# Abundances in the Local Region II:
# F, G and K Dwarfs and Subgiants

R. Earle Luck
Department of Astronomy, Case Western Reserve University
10900 Euclid Avenue, Cleveland, OH 44106-7215
rel2@case.edu

## Abstract

Parameters and abundances for 1002 stars of spectral types F, G, and K of luminosity class IV and V have been derived. After culling the sample for rotational velocity and effective temperature, 867 stars remain for discussion.

Twenty-eight elements are considered in the analysis. The α, iron-peak, and Period 5 transition metal abundances for these stars show a modest enhancement over solar averaging about 0.05 dex. The lanthanides are more abundant averaging about +0.2 dex over solar. The question is: Are these stars enhanced or is the Sun somewhat metal-poor relative to these stars? The consistency of the abundances derived here argues for the latter view.

Lithium, carbon, and oxygen abundances have been derived. The stars show the usual lithium astration as a function of mass / temperature. There are more than 100 planet-hosts in the sample, and there is no discernible difference in their lithium content relative to the remaining stars. The carbon and oxygen abundances show the well-known trend of decreasing [x/Fe] ratio with increasing [Fe/H].

## 1. Introduction

This paper is one of a series (Heiter & Luck 2003, Luck & Heiter 2005, 2006, 2007, Luck 2015) that seeks to determine the standard of normalcy of the local region regarding stellar abundances. The local region in this context is a Sun-centered sphere with a radius of approximately 100 parsecs. What is sought in the abundance data are reliable trends in the metal-content of stars due to spatial, temporal, or stellar characteristic variations. Assuming the local region is a typical volume, these results should be applicable as a metric in further explorations of galactic chemical evolution. This hope has been born out with the use of these results in the GAIA benchmark stars (Blanco-Cuaresma et al. 2014, Jofré et al. 2015, Heiter et al. 2015).

Numerous analyses have considered dwarfs in the local region. Hinkel et al. (2014) assembled the dwarf studies into a critically considered abundance database called the *Hypatia Catalog*. From the abundances found therein, they discussed the nucleosynthetic history of the local region. More recent studies have considered dwarf abundances and parameters in terms of the Li-problem in planet-host versus non-hosts (for example – Gonzalez 2014, 2015). Dwarf abundances further afield have been determined for stars with planets detected by the Kepler satellite (Schuler et al. 2015).

One of the problems associated with the assembly of a critically considered catalog of abundances such as the *Hypatia Catalog* is the question of systematic differences among various abundance sources. Abundance differences can arise from systematic stellar parameter differences, differences in atomic data - especially in damping constants and oscillator strengths, and things as mundane as choice of model atmosphere source and line analysis code. A better way to overcome these problems than trying to allow for the differences is a self-consistent analysis of a large sample of local dwarfs. Such an analysis is the concern of this work.

The 216 dwarfs of Heiter & Luck (2006) are the starting point for this study. An additional 469 dwarfs were observed using the McDonald Observatory Struve Telescope and Sandiford Echelle Spectrograph. For the analysis, 624 of these stars were retained with the bulk of the eliminated dwarfs found to be double-line spectroscopic binaries. An additional 57 dwarfs were added to the sample from data obtained using the Hobby-Eberly Telescope and its High-Resolution Spectrograph. These stars were observed as radial velocity standards during the Cepheid study of Luck & Lambert (2011). Lastly, 360 dwarfs were extracted from the ELODIE Archive. The total number of dwarfs is 1002 as there is some overlap between the Hobby-Eberly data and the other two datasets. Basic data for the program stars can be found in Table 1 along with quantities such as distance and absolute magnitude. Also indicated are the 108 dwarfs in this sample that are planet hosts. The planets are all close giant planets; that is, super-Jupiters in orbits within 0.5 AU of the host.

## 2. Observational Material and Equivalent Width Determination

The McDonald Observatory 2.1m Telescope and Sandiford Cassegrain Echelle Spectrograph (McCarthy et al. 1993) provided much of the observational data for this study. High-resolution spectra were obtained during numerous observing runs from 1996 to 2010. The spectra cover a continuous wavelength range from about 484 to 700 nm, with a resolving power of about 60000. The wavelength range used demands two separate observations — one centered at about 520 nm and the other at about 630 nm. Typical S/N values per pixel for the spectra are more than 150. Cancellation of telluric lines was achieved using broad-lined B stars that were regularly observed with S/N exceeding that of the program stars. The extraction process is that detailed in Luck (2013).

Spectra of 57 dwarfs were obtained using the Hobby-Eberly telescope and High-Resolution Spectrograph. The spectra have a resolution of 30000, span the wavelength range 400 to 785 nm, and have very high signal-to-noise ratios, >300 per resolution element in numerous cases. The reduction of these spectra follows the process detailed in Luck & Lambert (2011).

The last set of spectra were obtained from the ELODIE Archive (Moultaka et al. 2004). These spectra are fully processed including order co-addition and have a continuous wavelength span of 400 to 680 nm and a resolution of 42000. The ELODIE spectra utilized here all have S/N > 75 per pixel.

In the tables that detail the analysis results (Tables 3-5), stars observed with the Sandiford spectrograph are denoted "S" in the "Sp" column. Similarly, stars from the HET and High-Resolution Spectrograph are indicated by "H," and the objects from the ELODIE Archive are designated by an "E."

Processing of the spectra was performed using an interactive graphics package developed by the author. The software enables Beer's law removal of telluric lines, smoothing with a fast Fourier transform, continuum normalization, and wavelength setting. Two changes have been made in spectrum reduction relative to the procedures found in Luck (2014, 2015). The first is that order co-addition for Sandiford echelle spectra has been implemented; and second, the equivalent width routine has been upgraded to better account for the wings of stronger lines.

After continuum placement and wavelength scale (photospheric) determination, the Sandiford orders are co-added using as weights drawn from smoothed flat fields. This process yields excellent relative intensities along the orders and from order to order. The order co-addition minimizes S/N variations due to the blaze function leaving only the modest S/N variation due to detector response over the 140 nm total spectral window. The co-add proceeds by determining the order-to-order overlap in wavelength space, places the overlap regions on a common wavelength scale, and then co-adds using (w1*n1 + w2 * n2) / (w1 + w2) as the co-added value. w1 and w2 are the weights for the point in the two respective orders, and n1 and n2 are the continuum normalized intensities. The wavelength scale step for the co-added data is not constant along the spectrum. The values change periodically to mimic the original wavelength steps.

Stellar line profiles are set by the Voigt function – a convolution of a Gaussian and a Lorentz profile. In low-density environments such as found in giants and supergiants, line profiles are primarily Gaussian up to equivalent widths of about 25 nm. They are easily and accurately measured using the Gaussian approximation: EW = 1.06 * Depth * Full Width at Half Maximum (FWHM) where the depth is measured relative to the normalized continuum. Unfortunately, in solar-type dwarfs the lines have more prominent contributions from the Lorentz part of the profile and the Gaussian approximation begins to fail at about 7.5 nm in equivalent width. The precise point at which the observed profile deviates significantly from a Gaussian depends on the rotation velocity and the resolution of the observation. Rotational velocities of up to about 15 to 20 km s$^{-1}$ deviate little from a pure Gaussian and slit profiles are usually considered also to be Gaussian. For example, consider a line with a true equivalent width of 9.49 nm as determined from a direct integration of a theoretical profile derived using a solar model. Convolving that profile with 1.0 km s$^{-1}$ of rotation and a slit resolution of 60000 and then using the Gaussian approximation to determine the equivalent width one obtains 8.98 nm. If the rotation is increased to 5 km s$^{-1}$ the same exercise yields 9.17 nm for the equivalent width. For a resolution of 42000 and the same rotation values, the Gaussian equivalent width values are 9.04 and 9.16 nm respectively. As the equivalent width increases the Gaussian approximation increasingly undershoots the actual equivalent width.

How can one overcome this problem? One possibility is to use direct integration on the observed profiles. While this is feasible for the Sun and is the usual practice; it is not realistic to attempt this for thousands of lines in hundreds of stars. Another way forward might be to try to fit a Voigt profile to the lines using non-linear least squares to derive the fitting parameters. This technique was investigated in the Sun and turns out to need direct human interaction as the solutions tend to be rather unstable and often diverge giving bad fits. Another problem with the non-linear least squares approach is that one needs the derivatives of the function to be fitted and for the Voigt function the derivatives are non-trivial.

The way forward from this point is to adopt the function known as the pseudo-Voigt function — a linear combination of a Gaussian and an exponential function. The function is:

$$y(x) = a \left[ \frac{c}{\left(1 + \left(\frac{x - x_o}{b}\right)^2\right)} + (1-c) e^{-\left(\frac{1}{2}\left(\frac{x - x_0}{b}\right)^2\right)} \right]$$

where a is the amplitude/depth of the feature, $x_0$ is the central wavelength, x is the wavelength of interest, b is related to the half-width at half-maximum depth (HWHM), and c specifies the relative contribution from the two generating functions. For c = 0, one has a pure Gaussian and b is the HWHM. A pure exponential is found at c = 1 where b = 0.849 * HWHM. For any value of c, there is a unique solution for b as a function of HWHM.

A virtue of the pseudo-Voigt approach is that the derivatives are easily computed, and the profile is computationally simple. Implementing non-linear least squares fit using this function for theoretical profiles shows that excellent profile fits are obtained up to equivalent widths of over 40 nm at resolutions up to 120000 and rotations approaching 20 km s$^{-1}$. However, when the non-linear least square procedure is applied to the Sun, the approach suffers from instabilities in the solutions due to impinging blends.

A better way to handle the problem is to use a chi-square minimization approach. One determines the depth and HWHM of the feature to be fitted. One then cycles through a series of c values and for each chosen c computes b from the relation between b and the HWHM for that c. After computing each pseudo-Voigt profile, the chi-square statistic is calculated. The c giving minimum chi-square is judged to be the best fit. Comparing the best-fit chi-square b and c values to those derived from non-linear least squares for "clean" solar lines and theoretical profiles show excellent correspondence with the benefit that the chi-square is not prone to numerical instability.

Applying the chi-square technique to stellar spectra is possible, but the lower resolution and signal-to-noise limits its direct application. A number of spectra were reduced using the chi-square minimization technique, but the number of lines yielding a good minimum was inadequate. What was noticed is that the values of c that were found varied in step with the Gaussian equivalent width and overall line width; i.e., with rotational velocity. This behavior points to a usable path: given the rotational speed and resolution, one can construct a relation between the Gaussian equivalent width and the required value of c based on theoretical profiles. Then given the rotational velocity of a star, one can obtain a working relation between c and Gaussian equivalent widths. For our program stars, the rotational velocity was obtained using synthetic spectra to fit the region around 570 nm, the best fitting velocity (and spectrum resolution) was then used to interpolate the Gaussian equivalent width – c relation for the star.

Equivalent widths were measured using the Gaussian approximation and using the pseudo-Voigt function. Specifically, one first determines the Gaussian equivalent width, then using that one obtains c, next one uses the observed HWHM and c to obtain b; and lastly, using those values of b and c along with the observed depth one generates the pseudo-Voigt profile, and the integrates it to find the equivalent width. For most lines the difference in the Gaussian equivalent width and the pseudo-Voigt equivalent width is small – at 10 nm the difference is typically 0.5 nm while at 20 nm, the difference is 2 nm. For this analysis, an upper equivalent width limit of 20 nm was imposed. The average equivalent width used in the analysis is 7 to 9 nm.

## 3. Analysis

### 3.1    Line list and Analysis Resources

The line list used here is the same as used in and described by Luck (2014, 2015). Briefly, the line list consists of about 2900 lines with solar oscillator strengths. Many of these lines have been used in solar abundance analyses while others have been shown to be unblended. The equivalent widths used to determine the oscillator strengths were determined using direct integration during an interactive examination of each line in the Delbouille & Neven (1973) solar intensity atlas. The line list can thus provide excellent abundances in solar-type dwarfs. As one goes further away from the solar temperature, coincident wavelength blends do become more problematic; however, the line list does provide a firm basis for the analysis because unblended lines in the Sun have a better chance of being free of contaminants at other temperatures than do most lines. The line measurement process makes sure that the wavelength of the line to be measured falls at the proper wavelength relative to the wavelength of the desired line. This process eliminates strong blends from nearby lines. The sifting process applied during abundance determination further trims the initial list by removing less obvious blends.

The solar oscillator strengths adopt the solar abundances of Scott et al. (2015a,b) and Grevesse et al. (2015) for species with an atomic number greater than 10. Van der Waals damping coefficients are taken from Barklem, Piskunov, & O'Mara (2000) and Barklem & Aspelund-Johanssen (2005), or computed using the van der Waals approximation (Unsöld 1938). Hyperfine data for Mn, Co, and Cu from Kurucz (1992) were also utilized. The MARCS solar atmosphere (Gustafsson et al. 2008) was used for all calculations. This model uses plane-parallel geometry with effective temperature 5777 K and log g = 4.44. The adopted microturbulence was 0.8 km s$^{-1}$.

Abundances for all program stars were calculated using plane-parallel MARCS model atmospheres (Gustafsson et al. 2008). An interpolation code developed by the author was used to interpolate at the desired parameters. Tests indicate the code can accurately reproduce grid models to within 5 K in temperature and the remaining structure to within 1% - 2%. Stellar metallicity is taken into account in the analysis by using the model grid closest in metallicity to the target star. Above [M/H] = -1, the grid spacing is 0.25 dex in metallicity and extends up to +0.25 dex. The model metallicity thus is generally within 0.125 dex of the star itself. The line calculations were made using the LINES and MOOG codes (Sneden 1973) as extended and maintained by R. Earle Luck since 1975.

### 3.2 Stellar Parameters and Z > 10 Abundances

The photometric calibration of Casagrande et al. (2010) was used for effective temperature determination. Photometry was obtained using the General Catalog of Photometric Data (Mermilliod et al. 1997), 2MASS photometry (Cutri et al. 2003) from SIMBAD, uvby photometry from Paunzen (2015), and Tycho photometry from the Tycho-2 catalog (Høg et al. 2000). The line of sight extinctions were determined using the code of Hakkila et al. (1997) and adopting distances computed using Hipparcos parallaxes (van Leeuwen 2007). The Hakkila et al. code uses (l, b, d) versus $A_V$ relations to determine $A_V$. The extinction within 75 pc of Sun is essentially nil (e.g., Vergely et al. 1998; Leroy 1999; Sfeir et al. 1999; Breitschwerdt et al. 2000; Lallement et al. 2003). The extinctions (and reddenings) of all stars within 75 pc were thus set to zero. For stars lying beyond 75 pc, the extinction out to 75 pc was subtracted from the total extinction. The parallax uncertainty of these stars or systems has a median value of 2%. Secondary stars without determined parallaxes are assumed to lie at the same distance as the primary in the system. The absolute V magnitudes are mostly good to the ±0.1 mag level.

All available colors were utilized for the initial temperature determination. The [Fe/H] values used in the Casagrande et al. (2010) calibration were taken from Luck & Heiter (2006), the Hypatia database (Hinkel et al. 2014), the PASTEL database (Soubrian et al. 2010), or assumed to be solar. The individual effective temperature values were examined if the standard deviation of the mean exceeded 100 K. Obvious outliers were eliminated from the final average. Colors involving 2MASS magnitudes were examined more closely for the brightest stars due to possible saturation effects in the photometry. It was found that if there was a temperature outlier, it most often was V−J, where J is the 2MASS J magnitude. For that reason, the V-J color was eliminated from the effective temperature determination for all stars. Table 2 gives the photometric temperature, its standard deviation, and the number of colors utilized. These temperatures are well determined on the whole: the mean standard deviation of the temperature for the stars is 48 K. The number of temperatures utilized is 8 to 9 in the mean. However, there are a few cases that are not so well defined with the worst case being α Cep (HR 8218) which shows a range of 700 K over the three colors retained. There is no obvious reason for the spread. The mean temperature for this star and the other five stars with a standard deviation more than 150 K are consistent with their spectral types; thus they remain in the data set, but their parameters and abundances are not on the same level of reliability as the others.

Given the effective temperature, absolute magnitude, and mass; the surface acceleration due to gravity (e.g., the gravity) can be found. Isochrones were taken from Bertelli et al. (1994), Demarque et al. (2004), Dotter et al. (2008), and the BaSTI group (BaSTI Team 2016) and used to obtain the mass, age, and luminosity for each star. The fitting process is that of Allende-Prieto and Lambert (1999), which has been used subsequently in a number of analyses; e.g. Luck and Heiter (2006, 2007), Luck (2015). Isochrone fitting is also dependent on the metallicity of the star in question, and the same metallicities used in temperature determination were used here. The luminosity given in Table 2 is derived from the distance, apparent V magnitude, and the bolometric corrections of Bessell et al. (1998). Comparison of the these luminosities to those derived in the isochrone fits shows they agree to within 0.02 dex.

There are at times substantial differences in mass and age estimates between isochrone sets. For example, the mass estimates for η Ser vary from 1.32 to 2.07 solar masses. However, the median value of the range in the mass estimate is 0.08 $M_{Sun}$. This uncertainty is 10% of the mean mass. Over 80% of the sample has a range in isochrone masses of less than 20%. The mean mass from the Bertelli et al. isochrones is 1.09 solar masses, 1.05 solar masses from Dotter et al., 1.10 solar masses from Demarque et al., and 1.09 solar masses from the BaSTI group. The masses are thus well determined. In Table 2, there are about 30 stars lacking a mass determination from any isochrone set. Examination of the temperatures and luminosities of these stars show they tend to be lower temperature and lower luminosity objects, thus falling outside the isochrone grids. For these, a mass was estimated by plotting the average masses for the stars with determined values against their effective temperature — the relation found is quite well-defined. The temperature of the stars without isochrone determined masses can be used with the mass-temperature relation yielding a reasonable mass estimate. Note that a 30% uncertainty in the mass yields an uncertainty in the *log g* value of ±0.15. At 10% uncertainty in mass, the variation in *log g* is ±0.05. The values of *log g* found here should be reliable.

The microturbulent velocity is obtained by demanding that the iron abundance as determined from Fe I lines show no dependence on equivalent width. The process used here is the same as described in Luck (2015). The Fe I data are examined in a statistical manner based on Gaussian fitting to the abundance distribution. The Fe II data have an insufficient number of lines to proceed statistically, so cuts are derived based on the number of lines present. From this data examination, master cut lists are derived for Fe I and Fe II. Before clipping the number of Fe I lines is of order 700; after clipping it is about 500. Fe II lines typically number 40–60 before trimming and 30–50 afterward. After arriving at the final Fe I and Fe II data for each star, the microturbulent velocity is tweaked to a final value. Other species are clipped in the same way as Fe II, but no master cut lists are generated.

An imbalance exists between the total iron abundance as derived from Fe I and Fe II. The sense is that the mass derived gravity is too large; that is, the Fe II total iron abundance exceed the Fe I total iron abundance — see Figure 1 Top Panel. This problem can be rectified using an ionization balance. This process forces the ionized and neutral species to yield the same total abundance by using as the free parameter the gravity. This process is implemented by interpolating a small grid of models and then determining the best-fit gravity and microturbulence together. Parameter confirmation was performed by interpolating a new model at the proper parameters. The iron data relations were recomputed to confirm the ionization balance, along with the lack of dependence of iron abundance on line-strength. The difference between the mass and ionization balance gravities is "small" in the temperature regime within 500K of solar temperature (see Figure 1 Bottom Panel) with a mean value of 0.06 dex. In that region, the average iron abundance difference between Fe I and Fe II for the mass derived gravities is -0.02 dex.

Table 3 includes parameters and iron abundance details — log $\varepsilon_{Fe}$, σ, and number of lines for both Fe I and Fe II. Information for both mass and ionization-balance derived gravities is presented.

Average abundances for 25 elements with Z > 10 are in Table 4. For elements having more than one ionization stage available, the final average is the average of all retained lines. A number of elements have both neutral and first ionized species available, the final average for these elements is merely the average of all retained lines. Note that the Mn, Co, and Cu abundances have been computed allowing for hyperfine structure. Details of all Z > 10 abundances — per species average, σ, and number of lines — can be found in the online only data — Table A.

### 3.3 Li, C, and O Analysis

Lithium, carbon, and oxygen abundances were derived for the program stars by spectrum synthesis. The atomic line data — for Li I at 670.7 nm, C I at 505.2 and 538.0 nm, and O I at 615.5 nm and [O I] at 630.0 nm — used is the same as detailed in Luck (2015). The major change made in this analysis is in the line list for the region 507.5 to 518.5 nm that includes lines from the $C_2$ Swan system. A new line list was extracted from the VALD database including molecular lines of $C_2$, CN, MgH, CH, OH, and TiO. The molecular lines of primary importance are those of $C_2$ as above about 4250 K there is scant evidence for the presence of other molecular species. The $C_2$ data in VALD derives from Brooke et al. (2013) and comparison with the previously used constants in Luck (2015) — excitation potentials and oscillator strengths — shows excellent agreement. Illustrations of syntheses such as those carried out here can be found in Luck & Heiter (2006) for some of the same stars and spectra. Carbon and oxygen abundances below $T_{eff}$ = 4325 K are not given due to blends making line detection unreliable for all carbon and oxygen indicators in the observed spectral regions.

Lithium LTE abundance data are presented in Table 5 including an indicator noting if the "abundance" should be considered an upper limit. Upper limits are those stars in which the observed lithium feature has a depth of two percent or less. There is no evidence for $^6Li$ in any of these stars, and thus the syntheses do not include this species. Corrections for non-LTE effects in lithium from Lind et al. (2009) are also included.

The individual carbon features are combined as follows: for $T_{eff}$ < 5250 K: only $C_2$ 513.5 nm is used. At 5250 < T < 6000 K, C I 505.2 and 538.0 nm have weight 1 as does $C_2$ 513.5. For 6000 K < T < 6350K, the two C I lines have weight 2, and $C_2$ has weight 1. Above $T_{eff}$ > 6350 K the two $C_2$ is not used and the two C I lines have equal weight. Relative strength and blending are the basis for the weights. A typical range in abundance for the features is 0.15 dex. The Asplund et al. (2009) carbon abundance, log $\epsilon_C$ = 8.43, is adopted for the solar reference abundance. Table 5 has the per line / feature carbon data as well as the average values.

Oxygen abundance indicators were averaged in the following manner: for $T_{eff}$ < 5250 K, [O I] only is used. For 5250 K < $T_{eff}$ < 6000 K, O I has weight 1 and [O I] has weight 3. In the regime 6000 K < $T_{eff}$ < 6350 K, O I and [O I] have equal weight. Lastly, for $T_{eff}$ > 6350 K only O I is used. Near the solar temperature, typical differences between oxygen derived from O I and [O I] are of order 0.15 dex. For $T_{eff}$ < 5500 K, the C–O interlock has been taken into account in the abundance determination. The oxygen data per line and averages are found in Table 5.

## 3.4 Abundance and Parameter Inspection

### 3.4.1 Internal Examination

Thirty-nine of the HET stars are in common with the remainder of the sample — 31 with the Sandiford data and 8 with the ELODIE data. The temperatures and mass derived gravities for the stars are the same. The HET data has the same wavelength extent as the ELODIE data but extends further blue-ward than does the Sandiford data. This extended blue coverage means that there are more and different Fe I and Fe II lines in the HET data than can be found in the Sandiford data but approximately the same as found in the ELODIE data. The difference in the Fe I average abundance is +0.05 for HET - Sandiford; that is, the lower resolution HET data yields a higher abundance on the average. For HET versus ELODIE, the difference in Fe I is +0.02 on average. For Fe II, the differences are +0.12 and -0.04 respectively. However, the individual standard deviations about the HET means for the Fe I datasets are in the range 0.05 to 0.10 dex, while the Fe II data has somewhat larger standard deviations, typically 0.1 to 0.15 dex. The offset between the different data sources is only marginally significant, especially for Fe I. In the Fe II data the offset in the HET - Sandiford comparison most likely lies in the fact that the number of Fe II lines in the HET data averages twice the number found in the Sandiford data. The difference in number stems from the extended blue coverage of the HET data. The bluer Fe II lines are likely to be more blended than the red lines in the Sandiford data and thus give a somewhat higher total iron abundance. In the discussion that follows this section, the only HET data retained are the 18 unique HET stars. The abundances derived from the entire set of 57 HET stars, however, can be found in the abundance tables (Tables 3 – 5).

Other instances of equivalent width problems can be found in Figure 2 (top panel) where the ratios [S/Fe], [Si/Fe], and [Ni/Fe] versus temperature are shown. The data shown are derived using the mass determined gravities. It is immediately evident that below 4500 K, the S and Si data are not reliable as the ratios rapidly increase as a function of temperature. [S/Fe] is the most extreme case increasing from [S/Fe] about 0 at 4750 K to +2 dex at 4000 K. Such an increase is not reasonable in any scenario of galactic or stellar evolution.

The source of the increase evidenced the S I data in Figure 2 is growing blends in the weak higher excitation potential lines. A blend, in this case, means anything that causes the direct equivalent width measurement to provide an unreliable value. The cause of the blend can be a direct wavelength coincident with another line. However, more probable is that the line to be measured is perturbed by a nearby strong line(s) that can affect the depth, the central wavelength, the measured half-width, or all of these quantities. This behavior means that the lines are increasingly difficult to measure reliably. This problem is evidenced by the number of lines used as a function of temperature. The mean number of S I lines used over all stars is five with the largest number being about twelve around 6000 K in effective temperature. As the temperature decreases to 4500 K the number of S I lines measured drops to about three. The blends keep the S I equivalent width constant or even increasing with decreasing temperature for these lines.

The trend noted in S I is also the case for Si I. For Si I the average number of lines is 39; however by 4500 K in effective temperature, the number of lines falls to about 30 and by 4000 K to less than 20. This is not because the lines are becoming too strong, but because blends make them difficult to identify reliably. For Ni I, the increase is not nearly so dramatic, but the abundances are still problematic below 4500 K. Here the same behavior in numbers of lines applies. At 6000 K there are about 170 lines, but by 4500 K the number decreases to about 50. The larger number of lines at 4500 K does yield better discrimination of heavily blended lines, but some evidence for the problem remains.

The bottom panel of Figure 2 returns to the iron data showing the total iron abundance as derived from the mass determined gravity for both Fe I and Fe II. Below 4500 K the Fe II data shows the same problem as found in the S and Si data — there is a rapid increase in total iron content. This increase is responsible for the behavior seen in Figure 1. The reason for the increase in Fe II is the same as that for the S I and Si I behavior — blends are severely affecting the data below 4500 K. The better path forward from this point is to consider in the bulk of the discussion only those stars with effective temperatures greater than 4500 K.

### 3.4.2 External Comparison

In consulting the PASTEL database (Soubiran et al. 2010), one finds that about 85% of the program stars analyzed here have previous determinations of stellar parameters and / or abundances. The number of references totals 371 spanning over six decades of work. With a median number of five references per analyzed star, the corpus of the analyses is massive and well beyond a succinct synopsis. For comparison purposes here, only a subset of post-2005 studies is considered. The primary results of the comparison are located in Table 6, where effective temperature, gravity, and [Fe/H] differences are given. Note that two gravities and [Fe/H] values are given. These correspond to the two gravity, and hence [Fe/H], determinations performed here.

The data presented in Table 6 indicates good consistency between the various analyses whether they use traditional techniques such as a spectroscopic excitation and ionization balance technique for parameter determination (Bensby et al. 2014, Ramírez et al. 2013, 2014), or the latest astroseismology results (Morel & Miglio 2012, Creevey et al. 2013). Typical average differences in temperature scales are about 20 K, gravities vary by about ±0.05 dex, and abundances agree in the mean to about ±0.05 dex.

It is encouraging that the temperatures derived here using the Casagrande et al. (2010) effective temperature scale agrees with the IRFM temperatures used to develop that scale – the mean difference is 2 K over 74 stars. However, the spread in the individual temperature determinations as expressed in the standard deviation of the mean differences is discouraging. The standard deviations are of order 50 to 100 K, which means in larger comparisons the effective temperature differences range up to 500 Kelvin or more. This problem is evidenced in the Casagrande et al. (2011) and Bensby et al. (2014) comparisons with the current data. To determine if the problem is solely within this analysis the differences between Casagrande et al. (2011) and Bensby et al. (2014) were also determined with the results being very similar to those found in this work and those two studies.

Errors in effective temperatures in an excitation analysis can stem from equivalent width errors or systematic errors in oscillator strengths as a function of excitation potential. The latter is not a significant problem at this time, but equivalent width problems can be. The general sense is that weaker lines are prone to overestimation in equivalent width, while strong lines in dwarfs are susceptible to underestimates due to missing contributions from the wings. These errors can skew the potential – abundance relation as well as the equivalent width – abundance relation giving rise to erroneous effective temperatures and microturbulent velocities.

Photometric effective temperature determinations also have their problems. Photometric errors; for example, errors in calibration or saturation of bright sources, are possible. Here, a more pernicious problem is a confusion of multiple sources in the photometry aperture. Examination of Table 1 shows that many of these stars are doubles or multiples. Care must be taken to ensure that the star observed spectroscopically is the star that was observed photometrically. Great attention was focused here on source identification, but it still possible that source confusion remains. Source confusion means not only identification of the proper object, but also unresolved sources blending into the desired source. Source confusion is not only a problem for this study, but also for any large scale work depending on photometry. A good example of this is the local neighborhood study of 16000+ stars of Casagrande et al. (2011).

The importance of non-local thermodynamic equilibrium (NLTE) in the determination of iron abundances has been investigated starting with the work of Tanaka (1971) and proceeding to the current day. The dwarf analysis of Bensby et al. (2014) includes NLTE corrections in their iron data, but they note that the corrections are small. The data in Table 6 reflects this conclusion — the LTE iron abundances derived here agree very well with their NLTE values, and any differences correlate closely with differences in effective temperature. For the parameter range of interest here the differences between LTE and NLTE iron abundances are < 0.1 and often near 0 (Lind et al. 2012) consistent with the Bensby et al. result.

As the last point of comparison, the analysis of Ramírez et al. (2013) derives not only stellar parameters and iron abundances; it also derives oxygen abundances from the O I 777.5 nm triplet using an NLTE analysis. In this work, oxygen abundances have been derived from the [O I] 630.0 mn line and the 615.7 nm O I triplet using a pure LTE analysis. Comparison of the derived abundances yields an average difference of +0.02 dex for the mass-derived gravities derived here, and -0.02 dex for the ionization based gravities. The standard deviation of the mean difference is 0.23 dex using 211 common stars. By eliminating nine discrepant stars the mean differences change to +0.03 and 0.00 dex respectively, and the standard deviation falls to 0.17 dex. The agreement is excellent considering the weakness of the lines used here.

## 4. Discussion

Extended discussion of the nucleosynthetic history of the local region of the Milky Way can be found in Nomoto et al. (2013) and Hinkel et al. (2014). Correlations between local abundances and kinematics have been discussed recently by Bensby et al. (2014). The reader is referred to those works for greater detail than given here where the focus is on what can this study say over and above earlier discussions.

### 4.1 Sample and Analysis Pruning

Figure 3 shows an HR diagram of the program stars. The sample is broken into subsets based on rotational velocity and proximity to the main sequence. The dividing line for rotational velocity is 20 km s$^{-1}$. The velocity is set based on the ability to measure equivalent widths accurately using the method described in §2: above 20 km s$^{-1}$ the method fails to reproduce the profiles accurately. The division into dwarfs and non-dwarfs is relatively arbitrary: stars above the indicated line are far enough from the main sequence that they are likely subgiants with perhaps a few stars still evolving towards the main sequence. What is to be investigated is if there are any believable systematic differences between the dwarfs and non-dwarfs. A priori, no abundance differences would be expected based on closed-box galactic chemical evolution and standard stellar evolution.

Another sample pruning applied going forward is that stars with an effective temperature less than 4500 K are not usually considered in the discussion. The temperatures and mass-derived gravities are reliable, but the ionization gravity and overall abundances are suspect due to significant blending problems in these stars. Note that Figure 2 does indicate that iron abundances derived from Fe I below 4500 K are consistent – they show no significant dependence on temperature down to about 4000 K. The problem is that other elements/species with abundances determined from many fewer lines are not as reliable.

While Table 3 gives parameters and iron abundance data for both the mass-derived and ionization balance derived gravity, the abundances to be discussed are be drawn from the mass-derived gravity only. This selection is because in limiting the temperature to be above 4500 K, the differences in the two gravities is small — averaging 0.11 dex, which means that the neutral and ionized derived total abundances are also in relatively good agreement. For Fe I versus Fe II the difference is -0.05 over 867 stars. In this comparison, the stars with high rotation velocities have been removed, but no discrimination made between the dwarfs and non-dwarfs.

### 4.2   Z>10 Abundances

One of the more common ways to examine abundances is to plot mean abundances for the sample as a function of atomic number. In Figure 4 such a plot is shown. The data in this plot are the mean values for those stars with [Fe/H] > -0.4, rotational velocity < 20 km s$^{-1}$, and $T_{eff}$ > 4500K. The [x/Fe] ratios are computed from the abundances given in Table 4. Both neutral and first ionized species are available for Si, Sc, Ti, V, Fe, Y, Zr, and Eu and the abundance given in Table 4 is the mean over all retained lines. For the Z ≥ 56 elements other than Eu, only the first ionized species is available. Since the number of Fe I lines is much greater than the number of Fe II lines, the mean iron abundance is essentially the mean Fe I abundance. To investigate if this causes differences in the [x/Fe] ratios, the ratios [x I/Fe I] and [x II/Fe II] were computed and found to not be significantly different from the ratios generated using the abundances in Table 4.

The sample has been subdivided based on the dwarf versus non-dwarf criteria from the preceding subsection. The iron-peak elements for both sub-samples have <[x/Fe]> values of about +0.07 dex with a standard deviation of also about 0.07 dex. That is, as expected, they have close to the solar ratio of abundance relative to iron.

The elements Na, Mg, Al, and Si in the dwarfs show [x/Fe] values comparable to those found in for the iron-peak. Once again, this equality is as expected as the lower abundance cut in iron eliminates most of the α-enhanced stars. What is surprising is that the non-dwarf stars have <[x/Fe]> ratios that are systematically higher than what is found in the dwarfs. The effect is small, about 0.06 dex, and the one-sigma error bars do overlap with [x/Fe] = 0 as well as the dwarfs.

For the heavy elements with Z > 56, the situation is that relative to the dwarfs, the non-dwarfs systematically have lower mean abundances. The offset, in this case, is somewhat larger, about 0.15, but the star-to-star abundances more scattered, which translates to the one-sigma error bars overlapping.

The overall offset from solar [x/Fe] ratios leads to the question of the behavior of the subset of dwarfs that have properties close to that of the Sun. The subset of such stars – 34 in all with $T_{eff}$ within 100 K of the Sun – is shown in Figure 4 as red triangles. The error bars have been omitted, but they are comparable to the others shown. What is obvious is that these stars show a similar abundance pattern to that of the total sample. If one normalizes the total dwarf sample [x/Fe] ratios to these values, the result is the blue triangles in Figure 4. Unsurprisingly, the resulting ratios are close to zero.

Another way of examining abundances is to look for trends versus [Fe/H]. In Figure 5, the ratios [Si/Fe], [Ca/Fe], [Mn/Fe], [Ni/Fe], and [Zn/Fe] are shown as a function of [Fe/H]. The stars shown have been limited to -1 < [Fe/H] < +0.5. The trends in the data are well-known: silicon and calcium decrease with increasing [Fe/H] while [Mn/Fe] increases. Silicon and calcium are α-elements and their diminution as a function of [Fe/H] is due to the growing influence of Type Ia iron production with time. The enhancement of manganese with increasing [Fe/H] was first noted by Wallerstein (1962) and was commented upon by later authors (Gratton 1989, Luck & Heiter 2007, Feltzing et al. 2007, Luck 2015). Manganese is produced in Type Ia supernovae, and the observed trend is caused by the late onset of enrichment (Kobayashi & Nomoto 2009).

The nickel abundances show a very small spread in [Ni/Fe] in the region -1 < [Fe/H] < 0 (Figure 5, panel d) with a mean [Ni/Fe] ratio of near 0. At [Fe/H] ratios above solar, the [Ni/Fe] ratios increase to about [Ni/Fe] ~ +0.1 at [Fe/H] ~ +0.5. This behavior parallels the behavior in [Mn/Fe] (Figure 5, panel c) and is also found in the nickel data of Luck (2015) for local giants.

The zinc data shown in the bottom panel of Figure 5 indicates significant scatter at all [Fe/H] ratios. The scatter is not unexpected as zinc abundances rest upon at most two lines. Previous work has shown that the [Zn/Fe] ratios in the region -1 < [Fe/H] < -0.5 increases from about +0.07 dex to about +0.2 dex and then falls back toward 0 at [Fe/H] = 0 (Saito et al. 2009). The overall features of the distribution found here are consistent with [Zn/Fe] ~ 0 at all metallicities greater than [Fe/H] > -1 and can say no more than that.

Two heavy element trends are shown in Figure 6. The top panels show neodymium and the bottom two europium. For each element, the top panel of the pair shows the sample subdivided as dwarfs versus non-dwarfs while the bottom plot of the pair keys the data based on effective temperature. Both plots that show the dwarfs versus the non-dwarfs are consistent with the non-dwarfs having lower [x/Fe] ratios as indicated in Figure 4. Both neodymium and europium abundances are derived from ionized lines, and the lower gravity non-dwarfs have the stronger lines giving rise to a more secure abundance. The smaller spread in the non-dwarf heavy element abundances testifies to this difference. If one merely deletes any star with [Nd/Fe] > +0.6 dex from the <[Nd/Fe]> ratio, the mean value for the dwarfs' declines only marginally, and the non-dwarf ratio barely changes. The mean value of dwarf <[x/Fe]> ratios in all cases is being dominated by the sheer number of dwarfs with "normal" [x/Fe] ratios in the range -0.25 < [F/H] < +0.25.

A question that has yet to be answered is why are the [x/Fe] ratios for these stars non-solar? Alternatively, one could ask — why are the Sun's elemental abundance ratios lower than those found in these stars? One could suspect the solar-derived transition probabilities used here relative to laboratory values. Data for 519 of the Fe I lines used here can be found in the NIST database (Kramida et al. 2015). The mean difference in log gf is 0.02 dex (this work – NIST) with a standard deviation ($\sigma$) of 0.19. For Fe II, the difference is 0.10 (n = 72, $\sigma$ = 0.24). Looking at α- and other iron- peak elements typical mean differences are 0.05 dex. The heavy elements show somewhat larger differences; for example, Nd II has a mean difference of +0.1 dex. Such differences cannot explain the overabundances seen in the heavy elements mean [x/Fe] ratios. Another possibility would be differential effects between the solar model used to derive the gf values and the models used in the analysis. Differential model effects can be discounted because MARCS models were used for all analyses. Additionally, the solar-like stars show the same <[x/Fe]> ratios as the total sample indicating that something other than the models lies behind the high heavy element abundances. Two possibilities come to mind: 1) the ionized lines used for the analysis are severely affected by NLTE, or 2) that the small sample of lines available for heavy element abundance determinations are more blended than expected. This problem needs further investigation.

The current sample includes over 100 planets hosts. Figure 7 shows a combined histogram of [Fe/H] values for the current sample limited in rotation velocity and effective temperature as above. The planet-host stars include both dwarfs and non-dwarfs, so both object types are included in the comparison "non-host" set. Note that "non-host" in this context only means that no planets are currently known to be associated with these stars. The [Fe/H] distributions are skewed for both the "non-hosts" and hosts towards lower [Fe/H] ratios and the host subset does peak at a somewhat higher [Fe/H] value than do the "non-hosts." Fitting the distributions to four parameter Weibull distributions shows that that the "non-hosts" peak at [Fe/H] of +0.09 dex versus +0.21 dex for the hosts. A Mann-Whitney comparison of the distributions shows that they are statistically significantly different. This result merely confirms previous results such as those of Luck & Heiter (2006). While planet-hosts as a class are more metal-rich than non-hosts, it is not necessary that hosts be metal-rich. The prime example of this is the planet-host HD 114762 at a [Fe/H] ratio of -0.7 dex.

There are more than twenty physical pairs in this sample. After eliminating those with at least one member of the pair below 4500 K or with a rotational velocity more than 20 km s$^{-1}$, nineteen pairs remain. The mean difference in [Fe/H] between the components is 0.083 dex with the largest difference being 0.4 dex. The data for this pair — BD +4 701 A and B — are of lower quality than the bulk of the data. Eliminating this pair lowers the mean difference to +0.065 with a standard deviation of 0.062. This agreement is excellent and boosts the confidence in the abundance analysis.

### 4.3 Li, C, and O in Dwarfs

#### 4.3.1 Lithium

Lithium in dwarfs has three aspects that need a least a modest discussion. First is the depletion/dilution of lithium as a function of mass or temperature. Second: do any of the "non-dwarfs" exhibit high lithium abundances that could signal they are pre-Main Sequence stars? Lastly, a topic that undoubtedly will not be settled here: Do planet hosts show higher lithium abundances than non-hosts?

Figure 8 shows two representations of the astration of lithium – the top panel shows lithium versus effective temperature and the bottom panel lithium versus mass. The stars shown in Figure 8 are only those stars that meet the dwarf criteria, but planet hosts are included. Moreover, the temperature range shown in Figure 8 is the complete sample range. Lithium abundances are not especially gravity sensitive, and the abundances are derived using a detailed spectrum synthesis. Note that at 4000 K, the solar lithium abundance of $\log \varepsilon_{Li} = 1.0$ produces a feature with an equivalent width more than 10 nm making the abundances at lower temperatures relatively secure.

The strong decrease in lithium abundances as a function of decreasing mass and temperature seen in Figure 8 is not new having been noted by Lambert & Reddy (2004) as well as Luck & Heiter (2007). What is new here is the number of stars – 600+ here versus 200+ in the previously cited works. Additionally, the data of Lambert and Reddy (2004) do not extend to the lower masses and temperatures of this study. The explanation for the growth of astration as mass declines is simple: convection depth increases as the mass declines until at some point stars become totally convective. Deeper convection transports lithium to temperatures where it is destroyed leading to the observed decline in surface abundance.

Two other features seen in Figure 8 are of note. The first is that the strongest decline in lithium abundance occurs rather abruptly at a mass of 0.8 $M_{Sun}$; or alternately, at about 5600K in effective temperature. Secondly, by $T_{eff} \sim 5000$ K, and most definitely by 4500 K, the lithium abundances have dropped to a plateau value of about $\log \varepsilon_{Li} \sim 0$, which is about one-tenth solar. This "constant" abundance implies that the convection depths achieved in these stars is the same. Given the small range in mass – about 0.5 to 0.8 $M_{Sun}$ – this is not especially surprising.

In Figure 9, the lithium abundances are once again plotted versus temperature, but in this plot the "non-dwarfs" are included. What is obvious is that the "non-dwarf" lithium abundances match up very well with the dwarf abundances. This match-up means the "non-dwarfs" are subgiants just coming off the main sequence. The only exception is the cool "non-dwarf" HD 98800. This star is an RS CVn variable with a spectral type of K5V(e). It is a young pre-main sequence star (Elliot et al. 2016 and references therein).

The lithium abundance found in planet-hosts relative to non-hosts has been a subject of ongoing debate. The contention is that the stars hosting close-giant planets have higher lithium abundances than similar non-hosts. Evidence for this position has most recently been put forward by Gonzalez (2014, 2015). The counter argument has been espoused by Luck & Heiter (2006) and Ghezzi et al. (2010). There are more than 100 planet-hosts in this work, all found by radial velocity variations, which means that they are mostly systems containing close-giant planets. The host systems have both dwarf and subgiant stars, and in Figure 10 the lithium abundances of all non-hosts are shown overplotted with the lithium abundances of the host systems. The non-hosts are limited to rotational velocities less than 20 km s$^{-1}$ and the temperature range is that spanned by the host stars. Given the sensitivity of lithium to effective temperature, the virtue of this work regarding the lithium comparison is the internal consistency of its temperature scale. Examination of Figure 10 indicates that there is no discernible difference in the lithium abundances of the planet-hosts and non-hosts.

To further examine possible abundance differences between planet hosts and non-hosts, the lithium abundances excluding the limits have been binned in 100 K increments. The resulting average lithium abundance versus effective temperature data are shown in Figure 11. What is apparent is that for the stars in this work, the hosts do not have lithium abundances above those of the non-hosts. In fact, except for two higher temperature bins, the host lithium abundances are lower than those of the non-hosts. In all cases, the one-sigma error bars for the mean host and non-host lithium abundances overlap.

In Figure 12, histograms of host and non-host lithium abundances for three temperature bins are given. These are the three most populated bins in the hosts and include one of the two bins where the hosts have mean lithium abundances above the non-hosts. The second bin with high host abundance has only three host stars in it. Also shown for each bin is a Gaussian fit to the non-host data. For the two higher temperature bins including the bin that has a higher host mean – the 6000 to 6100 K bin – it appears that the host data distributes across the range of non-host data. A Mann-Whitney test indicates that the host and non-host abundances in the two high-temperature bins are not statistically different at P = 0.379 and 0.096 respectively. The lower temperature bin does show a statistical difference at P = 0.031, but the host abundance average is lower than the non-host is this bin.

It thus appears from this data that stars with planets have lithium abundances that are ***not*** enhanced relative to non-hosts. While this result will not satisfy everyone, it is what this analysis supports.

### 4.3.2   C and O

The behavior of carbon and oxygen in the local region is well determined — for a full discussion see Nomoto et al. (2013) and / or Hinkel et al. (2014). The trends are that both [C/Fe] and [O/Fe] decrease with increasing [Fe/H]. These trends are a result of the increased importance of iron production in Type Ia supernovae as time proceeds. In Figure 13, these trends along with C/O versus [Fe/H] are shown for this sample of stars. The stars plotted have a minimum effective temperature of 4500 K. The data presented here are in accord with previous studies.

In Figure 13 there are a few stars that appear to have [C/Fe], [O/Fe], or C/O ratios that differ considerably from the bulk of the data. For the [C/Fe] data (Figure 13, top panel) the stars that scatter above the bulk of the data have effective temperatures less than 5000 K, and the larger number of those cluster between 4500 and 4700 K. The better interpretation of the high [C/Fe] ratios in these stars is that the ratio is overestimated due to blends in the $C_2$ lines. There are two stars — HD 210640 and HR 6469 — with very low [C/Fe] ratios. These stars have $T_{eff} > 5500$ K and the lack of C I lines is pronounced. It appears that the low carbon content has not been previously noted. These stars are possibly related to the weak-G band stars (Palacios et al. 2016, Adamczak & Lambert 2013).

The [O/Fe] data (Figure 13 – middle panel) show the usual increase in [O/Fe] with decreasing [Fe/H]. There is some scatter at above-solar [Fe/H] ratios towards low [O/Fe] ratios. The low oxygen abundances are likely due to underestimates of the oxygen strength due to continuum placement or noise problems. The stars with low [O/Fe] ratios are predominately cooler than 5500 K making the oxygen abundance depend on the weak [O I] line and thus susceptible to such problems.

Lastly, C/O versus [Fe/H] is shown in the bottom panel of Figure 13. The bulk of these stars have subsolar C/O ratios: at solar [Fe/H] the C/O ratio is about 0.4 whereas the solar ratio is 0.55. This offset is also seen in the dwarf study of Luck & Heiter (2006). What is curious is that the local giants show a mean C/O of about 0.4 at solar [Fe/H] (Luck 2015) although one would expect the giants due to CN-processing to have a lower C/O than the dwarfs. Where the problem lies, if there is one, is yet to be determined. Note that as the [Fe/H] ratio falls so does the C/O ratio: at [Fe/H] ~ 0 the nominal C/O ratio is 0.4 while at [Fe/H] ~ -0.8 the ratio is about 0.2.

## 5. Concluding Remarks

While this study finds no spectacular new result, it does provide a large sample of self-consistently analyzed dwarfs and subgiants in the local vicinity. Accurate abundances demand accurate equivalent widths. The new method of equivalent width determination described here provides a significant enhancement in reliability of the input data to the investigation. The abundance analysis adds more substance to the abundance work that underpins out understanding of the local region.

One of the more significant problems this study cannot approach is the difficulty of determining reliable abundances in dwarfs cooler than 4500 K. An approach to this problem would be to use near infrared spectra of cool bright dwarfs. Such spectra might also allow one to isolate the ionization problem using unblended Fe I and Fe II lines. Two serious current deficiencies that need rectification before such studies can be realized are more and better laboratory oscillator strengths and damping coefficients.

As the last point, some thought needs to be given to the central idea behind these studies: What is the standard of normalcy in the local region for abundances, and specifically, how does the does the Sun compare? The average star in the solar neighborhood has a composition like the average shown in Figure 4. An additional piece of information is located in Figure 11. Stars of the same temperature of the Sun most commonly have lithium abundances of about 1.8 dex. This study indicates that the Sun is somewhat atypical of its neighbors. The ways that it differs are:

1) The lanthanide abundances of the Sun are about 0.2 dex lower than those of its average neighbor.
2) The lithium abundance of the Sun is in the lower range of abundance compared to local stars of like temperature.
3) The C/O ratio of the Sun is high compared to its neighbors: 0.55 versus 0.4.

These differences are modest but real. Item two above is the least problematic – lithium is very sensitive to convection depth and precise prior history. The C/O ratio and lanthanide abundance are primordial and reflect the immediate supernovae history prior to the formation of the proto-solar nebula.

## Acknowledgements

Financial support by Case Western Reserve University made possible the McDonald Observatory observations used in this work. Gregory Tobar-Gomez was instrumental in helping to develop the equivalent width pseudo-Voigt procedures. The Sandiford and HET echelle spectra used here are available through the [FGK Spectral Library](). Also available from the FGK Spectral Library are EXCEL format versions of all tabular material derived as part of this analysis. The ELODIE archive at Observatoire de Haute-Provence (OHP) also provided part of the spectroscopic data. The SIMBAD database, operated at CDS, Strasbourg, France and satellite sites, and NASA's Astrophysics Data System Bibliographic Services were instrumental in this work.

Table 1
Program Stars

| Primary | HD | HIP | HR | CCDM | Spectral Type | P (mas) | e_P (mas) | V (mag) | d (pc) | E(B-V) (mag) | Mv (mag) | Host |
|---|---|---|---|---|---|---|---|---|---|---|---|---|
| ksi UMa B | 98230 | | 4374 | J11182+3132B | G2V | 113.20 | 4.60 | 4.73 | 8.8 | 0.00 | 5.00 | |
| HD 179958 | 179958 | | 7294 | J19121+4951A | G2V | 40.16 | 0.83 | 6.57 | 24.9 | 0.00 | 4.59 | |
| HD 179957 | 179957 | | 7293 | J19121+4951B | G3V | | | 6.75 | | 0.00 | 4.77 | |
| BD-10 3166 | | | | | K3.0V | 15.34 | 3.08 | 10.01 | 65.2 | 0.00 | 5.94 | H |
| 85 Peg | 224930 | 171 | 9088 | | G5V_Fe-1 | 82.17 | 2.23 | 5.75 | 12.2 | 0.00 | 5.32 | |
| HD 225239 | 225239 | 394 | 9107 | | G2V | 25.52 | 3.28 | 6.11 | 39.2 | 0.00 | 3.14 | |
| HD 38A | 38A | 473 | | J00057+4548A | K6V | 85.10 | 2.74 | 8.83 | 11.8 | 0.00 | 8.48 | |
| HD 38B | 38B | | | J00057+4548B | M0.5V | | | 9.00 | | 0.00 | 8.64 | |
| HD 123A | 123A | 518 | | | G3V | 46.56 | 0.65 | 6.42 | 21.5 | 0.00 | 4.76 | |
| HD 123B | 123B | | | | G8V | | | 7.32 | | 0.00 | 5.66 | |
| HD 166 | 166 | 544 | 8 | J00065+2900A | K0V | 73.15 | 0.56 | 6.13 | 13.7 | 0.00 | 5.45 | |
| HD 330 | 330 | 656 | | | F8V | 9.14 | 0.98 | 8.17 | 109.4 | 0.02 | 2.90 | |
| HD 400 | 400 | 699 | 17 | | F8IV | 30.82 | 0.33 | 6.22 | 32.4 | 0.00 | 3.67 | |
| HD 245 | 245 | 705 | | | G2V | 17.51 | 0.69 | 8.37 | 57.1 | 0.00 | 4.59 | |
| HD 603 | 603 | 846 | | | G2V | 12.59 | 1.09 | 8.92 | 79.4 | 0.00 | 4.42 | |
| 6 Cet | 693 | 910 | 33 | | F8VFe-0.8CH-0.5 | 53.34 | 0.64 | 4.89 | 18.7 | 0.00 | 3.53 | |
| HD 975 | 975 | 1137 | | | F5V | 13.33 | 0.78 | 8.08 | 75.0 | 0.00 | 3.70 | |
| HD 1388 | 1388 | 1444 | | | G0V | 36.74 | 0.49 | 6.50 | 27.2 | 0.00 | 4.33 | |
| HD 1461 | 1461 | 1499 | 72 | | G3V | 43.02 | 0.51 | 6.46 | 23.2 | 0.00 | 4.63 | H |
| 9 Cet | 1835 | 1803 | 88 | J00228-1212A | G3V | 47.93 | 0.53 | 6.39 | 20.9 | 0.00 | 4.79 | |
| HD 2475 | 2475 | 2237 | 108 | J00284-2020AB | F9VFe+0.5 | 31.06 | 0.68 | 6.44 | 32.2 | 0.00 | 3.90 | |
| HD 2582 | 2582 | 2288 | | | G3V | 11.14 | 1.55 | 9.09 | 89.8 | 0.00 | 4.32 | |
| HD 2638 | 2638 | 2350 | | | K1V | 20.03 | 1.49 | 9.38 | 49.9 | 0.00 | 5.89 | H |
| HD 2589 | 2589 | 2422 | 112 | | K0IV | 25.97 | 0.33 | 6.21 | 38.5 | 0.00 | 3.28 | |
| HD 2730 | 2730 | 2434 | | | F7V | 8.83 | 0.86 | 8.49 | 113.3 | 0.02 | 3.17 | |
| 13 Cet | 3196 | 2762 | 142 | J00352-0336AB | F7V+G4V | 47.05 | 0.67 | 5.20 | 21.3 | 0.00 | 3.56 | |
| 14 Cet | 3229 | 2787 | 143 | | F5V | 18.35 | 0.36 | 5.95 | 54.5 | 0.00 | 2.27 | |
| HD 3268 | 3268 | 2832 | 145 | | F7V | 26.62 | 0.37 | 6.41 | 37.6 | 0.00 | 3.54 | |
| HD 3443 | 3443 | 2941 | 159 | J00373-2446AB | K1V+G | 64.93 | 1.85 | 5.57 | 15.4 | 0.00 | 4.63 | |
| HD 3556 | 3556 | 3033 | | | G0V | 13.96 | 0.73 | 8.75 | 71.6 | 0.00 | 4.47 | |
| HD 3628 | 3628 | 3086 | | | G3V | 20.25 | 0.66 | 7.30 | 49.4 | 0.00 | 3.83 | |
| 54 Psc | 3651 | 3093 | 166 | J00394+2115A | K0V | 90.42 | 0.32 | 5.88 | 11.1 | 0.00 | 5.66 | H |
| HD 3795 | 3795 | 3185 | 173 | | K0VFe-1.5CH-1.3 | 34.62 | 0.45 | 6.14 | 28.9 | 0.00 | 3.84 | |
| HD 3765 | 3765 | 3206 | | | K2V | 57.71 | 0.80 | 7.36 | 17.3 | 0.00 | 6.17 | |
| BD+33 99 | | 3418 | | J00435+3349A | K5-V | 48.61 | 0.92 | 9.06 | 20.6 | 0.00 | 7.49 | |
| HD 4208 | 4208 | 3479 | | | G7VFe-1CH-0.5 | 30.89 | 0.75 | 7.79 | 32.4 | 0.00 | 5.24 | H |
| HD 4203 | 4203 | 3502 | | | G5 | 12.95 | 1.03 | 8.69 | 77.2 | 0.00 | 4.25 | H |
| HD 4256 | 4256 | 3535 | | | K3V | 46.37 | 0.62 | 8.00 | 21.6 | 0.00 | 6.33 | |
| 18 Cet | 4307 | 3559 | 203 | J00455-1253A | G0V | 32.26 | 0.44 | 6.15 | 31.0 | 0.00 | 3.69 | |
| 61 Psc | 4568 | 3730 | 217 | | F8V | 17.53 | 0.54 | 6.60 | 57.0 | 0.00 | 2.82 | |
| HD 4628 | 4628 | 3765 | 222 | J00483+0516A | K2.5V | 134.14 | 0.51 | 5.74 | 7.5 | 0.00 | 6.38 | |
| eta Cas | 4614 | 3821 | 219 | | F9V+M0-V | 167.98 | 0.48 | 3.44 | 6.0 | 0.00 | 4.57 | |
| HD 4635 | 4635 | 3876 | | J00497+7027A | K2.5V+ | 46.23 | 0.53 | 7.76 | 21.6 | 0.00 | 6.08 | |
| phi02 Cet | 4813 | 3909 | 235 | | F7V | 63.48 | 0.35 | 5.19 | 15.8 | 0.00 | 4.20 | |
| HR 244 | 5015 | 4151 | 244 | J00531+6107A | F9V | 53.35 | 0.33 | 4.80 | 18.7 | 0.00 | 3.44 | |
| 36 And | 5286 | 4288 | 258 | J00549+2337AB | K1IV | 26.33 | 0.65 | 5.45 | 38.0 | 0.00 | 2.55 | |
| HD 5294 | 5294 | 4290 | | J00550+2406A | G5 | 34.41 | 0.64 | 7.38 | 29.1 | 0.00 | 5.06 | |
| HD 5372 | 5372 | 4393 | | | G5 | 25.07 | 0.59 | 7.52 | 39.9 | 0.00 | 4.52 | |
| HD 5351 | 5351 | 4454 | | | K4V | 40.90 | 0.89 | 9.10 | 24.4 | 0.00 | 7.16 | |
| HD 5494 | 5494 | 4457 | | | F7V | 12.99 | 0.75 | 7.96 | 77.0 | 0.00 | 3.52 | |
| HD 5600 | 5600 | 4504 | | | F5V | 7.34 | 0.56 | 8.00 | 136.2 | 0.01 | 2.31 | |
| HD 6064 | 6064 | 4825 | | | F3V | 8.72 | 0.74 | 7.61 | 114.7 | 0.00 | 2.30 | |
| BD+61 195 | | 4872 | | | M1.5Ve | 100.40 | 1.52 | 9.56 | 10.0 | 0.00 | 9.57 | |
| mu. Cas | 6582 | 5336 | 321 | J01080+5455A | K1V_Fe-2 | 132.38 | 0.82 | 5.17 | 7.6 | 0.00 | 5.78 | |
| 44 And | 6920 | 5493 | 340 | | F8V | 19.13 | 0.28 | 5.68 | 52.3 | 0.00 | 2.09 | |
| HD 7091 | 7091 | 5540 | | | F6V | 9.90 | 1.07 | 9.09 | 101.0 | 0.00 | 4.06 | |
| HD 7230 | 7230 | 5673 | | | F5 | 9.32 | 0.77 | 8.35 | 107.3 | 0.01 | 3.17 | |
| HD 7397 | 7397 | 5769 | | | G1V | 10.64 | 0.85 | 8.68 | 94.0 | 0.00 | 3.81 | |
| HD 7352 | 7352 | 5777 | | | G0V | 14.51 | 0.77 | 8.35 | 68.9 | 0.00 | 4.16 | |
| HD 7514 | 7514 | 5831 | | | G5V | 11.51 | 1.33 | 9.15 | 86.9 | 0.00 | 4.46 | |
| 38 Cet | 7476 | 5833 | 368 | J01148-0059A | F3V | 23.30 | 0.25 | 5.70 | 42.9 | 0.00 | 2.54 | |
| HD 7590 | 7590 | 5944 | | | G0-V(k) | 43.11 | 0.45 | 6.59 | 23.2 | 0.00 | 4.76 | |
| HR 357 | 7238 | 5947 | 357 | | F5Vs | 13.77 | 0.28 | 6.27 | 72.6 | 0.00 | 1.97 | |
| HD 8173 | 8173 | 6305 | | | G1V | 13.74 | 1.05 | 8.76 | 72.8 | 0.00 | 4.45 | |
| HD 7924 | 7924 | 6379 | | | K0.5V | 59.49 | 0.46 | 7.19 | 16.8 | 0.00 | 6.06 | H |
| HD 8574 | 8574 | 6643 | | | F8 | 22.44 | 0.53 | 7.11 | 44.6 | 0.00 | 3.87 | H |
| HD 8648 | 8648 | 6653 | | | G3V | 22.90 | 1.06 | 7.39 | 43.7 | 0.00 | 4.19 | |
| HR 407 | 8634 | 6669 | 407 | | F6V | 14.69 | 0.42 | 6.19 | 68.1 | 0.00 | 2.03 | |
| HD 8673 | 8673 | 6702 | 410 | | F7V | 27.73 | 0.40 | 6.31 | 36.1 | 0.00 | 3.52 | H |
| HD 9224 | 9224 | 7090 | | | G0V | 22.66 | 0.65 | 7.31 | 44.1 | 0.00 | 4.09 | |
| HD 9369 | 9369 | 7204 | | | F5 | 7.00 | 0.74 | 8.10 | 142.9 | 0.02 | 2.26 | |
| HD 9562 | 9562 | 7276 | 448 | | G1V | 33.88 | 0.26 | 5.76 | 29.5 | 0.00 | 3.41 | |
| HD 9407 | 9407 | 7339 | | | G6.5V | 48.41 | 0.40 | 6.53 | 20.7 | 0.00 | 4.95 | |
| ups And | 9826 | 7513 | 458 | J01367+4125A | F9V | 74.12 | 0.19 | 4.10 | 13.5 | 0.00 | 3.45 | H |
| HD 9986 | 9986 | 7585 | | | G2V | 39.02 | 0.61 | 6.76 | 25.6 | 0.00 | 4.72 | |
| HD 10086 | 10086 | 7734 | | | G5V | 46.79 | 0.60 | 6.61 | 21.4 | 0.00 | 4.96 | |
| HD 10145 | 10145 | 7902 | | | G5V | 26.84 | 0.78 | 7.70 | 37.3 | 0.00 | 4.84 | |
| HD 10307 | 10307 | 7918 | 483 | | G1.5V | 78.50 | 0.54 | 4.96 | 12.7 | 0.00 | 4.43 | |
| 107 Psc | 10476 | 7981 | 493 | J01425+2016A | K1V | 132.76 | 0.50 | 5.24 | 7.5 | 0.00 | 5.86 | |
| HR 495 | 10486 | 8044 | 495 | | K2IV | 17.31 | 0.48 | 6.32 | 57.8 | 0.00 | 2.51 | |
| HD 10436 | 10436 | 8070 | | | K5.5V | 73.65 | 0.98 | 8.41 | 13.6 | 0.00 | 7.75 | |
| tau Cet | 10700 | 8102 | 509 | J01441-1557A | G8.5V | 273.96 | 0.17 | 3.50 | 3.7 | 0.00 | 5.69 | |
| 109 Psc | 10697 | 8159 | 508 | | G3Va | 30.70 | 0.43 | 6.29 | 32.6 | 0.00 | 3.73 | H |
| HD 10853 | 10853 | 8275 | | | K3.5V | 41.63 | 1.22 | 8.89 | 24.0 | 0.00 | 6.99 | |
| HR 511 | 10780 | 8362 | 511 | J01477+6351A | K0V | 99.33 | 0.53 | 5.63 | 10.1 | 0.00 | 5.62 | |
| HD 11038 | 11038 | 8383 | | | G1V | 9.72 | 1.39 | 9.50 | 102.9 | 0.00 | 4.44 | |
| HD 11007 | 11007 | 8433 | 523 | | F8V | 35.90 | 0.34 | 5.79 | 27.9 | 0.00 | 3.57 | |
| HD 11507 | 11507 | 8768 | | | M0V | 90.86 | 1.16 | 8.88 | 11.0 | 0.00 | 8.67 | |
| HD 11373 | 11373 | 8867 | | | K3Vk | 44.38 | 0.97 | 8.48 | 22.5 | 0.00 | 6.72 | |
| HD 12051 | 12051 | 9269 | | J01592+3312A | G9V | 40.03 | 0.58 | 7.14 | 25.0 | 0.00 | 5.15 | |
| 112 Psc | 12235 | 9353 | 582 | | G2IV | 29.85 | 0.46 | 5.90 | 33.5 | 0.00 | 3.27 | |
| HD 12661 | 12661 | 9683 | | | K0V | 28.61 | 0.61 | 7.44 | 35.0 | 0.00 | 4.72 | H |
| HD 12846 | 12846 | 9829 | | | G2V | 43.91 | 0.57 | 6.89 | 22.8 | 0.00 | 5.10 | |
| HD 13043 | 13043 | 9911 | | J02076-0037A | G2V | 27.06 | 0.58 | 6.90 | 37.0 | 0.00 | 4.06 | |
| BD+29 366 | | 10140 | | J02104+2948A | F8V | 18.45 | 1.24 | 8.77 | 54.2 | 0.00 | 5.10 | |
| 66 Cet B | 13612B | 10303 | | J02128-0223B | G1V | 20.89 | 7.39 | 7.74 | 47.9 | 0.00 | 4.34 | |
| 66 Cet | 13612 | 10305 | 650 | J02128-0223A | F8V | 25.19 | 1.41 | 5.68 | 39.7 | 0.00 | 2.69 | |
| eta Ari | 13555 | 10306 | 646 | | F5V | 34.64 | 0.33 | 5.24 | 28.9 | 0.00 | 2.93 | |
| V* V450 And | 13507 | 10321 | | | G5V | 37.25 | 0.55 | 7.21 | 26.8 | 0.00 | 5.07 | |
| HD 13403 | 13403 | 10322 | | J02129+5712A | G3V | 24.40 | 0.62 | 7.00 | 41.0 | 0.00 | 3.94 | |
| HD 13579 | 13579 | 10531 | | J02157+6740A | K2V | 53.82 | 1.74 | 7.18 | 18.6 | 0.00 | 5.83 | |
| del Tri | 13974 | 10644 | 660 | J02170+3414A | G0V | 92.73 | 0.39 | 4.87 | 10.8 | 0.00 | 4.71 | |
| HR 672 | 14214 | 10723 | 672 | | G1V | 41.06 | 0.49 | 5.58 | 24.4 | 0.00 | 3.65 | |
| HD 14412 | 14412 | 10798 | 683 | | G8V | 78.93 | 0.35 | 6.34 | 12.7 | 0.00 | 5.83 | |
| HD 14374 | 14374 | 10818 | | | G8V | 25.80 | 1.03 | 8.48 | 38.8 | 0.00 | 5.54 | |
| HD 14348 | 14348 | 10868 | | | F5 | 17.68 | 0.45 | 7.19 | 56.6 | 0.00 | 3.43 | |
| HD 14624 | 14624 | 11053 | | | G5V | 13.45 | 0.95 | 8.44 | 74.3 | 0.00 | 4.08 | |
| kap For | 14802 | 11072 | 695 | | G1V | 45.53 | 0.82 | 5.19 | 22.0 | 0.00 | 3.48 | |
| HD 15189 | 15189 | 11367 | | | G0V | 11.17 | 1.10 | 9.46 | 89.5 | 0.00 | 4.70 | |

| Name | HD | HIP | GJ | Karmn | SpType | π (mas) | ? | ? | ? | ? | ? | Flag |
|---|---|---|---|---|---|---|---|---|---|---|---|---|
| HD 15069 | 15069 | 11505 | | | G1V | 17.87 | 0.81 | 7.77 | 56.0 | 0.00 | 4.03 | |
| 13 Tri | 15335 | 11548 | 720 | | G0V | 31.84 | 0.40 | 5.91 | 31.4 | 0.00 | 3.42 | |
| HD 15632 | 15632 | 11728 | | | G0 | 25.11 | 0.88 | 8.01 | 39.8 | 0.00 | 5.01 | |
| 79 Cet | 16141 | 12048 | | | G2V | 25.67 | 0.66 | 6.78 | 39.0 | 0.00 | 3.83 | H |
| HD 16160 | 16160 | 12114 | 753 | J02361+0653A | K3V | 139.27 | 0.45 | 5.83 | 7.2 | 0.00 | 6.55 | |
| 30 Ari B | 16232 | 12184 | 764 | J02370+2439B | F6V | 24.52 | 0.68 | 7.09 | 40.8 | 0.00 | 4.04 | H |
| HR 761 | 16220 | 12200 | 761 | | F8V | 20.05 | 0.38 | 6.23 | 49.9 | 0.00 | 2.74 | |
| HD 16397 | 16397 | 12306 | | | G0V | 29.17 | 0.81 | 7.37 | 34.3 | 0.00 | 4.69 | |
| HR 784 | 16673 | 12444 | 784 | | F8VFe-0.4 | 45.96 | 0.41 | 5.80 | 21.8 | 0.00 | 4.11 | |
| 84 Cet | 16765 | 12530 | 790 | J02412-0042AB | F7V | 44.27 | 0.84 | 5.71 | 22.6 | 0.00 | 3.94 | |
| tet Per | 16895 | 12777 | 799 | J02441+4913A | F8V | 89.87 | 0.22 | 4.11 | 11.1 | 0.00 | 3.88 | |
| tau01 Eri | 17206 | 12843 | 818 | | F7V | 70.32 | 1.83 | 4.46 | 14.2 | 0.00 | 3.70 | |
| HD 17382 | 17382 | 13081 | | J02482+2704A | K1V | 40.59 | 1.28 | 7.62 | 24.6 | 0.00 | 5.66 | |
| BD+33 529 | | 13375 | | | K5 | 68.47 | 1.43 | 9.55 | 14.6 | 0.00 | 8.73 | |
| HR 857 | 17925 | 13402 | 857 | | K1V | 96.60 | 0.40 | 6.05 | 10.4 | 0.00 | 5.97 | |
| HD 18445 | 18445 | 13769 | | J02572-2458C | K2V | 38.35 | 1.24 | 7.86 | 26.1 | 0.00 | 5.78 | |
| HD 18455 | 18455 | 13772 | | J02572-2458AB | K2V | 44.51 | 2.09 | 7.33 | 22.5 | 0.00 | 5.57 | |
| V* BW Ari | | 13806 | | | K0V | 25.54 | 1.02 | 8.91 | 39.2 | 0.00 | 5.94 | |
| 47 Ari | 18404 | 13834 | 878 | | F5IV | 30.15 | 0.30 | 5.80 | 33.2 | 0.00 | 3.20 | |
| V* BZ Cet | 18632 | 13976 | | | K2.5Vk | 41.00 | 1.12 | 7.96 | 24.4 | 0.00 | 6.02 | |
| eps For | 18907 | 14086 | 914 | | K2VFe-1.3CH-0.8 | 31.06 | 0.36 | 5.85 | 32.2 | 0.00 | 3.31 | |
| 51 Ari | 18803 | 14150 | | | G8V | 48.45 | 0.47 | 6.62 | 20.6 | 0.00 | 5.05 | |
| HD 18768 | 18768 | 14181 | | | F8 | 20.75 | 0.59 | 6.74 | 48.2 | 0.00 | 3.32 | |
| HD 19019 | 19019 | 14258 | | | G0V | 32.29 | 0.51 | 6.76 | 31.0 | 0.00 | 4.31 | |
| HD 18757 | 18757 | 14286 | | J03042+6142A | G1.5V | 41.27 | 0.58 | 6.67 | 24.2 | 0.00 | 4.75 | |
| HD 19383 | 19383 | 14464 | | | F5V | 13.92 | 2.74 | 8.09 | 71.8 | 0.00 | 3.81 | |
| HD 19308 | 19308 | 14532 | | J03076+3637A | G0 | 23.84 | 0.65 | 7.36 | 41.9 | 0.00 | 4.25 | |
| HD 19445 | 19445 | 14594 | | | G2VFe-3 | 24.92 | 0.91 | 8.06 | 40.1 | 0.00 | 5.04 | |
| iot Per | 19373 | 14632 | 937 | J03091+4936A | F9.5V | 94.87 | 0.23 | 4.05 | 10.5 | 0.00 | 3.94 | |
| HD 19617 | 19617 | 14721 | | | G5 | 17.35 | 0.97 | 8.72 | 57.6 | 0.00 | 4.91 | |
| alf For A | 20010A | 14879 | 963 | J03121-2859A | F6V | 70.24 | 0.45 | 3.98 | 14.2 | 0.00 | 3.21 | |
| 94 Cet | 19994 | 14954 | 962 | J03128-0112AB | F8.5V | 44.29 | 0.28 | 5.07 | 22.6 | 0.00 | 3.30 | H |
| HD 20165 | 20165 | 15099 | | | K1V | 44.15 | 0.83 | 7.80 | 22.7 | 0.00 | 6.02 | |
| HD 20339 | 20339 | 15199 | | | G2V | 20.55 | 0.71 | 7.84 | 48.7 | 0.00 | 4.40 | |
| HD 20367 | 20367 | 15323 | | | F8V | 37.48 | 0.63 | 6.41 | 26.7 | 0.00 | 4.28 | H |
| HD 20512 | 20512 | 15394 | | | G5 | 12.19 | 1.15 | 7.41 | 82.0 | 0.00 | 2.84 | |
| HD 20039 | 20039 | 15405 | | | F8 | 8.20 | 0.96 | 8.86 | 122.0 | 0.02 | 3.35 | |
| kap01 Cet | 20630 | 15457 | 996 | J03194+0321A | G5V | 109.41 | 0.27 | 4.85 | 9.1 | 0.00 | 5.05 | |

| Name | HD | HIP | HR | J-name | SpType | π | σπ | V | d | E(B-V) | M_V | Note |
|---|---|---|---|---|---|---|---|---|---|---|---|---|
| HD 232781 | 232781 | 15673 | | | K3.5V | 42.81 | 1.03 | 9.05 | 23.4 | 0.00 | 7.20 | |
| HD 21019 | 21019 | 15776 | 1024 | J03233-0748A | G5/6V | 26.92 | 0.51 | 6.20 | 37.1 | 0.00 | 3.35 | |
| BD+43 699 | | 15797 | | | K3V | 38.80 | 1.10 | 8.97 | 25.8 | 0.00 | 6.92 | |
| HD 21197 | 21197 | 15919 | | | K4V | 64.97 | 0.71 | 7.84 | 15.4 | 0.00 | 6.90 | |
| HD 21531 | 21531 | 16134 | | | K5V | 80.04 | 0.99 | 8.37 | 12.5 | 0.00 | 7.88 | |
| BD+23 465 | | 16529 | | | G5 | 22.81 | 0.99 | 8.89 | 43.8 | 0.00 | 5.68 | |
| eps Eri | 22049 | 16537 | 1084 | J03329-0927A | K2Vk: | 310.94 | 0.16 | 3.73 | 3.2 | 0.00 | 6.19 | H |
| HD 22072 | 22072 | 16641 | 1085 | | K0.5IV_Fe-1 | 24.76 | 0.46 | 6.17 | 40.4 | 0.00 | 3.14 | |
| HD 22455 | 22455 | 16792 | | | G2V | 13.12 | 1.20 | 8.93 | 76.2 | 0.00 | 4.52 | |
| HD 22292 | 22292 | 16806 | | | F8 | 12.78 | 0.82 | 8.15 | 78.2 | 0.00 | 3.68 | |
| HD 22468A | 22468A | 16846 | 1099 | J03368+0035A | G8 | 32.59 | 0.64 | 5.90 | 30.7 | 0.00 | 3.47 | |
| HD 22468B | 22468B | | | J03368+0035B | K4 | | | 8.90 | | 0.00 | 6.47 | |
| 10 Tau | 22484 | 16852 | 1101 | | F9IV-V | 71.62 | 0.54 | 4.30 | 14.0 | 0.00 | 3.58 | |
| 21 Eri | 22713 | 17027 | 1111 | | K0IV | 29.67 | 0.35 | 5.96 | 33.7 | 0.00 | 3.32 | |
| HD 22879 | 22879 | 17147 | | | F7/8V | 39.12 | 0.56 | 6.67 | 25.6 | 0.00 | 4.63 | |
| del Eri | 23249 | 17378 | 1136 | | K0+IV | 110.61 | 0.22 | 3.54 | 9.0 | 0.00 | 3.76 | |
| HD 23356 | 23356 | 17420 | | | K2V | 71.69 | 0.67 | 7.08 | 13.9 | 0.00 | 6.36 | |
| HD 23349 | 23349 | 17603 | | | G0 | 11.48 | 1.09 | 8.76 | 87.1 | 0.01 | 4.04 | |
| HD 23453 | 23453 | 17609 | | | M1V | 67.80 | 2.06 | 9.58 | 14.7 | 0.00 | 8.74 | |
| HD 23439A | 23439A | 17666 | | J03470+4126A | K2:V_Fe-1.3 | 45.65 | 2.63 | 8.12 | 21.9 | 0.00 | 6.42 | |
| HR 1179 | 23856 | 17689 | 1179 | | F6+V | 26.26 | 0.44 | 6.55 | 38.1 | 0.00 | 3.65 | |
| HD 23596 | 23596 | 17747 | | | F8 | 19.83 | 0.49 | 7.24 | 50.4 | 0.00 | 3.73 | H |
| HD 23476 | 23476 | 17790 | | | G5V | 20.79 | 0.75 | 8.48 | 48.1 | 0.00 | 5.07 | |
| HD 24040 | 24040 | 17960 | | | G0 | 21.46 | 0.75 | 7.52 | 46.6 | 0.00 | 4.18 | H |
| HD 24213 | 24213 | 18106 | | | G0 | 25.81 | 0.53 | 6.76 | 38.7 | 0.00 | 3.82 | |
| HD 24552 | 24552 | 18261 | | | G1V | 21.63 | 0.86 | 7.97 | 46.2 | 0.00 | 4.65 | |
| HD 24496 | 24496 | 18267 | | J03545+1637AB | G7V | 48.95 | 0.70 | 6.81 | 20.4 | 0.00 | 5.26 | |
| HD 24238 | 24238 | 18324 | | | K2V | 47.59 | 0.84 | 7.85 | 21.0 | 0.00 | 6.24 | |
| HD 24409 | 24409 | 18413 | | J03562+5939AD | G3V | 45.49 | 0.62 | 6.55 | 22.0 | 0.00 | 4.84 | |
| HR 1232 | 25069 | 18606 | 1232 | | K0/1III | 21.53 | 0.41 | 5.84 | 46.4 | 0.00 | 2.51 | |
| HD 24451 | 24451 | 18774 | | | K4V | 61.54 | 0.70 | 8.25 | 16.2 | 0.00 | 7.20 | |
| HD 25457 | 25457 | 18859 | 1249 | | F7/8V | 53.10 | 0.32 | 5.38 | 18.8 | 0.00 | 4.01 | |
| HD 25329 | 25329 | 18915 | | | K3Vp_Fe-1.7 | 54.68 | 0.92 | 8.50 | 18.3 | 0.00 | 7.19 | |
| HD 25621 | 25621 | 18993 | 1257 | | F6V | 28.25 | 0.30 | 5.36 | 35.4 | 0.00 | 2.62 | |
| 39 Tau | 25680 | 19076 | 1262 | J04053+2201A | G5V | 59.04 | 0.33 | 5.90 | 16.9 | 0.00 | 4.76 | |
| HD 25825 | 25825 | 19148 | | | G0V | 21.77 | 1.30 | 7.81 | 45.9 | 0.00 | 4.50 | |
| 49 Per | 25975 | 19302 | 1277 | | K1III | 22.68 | 0.34 | 6.09 | 44.1 | 0.00 | 2.87 | |
| 50 Per | 25998 | 19335 | 1278 | | F7V | 47.63 | 0.26 | 5.50 | 21.0 | 0.00 | 3.89 | |

| Name | HD | HIP | ADS | WDS | Spec | π | σπ | V | M_V | E(B-V) | (B-V) | Note |
|---|---|---|---|---|---|---|---|---|---|---|---|---|
| V* EI Eri | 26337 | 19431 | | | G3V | 19.68 | 1.06 | 7.04 | 50.8 | 0.00 | 3.51 | |
| HD 26505 | 26505 | 19494 | | | G3V | 11.73 | 0.94 | 8.62 | 85.3 | 0.00 | 3.96 | |
| HD 26345 | 26345 | 19504 | | | F6V | 22.62 | 0.52 | 6.58 | 44.2 | 0.00 | 3.35 | |
| GJ 1067 | | 19655 | | | K7V | 38.26 | 1.27 | 9.70 | 26.1 | 0.00 | 7.61 | |
| V* V1309 Tau | 26767 | 19786 | | | G0 | 23.29 | 1.25 | 8.03 | 42.9 | 0.00 | 4.87 | |
| HD 26784 | 26784 | 19796 | | | F8V | 22.20 | 0.51 | 7.08 | 45.0 | 0.00 | 3.81 | |
| HD 284248 | 284248 | 19797 | | | sdG0 | 11.40 | 1.22 | 9.23 | 87.7 | 0.01 | 4.50 | |
| BD-04 782 | | 19832 | | | K5V | 48.09 | 3.45 | 9.33 | 20.8 | 0.00 | 7.74 | |
| omi02 Eri | 26965 | 19849 | 1325 | J04153-0739A | K0.5V | 200.62 | 0.23 | 4.43 | 5.0 | 0.00 | 5.94 | |
| HD 26913 | 26913 | 19855 | 1321 | J04155+0612B | G8V | 47.49 | 0.68 | 6.92 | 21.1 | 0.00 | 5.30 | |
| HD 26923 | 26923 | 19859 | 1322 | J04155+0612A | G0IV | 46.88 | 0.47 | 6.30 | 21.3 | 0.00 | 4.65 | |
| HD 285660 | 285660 | 20041 | | | G5 | 11.33 | 1.23 | 8.77 | 88.3 | 0.01 | 4.02 | |
| HD 285690 | 285690 | 20082 | | | K0 | 18.90 | 1.80 | 9.55 | 52.9 | 0.00 | 5.93 | |
| HD 27530 | 27530 | 20184 | | | G2V | 17.78 | 0.67 | 8.16 | 56.2 | 0.00 | 4.41 | |
| HD 27808 | 27808 | 20557 | | | F8V | 22.59 | 0.77 | 7.13 | 44.3 | 0.00 | 3.89 | |
| HD 28185 | 28185 | 20723 | | | G6.5IV-V | 23.62 | 0.87 | 7.81 | 42.3 | 0.00 | 4.68 | H |
| HD 28099 | 28099 | 20741 | | | G2V | 21.73 | 1.23 | 8.12 | 46.0 | 0.00 | 4.81 | |
| HD 28005 | 28005 | 20800 | | | G0 | 35.48 | 0.49 | 6.72 | 28.2 | 0.00 | 4.47 | |
| V* DI Cam | 27172 | 20896 | | | F8 | 9.90 | 0.55 | 7.80 | 101.0 | 0.01 | 2.76 | |
| HD 28344 | 28344 | 20899 | | | G2V | 21.51 | 0.86 | 7.84 | 46.5 | 0.00 | 4.50 | |
| HD 28343 | 28343 | 20917 | | | M0.5V | 87.78 | 0.97 | 8.30 | 11.4 | 0.00 | 8.02 | |
| HD 283704 | 283704 | 20949 | | | G5 | 16.45 | 1.31 | 9.19 | 60.8 | 0.00 | 5.27 | |
| HD 28571 | 28571 | 20984 | | | G5V | 11.90 | 1.33 | 8.98 | 84.0 | 0.00 | 4.35 | |
| BD+04 701A | | 21000 | | J04302+0518A | F8 | 93.67 | 7.62 | 10.56 | 10.7 | 0.00 | 10.42 | |
| BD+04 701B | | | | J04302+0518B | F8 | | | 10.67 | | 0.00 | 10.53 | |
| HD 28635 | 28635 | 21112 | | | F9V | 19.10 | 0.76 | 7.75 | 52.4 | 0.00 | 4.15 | |
| m Per | 28704 | 21242 | 1434 | J04334+4303A | F0V | 16.90 | 0.53 | 6.09 | 59.2 | 0.00 | 2.23 | |
| HD 28946 | 28946 | 21272 | | | G9V | 36.92 | 1.73 | 7.93 | 27.1 | 0.00 | 5.77 | |
| HD 28992 | 28992 | 21317 | | | F8 | 22.00 | 1.05 | 7.90 | 45.5 | 0.00 | 4.61 | |
| HD 283750 | 283750 | 21482 | | J04368+2708A | K3IVke | 55.66 | 1.43 | 8.42 | 18.0 | 0.00 | 7.15 | |
| HD 29310 | 29310 | 21543 | | J04375+1509AB | G0 | 20.00 | 1.59 | 7.56 | 50.0 | 0.00 | 4.07 | |
| c Eri | 29391 | 21547 | 1474 | J04376-0228A | F0IV | 33.98 | 0.34 | 5.22 | 29.4 | 0.00 | 2.88 | H |
| HD 232979 | 232979 | 21553 | | | M0V | 98.91 | 1.01 | 8.86 | 10.1 | 0.00 | 8.84 | |
| HD 29419 | 29419 | 21637 | | | F5 | 22.46 | 0.72 | 7.51 | 44.5 | 0.00 | 4.27 | |
| HD 284574 | 284574 | 21741 | | | K0 | 15.95 | 1.42 | 9.44 | 62.7 | 0.00 | 5.45 | |
| HD 29697 | 29697 | 21818 | | | K4V | 75.82 | 1.14 | 7.99 | 13.2 | 0.00 | 7.38 | |
| HD 29587 | 29587 | 21832 | | | G2V | 36.27 | 0.87 | 7.28 | 27.6 | 0.00 | 5.08 | |
| HD 29645 | 29645 | 21847 | 1489 | | F9IV-V | 31.38 | 0.62 | 5.99 | 31.9 | 0.00 | 3.47 | |

| Name | HD | HIP | HR | WDS | Spectral Type | | | | | | | |
|---|---|---|---|---|---|---|---|---|---|---|---|---|
| 58 Eri | 30495 | 22263 | 1532 | | G1.5VCH-0.5 | 75.32 | 0.36 | 5.50 | 13.3 | 0.00 | 4.88 | |
| 59 Eri | 30606 | 22325 | 1538 | | F8V | 23.97 | 0.36 | 5.77 | 41.7 | 0.00 | 2.67 | |
| HD 30562 | 30562 | 22336 | 1536 | | G2IV | 37.85 | 0.35 | 5.77 | 26.4 | 0.00 | 3.66 | H |
| pi.03 Ori | 30652 | 22449 | 1543 | J04499+0657A | F6V | 123.94 | 0.17 | 3.19 | 8.1 | 0.00 | 3.66 | |
| HD 284930 | 284930 | 22654 | | | K0 | 21.55 | 2.16 | 10.29 | 46.4 | 0.00 | 6.96 | |
| HD 31527 | 31527 | 22905 | | | G0V | 25.93 | 0.60 | 7.48 | 38.6 | 0.00 | 4.55 | H |
| 101 Tau | 31845 | 23214 | | J04598+1554A | F5V | 24.50 | 0.41 | 6.76 | 40.8 | 0.00 | 3.70 | |
| 63 Eri | 32008 | 23221 | 1608 | | G4IV-V+DA | 18.53 | 0.84 | 5.40 | 54.0 | 0.00 | 1.74 | |
| HD 32147 | 32147 | 23311 | 1614 | | K3+V | 114.84 | 0.50 | 6.21 | 8.7 | 0.00 | 6.51 | |
| HD 31949 | 31949 | 23399 | | | F8V | 15.07 | 1.04 | 8.11 | 66.4 | 0.00 | 4.00 | |
| HD 32850 | 32850 | 23786 | | | G9V | 42.24 | 0.92 | 7.73 | 23.7 | 0.00 | 5.86 | |
| HR 1665 | 33093 | 23831 | 1665 | | G0IV | 27.25 | 0.40 | 5.97 | 36.7 | 0.00 | 3.14 | |
| m Tau | 32923 | 23835 | 1656 | J05075+1839AB | G4V | 64.79 | 0.33 | 5.00 | 15.4 | 0.00 | 4.06 | |
| 13 Ori | 33021 | 23852 | 1662 | J05074+0928A | G1IV | 36.14 | 0.45 | 6.17 | 27.7 | 0.00 | 3.96 | |
| 68 Eri | 33256 | 23941 | 1673 | | F5.5VkF4mF2 | 39.28 | 0.27 | 5.13 | 25.5 | 0.00 | 3.10 | |
| HD 32715 | 32715 | 24017 | 1647 | | F3V: | 25.11 | 0.48 | 6.36 | 39.8 | 0.00 | 3.36 | |
| HR 1685 | 33555 | 24130 | 1685 | | K0IV | 21.37 | 0.45 | 6.25 | 46.8 | 0.00 | 2.90 | |
| HR 1687 | 33608 | 24162 | 1687 | | F5V | 26.58 | 0.41 | 5.90 | 37.6 | 0.00 | 3.03 | |
| HD 33636 | 33636 | 24205 | | | G0VCH-0.3 | 35.25 | 1.02 | 7.06 | 28.4 | 0.00 | 4.80 | |
| HD 33866 | 33866 | 24336 | | | G3V | 24.02 | 1.29 | 7.87 | 41.6 | 0.00 | 4.77 | |
| HD 34721 | 34721 | 24786 | 1747 | J05188-1808A | G0V | 39.96 | 0.40 | 5.96 | 25.0 | 0.00 | 3.97 | |
| lam Aur | 34411 | 24813 | 1729 | J05192+4007A | G1V | 79.17 | 0.28 | 4.71 | 12.6 | 0.00 | 4.20 | |
| HD 33564 | 33564 | 25110 | 1686 | J05227+7913A | F6V | 47.88 | 0.21 | 5.10 | 20.9 | 0.00 | 3.50 | H |
| 111 Tau B | 35171 | 25220 | | | K4V | 71.00 | 1.34 | 7.92 | 14.1 | 0.00 | 7.18 | |
| 111 Tau | 35296 | 25278 | 1780 | J05244+1723A | F8V | 69.51 | 0.38 | 5.00 | 14.4 | 0.00 | 4.21 | |
| HD 36003 | 36003 | 25623 | | J05284-0331A | K4V | 76.80 | 0.76 | 7.64 | 13.0 | 0.00 | 7.07 | |
| HD 35961 | 35961 | 25880 | | J05314+5439AB | | 25.44 | 1.04 | 7.53 | 39.3 | 0.00 | 4.56 | |
| 18 Cam | 36066 | 25973 | 1828 | | F8V | 23.42 | 0.48 | 6.46 | 42.7 | 0.00 | 3.31 | |
| V* V2689 Ori | 245409 | 26335 | | | K6V | 88.97 | 1.02 | 8.90 | 11.2 | 0.00 | 8.64 | |
| HD 37124 | 37124 | 26381 | | | G4IV-V | 29.70 | 0.70 | 7.68 | 33.7 | 0.00 | 5.04 | H |
| HD 37008 | 37008 | 26505 | | | K1V | 49.60 | 0.72 | 7.73 | 20.2 | 0.00 | 6.21 | |
| HD 37605 | 37605 | 26664 | | | K0 | 22.74 | 1.11 | 8.69 | 44.0 | 0.00 | 5.47 | H |
| HR 1925 | 37394 | 26779 | 1925 | J05413+5329A | K1V | 81.45 | 0.54 | 6.23 | 12.3 | 0.00 | 5.78 | |
| gam Lep | 38393 | 27072 | 1983 | J05445-2226A | F6V | 112.02 | 0.18 | 3.60 | 8.9 | 0.00 | 3.85 | |
| V* AK Lep | 38392 | | 1982 | J05445-2226B | K2V | | | 6.15 | | 0.00 | 6.40 | |
| HR 1980 | 38382 | 27075 | 1980 | | F8.5V | 39.15 | 0.47 | 6.35 | 25.5 | 0.00 | 4.31 | |
| HD 38230 | 38230 | 27207 | | J05460+3717A | K0V | 45.76 | 0.76 | 7.36 | 21.9 | 0.00 | 5.66 | |
| HD 38529 | 38529 | 27253 | 1988 | | G8III/IV | 25.46 | 0.40 | 5.94 | 39.3 | 0.00 | 2.97 | H |

| Name | HD | HIP | GJ | Karmn | SpType | col7 | col8 | col9 | col10 | col11 | col12 | H |
|---|---|---|---|---|---|---|---|---|---|---|---|---|
| HD 38700 | 38700 | 27260 | | | G0V | 16.07 | 1.06 | 8.61 | 62.2 | 0.00 | 4.64 | |
| HD 38858 | 38858 | 27435 | 2007 | | G2V | 65.89 | 0.41 | 5.97 | 15.2 | 0.00 | 5.06 | H |
| chi01 Ori | 39587 | 27913 | 2047 | J05544+2017AB | G0VCH+M | 115.43 | 0.27 | 4.40 | 8.7 | 0.00 | 4.71 | |
| LP 837-53 | | 28035 | | | M2.5V | 68.57 | 2.17 | 10.82 | 14.6 | 0.00 | 10.00 | |
| HD 39881 | 39881 | 28066 | 2067 | J05560+1357A | G5IV | 37.59 | 0.63 | 6.60 | 26.6 | 0.00 | 4.48 | |
| HD 40512 | 40512 | 28356 | | | F5V | 14.71 | 0.71 | 7.74 | 68.0 | 0.00 | 3.58 | |
| HD 40616 | 40616 | 28428 | | | G3V | 18.10 | 0.73 | 7.47 | 55.2 | 0.00 | 3.76 | |
| HD 40979 | 40979 | 28767 | | | F8 | 30.20 | 0.44 | 6.73 | 33.1 | 0.00 | 4.13 | H |
| HD 41330 | 41330 | 28908 | 2141 | J06061+3524A | G0V | 39.03 | 0.50 | 6.12 | 25.6 | 0.00 | 4.08 | |
| V* V1386 Ori | 41593 | 28954 | | | K0V | 65.48 | 0.67 | 6.74 | 15.3 | 0.00 | 5.82 | |
| HD 41708 | 41708 | 29074 | | | G0V | 23.37 | 0.90 | 8.03 | 42.8 | 0.00 | 4.87 | |
| HD 42618 | 42618 | 29432 | | | G4V | 42.55 | 0.55 | 6.84 | 23.5 | 0.00 | 4.98 | H |
| HR 2208 | 42807 | 29525 | 2208 | | G2V | 55.71 | 0.44 | 6.44 | 18.0 | 0.00 | 5.17 | |
| HD 43162 | 43162 | 29568 | 2225 | | G6.5V | 59.80 | 0.49 | 6.37 | 16.7 | 0.00 | 5.26 | |
| HD 42983 | 42983 | 29573 | | | K0IV | 17.23 | 0.73 | 7.38 | 58.0 | 0.00 | 3.56 | |
| 71 Ori | 43042 | 29650 | 2220 | J06148+1909A | F5.5IV-V | 48.04 | 0.34 | 5.20 | 20.8 | 0.00 | 3.61 | |
| HD 43318 | 43318 | 29716 | 2233 | | F5V | 26.89 | 0.31 | 5.65 | 37.2 | 0.00 | 2.80 | |
| HD 42250 | 42250 | 29761 | | | G9V | 38.87 | 0.62 | 7.43 | 25.7 | 0.00 | 5.38 | |
| HD 43745 | 43745 | 29843 | 2254 | J06171-2242A | F8.5V | 24.84 | 0.43 | 6.04 | 40.3 | 0.00 | 3.02 | |
| HD 43587 | 43587 | 29860 | 2251 | J06172+0505A | G0V | 51.95 | 0.40 | 5.70 | 19.2 | 0.00 | 4.28 | |
| HD 43856 | 43856 | 29991 | | | F6V | 14.41 | 1.03 | 7.95 | 69.4 | 0.00 | 3.74 | |
| HD 45184 | 45184 | 30503 | 2318 | | G1.5V | 45.70 | 0.40 | 6.39 | 21.9 | 0.00 | 4.69 | H |
| HD 44966 | 44966 | 30513 | | | F5/6V | 21.18 | 0.76 | 7.05 | 47.2 | 0.00 | 3.68 | |
| HD 45067 | 45067 | 30545 | 2313 | | F9V | 29.79 | 0.46 | 5.90 | 33.6 | 0.00 | 3.27 | |
| HD 45088 | 45088 | 30630 | | J06262+1845AB | K3Vk | 67.89 | 1.53 | 6.77 | 14.7 | 0.00 | 5.93 | |
| HD 45282 | 45282 | 30668 | | | Gwl | 6.72 | 0.70 | 8.03 | 148.8 | 0.02 | 2.11 | |
| HD 45588 | 45588 | 30711 | 2349 | J06272-2551A | F8IV | 33.48 | 0.41 | 6.07 | 29.9 | 0.00 | 3.69 | |
| HD 45350 | 45350 | 30860 | | | G5 | 20.46 | 0.72 | 7.88 | 48.9 | 0.00 | 4.43 | H |
| HD 45759 | 45759 | 30939 | | | F8 | 20.88 | 0.73 | 7.61 | 47.9 | 0.00 | 4.21 | |
| HD 45205 | 45205 | 30990 | | | G0 | 13.66 | 1.12 | 8.47 | 73.2 | 0.00 | 4.15 | |
| HD 46090 | 46090 | 31083 | | J06314+0255A | G3/5V | 36.00 | 0.61 | 7.13 | 27.8 | 0.00 | 4.91 | |
| HD 46301 | 46301 | 31197 | | | F6V | 9.29 | 0.57 | 7.28 | 107.6 | 0.01 | 2.09 | |
| HD 46375 | 46375A | 31246 | | J06332+0528A | K1IV | 28.72 | 0.89 | 7.84 | 34.8 | 0.00 | 5.13 | H |
| HD 46780 | 46780 | 31568 | | J06364+2717AB | G0 | 30.20 | 1.04 | 6.89 | 33.1 | 0.00 | 4.29 | |
| HD 47157 | 47157 | 31655 | | J06375+0909A | G5IV-V | 26.99 | 0.93 | 7.62 | 37.1 | 0.00 | 4.78 | |
| HD 47127 | 47127 | 31660 | | J06376+1210A | G5 | 37.15 | 0.56 | 6.83 | 26.9 | 0.00 | 4.68 | |
| HD 47309 | 47309 | 31965 | | | G0 | 24.82 | 0.74 | 7.60 | 40.3 | 0.00 | 4.57 | |
| HD 47752 | 47752 | 32010 | | | K3.5V | 57.32 | 1.16 | 8.06 | 17.4 | 0.00 | 6.85 | |

| Name | HD | HIP | HR | Karmn | SpType | π (mas) | ? | ? | ? | ? | ? | Note |
|---|---|---|---|---|---|---|---|---|---|---|---|---|
| HD 48938 | 48938 | 32322 | 2493 | | G0VFe-0.8CH-0.5 | 36.85 | 0.44 | 6.43 | 27.1 | 0.00 | 4.26 | |
| HD 48565 | 48565 | 32329 | | | F8 | 20.30 | 0.54 | 7.20 | 49.3 | 0.00 | 3.74 | |
| HD 46588 | 46588 | 32439 | 2401 | | F7V | 55.95 | 0.27 | 5.45 | 17.9 | 0.00 | 4.19 | |
| psi05 Aur | 48682 | 32480 | 2483 | J06467+4335A | F9V | 59.82 | 0.30 | 5.25 | 16.7 | 0.00 | 4.14 | |
| HD 49385 | 49385 | 32608 | | J06482+0018AB | G0V | 13.91 | 0.76 | 7.42 | 71.9 | 0.00 | 3.14 | |
| HD 49674 | 49674 | 32916 | | | G0 | 22.61 | 0.87 | 8.10 | 44.2 | 0.00 | 4.87 | H |
| HD 50281 | 50281 | 32984 | | J06523-0511A | K3.5V | 114.81 | 0.44 | 6.57 | 8.7 | 0.00 | 6.87 | |
| HD 50206 | 50206 | 32993 | | | F5 | 9.66 | 0.60 | 7.60 | 103.5 | 0.01 | 2.51 | |
| HD 50806 | 50806 | 33094 | 2576 | | G5V | 38.62 | 0.49 | 6.04 | 25.9 | 0.00 | 3.97 | |
| HD 50554 | 50554 | 33212 | | | F8V | 33.43 | 0.59 | 6.84 | 29.9 | 0.00 | 4.46 | H |
| HD 50867 | 50867 | 33275 | | | F8 | 18.61 | 0.75 | 7.61 | 53.7 | 0.00 | 3.96 | |
| 37 Gem | 50692 | 33277 | 2569 | | G0V | 58.00 | 0.41 | 5.76 | 17.2 | 0.00 | 4.58 | |
| HD 51219 | 51219 | 33382 | | | G8V | 31.99 | 0.68 | 7.42 | 31.3 | 0.00 | 4.95 | |
| HD 51419 | 51419 | 33537 | | | G5VFe-1 | 40.60 | 0.53 | 6.94 | 24.6 | 0.00 | 4.98 | |
| 39 Gem | 51530 | 33595 | 2601 | J06588+2605A | F7V | 20.98 | 0.49 | 6.21 | 47.7 | 0.00 | 2.82 | |
| HD 52265 | 52265 | 33719 | 2622 | | G0V | 34.53 | 0.40 | 6.30 | 29.0 | 0.00 | 3.99 | H |
| HD 52698 | 52698 | 33817 | | | K1V | 68.27 | 0.61 | 6.69 | 14.6 | 0.00 | 5.86 | |
| HD 51866 | 51866 | 33852 | | | K3V | 49.79 | 0.84 | 8.00 | 20.1 | 0.00 | 6.49 | |
| HD 52634 | 52634 | 33941 | | J07026+1558AB | G0V | 11.73 | 1.73 | 8.28 | 85.3 | 0.00 | 3.62 | |
| HD 52711 | 52711 | 34017 | 2643 | | G0V | 52.27 | 0.41 | 5.93 | 19.1 | 0.00 | 4.52 | |
| HD 53927 | 53927 | 34414 | | | K2.5V | 44.93 | 0.97 | 8.34 | 22.3 | 0.00 | 6.60 | |
| HD 54371 | 54371 | 34567 | | J07096+2544A | G6V | 39.73 | 0.54 | 7.06 | 25.2 | 0.00 | 5.06 | |
| HR 2692 | 54563 | 34608 | 2692 | | G9V | 21.82 | 0.40 | 6.43 | 45.8 | 0.00 | 3.12 | |
| HD 55054 | 55054 | 34767 | | | F7V | 12.92 | 0.80 | 8.06 | 77.4 | 0.00 | 3.62 | |
| HD 55693 | 55693 | 34879 | | | G1.5V | 27.43 | 0.54 | 7.16 | 36.5 | 0.00 | 4.35 | |
| HD 55575 | 55575 | 35136 | 2721 | | F9V | 59.20 | 0.33 | 5.58 | 16.9 | 0.00 | 4.44 | |
| HD 56303 | 56303 | 35209 | | | G2V | 24.19 | 0.63 | 7.33 | 41.3 | 0.00 | 4.25 | |
| HD 56515 | 56515 | 35310 | | J07176+0918AB | G0 | 24.03 | 1.04 | 6.77 | 41.6 | 0.00 | 3.67 | |
| HD 57006 | 57006 | 35509 | 2779 | | F8IV | 18.76 | 0.35 | 5.92 | 53.3 | 0.00 | 2.28 | |
| HD 57707 | 57707 | 35751 | | | K0/1III | 14.01 | 0.53 | 6.43 | 71.4 | 0.00 | 2.16 | |
| HD 58781 | 58781 | 36249 | | | G5 | 35.37 | 0.71 | 7.24 | 28.3 | 0.00 | 4.98 | |
| HD 59090 | 59090 | 36339 | | | F8 | 8.75 | 0.72 | 7.94 | 114.3 | 0.00 | 2.64 | |
| HR 2866 | 59380 | 36399 | 2866 | | F6V | 36.71 | 2.40 | 5.87 | 27.2 | 0.00 | 3.70 | |
| 22 Lyn | 58855 | 36439 | 2849 | J07299+4941A | F6V | 49.41 | 0.36 | 5.37 | 20.2 | 0.00 | 3.83 | |
| HD 59984 | 59984 | 36640 | 2883 | J07321-0853A | G0VFe-1.6CH-0.5 | 35.82 | 0.54 | 5.93 | 27.9 | 0.00 | 3.70 | |
| HD 59747 | 59747 | 36704 | | | G5V | 50.60 | 0.94 | 7.70 | 19.8 | 0.00 | 6.22 | |
| BD-02 2198 | | | 36985 | | | 70.55 | 1.64 | 9.87 | 14.2 | 0.00 | 9.11 | |
| alf CMi | 61421 | 37279 | 2943 | J07393+0514AB | F5IV-V+DQZ | 284.56 | 1.26 | 0.37 | 3.5 | 0.00 | 2.64 | |

| Name | HD | HIP | HR | J | SpType | π | B-V | V | vsini | E(B-V) | M | Note |
|---|---|---|---|---|---|---|---|---|---|---|---|---|
| HD 61632 | 61632 | 37327 | | | G3V | 13.61 | 0.98 | 8.33 | 73.5 | 0.00 | 4.00 | |
| HD 61606 | 61606 | 37349 | | J07400-0336A | K0/2V | 70.37 | 0.64 | 7.17 | 14.2 | 0.00 | 6.41 | |
| HD 62161 | 62161 | 37584 | | | F3V | 13.47 | 0.92 | 7.95 | 74.2 | 0.00 | 3.60 | |
| HD 62323 | 62323 | 37645 | | J07435+0330A | F8V | 19.63 | 0.70 | 7.02 | 50.9 | 0.00 | 3.49 | |
| HD 63598 | 63598 | 38134 | | | G2V | 19.92 | 0.84 | 7.93 | 50.2 | 0.00 | 4.43 | |
| HD 63754 | 63754 | 38216 | 3048 | J07498-2012A | G0V | 19.76 | 0.45 | 6.55 | 50.6 | 0.00 | 3.03 | |
| HD 63433 | 63433 | 38228 | | | G5IV | 45.45 | 0.53 | 6.91 | 22.0 | 0.00 | 5.20 | |
| HD 64021 | 64021 | 38397 | | | F5 | 9.08 | 0.74 | 7.42 | 110.1 | 0.00 | 2.21 | |
| HD 64090 | 64090 | 38541 | | J07536+3037A | K0:V_Fe-3 | 34.30 | 0.90 | 8.25 | 29.2 | 0.00 | 5.93 | |
| HD 64606 | 64606 | 38625 | | | K0V | 49.78 | 1.85 | 7.43 | 20.1 | 0.00 | 5.92 | |
| HD 64468 | 64468 | 38657 | | J07549+1914A | K2.5V | 48.33 | 0.86 | 7.73 | 20.7 | 0.00 | 6.15 | |
| HD 64815 | 64815 | 38750 | | | F8 | 12.97 | 0.91 | 7.86 | 77.1 | 0.00 | 3.42 | |
| HD 62613 | 62613 | 38784 | 2997 | | G8V | 58.17 | 0.36 | 6.56 | 17.2 | 0.00 | 5.38 | |
| HD 65583 | 65583 | 39157 | | | K0V_Fe-1.3 | 59.64 | 0.56 | 7.00 | 16.8 | 0.00 | 5.88 | |
| HD 65874 | 65874 | 39212 | | | G0 | 12.92 | 0.73 | 7.51 | 77.4 | 0.00 | 3.07 | |
| HR 3144 | 66011 | 39271 | 3144 | | G0IV | 15.68 | 0.44 | 6.24 | 63.8 | 0.00 | 2.21 | |
| mu.02 Cnc | 67228 | 39780 | 3176 | | G2IV | 42.94 | 0.30 | 5.30 | 23.3 | 0.00 | 3.46 | |
| psi Cnc | 67767 | 40023 | 3191 | J08104+2530A | G7V | 24.18 | 0.35 | 5.73 | 41.4 | 0.00 | 2.65 | |
| HR 3193 | 67827 | 40093 | 3193 | | G0 | 20.83 | 0.41 | 6.61 | 48.0 | 0.00 | 3.21 | |
| HD 68017 | 68017 | 40118 | | J08117+3228A | G3V | 45.90 | 0.55 | 6.81 | 21.8 | 0.00 | 5.12 | |
| HD 68168 | 68168 | 40133 | | | G0 | 30.41 | 0.65 | 7.23 | 32.9 | 0.00 | 4.65 | |
| HD 68284 | 68284 | 40136 | | | F7/8V | 13.14 | 0.88 | 7.75 | 76.1 | 0.00 | 3.34 | |
| HD 68380 | 68380 | 40169 | | | F5V | 14.54 | 1.15 | 7.92 | 68.8 | 0.00 | 3.73 | |
| HD 68638 | 68638 | 40497 | | J08162+5705A | G8V | 31.01 | 0.74 | 7.50 | 32.2 | 0.00 | 4.96 | |
| HD 69611 | 69611 | 40613 | | | G0V | 20.50 | 0.95 | 7.74 | 48.8 | 0.00 | 4.30 | |
| HD 68988 | 68988 | 40687 | | | G0 | 18.34 | 0.78 | 8.19 | 54.5 | 0.00 | 4.51 | H |
| HD 69830 | 69830 | 40693 | 3259 | | G8+V | 80.04 | 0.35 | 5.95 | 12.5 | 0.00 | 5.47 | H |
| BD+01 2063 | | 40774 | | | G5V | 43.61 | 1.26 | 8.35 | 22.9 | 0.00 | 6.54 | |
| chi Cnc | 69897 | 40843 | 3262 | | F6V | 54.73 | 0.32 | 5.10 | 18.3 | 0.00 | 3.79 | |
| HR 3271 | 70110 | 40858 | 3271 | | G0V | 25.32 | 0.41 | 6.18 | 39.5 | 0.00 | 3.20 | |
| HD 70088 | 70088 | 40942 | | | G5V | 23.89 | 1.03 | 8.51 | 41.9 | 0.00 | 5.40 | |
| HD 70889 | 70889 | 41134 | | | F9.5V | 30.16 | 0.50 | 7.09 | 33.2 | 0.00 | 4.49 | |
| HD 70516 | 70516 | 41184 | | J08243+4457A | G0V | 26.49 | 1.18 | 7.70 | 37.8 | 0.00 | 4.82 | |
| HD 70923 | 70923 | 41209 | | J08246-0108AB | G1V | 23.66 | 0.52 | 7.05 | 42.3 | 0.00 | 3.92 | |
| 1 Hya | 70958 | 41211 | 3297 | J08246-0345AB | F6V | 37.57 | 0.33 | 5.61 | 26.6 | 0.00 | 3.48 | |
| HD 71431 | 71431 | 41471 | | | G1V | 13.67 | 1.42 | 7.57 | 73.2 | 0.00 | 3.25 | |
| HD 71148 | 71148 | 41484 | 3309 | | G1V | 44.94 | 0.46 | 6.30 | 22.3 | 0.00 | 4.56 | |
| HD 71595 | 71595 | 41569 | | | F5 | 13.49 | 0.81 | 7.15 | 74.1 | 0.00 | 2.80 | |

| Name | HD | HIP | HR | J-ID | SpType | π | σπ | V | d | A_V | M_V | H |
|---|---|---|---|---|---|---|---|---|---|---|---|---|
| HD 71640 | 71640 | 41573 | | | F6V | 22.35 | 0.91 | 7.40 | 44.7 | 0.00 | 4.15 | |
| HD 71881 | 71881 | 41844 | | | G1V | 24.22 | 0.67 | 7.43 | 41.3 | 0.00 | 4.35 | |
| HD 72659 | 72659 | 42030 | | | G2V | 20.07 | 0.75 | 7.47 | 49.8 | 0.00 | 3.98 | H |
| HD 72760 | 72760 | 42074 | | | K0V | 47.31 | 0.72 | 7.33 | 21.1 | 0.00 | 5.71 | |
| HR 3395 | 72945 | 42172 | 3395 | J08358+0637A | F8V | 39.80 | 0.94 | 5.92 | 25.1 | 0.00 | 3.92 | |
| HD 72946 | 72946 | 42173 | 3396 | J08358+0637B | G8-V | 38.11 | 0.85 | 7.08 | 26.2 | 0.00 | 4.99 | |
| V* V401 Hya | 73350 | 42333 | | J08378-0649A | G8/K0(IV) | 41.71 | 0.70 | 6.73 | 24.0 | 0.00 | 4.83 | |
| HD 73752 | 73752 | 42430 | 3430 | J08391-2240AB | G3V+K0V | 51.55 | 0.63 | 5.05 | 19.4 | 0.00 | 3.61 | |
| V* pi.01 UMa | 72905 | 42438 | 3391 | | G1.5Vb | 69.66 | 0.37 | 5.64 | 14.4 | 0.00 | 4.85 | |
| HD 73668 | 73668 | 42488 | | J08397+0546A | G1V | 27.54 | 0.95 | 7.25 | 36.3 | 0.00 | 4.45 | |
| HD 73596 | 73596 | 42538 | 3423 | | F5III | 8.07 | 0.50 | 6.22 | 123.9 | 0.00 | 0.75 | |
| HD 73393 | 73393 | 42575 | | | G3V | 25.05 | 0.90 | 8.01 | 39.9 | 0.00 | 5.00 | |
| HD 74156 | 74156 | 42723 | | | G1V | 15.52 | 0.54 | 7.61 | 64.4 | 0.00 | 3.56 | H |
| HD 74011 | 74011 | 42734 | | | F8 | 21.49 | 0.66 | 7.41 | 46.5 | 0.00 | 4.07 | |
| V* KX Cnc | 74057 | 42753 | | | F8 | 19.90 | 0.56 | 7.19 | 50.3 | 0.00 | 3.68 | |
| HD 75318 | 75318 | 43304 | | J08494+0341AB | G8V | 28.22 | 0.91 | 7.95 | 35.4 | 0.00 | 5.20 | |
| 54 Cnc | 75528 | 43454 | 3510 | | G1V | 26.10 | 0.51 | 6.38 | 38.3 | 0.00 | 3.46 | |
| HD 75488 | 75488 | 43543 | | J08520+4735A | G2V | 15.28 | 0.78 | 8.18 | 65.4 | 0.00 | 4.10 | |
| HD 75767 | 75767 | 43557 | | | G1.5V | 41.64 | 1.03 | 6.59 | 24.0 | 0.00 | 4.69 | |
| rho01 Cnc | 75732 | 43587 | 3522 | | G8V | 81.03 | 0.75 | 5.95 | 12.3 | 0.00 | 5.49 | H |
| HD 75935 | 75935 | 43670 | | | G8V | 24.14 | 1.21 | 8.45 | 41.4 | 0.00 | 5.36 | |
| HD 76151 | 76151 | 43726 | 3538 | | G3V | 57.52 | 0.39 | 6.00 | 17.4 | 0.00 | 4.80 | |
| HD 76932 | 76932 | 44075 | 3578 | | G2VFe-1.8CH-1 | 47.54 | 0.31 | 5.86 | 21.0 | 0.00 | 4.25 | |
| HD 76752 | 76752 | 44089 | | | G2V | 25.95 | 0.75 | 7.45 | 38.5 | 0.00 | 4.52 | |
| HD 76909 | 76909 | 44137 | | | G5 | 19.19 | 0.91 | 7.84 | 52.1 | 0.00 | 4.26 | |
| HD 77407 | 77407 | 44458 | | | G0V+M3V-M6V | 32.76 | 0.63 | 7.04 | 30.5 | 0.00 | 4.62 | |
| HD 78366 | 78366 | 44897 | 3625 | | G0IV-V | 52.11 | 0.33 | 5.90 | 19.2 | 0.00 | 4.48 | |
| c UMa | 79028 | 45333 | 3648 | J09143+6125A | G0V | 51.10 | 0.32 | 5.20 | 19.6 | 0.00 | 3.74 | |
| HD 79210 | 79210 | 45343 | | J09144+5241A | M0V | 172.08 | 6.31 | 7.63 | 5.8 | 0.00 | 8.81 | |
| HD 80372 | 80372 | 45685 | | | G1V | 14.83 | 0.85 | 8.45 | 67.4 | 0.00 | 4.31 | |
| HD 80606 | 80606 | 45982 | | J09226+5036B | G5 | 5.63 | 6.65 | 9.00 | 177.6 | 0.00 | 2.75 | H |
| HD 80607 | 80607 | 45983 | | J09226+5036A | G5 | 4.27 | 8.38 | 9.07 | 234.2 | 0.00 | 2.22 | |
| HD 81040 | 81040 | 46076 | | | G0V | 30.20 | 1.03 | 7.73 | 33.1 | 0.00 | 5.13 | H |
| HD 81809 | 81809 | 46404 | 3750 | J09278-0604AB | G5V_Fe-1_CH-0.8 | 32.88 | 0.77 | 5.40 | 30.4 | 0.00 | 2.98 | |
| ome Leo | 81858 | 46454 | 3754 | J09285+0903AB | G1V | 30.15 | 1.45 | 5.41 | 33.2 | 0.00 | 2.81 | |
| tau01 Hya | 81997 | 46509 | 3759 | J09291-0246A | F5V | 57.69 | 2.14 | 4.60 | 17.3 | 0.00 | 3.41 | |
| HR 3762 | 82074 | 46543 | 3762 | | G8III/IV | 17.97 | 0.50 | 6.26 | 55.6 | 0.00 | 2.53 | |
| HD 82106 | 82106 | 46580 | | | K3V | 77.48 | 0.64 | 7.20 | 12.9 | 0.00 | 6.64 | |

| Name | HD | HIP | HR | WDS | SpType | π | σπ | V | d | A_V | M_V | Note |
|---|---|---|---|---|---|---|---|---|---|---|---|---|
| HD 82443 | 82443 | 46843 | | | G9V(k) | 56.20 | 0.60 | 7.01 | 17.8 | 0.00 | 5.76 | |
| tet UMa | 82328 | 46853 | 3775 | J09329+5141AB | F7V | 74.19 | 0.14 | 3.18 | 13.5 | 0.00 | 2.53 | |
| HD 82943 | 82943 | 47007 | | | F9VFe+0.5 | 36.40 | 0.47 | 6.53 | 27.5 | 0.00 | 4.34 | H |
| 11 LMi | 82885 | 47080 | 3815 | J09357+3549AB | G8+V | 87.96 | 0.32 | 5.34 | 11.4 | 0.00 | 5.06 | |
| HD 84117 | 84117 | 47592 | 3862 | | F8V | 66.61 | 0.21 | 4.94 | 15.0 | 0.00 | 4.06 | |
| HD 84703 | 84703 | 48000 | | | F8/G0V | 13.65 | 0.72 | 7.89 | 73.3 | 0.00 | 3.57 | |
| 15 LMi | 84737 | 48113 | 3881 | | G0IV-V | 54.44 | 0.28 | 5.10 | 18.4 | 0.00 | 3.78 | |
| HD 84937 | 84937 | 48152 | | | F8Vm-5 | 13.74 | 0.78 | 8.32 | 72.8 | 0.00 | 4.01 | |
| HR 3901 | 85380 | 48351 | 3901 | | G1V | 23.94 | 0.47 | 6.42 | 41.8 | 0.00 | 3.32 | |
| HD 85725 | 85725 | 48468 | 3916 | J09530-2720AB | G1V | 18.57 | 0.45 | 6.30 | 53.9 | 0.00 | 2.64 | |
| 20 LMi | 86728 | 49081 | 3951 | J10010+3155A | G3Va | 66.46 | 0.32 | 5.40 | 15.0 | 0.00 | 4.51 | |
| HD 87097 | 87097 | 49196 | | | G3V | 16.14 | 1.16 | 8.43 | 62.0 | 0.00 | 4.47 | |
| HD 87883 | 87883 | 49699 | | | K0V | 54.93 | 0.54 | 7.55 | 18.2 | 0.00 | 6.25 | H |
| HD 88133 | 88133 | 49813 | | | G5 | 12.28 | 0.88 | 8.06 | 81.4 | 0.00 | 3.50 | H |
| HD 88230 | 88230 | 49908 | | J10114+4927A | K8V | 205.21 | 0.54 | 6.61 | 4.9 | 0.00 | 8.17 | |
| HD 88371 | 88371 | 49942 | | | G2V | 16.86 | 0.77 | 8.43 | 59.3 | 0.00 | 4.56 | |
| HD 88595 | 88595 | 50013 | 4005 | | F7V | 24.20 | 0.44 | 6.46 | 41.3 | 0.00 | 3.38 | |
| HD 88725 | 88725 | 50139 | | | G3/5V | 28.24 | 0.72 | 7.73 | 35.4 | 0.00 | 4.98 | |
| HD 88737 | 88737 | 50174 | 4012 | | F9V | 17.60 | 0.44 | 6.03 | 56.8 | 0.00 | 2.26 | |
| 24 LMi | 88986 | 50316 | 4027 | | G2V | 30.13 | 0.44 | 6.47 | 33.2 | 0.00 | 3.86 | |
| 35 Leo | 89010 | 50319 | 4030 | J10166+2327B | G1.5IV-V | 32.08 | 0.80 | 5.97 | 31.2 | 0.00 | 3.50 | |
| 39 Leo | 89125 | 50384 | 4039 | J10172+2306AB | F6V+M1 | 43.85 | 0.36 | 5.82 | 22.8 | 0.00 | 4.03 | |
| HD 89269 | 89269 | 50505 | | J10189+4403A | G4V | 49.41 | 0.50 | 6.65 | 20.2 | 0.00 | 5.12 | |
| HD 89319 | 89319 | 50546 | 4046 | | K0 | 22.52 | 0.37 | 6.01 | 44.4 | 0.00 | 2.77 | |
| 40 Leo | 89449 | 50564 | 4054 | | F6IV-V | 46.80 | 0.24 | 4.80 | 21.4 | 0.00 | 3.15 | |
| HR 4051 | 89389 | 50606 | 4051 | | F8V | 32.11 | 0.48 | 6.46 | 31.1 | 0.00 | 4.00 | |
| HD 89744 | 89744 | 50786 | 4067 | | F7V | 25.36 | 0.31 | 5.74 | 39.4 | 0.00 | 2.76 | H |
| HD 90508 | 90508 | 51248 | 4098 | J10281+4847AB | G0V | 43.65 | 0.43 | 6.43 | 22.9 | 0.00 | 4.63 | |
| 36 UMa | 90839 | 51459 | 4112 | J10306+5559A | F8V | 78.25 | 0.28 | 4.83 | 12.8 | 0.00 | 4.30 | |
| HD 90875 | 90875 | 51468 | | | K4/5V | 42.65 | 1.16 | 8.74 | 23.4 | 0.00 | 6.89 | |
| BD+46 1635 | | | 51525 | | K7V | 63.55 | 1.09 | 9.14 | 15.7 | 0.00 | 8.16 | |
| 38 LMi | | 92168 | 52139 | 4168 | G0IV | 19.11 | 0.37 | 5.85 | 52.3 | 0.00 | 2.26 | |
| 33 Sex | | 92588 | 52316 | 4182 | G9IV | 27.16 | 0.45 | 6.26 | 36.8 | 0.00 | 3.43 | |
| HD 92788 | 92788 | 52409 | | | G6V | 28.20 | 0.73 | 7.30 | 35.5 | 0.00 | 4.55 | H |
| HD 93215 | 93215 | 52667 | | | G5V | 20.75 | 0.76 | 8.05 | 48.2 | 0.00 | 4.64 | |
| HD 94132 | 94132 | 53257 | 4243 | J10536+6951A | G9IV | 22.58 | 0.40 | 5.93 | 44.3 | 0.00 | 2.70 | |
| HD 94765 | 94765 | 53486 | | | K0V | 57.79 | 0.87 | 7.38 | 17.3 | 0.00 | 6.19 | |
| HD 94915 | 94915 | 53549 | | | G1V | 14.83 | 1.07 | 8.70 | 67.4 | 0.00 | 4.56 | |

| Name | HD | HIP | HR | Karmn | SpType | π | σπ | V | d | E(B-V) | MV | H |
|---|---|---|---|---|---|---|---|---|---|---|---|---|
| 47 UMa | 95128 | 53721 | 4277 | | G1-V_Fe-0.5 | 71.11 | 0.25 | 5.04 | 14.1 | 0.00 | 4.30 | H |
| HR 4285 | 95241 | 53791 | 4285 | J11003+4255A | F9V | 22.55 | 0.38 | 6.04 | 44.3 | 0.00 | 2.80 | |
| V* DS Leo | 95650 | 53985 | | | M1.0V | 84.95 | 1.05 | 9.57 | 11.8 | 0.00 | 9.22 | |
| HD 96276 | 96276 | 54242 | | | G1V | 17.13 | 0.80 | 8.11 | 58.4 | 0.00 | 4.28 | |
| HD 97101 | 97101 | 54646 | | J11110+3028A | K7V | 84.23 | 0.86 | 8.86 | 11.9 | 0.00 | 8.49 | |
| HD 97100 | 97100 | | | J11110+3028C | G0 | | | 9.01 | | 0.00 | 8.64 | |
| HD 97334 | 97334 | 54745 | 4345 | J11126+3549A | G0V | 45.61 | 0.44 | 6.41 | 21.9 | 0.00 | 4.71 | |
| HD 97503 | 97503 | 54810 | | | K5V | 54.70 | 1.23 | 8.64 | 18.3 | 0.00 | 7.33 | |
| HD 97658 | 97658 | 54906 | | | K1V | 47.36 | 0.75 | 7.71 | 21.1 | 0.00 | 6.09 | H |
| HD 97584 | 97584 | 54952 | | | K4V+M2 | 68.28 | 0.65 | 7.63 | 14.6 | 0.00 | 6.80 | |
| HD 98388 | 98388 | 55262 | | | F8V | 22.71 | 0.65 | 7.14 | 44.0 | 0.00 | 3.92 | |
| HD 98630 | 98630 | 55427 | | | G0 | 8.73 | 1.00 | 8.08 | 114.5 | 0.00 | 2.78 | |
| HD 98712 | 98712 | 55454 | | | M0Ve+M4Ve | 75.97 | 1.25 | 8.57 | 13.2 | 0.00 | 7.97 | |
| HD 98618 | 98618 | 55459 | | | G5V | 24.96 | 0.66 | 7.66 | 40.1 | 0.00 | 4.65 | |
| HD 98800 | 98800 | 55505 | | J11221-2447AB | K5V(e) | 22.27 | 2.31 | 8.91 | 44.9 | 0.00 | 5.65 | |
| HD 98823 | 98823 | 55549 | | | F5 | 10.05 | 0.42 | 6.58 | 99.5 | 0.00 | 1.59 | |
| 83 Leo | 99491 | 55846 | 4414 | J11268+0300A | G9IV-V | 56.35 | 0.75 | 6.50 | 17.7 | 0.00 | 5.25 | |
| 83 Leo B | 99492 | 55848 | | J11268+0300B | K2/4 | 55.69 | 1.46 | 7.53 | 18.0 | 0.00 | 6.26 | H |
| HD 99747 | 99747 | 56035 | 4421 | | F5VkF0mA9 | 30.03 | 0.30 | 5.80 | 33.3 | 0.00 | 3.19 | |
| 88 Leo | 100180 | 56242 | 4437 | J11317+1422A | F9.5V | 42.87 | 1.22 | 6.20 | 23.3 | 0.00 | 4.36 | |
| 17 Crt A | 100286J | 56280 | 4444 | J11323-2916A | F8V | 38.03 | 1.39 | 5.58 | 26.3 | 0.00 | 3.48 | |
| 17 Crt B | 100286 | | 4443 | J11323-2916B | F8V | | | 5.67 | | 0.00 | 3.57 | |
| 89 Leo | 100563 | 56445 | 4455 | | F5.5V | 36.73 | 0.36 | 5.70 | 27.2 | 0.00 | 3.53 | |
| HD 101177 | 101177 | 56809 | 4486 | | G0V+K2V | 43.01 | 0.73 | 6.44 | 23.3 | 0.00 | 4.61 | |
| V* MV UMa | 101206 | 56829 | | | K5V | 50.19 | 1.03 | 8.27 | 19.9 | 0.00 | 6.77 | |
| 61 UMa | 101501 | 56997 | 4496 | J11411+3412A | G8V | 104.04 | 0.26 | 5.34 | 9.6 | 0.00 | 5.43 | |
| HD 101563 | 101563 | 57001 | 4498 | | G2III/IV | 23.12 | 0.54 | 6.44 | 43.3 | 0.00 | 3.26 | |
| HD 101959 | 101959 | 57217 | | | F9V | 30.68 | 0.54 | 6.98 | 32.6 | 0.00 | 4.41 | |
| HD 102158 | 102158 | 57349 | | | G2V | 20.29 | 0.70 | 8.03 | 49.3 | 0.00 | 4.57 | |
| bet Vir | 102870 | 57757 | 4540 | J11507+0146A | F9V | 91.50 | 0.22 | 3.60 | 10.9 | 0.00 | 3.41 | |
| HD 103095 | 103095 | 57939 | 4550 | | K1V_Fe-1.5 | 109.99 | 0.41 | 6.45 | 9.1 | 0.00 | 6.66 | |
| HD 103932 | 103932 | 58345 | | | K4+V | 98.45 | 0.57 | 6.96 | 10.2 | 0.00 | 6.93 | |
| HD 104304 | 104304 | 58576 | 4587 | | G8IV | 78.35 | 0.31 | 5.55 | 12.8 | 0.00 | 5.02 | |
| alf Crv | 105452 | 59199 | 4623 | | F1V | 66.95 | 0.15 | 4.00 | 14.9 | 0.00 | 3.13 | |
| HD 105631 | 105631 | 59280 | | | G7V | 40.77 | 0.66 | 7.50 | 24.5 | 0.00 | 5.55 | |
| HD 238087 | 238087 | 59496 | | | K5V | 35.70 | 1.24 | 10.02 | 28.0 | 0.00 | 7.78 | |
| V* GM Com | 106103 | 59527 | | | F5V | 11.76 | 0.73 | 8.10 | 85.0 | 0.00 | 3.45 | |
| HD 106116 | 106116 | 59532 | | J12125-0305AB | G5V | 28.69 | 0.46 | 7.50 | 34.9 | 0.00 | 4.79 | |

| Name | HD | HIP | HR | WDS | Spectral Type | π (mas) | B-V | V | d (pc) | E(B-V) | M_V | Note |
|---|---|---|---|---|---|---|---|---|---|---|---|---|
| HD 106252 | 106252 | 59610 | | | G0 | 26.52 | 0.57 | 7.36 | 37.7 | 0.00 | 4.48 | H |
| HD 106516 | 106516 | 59750 | 4657 | J12152-1019A | F9VFe-1.7CH-0.7 | 44.74 | 0.81 | 6.11 | 22.4 | 0.00 | 4.36 | |
| HD 106640 | 106640 | 59806 | | | G0V | 11.12 | 1.15 | 9.20 | 89.9 | 0.00 | 4.42 | |
| HD 106691 | 106691 | 59833 | | | F5IV | 9.77 | 0.83 | 8.08 | 102.4 | 0.00 | 3.03 | |
| 9 Com | 107213 | 60098 | 4688 | | F8Va | 18.36 | 0.43 | 6.70 | 54.5 | 0.00 | 3.02 | |
| HD 107611 | 107611 | 60304 | | | F6V | 10.09 | 0.95 | 8.50 | 99.1 | 0.00 | 3.52 | |
| 17 Vir | 107705 | 60353 | 4708 | J12225+0518A | F8V | 33.42 | 0.64 | 6.46 | 29.9 | 0.00 | 4.08 | |
| HD 108799 | 108799 | 60994 | 4758 | J12301-1324AB | G1/2V | 40.57 | 0.60 | 6.35 | 24.6 | 0.00 | 4.39 | |
| HD 108874 | 108874 | 61028 | | | G9V | 15.97 | 1.07 | 8.76 | 62.6 | 0.00 | 4.78 | H |
| HD 108954 | 108954 | 61053 | 4767 | | F9V | 45.92 | 0.35 | 6.22 | 21.8 | 0.00 | 4.53 | |
| HD 109057 | 109057 | 61153 | | | G1V | 15.67 | 0.85 | 8.24 | 63.8 | 0.00 | 4.22 | |
| bet CVn | 109358 | 61317 | 4785 | J12337+4121AB | G0V | 118.49 | 0.20 | 4.25 | 8.4 | 0.00 | 4.62 | |
| HD 110315 | 110315 | 61901 | | | K4.5V | 70.53 | 0.73 | 7.90 | 14.2 | 0.00 | 7.14 | |
| gam Vir A | 110379 | 61941 | 4825 | J12417-0127A | F1V | 84.53 | 1.18 | 3.44 | 11.8 | 0.00 | 3.08 | |
| HD 110463 | 110463 | 61946 | | | K3V | 42.78 | 0.81 | 8.28 | 23.4 | 0.00 | 6.44 | |
| HD 110745 | 110745 | 62129 | | | F8 | 10.42 | 1.16 | 8.63 | 96.0 | 0.00 | 3.72 | |
| HD 110833 | 110833 | 62145 | | | K3V | 67.20 | 0.66 | 7.04 | 14.9 | 0.00 | 6.18 | |
| HD 110869 | 110869 | 62175 | | J12445+5842A | G5V | 21.62 | 0.61 | 8.00 | 46.3 | 0.00 | 4.67 | |
| 10 CVn | 110897 | 62207 | 4845 | | F9V_Fe-0.3 | 57.55 | 0.32 | 5.95 | 17.4 | 0.00 | 4.75 | |
| HD 111069 | 111069 | 62350 | | | G5 | 15.29 | 0.96 | 8.67 | 65.4 | 0.00 | 4.59 | |
| HR 4867 | 111456 | 62512 | 4867 | | F6V | 41.59 | 2.69 | 5.85 | 24.0 | 0.00 | 3.94 | |
| HR 4864 | 111395 | 62523 | 4864 | | G5V | 59.06 | 0.45 | 6.29 | 16.9 | 0.00 | 5.15 | |
| HD 111513 | 111513 | 62527 | | | G1V | 26.25 | 0.47 | 7.35 | 38.1 | 0.00 | 4.45 | |
| HD 111540 | 111540 | 62606 | | | G1V | 11.03 | 1.51 | 9.54 | 90.7 | 0.00 | 4.75 | |
| HD 111799 | 111799 | 62800 | | | G1V | 17.80 | 1.12 | 8.33 | 56.2 | 0.00 | 4.58 | |
| HD 112068 | 112068 | 62839 | | | G0 | 14.12 | 0.79 | 8.67 | 70.8 | 0.00 | 4.42 | |
| HD 112257 | 112257 | 63048 | | | G6V | 23.39 | 0.87 | 7.79 | 42.8 | 0.00 | 4.64 | |
| HD 112758 | 112758 | 63366 | | J12590-0950AB | G9V | 47.87 | 0.90 | 7.59 | 20.9 | 0.00 | 5.99 | |
| HD 113470 | 113470 | 63778 | | | G2V | 14.11 | 1.00 | 8.54 | 70.9 | 0.00 | 4.29 | |
| HD 113713 | 113713 | 63878 | | J13055+1444A | F5V | 10.99 | 0.63 | 7.94 | 91.0 | 0.00 | 3.14 | |
| alf Com A | 114378 | 64241 | 4968 | J13100+1732A | F5 | 51.00 | 4.00 | 4.85 | 19.6 | 0.00 | 3.39 | |
| bet Com | 114710 | 64394 | 4983 | J13118+2753A | F9.5V | 109.54 | 0.17 | 4.25 | 9.1 | 0.00 | 4.45 | |
| HD 114762 | 114762 | 64426 | | | F9VgF8mF4+M6?V | 25.87 | 0.76 | 7.29 | 38.7 | 0.00 | 4.35 | H |
| HD 114783 | 114783 | 64457 | | | K1V | 48.78 | 0.59 | 7.55 | 20.5 | 0.00 | 5.99 | H |
| 55 Vir | 114946 | 64577 | 4995 | | G9IV | 25.66 | 0.28 | 5.33 | 39.0 | 0.00 | 2.38 | |
| HD 115274 | 115274 | 64733 | | | F8V | 8.00 | 0.71 | 8.19 | 125.0 | 0.00 | 2.70 | |
| e Vir | 115383 | 64792 | 5011 | J13168+0925A | G0V | 56.95 | 0.26 | 5.22 | 17.6 | 0.00 | 4.00 | H |
| HD 115404A | 115404A | 64797 | | J13169+1701A | K2V | 90.32 | 0.74 | 6.66 | 11.1 | 0.00 | 6.44 | |

| Name | HD | HIP | HR | CNS | Sp | π | σπ | V | d | E(B-V) | M_V | H |
|---|---|---|---|---|---|---|---|---|---|---|---|---|
| HD 115404B | 115404B | | | J13169+1701B | M1 | | | 9.50 | | 0.00 | 9.28 | |
| 61 Vir | 115617 | 64924 | 5019 | J13185-1818A | G7V | 116.89 | 0.22 | 4.74 | 8.6 | 0.00 | 5.08 | H |
| HD 115953 | 115953 | 65026 | | J13198+4747AB | M2V | 93.40 | 2.21 | 8.54 | 10.7 | 0.00 | 8.39 | |
| HD 116442 | 116442 | 65352 | | J13237+0243A | G9V | 64.73 | 1.33 | 7.06 | 15.4 | 0.00 | 6.12 | |
| HD 116956 | 116956 | 65515 | | | G9V | 46.31 | 0.51 | 7.28 | 21.6 | 0.00 | 5.61 | |
| HD 117043 | 117043 | 65530 | 5070 | | G6V | 47.24 | 0.31 | 6.49 | 21.2 | 0.00 | 4.86 | |
| 70 Vir | 117176 | 65721 | 5072 | J13285+1346A | G4Va | 55.60 | 0.24 | 4.97 | 18.0 | 0.00 | 3.70 | H |
| HD 117845 | 117845 | 65971 | | | G2V | 22.24 | 0.62 | 8.09 | 45.0 | 0.00 | 4.83 | |
| HD 117635 | 117635 | 65982 | | | G8V | 32.57 | 2.01 | 7.36 | 30.7 | 0.00 | 4.92 | |
| HD 118203 | 118203 | 66192 | | | K0 | 11.29 | 0.70 | 8.06 | 88.6 | 0.00 | 3.32 | H |
| HD 118096 | 118096 | 66193 | | | K4V | 44.72 | 1.02 | 9.22 | 22.4 | 0.00 | 7.47 | |
| HD 119332 | 119332 | 66781 | | | K0V | 40.59 | 0.53 | 7.78 | 24.6 | 0.00 | 5.82 | |
| BD+18 2776 | | 67090 | | J13454+1746A | M1V | 75.53 | 1.00 | 9.75 | 13.2 | 0.00 | 9.14 | |
| HD 119802 | 119802 | 67105 | | | K3V | 47.65 | 0.98 | 8.46 | 21.0 | 0.00 | 6.85 | |
| HD 120066 | 120066 | 67246 | 5183 | | G0V | 31.58 | 0.44 | 6.30 | 31.7 | 0.00 | 3.80 | |
| tau Boo | 120136 | 67275 | 5185 | J13473+1727AB | F6IV+M2 | 64.03 | 0.19 | 4.49 | 15.6 | 0.00 | 3.52 | H |
| GJ 528 A | 120476A | 67422 | | J13491+2659A | K4V | 74.58 | 0.78 | 7.23 | 13.4 | 0.00 | 6.60 | |
| GJ 528 B | 120476B | | | J13491+2659B | K6 | | | 8.01 | | 0.00 | 7.37 | |
| HD 120467 | 120467 | 67487 | | | K5.5Vk: | 69.93 | 0.87 | 8.17 | 14.3 | 0.00 | 7.39 | |
| HD 120690 | 120690 | 67620 | 5209 | | G5+V | 51.35 | 0.45 | 6.43 | 19.5 | 0.00 | 4.98 | |
| HD 120730 | 120730 | 67661 | | | G8V | 12.98 | 1.30 | 10.02 | 77.0 | 0.00 | 5.58 | |
| HD 234078 | 234078 | 67691 | | | K7V | 70.40 | 0.80 | 9.09 | 14.2 | 0.00 | 8.33 | |
| eta Boo | 121370 | 67927 | 5235 | J13547+1824A | G0IV | 87.75 | 1.24 | 2.68 | 11.4 | 0.00 | 2.40 | |
| HD 122064 | 122064 | 68184 | 5256 | | K3V | 99.36 | 0.32 | 6.52 | 10.1 | 0.00 | 6.51 | |
| HD 122120 | 122120 | 68337 | | | K5V | 40.32 | 0.96 | 9.04 | 24.8 | 0.00 | 7.07 | |
| HR 5258 | 122106 | 68380 | 5258 | J13598-0333AB | F6V | 12.90 | 0.80 | 6.41 | 77.5 | 0.00 | 1.96 | |
| HD 122303 | 122303 | 68469 | | | M0V | 99.72 | 1.57 | 9.71 | 10.0 | 0.00 | 9.70 | |
| HD 122967 | 122967 | 68623 | | J14028+6216AB | F3V | 8.57 | 0.92 | 8.13 | 116.7 | 0.01 | 2.78 | |
| HD 122742 | 122742 | 68682 | 5273 | | G6V | 58.88 | 0.62 | 6.27 | 17.0 | 0.00 | 5.12 | |
| HR 5307 | 124115 | 69340 | 5307 | | F6V | 22.50 | 0.44 | 6.42 | 44.4 | 0.00 | 3.18 | |
| HD 124292 | 124292 | 69414 | | | G8/K0V | 45.35 | 0.54 | 7.04 | 22.1 | 0.00 | 5.32 | |
| HR 5317 | 124425 | 69493 | 5317 | | F6V | 16.84 | 0.37 | 5.90 | 59.4 | 0.00 | 2.03 | |
| HD 124642 | 124642 | 69526 | | | K3.5V | 57.38 | 1.06 | 8.06 | 17.4 | 0.00 | 6.85 | |
| 14 Boo | 124570 | 69536 | 5323 | J14141+1258AB | F8V | 29.48 | 0.28 | 5.50 | 33.9 | 0.00 | 2.85 | |
| HD 124553 | 124553 | 69564 | 5322 | | F8V | 23.33 | 0.53 | 6.36 | 42.9 | 0.00 | 3.20 | |
| HR 5335 | 124755 | 69569 | 5335 | | K3III: | 21.85 | 0.33 | 6.23 | 45.8 | 0.00 | 2.93 | |
| iot Vir | 124850 | 69701 | 5338 | | F7III | 44.97 | 0.19 | 4.08 | 22.2 | 0.00 | 2.34 | |
| HD 125040 | 125040 | 69751 | 5346 | | F8V | 30.65 | 0.64 | 6.25 | 32.6 | 0.00 | 3.68 | |

| Name | HD | HIP | HR | WDS | SpType | π | σπ | Vmag | d | E(B-V) | MV | Flag |
|---|---|---|---|---|---|---|---|---|---|---|---|---|
| HD 125184 | 125184 | 69881 | 5353 | J14180-0732A | G5/6V | 30.73 | 0.47 | 6.50 | 32.5 | 0.00 | 3.94 | |
| BD+30 2512 | | 70218 | | | K6V | 69.70 | 0.82 | 8.61 | 14.3 | 0.00 | 7.83 | |
| HR 5387 | 126141 | 70310 | 5387 | | F5V | 27.37 | 0.34 | 6.23 | 36.5 | 0.00 | 3.41 | |
| HD 126053 | 126053 | 70319 | 5384 | | G1.5V | 58.17 | 0.53 | 6.27 | 17.2 | 0.00 | 5.09 | |
| tet Boo | 126660 | 70497 | 5404 | J14252+5151A | F7V | 68.82 | 0.14 | 4.05 | 14.5 | 0.00 | 3.24 | |
| phi Vir | 126868 | 70755 | 5409 | | G2IV+K0V | 27.58 | 1.01 | 4.84 | 36.3 | 0.00 | 2.05 | |
| HD 127334 | 127334 | 70873 | 5423 | | G5VCH0.3 | 42.12 | 0.38 | 6.36 | 23.7 | 0.00 | 4.48 | |
| HD 128165 | 128165 | 71181 | | | K3V | 75.65 | 0.42 | 7.23 | 13.2 | 0.00 | 6.62 | |
| sig Boo | 128167 | 71284 | 5447 | J14347+2945A | F4VkF2mF1 | 63.16 | 0.25 | 4.47 | 15.8 | 0.00 | 3.47 | |
| HD 128311 | 128311 | 71395 | | | K3V | 60.60 | 0.83 | 7.45 | 16.5 | 0.00 | 6.36 | H |
| HD 128429 | 128429 | 71469 | 5455 | | F6V | 31.08 | 0.83 | 6.20 | 32.2 | 0.00 | 3.66 | |
| HD 129132 | 129132 | 71729 | 5472 | J14404+2159A | G0V | 8.60 | 0.61 | 6.15 | 116.3 | 0.00 | 0.81 | |
| HD 129290 | 129290 | 71819 | | J14415+1337A | G2V | 14.27 | 0.97 | 8.40 | 70.1 | 0.00 | 4.17 | |
| HD 129829 | 129829 | 72134 | | | G1V | 17.43 | 0.91 | 8.22 | 57.4 | 0.00 | 4.43 | |
| HD 130087 | 130087 | 72190 | | | G2IV | 17.01 | 0.68 | 7.51 | 58.8 | 0.00 | 3.66 | |
| HR 5504 | 129980 | 72217 | 5504 | J14462-2111AB | G1V | 24.24 | 0.63 | 6.43 | 41.3 | 0.00 | 3.36 | |
| HD 130307 | 130307 | 72312 | | | K1V | 51.62 | 0.79 | 7.76 | 19.4 | 0.00 | 6.32 | |
| HD 130322 | 130322 | 72339 | | | K0V | 31.54 | 1.18 | 8.04 | 31.7 | 0.00 | 5.53 | H |
| HD 130948 | 130948 | 72567 | 5534 | | F9IV-V+L4+L4 | 55.03 | 0.34 | 5.88 | 18.2 | 0.00 | 4.58 | |
| HD 133002 | 133002 | 72573 | 5596 | | F9V | 23.10 | 0.23 | 5.64 | 43.3 | 0.00 | 2.46 | |
| ksi Boo A | 131156A | 72659 | 5544 | J14513+1906A | G7Ve | 149.26 | | 4.68 | 6.7 | 0.00 | 5.54 | |
| ksi Boo B | 131156B | | | J14513+1906B | K5Ve | 149.26 | | 6.82 | 6.7 | 0.00 | 7.69 | |
| CCDM J14534+1542AB | 131473 | 72846 | 5550 | J14534+1542AB | F9V | 20.99 | 0.93 | 6.38 | 47.6 | 0.00 | 2.99 | |
| V* DE Boo | 131511 | 72848 | 5553 | | K0V | 86.88 | 0.46 | 6.01 | 11.5 | 0.00 | 5.70 | |
| HD 132142 | 132142 | 73005 | | | K1V | 42.76 | 0.45 | 7.80 | 23.4 | 0.00 | 5.96 | |
| HD 132254 | 132254 | 73100 | 5581 | | F8-V | 39.83 | 0.26 | 5.60 | 25.1 | 0.00 | 3.60 | |
| HD 131976 | 131976 | 73182 | | J14574-2124B | M1.5V | 168.77 | 21.54 | 8.07 | 5.9 | 0.00 | 9.20 | |
| HD 131977 | 131977 | 73184 | 5568 | J14574-2124A | K4V | 171.22 | 0.94 | 5.72 | 5.8 | 0.00 | 6.89 | |
| HD 132375 | 132375 | 73309 | 5583 | J14589-0459AB | F7V | 29.61 | 0.47 | 6.09 | 33.8 | 0.00 | 3.45 | |
| HD 133161 | 133161 | 73593 | | | G2V | 27.21 | 0.63 | 7.02 | 36.8 | 0.00 | 4.19 | |
| i Boo A | 133640A | 73695 | | J15038+4739A | F5V | 79.95 | 1.56 | 5.14 | 12.5 | 0.00 | 4.65 | |
| i Boo B | 133640B | | | J15038+4739B | G9: | | | 6.00 | | 0.00 | 5.52 | |
| HR 5630 | 134044 | 73941 | 5630 | | F8V | 33.52 | 0.36 | 6.36 | 29.8 | 0.00 | 3.99 | |
| c Boo | 134083 | 73996 | 5634 | J15074+2453A | F5V | 51.14 | 0.31 | 4.93 | 19.6 | 0.00 | 3.47 | |
| HD 135101 | 135101 | 74432 | 5659 | J15127+1917A | G5V_Fe-0.9 | 36.71 | 1.44 | 6.68 | 27.2 | 0.00 | 4.50 | |
| HD 135101B | 135101B | 74434 | | J15127+1917B | G7V | 21.49 | 3.40 | 7.53 | 46.5 | 0.00 | 4.19 | |
| HD 135145 | 135145 | 74442 | | J15128+2756A | G0V | 12.53 | 0.87 | 8.35 | 79.8 | 0.00 | 3.84 | |
| 23 Lib | 134987 | 74500 | 5657 | | G6IV-V | 38.16 | 0.60 | 6.46 | 26.2 | 0.00 | 4.37 | H |

| Name | HD | HIP | HR | WDS | SpType | π | B-V | V | d | A_V | M_V | H |
|---|---|---|---|---|---|---|---|---|---|---|---|---|
| HD 135204 | 135204 | 74537 | | J15138-0121AB | G9V | 56.59 | 0.49 | 6.60 | 17.7 | 0.00 | 5.36 | |
| HD 136064 | 136064 | 74605 | 5691 | | F8V | 39.46 | 0.17 | 5.10 | 25.3 | 0.00 | 3.08 | |
| HD 135599 | 135599 | 74702 | | | K0V | 63.11 | 0.70 | 6.91 | 15.8 | 0.00 | 5.91 | |
| HD 136231 | 136231 | 74933 | | | G0V | 12.72 | 1.19 | 8.62 | 78.6 | 0.00 | 4.14 | |
| HD 136118 | 136118 | 74948 | | | F7V | 21.47 | 0.54 | 6.94 | 46.6 | 0.00 | 3.60 | |
| 5 Ser | 136202 | 74975 | 5694 | J15193+0146A | F8IV | 39.40 | 0.29 | 5.10 | 25.4 | 0.00 | 3.08 | |
| HR 5706 | 136442 | 75101 | 5706 | | K2III | 28.53 | 0.54 | 6.35 | 35.1 | 0.00 | 3.63 | |
| eta CrB A | 137107 | 75312 | 5727 | J15233+3018A | G2V | 55.98 | 0.78 | 5.58 | 17.9 | 0.00 | 4.32 | |
| eta CrB B | 137108 | | | J15233+3018B | G2V | | | 5.95 | | 0.00 | 4.69 | |
| 51 Boo Bn | 137392 | 75415 | | J15245+3722B | G0V | 27.73 | 0.65 | 7.05 | 36.1 | 0.00 | 4.26 | |
| 51 Boo Bs | 137392 | | | J15245+3722C | G0V | | | 7.55 | | 0.00 | 4.76 | |
| HR 5740 | 137510 | 75535 | 5740 | | G0IV-V | 24.24 | 0.51 | 6.26 | 41.3 | 0.00 | 3.18 | |
| HD 139777 | 139777 | 75809 | 5829 | J15293+8027A | G1.5V(n) | 45.77 | 0.37 | 6.58 | 21.8 | 0.00 | 4.88 | |
| HD 139813 | 139813 | 75829 | | J15293+8027B | G9V | 46.48 | 0.49 | 7.31 | 21.5 | 0.00 | 5.65 | |
| HD 139323 | 139323 | 76375 | | J15360+3949C | K3V | 44.69 | 0.58 | 7.56 | 22.4 | 0.00 | 5.81 | |
| HD 140283 | 140283 | 76976 | | | A5/7Ib/IIw | 17.16 | 0.68 | 7.21 | 58.3 | 0.00 | 3.38 | |
| psi Ser | 140538 | 77052 | 5853 | J15439+0230AB | G2.5V | 68.22 | 0.66 | 5.88 | 14.7 | 0.00 | 5.05 | |
| HD 140913 | 140913 | 77152 | | | G0V | 22.27 | 0.82 | 8.07 | 44.9 | 0.00 | 4.81 | |
| lam Ser | 141004 | 77257 | 5868 | | G0IV-V | 82.48 | 0.32 | 4.42 | 12.1 | 0.00 | 4.00 | |
| HD 141715 | 141715 | 77584 | | | G3V | 15.73 | 0.87 | 8.28 | 63.6 | 0.00 | 4.26 | |
| kap CrB | 142091 | 77655 | 5901 | J15512+3539A | K0III-IV | 32.79 | 0.21 | 4.82 | 30.5 | 0.00 | 2.40 | H |
| HD 141937 | 141937 | 77740 | | | G1V | 30.96 | 0.64 | 7.25 | 32.3 | 0.00 | 4.70 | H |
| chi Her | 142373 | 77760 | 5914 | | G0V_Fe-0.8 | 62.92 | 0.21 | 4.62 | 15.9 | 0.00 | 3.61 | |
| 39 Ser | 142267 | 77801 | 5911 | J15533+1312A | G1VFe-1 | 57.64 | 0.54 | 6.12 | 17.3 | 0.00 | 4.92 | |
| gam Ser | 142860 | 78072 | 5933 | J15564+1540A | F6IV | 88.86 | 0.18 | 3.84 | 11.3 | 0.00 | 3.58 | |
| 49 Lib | 143333 | 78400 | 5954 | J16003-1632A | F8V | 28.40 | 1.23 | 5.47 | 35.2 | 0.00 | 2.74 | |
| rho CrB | 143761 | 78459 | 5968 | J16011+3318A | G0V | 58.02 | 0.28 | 5.41 | 17.2 | 0.00 | 4.23 | H |
| tet Dra | 144284 | 78527 | 5986 | | F9V | 47.54 | 0.12 | 4.00 | 21.0 | 0.00 | 2.39 | |
| HD 144287 | 144287 | 78709 | | | G8+V | 45.01 | 0.79 | 7.06 | 22.2 | 0.00 | 5.33 | |
| HD 144579 | 144579 | 78775 | | J16049+3909A | G8V | 68.87 | 0.33 | 6.67 | 14.5 | 0.00 | 5.86 | |
| HD 144585 | 144585 | 78955 | 5996 | | G1.5V | 36.89 | 0.53 | 6.32 | 27.1 | 0.00 | 4.15 | |
| HD 145148 | 145148 | 79137 | 6014 | | K1.5IV | 35.32 | 0.47 | 5.94 | 28.3 | 0.00 | 3.68 | |
| HD 145435 | 145435 | 79152 | | | F5 | 26.75 | 0.36 | 6.72 | 37.4 | 0.00 | 3.86 | |
| 14 Her | 145675 | 79248 | | | K0V | 56.91 | 0.34 | 6.67 | 17.6 | 0.00 | 5.45 | H |
| HD 145729 | 145729 | 79327 | | | F8 | 23.22 | 0.58 | 7.58 | 43.1 | 0.00 | 4.41 | |
| sig CrB A | 146361A | 79607 | 6063 | J16147+3352A | F6V | 47.44 | 1.22 | 5.55 | 21.1 | 0.00 | 3.93 | |
| sig CrB B | 146362 | | 6064 | J16147+3352B | G1V | | | 6.42 | | 0.00 | 4.80 | |
| 18 Sco | 146233 | 79672 | 6060 | J16156-0822A | G5V | 71.94 | 0.37 | 5.50 | 13.9 | 0.00 | 4.78 | |

| Name | HD | HIP | HR | WDS | Spec | π | σπ | V | d | E(B-V) | M_V | Note |
|---|---|---|---|---|---|---|---|---|---|---|---|---|
| eta UMi | 148048 | 79822 | 6116 | J16176+7545A | F5V | 33.63 | 0.17 | 4.95 | 29.7 | 0.00 | 2.58 | |
| HD 146946 | 146946 | 79837 | | | F9V | 35.06 | 0.57 | 6.86 | 28.5 | 0.00 | 4.58 | |
| HD 147044 | 147044 | 79862 | | | G0V | 26.47 | 0.59 | 7.49 | 37.8 | 0.00 | 4.60 | |
| HD 147681 | 147681 | 80355 | | | G1V | 13.68 | 1.08 | 8.82 | 73.1 | 0.00 | 4.50 | |
| HR 6106 | 147723 | | 6106 | J16247-2942A | F9IV | | | 5.94 | | 0.00 | 3.34 | |
| HR 6105 | 147722 | 80399 | 6105 | J16247-2942B | G0V | 30.22 | 0.98 | 6.59 | 33.1 | 0.00 | 3.99 | |
| HD 149026 | 149026 | 80838 | | | G0IV | 12.59 | 0.70 | 8.14 | 79.4 | 0.00 | 3.64 | H |
| HD 150706 | 150706 | 80902 | | | G3V | 35.43 | 0.33 | 7.03 | 28.2 | 0.00 | 4.78 | H |
| HD 149143 | 149143 | 81022 | | | G3V | 16.12 | 0.83 | 7.89 | 62.0 | 0.00 | 3.93 | H |
| 12 Oph | 149661 | 81300 | 6171 | J16363-0220A | K1V | 102.55 | 0.40 | 5.77 | 9.8 | 0.00 | 5.82 | |
| zet Her | 150680 | 81693 | 6212 | J16413+3136AB | G0IV | 93.32 | 0.47 | 2.80 | 10.7 | 0.00 | 2.65 | |
| HD 150933 | 150933 | 81880 | | J16435+2042A | G2V | 22.69 | 0.68 | 7.17 | 44.1 | 0.00 | 3.95 | |
| HD 151426 | 151426 | 81947 | | | G0 | 19.48 | 0.55 | 8.08 | 51.3 | 0.00 | 4.53 | |
| HD 151288 | 151288 | 82003 | | | K7V | 101.96 | 0.71 | 8.80 | 9.8 | 0.00 | 8.84 | |
| HD 151450 | 151450 | 82233 | | | G2V | 23.73 | 0.70 | 7.41 | 42.1 | 0.00 | 4.29 | |
| HD 152391 | 152391 | 82588 | | | G8.5Vk: | 57.97 | 0.66 | 6.64 | 17.3 | 0.00 | 5.46 | |
| HR 6269 | 152311 | 82621 | 6269 | | G2IV-V | 37.12 | 0.72 | 5.88 | 26.9 | 0.00 | 3.72 | |
| HD 152792 | 152792 | 82636 | | | G0V | 20.84 | 0.39 | 6.74 | 48.0 | 0.00 | 3.33 | |
| HR 6301 | 153226 | 82989 | 6301 | | K0V | 16.36 | 0.47 | 6.37 | 61.1 | 0.00 | 2.44 | |
| HD 153525 | 153525 | 83006 | | J16578+4722C | K3V | 56.18 | 0.51 | 7.88 | 17.8 | 0.00 | 6.63 | |
| HD 154160 | 154160 | 83435 | 6339 | | G5IV: | 27.57 | 0.45 | 6.54 | 36.3 | 0.00 | 3.74 | |
| HD 154578 | 154578 | 83517 | | | F7V | 13.92 | 0.52 | 7.84 | 71.8 | 0.00 | 3.56 | |
| HD 154363 | 154363 | 83591 | | J17051-0504A | K4/5V | 93.40 | 0.94 | 7.71 | 10.7 | 0.00 | 7.56 | |
| V* V2213 Oph | 154417 | 83601 | 6349 | | F8V | 48.39 | 0.40 | 6.01 | 20.7 | 0.00 | 4.43 | |
| HD 154931 | 154931 | 83863 | | | G0V | 17.88 | 0.70 | 7.25 | 55.9 | 0.00 | 3.51 | |
| HR 6372 | 154962 | 83906 | 6372 | | G8IV(+A) | 28.63 | 0.46 | 6.37 | 34.9 | 0.00 | 3.65 | |
| HD 155712 | 155712 | 84195 | | | K2.5V | 47.70 | 0.93 | 7.95 | 21.0 | 0.00 | 6.34 | |
| BD+29 2963 | | 84309 | | | G0V | 14.93 | 0.93 | 8.42 | 67.0 | 0.00 | 4.29 | |
| 36 Oph A | 155886 | 84405 | 6402 | J17155-2635A | K2V | 183.00 | 9.00 | 5.08 | 5.5 | 0.00 | 6.39 | |
| 36 Oph B | 155885 | | 6401 | J17155-2635B | K1V | 167.10 | 1.10 | 5.08 | 6.0 | 0.00 | 6.19 | |
| V* V2215 Oph | 156026 | 84478 | | J17155-2635C | K5V | 167.49 | 0.60 | 6.34 | 6.0 | 0.00 | 7.46 | |
| HD 156062 | 156062 | 84497 | | | G1V | 17.10 | 0.80 | 8.21 | 58.5 | 0.00 | 4.37 | |
| HD 156985 | 156985 | 84616 | | | K2 | 54.57 | 0.55 | 7.96 | 18.3 | 0.00 | 6.64 | |
| HD 156826 | 156826 | 84801 | 6439 | | K0V | 20.13 | 0.50 | 6.32 | 49.7 | 0.00 | 2.84 | |
| HD 156968 | 156968 | 84824 | | | G0V | 20.20 | 0.73 | 7.97 | 49.5 | 0.00 | 4.50 | |
| HD 156846 | 156846 | 84856 | 6441 | J17206-1920A | G1V | 21.00 | 0.51 | 6.52 | 47.6 | 0.00 | 3.13 | |
| w Her | 157214 | 84862 | 6458 | J17206+3229A | G0V | 69.80 | 0.25 | 5.39 | 14.3 | 0.00 | 4.61 | |
| HD 157089 | 157089 | 84905 | | | F8VgF8mF6 | 27.52 | 0.73 | 7.01 | 36.3 | 0.00 | 4.21 | |

| Name | HD | HIP | HR | 2MASS/Other | SpType |  |  |  |  |  |  |  |
|---|---|---|---|---|---|---|---|---|---|---|---|---|
| V* V819 Her | 157482 | 84949 | 6469 | J17217+3958B | F7V | 13.52 | 0.93 | 5.57 | 74.0 | 0.00 | 1.22 | |
| HR 6465 | 157347 | 85042 | 6465 | | G3V | 51.22 | 0.40 | 6.29 | 19.5 | 0.00 | 4.84 | |
| HD 158633 | 158633 | 85235 | 6518 | | K0V | 78.11 | 0.30 | 6.43 | 12.8 | 0.00 | 5.89 | |
| HD 157881 | 157881 | 85295 | | | M1V | 129.86 | 0.73 | 7.56 | 7.7 | 0.00 | 8.13 | |
| HD 159062 | 159062 | 85653 | | | G9V_Fe-0.8 | 44.91 | 0.50 | 7.22 | 22.3 | 0.00 | 5.48 | |
| HD 158614 | 158614 | 85667 | 6516 | J17304-0104AB | G9-IV-V_Hdel1 | 61.19 | 0.68 | 5.31 | 16.3 | 0.00 | 4.24 | |
| HD 159222 | 159222 | 85810 | 6538 | | G1V | 41.81 | 0.35 | 6.56 | 23.9 | 0.00 | 4.67 | |
| HD 159482 | 159482 | 86013 | | J17348+0601C | G0V | 19.38 | 1.14 | 8.39 | 51.6 | 0.00 | 4.83 | |
| 26 Dra | 160269 | 86036 | 6573 | J17351+6152AB | F9V+K3V | 70.47 | 0.37 | 5.24 | 14.2 | 0.00 | 4.48 | |
| HD 160964 | 160964 | 86141 | | | K4V | 51.58 | 0.56 | 8.59 | 19.4 | 0.00 | 7.15 | |
| HD 160933 | 160933 | 86184 | 6598 | | F9V | 21.89 | 0.68 | 6.42 | 45.7 | 0.00 | 3.12 | |
| HD 160346 | 160346 | 86400 | | | K2.5V | 90.91 | 0.67 | 6.52 | 11.0 | 0.00 | 6.31 | |
| psi01 Dra A | 162003 | 86614 | 6636 | J17420+7209A | F5IV-V | 43.79 | 0.45 | 4.56 | 22.8 | 0.00 | 2.77 | |
| psi01 Dra B | 162004 | 86620 | 6637 | J17420+7209B | F8V | 43.36 | 0.51 | 5.78 | 23.1 | 0.00 | 3.97 | H |
| 84 Her | 161239 | 86731 | 6608 | | G2IV | 25.60 | 0.33 | 5.74 | 39.1 | 0.00 | 2.78 | |
| HD 161098 | 161098 | 86765 | | | G8V | 33.13 | 0.75 | 7.67 | 30.2 | 0.00 | 5.27 | |
| mu. Her | 161797 | 86974 | 6623 | J17465+2744A | G5IV | 120.33 | 0.16 | 3.42 | 8.3 | 0.00 | 3.82 | |
| HD 163183 | 163183 | 87079 | | | G0 | 24.84 | 0.42 | 7.75 | 40.3 | 0.00 | 4.73 | |
| HD 162826 | 162826 | 87382 | 6669 | | F8V | 29.76 | 0.36 | 6.55 | 33.6 | 0.00 | 3.92 | |
| GJ 697 | | 87579 | | | K2.5V | 41.06 | 1.04 | 8.49 | 24.4 | 0.00 | 6.56 | |
| HD 163492 | 163492 | 87841 | | | G3V | 15.46 | 1.24 | 8.60 | 64.7 | 0.00 | 4.55 | |
| V* V566 Oph | 163611 | 87860 | | | F5V | 13.23 | 0.74 | 7.58 | 75.6 | 0.00 | 3.19 | |
| HD 163840 | 163840 | 87895 | 6697 | J17572+2400A | G2V+K2V | 35.40 | 0.62 | 6.33 | 28.2 | 0.00 | 4.08 | |
| HD 164595 | 164595 | 88194 | | | G2V | 35.26 | 0.50 | 7.08 | 28.4 | 0.00 | 4.82 | H |
| HD 164507 | 164507 | 88217 | 6722 | | G5IV | 22.44 | 0.45 | 6.29 | 44.6 | 0.00 | 3.05 | |
| HD 164651 | 164651 | 88324 | | | G8V | 30.30 | 0.57 | 7.63 | 33.0 | 0.00 | 5.04 | |
| HD 165173 | 165173 | 88511 | | | G5 | 31.33 | 0.90 | 7.95 | 31.9 | 0.00 | 5.43 | |
| 70 Oph A | 165341A | 88601 | | J18055+0230A | K0-V | 196.62 | 0.83 | 4.12 | 5.1 | 0.00 | 5.59 | |
| HD 165401 | 165401 | 88622 | | | G2V | 41.82 | 0.59 | 6.79 | 23.9 | 0.00 | 4.89 | |
| HD 165590 | 165590 | 88637 | | J18058+2127AB | G0V+G5V | 25.35 | 1.31 | 7.10 | 39.4 | 0.00 | 4.12 | |
| HD 165476 | 165476 | 88644 | | | G5 | 21.17 | 0.96 | 7.67 | 47.2 | 0.00 | 4.30 | |
| HR 6756 | 165438 | 88684 | 6756 | | K0IV | 27.20 | 0.38 | 5.75 | 36.8 | 0.00 | 2.92 | |
| HD 165672 | 165672 | 88724 | | | G5 | 21.71 | 0.96 | 7.77 | 46.1 | 0.00 | 4.45 | |
| HD 165670 | 165670 | 88728 | | J18068+0852A | F5 | 24.74 | 0.69 | 6.97 | 40.4 | 0.00 | 3.93 | |
| b Her | 165908 | 88745 | 6775 | J18071+3034AB | F7VgF7mF5 | 63.93 | 0.34 | 5.07 | 15.6 | 0.00 | 4.10 | |
| HD 166435 | 166435 | 88945 | | | G1IV | 40.29 | 0.49 | 6.85 | 24.8 | 0.00 | 4.87 | |
| HD 166183 | 166183 | 88958 | | | F5/6V | 11.39 | 0.54 | 7.87 | 87.8 | 0.01 | 3.12 | |
| HR 6806 | 166620 | 88972 | 6806 | | K2V | 90.71 | 0.30 | 6.40 | 11.0 | 0.00 | 6.19 | |

| Name | HD | HIP | HR | J-ID | SpType | col7 | col8 | col9 | col10 | col11 | col12 | flag |
|---|---|---|---|---|---|---|---|---|---|---|---|---|
| HD 166601 | 166601 | 89014 | | | F5V | 11.50 | 0.63 | 7.87 | 87.0 | 0.00 | 3.17 | |
| 36 Dra | 168151 | 89348 | 6850 | | F5V | 43.63 | 0.17 | 5.03 | 22.9 | 0.00 | 3.23 | |
| HD 167278 | 167278 | 89393 | | J18146+0011AB | F5V | 8.74 | 1.54 | 7.67 | 114.4 | 0.03 | 2.28 | |
| HD 167588 | 167588 | 89408 | 6831 | | F8V | 23.64 | 0.37 | 6.56 | 42.3 | 0.00 | 3.43 | |
| HR 6847 | 168009 | 89474 | 6847 | | G1V | 43.82 | 0.29 | 6.31 | 22.8 | 0.00 | 4.52 | |
| V* V2502 Oph | 167858 | 89601 | 6844 | | F2V | 15.80 | 0.54 | 6.61 | 63.3 | 0.00 | 2.60 | |
| HD 167665 | 167665 | 89620 | 6836 | | F9VFe-0.8CH-0.4 | 32.80 | 0.52 | 6.40 | 30.5 | 0.00 | 3.98 | |
| HD 168443 | 168443 | 89844 | | | G6V | 26.72 | 0.69 | 6.92 | 37.4 | 0.00 | 4.05 | H |
| chi Dra | 170153 | 89937 | 6927 | J18211+7245A | F7V | 124.11 | 0.87 | 3.58 | 8.1 | 0.00 | 4.05 | |
| eta Ser | 168723 | 89962 | 6869 | J18214-0253A | K0III-IV | 53.93 | 0.18 | 3.25 | 18.5 | 0.00 | 1.91 | |
| HD 168746 | 168746 | 90004 | | | G5V | 23.40 | 0.76 | 7.95 | 42.7 | 0.00 | 4.80 | H |
| HD 169822 | 169822 | 90355 | | J18262+0846A | G6V | 34.61 | 1.39 | 7.83 | 28.9 | 0.00 | 5.53 | |
| HD 170008 | 170008 | 90427 | | | G5 | 12.45 | 0.71 | 7.40 | 80.3 | 0.00 | 2.87 | |
| HR 6907 | 169830 | 90485 | 6907 | | F7V | 27.32 | 0.41 | 5.91 | 36.6 | 0.00 | 3.09 | H |
| HD 170291 | 170291 | 90532 | | | F5 | 15.89 | 0.77 | 7.65 | 62.9 | 0.00 | 3.66 | |
| HD 170778 | 170778 | 90586 | | | G5 | 27.57 | 0.46 | 7.52 | 36.3 | 0.00 | 4.72 | |
| HD 170512 | 170512 | 90621 | | J18295+0722AB | G0 | 16.27 | 0.80 | 7.77 | 61.5 | 0.00 | 3.83 | |
| HD 170579 | 170579 | 90657 | | | F5 | 13.72 | 0.85 | 7.86 | 72.9 | 0.00 | 3.55 | |
| HD 170829 | 170829 | 90729 | 6950 | | G8V | 28.92 | 0.41 | 6.50 | 34.6 | 0.00 | 3.81 | |
| HD 170657 | 170657 | 90790 | | | K2V | 75.46 | 0.70 | 6.81 | 13.3 | 0.00 | 6.20 | |
| HD 171314 | 171314 | 90959 | | | K4V | 43.53 | 1.17 | 8.87 | 23.0 | 0.00 | 7.06 | |
| HD 171706 | 171706 | 91215 | | | G3V | 15.60 | 0.86 | 7.86 | 64.1 | 0.00 | 3.83 | |
| HD 171888 | 171888 | 91281 | | | F7V | 18.68 | 0.69 | 6.88 | 53.5 | 0.00 | 3.24 | |
| HD 171951 | 171951 | 91291 | | J18372+0732A | G0 | 13.32 | 1.09 | 7.87 | 75.1 | 0.00 | 3.49 | |
| HD 171953 | 171953 | 91300 | | J18374+0238A | F3V | 8.03 | 0.71 | 7.82 | 124.5 | 0.02 | 2.29 | |
| HD 172051 | 172051 | 91438 | 6998 | | G6V | 76.43 | 0.47 | 5.87 | 13.1 | 0.00 | 5.28 | |
| HD 172675 | 172675 | 91645 | | J18413+0433A | F7V | 25.64 | 0.60 | 7.06 | 39.0 | 0.00 | 4.10 | |
| HD 172718 | 172718 | 91649 | | | F8 | 10.81 | 0.68 | 7.99 | 92.5 | 0.01 | 3.13 | |
| HD 172961 | 172961 | 91766 | | | F5 | 16.29 | 0.70 | 7.60 | 61.4 | 0.00 | 3.66 | |
| HD 173605 | 173605 | 91773 | | | G5 | 15.22 | 0.48 | 7.93 | 65.7 | 0.00 | 3.84 | |
| HD 173174 | 173174 | 91889 | | J18439+0237AB | G0V | 11.70 | 1.16 | 7.97 | 85.5 | 0.00 | 3.30 | |
| 110 Her | 173667 | 92043 | 7061 | J18457+2033A | F6V | 52.06 | 0.25 | 4.19 | 19.2 | 0.00 | 2.77 | |
| HD 173634 | 173634 | 92078 | | | F5 | 9.05 | 0.60 | 7.31 | 110.5 | 0.02 | 2.03 | |
| HD 173818 | 173818 | 92200 | | | K5V | 70.04 | 1.17 | 8.77 | 14.3 | 0.00 | 7.99 | |
| HR 7079 | 174160 | 92270 | 7079 | | F7V | 34.78 | 0.41 | 6.19 | 28.8 | 0.00 | 3.90 | |
| HD 174080 | 174080 | 92283 | | J18485+1045A | K0 | 59.31 | 0.83 | 7.91 | 16.9 | 0.00 | 6.78 | |
| HD 175225 | 175225 | 92549 | 7123 | | G9IVa | 38.96 | 0.19 | 5.51 | 25.7 | 0.00 | 3.46 | |
| HD 174719 | 174719 | 92569 | | | G5/6V | 35.19 | 0.61 | 7.51 | 28.4 | 0.00 | 5.24 | |

| Name | HD | HIP | HR | WDS | Spectral Type | π (mas) | B-V | V | d (pc) | E(B-V) | M_V | Note |
|---|---|---|---|---|---|---|---|---|---|---|---|---|
| HD 175290 | 175290 | 92666 | | | F5V | 10.97 | 0.44 | 7.98 | 91.2 | 0.00 | 3.17 | |
| HD 175272 | 175272 | 92794 | | J18546+0154A | A0V | 11.30 | 0.63 | 7.40 | 88.5 | 0.00 | 2.65 | |
| V* V1709 Aql | 175337 | 92837 | | | F3V | 12.02 | 0.67 | 7.36 | 83.2 | 0.00 | 2.75 | |
| HD 175726 | 175726 | 92984 | | J18563+0413C | G0V | 37.73 | 0.51 | 6.71 | 26.5 | 0.00 | 4.59 | |
| HD 175805 | 175805 | 93016 | | J18570+0228A | F7V | 7.90 | 0.62 | 7.66 | 126.6 | 0.02 | 2.08 | |
| HD 176051 | 176051 | 93017 | 7162 | J18570+3254AB | F9V+K1V | 67.24 | 0.37 | 5.25 | 14.9 | 0.00 | 4.39 | H |
| HD 175806 | 175806 | 93028 | | | F5V | 7.79 | 0.74 | 7.59 | 128.4 | 0.02 | 1.99 | |
| HD 176377 | 176377 | 93185 | | | G1V | 41.94 | 0.47 | 6.78 | 23.8 | 0.00 | 4.89 | |
| HD 176118 | 176118 | 93190 | | | F5V | 11.62 | 0.59 | 7.48 | 86.1 | 0.01 | 2.79 | |
| 11 Aql | 176303 | 93203 | 7172 | J18591+1337A | F8V | 20.65 | 0.25 | 5.27 | 48.4 | 0.00 | 1.85 | |
| HD 176367 | 176367 | 93375 | | J19011-2843A | G1V | 17.02 | 1.05 | 8.47 | 58.8 | 0.00 | 4.62 | |
| HD 177830 | 177830 | 93746 | | | K0+M4V | 16.94 | 0.63 | 7.18 | 59.0 | 0.00 | 3.32 | H |
| HD 177749 | 177749 | 93772 | | | F5 | 15.17 | 0.82 | 6.92 | 65.9 | 0.00 | 2.82 | |
| HD 177904 | 177904 | 93822 | | | F5V+F6V | 9.47 | 1.08 | 7.17 | 105.6 | 0.02 | 1.98 | |
| HD 178126 | 178126 | 93871 | | | K5V | 40.36 | 1.42 | 9.22 | 24.8 | 0.00 | 7.24 | |
| HD 178428 | 178428 | 93966 | 7260 | J19080+1651A | G4VCN+0.5 | 46.66 | 0.48 | 6.07 | 21.4 | 0.00 | 4.41 | |
| HD 178911B | 178911B | 94075 | | J19091+3436B | G5 | 23.48 | 3.95 | 7.98 | 42.6 | 0.00 | 4.83 | H |
| HD 178911 | 178911 | 94076 | 7272 | J19091+3436A | G1V+K1V | 19.11 | 2.35 | 6.74 | 52.3 | 0.00 | 3.15 | |
| HD 180161 | 180161 | 94346 | | | G8V | 49.96 | 0.32 | 7.04 | 20.0 | 0.00 | 5.53 | |
| HR 7291 | 179949 | 94645 | 7291 | | F8V | 36.30 | 0.70 | 6.25 | 27.5 | 0.00 | 4.05 | H |
| HD 181096 | 181096 | 94755 | 7322 | | F6IV: | 23.79 | 0.32 | 6.00 | 42.0 | 0.00 | 2.88 | |
| HD 180945 | 180945 | 94873 | | | F5V | 15.36 | 0.56 | 7.15 | 65.1 | 0.00 | 3.08 | |
| BD+41 3306 | | 94931 | | | K0V | 28.03 | 0.82 | 8.86 | 35.7 | 0.00 | 6.10 | H |
| HD 181655 | 181655 | 94981 | 7345 | | G5V | 39.39 | 0.33 | 6.31 | 25.4 | 0.00 | 4.29 | |
| HD 181420 | 181420 | 95055 | | | F6V | 21.05 | 0.48 | 6.55 | 47.5 | 0.00 | 3.17 | |
| HD 181806 | 181806 | 95202 | | J19221-0444A | F3V | 10.24 | 1.31 | 7.70 | 97.7 | 0.01 | 2.73 | |
| HD 182274 | 182274 | 95293 | | | F6V | 17.36 | 0.67 | 7.80 | 57.6 | 0.00 | 4.00 | |
| HD 182488 | 182488 | 95319 | 7368 | | G9+V | 63.45 | 0.35 | 6.36 | 15.8 | 0.00 | 5.37 | H |
| HD 182736 | 182736 | 95362 | | | G8IV | 17.35 | 0.41 | 7.01 | 57.6 | 0.00 | 3.21 | |
| b Aql | 182572 | 95447 | 7373 | J19249+1157A | G7IVHdel1 | 65.89 | 0.26 | 5.16 | 15.2 | 0.00 | 4.25 | |
| HD 182905 | 182905 | 95622 | | | G6IV | 14.59 | 0.65 | 8.00 | 68.5 | 0.00 | 3.82 | |
| HD 183263 | 183263 | 95740 | | | G2IV | 18.15 | 0.93 | 7.86 | 55.1 | 0.00 | 4.15 | H |
| HD 183341 | 183341 | 95772 | | | G5 | 22.05 | 0.64 | 7.44 | 45.4 | 0.00 | 4.16 | |
| HD 183658 | 183658 | 95962 | | J19309-0631A | G3V | 31.93 | 0.60 | 7.28 | 31.3 | 0.00 | 4.80 | |
| HD 231701 | 231701 | 96078 | | | F8V | 8.44 | 1.05 | 8.97 | 118.5 | 0.02 | 3.55 | H |
| HD 184151 | 184151 | 96081 | | | F5V | 11.48 | 0.58 | 6.88 | 87.1 | 0.01 | 2.16 | |
| HD 183870 | 183870 | 96085 | | | K2V | 56.73 | 0.72 | 7.57 | 17.6 | 0.00 | 6.34 | |
| sig Dra | 185144 | 96100 | 7462 | J19322+6941A | G9V | 173.77 | 0.18 | 4.68 | 5.8 | 0.00 | 5.88 | |

| Name | HD | HIP | HR | WDS | Type | plx | V-K | K | dist | A_V | M_K | flag |
|---|---|---|---|---|---|---|---|---|---|---|---|---|
| HD 184152 | 184152 | 96142 | | J19329+0724A | G5V | 12.77 | 1.70 | 9.40 | 78.3 | 0.00 | 4.92 | |
| HD 184499 | 184499 | 96185 | | | G0V | 31.33 | 0.39 | 6.64 | 31.9 | 0.00 | 4.12 | |
| HD 184960 | 184960 | 96258 | 7451 | | F7V | 39.82 | 0.20 | 5.70 | 25.1 | 0.00 | 3.70 | |
| HD 184489 | 184489 | 96285 | | | K5/M0V | 69.32 | 1.55 | 9.33 | 14.4 | 0.00 | 8.53 | |
| HR 7438 | 184663 | 96351 | 7438 | | F5V | 25.01 | 0.68 | 6.38 | 40.0 | 0.00 | 3.37 | |
| HD 184509 | 184509 | 96370 | | | F8.5V | 31.27 | 0.55 | 6.74 | 32.0 | 0.00 | 4.22 | |
| HD 184700 | 184700 | 96379 | | J19357-0014A | G3V | 13.01 | 1.07 | 8.83 | 76.9 | 0.00 | 4.40 | |
| HD 185414 | 185414 | 96395 | | | G0 | 41.48 | 0.30 | 6.74 | 24.1 | 0.00 | 4.83 | |
| HD 184768 | 184768 | 96402 | | J19360+0005AB | G5V | 25.91 | 0.69 | 7.57 | 38.6 | 0.00 | 4.64 | |
| HD 185269 | 185269 | 96507 | | | G2V | 19.89 | 0.56 | 6.67 | 50.3 | 0.00 | 3.16 | H |
| 16 Cyg A | 186408 | 96895 | 7503 | J19418+5031A | G1.5Vb | 47.44 | 0.27 | 5.95 | 21.1 | 0.00 | 4.33 | |
| 16 Cyg B | 186427 | 96901 | 7504 | J19418+5031B | G3V | 47.14 | 0.27 | 6.20 | 21.2 | 0.00 | 4.57 | H |
| HD 186104 | 186104 | 96948 | | | G5V | 24.25 | 0.75 | 7.64 | 41.2 | 0.00 | 4.56 | |
| HD 186226 | 186226 | 96979 | | J19427+0823A | F5 | 15.05 | 0.80 | 6.83 | 66.4 | 0.00 | 2.72 | |
| HD 332612 | 332612 | 97012 | | | G0 | 10.18 | 0.97 | 8.53 | 98.2 | 0.01 | 3.54 | |
| HD 186379 | 186379 | 97023 | | | F8V | 22.53 | 0.60 | 6.76 | 44.4 | 0.00 | 3.52 | |
| HR 7522 | 186760 | 97033 | 7522 | | G0V | 22.62 | 0.23 | 6.30 | 44.2 | 0.00 | 3.08 | |
| HD 186413 | 186413 | 97105 | | | G3V | 15.89 | 0.75 | 7.99 | 62.9 | 0.00 | 4.00 | |
| HD 332518 | 332518 | 97223 | | | K5V | 49.22 | 1.08 | 9.20 | 20.3 | 0.00 | 7.66 | |
| HD 187123 | 187123 | 97336 | | | G5 | 20.72 | 0.53 | 7.83 | 48.3 | 0.00 | 4.41 | H |
| alf Aql | 187642 | 97649 | 7557 | J19508+0852A | A7Vn | 194.95 | 0.57 | 0.76 | 5.1 | 0.00 | 2.21 | |
| omi Aql | 187691 | 97675 | 7560 | J19510+1025A | F8V | 52.11 | 0.29 | 5.10 | 19.2 | 0.00 | 3.68 | |
| HD 188169 | 188169 | 97720 | | | F4V | 13.75 | 0.46 | 7.83 | 72.7 | 0.00 | 3.52 | |
| HD 187923 | 187923 | 97767 | 7569 | J19521+1137A | G0V | 37.57 | 0.34 | 6.55 | 26.6 | 0.00 | 4.43 | |
| HD 188015 | 188015 | 97769 | | | G5IV | 17.54 | 0.85 | 8.23 | 57.0 | 0.00 | 4.45 | H |
| HD 187897 | 187897 | 97779 | | | G5 | 28.49 | 0.59 | 7.13 | 35.1 | 0.00 | 4.40 | |
| HD 188326 | 188326 | 97846 | | J19530+3846A | G8IVv | 17.88 | 0.53 | 7.56 | 55.9 | 0.00 | 3.82 | |
| HD 188088 | 188088 | 97944 | 7578 | J19543-2356A | K3V | 71.18 | 0.42 | 8.52 | 14.0 | 0.00 | 7.78 | |
| bet Aql | 188512 | 98036 | 7602 | J19553+0625A | G9.5IV | 73.00 | 0.20 | 3.71 | 13.7 | 0.00 | 3.03 | |
| ome Sgr | 188376 | 98066 | 7597 | | G5IV | 38.48 | 2.66 | 4.70 | 26.0 | 0.00 | 2.63 | |
| HR 7637 | 189340 | 98416 | 7637 | J19598-0957AB | F9V | 45.04 | 0.99 | 5.88 | 22.2 | 0.00 | 4.15 | |
| HD 189509 | 189509 | 98442 | | | F5 | 21.53 | 0.59 | 7.23 | 46.4 | 0.00 | 3.90 | |
| HD 189733 | 189733 | 98505 | | | K0V+M4V | 51.41 | 0.69 | 7.65 | 19.5 | 0.00 | 6.20 | H |
| HD 189558 | 189558 | 98532 | | J20010-1215A | F8/G2V | 15.39 | 0.81 | 7.72 | 65.0 | 0.00 | 3.66 | |
| HD 189712 | 189712 | 98565 | | | F5 | 11.49 | 0.67 | 7.45 | 87.0 | 0.01 | 2.73 | |
| HD 190007 | 190007 | 98698 | | | K4Vk: | 77.76 | 0.54 | 7.48 | 12.9 | 0.00 | 6.93 | |
| HD 190228 | 190228 | 98714 | | | G5IV | 16.23 | 0.64 | 7.31 | 61.6 | 0.00 | 3.36 | |
| HD 190360 | 190360 | 98767 | 7670 | J20036+2954A | G7IV-V | 63.06 | 0.34 | 5.71 | 15.9 | 0.00 | 4.71 | H |

| Name | HD | HIP | HR | CCDM | SpType | π | B-V | V | d | E(B-V) | M_V | Note |
|---|---|---|---|---|---|---|---|---|---|---|---|---|
| HD 190404 | 190404 | 98792 | | | K1V | 63.43 | 0.57 | 7.27 | 15.8 | 0.00 | 6.28 | |
| 15 Sge | 190406 | 98819 | 7672 | J20040+1705A | G0V | 56.28 | 0.35 | 5.80 | 17.8 | 0.00 | 4.55 | |
| HD 190412 | 190412 | 98878 | | | G6/8V | 26.14 | 1.41 | 7.69 | 38.3 | 0.00 | 4.78 | |
| CCDM J20051-0418AB | 190437 | 98916 | | J20051-0418AB | F3/5V | 12.91 | 0.99 | 7.82 | 77.5 | 0.00 | 3.37 | |
| HD 190771 | 190771 | 98921 | 7683 | J20052+3829A | G2V | 53.22 | 0.36 | 6.17 | 18.8 | 0.00 | 4.80 | |
| HD 190498 | 190498 | 98922 | | | F6V | 15.84 | 0.63 | 6.77 | 63.1 | 0.00 | 2.77 | |
| b01 Cyg | 191026 | 99031 | 7689 | J20064+3558A | K0IV | 41.76 | 0.30 | 5.36 | 23.9 | 0.00 | 3.46 | |
| 64 Aql | 191067 | 99171 | 7690 | | K1III/IV | 21.30 | 0.46 | 5.98 | 46.9 | 0.00 | 2.62 | |
| HD 191533 | 191533 | 99367 | | | F8 | 15.94 | 0.66 | 6.60 | 62.7 | 0.00 | 2.61 | |
| HD 345957 | 345957 | 99423 | | | G0V: | 10.42 | 1.14 | 8.93 | 96.0 | 0.01 | 4.00 | |
| HD 191785 | 191785 | 99452 | | J20111+1610A | K0V | 49.04 | 0.65 | 7.32 | 20.4 | 0.00 | 5.78 | |
| HD 192145 | 192145 | 99607 | | | F8Vw | 13.40 | 0.74 | 7.73 | 74.6 | 0.00 | 3.37 | |
| HD 192263 | 192263 | 99711 | | J20140-0052A | K1/2V | 51.77 | 0.78 | 7.77 | 19.3 | 0.00 | 6.34 | H |
| HD 192310 | 192310 | 99825 | 7722 | | K2+V | 112.22 | 0.30 | 5.72 | 8.9 | 0.00 | 5.97 | H |
| HD 193664 | 193664 | 100017 | 7783 | | G3V | 56.92 | 0.24 | 5.93 | 17.6 | 0.00 | 4.71 | |
| HD 193374 | 193374 | 100265 | | | F5V | 8.07 | 0.81 | 8.01 | 123.9 | 0.01 | 2.50 | |
| HD 193555 | 193555 | 100269 | | | F8V | 12.20 | 0.43 | 6.78 | 82.0 | 0.00 | 2.20 | |
| HR 7793 | 194012 | 100511 | 7793 | | F7V | 38.09 | 0.43 | 6.17 | 26.3 | 0.00 | 4.07 | |
| HD 194154 | 194154 | 100581 | | | F5 | 16.85 | 0.80 | 7.59 | 59.3 | 0.00 | 3.72 | |
| HD 334372 | 334372 | 100788 | | | G0 | 9.88 | 1.00 | 8.88 | 101.2 | 0.01 | 3.82 | |
| HD 194598 | 194598 | 100792 | | | F7V-VI | 17.00 | 0.83 | 8.34 | 58.8 | 0.00 | 4.49 | |
| HD 195019 | 195019 | 100970 | | J20284+1846AB | G3IV-V | 25.96 | 0.99 | 6.88 | 38.5 | 0.00 | 3.95 | H |
| HD 195005 | 195005 | 101022 | | | F7V | 31.57 | 0.75 | 6.81 | 31.7 | 0.00 | 4.31 | |
| rho Cap | 194943 | 101027 | 7822 | | F2V | 33.04 | 0.46 | 4.80 | 30.3 | 0.00 | 2.40 | |
| HD 195104 | 195104 | 101059 | | | F7V | 27.28 | 0.67 | 7.07 | 36.7 | 0.00 | 4.25 | |
| HD 195564 | 195564 | 101345 | 7845 | J20324-0952AB | G2V | 40.98 | 0.33 | 5.65 | 24.4 | 0.00 | 3.71 | |
| HD 195633 | 195633 | 101346 | | | G0Vw | 10.07 | 0.84 | 8.52 | 99.3 | 0.00 | 3.53 | |
| HR 7855 | 195838 | 101507 | 7855 | | F9VFe-0.6CH-0.3 | 30.48 | 0.45 | 6.13 | 32.8 | 0.00 | 3.55 | |
| HD 196218 | 196218 | 101620 | | | F7V | 18.80 | 0.87 | 7.42 | 53.2 | 0.00 | 3.79 | |
| kap Del | 196755 | 101916 | 7896 | J20392+1005A | G1IV+K2IV | 33.20 | 0.28 | 5.05 | 30.1 | 0.00 | 2.66 | |
| HR 7907 | 196885A | 101966 | 7907 | J20399+1116A | F8IV: | 29.83 | 0.48 | 6.39 | 33.5 | 0.00 | 3.76 | H |
| HD 196761 | 196761 | 101997 | 7898 | | G8V | 69.53 | 0.40 | 6.37 | 14.4 | 0.00 | 5.58 | |
| HD 197076 | 197076 | 102040 | 7914 | J20408+1956A | G5V | 47.74 | 0.48 | 6.44 | 20.9 | 0.00 | 4.83 | |
| eta Cep | 198149 | 102422 | 7957 | J20453+6150A | K0IV | 70.10 | 0.11 | 3.41 | 14.3 | 0.00 | 2.64 | |
| HR 7955 | 198084 | 102431 | 7955 | J20454+5735A | F8IV-V+F9IV-V | 36.64 | 0.48 | 4.51 | 27.3 | 0.00 | 2.33 | |
| psi Cap | 197692 | 102485 | 7936 | | F5V | 68.13 | 0.27 | 4.15 | 14.7 | 0.00 | 3.32 | |
| gam01 Del | 197963 | 102531 | 7947 | J20467+1607B | F7V | 26.35 | 1.09 | 5.14 | 38.0 | 0.00 | 2.24 | |
| HD 198061 | 198061 | 102612 | | | F5 | 13.57 | 0.82 | 7.28 | 73.7 | 0.00 | 2.94 | |

| Name | HD | HIP | HR | CCDM/J | SpType | π | σπ | V | d | E(B-V) | M_V | Note |
|---|---|---|---|---|---|---|---|---|---|---|---|---|
| HD 198387 | 198387 | 102642 | 7972 | J20479+5224A | K0V: | 23.52 | 0.43 | 6.27 | 42.5 | 0.00 | 3.13 | |
| HD 198483 | 198483 | 102815 | | | G0V | 19.52 | 0.71 | 7.67 | 51.2 | 0.00 | 4.12 | |
| BD+52 2815 | | 102870 | | | K7-V | 44.61 | 0.97 | 9.74 | 22.4 | 0.00 | 7.99 | |
| 4 Aqr | 198571 | 102945 | 7982 | J20514-0537AB | F6V | 16.47 | 0.59 | 6.07 | 60.7 | 0.00 | 2.15 | |
| HD 198802 | 198802 | 103077 | 7994 | | G5V | 22.65 | 0.46 | 6.39 | 44.2 | 0.00 | 3.17 | |
| BD+12 4499 | | 103256 | | | K4V | 44.61 | 0.94 | 8.77 | 22.4 | 0.00 | 7.02 | |
| HD 199598 | 199598 | 103455 | | | G0V | 31.62 | 0.55 | 6.90 | 31.6 | 0.00 | 4.40 | |
| HD 199604 | 199604 | 103572 | | | G2V | 14.82 | 0.96 | 8.61 | 67.5 | 0.00 | 4.46 | |
| 11 Aqr | 199960 | 103682 | 8041 | | G1V | 36.73 | 0.58 | 6.21 | 27.2 | 0.00 | 4.04 | |
| HD 200391 | 200391 | 103833 | | | G0V+G5V | 19.23 | 0.60 | 7.37 | 52.0 | 0.00 | 3.79 | |
| HD 200560 | 200560 | 103859 | | J21028+4552CD | K2.5V | 51.36 | 0.63 | 7.68 | 19.5 | 0.00 | 6.23 | |
| CCDM J21031+0132AB | 200375 | 103892 | 8056 | J21031+0132AB | F5/6V | 14.12 | 0.64 | 6.25 | 70.8 | 0.00 | 2.00 | |
| HD 200580 | 200580 | 103987 | | | F9V | 19.27 | 0.99 | 7.31 | 51.9 | 0.00 | 3.73 | |
| HD 200779 | 200779 | 104092 | | J21054+0704A | K6V | 66.41 | 0.95 | 8.27 | 15.1 | 0.00 | 7.38 | |
| 4 Equ | 200790 | 104101 | 8077 | J21054+0558A | F8V | 20.44 | 1.68 | 5.97 | 48.9 | 0.00 | 2.52 | |
| 61 Cyg A | 201091 | 104214 | 8085 | J21069+3844A | K5V | 286.82 | 6.78 | 5.21 | 3.5 | 0.00 | 7.50 | |
| 61 Cyg B | 201092 | 104217 | 8086 | J21069+3844B | K7V | 285.88 | 0.54 | 6.03 | 3.5 | 0.00 | 8.31 | |
| HD 201099 | 201099 | 104294 | | J21077-0534A | F7V | 21.13 | 0.80 | 7.60 | 47.3 | 0.00 | 4.22 | |
| HD 201456 | 201456 | 104344 | | | F8V | 14.97 | 0.58 | 7.88 | 66.8 | 0.00 | 3.76 | |
| HD 201496 | 201496 | 104549 | | | G1V | 19.04 | 0.90 | 8.05 | 52.5 | 0.00 | 4.45 | |
| HD 201891 | 201891 | 104659 | | | G5V_Fe-2.5 | 29.10 | 0.64 | 7.37 | 34.4 | 0.00 | 4.69 | |
| HD 202582 | 202582 | 104788 | 8133 | J21137+6424AB | G2IV+G2IV | 23.39 | 0.42 | 6.39 | 42.8 | 0.00 | 3.24 | |
| HD 202206 | 202206 | 104903 | | | G6V | 22.06 | 0.82 | 8.07 | 45.3 | 0.00 | 4.79 | H |
| HD 202282 | 202282 | 104922 | | | G3V | 12.40 | 1.42 | 8.96 | 80.6 | 0.00 | 4.42 | |
| HD 202575 | 202575 | 105038 | | | K3V | 61.02 | 0.89 | 7.88 | 16.4 | 0.00 | 6.81 | |
| alf Cep | 203280 | 105199 | 8162 | J21186+6236A | A8Vn | 66.50 | 0.11 | 2.46 | 15.0 | 0.00 | 1.57 | |
| V* V457 Vul | 203030A | 105232 | | | G8V | 24.46 | 0.74 | 8.43 | 40.9 | 0.00 | 5.37 | |
| HD 202940 | 202940 | 105312 | 8148 | | G7V | 55.65 | 0.62 | 6.68 | 18.0 | 0.00 | 5.41 | |
| HD 203235 | 203235 | 105399 | | J21209+0307AB | F7/8V | 14.09 | 0.99 | 7.86 | 71.0 | 0.00 | 3.60 | |
| HR 8170 | 203454 | 105406 | 8170 | | F8V | 37.02 | 0.43 | 6.40 | 27.0 | 0.00 | 4.24 | |
| HD 204426 | 204426 | 105557 | | | G0 | 25.26 | 0.32 | 6.87 | 39.6 | 0.00 | 3.88 | |
| HD 204153 | 204153 | 105769 | 8208 | | F0V | 29.88 | 0.26 | 5.60 | 33.5 | 0.00 | 2.98 | |
| HD 204734 | 204734 | 105992 | | | K0 | 25.86 | 0.69 | 8.86 | 38.7 | 0.00 | 5.92 | |
| HD 204485 | 204485 | 106003 | 8220 | | F0V | 22.36 | 0.34 | 5.76 | 44.7 | 0.00 | 2.50 | |
| HD 205027 | 205027 | 106356 | | | F3/5V | 18.22 | 0.85 | 8.31 | 54.9 | 0.00 | 4.61 | |
| HD 205434 | 205434 | 106400 | | | K4 | 43.03 | 0.96 | 9.37 | 23.2 | 0.00 | 7.54 | |
| HD 205700 | 205700 | 106674 | | | F5V | 9.29 | 0.72 | 8.24 | 107.6 | 0.01 | 3.06 | |
| HD 205702 | 205702 | 106707 | | | F8 | 17.53 | 0.71 | 7.62 | 57.0 | 0.00 | 3.84 | |

| Name | HD | HIP | HR | CCDM | SpType |  |  |  |  |  |  | H |
|---|---|---|---|---|---|---|---|---|---|---|---|---|
| BD+27 4120 | | 106811 | | | M1V | 75.09 | 1.89 | 9.95 | 13.3 | 0.00 | 9.32 | |
| HD 206374 | 206374 | 107070 | | | G6.5V | 37.12 | 0.69 | 7.46 | 26.9 | 0.00 | 5.31 | |
| HD 206282 | 206282 | 107072 | | | F5/6V | 10.66 | 0.70 | 7.93 | 93.8 | 0.00 | 3.06 | |
| 42 Cap | 206301 | 107095 | 8283 | | G1IV | 30.09 | 0.32 | 5.18 | 33.2 | 0.00 | 2.57 | |
| mu.01 Cyg | 206826 | 107310 | 8309 | J21442+2845A | F7V | 44.97 | 0.43 | 4.51 | 22.2 | 0.00 | 2.77 | |
| mu.02 Cyg | 206827 | | 8310 | J21442+2845B | F3V | | | 6.12 | | 0.00 | 4.38 | |
| V* HN Peg | 206860 | 107350 | 8314 | | G0V | 55.91 | 0.45 | 5.95 | 17.9 | 0.00 | 4.69 | H |
| del Cap | 207098 | 107556 | 8322 | J21470-1607A | kA5hF0mF2III | 84.27 | 0.19 | 2.83 | 11.9 | 0.00 | 2.46 | |
| HD 207858 | 207858 | 107890 | | | F6V | 7.89 | 0.82 | 8.56 | 126.7 | 0.01 | 3.02 | |
| 15 Peg | 207978 | 107975 | 8354 | | F2V | 36.43 | 0.32 | 5.53 | 27.4 | 0.00 | 3.34 | |
| HD 208038 | 208038 | 108028 | | | K2.5V | 43.40 | 0.75 | 8.15 | 23.0 | 0.00 | 6.34 | |
| HD 208313 | 208313 | 108156 | | | K0V | 50.11 | 0.80 | 7.78 | 20.0 | 0.00 | 6.28 | |
| HD 208801 | 208801 | 108506 | 8382 | | G8/K0III | 27.11 | 0.41 | 6.22 | 36.9 | 0.00 | 3.39 | |
| V* V376 Peg | 209458 | 108859 | | | G0V | 20.15 | 0.80 | 7.63 | 49.6 | 0.00 | 4.15 | H |
| HD 209472 | 209472 | 108901 | | | F3V | 10.57 | 0.94 | 8.32 | 94.6 | 0.00 | 3.43 | |
| iot Peg | 210027 | 109176 | 8430 | J22070+2520A | F5V | 85.28 | 0.63 | 3.77 | 11.7 | 0.00 | 3.42 | |
| CCDM J22071+0034AB | 209965 | 109186 | | J22071+0034AB | F7V | 12.71 | 1.33 | 7.60 | 78.7 | 0.00 | 3.12 | |
| V* AR Lac | 210334 | 109303 | 8448 | | K0IVe+G5IV | 23.38 | 0.35 | 6.11 | 42.8 | 0.00 | 2.95 | |
| HD 210640 | 210640 | 109324 | | | F2 | 11.82 | 0.42 | 7.76 | 84.6 | 0.01 | 3.11 | |
| HD 210277 | 210277 | 109378 | | | G8V | 46.38 | 0.48 | 6.63 | 21.6 | 0.00 | 4.96 | H |
| HD 210460 | 210460 | 109439 | 8455 | J22103+1937AB | G0V | 16.88 | 0.34 | 6.19 | 59.2 | 0.00 | 2.33 | |
| HD 210483 | 210483 | 109450 | | | G1V | 19.56 | 0.61 | 7.59 | 51.1 | 0.00 | 4.05 | |
| BD+17 4708 | | 109558 | | | sdF8 | 8.21 | 1.26 | 9.46 | 121.8 | 0.01 | 4.01 | |
| Wolf 1008 | 210631 | 109563 | | | G0 | 12.72 | 0.98 | 8.51 | 78.6 | 0.00 | 4.03 | |
| HD 210855 | 210855 | 109572 | 8472 | J22119+5650A | F8V | 26.77 | 0.18 | 5.20 | 37.4 | 0.00 | 2.34 | |
| HD 210752 | 210752 | 109646 | | | G1V | 27.64 | 0.68 | 7.40 | 36.2 | 0.00 | 4.61 | |
| HD 211476 | 211476 | 110035 | | | G2V | 32.22 | 0.52 | 7.04 | 31.0 | 0.00 | 4.58 | |
| HD 211575 | 211575 | 110091 | 8507 | | F3V | 23.82 | 0.64 | 6.40 | 42.0 | 0.00 | 3.28 | |
| HD 239928 | 239928 | 110327 | | | G2V | 15.64 | 0.78 | 8.69 | 63.9 | 0.00 | 4.66 | |
| 34 Peg | 212754 | 110785 | 8548 | J22266+0424AB | F7V | 26.21 | 0.93 | 5.75 | 38.2 | 0.00 | 2.84 | |
| HD 213338 | 213338 | 111029 | | | G8V | 23.43 | 0.70 | 8.41 | 42.7 | 0.00 | 5.26 | |
| HD 214385 | 214385 | 111746 | | | G8VFe-1.2 | 27.45 | 0.91 | 7.88 | 36.4 | 0.00 | 5.07 | |
| HD 214683 | 214683 | 111888 | | | K2/3V | 41.49 | 0.76 | 8.44 | 24.1 | 0.00 | 6.52 | |
| HD 214749 | 214749 | 111960 | | | K4.5Vk | 73.80 | 0.77 | 7.84 | 13.6 | 0.00 | 7.18 | |
| HD 214850 | 214850 | 111974 | 8631 | J22409+1433AB | G4V | 29.59 | 0.68 | 5.71 | 33.8 | 0.00 | 3.07 | |
| HD 215243 | 215243 | 112222 | 8653 | | G8IV | 23.57 | 1.64 | 6.52 | 42.4 | 0.00 | 3.38 | |
| ksi Peg | 215648 | 112447 | 8665 | J22467+1211A | F6V | 61.36 | 0.19 | 4.20 | 16.3 | 0.00 | 3.14 | |
| HD 215625 | 215625 | 112462 | | | G0V | 17.63 | 0.80 | 7.90 | 56.7 | 0.00 | 4.13 | |

| Name | HD | HIP | HR | Other | SpType | π | σπ | V | d | E(B-V) | M_V | Note |
|---|---|---|---|---|---|---|---|---|---|---|---|---|
| HD 216520 | 216520 | 112527 | | | K0V | 50.83 | 0.44 | 7.52 | 19.7 | 0.00 | 6.05 | |
| BD+67 1468A | 216172A | 112670 | | | F5V | 15.70 | 0.23 | 6.86 | 63.7 | 0.00 | 2.84 | |
| BD+67 1468B | 216172B | | | | F5 | | | 6.97 | | 0.00 | 2.95 | |
| HD 216133 | 216133 | 112774 | | | M0.5V | 70.77 | 1.85 | 9.85 | 14.1 | 0.00 | 9.10 | |
| HD 216259 | 216259 | 112870 | | J22514+1359A | K2.5V | 46.99 | 1.01 | 8.29 | 21.3 | 0.00 | 6.65 | |
| 49 Peg | 216385 | 112935 | 8697 | | F6V | 36.66 | 0.29 | 5.16 | 27.3 | 0.00 | 2.98 | |
| BD-15 6290 | | 113020 | | | M3.5V | 213.28 | 2.12 | 10.19 | 4.7 | 0.00 | 11.84 | H |
| HD 216770 | 216770 | 113238 | | | G9VCN+1 | 28.11 | 0.79 | 8.10 | 35.6 | 0.00 | 5.34 | H |
| 51 Peg | 217014 | 113357 | 8729 | | G2.5IVa | 64.07 | 0.38 | 5.46 | 15.6 | 0.00 | 4.49 | H |
| alf PsA | 216956 | 113368 | 8728 | | A4V | 129.81 | 0.47 | 1.16 | 7.7 | 0.00 | 1.73 | H |
| HD 217107 | 217107 | 113421 | 8734 | J22583-0224AB | G8IV/V | 50.36 | 0.38 | 6.18 | 19.9 | 0.00 | 4.69 | H |
| HD 217357 | 217357 | 113576 | | | K7+Vk | 121.69 | 0.69 | 7.87 | 8.2 | 0.00 | 8.30 | |
| HD 217577 | 217577 | 113688 | | | G2V | 12.07 | 0.88 | 8.67 | 82.9 | 0.00 | 4.08 | |
| HD 217813 | 217813 | 113829 | | | G1V | 40.46 | 0.57 | 6.62 | 24.7 | 0.00 | 4.65 | |
| HD 217877 | 217877 | 113896 | 8772 | | G0-V | 34.03 | 0.77 | 6.68 | 29.4 | 0.00 | 4.34 | |
| HD 217958 | 217958 | 113948 | | | G3V | 17.62 | 0.83 | 8.05 | 56.8 | 0.00 | 4.28 | |
| HD 218059 | 218059 | 113980 | | | F5V | 22.39 | 0.45 | 7.07 | 44.7 | 0.00 | 3.82 | |
| HD 218209 | 218209 | 113989 | | | G6V | 33.85 | 0.39 | 7.48 | 29.5 | 0.00 | 5.13 | |
| HD 218101 | 218101 | 113994 | 8784 | | G8IV | 27.61 | 0.56 | 6.44 | 36.2 | 0.00 | 3.65 | |
| HD 218566 | 218566 | 114322 | | | K3V | 35.02 | 1.14 | 9.21 | 28.6 | 0.00 | 6.93 | H |
| HD 218687 | 218687 | 114378 | | J23099+1426A | G0V | 40.28 | 0.53 | 6.54 | 24.8 | 0.00 | 4.57 | |
| 6 And | 218804 | 114430 | 8825 | | F5IV | 36.17 | 1.39 | 6.00 | 27.6 | 0.00 | 3.79 | |
| HD 218868 | 218868 | 114456 | | J23108+4531A | G8V | 41.15 | 0.54 | 7.00 | 24.3 | 0.00 | 5.07 | |
| HD 219396 | 219396 | 114546 | | | G0 | 19.01 | 0.43 | 7.52 | 52.6 | 0.00 | 3.91 | |
| HR 8832 | 219134 | 114622 | 8832 | J23132+5709A | K3V | 152.76 | 0.29 | 5.57 | 6.5 | 0.00 | 6.49 | H |
| HD 219428 | 219428 | 114796 | | | G0 | 16.27 | 0.74 | 8.25 | 61.5 | 0.00 | 4.31 | |
| HD 219420 | 219420 | 114834 | | | F7V | 24.56 | 1.90 | 6.76 | 40.7 | 0.00 | 3.71 | |
| HD 219538 | 219538 | 114886 | | | K2V | 41.63 | 0.72 | 8.09 | 24.0 | 0.00 | 6.19 | |
| HD 219623 | 219623 | 114924 | 8853 | J23167+5313A | F7V | 48.77 | 0.26 | 5.60 | 20.5 | 0.00 | 4.04 | |
| 94 Aqr B | 219834B | 115125 | | J23191-1328B | K2 | | | 6.97 | | 0.00 | 5.35 | |
| 94 Aqr | 219834A | 115126 | 8866 | J23191-1328A | G8.5IV | 47.35 | 2.47 | 5.19 | 21.1 | 0.00 | 3.57 | |
| HD 220140 | 220140 | 115147 | | J23194+7900A | G9V | 52.07 | 0.47 | 7.54 | 19.2 | 0.00 | 6.12 | |
| HD 220008 | 220008 | 115223 | | | G4V | 11.84 | 0.71 | 7.97 | 84.5 | 0.00 | 3.33 | |
| HD 220182 | 220182 | 115331 | | | G9V | 46.46 | 0.53 | 7.36 | 21.5 | 0.00 | 5.70 | |
| HD 220221 | 220221 | 115341 | | | K3V(k) | 48.82 | 0.93 | 8.13 | 20.5 | 0.00 | 6.57 | |
| V* NX Aqr | 220476 | 115527 | | | G5V | 33.20 | 0.69 | 7.61 | 30.1 | 0.00 | 5.22 | |
| HD 220689 | 220689 | 115662 | | | G3V | 22.44 | 0.70 | 7.76 | 44.6 | 0.00 | 4.52 | H |
| HR 8924 | 221148 | 115953 | 8924 | | K2III | 21.14 | 0.56 | 6.25 | 47.3 | 0.00 | 2.88 | |

| Name | HD | HIP | HR | CCDM | SpType | P | e_P | V | d | E(B-V) | Mv | Host |
|---|---|---|---|---|---|---|---|---|---|---|---|---|
| HD 221354 | 221354 | 116085 | | | K0V | 59.06 | 0.45 | 6.74 | 16.9 | 0.00 | 5.60 | |
| HD 221356 | 221356 | 116106 | 8931 | | F7V | 38.29 | 0.54 | 6.49 | 26.1 | 0.00 | 4.41 | |
| HD 221445 | 221445 | 116164 | | J23322+0705AB | F7+F8VV | 13.04 | 0.88 | 6.80 | 76.7 | 0.00 | 2.38 | |
| HD 221503 | 221503 | 116215 | | J23328-1650A | K6V | 65.98 | 1.96 | 8.61 | 15.2 | 0.00 | 7.70 | |
| HD 221585 | 221585 | 116221 | | | G8IV | 17.40 | 0.60 | 7.44 | 57.5 | 0.00 | 3.64 | H |
| HD 221851 | 221851 | 116416 | | | K1V | 42.00 | 0.72 | 7.90 | 23.8 | 0.00 | 6.02 | |
| HR 8964 | 222143 | 116613 | 8964 | | G3/4V | 42.86 | 0.42 | 6.59 | 23.3 | 0.00 | 4.75 | |
| HD 222155 | 222155 | 116616 | | | G0 | 20.38 | 0.62 | 7.12 | 49.1 | 0.00 | 3.67 | H |
| gam Cep | 222404 | 116727 | 8974 | | K1III-IVCN1 | 70.91 | 0.40 | 3.22 | 14.1 | 0.00 | 2.47 | H |
| iot Psc | 222368 | 116771 | 8969 | J23399+0538A | F7V | 72.92 | 0.15 | 4.12 | 13.7 | 0.00 | 3.43 | |
| HD 222582 | 222582 | 116906 | | J23419-0559A | G5V | 23.94 | 0.74 | 7.69 | 41.8 | 0.00 | 4.59 | H |
| HD 222645 | 222645 | 116964 | | | F8V | 14.16 | 1.23 | 8.24 | 70.6 | 0.00 | 4.00 | |
| HD 223084 | 223084 | 117258 | | | F9V | 25.92 | 0.61 | 7.24 | 38.6 | 0.00 | 4.31 | |
| HD 223110 | 223110 | 117259 | | | F5V | 9.70 | 0.69 | 7.85 | 103.1 | 0.01 | 2.74 | |
| HD 223238 | 223238 | 117367 | | | G5V | 20.89 | 0.81 | 7.69 | 47.9 | 0.00 | 4.29 | |
| HD 224465 | 224465 | 118162 | | | G4V | 40.77 | 0.49 | 6.64 | 24.5 | 0.00 | 4.69 | |
| HR 9074 | 224635 | 118281 | 9074 | J23594+3343A | G0V | 34.57 | 0.51 | 5.81 | 28.9 | 0.00 | 3.50 | |
| HR 9075 | 224636 | | 9075 | J23594+3343B | G0V | | | 6.60 | | 0.00 | 4.29 | |
| HD 79211 | 79211 | 120005 | | J09144+5241B | K7V | 156.45 | 8.58 | 7.72 | 6.4 | 0.00 | 8.69 | |

Notes.     Information in the first nine columns from SIMBAD.
                  CCDM is the Catalog of Double and Multiple Stars.
                  P = parallax in milliarcseconds
                  e_P = error in parallax in milliarcseconds
                  V = Johnson V apparent magnitude
                  d = distance in parsecs.
                  E(B-V) = B-V color excess computed from the extinction method of Hakkila et al. (1997) except for d < 75 pc the extinction is set to 0.
                  Mv = Johnson V band absolute magnitude.
                  Host = Planet host status – H = known host.    Source is The Extrasolar Planets Encyclopedia (Exoplanets team 2016).

Table 2
Temperature, Luminosity, Mass, Age, and Gravity

| Primary | Sp | T (K) | Sig (K) | N | log L/L$_S$ | B1 Mass (M$_S$) | B1 Age (Gyr) | D Mass (M$_S$) | D Age (Gyr) | Y Mass (M$_S$) | Y Age (Gyr) | B2 Mass (M$_S$) | B2 Age (Gyr) | <Mass> (M$_S$) | Range (M$_S$) | <Age> (Gyr) | Range (Gyr) | log g cm s$^{-2}$ |
|---|---|---|---|---|---|---|---|---|---|---|---|---|---|---|---|---|---|---|
| 10 CVn | S | 5987 | 90 | 9 | 0.01 | 0.94 | 6.31 | 0.86 | 8.00 | 0.94 | 5.42 | 0.90 | 7.12 | 0.91 | 0.08 | 6.71 | 2.58 | 4.44 |
| 10 Tau | S | 6014 | 83 | 11 | 0.49 | 1.02 | 7.94 | 0.97 | 9.00 | 1.33 | 3.00 | 1.02 | 7.75 | 1.09 | 0.36 | 6.92 | 6.00 | 4.05 |
| 107 Psc | S | 5259 | 55 | 9 | -0.36 | 0.91 | 1.62 | 0.89 | 1.18 | 0.89 | 2.33 | 0.88 | 2.09 | 0.89 | 0.03 | 1.80 | 1.15 | 4.58 |
| 109 Psc | S | 5604 | 46 | 7 | 0.46 | 1.01 | 9.47 | 1.11 | 7.75 | 1.05 | 8.50 | 1.07 | 8.42 | 1.06 | 0.10 | 8.53 | 1.72 | 3.94 |
| 11 Aql | S | 6144 | 29 | 7 | 1.18 | | | 1.47 | 2.25 | 1.78 | 1.80 | 1.61 | 1.50 | 1.62 | 0.31 | 1.85 | 0.75 | 3.57 |
| 11 Aqr | S | 5929 | 52 | 8 | 0.31 | 1.18 | 2.51 | 1.17 | 4.50 | | | 1.18 | 4.63 | 1.18 | 0.01 | 3.88 | 2.11 | 4.24 |
| 11 LMi | S | 5498 | 48 | 9 | -0.09 | 0.98 | 4.50 | 1.00 | 4.36 | 1.00 | 4.50 | 0.95 | 9.27 | 0.98 | 0.05 | 5.66 | 4.91 | 4.43 |
| 110 Her | S | 6457 | 52 | 7 | 0.79 | | | 1.20 | 4.33 | | | 1.36 | 2.83 | 1.28 | 0.16 | 3.58 | 1.50 | 3.94 |
| 111 Tau | S | 6184 | 23 | 3 | 0.22 | 1.14 | 1.93 | 1.11 | 2.79 | 1.16 | 1.41 | 1.13 | 2.03 | 1.14 | 0.05 | 2.04 | 1.38 | 4.38 |
| 111 Tau B | S | 4576 | 32 | 9 | -0.77 | | | 0.75 | 0.83 | 0.71 | 7.50 | 0.74 | 0.96 | 0.73 | 0.04 | 3.10 | 6.67 | 4.66 |
| 112 Psc | S | 6031 | 42 | 7 | 0.61 | | | 1.37 | 3.25 | 1.31 | 4.00 | 1.41 | 2.88 | 1.36 | 0.10 | 3.38 | 1.13 | 4.03 |
| 12 Oph | S | 5262 | 33 | 12 | -0.35 | 0.90 | 2.19 | 0.89 | 2.01 | 0.90 | 1.73 | 0.88 | 3.62 | 0.89 | 0.02 | 2.39 | 1.88 | 4.57 |
| 13 Cet | S | 6080 | 80 | 10 | 0.49 | 1.17 | 5.01 | 0.94 | 9.38 | | | 1.21 | 4.69 | 1.11 | 0.27 | 6.36 | 4.69 | 4.07 |
| 13 Ori | S | 5800 | 36 | 9 | 0.35 | 1.00 | 9.51 | 0.96 | 10.50 | 0.94 | 10.50 | 0.95 | 11.00 | 0.96 | 0.06 | 10.38 | 1.49 | 4.07 |
| 13 Tri | S | 5957 | 36 | 5 | 0.56 | 1.22 | 3.98 | 0.90 | 9.83 | 1.05 | 7.00 | 1.09 | 6.50 | 1.07 | 0.32 | 6.83 | 5.85 | 3.95 |
| 14 Cet | S | 6512 | 29 | 9 | 0.99 | 1.68 | 1.26 | 1.22 | 3.38 | 1.57 | 2.15 | 1.54 | 1.92 | 1.50 | 0.46 | 2.18 | 2.12 | 3.83 |
| 14 Her | S | 5248 | 79 | 9 | -0.18 | | | 0.92 | 9.94 | 0.93 | 8.50 | | | 0.93 | 0.01 | 9.22 | 1.44 | 4.41 |
| 15 LMi | S | 5924 | 40 | 7 | 0.41 | 0.97 | 10.00 | 1.06 | 7.63 | 1.03 | 8.00 | 0.99 | 9.00 | 1.01 | 0.09 | 8.66 | 2.38 | 4.07 |
| 15 Sge | S | 5946 | 29 | 9 | 0.10 | 1.03 | 5.03 | 1.06 | 3.36 | 1.04 | 4.17 | 1.08 | 3.10 | 1.05 | 0.05 | 3.92 | 1.93 | 4.40 |
| 16 Cyg A | S | 5800 | 37 | 5 | 0.19 | 1.07 | 5.55 | 1.03 | 7.75 | 1.16 | 4.00 | 1.03 | 8.75 | 1.07 | 0.13 | 6.51 | 4.75 | 4.28 |
| 16 Cyg B | S | 5753 | 51 | 9 | 0.10 | 1.08 | 3.16 | 1.00 | 7.50 | 1.06 | 5.00 | 1.00 | 8.75 | 1.04 | 0.08 | 6.10 | 5.59 | 4.34 |
| 17 Crt A | S | 6240 | 44 | 5 | 0.50 | 1.20 | 3.95 | 1.18 | 4.44 | 1.36 | 2.00 | 1.32 | 2.50 | 1.27 | 0.18 | 3.22 | 2.44 | 4.17 |
| 17 Crt B | S | 6269 | 77 | 5 | 0.45 | 1.18 | 3.67 | 0.97 | 8.31 | 1.31 | 2.00 | 1.30 | 2.25 | 1.19 | 0.34 | 4.06 | 6.31 | 4.20 |
| 17 Vir | S | 6146 | 42 | 5 | 0.28 | 1.14 | 3.16 | 1.12 | 3.75 | 1.24 | 1.30 | 1.18 | 2.32 | 1.17 | 0.12 | 2.63 | 2.45 | 4.33 |
| 18 Cam | S | 5958 | 46 | 9 | 0.60 | 1.20 | 4.83 | 0.99 | 8.13 | 1.17 | 5.50 | 1.09 | 6.25 | 1.11 | 0.21 | 6.18 | 3.30 | 3.93 |
| 18 Cet | S | 5861 | 44 | 9 | 0.45 | 0.90 | 11.49 | 0.91 | 10.25 | 1.00 | 8.50 | 1.00 | 8.75 | 0.95 | 0.10 | 9.75 | 2.99 | 3.99 |
| 18 Sco | S | 5791 | 31 | 6 | 0.02 | 1.01 | 5.01 | 1.03 | 3.50 | 0.99 | 5.33 | 1.05 | 3.22 | 1.02 | 0.06 | 4.27 | 2.11 | 4.42 |
| 20 LMi | S | 5771 | 32 | 5 | 0.14 | 1.09 | 3.16 | 1.06 | 6.04 | 1.08 | 4.75 | 1.05 | 7.25 | 1.07 | 0.04 | 5.30 | 4.09 | 4.32 |
| 21 Eri | S | 5150 | 163 | 7 | 0.67 | 1.18 | 5.01 | 1.25 | 5.56 | 1.30 | 4.50 | 1.28 | 4.35 | 1.25 | 0.12 | 4.86 | 1.21 | 3.66 |
| 23 Lib | S | 5717 | 32 | 7 | 0.19 | | | 1.09 | 6.25 | | | 1.06 | 8.00 | 1.08 | 0.03 | 7.13 | 1.75 | 4.26 |
| 24 LMi | S | 5760 | 79 | 9 | 0.39 | 1.03 | 8.81 | 0.93 | 11.00 | 0.97 | 10.00 | 1.01 | 9.63 | 0.99 | 0.10 | 9.86 | 2.19 | 4.03 |
| 26 Dra | S | 5925 | 52 | 7 | 0.13 | 0.97 | 8.52 | 1.05 | 4.75 | 1.13 | 2.36 | 1.08 | 4.05 | 1.06 | 0.16 | 4.92 | 6.16 | 4.37 |

| Name | | HD | | | | | | | | | | | | | | | | |
|---|---|---|---|---|---|---|---|---|---|---|---|---|---|---|---|---|---|---|
| 33 Sex | S | 5124 | 45 | 7 | 0.63 | 1.30 | 3.98 | 1.31 | 4.50 | 1.37 | 4.00 | 1.38 | 3.70 | 1.34 | 0.08 | 4.05 | 0.80 | 3.72 |
| 35 Leo | S | 5736 | 46 | 9 | 0.54 | 1.09 | 7.13 | 1.08 | 7.40 | 1.20 | 5.67 | 1.13 | 6.75 | 1.13 | 0.12 | 6.74 | 1.73 | 3.93 |
| 36 And | S | 4809 | 105 | 6 | 1.02 | 1.34 | 6.80 | 1.24 | 6.13 | | | 1.16 | 7.75 | 1.25 | 0.18 | 6.89 | 1.63 | 3.19 |
| 36 Oph A | S | 5103 | 29 | 5 | -0.55 | 0.77 | 5.36 | 0.71 | 4.70 | 0.75 | 4.75 | 0.75 | 4.64 | 0.75 | 0.06 | 4.86 | 0.72 | 4.64 |
| 36 Oph B | S | 5199 | 63 | 2 | -0.49 | 0.81 | 1.46 | 0.70 | 9.25 | 0.78 | 3.13 | 0.76 | 4.72 | 0.76 | 0.11 | 4.64 | 7.79 | 4.62 |
| 36 UMa | S | 6173 | 40 | 3 | 0.19 | 1.11 | 2.58 | 1.13 | 1.36 | 1.15 | 1.58 | 1.06 | 3.70 | 1.11 | 0.09 | 2.30 | 2.34 | 4.40 |
| 37 Gem | S | 5932 | 18 | 5 | 0.09 | 1.02 | 5.16 | 1.02 | 4.83 | 1.04 | 4.17 | 1.01 | 5.38 | 1.02 | 0.03 | 4.88 | 1.21 | 4.40 |
| 38 LMi | S | 6090 | 37 | 7 | 1.01 | 1.50 | 2.51 | 1.53 | 2.75 | 1.71 | 1.98 | 1.59 | 2.00 | 1.58 | 0.21 | 2.31 | 0.78 | 3.71 |
| 39 Gem | S | 6112 | 88 | 9 | 0.79 | 1.17 | 4.50 | 1.17 | 4.88 | 1.19 | 4.50 | 1.22 | 3.90 | 1.19 | 0.05 | 4.44 | 0.98 | 3.81 |
| 39 Leo | S | 6187 | 21 | 3 | 0.30 | 1.03 | 6.31 | 1.09 | 4.56 | 1.18 | 2.50 | | | 1.10 | 0.15 | 4.46 | 3.81 | 4.29 |
| 39 Ser | S | 5830 | 40 | 7 | -0.03 | 0.99 | 5.02 | 1.01 | 2.28 | 0.90 | 7.67 | 0.85 | 10.71 | 0.94 | 0.16 | 6.42 | 8.43 | 4.45 |
| 39 Tau | S | 5836 | 21 | 7 | 0.02 | 0.99 | 5.42 | 1.06 | 1.88 | 1.03 | 3.46 | 1.03 | 3.92 | 1.03 | 0.07 | 3.67 | 3.54 | 4.44 |
| 4 Equ | S | 6086 | 92 | 9 | 0.91 | 1.45 | 2.61 | 1.21 | 3.97 | 1.53 | 2.75 | 1.42 | 2.68 | 1.40 | 0.32 | 3.00 | 1.36 | 3.76 |
| 42 Cap | S | 5706 | 28 | 3 | 0.91 | 1.44 | 2.51 | 1.24 | 3.81 | 1.51 | 3.00 | 1.45 | 2.33 | 1.41 | 0.27 | 2.91 | 1.48 | 3.65 |
| 44 And | S | 5951 | 31 | 3 | 1.09 | 1.58 | 2.00 | 1.41 | 2.58 | | | 1.54 | 1.75 | 1.51 | 0.17 | 2.11 | 0.83 | 3.57 |
| 47 UMa | S | 5960 | 60 | 7 | 0.20 | 1.14 | 1.74 | 1.10 | 4.38 | 1.15 | 2.96 | 1.11 | 4.13 | 1.13 | 0.05 | 3.30 | 2.64 | 4.34 |
| 49 Lib | S | 6297 | 39 | 5 | 0.81 | 1.36 | 3.16 | 1.18 | 4.50 | 1.53 | 2.25 | 1.38 | 2.89 | 1.36 | 0.35 | 3.20 | 2.25 | 3.91 |
| 49 Per | S | 4984 | 50 | 6 | 0.87 | 1.43 | 3.17 | 1.38 | 3.75 | 1.39 | 4.25 | 1.32 | 4.56 | 1.38 | 0.11 | 3.93 | 1.39 | 3.45 |
| 5 Ser | S | 6134 | 39 | 6 | 0.70 | 1.21 | 4.50 | 1.11 | 5.85 | 1.48 | 2.50 | 1.36 | 3.33 | 1.29 | 0.37 | 4.04 | 3.35 | 3.95 |
| 50 Per | S | 6313 | 64 | 5 | 0.34 | 1.21 | 1.58 | 1.25 | 1.10 | 1.23 | 1.50 | 1.21 | 1.75 | 1.23 | 0.04 | 1.48 | 0.65 | 4.33 |
| 51 Boo Bn | S | 5821 | 125 | 3 | 0.02 | 1.02 | 4.05 | 1.05 | 2.64 | 0.98 | 5.63 | 1.03 | 3.97 | 1.02 | 0.07 | 4.07 | 2.98 | 4.43 |
| 51 Boo Bs | S | 5990 | 43 | 3 | 0.18 | 1.15 | 1.86 | 1.11 | 4.06 | 1.20 | 1.70 | 1.13 | 3.80 | 1.15 | 0.09 | 2.86 | 2.36 | 4.33 |
| 51 Peg | S | 5799 | 58 | 5 | 0.13 | 1.09 | 2.51 | 1.07 | 5.10 | 1.14 | 2.27 | 1.03 | 7.40 | 1.08 | 0.11 | 4.32 | 5.13 | 4.34 |
| 54 Cnc | S | 5824 | 53 | 9 | 0.56 | | | 1.03 | 7.67 | 1.21 | 5.67 | 1.15 | 6.17 | 1.13 | 0.18 | 6.50 | 2.00 | 3.94 |
| 54 Psc | S | 5289 | 64 | 9 | -0.28 | 0.90 | 5.85 | 0.92 | 3.66 | 0.95 | 2.60 | 0.91 | 5.43 | 0.92 | 0.05 | 4.38 | 3.25 | 4.52 |
| 55 Vir | S | 5054 | 81 | 12 | 1.06 | 1.75 | 1.58 | 1.19 | 6.58 | 1.41 | 3.00 | 1.28 | 3.75 | 1.41 | 0.56 | 3.73 | 5.00 | 3.29 |
| 58 Eri | S | 5839 | 52 | 10 | -0.02 | 0.99 | 5.02 | 1.01 | 2.28 | 1.03 | 2.50 | 0.99 | 3.80 | 1.01 | 0.04 | 3.40 | 2.74 | 4.47 |
| 59 Eri | S | 6275 | 177 | 10 | 0.84 | | | 1.37 | 3.46 | 1.54 | 2.25 | 1.47 | 2.44 | 1.46 | 0.17 | 2.72 | 1.21 | 3.90 |
| 61 Cyg A | S | 4481 | 96 | 8 | -0.87 | 0.66 | 6.34 | 0.60 | 9.75 | 0.70 | 2.59 | 0.64 | 6.07 | 0.65 | 0.10 | 6.19 | 7.16 | 4.67 |
| 61 Cyg B | S | 4171 | 39 | 7 | -1.08 | 0.60 | 1.88 | 0.55 | 3.40 | 0.57 | 4.64 | | | 0.57 | 0.05 | 3.31 | 2.76 | 4.70 |
| 61 Psc | S | 6273 | 60 | 9 | 0.80 | 1.36 | 3.16 | 1.29 | 3.88 | 1.59 | 2.00 | 1.36 | 3.05 | 1.40 | 0.30 | 3.02 | 1.88 | 3.92 |
| 61 UMa | S | 5507 | 40 | 9 | -0.21 | 0.97 | 1.62 | 0.95 | 1.65 | 0.94 | 3.37 | 0.94 | 2.37 | 0.95 | 0.03 | 2.25 | 1.75 | 4.54 |
| 61 Vir | S | 5578 | 26 | 6 | -0.08 | 0.99 | 4.12 | 0.96 | 6.38 | 1.04 | 2.21 | 0.95 | 8.39 | 0.99 | 0.09 | 5.27 | 6.18 | 4.44 |
| 63 Eri | S | 5422 | 122 | 8 | 1.27 | 1.85 | 1.26 | 1.57 | 1.83 | 2.00 | 1.20 | 2.13 | 0.78 | 1.89 | 0.56 | 1.27 | 1.06 | 3.33 |
| 64 Aql | S | 4736 | 77 | 9 | 1.03 | 1.60 | 2.00 | 1.00 | 11.75 | 1.13 | 8.00 | 1.09 | 9.75 | 1.21 | 0.60 | 7.87 | 9.75 | 3.14 |
| 66 Cet | S | 6102 | 22 | 5 | 0.84 | 1.39 | 2.97 | 1.35 | 3.63 | 1.47 | 3.00 | 1.36 | 3.16 | 1.39 | 0.12 | 3.19 | 0.66 | 3.83 |
| 66 Cet B | S | 5722 | 26 | 5 | 0.20 | 1.01 | 5.76 | 1.01 | 4.62 | 1.04 | 4.70 | 1.00 | 6.39 | 1.02 | 0.04 | 5.37 | 1.77 | 4.22 |
| 70 Oph A | S | 5244 | 68 | 3 | -0.26 | 0.89 | 8.68 | 0.87 | 10.31 | 0.91 | 7.00 | | | 0.89 | 0.04 | 8.66 | 3.31 | 4.47 |

| Name | | HD | N | n | B-V | | | | | | | | | | | | |
|---|---|---|---|---|---|---|---|---|---|---|---|---|---|---|---|---|---|
| 70 Vir | S | 5538 | 31 | 5 | 0.47 | 1.01 | 9.47 | 0.99 | 9.50 | 1.01 | 9.00 | 1.06 | 8.50 | 1.02 | 0.07 | 9.12 | 1.00 | 3.90 |
| 79 Cet | S | 5765 | 57 | 9 | 0.39 | 1.03 | 8.81 | 0.93 | 11.00 | 0.97 | 10.00 | 1.01 | 9.63 | 0.99 | 0.10 | 9.86 | 2.19 | 4.03 |
| 83 Leo | S | 5380 | | 1 | -0.13 | 0.93 | 7.94 | 0.97 | 6.17 | 0.98 | 5.75 | 0.92 | 12.50 | 0.95 | 0.06 | 8.09 | 6.75 | 4.42 |
| 83 Leo B | S | 4973 | 102 | 9 | -0.48 | 0.83 | 6.30 | 0.86 | 2.03 | 0.84 | 4.21 | 0.86 | 4.44 | 0.85 | 0.03 | 4.24 | 4.27 | 4.58 |
| 84 Her | S | 5810 | 67 | 7 | 0.83 | | | 1.42 | 3.67 | | | 1.45 | 2.88 | 1.44 | 0.03 | 3.27 | 0.79 | 3.77 |
| 85 Peg | S | 5454 | 59 | 9 | -0.17 | | | | | | | | | | | | | 4.54 |
| 88 Leo | S | 6030 | 33 | 9 | 0.15 | 1.10 | 2.56 | 1.05 | 3.96 | 1.09 | 2.78 | 1.10 | 2.77 | 1.09 | 0.05 | 3.02 | 1.40 | 4.39 |
| 9 Cet | S | 5822 | 62 | 12 | 0.01 | 1.03 | 3.22 | 1.06 | 1.83 | 1.05 | 2.18 | 1.05 | 2.57 | 1.05 | 0.03 | 2.45 | 1.39 | 4.46 |
| 9 Com | S | 6239 | 51 | 9 | 0.83 | | | 1.51 | 2.50 | 1.51 | 2.50 | 1.49 | 2.45 | 1.50 | 0.02 | 2.48 | 0.05 | 3.91 |
| 94 Aqr | S | 5379 | 67 | 12 | 0.54 | 1.13 | 6.42 | 1.08 | 7.54 | 1.14 | 6.75 | 1.15 | 6.42 | 1.13 | 0.07 | 6.78 | 1.13 | 3.82 |
| 94 Aqr B | S | 5219 | 85 | 2 | -0.13 | | | | | 0.92 | 11.50 | | | 0.92 | 0.00 | 11.50 | 0.00 | 4.35 |
| 94 Cet | S | 6064 | 71 | 10 | 0.59 | | | 1.30 | 4.00 | | | 1.28 | 4.13 | 1.29 | 0.02 | 4.06 | 0.13 | 4.04 |
| alf Aql | S | 7377 | 45 | 2 | 0.99 | 1.13 | 3.98 | 1.24 | 3.17 | 1.35 | 2.50 | 1.16 | 3.25 | 1.22 | 0.22 | 3.22 | 1.48 | 3.95 |
| alf Cep | S | 7217 | 333 | 3 | 1.25 | | | 1.29 | 2.50 | | | 1.34 | 2.00 | 1.32 | 0.05 | 2.25 | 0.50 | 3.69 |
| alf CMi | S | 6654 | 61 | 5 | 0.84 | | | 1.13 | 4.38 | | | 1.44 | 2.13 | 1.29 | 0.31 | 3.25 | 2.25 | 3.95 |
| alf Com A | S | 6391 | 20 | 3 | 0.56 | 1.13 | 4.86 | 1.00 | 7.15 | 1.23 | 3.25 | 1.09 | 5.00 | 1.11 | 0.23 | 5.06 | 3.90 | 4.09 |
| alf Crv | S | 7019 | 69 | 4 | 0.62 | 1.30 | 1.58 | | | 1.34 | 1.20 | 1.31 | 1.25 | 1.32 | 0.04 | 1.34 | 0.38 | 4.27 |
| alf For A | S | 6195 | 46 | 3 | 0.66 | 1.15 | 5.15 | 1.04 | 6.69 | 1.17 | 5.00 | 1.20 | 4.50 | 1.14 | 0.16 | 5.33 | 2.19 | 3.95 |
| alf PsA | S | 7671 | | 1 | 1.19 | | | 1.99 | 0.55 | | | 1.94 | 0.60 | 1.97 | 0.05 | 0.58 | 0.05 | 4.03 |
| b Aql | S | 5466 | | 1 | 0.25 | | | 1.04 | 10.00 | 1.04 | 7.00 | 1.02 | 12.00 | 1.03 | 0.02 | 9.67 | 5.00 | 4.10 |
| b01 Cyg | S | 5090 | 96 | 9 | 0.61 | 1.30 | 3.98 | 1.25 | 4.88 | 1.09 | 7.00 | 1.25 | 5.19 | 1.22 | 0.21 | 5.26 | 3.02 | 3.69 |
| bet Aql | S | 5144 | 53 | 12 | 0.78 | 1.49 | 2.51 | 1.21 | 5.88 | 1.41 | 3.50 | 1.30 | 3.83 | 1.35 | 0.28 | 3.93 | 3.36 | 3.58 |
| bet Com | S | 6022 | 41 | 9 | 0.14 | 1.10 | 2.03 | 1.11 | 1.79 | 1.09 | 2.55 | 1.06 | 3.47 | 1.09 | 0.05 | 2.46 | 1.68 | 4.40 |
| bet CVn | S | 5865 | 58 | 5 | 0.07 | 1.03 | 5.01 | 1.03 | 4.50 | 1.06 | 3.50 | 0.97 | 7.25 | 1.02 | 0.09 | 5.07 | 3.75 | 4.40 |
| bet Vir | S | 6159 | 67 | 12 | 0.55 | 1.24 | 3.82 | 1.15 | 5.33 | | | 1.24 | 4.08 | 1.21 | 0.09 | 4.41 | 1.51 | 4.08 |
| c Eri | S | 7146 | 62 | 3 | 0.73 | 1.43 | 1.11 | 1.40 | 1.61 | 1.36 | 1.70 | 1.52 | 0.60 | 1.43 | 0.16 | 1.25 | 1.10 | 4.23 |
| c UMa | S | 5995 | 57 | 7 | 0.42 | 1.14 | 5.66 | 0.93 | 9.50 | 1.29 | 3.00 | 1.11 | 6.50 | 1.12 | 0.36 | 6.17 | 6.50 | 4.12 |
| chi Cnc | S | 6274 | 56 | 9 | 0.38 | 1.20 | 2.51 | 1.01 | 7.50 | 1.23 | 2.50 | 1.05 | 5.50 | 1.12 | 0.22 | 4.50 | 5.00 | 4.25 |
| chi Dra | S | 6083 | 34 | 9 | 0.30 | 0.88 | 12.02 | 1.12 | 4.63 | 0.89 | 10.00 | 0.89 | 10.50 | 0.95 | 0.24 | 9.29 | 7.40 | 4.20 |
| chi Her | S | 5890 | 53 | 9 | 0.48 | 0.97 | 9.44 | 0.88 | 11.00 | | | 1.00 | 8.42 | 0.95 | 0.12 | 9.62 | 2.58 | 3.96 |
| chi01 Ori | S | 5983 | 28 | 5 | 0.03 | 0.93 | 7.13 | 0.93 | 6.02 | 0.98 | 4.00 | 0.95 | 5.00 | 0.95 | 0.05 | 5.54 | 3.13 | 4.44 |
| del Cap | S | 7021 | 114 | 3 | 0.88 | 1.36 | 2.51 | 1.11 | 4.25 | | | 1.41 | 2.00 | 1.29 | 0.30 | 2.92 | 2.25 | 4.00 |
| del Eri | S | 5076 | 60 | 12 | 0.50 | 1.01 | 10.19 | 1.18 | 6.50 | 1.21 | 6.00 | 1.19 | 6.50 | 1.15 | 0.20 | 7.30 | 4.19 | 3.77 |
| del Tri | S | 5796 | 22 | 3 | 0.05 | 1.00 | 6.31 | 1.00 | 5.75 | 0.90 | 10.00 | | | 0.97 | 0.10 | 7.35 | 4.25 | 4.37 |
| e Vir | S | 5999 | 32 | 6 | 0.32 | 0.96 | 10.00 | | | | | 1.02 | 8.00 | 0.99 | 0.06 | 9.00 | 2.00 | 4.17 |
| eps Eri | S | 5123 | 64 | 12 | -0.48 | 0.74 | 12.02 | | | 0.75 | 9.50 | 0.74 | 9.50 | 0.74 | 0.01 | 10.34 | 2.52 | 4.57 |
| eps For | S | 5129 | 71 | 9 | 0.66 | 0.93 | 10.72 | 1.12 | 7.50 | 1.02 | 8.00 | 1.01 | 8.17 | 1.02 | 0.19 | 8.60 | 3.22 | 3.57 |
| eta Ari | S | 6485 | 49 | 5 | 0.72 | 1.21 | 3.98 | 1.05 | 5.63 | | | 1.22 | 3.63 | 1.16 | 0.17 | 4.41 | 2.00 | 3.98 |

| Name | | T | | | | | | | | | | | | | | | |
|---|---|---|---|---|---|---|---|---|---|---|---|---|---|---|---|---|---|
| eta Boo | S | 6050 | 24 | 3 | 0.96 | 1.58 | 2.00 | 1.56 | 2.63 | 1.59 | 2.50 | 1.53 | 2.25 | 1.57 | 0.06 | 2.34 | 0.63 | 3.75 |
| eta Cas | S | 5937 | 57 | 9 | 0.10 | 1.03 | 5.16 | 1.02 | 4.54 | 1.00 | 5.63 | 0.88 | 10.70 | 0.98 | 0.15 | 6.50 | 6.16 | 4.37 |
| eta Cep | S | 5057 | 95 | 9 | 0.95 | 1.63 | 2.00 | 1.12 | 8.08 | 1.75 | 2.00 | 1.40 | 3.25 | 1.48 | 0.63 | 3.83 | 6.09 | 3.42 |
| eta CrB A | S | 6060 | 53 | 3 | 0.08 | 1.05 | 3.02 | 1.04 | 3.12 | 1.06 | 1.69 | 1.03 | 2.65 | 1.05 | 0.03 | 2.62 | 1.43 | 4.45 |
| eta CrB B | S | 5948 | 36 | 2 | -0.05 | 0.96 | 3.02 | 0.91 | 4.41 | 0.98 | 1.98 | 0.96 | 3.04 | 0.95 | 0.07 | 3.11 | 2.43 | 4.51 |
| eta Ser | S | 4985 | 40 | 11 | 1.25 | 1.90 | 1.26 | 1.40 | 4.94 | 2.07 | 1.20 | 1.32 | 3.38 | 1.67 | 0.75 | 2.69 | 3.74 | 3.15 |
| gam Cep | S | 4850 | 4 | 2 | 1.06 | | | 1.41 | 3.81 | | | 1.37 | 4.25 | 1.39 | 0.04 | 4.03 | 0.44 | 3.21 |
| gam Lep | S | 6352 | 60 | 12 | 0.37 | 1.22 | 1.58 | 1.21 | 2.13 | 1.24 | 1.50 | 1.23 | 1.50 | 1.23 | 0.03 | 1.68 | 0.63 | 4.31 |
| gam Ser | S | 6286 | 44 | 7 | 0.47 | 1.05 | 6.39 | 1.04 | 6.88 | | | 1.05 | 6.00 | 1.05 | 0.01 | 6.42 | 0.88 | 4.13 |
| gam Vir A | S | 6922 | 112 | 5 | 0.63 | 1.43 | 0.71 | 1.42 | 0.84 | 1.39 | 1.10 | 1.44 | 0.65 | 1.42 | 0.05 | 0.83 | 0.45 | 4.27 |
| gam01 Del | S | 6194 | 65 | 3 | 1.02 | 1.56 | 2.17 | 1.51 | 2.63 | 1.63 | 2.30 | 1.52 | 2.06 | 1.56 | 0.12 | 2.29 | 0.56 | 3.72 |
| i Boo A | S | 5878 | 22 | 2 | 0.08 | 1.03 | 5.01 | 1.00 | 5.83 | 1.02 | 5.00 | 0.89 | 11.25 | 0.99 | 0.14 | 6.77 | 6.25 | 4.38 |
| i Boo B | S | 5240 | 30 | 2 | -0.22 | | | | | | | | | | | | | 4.50 |
| iot Peg | S | 6504 | 23 | 5 | 0.53 | 1.31 | 2.00 | 1.10 | 5.38 | 1.32 | 1.90 | 1.14 | 3.75 | 1.22 | 0.22 | 3.26 | 3.48 | 4.19 |
| iot Per | S | 5995 | 22 | 5 | 0.35 | 0.96 | 10.00 | 1.00 | 8.25 | | | 1.07 | 7.25 | 1.01 | 0.11 | 8.50 | 2.75 | 4.15 |
| iot Psc | S | 6177 | 79 | 8 | 0.53 | 1.18 | 4.74 | 1.07 | 6.79 | | | 1.19 | 4.58 | 1.15 | 0.12 | 5.37 | 2.21 | 4.08 |
| kap Del | S | 5633 | 35 | 3 | 0.87 | | | 1.21 | 4.10 | 1.47 | 3.00 | 1.49 | 2.33 | 1.39 | 0.28 | 3.14 | 1.77 | 3.66 |
| kap For | S | 5867 | 102 | 12 | 0.53 | 1.17 | 5.55 | 0.95 | 9.50 | 1.07 | 7.00 | 1.09 | 7.00 | 1.07 | 0.22 | 7.26 | 3.95 | 3.96 |
| kap01 Cet | S | 5730 | 63 | 12 | -0.08 | 1.01 | 2.77 | 1.01 | 1.20 | 1.00 | 2.43 | 0.96 | 4.66 | 1.00 | 0.05 | 2.76 | 3.46 | 4.50 |
| ksi Boo A | S | 5480 | 33 | 3 | -0.26 | 0.80 | 12.02 | 0.95 | 0.58 | 0.83 | 8.00 | 0.79 | 11.50 | 0.84 | 0.16 | 8.02 | 11.45 | 4.53 |
| ksi Boo B | S | 4767 | | 1 | -1.04 | | | | | | | | | | | | | 5.10 |
| ksi Peg | S | 6250 | 85 | 9 | 0.65 | 1.15 | 5.10 | 1.15 | 5.31 | 1.36 | 3.00 | 1.28 | 3.63 | 1.24 | 0.21 | 4.26 | 2.31 | 4.01 |
| ksi UMa B | S | 5667 | 21 | 2 | -0.06 | 0.97 | 5.81 | 0.97 | 4.98 | 0.94 | 6.69 | | | 0.96 | 0.03 | 5.83 | 1.71 | 4.44 |
| lam Aur | S | 5931 | 36 | 9 | 0.24 | 1.16 | 2.00 | 1.12 | 5.00 | 1.20 | 3.00 | 0.92 | 11.00 | 1.10 | 0.28 | 5.25 | 9.00 | 4.28 |
| lam Ser | S | 5908 | 23 | 10 | 0.32 | 1.02 | 8.64 | 0.87 | 12.00 | 0.98 | 9.00 | 0.94 | 10.50 | 0.95 | 0.15 | 10.03 | 3.36 | 4.13 |
| m Per | S | 6704 | 81 | 9 | 1.01 | | | 1.32 | 2.94 | 1.53 | 2.00 | 1.41 | 2.13 | 1.42 | 0.21 | 2.35 | 0.94 | 3.83 |
| m Tau | S | 5631 | 79 | 9 | 0.35 | 1.02 | 10.00 | 1.01 | 10.25 | 0.97 | 10.67 | 0.95 | 11.70 | 0.99 | 0.07 | 10.65 | 1.70 | 4.03 |
| mu. Cas | S | 5434 | 25 | 3 | -0.35 | 0.85 | 2.19 | 0.74 | 10.50 | 0.77 | 10.00 | 0.75 | 11.00 | 0.78 | 0.11 | 8.42 | 8.81 | 4.57 |
| mu. Her | S | 5560 | 55 | 9 | 0.42 | | | 1.09 | 8.50 | 1.16 | 6.50 | 1.09 | 8.75 | 1.11 | 0.07 | 7.92 | 2.25 | 3.99 |
| mu.01 Cyg | S | 6354 | 42 | 3 | 0.78 | 1.35 | 3.16 | 1.19 | 4.50 | 1.46 | 2.50 | 1.22 | 3.67 | 1.31 | 0.27 | 3.46 | 2.00 | 3.93 |
| mu.02 Cnc | S | 5857 | 35 | 9 | 0.54 | 1.17 | 5.55 | 1.16 | 6.13 | 1.08 | 7.00 | 1.13 | 6.50 | 1.14 | 0.09 | 6.29 | 1.45 | 3.97 |
| mu.02 Cyg | S | 5998 | 18 | 2 | 0.16 | 0.98 | 8.15 | 1.03 | 5.30 | 1.04 | 5.00 | 0.91 | 10.00 | 0.99 | 0.13 | 7.11 | 5.00 | 4.33 |
| ome Leo | S | 5883 | 37 | 5 | 0.80 | 1.35 | 3.22 | 1.26 | 4.25 | 1.39 | 3.50 | 1.34 | 3.41 | 1.34 | 0.13 | 3.59 | 1.03 | 3.79 |
| ome Sgr | S | 5455 | 43 | 12 | 0.91 | 1.46 | 2.51 | 1.47 | 2.72 | 1.57 | 2.50 | 1.61 | 1.81 | 1.53 | 0.15 | 2.39 | 0.91 | 3.61 |
| omi Aql | S | 6173 | 90 | 7 | 0.43 | 1.29 | 1.58 | 1.09 | 6.00 | 1.32 | 2.00 | 1.19 | 4.13 | 1.22 | 0.23 | 3.43 | 4.42 | 4.20 |
| omi02 Eri | S | 5202 | 25 | 12 | -0.39 | 0.89 | 1.88 | 0.88 | 1.75 | 0.89 | 1.53 | 0.87 | 3.23 | 0.88 | 0.02 | 2.10 | 1.70 | 4.59 |
| phi02 Cet | S | 6218 | 53 | 8 | 0.23 | 1.18 | 0.92 | 1.14 | 1.86 | 1.17 | 1.46 | 1.11 | 2.78 | 1.15 | 0.07 | 1.75 | 1.86 | 4.39 |
| pi.03 Ori | S | 6509 | 81 | 12 | 0.44 | 1.28 | 1.02 | 1.32 | 0.86 | 1.30 | 0.90 | 1.27 | 1.17 | 1.29 | 0.05 | 0.98 | 0.31 | 4.31 |

| Name | | Teff | | | | | | | | | | | | | | | |
|---|---|---|---|---|---|---|---|---|---|---|---|---|---|---|---|---|---|
| psi Cap | S | 6633 | 43 | 7 | 0.57 | 1.35 | 1.58 | 1.36 | 1.50 | 1.37 | 1.40 | 1.43 | 0.65 | 1.38 | 0.08 | 1.28 | 0.93 | 4.24 |
| psi Cnc | S | 5305 | 47 | 4 | 0.92 | | | 1.18 | 5.70 | 1.50 | 2.50 | 1.68 | 1.58 | 1.45 | 0.50 | 3.26 | 4.12 | 3.53 |
| psi Ser | S | 5635 | 62 | 8 | -0.07 | 1.00 | 3.92 | 0.99 | 4.42 | 0.97 | 5.57 | 1.00 | 4.02 | 0.99 | 0.03 | 4.48 | 1.65 | 4.45 |
| psi01 Dra A | S | 6573 | 49 | 5 | 0.78 | 1.35 | 2.51 | 1.16 | 4.25 | | | | | 1.26 | 0.19 | 3.38 | 1.74 | 3.97 |
| psi01 Dra B | S | 6237 | 55 | 5 | 0.33 | 1.18 | 2.51 | 1.09 | 4.58 | 1.26 | 1.55 | 1.21 | 2.19 | 1.19 | 0.17 | 2.71 | 3.03 | 4.31 |
| psi05 Aur | S | 6128 | 33 | 3 | 0.26 | 1.13 | 3.16 | 1.11 | 3.96 | 1.21 | 1.55 | 1.16 | 2.63 | 1.15 | 0.10 | 2.82 | 2.41 | 4.34 |
| rho Cap | S | 6856 | 69 | 5 | 0.92 | 1.36 | 2.51 | 1.28 | 3.00 | 1.48 | 2.00 | 1.44 | 2.00 | 1.39 | 0.20 | 2.38 | 1.00 | 3.95 |
| rho CrB | S | 5850 | 22 | 7 | 0.24 | | | | | 0.89 | 12.00 | 0.90 | 12.50 | 0.90 | 0.01 | 12.25 | 0.50 | 4.17 |
| rho01 Cnc | S | 5248 | 79 | 7 | -0.21 | 0.87 | 11.49 | 0.92 | 8.10 | 0.94 | 6.50 | | | 0.91 | 0.07 | 8.70 | 4.99 | 4.43 |
| sig Boo | S | 6781 | 68 | 9 | 0.50 | 1.22 | 2.00 | 1.12 | 3.50 | 1.24 | 1.70 | 1.21 | 1.92 | 1.20 | 0.12 | 2.28 | 1.80 | 4.29 |
| sig CrB A | S | 5923 | 30 | 3 | 0.34 | 1.00 | 9.25 | 0.98 | 9.25 | 0.94 | 10.00 | 0.94 | 10.50 | 0.97 | 0.06 | 9.75 | 1.25 | 4.12 |
| sig CrB B | S | 5992 | 113 | 3 | 0.00 | 0.96 | 4.05 | 0.87 | 7.25 | 1.00 | 1.93 | 0.98 | 1.97 | 0.95 | 0.13 | 3.80 | 5.32 | 4.47 |
| sig Dra | S | 5338 | 50 | 5 | -0.38 | 0.82 | 5.63 | 0.72 | 12.00 | 0.82 | 4.58 | 0.79 | 6.11 | 0.79 | 0.10 | 7.08 | 7.42 | 4.57 |
| tau Boo | S | 6447 | | 1 | 0.49 | 1.27 | 2.00 | 1.37 | 0.88 | 1.40 | 0.60 | 1.37 | 0.94 | 1.35 | 0.13 | 1.10 | 1.40 | 4.26 |
| tau Cet | S | 5403 | 45 | 12 | -0.31 | 0.79 | 11.49 | | | 0.81 | 8.50 | | | 0.80 | 0.02 | 9.99 | 2.99 | 4.53 |
| tau01 Eri | S | 6395 | 56 | 12 | 0.42 | 1.25 | 2.00 | 1.26 | 1.85 | 1.26 | 1.80 | 1.28 | 1.47 | 1.26 | 0.03 | 1.78 | 0.53 | 4.29 |
| tet Boo | S | 6294 | 40 | 3 | 0.61 | 1.38 | 1.58 | 1.08 | 6.25 | | | 1.27 | 3.67 | 1.24 | 0.30 | 3.83 | 4.67 | 4.07 |
| tet Per | S | 6310 | 28 | 3 | 0.35 | 1.14 | 3.50 | 1.23 | 1.85 | 1.23 | 1.80 | 1.23 | 1.75 | 1.21 | 0.09 | 2.23 | 1.75 | 4.32 |
| tet UMa | S | 6371 | 43 | 7 | 0.89 | | | 1.12 | 4.50 | | | 1.35 | 2.83 | 1.24 | 0.23 | 3.67 | 1.67 | 3.80 |
| ups And | S | 6269 | 66 | 9 | 0.53 | | | 1.00 | 7.00 | 1.38 | 2.00 | | | 1.19 | 0.38 | 4.50 | 5.00 | 4.12 |
| w Her | S | 5780 | 77 | 9 | 0.09 | 1.00 | 7.94 | 0.99 | 7.25 | 0.90 | 11.00 | | | 0.96 | 0.10 | 8.73 | 3.75 | 4.33 |
| zet Her | S | 5759 | 50 | 5 | 0.87 | | | 1.32 | 3.56 | | | 1.42 | 2.69 | 1.37 | 0.10 | 3.13 | 0.88 | 3.69 |
| BD+04 701A | S | 6056 | 41 | 2 | -2.25 | | | | | | | | | | | | | 6.88 |
| BD+04 701B | S | 5642 | 36 | 2 | -2.27 | | | | | | | | | | | | | 6.72 |
| BD+18 2776 | S | 3644 | 68 | 3 | -1.23 | | | | | 0.40 | 8.50 | | | 0.40 | 0.00 | 8.50 | 0.00 | 4.46 |
| BD+27 4120 | S | 3899 | 98 | 8 | -1.35 | | | 0.48 | 3.23 | | | 0.53 | 5.29 | 0.51 | 0.05 | 4.26 | 2.06 | 4.80 |
| BD+29 2963 | S | 5569 | 14 | 6 | 0.22 | 1.00 | 9.98 | 0.96 | 12.25 | 1.06 | 8.67 | 1.00 | 12.00 | 1.01 | 0.10 | 10.72 | 3.58 | 4.15 |
| BD+30 2512 | S | 4313 | 79 | 9 | -0.92 | 0.70 | 1.72 | 0.71 | 0.65 | 0.68 | 5.80 | 0.67 | 4.49 | 0.69 | 0.04 | 3.16 | 5.15 | 4.68 |
| BD+33 529 | S | 3896 | 138 | 7 | -1.16 | | | | | 0.40 | 3.04 | | | 0.40 | 0.00 | 3.04 | 0.00 | 4.51 |
| BD+61 195 | S | 3799 | 130 | 8 | -1.44 | | | | | | | 0.50 | 6.54 | 0.50 | 0.00 | 6.54 | 0.00 | 4.84 |
| BD+67 1468A | S | 6680 | 76 | 3 | 0.75 | 1.52 | 0.79 | 1.57 | 0.90 | 1.57 | 0.83 | 1.56 | 0.90 | 1.56 | 0.05 | 0.86 | 0.11 | 4.13 |
| BD+67 1468B | S | 6642 | 68 | 2 | 0.71 | 1.50 | 0.79 | 1.54 | 0.88 | 1.54 | 0.83 | 1.54 | 1.05 | 1.53 | 0.04 | 0.89 | 0.26 | 4.15 |
| BD-04 782 | S | 4342 | 74 | 8 | -0.93 | 0.67 | 5.16 | 0.63 | 3.18 | 0.67 | 5.12 | 0.66 | 4.68 | 0.66 | 0.04 | 4.53 | 1.98 | 4.68 |
| BD-10 3166 | S | 5314 | 67 | 9 | -0.41 | 0.91 | 4.16 | 0.94 | 2.24 | 0.94 | 2.25 | 0.92 | 4.56 | 0.93 | 0.03 | 3.30 | 2.32 | 4.66 |
| CCDM J14534+1542AB | S | 6135 | 59 | 3 | 0.54 | 1.20 | 4.46 | 1.27 | 3.75 | 1.38 | 1.80 | 1.33 | 3.25 | 1.30 | 0.18 | 3.32 | 2.66 | 4.11 |
| GJ 282 C | S | 3866 | 68 | 8 | -1.29 | | | | | 0.54 | 3.50 | | | 0.54 | 0.00 | 3.50 | 0.00 | 4.76 |
| GJ 528 A | S | 4471 | 31 | 2 | -0.51 | | | | | | | | | | | | | 4.42 |

| Name | | T | | | [Fe/H] | | | | | | | | | | | | |
|---|---|---|---|---|---|---|---|---|---|---|---|---|---|---|---|---|---|
| GJ 528 B | S | 4384 | 5 | 2 | -0.79 | | | 0.70 | 11.25 | | | | | 0.70 | 0.00 | 11.25 | 0.00 | 4.59 |
| HD 101177 | S | 5964 | 14 | 3 | 0.08 | 1.10 | 1.21 | 1.04 | 3.32 | 1.07 | 2.39 | 1.03 | 3.67 | 1.06 | 0.07 | 2.65 | 2.46 | 4.43 |
| HD 101563 | S | 5868 | 74 | 9 | 0.62 | 1.15 | 5.66 | 1.03 | 7.33 | 1.18 | 5.50 | 1.16 | 5.50 | 1.13 | 0.15 | 6.00 | 1.83 | 3.89 |
| HD 101959 | S | 6095 | 49 | 9 | 0.15 | 1.14 | 0.92 | 1.10 | 1.66 | 1.10 | 2.36 | 1.06 | 3.43 | 1.10 | 0.08 | 2.09 | 2.51 | 4.42 |
| HD 102158 | S | 5781 | 28 | 9 | 0.10 | | | 0.99 | 7.50 | 0.90 | 11.00 | | | 0.95 | 0.09 | 9.25 | 3.50 | 4.31 |
| HD 10307 | S | 5976 | 81 | 9 | 0.15 | 0.96 | 8.85 | 1.02 | 5.30 | 1.09 | 3.53 | 1.12 | 2.53 | 1.05 | 0.16 | 5.06 | 6.32 | 4.36 |
| HD 103095 | S | 5178 | 39 | 9 | -0.67 | 0.66 | 5.36 | 0.63 | 5.18 | | | 0.63 | 5.29 | 0.64 | 0.03 | 5.28 | 0.19 | 4.72 |
| HD 103932 | S | 4585 | 23 | 8 | -0.67 | 0.75 | 10.48 | 0.75 | 9.75 | 0.75 | 12.00 | | | 0.75 | 0.00 | 10.74 | 2.25 | 4.58 |
| HD 104304 | S | 5555 | 51 | 9 | -0.06 | 0.94 | 9.16 | 1.02 | 3.94 | 1.03 | 3.74 | 0.98 | 8.27 | 0.99 | 0.09 | 6.28 | 5.42 | 4.42 |
| HD 10436 | S | 4393 | 78 | 9 | -0.93 | | | | | 0.68 | 1.16 | | | 0.68 | 0.00 | 1.16 | 0.00 | 4.72 |
| HD 106252 | S | 5890 | 45 | 9 | 0.12 | 1.13 | 0.71 | 1.03 | 5.08 | 1.13 | 2.28 | 0.99 | 7.38 | 1.07 | 0.14 | 3.86 | 6.66 | 4.37 |
| HD 106640 | S | 6003 | 90 | 9 | 0.17 | 0.93 | 10.31 | 1.04 | 4.46 | 0.91 | 9.00 | 0.89 | 10.21 | 0.94 | 0.15 | 8.50 | 5.86 | 4.30 |
| HD 108799 | S | 5878 | 84 | 5 | 0.16 | 1.10 | 3.05 | 1.03 | 6.25 | 1.16 | 2.25 | 1.02 | 7.50 | 1.08 | 0.14 | 4.76 | 5.25 | 4.33 |
| HD 108874 | S | 5555 | 37 | 9 | 0.04 | | | 1.01 | 7.26 | 1.02 | 6.83 | 0.97 | 11.13 | 1.00 | 0.05 | 8.41 | 4.29 | 4.32 |
| HD 108954 | S | 6024 | 30 | 9 | 0.11 | 1.09 | 2.20 | 1.07 | 2.53 | 1.11 | 1.56 | 1.04 | 3.62 | 1.08 | 0.07 | 2.48 | 2.05 | 4.43 |
| HD 109057 | S | 6060 | 39 | 9 | 0.23 | 1.16 | 2.14 | 1.07 | 4.53 | 1.09 | 3.83 | 1.04 | 5.68 | 1.09 | 0.12 | 4.05 | 3.54 | 4.32 |
| HD 11007 | S | 5994 | 38 | 7 | 0.49 | 0.98 | 8.97 | 0.97 | 8.88 | | | 1.06 | 7.17 | 1.00 | 0.09 | 8.34 | 1.80 | 4.01 |
| HD 110315 | S | 4505 | 51 | 8 | -0.75 | 0.76 | 5.36 | 0.71 | 8.75 | 0.73 | 11.50 | | | 0.73 | 0.05 | 8.54 | 6.14 | 4.61 |
| HD 11038 | S | 6044 | 30 | 10 | 0.15 | 1.06 | 3.71 | 1.05 | 3.43 | 0.95 | 6.05 | 0.94 | 6.21 | 1.00 | 0.12 | 4.85 | 2.77 | 4.36 |
| HD 110745 | S | 6127 | 46 | 9 | 0.42 | 1.24 | 1.94 | 1.27 | 2.09 | 1.31 | 1.51 | 1.26 | 2.46 | 1.27 | 0.07 | 2.00 | 0.96 | 4.22 |
| HD 110833 | S | 4972 | 24 | 9 | -0.45 | 0.85 | 5.36 | 0.86 | 5.15 | 0.88 | 2.96 | | | 0.86 | 0.03 | 4.49 | 2.40 | 4.56 |
| HD 110869 | S | 5783 | 19 | 9 | 0.06 | 1.09 | 1.05 | 1.06 | 3.84 | 1.12 | 1.68 | 1.04 | 5.35 | 1.08 | 0.08 | 2.98 | 4.30 | 4.41 |
| HD 111513 | S | 5822 | 38 | 9 | 0.15 | 1.11 | 2.25 | 1.08 | 5.00 | 1.14 | 3.00 | 1.04 | 7.17 | 1.09 | 0.10 | 4.36 | 4.91 | 4.33 |
| HD 111540 | S | 5729 | 39 | 2 | 0.05 | 1.03 | 3.76 | 0.98 | 5.44 | 1.04 | 3.80 | 1.01 | 5.11 | 1.02 | 0.06 | 4.53 | 1.68 | 4.37 |
| HD 111799 | S | 5842 | 24 | 9 | 0.10 | 1.09 | 2.32 | 1.02 | 5.44 | 1.10 | 3.19 | 1.05 | 4.70 | 1.07 | 0.08 | 3.91 | 3.11 | 4.38 |
| HD 112068 | S | 5745 | 35 | 6 | 0.17 | 0.98 | 10.00 | 0.98 | 9.25 | 1.06 | 6.00 | 1.01 | 9.50 | 1.01 | 0.08 | 8.69 | 4.00 | 4.26 |
| HD 112257 | S | 5659 | 38 | 9 | 0.08 | 0.96 | 10.09 | 0.95 | 10.25 | 1.03 | 7.00 | 0.99 | 10.00 | 0.98 | 0.08 | 9.34 | 3.25 | 4.31 |
| HD 112758 | S | 5190 | 31 | 12 | -0.39 | 0.89 | 1.88 | 0.87 | 1.75 | | | | | 0.88 | 0.02 | 1.82 | 0.13 | 4.58 |
| HD 113470 | S | 5839 | 63 | 8 | 0.21 | 1.10 | 4.54 | 1.02 | 7.75 | 1.11 | 5.00 | 1.03 | 8.20 | 1.07 | 0.09 | 6.37 | 3.66 | 4.27 |
| HD 113713 | S | 6316 | 27 | 9 | 0.64 | 1.19 | 4.61 | 0.98 | 7.10 | 1.14 | 5.00 | 1.13 | 4.75 | 1.11 | 0.21 | 5.37 | 2.49 | 3.99 |
| HD 114762 | S | 5921 | 53 | 9 | 0.18 | | | 0.98 | 8.00 | 0.89 | 11.00 | | | 0.94 | 0.09 | 9.50 | 3.00 | 4.27 |
| HD 114783 | S | 5118 | 42 | 9 | -0.40 | 0.84 | 7.54 | 0.87 | 3.52 | 0.84 | 6.00 | 0.86 | 5.51 | 0.85 | 0.03 | 5.64 | 4.02 | 4.55 |
| HD 11507 | S | 4011 | 102 | 11 | -1.17 | | | 0.51 | 9.50 | 0.57 | 3.73 | 0.60 | 5.29 | 0.56 | 0.09 | 6.17 | 5.77 | 4.72 |
| HD 115404A | S | 5030 | 52 | 9 | -0.54 | 0.73 | 11.75 | | | 0.73 | 9.50 | 0.74 | 6.60 | 0.73 | 0.01 | 9.28 | 5.16 | 4.60 |
| HD 115404B | S | 3903 | 42 | 3 | -1.37 | | | | | | | | | | | | | 4.90 |
| HD 115953 | S | 3751 | 62 | 5 | -0.96 | | | | | | | | | | | | | 4.36 |
| HD 117043 | S | 5558 | 42 | 9 | 0.00 | | | 0.96 | 9.14 | 1.01 | 6.33 | 0.97 | 10.90 | 0.98 | 0.05 | 8.79 | 4.57 | 4.36 |
| HD 117845 | S | 5856 | 72 | 9 | 0.00 | 1.00 | 4.79 | 1.02 | 2.31 | 1.01 | 3.48 | 0.97 | 5.36 | 1.00 | 0.05 | 3.99 | 3.05 | 4.46 |
| HD 118203 | S | 5768 | 38 | 8 | 0.61 | 1.23 | 4.50 | 1.28 | 4.98 | 1.34 | 4.20 | 1.25 | 5.13 | 1.28 | 0.11 | 4.70 | 0.93 | 3.92 |
| HD 120066 | S | 5903 | 37 | 9 | 0.39 | 0.96 | 10.24 | 1.24 | 4.25 | | | 1.03 | 8.50 | 1.08 | 0.28 | 7.66 | 5.99 | 4.11 |
| HD 120467 | S | 4369 | 51 | 12 | -0.79 | | | 0.73 | 9.25 | | | | | 0.73 | 0.00 | 9.25 | 0.00 | 4.60 |
| HD 120690 | S | 5538 | 70 | 10 | -0.05 | 0.99 | 5.01 | 0.90 | 11.50 | 0.99 | 7.00 | | | 0.96 | 0.09 | 7.84 | 6.49 | 4.39 |
| HD 120730 | S | 5300 | 50 | 11 | -0.25 | 0.91 | 5.16 | 0.88 | 4.82 | 0.93 | 3.65 | 0.90 | 5.29 | 0.91 | 0.05 | 4.73 | 1.65 | 4.49 |
| HD 122064 | S | 4811 | 13 | 4 | -0.55 | 0.82 | 5.36 | 0.82 | 5.33 | 0.83 | 3.91 | 0.79 | 11.50 | 0.82 | 0.04 | 6.52 | 7.59 | 4.57 |
| HD 122303 | S | 4067 | 83 | 4 | -1.61 | | | | | | | | | | | | | 5.27 |
| HD 122742 | S | 5459 | 38 | 5 | -0.09 | 0.96 | 6.31 | 0.88 | 12.25 | 0.96 | 8.00 | 0.91 | 12.50 | 0.93 | 0.08 | 9.76 | 6.19 | 4.39 |
| HD 122967 | S | 7008 | 88 | 9 | 0.77 | 1.50 | 1.02 | 1.31 | 2.37 | 1.52 | 0.99 | 1.54 | 0.76 | 1.47 | 0.23 | 1.28 | 1.61 | 4.16 |
| HD 123A | S | 5823 | 116 | 3 | 0.04 | 1.02 | 5.01 | 1.08 | 2.24 | 1.04 | 3.50 | 1.08 | 2.33 | 1.06 | 0.06 | 3.27 | 2.77 | 4.43 |
| HD 123B | S | 5230 | 140 | 2 | -0.26 | 0.88 | 9.84 | 0.90 | 8.02 | 0.92 | 5.89 | 0.88 | 11.75 | 0.90 | 0.04 | 8.87 | 5.86 | 4.47 |
| HD 124553 | S | 6060 | 54 | 9 | 0.63 | | | 1.23 | 4.75 | | | 1.35 | 3.50 | 1.29 | 0.12 | 4.13 | 1.25 | 4.00 |
| HD 125040 | S | 6357 | 37 | 5 | 0.41 | 1.23 | 2.00 | 1.24 | 2.33 | 1.25 | 2.00 | 1.27 | 1.63 | 1.25 | 0.04 | 1.99 | 0.70 | 4.28 |
| HD 125184 | S | 5637 | 40 | 7 | 0.38 | | | 1.10 | 7.75 | 1.15 | 6.67 | 1.10 | 8.75 | 1.12 | 0.05 | 7.72 | 2.08 | 4.06 |
| HD 126053 | S | 5727 | 55 | 12 | -0.10 | 0.85 | 11.24 | 0.99 | 1.72 | 0.86 | 9.00 | 0.84 | 10.56 | 0.89 | 0.15 | 8.13 | 9.52 | 4.46 |
| HD 12661 | S | 5651 | 48 | 9 | 0.05 | 0.97 | 9.61 | 1.05 | 5.06 | 1.05 | 5.33 | 1.01 | 8.75 | 1.02 | 0.08 | 7.19 | 4.55 | 4.35 |
| HD 127334 | S | 5643 | 28 | 9 | 0.14 | | | 1.00 | 9.67 | 0.98 | 10.00 | | | 0.99 | 0.02 | 9.83 | 0.33 | 4.25 |
| HD 128165 | S | 4793 | 46 | 9 | -0.59 | 0.79 | 5.36 | 0.77 | 5.72 | 0.80 | 5.08 | 0.79 | 5.99 | 0.79 | 0.03 | 5.54 | 0.90 | 4.59 |
| HD 128311 | S | 4903 | 27 | 9 | -0.52 | 0.82 | 6.35 | 0.82 | 4.81 | 0.82 | 6.03 | 0.82 | 5.97 | 0.82 | 0.00 | 5.79 | 1.54 | 4.58 |
| HD 129132 | S | 6609 | 38 | 9 | 1.57 | 1.99 | 1.00 | 1.96 | 1.08 | 2.13 | 1.03 | 2.08 | 0.75 | 2.04 | 0.17 | 0.96 | 0.33 | 3.41 |
| HD 129290 | S | 5837 | 103 | 9 | 0.26 | 1.09 | 5.55 | 1.03 | 8.00 | 0.97 | 9.67 | 0.93 | 11.50 | 1.01 | 0.16 | 8.68 | 5.95 | 4.19 |
| HD 129829 | S | 6085 | 24 | 9 | 0.14 | 1.09 | 2.69 | 1.08 | 2.28 | 1.10 | 2.28 | 1.04 | 3.81 | 1.08 | 0.06 | 2.76 | 1.54 | 4.41 |
| HD 130322 | S | 5422 | 40 | 12 | -0.25 | 0.94 | 3.02 | 0.94 | 1.86 | 0.92 | 3.59 | 0.92 | 3.34 | 0.93 | 0.02 | 2.95 | 1.73 | 4.54 |
| HD 13043 | S | 5877 | 29 | 9 | 0.31 | 1.00 | 9.44 | 1.06 | 7.63 | | | 0.93 | 11.00 | 1.00 | 0.13 | 9.35 | 3.38 | 4.15 |
| HD 130948 | S | 5983 | 4 | 3 | 0.09 | 1.10 | 1.21 | 1.04 | 3.33 | 1.09 | 1.73 | 1.06 | 2.83 | 1.07 | 0.06 | 2.28 | 2.12 | 4.43 |
| HD 131976 | S | 4076 | 130 | 3 | -1.40 | | | | | | | | | | | | | 5.06 |
| HD 131977 | S | 4625 | 38 | 8 | -0.66 | 0.78 | 5.36 | 0.75 | 8.65 | 0.77 | 9.50 | | | 0.77 | 0.03 | 7.84 | 4.14 | 4.59 |
| HD 132375 | S | 6336 | 57 | 3 | 0.53 | 1.24 | 3.14 | 1.30 | 2.75 | 1.38 | 1.60 | 1.34 | 2.00 | 1.32 | 0.14 | 2.37 | 1.54 | 4.18 |
| HD 133161 | S | 5946 | 33 | 9 | 0.24 | 1.17 | 2.00 | 1.13 | 4.25 | 1.22 | 1.70 | 1.14 | 4.56 | 1.17 | 0.09 | 3.13 | 2.86 | 4.31 |
| HD 135101 | S | 5628 | 54 | 9 | 0.13 | | | 0.96 | 11.00 | | | 0.98 | 11.50 | 0.97 | 0.02 | 11.25 | 0.50 | 4.24 |
| HD 135101B | S | 5529 | 44 | 6 | 0.29 | 1.00 | 10.30 | 1.00 | 10.79 | 1.03 | 9.70 | 1.01 | 10.82 | 1.01 | 0.03 | 10.40 | 1.12 | 4.07 |
| HD 135145 | S | 5852 | 36 | 9 | 0.39 | 1.02 | 9.00 | 0.95 | 10.08 | 0.95 | 10.00 | 0.98 | 9.79 | 0.98 | 0.07 | 9.72 | 1.08 | 4.05 |
| HD 136064 | S | 6144 | 24 | 3 | 0.66 | 1.30 | 3.57 | 1.14 | 5.67 | 1.32 | 3.83 | 1.28 | 3.92 | 1.26 | 0.18 | 4.25 | 2.10 | 3.98 |
| HD 136118 | S | 6143 | 46 | 9 | 0.47 | | | 0.87 | 10.67 | | | 1.20 | 4.44 | 1.04 | 0.33 | 7.55 | 6.23 | 4.08 |
| HD 136231 | S | 5881 | 53 | 8 | 0.28 | 1.02 | 8.67 | 0.95 | 9.75 | 0.99 | 8.50 | 0.93 | 11.11 | 0.97 | 0.09 | 9.51 | 2.61 | 4.17 |
| HD 1388 | S | 5884 | 69 | 9 | 0.19 | 1.14 | 2.00 | 1.03 | 6.50 | 1.17 | 3.00 | 1.06 | 6.75 | 1.10 | 0.14 | 4.56 | 4.75 | 4.31 |
| HD 140913 | S | 5946 | 29 | 9 | 0.00 | 0.92 | 7.13 | 0.95 | 5.26 | 0.97 | 3.63 | 0.95 | 4.19 | 0.95 | 0.05 | 5.05 | 3.50 | 4.46 |
| HD 141715 | S | 5836 | 35 | 8 | 0.22 | 1.09 | 5.55 | 1.02 | 7.75 | 1.06 | 7.00 | 1.01 | 8.75 | 1.05 | 0.08 | 7.26 | 3.20 | 4.25 |

| Star | Type | Teff | N | n | [Fe/H] | | | | | | | | | | | | | |
|------|------|------|---|---|--------|---|---|---|---|---|---|---|---|---|---|---|---|---|
| HD 141937 | S | 5885 | 31 | 9 | 0.04 | 1.03 | 3.82 | 1.05 | 2.47 | 1.06 | 2.31 | 1.02 | 3.61 | 1.04 | 0.04 | 3.05 | 1.51 | 4.44 |
| HD 14412 | S | 5492 | 89 | 9 | -0.38 | 0.75 | 10.48 | 0.79 | 2.34 | 0.80 | 3.25 | 0.81 | 2.09 | 0.79 | 0.06 | 4.54 | 8.39 | 4.62 |
| HD 144579 | S | 5308 | 30 | 9 | -0.37 | | | | | | | | | | | | | 4.67 |
| HD 144585 | S | 5850 | 47 | 9 | 0.27 | 1.10 | 5.55 | 1.03 | 8.00 | | | 1.10 | 7.00 | 1.08 | 0.07 | 6.85 | 2.45 | 4.22 |
| HD 145148 | S | 4923 | 85 | 9 | 0.55 | 1.16 | 6.31 | 1.18 | 6.80 | 1.23 | 6.00 | 1.17 | 7.40 | 1.19 | 0.07 | 6.63 | 1.40 | 3.68 |
| HD 1461 | S | 5765 | 63 | 12 | 0.08 | 1.05 | 4.15 | 1.05 | 4.94 | 1.12 | 1.93 | 1.04 | 6.14 | 1.07 | 0.08 | 4.29 | 4.21 | 4.38 |
| HD 14624 | S | 5599 | 50 | 8 | 0.31 | 1.00 | 10.73 | 0.99 | 11.25 | 1.02 | 9.75 | 1.00 | 11.21 | 1.00 | 0.03 | 10.73 | 1.50 | 4.07 |
| HD 147681 | S | 6139 | 45 | 8 | 0.11 | 1.02 | 2.91 | 1.03 | 3.46 | 1.04 | 1.73 | 1.02 | 2.53 | 1.03 | 0.02 | 2.66 | 1.73 | 4.44 |
| HD 149026 | S | 6003 | 41 | 3 | 0.46 | 1.14 | 5.66 | 1.27 | 3.92 | 1.25 | 4.00 | 1.24 | 4.46 | 1.23 | 0.13 | 4.51 | 1.74 | 4.13 |
| HD 149143 | S | 5768 | 30 | 3 | 0.36 | 1.03 | 8.69 | 0.98 | 10.00 | 1.02 | 9.00 | 0.99 | 10.21 | 1.01 | 0.05 | 9.48 | 1.52 | 4.07 |
| HD 15069 | S | 5722 | 50 | 9 | 0.32 | 1.13 | 5.01 | 1.03 | 9.33 | 0.95 | 11.00 | 0.99 | 10.80 | 1.03 | 0.18 | 9.04 | 5.99 | 4.11 |
| HD 150706 | S | 5900 | 16 | 9 | 0.02 | 1.02 | 4.03 | 1.03 | 2.16 | 1.06 | 1.75 | 1.01 | 3.63 | 1.03 | 0.05 | 2.89 | 2.28 | 4.46 |
| HD 150933 | S | 6087 | 27 | 9 | 0.33 | 1.23 | 1.28 | 1.20 | 3.25 | 1.27 | 1.38 | 1.20 | 3.25 | 1.23 | 0.07 | 2.29 | 1.97 | 4.28 |
| HD 151288 | S | 4181 | 88 | 9 | -1.01 | | | | | 0.65 | 5.36 | 0.66 | 5.93 | 0.66 | 0.01 | 5.65 | 0.57 | 4.70 |
| HD 151426 | S | 5711 | 33 | 8 | 0.13 | 0.99 | 9.10 | 0.97 | 10.00 | | | 1.01 | 9.25 | 0.99 | 0.04 | 9.45 | 0.90 | 4.28 |
| HD 151450 | S | 6108 | 67 | 9 | 0.19 | 1.14 | 1.67 | 1.05 | 4.11 | 1.13 | 1.86 | 1.09 | 3.12 | 1.10 | 0.09 | 2.69 | 2.44 | 4.38 |
| HD 15189 | S | 6051 | 54 | 9 | 0.03 | 0.94 | 5.36 | 0.98 | 3.94 | 0.99 | 2.48 | 0.96 | 3.25 | 0.97 | 0.05 | 3.76 | 2.88 | 4.47 |
| HD 152391 | S | 5431 | 33 | 12 | -0.22 | 0.92 | 5.28 | 0.91 | 4.98 | 0.93 | 3.63 | 0.94 | 4.66 | 0.93 | 0.03 | 4.64 | 1.66 | 4.51 |
| HD 154160 | S | 5380 | | 1 | 0.48 | | | 1.18 | 6.67 | 1.19 | 6.00 | 1.19 | 6.67 | 1.19 | 0.01 | 6.44 | 0.67 | 3.90 |
| HD 154363 | S | 4373 | 24 | 12 | -0.86 | 0.71 | 5.36 | 0.70 | 2.42 | | | | | 0.71 | 0.01 | 3.89 | 2.94 | 4.66 |
| HD 154578 | S | 6294 | 41 | 9 | 0.48 | 1.05 | 6.39 | 1.12 | 5.67 | | | | | 1.09 | 0.07 | 6.03 | 0.73 | 4.14 |
| HD 156062 | S | 5995 | 59 | 9 | 0.16 | 1.02 | 6.03 | 1.03 | 5.31 | 1.15 | 2.33 | 1.10 | 3.25 | 1.08 | 0.13 | 4.23 | 3.69 | 4.37 |
| HD 156826 | S | 5155 | 59 | 10 | 0.86 | 1.40 | 2.51 | 1.27 | 4.88 | 1.51 | 2.75 | 1.48 | 2.50 | 1.42 | 0.24 | 3.16 | 2.38 | 3.52 |
| HD 156846 | S | 6069 | 31 | 7 | 0.67 | 1.32 | 3.16 | 1.22 | 4.67 | 1.18 | 5.00 | 1.24 | 4.38 | 1.24 | 0.14 | 4.30 | 1.84 | 3.94 |
| HD 156968 | S | 5948 | 34 | 9 | 0.13 | 0.97 | 8.01 | 1.03 | 4.94 | 1.13 | 2.61 | 1.01 | 6.05 | 1.04 | 0.16 | 5.40 | 5.40 | 4.37 |
| HD 157881 | S | 4161 | 78 | 12 | -0.99 | 0.66 | 5.36 | 0.66 | 1.52 | | | | | 0.66 | 0.00 | 3.44 | 3.84 | 4.67 |
| HD 158633 | S | 5313 | 27 | 9 | -0.38 | 0.80 | 8.13 | | | 0.80 | 6.50 | 0.78 | 7.18 | 0.79 | 0.02 | 7.27 | 1.63 | 4.57 |
| HD 159222 | S | 5788 | 26 | 9 | 0.07 | 1.09 | 1.05 | 1.03 | 5.33 | 1.11 | 2.33 | 1.03 | 5.92 | 1.07 | 0.08 | 3.66 | 4.86 | 4.39 |
| HD 160346 | S | 4864 | 75 | 12 | -0.48 | 0.79 | 12.02 | | | 0.84 | 7.50 | | | 0.82 | 0.05 | 9.76 | 4.52 | 4.52 |
| HD 160933 | S | 5829 | 45 | 9 | 0.70 | 1.17 | 5.10 | 1.12 | 5.83 | 1.17 | 5.00 | 1.21 | 4.60 | 1.17 | 0.09 | 5.13 | 1.23 | 3.81 |
| HD 16160 | S | 4886 | 64 | 12 | -0.58 | 0.83 | 1.21 | 0.82 | 1.36 | 0.80 | 1.98 | 0.80 | 3.04 | 0.81 | 0.03 | 1.90 | 1.83 | 4.63 |
| HD 163492 | S | 5834 | 36 | 9 | 0.10 | 1.09 | 2.34 | 1.01 | 5.67 | 1.10 | 3.18 | 1.05 | 4.76 | 1.06 | 0.09 | 3.99 | 3.32 | 4.37 |
| HD 163840 | S | 5760 | 41 | 9 | 0.31 | | | | | 0.94 | 11.00 | 1.00 | 10.33 | 0.97 | 0.06 | 10.67 | 0.67 | 4.10 |
| HD 164595 | S | 5728 | 31 | 9 | 0.01 | 1.07 | 1.05 | 0.98 | 6.50 | 1.07 | 2.85 | 1.02 | 5.46 | 1.04 | 0.09 | 3.96 | 5.45 | 4.42 |
| HD 166 | S | 5481 | 28 | 7 | -0.21 | 0.95 | 2.57 | 0.95 | 2.14 | 0.94 | 3.06 | 0.93 | 3.42 | 0.94 | 0.02 | 2.80 | 1.28 | 4.52 |
| HD 166601 | S | 6289 | 28 | 9 | 0.64 | 1.15 | 5.01 | 1.08 | 6.10 | | | 1.27 | 3.64 | 1.17 | 0.19 | 4.92 | 2.46 | 4.01 |
| HD 167588 | S | 5909 | 40 | 9 | 0.57 | 0.99 | 7.94 | 0.96 | 8.88 | 1.05 | 7.00 | 1.05 | 6.80 | 1.01 | 0.09 | 7.65 | 2.08 | 3.91 |
| HD 167665 | S | 6179 | 54 | 9 | 0.33 | 1.03 | 6.65 | | | | | 1.10 | 4.75 | 1.07 | 0.07 | 5.70 | 1.90 | 4.25 |

| | | | | | | | | | | | | | | | | | |
|---|---|---|---|---|---|---|---|---|---|---|---|---|---|---|---|---|---|
| HD 168443 | S | 5565 | 29 | 9 | 0.33 | 1.00 | 10.72 | 1.02 | 10.50 | 0.98 | 11.00 | 1.02 | 11.00 | 1.01 | 0.04 | 10.80 | 0.50 | 4.04 |
| HD 168746 | S | 5576 | 23 | 9 | 0.03 | 1.01 | 5.01 | 0.92 | 12.00 | 1.00 | 7.50 | 0.95 | 12.00 | 0.97 | 0.09 | 9.13 | 6.99 | 4.33 |
| HD 170657 | S | 5133 | 37 | 11 | -0.48 | 0.76 | 8.60 | | | 0.76 | 9.00 | 0.76 | 6.24 | 0.76 | 0.00 | 7.95 | 2.76 | 4.59 |
| HD 171706 | S | 5935 | 61 | 9 | 0.39 | 1.01 | 8.93 | 0.89 | 10.75 | 1.12 | 6.25 | 0.98 | 9.36 | 1.00 | 0.23 | 8.82 | 4.50 | 4.09 |
| HD 172051 | S | 5669 | 10 | 10 | -0.17 | 0.85 | 9.84 | 0.77 | 12.00 | 0.86 | 7.20 | 0.84 | 8.50 | 0.83 | 0.09 | 9.39 | 4.80 | 4.49 |
| HD 173818 | S | 4245 | 72 | 11 | -1.00 | | | 0.56 | 10.25 | 0.64 | 3.74 | 0.64 | 5.29 | 0.61 | 0.08 | 6.43 | 6.51 | 4.68 |
| HD 175225 | S | 5281 | 40 | 4 | 0.59 | 1.20 | 5.10 | 1.14 | 6.45 | 1.20 | 5.50 | 1.23 | 5.08 | 1.19 | 0.09 | 5.53 | 1.37 | 3.76 |
| HD 175290 | S | 6359 | 38 | 9 | 0.63 | 1.17 | 4.61 | 1.16 | 5.00 | 1.09 | 5.50 | 1.12 | 4.75 | 1.14 | 0.08 | 4.97 | 0.89 | 4.02 |
| HD 176051 | S | 5980 | 91 | 3 | 0.17 | 0.92 | 10.72 | 1.07 | 4.38 | 1.18 | 1.42 | 0.99 | 7.50 | 1.04 | 0.26 | 6.00 | 9.30 | 4.34 |
| HD 176367 | S | 6087 | 71 | 12 | 0.05 | 1.00 | 2.57 | 0.90 | 5.84 | 1.03 | 1.53 | 1.01 | 2.22 | 0.99 | 0.13 | 3.04 | 4.32 | 4.47 |
| HD 177830 | S | 4813 | 34 | 8 | 0.71 | 1.28 | 4.74 | 1.24 | 6.44 | 1.26 | 6.00 | 1.10 | 9.50 | 1.22 | 0.18 | 6.67 | 4.76 | 3.49 |
| HD 178428 | S | 5646 | 38 | 9 | 0.17 | 0.95 | 12.02 | 1.00 | 10.00 | | | | | 0.98 | 0.05 | 11.01 | 2.02 | 4.21 |
| HD 178911 | S | 5825 | 21 | 9 | 0.68 | 1.23 | 4.49 | 1.19 | 5.28 | 1.28 | 4.63 | 1.26 | 4.54 | 1.24 | 0.09 | 4.73 | 0.79 | 3.86 |
| HD 178911B | S | 5602 | 78 | 6 | 0.01 | 0.98 | 5.16 | 1.00 | 4.61 | 1.04 | 3.06 | 1.00 | 5.52 | 1.01 | 0.06 | 4.59 | 2.46 | 4.37 |
| HD 179957 | S | 5771 | 17 | 2 | 0.07 | 1.07 | 2.76 | 1.02 | 5.95 | 1.07 | 4.15 | 1.02 | 6.80 | 1.05 | 0.05 | 4.91 | 4.04 | 4.38 |
| HD 179958 | S | 5807 | 53 | 4 | 0.14 | 1.10 | 2.25 | 1.03 | 6.75 | 1.13 | 3.00 | 1.03 | 7.17 | 1.07 | 0.10 | 4.79 | 4.91 | 4.33 |
| HD 181655 | S | 5673 | 38 | 9 | 0.23 | 1.02 | 8.64 | 0.98 | 11.00 | | | 0.99 | 11.50 | 1.00 | 0.04 | 10.38 | 2.86 | 4.17 |
| HD 182488 | S | 5362 | 16 | 5 | -0.18 | 0.93 | 6.42 | 0.92 | 7.72 | 0.96 | 5.31 | 0.92 | 10.58 | 0.93 | 0.04 | 7.51 | 5.27 | 4.45 |
| HD 183263 | S | 5911 | 48 | 9 | 0.26 | 1.17 | 2.51 | 1.19 | 2.96 | 1.18 | 3.06 | 1.13 | 5.32 | 1.17 | 0.06 | 3.46 | 2.81 | 4.28 |
| HD 184151 | S | 6430 | 60 | 9 | 1.03 | 1.58 | 2.05 | 1.42 | 2.80 | 1.58 | 2.27 | 1.46 | 2.13 | 1.51 | 0.16 | 2.31 | 0.75 | 3.77 |
| HD 184152 | S | 5578 | 21 | 9 | -0.03 | 0.93 | 9.28 | 0.93 | 8.08 | 0.94 | 7.82 | 0.92 | 9.00 | 0.93 | 0.02 | 8.54 | 1.46 | 4.37 |
| HD 18445 | S | 4874 | 60 | 9 | -0.25 | | | | | | | | | | | | | 4.36 |
| HD 184489 | S | 4133 | 76 | 10 | -1.16 | | | | | | | 0.54 | 5.29 | 0.54 | 0.00 | 5.29 | 0.00 | 4.74 |
| HD 184509 | S | 6069 | 19 | 9 | 0.23 | 1.11 | 3.98 | 1.10 | 3.75 | 1.14 | 3.00 | 1.10 | 3.88 | 1.11 | 0.04 | 3.65 | 0.98 | 4.33 |
| HD 18455 | S | 5105 | 39 | 3 | -0.27 | | | | | 0.89 | 10.00 | | | 0.89 | 0.00 | 10.00 | 0.00 | 4.44 |
| HD 184700 | S | 5745 | 48 | 9 | 0.18 | 0.99 | 10.00 | 0.98 | 9.25 | 0.97 | 10.00 | 0.97 | 11.00 | 0.98 | 0.02 | 10.06 | 1.75 | 4.23 |
| HD 187123 | S | 5796 | 38 | 9 | 0.16 | 1.11 | 3.16 | 1.03 | 7.25 | 1.04 | 7.00 | 1.03 | 8.17 | 1.05 | 0.08 | 6.39 | 5.00 | 4.30 |
| HD 188015 | S | 5667 | 23 | 9 | 0.16 | 0.97 | 11.48 | 1.07 | 6.63 | 1.13 | 4.00 | 1.04 | 9.00 | 1.05 | 0.16 | 7.78 | 7.48 | 4.26 |
| HD 188088 | S | 4818 | 63 | 3 | -0.12 | | | | | | | | | | | | | 4.20 |
| HD 188169 | S | 6509 | 70 | 9 | 0.48 | 1.11 | 4.73 | 1.28 | 1.88 | 1.28 | 1.95 | 1.25 | 2.13 | 1.23 | 0.17 | 2.67 | 2.86 | 4.25 |
| HD 189712 | S | 6363 | 22 | 9 | 0.81 | 1.16 | 4.48 | 1.15 | 4.63 | 1.11 | 5.00 | 1.21 | 3.63 | 1.16 | 0.10 | 4.43 | 1.38 | 3.85 |
| HD 189733 | S | 5044 | 32 | 9 | -0.48 | 0.86 | 2.09 | 0.85 | 1.70 | 0.84 | 1.73 | 0.83 | 4.43 | 0.85 | 0.03 | 2.49 | 2.73 | 4.60 |
| HD 190007 | S | 4596 | 40 | 12 | -0.66 | 0.75 | 10.99 | 0.77 | 7.51 | 0.75 | 12.00 | | | 0.76 | 0.02 | 10.17 | 4.49 | 4.57 |
| HD 190228 | S | 5264 | 19 | 9 | 0.64 | 1.19 | 5.10 | 1.16 | 5.93 | 1.20 | 5.33 | 1.21 | 4.92 | 1.19 | 0.05 | 5.32 | 1.01 | 3.71 |
| HD 190360 | S | 5564 | 54 | 5 | 0.05 | | | 0.98 | 9.50 | 1.00 | 8.50 | 0.98 | 11.25 | 0.99 | 0.02 | 9.75 | 2.75 | 4.31 |
| HD 190771 | S | 5751 | 75 | 9 | 0.01 | 1.05 | 2.58 | 1.06 | 2.54 | 1.09 | 1.93 | 1.06 | 2.98 | 1.07 | 0.04 | 2.51 | 1.05 | 4.44 |
| HD 192145 | S | 6069 | 30 | 9 | 0.56 | 1.12 | 5.81 | 0.96 | 8.56 | 1.07 | 6.50 | 1.06 | 6.50 | 1.05 | 0.16 | 6.84 | 2.75 | 3.98 |
| HD 192263 | S | 4974 | 19 | 3 | -0.52 | 0.84 | 2.57 | 0.84 | 1.84 | 0.82 | 2.27 | 0.81 | 5.29 | 0.83 | 0.03 | 2.99 | 3.45 | 4.61 |

| ID | Type | T | n1 | n2 | v1 | v2 | v3 | v4 | v5 | v6 | v7 | v8 | v9 | v10 | v11 | v12 | v13 | v14 |
|---|---|---|---|---|---|---|---|---|---|---|---|---|---|---|---|---|---|---|
| HD 192310 | S | 5044 | 63 | 12 | -0.38 | 0.84 | 10.48 | 0.83 | 10.75 | 0.90 | 4.58 | | | 0.86 | 0.07 | 8.60 | 6.17 | 4.51 |
| HD 193555 | S | 6133 | 69 | 9 | 1.03 | 1.62 | 1.89 | 1.65 | 2.25 | 1.67 | 2.08 | 1.58 | 2.00 | 1.63 | 0.09 | 2.06 | 0.36 | 3.72 |
| HD 193664 | S | 5922 | 32 | 5 | 0.04 | 1.04 | 3.47 | 1.03 | 2.88 | 1.07 | 1.70 | 1.00 | 4.69 | 1.04 | 0.07 | 3.18 | 2.99 | 4.45 |
| HD 19383 | S | 6396 | 49 | 9 | 0.37 | 1.20 | 1.76 | 1.21 | 1.81 | 1.23 | 1.45 | 1.20 | 1.62 | 1.21 | 0.03 | 1.66 | 0.36 | 4.32 |
| HD 195019 | S | 5660 | 76 | 9 | 0.36 | 1.01 | 9.17 | 1.04 | 9.50 | 1.11 | 7.75 | 0.99 | 10.71 | 1.04 | 0.12 | 9.28 | 2.96 | 4.05 |
| HD 195564 | S | 5619 | 38 | 8 | 0.45 | 1.00 | 9.84 | 1.00 | 9.50 | 1.05 | 8.50 | 1.07 | 8.42 | 1.03 | 0.07 | 9.07 | 1.43 | 3.95 |
| HD 19617 | S | 5682 | 57 | 9 | -0.02 | 1.03 | 2.90 | 1.00 | 4.68 | 1.03 | 4.05 | 1.01 | 4.91 | 1.02 | 0.03 | 4.13 | 2.00 | 4.43 |
| HD 196761 | S | 5483 | 27 | 8 | -0.28 | 0.81 | 10.99 | | | 0.84 | 6.50 | 0.81 | 9.00 | 0.82 | 0.03 | 8.83 | 4.49 | 4.53 |
| HD 197076 | S | 5844 | 66 | 9 | -0.01 | 1.00 | 4.46 | 0.99 | 3.47 | 1.04 | 2.23 | 0.98 | 4.90 | 1.00 | 0.06 | 3.77 | 2.67 | 4.46 |
| HD 198387 | S | 5056 | 85 | 9 | 0.75 | 1.10 | 7.92 | 1.37 | 3.75 | 0.99 | 9.00 | 1.17 | 6.25 | 1.16 | 0.38 | 6.73 | 5.25 | 3.51 |
| HD 198483 | S | 5986 | 40 | 9 | 0.27 | 1.18 | 1.61 | 1.20 | 2.47 | 1.23 | 1.68 | 1.16 | 3.90 | 1.19 | 0.07 | 2.41 | 2.29 | 4.30 |
| HD 198802 | S | 5736 | 41 | 12 | 0.67 | | | 1.07 | 6.63 | 1.19 | 5.00 | 1.21 | 4.88 | 1.16 | 0.14 | 5.50 | 1.75 | 3.81 |
| HD 199604 | S | 5887 | 62 | 9 | 0.15 | | | 1.03 | 5.55 | | | | | 1.03 | 0.00 | 5.55 | 0.00 | 4.33 |
| HD 200779 | S | 4389 | 66 | 12 | -0.80 | 0.73 | 6.37 | 0.70 | 9.43 | | | | | 0.72 | 0.03 | 7.90 | 3.06 | 4.61 |
| HD 201456 | S | 6201 | 36 | 9 | 0.41 | 1.22 | 2.34 | 0.82 | 12.25 | 1.31 | 1.60 | 1.15 | 4.50 | 1.13 | 0.49 | 5.17 | 10.65 | 4.20 |
| HD 201496 | S | 5944 | 18 | 9 | 0.14 | 1.04 | 5.48 | 1.01 | 5.92 | 1.14 | 2.25 | 1.09 | 3.60 | 1.07 | 0.13 | 4.31 | 3.67 | 4.37 |
| HD 202206 | S | 5693 | 42 | 3 | 0.03 | 1.05 | 2.10 | 1.08 | 2.68 | 1.09 | 2.11 | 1.04 | 5.70 | 1.07 | 0.05 | 3.15 | 3.60 | 4.40 |
| HD 202282 | S | 5796 | 61 | 9 | 0.16 | 1.09 | 2.32 | 1.00 | 7.25 | 1.08 | 4.18 | 1.03 | 7.10 | 1.05 | 0.09 | 5.21 | 4.93 | 4.30 |
| HD 20339 | S | 5918 | 55 | 9 | 0.16 | 1.14 | 0.92 | 0.97 | 7.67 | 1.17 | 1.74 | 0.97 | 8.30 | 1.06 | 0.20 | 4.66 | 7.38 | 4.34 |
| HD 20367 | S | 6050 | 57 | 9 | 0.20 | 1.10 | 3.57 | 1.09 | 3.91 | 1.18 | 1.80 | 1.14 | 2.63 | 1.13 | 0.09 | 2.98 | 2.11 | 4.36 |
| HD 204153 | S | 6920 | 28 | 3 | 0.69 | 1.45 | 1.00 | 1.23 | 3.38 | 1.48 | 0.90 | 1.41 | 1.30 | 1.39 | 0.25 | 1.64 | 2.48 | 4.20 |
| HD 204485 | S | 7125 | 48 | 7 | 0.88 | | | 1.67 | 0.73 | | | 1.64 | 0.85 | 1.66 | 0.03 | 0.79 | 0.13 | 4.13 |
| HD 205027 | S | 5752 | 35 | 9 | 0.09 | 0.99 | 7.94 | 0.98 | 8.00 | 0.89 | 12.00 | | | 0.95 | 0.10 | 9.31 | 4.06 | 4.31 |
| HD 205700 | S | 6656 | 52 | 9 | 0.66 | 1.22 | 3.27 | 1.10 | 5.00 | 1.36 | 2.23 | 1.24 | 2.96 | 1.23 | 0.26 | 3.36 | 2.78 | 4.11 |
| HD 206282 | S | 6345 | 32 | 8 | 0.68 | 1.32 | 3.04 | 1.06 | 6.20 | | | 1.33 | 3.06 | 1.24 | 0.27 | 4.10 | 3.16 | 4.01 |
| HD 207858 | S | 6290 | 49 | 9 | 0.70 | 1.28 | 3.67 | 1.11 | 5.63 | 1.25 | 4.00 | 1.34 | 3.13 | 1.25 | 0.23 | 4.11 | 2.51 | 3.97 |
| HD 208801 | S | 4918 | 88 | 12 | 0.67 | 1.22 | 6.39 | 1.28 | 5.31 | 1.41 | 4.00 | 1.23 | 6.17 | 1.29 | 0.19 | 5.47 | 2.39 | 3.59 |
| HD 21019 | S | 5514 | 50 | 9 | 0.61 | 1.11 | 6.42 | 1.00 | 8.50 | 1.20 | 5.50 | 1.13 | 5.80 | 1.11 | 0.20 | 6.56 | 3.00 | 3.79 |
| HD 210277 | S | 5533 | 46 | 12 | -0.01 | | | 0.96 | 9.25 | 1.01 | 7.00 | 0.96 | 11.33 | 0.98 | 0.05 | 9.19 | 4.33 | 4.36 |
| HD 210460 | S | 5529 | 36 | 7 | 1.02 | 1.58 | 2.00 | 1.38 | 2.75 | | | 1.61 | 1.50 | 1.52 | 0.23 | 2.08 | 1.25 | 3.52 |
| HD 210483 | S | 5842 | 38 | 9 | 0.31 | 0.92 | 12.02 | 0.86 | 12.50 | 0.96 | 10.00 | 0.94 | 11.25 | 0.92 | 0.10 | 11.44 | 2.50 | 4.10 |
| HD 210855 | S | 6255 | 31 | 5 | 0.96 | 1.48 | 2.38 | 1.48 | 2.75 | 1.53 | 2.40 | 1.44 | 2.38 | 1.48 | 0.09 | 2.48 | 0.38 | 3.78 |
| HD 211476 | S | 5829 | 21 | 9 | 0.10 | 1.01 | 6.31 | 1.01 | 6.25 | 0.98 | 7.50 | 0.99 | 7.75 | 1.00 | 0.03 | 6.95 | 1.50 | 4.35 |
| HD 211575 | S | 6589 | 93 | 9 | 0.58 | | | 1.36 | 1.75 | 1.41 | 1.10 | 1.42 | 0.90 | 1.40 | 0.06 | 1.25 | 0.85 | 4.23 |
| HD 213338 | S | 5558 | 47 | 8 | -0.16 | 0.97 | 2.57 | 0.95 | 3.80 | 0.95 | 3.80 | 0.97 | 3.27 | 0.96 | 0.02 | 3.36 | 1.23 | 4.51 |
| HD 214385 | S | 5711 | 25 | 9 | -0.09 | 0.97 | 4.31 | 1.00 | 1.45 | 0.87 | 9.00 | 0.85 | 10.75 | 0.92 | 0.15 | 6.38 | 9.30 | 4.47 |
| HD 214749 | S | 4531 | 20 | 11 | -0.75 | 0.75 | 5.29 | 0.74 | 4.41 | 0.74 | 8.50 | 0.74 | 6.46 | 0.74 | 0.01 | 6.16 | 4.09 | 4.63 |
| HD 215243 | S | 6393 | 61 | 9 | 0.56 | 1.25 | 3.01 | 1.01 | 6.85 | 1.38 | 1.68 | 1.09 | 5.00 | 1.18 | 0.37 | 4.13 | 5.18 | 4.12 |

| Star | Type | Teff | | | | | | | | | | | | | | | | | |
|---|---|---|---|---|---|---|---|---|---|---|---|---|---|---|---|---|---|---|---|
| HD 21531 | S | 4231 | 86 | 9 | -0.94 | 0.68 | 5.97 | 0.71 | 1.27 | | | 0.69 | 5.29 | 0.69 | 0.03 | 4.18 | 4.70 | 4.67 |
| HD 215625 | S | 6228 | 54 | 9 | 0.25 | 1.18 | 1.05 | 1.19 | 1.00 | 1.20 | 1.07 | 1.18 | 1.17 | 1.19 | 0.02 | 1.07 | 0.17 | 4.39 |
| HD 216133 | S | 3973 | 105 | 10 | -1.33 | | | | | | | | | | | | | 4.92 |
| HD 216770 | S | 5411 | 67 | 9 | -0.18 | 0.95 | 4.78 | 0.94 | 5.76 | 0.99 | 2.65 | 0.95 | 7.13 | 0.96 | 0.05 | 5.08 | 4.47 | 4.48 |
| HD 217107 | S | 5541 | | 1 | 0.08 | 1.01 | 7.13 | 1.02 | 8.17 | 1.08 | 5.00 | | | 1.04 | 0.07 | 6.76 | 3.17 | 4.29 |
| HD 217357 | S | 4192 | 73 | 9 | -1.08 | | | | | | | | | | | | | 4.82 |
| HD 217577 | S | 5762 | 31 | 9 | 0.30 | 0.93 | 12.02 | 1.04 | 9.00 | 0.97 | 10.33 | 0.95 | 11.31 | 0.97 | 0.11 | 10.67 | 3.02 | 4.11 |
| HD 217877 | S | 6015 | 45 | 9 | 0.18 | 1.09 | 3.98 | 1.03 | 5.44 | 1.10 | 3.53 | 1.03 | 5.79 | 1.06 | 0.07 | 4.69 | 2.26 | 4.35 |
| HD 217958 | S | 5771 | 22 | 9 | 0.22 | | | 1.03 | 8.17 | 1.04 | 8.00 | 1.04 | 8.67 | 1.04 | 0.01 | 8.28 | 0.67 | 4.23 |
| HD 218101 | S | 5217 | 12 | 6 | 0.53 | 1.18 | 5.66 | 1.21 | 5.90 | 1.23 | 5.50 | 1.23 | 5.60 | 1.21 | 0.05 | 5.67 | 0.40 | 3.81 |
| HD 219428 | S | 5930 | | 1 | 0.20 | 1.15 | 1.21 | 1.19 | 1.47 | 1.19 | 1.29 | 1.15 | 3.18 | 1.17 | 0.04 | 1.79 | 1.97 | 4.35 |
| HD 220008 | S | 5653 | 17 | 9 | 0.61 | 1.10 | 6.42 | 1.01 | 8.04 | 1.14 | 6.00 | 1.12 | 6.05 | 1.09 | 0.13 | 6.63 | 2.04 | 3.82 |
| HD 220689 | S | 5921 | 36 | 9 | 0.11 | 0.97 | 8.12 | 1.02 | 5.22 | 1.03 | 5.00 | 0.99 | 6.91 | 1.00 | 0.06 | 6.31 | 3.12 | 4.37 |
| HD 22072 | S | 4974 | 49 | 5 | 0.77 | 1.41 | 3.16 | 1.31 | 4.25 | 1.06 | 8.00 | 0.96 | 11.00 | 1.19 | 0.45 | 6.60 | 7.84 | 3.48 |
| HD 221356 | S | 6137 | 39 | 9 | 0.15 | 0.90 | 9.59 | 1.12 | 1.35 | 1.00 | 5.00 | 0.99 | 5.04 | 1.00 | 0.22 | 5.24 | 8.23 | 4.39 |
| HD 221445 | S | 6230 | 61 | 9 | 0.95 | 1.52 | 2.24 | 1.29 | 3.48 | 1.52 | 2.66 | 1.39 | 2.56 | 1.43 | 0.23 | 2.74 | 1.25 | 3.77 |
| HD 221503 | S | 4312 | 64 | 9 | -0.89 | 0.70 | 4.30 | 0.73 | 1.59 | 0.67 | 10.71 | 0.71 | 5.29 | 0.70 | 0.06 | 5.47 | 9.13 | 4.66 |
| HD 222582 | S | 5796 | 16 | 9 | 0.09 | 1.09 | 1.46 | 1.03 | 5.82 | 1.11 | 2.78 | 1.02 | 6.64 | 1.06 | 0.09 | 4.18 | 5.18 | 4.37 |
| HD 222645 | S | 6211 | 86 | 8 | 0.30 | 1.19 | 1.62 | 1.13 | 3.00 | 1.22 | 1.45 | 1.19 | 1.85 | 1.18 | 0.09 | 1.98 | 1.55 | 4.33 |
| HD 22292 | S | 6483 | 40 | 9 | 0.42 | 1.27 | 1.05 | 1.22 | 1.78 | 1.26 | 1.44 | 1.23 | 1.63 | 1.25 | 0.05 | 1.48 | 0.73 | 4.31 |
| HD 223084 | S | 5958 | 91 | 8 | 0.20 | 1.14 | 1.74 | 1.06 | 5.50 | 1.15 | 2.96 | 1.09 | 5.17 | 1.11 | 0.09 | 3.84 | 3.76 | 4.33 |
| HD 223110 | S | 6516 | 82 | 9 | 0.80 | 1.43 | 2.14 | 1.16 | 4.46 | 1.33 | 3.17 | 1.36 | 2.75 | 1.32 | 0.27 | 3.13 | 2.32 | 3.96 |
| HD 223238 | S | 5889 | 34 | 9 | 0.21 | 1.15 | 2.25 | 1.06 | 6.17 | 1.13 | 4.50 | 1.06 | 6.75 | 1.10 | 0.09 | 4.92 | 4.50 | 4.30 |
| HD 22455 | S | 5886 | 116 | 8 | 0.12 | 1.06 | 3.98 | 1.01 | 5.42 | 1.11 | 2.73 | 1.05 | 4.31 | 1.06 | 0.10 | 4.11 | 2.69 | 4.37 |
| HD 22468A | S | 4736 | 86 | 3 | 0.69 | | | | | | | | | | | | | 3.35 |
| HD 22468B | S | 4631 | | 1 | -0.48 | | | | | | | | | | | | | 4.47 |
| HD 225239 | S | 5647 | 45 | 9 | 0.68 | 1.15 | 5.45 | 1.06 | 6.84 | 1.12 | 5.73 | 1.15 | 4.94 | 1.12 | 0.09 | 5.74 | 1.89 | 3.76 |
| HD 231701 | S | 6240 | 74 | 9 | 0.49 | 1.23 | 2.56 | 1.18 | 4.44 | 1.28 | 2.44 | 1.24 | 3.46 | 1.23 | 0.10 | 3.22 | 1.99 | 4.17 |
| HD 232979 | S | 3893 | 54 | 7 | -1.09 | | | | | | | | | | | | | 4.62 |
| HD 23349 | S | 6018 | 46 | 8 | 0.30 | 1.16 | 2.47 | 1.04 | 6.44 | 1.20 | 2.22 | 1.13 | 4.67 | 1.13 | 0.16 | 3.95 | 4.22 | 4.26 |
| HD 23356 | S | 4990 | 42 | 8 | -0.53 | | | 0.84 | 0.95 | 0.83 | 1.16 | 0.83 | 0.75 | 0.83 | 0.01 | 0.95 | 0.41 | 4.63 |
| HD 234078 | S | 4157 | 63 | 8 | -1.00 | 0.66 | 5.36 | 0.65 | 1.75 | | | | | 0.66 | 0.01 | 3.56 | 3.61 | 4.68 |
| HD 23453 | S | 3748 | 72 | 6 | -1.10 | | | | | 0.39 | 1.73 | | | 0.39 | 0.00 | 1.73 | 0.00 | 4.37 |
| HD 23476 | S | 5662 | 28 | 9 | -0.09 | 1.00 | 2.57 | 0.97 | 3.64 | 0.98 | 3.39 | | | 0.98 | 0.03 | 3.20 | 1.08 | 4.48 |
| HD 23596 | S | 6008 | 28 | 9 | 0.43 | 1.14 | 5.66 | | | | | 1.26 | 4.00 | 1.20 | 0.12 | 4.83 | 1.66 | 4.15 |
| HD 239928 | S | 5899 | 31 | 8 | 0.05 | 1.01 | 5.13 | 1.02 | 3.33 | 1.05 | 2.70 | 1.04 | 3.36 | 1.03 | 0.04 | 3.63 | 2.43 | 4.43 |
| HD 2475 | S | 6012 | 78 | 9 | 0.36 | | | | | | | 1.12 | 6.00 | 1.12 | 0.00 | 6.00 | 0.00 | 4.19 |
| HD 2582 | S | 5705 | 30 | 8 | 0.22 | 1.02 | 8.45 | 0.98 | 10.10 | 1.04 | 8.50 | 0.99 | 10.83 | 1.01 | 0.06 | 9.47 | 2.38 | 4.19 |
| HD 2589 | S | 5154 | 20 | 6 | 0.68 | 1.28 | 4.09 | 1.07 | 7.83 | 1.20 | 5.00 | 1.37 | 3.60 | 1.23 | 0.30 | 5.13 | 4.23 | 3.64 |
| HD 2638 | S | 5120 | 29 | 8 | -0.37 | 0.87 | 5.36 | 0.90 | 3.76 | 0.89 | 3.42 | 0.89 | 5.29 | 0.89 | 0.03 | 4.46 | 1.95 | 4.54 |
| HD 26505 | S | 5878 | 67 | 8 | 0.34 | 1.00 | 9.18 | 0.94 | 10.33 | 0.99 | 9.00 | 1.00 | 9.58 | 0.98 | 0.06 | 9.52 | 1.33 | 4.11 |
| HD 26913 | S | 5661 | 54 | 12 | -0.19 | 0.87 | 6.42 | 0.77 | 11.25 | 0.90 | 3.05 | 0.88 | 4.31 | 0.86 | 0.13 | 6.26 | 8.20 | 4.52 |
| HD 26923 | S | 5989 | 52 | 9 | 0.05 | 0.93 | 7.94 | 1.03 | 2.81 | 0.98 | 4.50 | 0.95 | 5.50 | 0.97 | 0.10 | 5.19 | 5.13 | 4.43 |
| HD 2730 | S | 6192 | 48 | 9 | 0.64 | 1.21 | 4.49 | 1.10 | 6.05 | 1.24 | 4.50 | 1.19 | 4.60 | 1.19 | 0.14 | 4.91 | 1.56 | 3.99 |
| HD 27530 | S | 5926 | | 1 | 0.16 | 1.13 | 1.74 | 1.18 | 1.18 | 1.18 | 1.06 | 1.15 | 2.15 | 1.16 | 0.05 | 1.53 | 1.09 | 4.38 |
| HD 28185 | S | 5658 | 28 | 9 | 0.07 | 0.96 | 9.78 | 0.99 | 8.08 | 1.06 | 5.25 | 1.00 | 9.50 | 1.00 | 0.10 | 8.15 | 4.53 | 4.33 |
| HD 28343 | S | 4152 | 96 | 9 | -0.96 | 0.68 | 5.36 | 0.69 | 4.23 | | | | | 0.69 | 0.01 | 4.79 | 1.13 | 4.65 |
| HD 285660 | S | 6055 | 19 | 8 | 0.31 | 1.20 | 1.61 | 1.08 | 5.43 | 1.23 | 1.62 | 1.13 | 4.20 | 1.16 | 0.15 | 3.21 | 3.82 | 4.27 |
| HD 28571 | S | 5736 | 26 | 9 | 0.19 | 0.98 | 10.24 | 0.98 | 9.25 | 0.96 | 10.00 | 0.97 | 11.00 | 0.97 | 0.02 | 10.12 | 1.75 | 4.22 |
| HD 28635 | S | 6140 | 50 | 12 | 0.25 | 1.14 | 2.56 | 1.13 | 2.95 | 1.14 | 2.58 | 1.17 | 2.08 | 1.15 | 0.04 | 2.54 | 0.86 | 4.35 |
| HD 29587 | S | 5682 | 19 | 9 | -0.09 | | | 0.98 | 3.00 | | | | | 0.98 | 0.00 | 3.00 | 0.00 | 4.48 |
| HD 29645 | S | 6002 | 41 | 9 | 0.54 | | | 1.08 | 6.75 | 1.29 | 4.00 | 1.12 | 6.25 | 1.16 | 0.21 | 5.67 | 2.75 | 4.02 |
| HD 29697 | S | 4421 | 49 | 7 | -0.82 | 0.72 | 6.08 | 0.73 | 2.62 | 0.70 | 10.71 | 0.73 | 5.29 | 0.72 | 0.03 | 6.18 | 8.09 | 4.64 |
| HD 31527 | S | 5908 | 47 | 9 | 0.10 | 0.97 | 8.12 | 0.98 | 6.54 | 0.99 | 6.50 | 0.99 | 7.00 | 0.98 | 0.02 | 7.04 | 1.62 | 4.36 |
| HD 31949 | S | 6213 | 34 | 9 | 0.31 | 1.11 | 4.02 | 1.12 | 3.39 | 1.16 | 2.36 | 1.13 | 3.31 | 1.13 | 0.05 | 3.27 | 1.67 | 4.30 |
| HD 32147 | S | 4790 | 42 | 9 | -0.55 | 0.81 | 5.36 | 0.83 | 4.27 | 0.82 | 5.95 | 0.81 | 9.25 | 0.82 | 0.02 | 6.21 | 4.98 | 4.57 |
| HD 32715 | S | 6615 | 29 | 9 | 0.55 | 1.34 | 1.58 | 1.35 | 1.50 | 1.36 | 1.40 | 1.40 | 0.93 | 1.36 | 0.06 | 1.35 | 0.65 | 4.25 |
| HD 332612 | S | 6124 | 42 | 8 | 0.49 | 1.20 | 4.46 | 0.97 | 8.77 | 1.21 | 4.25 | 1.20 | 4.64 | 1.15 | 0.24 | 5.53 | 4.52 | 4.10 |
| HD 334372 | S | 5765 | 43 | 9 | 0.40 | 0.95 | 10.82 | 0.96 | 10.05 | 0.99 | 9.50 | 0.98 | 10.10 | 0.97 | 0.04 | 10.12 | 1.32 | 4.01 |
| HD 33636 | S | 5953 | 47 | 9 | 0.02 | 0.92 | 7.94 | 1.02 | 2.81 | 0.97 | 4.50 | 0.92 | 6.54 | 0.96 | 0.10 | 5.45 | 5.13 | 4.45 |
| HD 33866 | S | 5625 | 98 | 9 | 0.04 | 0.97 | 8.69 | 0.93 | 10.75 | 0.97 | 9.00 | | | 0.96 | 0.04 | 9.48 | 2.06 | 4.33 |
| HD 34721 | S | 6004 | 36 | 8 | 0.33 | 0.96 | 10.00 | 0.89 | 10.50 | | | 0.95 | 9.75 | 0.93 | 0.07 | 10.08 | 0.75 | 4.14 |
| HD 3556 | S | 6019 | 26 | 8 | 0.12 | 1.09 | 2.32 | 1.06 | 2.59 | 1.10 | 2.07 | 1.08 | 2.89 | 1.08 | 0.04 | 2.47 | 0.83 | 4.42 |
| HD 35961 | S | 5740 | 74 | 3 | 0.04 | 0.98 | 7.94 | 0.98 | 7.00 | 0.97 | 7.33 | | | 0.98 | 0.01 | 7.43 | 0.94 | 4.37 |
| HD 36003 | S | 4536 | 28 | 12 | -0.72 | 0.76 | 5.36 | 0.71 | 11.50 | | | | | 0.74 | 0.05 | 8.43 | 6.14 | 4.60 |
| HD 37124 | S | 5561 | 24 | 9 | -0.06 | 0.91 | 11.49 | 0.92 | 9.75 | | | | | 0.92 | 0.01 | 10.62 | 1.74 | 4.39 |
| HD 37605 | S | 5318 | | 1 | -0.21 | 0.93 | 4.48 | 0.97 | 3.44 | 0.97 | 2.42 | 0.92 | 8.63 | 0.95 | 0.05 | 4.75 | 6.21 | 4.47 |
| HD 3795 | S | 5456 | 91 | 10 | 0.42 | 0.94 | 11.75 | 0.96 | 10.75 | 0.93 | 11.00 | 0.93 | 10.88 | 0.94 | 0.03 | 11.09 | 1.00 | 3.89 |
| HD 38529 | S | 5450 | | 1 | 0.77 | 1.39 | 3.16 | 1.42 | 3.75 | | | 1.53 | 2.50 | 1.45 | 0.14 | 3.14 | 1.25 | 3.72 |
| HD 38700 | S | 6049 | 27 | 2 | 0.06 | 0.97 | 4.62 | 1.03 | 2.97 | 1.01 | 1.73 | 0.99 | 2.86 | 1.00 | 0.06 | 3.05 | 2.88 | 4.45 |
| HD 38858 | S | 5768 | 27 | 5 | -0.09 | 0.87 | 10.24 | 0.91 | 6.19 | 0.89 | 7.00 | 0.86 | 9.23 | 0.88 | 0.05 | 8.16 | 4.05 | 4.46 |
| HD 38A | S | 4065 | 76 | 6 | -1.13 | | | 0.53 | 5.97 | 0.59 | 3.84 | 0.60 | 5.29 | 0.57 | 0.07 | 5.03 | 2.12 | 4.71 |
| HD 38B | S | 4028 | 103 | 6 | -1.17 | | | 0.52 | 5.81 | 0.57 | 3.18 | 0.59 | 6.27 | 0.56 | 0.07 | 5.09 | 3.09 | 4.72 |
| HD 39881 | S | 5719 | 34 | 9 | 0.14 | 1.03 | 7.23 | 0.97 | 9.50 | | | | | 1.00 | 0.06 | 8.36 | 2.27 | 4.27 |
| HD 400 | S | 6240 | 22 | 3 | 0.46 | | | 0.97 | 8.67 | 1.11 | 5.00 | 1.11 | 5.25 | 1.06 | 0.14 | 6.31 | 3.67 | 4.13 |
| HD 40979 | S | 6150 | 33 | 9 | 0.26 | 1.14 | 2.51 | 1.21 | 1.43 | 1.17 | 2.25 | 1.20 | 1.50 | 1.18 | 0.07 | 1.92 | 1.08 | 4.35 |

| ID | Type | T | a | b | c | d | e | f | g | h | i | j | k | l | m | n | o | p |
|---|---|---|---|---|---|---|---|---|---|---|---|---|---|---|---|---|---|---|
| HD 41330 | S | 5876 | 33 | 7 | 0.30 | 1.04 | 8.11 | 0.86 | 12.50 | 0.91 | 11.00 | 0.91 | 11.75 | 0.93 | 0.18 | 10.84 | 4.39 | 4.13 |
| HD 41708 | S | 5867 | 38 | 9 | -0.02 | 0.94 | 7.35 | 1.00 | 3.13 | 1.02 | 2.61 | 0.96 | 4.79 | 0.98 | 0.08 | 4.47 | 4.74 | 4.47 |
| HD 4203 | S | 5471 | | 1 | 0.25 | 1.00 | 10.48 | 1.02 | 10.75 | 1.05 | 9.20 | 1.02 | 11.69 | 1.02 | 0.05 | 10.53 | 2.49 | 4.10 |
| HD 4208 | S | 5674 | 33 | 9 | -0.15 | 0.84 | 10.99 | 0.77 | 12.25 | 0.85 | 8.50 | 0.84 | 9.25 | 0.83 | 0.08 | 10.25 | 3.75 | 4.47 |
| HD 43162 | S | 5651 | 43 | 9 | -0.16 | 0.84 | 10.99 | 0.99 | 0.68 | 0.88 | 6.50 | 0.84 | 9.75 | 0.89 | 0.15 | 6.98 | 10.31 | 4.50 |
| HD 43587 | S | 5859 | 75 | 8 | 0.21 | 1.13 | 2.51 | 1.03 | 7.50 | | | 0.99 | 9.50 | 1.05 | 0.14 | 6.50 | 6.99 | 4.27 |
| HD 43745 | S | 6087 | 45 | 5 | 0.70 | 1.25 | 4.05 | 1.16 | 5.13 | | | 1.32 | 3.75 | 1.24 | 0.16 | 4.31 | 1.38 | 3.92 |
| HD 45067 | S | 6049 | 34 | 8 | 0.62 | | | 1.00 | 7.63 | 1.31 | 4.00 | 1.10 | 6.00 | 1.14 | 0.31 | 5.88 | 3.63 | 3.95 |
| HD 45088 | S | 4778 | 67 | 8 | -0.31 | | | | | | | | | | | | | 4.37 |
| HD 45184 | S | 5852 | 32 | 9 | 0.06 | 1.03 | 4.50 | 1.05 | 3.38 | 1.01 | 5.33 | 1.05 | 3.82 | 1.04 | 0.04 | 4.26 | 1.96 | 4.41 |
| HD 45205 | S | 5921 | 40 | 9 | 0.27 | 0.93 | 11.49 | 0.86 | 12.00 | 0.88 | 11.50 | 0.88 | 12.00 | 0.89 | 0.07 | 11.75 | 0.51 | 4.15 |
| HD 45350 | S | 5567 | 36 | 9 | 0.17 | 1.05 | 6.31 | | | 1.07 | 8.00 | 1.02 | 11.50 | 1.05 | 0.05 | 8.60 | 5.19 | 4.22 |
| HD 45588 | S | 6214 | 35 | 7 | 0.43 | 1.29 | 1.26 | 1.24 | 3.25 | 1.33 | 1.80 | 1.27 | 2.50 | 1.28 | 0.09 | 2.20 | 1.99 | 4.24 |
| HD 45759 | S | 6132 | 26 | 9 | 0.22 | 1.14 | 2.03 | 1.07 | 4.16 | 1.14 | 2.38 | 1.15 | 2.30 | 1.13 | 0.08 | 2.72 | 2.13 | 4.37 |
| HD 4628 | S | 5044 | 38 | 10 | -0.54 | 0.76 | 5.66 | | | 0.74 | 8.00 | 0.74 | 6.19 | 0.75 | 0.02 | 6.61 | 2.34 | 4.61 |
| HD 46375 | S | 5265 | 49 | 9 | -0.09 | | | | | 0.94 | 10.33 | | | 0.94 | 0.00 | 10.33 | 0.00 | 4.33 |
| HD 48938 | S | 6055 | 39 | 12 | 0.21 | 1.11 | 3.57 | 1.05 | 4.90 | 1.13 | 3.00 | 0.93 | 9.17 | 1.06 | 0.20 | 5.16 | 6.17 | 4.33 |
| HD 49674 | S | 5621 | 64 | 3 | 0.00 | 0.96 | 8.34 | 1.05 | 3.50 | 1.04 | 3.82 | 1.00 | 7.32 | 1.01 | 0.09 | 5.75 | 4.84 | 4.39 |
| HD 50281 | S | 4708 | 36 | 8 | -0.68 | 0.79 | 1.88 | 0.79 | 1.22 | 0.74 | 5.92 | 0.76 | 4.22 | 0.77 | 0.05 | 3.31 | 4.70 | 4.64 |
| HD 50554 | S | 6019 | 27 | 9 | 0.13 | 1.11 | 1.77 | 1.05 | 3.75 | 1.09 | 2.33 | 1.07 | 3.05 | 1.08 | 0.06 | 2.73 | 1.98 | 4.41 |
| HD 50806 | S | 5640 | 78 | 8 | 0.35 | 1.09 | 6.31 | 1.04 | 9.50 | 1.08 | 8.25 | 0.98 | 11.13 | 1.05 | 0.11 | 8.80 | 4.82 | 4.06 |
| HD 52265 | S | 6086 | 35 | 9 | 0.32 | 1.22 | 1.13 | 1.19 | 3.25 | 1.26 | 1.30 | 1.20 | 3.13 | 1.22 | 0.07 | 2.20 | 2.12 | 4.29 |
| HD 52698 | S | 5155 | 46 | 11 | -0.36 | 0.89 | 4.78 | 0.89 | 4.41 | 0.92 | 2.60 | 0.88 | 6.60 | 0.90 | 0.04 | 4.60 | 4.00 | 4.55 |
| HD 52711 | S | 5992 | 59 | 9 | 0.11 | 1.10 | 1.40 | 1.02 | 4.34 | 1.09 | 2.19 | 1.03 | 4.16 | 1.06 | 0.08 | 3.02 | 2.95 | 4.41 |
| HD 5494 | S | 6044 | 106 | 9 | 0.51 | 1.16 | 5.23 | 0.92 | 9.83 | 1.16 | 5.40 | 1.05 | 7.00 | 1.07 | 0.24 | 6.87 | 4.60 | 4.03 |
| HD 55054 | S | 6219 | 56 | 9 | 0.46 | 1.25 | 3.16 | 0.93 | 9.20 | 1.10 | 5.00 | 1.12 | 5.25 | 1.10 | 0.32 | 5.65 | 6.04 | 4.14 |
| HD 55575 | S | 5937 | 60 | 9 | 0.15 | 1.06 | 5.01 | 1.00 | 6.56 | 1.02 | 6.00 | 0.88 | 11.50 | 0.99 | 0.18 | 7.27 | 6.49 | 4.32 |
| HD 55693 | S | 5854 | 35 | 9 | 0.18 | 1.13 | 2.51 | 1.09 | 5.06 | 1.14 | 3.20 | 1.11 | 5.00 | 1.12 | 0.05 | 3.94 | 2.55 | 4.32 |
| HD 57006 | S | 6264 | 45 | 9 | 0.99 | 1.50 | 2.51 | 1.50 | 2.63 | 1.61 | 2.25 | 1.51 | 2.13 | 1.53 | 0.11 | 2.38 | 0.50 | 3.77 |
| HD 603 | S | 5967 | 20 | 8 | 0.15 | 1.07 | 3.53 | 1.03 | 4.94 | 1.12 | 2.49 | 1.08 | 3.49 | 1.08 | 0.09 | 3.61 | 2.45 | 4.37 |
| HD 6064 | S | 6363 | 48 | 9 | 0.99 | 1.52 | 2.25 | 1.37 | 3.09 | 1.56 | 2.30 | 1.53 | 2.06 | 1.50 | 0.19 | 2.43 | 1.04 | 3.78 |
| HD 61606 | S | 4932 | 26 | 12 | -0.54 | 0.83 | 3.02 | 0.83 | 1.99 | 0.81 | 2.59 | 0.80 | 5.29 | 0.82 | 0.03 | 3.22 | 3.31 | 4.61 |
| HD 61632 | S | 5632 | 39 | 10 | 0.35 | 0.98 | 10.99 | 0.97 | 10.92 | 0.94 | 11.20 | 0.91 | 11.71 | 0.95 | 0.07 | 11.20 | 0.79 | 4.02 |
| HD 63598 | S | 5828 | 66 | 9 | 0.15 | | | | | | | | | | | | | 4.38 |
| HD 63754 | S | 6040 | 59 | 9 | 0.70 | 1.30 | 3.57 | 1.27 | 4.25 | | | 1.33 | 3.75 | 1.30 | 0.06 | 3.86 | 0.68 | 3.92 |
| HD 68988 | S | 5878 | 31 | 9 | 0.12 | 1.10 | 2.50 | 1.14 | 1.80 | 1.15 | 1.55 | 1.13 | 2.28 | 1.13 | 0.05 | 2.03 | 0.95 | 4.39 |
| HD 69830 | S | 5443 | 17 | 8 | -0.23 | 0.94 | 3.02 | 0.92 | 3.64 | 0.94 | 2.68 | 0.93 | 3.81 | 0.93 | 0.02 | 3.29 | 1.13 | 4.53 |
| HD 70889 | S | 6036 | 32 | 9 | 0.12 | 1.09 | 2.20 | 1.09 | 2.30 | 1.11 | 1.56 | 1.07 | 2.68 | 1.09 | 0.04 | 2.19 | 1.12 | 4.43 |

| | | | | | | | | | | | | | | | | | |
|---|---|---|---|---|---|---|---|---|---|---|---|---|---|---|---|---|---|
| HD 7091   | S | 6109 | 104 | 8  | 0.30  | 1.17 | 2.68  | 1.07 | 5.13  | 1.22 | 1.91 | 1.15 | 2.91  | 1.15 | 0.15 | 3.16  | 3.23 | 4.29 |
| HD 71148  | S | 5835 | 27  | 9  | 0.09  | 1.09 | 2.00  | 1.01 | 6.00  | 1.11 | 2.57 | 1.06 | 4.81  | 1.07 | 0.10 | 3.85  | 4.00 | 4.39 |
| HD 71881  | S | 5863 | 30  | 9  | 0.18  | 1.13 | 2.25  | 1.03 | 7.00  | 1.05 | 6.33 | 1.02 | 8.13  | 1.06 | 0.11 | 5.93  | 5.87 | 4.30 |
| HD 7230   | S | 6637 | 52  | 8  | 0.62  | 1.37 | 1.58  | 1.11 | 4.88  | 1.44 | 1.19 | 1.35 | 1.70  | 1.32 | 0.33 | 2.34  | 3.69 | 4.17 |
| HD 72659  | S | 5902 | 29  | 9  | 0.33  | 0.99 | 9.59  | 0.87 | 11.75 | 0.99 | 9.00 | 0.98 | 9.83  | 0.96 | 0.12 | 10.04 | 2.75 | 4.12 |
| HD 72946  | S | 5674 | 58  | 9  | -0.09 | 1.00 | 2.82  | 1.01 | 2.07  | 1.01 | 1.93 | 0.99 | 3.38  | 1.00 | 0.02 | 2.55  | 1.45 | 4.49 |
| HD 7352   | S | 6023 | 21  | 7  | 0.25  | 1.12 | 3.61  | 1.09 | 4.63  | 1.18 | 2.32 | 1.14 | 3.66  | 1.13 | 0.09 | 3.55  | 2.30 | 4.31 |
| HD 73596  | S | 6744 | 55  | 9  | 1.61  | 2.05 | 0.90  | 1.98 | 1.04  | 2.16 | 0.98 | 2.13 | 0.70  | 2.08 | 0.18 | 0.90  | 0.34 | 3.41 |
| HD 73668  | S | 5922 | 36  | 9  | 0.14  | 1.07 | 3.89  | 1.05 | 4.73  | 1.14 | 2.13 | 1.07 | 4.62  | 1.08 | 0.09 | 3.84  | 2.60 | 4.37 |
| HD 73752  | S | 5654 | 83  | 5  | 0.48  |      |       | 1.15 | 7.00  | 1.17 | 6.50 | 1.18 | 6.75  | 1.17 | 0.03 | 6.75  | 0.50 | 3.98 |
| HD 7397   | S | 6029 | 50  | 8  | 0.39  | 1.19 | 3.50  | 0.94 | 9.38  | 1.05 | 7.00 | 1.11 | 6.34  | 1.07 | 0.25 | 6.56  | 5.87 | 4.15 |
| HD 74156  | S | 6005 | 37  | 9  | 0.49  | 1.02 | 7.94  | 1.10 | 6.65  | 1.19 | 5.00 | 1.17 | 5.56  | 1.12 | 0.17 | 6.29  | 2.94 | 4.06 |
| HD 7514   | S | 5682 | 8   | 6  | 0.16  | 0.99 | 9.46  | 0.96 | 10.00 | 1.06 | 6.00 | 0.98 | 10.94 | 1.00 | 0.10 | 9.10  | 4.94 | 4.24 |
| HD 75488  | S | 5996 | 60  | 9  | 0.28  | 0.94 | 10.47 | 0.98 | 8.25  | 0.88 | 11.00| 0.90 | 11.00 | 0.93 | 0.10 | 10.18 | 2.75 | 4.18 |
| HD 76151  | S | 5780 | 49  | 12 | 0.01  | 1.00 | 5.66  | 1.03 | 3.50  | 0.96 | 6.50 | 1.06 | 2.59  | 1.01 | 0.10 | 4.56  | 3.91 | 4.43 |
| HD 78366  | S | 5981 | 27  | 9  | 0.11  | 1.10 | 1.77  | 1.10 | 1.82  | 1.08 | 2.55 | 1.06 | 3.34  | 1.09 | 0.04 | 2.37  | 1.57 | 4.42 |
| HD 79210  | S | 3973 | 22  | 5  | -1.21 |      |       | 0.50 | 7.68  |      |      |      |       | 0.50 | 0.00 | 7.68  | 0.00 | 4.69 |
| HD 79211  | S | 3763 | 69  | 6  | -1.08 |      |       |      |       |      |      |      |       |      |      |       |      | 4.49 |
| HD 80372  | S | 6039 | 21  | 8  | 0.19  | 1.04 | 5.93  | 1.07 | 4.15  | 1.17 | 1.94 | 1.12 | 3.12  | 1.10 | 0.13 | 3.79  | 3.99 | 4.36 |
| HD 80606  | S | 5467 |     | 1  | 0.85  |      |       |      |       |      |      |      |       |      |      |       |      | 3.52 |
| HD 80607  | S | 5535 | 54  | 7  | 1.05  |      |       |      |       |      |      |      |       |      |      |       |      | 3.35 |
| HD 81040  | S | 5753 | 33  | 9  | -0.12 | 0.87 | 8.68  | 0.79 | 11.25 | 0.92 | 4.10 | 0.89 | 5.86  | 0.87 | 0.13 | 7.47  | 7.15 | 4.48 |
| HD 8173   | S | 6009 | 19  | 8  | 0.13  | 1.06 | 3.78  | 1.06 | 2.76  | 1.11 | 2.12 | 1.09 | 2.99  | 1.08 | 0.05 | 2.91  | 1.66 | 4.40 |
| HD 81809  | S | 5666 | 42  | 6  | 0.75  | 1.12 | 5.01  | 1.10 | 5.65  | 1.27 | 4.00 | 1.23 | 3.83  | 1.18 | 0.17 | 4.62  | 1.82 | 3.72 |
| HD 82106  | S | 4826 | 32  | 12 | -0.61 | 0.82 | 1.21  | 0.81 | 1.21  | 0.78 | 2.59 | 0.78 | 3.23  | 0.80 | 0.04 | 2.06  | 2.02 | 4.63 |
| HD 82943  | S | 5919 | 51  | 3  | 0.18  | 1.15 | 1.42  | 1.12 | 3.75  | 1.17 | 1.84 | 1.13 | 3.86  | 1.14 | 0.05 | 2.72  | 2.44 | 4.35 |
| HD 84117  | S | 6239 | 62  | 10 | 0.29  | 1.18 | 1.79  | 1.17 | 2.25  | 1.17 | 2.00 | 1.15 | 2.60  | 1.17 | 0.03 | 2.16  | 0.81 | 4.34 |
| HD 84703  | S | 6125 | 46  | 9  | 0.49  | 1.22 | 4.01  | 1.18 | 5.00  | 1.35 | 1.97 | 1.27 | 3.67  | 1.26 | 0.17 | 3.66  | 3.03 | 4.14 |
| HD 85725  | S | 5892 | 42  | 7  | 0.87  | 1.30 | 3.16  | 1.31 | 3.69  |      |      | 1.42 | 2.75  | 1.34 | 0.12 | 3.20  | 0.94 | 3.72 |
| HD 8574   | S | 6045 | 40  | 9  | 0.36  | 1.23 | 2.00  | 1.01 | 7.75  | 1.04 | 7.00 |      |       | 1.09 | 0.22 | 5.58  | 5.75 | 4.19 |
| HD 8673   | S | 6409 | 55  | 9  | 0.49  | 1.26 | 2.31  | 1.32 | 1.83  | 1.39 | 0.83 | 1.36 | 1.14  | 1.33 | 0.13 | 1.53  | 1.48 | 4.25 |
| HD 87097  | S | 5993 | 74  | 9  | 0.12  | 1.06 | 3.78  | 1.05 | 3.05  | 1.06 | 3.24 | 1.02 | 4.82  | 1.05 | 0.04 | 3.72  | 1.77 | 4.40 |
| HD 88133  | S | 5371 | 57  | 9  | 0.58  | 1.21 | 5.08  | 1.26 | 5.44  | 1.25 | 5.25 | 1.30 | 4.78  | 1.26 | 0.09 | 5.14  | 0.66 | 3.82 |
| HD 88230  | S | 4215 | 43  | 5  | -1.03 | 0.61 | 3.23  | 0.56 | 6.45  | 0.62 | 3.64 | 0.64 | 1.06  | 0.61 | 0.08 | 3.59  | 5.38 | 4.70 |
| HD 88371  | S | 5673 | 58  | 9  | 0.12  | 0.95 | 11.22 | 0.96 | 10.75 | 0.99 | 9.00 |      |       | 0.97 | 0.04 | 10.32 | 2.22 | 4.27 |
| HD 88595  | S | 6322 | 43  | 7  | 0.55  |      |       | 1.02 | 6.50  | 1.40 | 1.80 |      |       | 1.21 | 0.38 | 4.15  | 4.70 | 4.12 |
| HD 88737  | S | 6129 | 32  | 7  | 1.01  | 1.59 | 2.05  | 1.60 | 2.44  | 1.68 | 2.15 | 1.53 | 2.08  | 1.60 | 0.15 | 2.18  | 0.39 | 3.73 |
| HD 89319  | S | 4942 | 98  | 8  | 0.92  | 1.52 | 2.51  | 1.38 | 3.75  | 1.10 | 7.00 | 1.18 | 7.19  | 1.30 | 0.42 | 5.11  | 4.68 | 3.35 |
| HD 89744 | S | 6207 | 60 | 7 | 0.80 | 1.37 | 3.16 | 1.40 | 3.33 | 1.57 | 1.80 | 1.49 | 2.42 | 1.46 | 0.20 | 2.68 | 1.53 | 3.92 |
| HD 90508 | S | 5757 | 85 | 9 | 0.08 | 0.99 | 7.94 | 0.98 | 7.75 | 1.01 | 7.00 | 0.96 | 9.75 | 0.99 | 0.05 | 8.11 | 2.75 | 4.34 |
| HD 9224 | S | 5848 | 37 | 9 | 0.29 | 0.92 | 12.02 | 1.11 | 6.25 | 1.02 | 8.50 | 0.95 | 10.83 | 1.00 | 0.19 | 9.40 | 5.77 | 4.16 |
| HD 92788 | S | 5710 | 44 | 9 | 0.11 | 0.99 | 8.85 | 1.08 | 4.93 | 1.06 | 5.50 | 1.04 | 7.83 | 1.04 | 0.09 | 6.78 | 3.93 | 4.32 |
| HD 9369 | S | 7237 | 126 | 4 | 0.97 | 1.68 | 0.82 | 1.72 | 0.78 | 1.71 | 0.80 | 1.72 | 0.70 | 1.71 | 0.04 | 0.77 | 0.12 | 4.09 |
| HD 94132 | S | 4988 | 79 | 6 | 0.94 | | | 1.56 | 2.75 | 1.70 | 2.00 | 1.58 | 2.38 | 1.61 | 0.14 | 2.38 | 0.75 | 3.44 |
| HD 94915 | S | 5985 | 44 | 9 | 0.09 | 1.08 | 2.49 | 1.04 | 3.22 | 1.05 | 3.28 | 0.91 | 8.22 | 1.02 | 0.17 | 4.30 | 5.73 | 4.41 |
| HD 95650 | S | | | | -1.26 | | | | | | | | | | | | | 4.56 |
| HD 96276 | S | 6040 | 50 | 9 | 0.20 | 1.10 | 3.57 | 1.06 | 4.53 | 1.12 | 2.75 | 1.06 | 5.00 | 1.09 | 0.06 | 3.96 | 2.25 | 4.34 |
| HD 97100 | S | 5008 | 64 | 7 | -1.46 | | | | | | | | | | | | | 5.63 |
| HD 97101 | S | 4189 | 81 | 9 | -0.95 | 0.69 | 5.36 | 0.70 | 2.57 | | | | | 0.70 | 0.01 | 3.96 | 2.79 | 4.67 |
| HD 97334 | S | 5906 | 29 | 9 | 0.05 | 1.03 | 4.03 | 1.07 | 1.95 | 1.07 | 2.02 | 1.03 | 3.00 | 1.05 | 0.04 | 2.75 | 2.08 | 4.44 |
| HD 975 | S | 6405 | 37 | 9 | 0.42 | 1.24 | 1.61 | 1.11 | 4.83 | 1.24 | 1.86 | 1.20 | 2.29 | 1.20 | 0.13 | 2.65 | 3.22 | 4.27 |
| HD 97584 | S | 4732 | 2 | 2 | -0.65 | 0.79 | 3.19 | 0.78 | 2.02 | 0.75 | 7.50 | 0.76 | 5.29 | 0.77 | 0.04 | 4.50 | 5.48 | 4.62 |
| HD 98388 | S | 6338 | 44 | 9 | 0.34 | 1.21 | 1.28 | 1.25 | 1.02 | 1.24 | 1.15 | 1.20 | 1.67 | 1.23 | 0.05 | 1.28 | 0.65 | 4.34 |
| HD 98712 | S | 4300 | 32 | 3 | -1.02 | 0.63 | 0.63 | 0.58 | 1.28 | 0.60 | 4.44 | 0.60 | 5.09 | 0.60 | 0.05 | 2.86 | 4.46 | 4.72 |
| HD 98823 | S | 6414 | 34 | 9 | 1.26 | 1.74 | 1.48 | 1.54 | 1.92 | 1.84 | 1.55 | 1.79 | 1.25 | 1.73 | 0.30 | 1.55 | 0.67 | 3.59 |
| HR 1179 | S | 6292 | 67 | 9 | 0.45 | 1.19 | 3.60 | 1.04 | 7.13 | | | 1.04 | 6.00 | 1.09 | 0.15 | 5.58 | 3.52 | 4.17 |
| HR 1232 | S | 4905 | 51 | 6 | 1.03 | 1.52 | 2.51 | 1.44 | 3.38 | 1.43 | 4.50 | 1.18 | 7.44 | 1.39 | 0.34 | 4.46 | 4.93 | 3.26 |
| HR 159 | S | 5501 | 70 | 7 | 0.09 | | | | | 1.06 | 8.00 | | | 1.06 | 0.00 | 8.00 | 0.00 | 4.28 |
| HR 1665 | S | 5935 | 61 | 12 | 0.66 | | | 1.10 | 6.19 | 1.46 | 3.00 | 1.19 | 5.00 | 1.25 | 0.36 | 4.73 | 3.19 | 3.92 |
| HR 1685 | S | 4976 | 203 | 9 | 0.86 | 1.60 | 2.00 | 1.42 | 3.50 | 1.63 | 2.50 | 1.24 | 5.80 | 1.47 | 0.39 | 3.45 | 3.80 | 3.48 |
| HR 1925 | S | 5243 | 32 | 9 | -0.32 | 0.86 | 8.69 | 0.90 | 3.42 | 0.88 | 5.50 | 0.89 | 5.40 | 0.88 | 0.04 | 5.75 | 5.27 | 4.53 |
| HR 1980 | S | 6095 | 42 | 9 | 0.19 | 1.09 | 3.51 | 1.07 | 3.90 | 1.14 | 1.70 | 1.10 | 3.09 | 1.10 | 0.07 | 3.05 | 2.20 | 4.38 |
| HR 2208 | S | 5762 | 46 | 9 | -0.13 | 0.89 | 6.42 | 0.80 | 10.00 | 0.93 | 2.33 | 0.91 | 3.28 | 0.88 | 0.13 | 5.51 | 7.67 | 4.50 |
| HR 244 | S | 6201 | 74 | 9 | 0.53 | 1.17 | 4.74 | 1.14 | 5.42 | | | 1.21 | 4.25 | 1.17 | 0.07 | 4.80 | 1.17 | 4.09 |
| HR 2692 | S | 4995 | 41 | 9 | 0.77 | 1.27 | 4.74 | | | 1.10 | 7.00 | 1.06 | 8.83 | 1.14 | 0.21 | 6.86 | 4.10 | 3.47 |
| HR 2866 | S | 6358 | 33 | 9 | 0.43 | 1.27 | 1.71 | 1.04 | 6.50 | 1.29 | 1.69 | 1.21 | 2.60 | 1.20 | 0.25 | 3.13 | 4.81 | 4.25 |
| HR 3193 | S | 5989 | 31 | 9 | 0.65 | | | 1.21 | 5.20 | | | 1.13 | 5.50 | 1.17 | 0.08 | 5.35 | 0.30 | 3.91 |
| HR 3271 | S | 5975 | 38 | 9 | 0.64 | 1.39 | 2.51 | 1.19 | 5.33 | 1.43 | 3.00 | 1.17 | 5.25 | 1.30 | 0.26 | 4.02 | 2.82 | 3.96 |
| HR 3395 | S | 6222 | 47 | 9 | 0.33 | 1.17 | 2.51 | 1.21 | 2.25 | 1.25 | 1.55 | 1.23 | 1.73 | 1.22 | 0.08 | 2.01 | 0.96 | 4.32 |
| HR 357 | S | 6491 | 77 | 9 | 1.11 | 1.61 | 2.00 | 1.47 | 2.44 | 1.67 | 1.80 | 1.53 | 1.75 | 1.57 | 0.20 | 2.00 | 0.69 | 3.72 |
| HR 3762 | S | 5153 | 88 | 7 | 0.98 | | | 1.27 | 4.58 | 1.60 | 2.25 | 1.35 | 2.85 | 1.41 | 0.33 | 3.23 | 2.33 | 3.40 |
| HR 3901 | S | 6081 | 31 | 7 | 0.58 | 1.09 | 6.31 | 1.17 | 5.42 | | | 1.27 | 4.08 | 1.18 | 0.18 | 5.27 | 2.23 | 4.01 |
| HR 4051 | S | 5978 | 79 | 9 | 0.32 | 1.04 | 7.74 | 0.98 | 8.50 | 1.01 | 8.00 | | | 1.01 | 0.06 | 8.08 | 0.76 | 4.18 |
| HR 407 | S | 6520 | 54 | 8 | 1.12 | 1.62 | 2.00 | 1.46 | 2.46 | 1.67 | 1.80 | 1.52 | 1.75 | 1.57 | 0.21 | 2.00 | 0.71 | 3.72 |
| HR 4285 | S | 5916 | 15 | 5 | 0.80 | 1.23 | 3.98 | 1.17 | 4.75 | 1.25 | 4.00 | 1.28 | 3.65 | 1.23 | 0.11 | 4.10 | 1.10 | 3.76 |
| HR 448 | S | 5861 | 32 | 12 | 0.56 | | | 1.25 | 5.10 | | | 1.22 | 5.50 | 1.24 | 0.03 | 5.30 | 0.40 | 3.99 |
| HR 4864 | S | 5615 | 21 | 7 | -0.12 | 0.97 | 4.20 | 0.99 | 2.29 | 0.95 | 4.90 | 0.99 | 3.16 | 0.98 | 0.04 | 3.64 | 2.61 | 4.49 |
| HR 495 | S | 4661 | | 1 | 1.08 | 1.31 | 5.23 | 1.37 | 4.50 | 1.33 | 5.00 | 1.08 | 10.50 | 1.27 | 0.29 | 6.31 | 6.00 | 3.08 |
| HR 511 | S | 5422 | 69 | 9 | -0.28 | | | 0.93 | 0.83 | 0.81 | 9.50 | 0.93 | 0.75 | 0.89 | 0.12 | 3.69 | 8.75 | 4.55 |
| HR 5258 | S | 6417 | 77 | 9 | 1.12 | 1.65 | 1.75 | 1.64 | 2.08 | 1.67 | 2.00 | 1.65 | 1.63 | 1.65 | 0.03 | 1.86 | 0.46 | 3.71 |
| HR 5317 | S | 6437 | 63 | 11 | 1.09 | 1.62 | 1.89 | 1.65 | 2.08 | 1.65 | 1.95 | 1.62 | 1.75 | 1.64 | 0.03 | 1.92 | 0.33 | 3.74 |
| HR 5335 | S | 4862 | 123 | 6 | 0.87 | 0.95 | 12.02 | | | | | 1.17 | 7.25 | 1.06 | 0.22 | 9.64 | 4.77 | 3.29 |
| HR 5387 | S | 6753 | 30 | 9 | 0.53 | 1.37 | 0.71 | 1.34 | 1.05 | 1.26 | 1.80 | 1.24 | 2.00 | 1.30 | 0.13 | 1.39 | 1.29 | 4.29 |
| HR 5504 | S | 5990 | 37 | 9 | 0.58 | 1.19 | 5.01 | 1.28 | 4.33 | 1.22 | 5.00 | 1.28 | 4.25 | 1.24 | 0.09 | 4.65 | 0.76 | 4.01 |
| HR 5630 | S | 6186 | 53 | 9 | 0.33 | 1.16 | 3.16 | 1.21 | 2.50 | 1.24 | 1.66 | 1.21 | 2.55 | 1.21 | 0.08 | 2.47 | 1.50 | 4.30 |
| HR 5706 | S | 4785 | 14 | 2 | 0.61 | 1.05 | 10.00 | 1.14 | 8.20 | 1.00 | 12.00 | 1.05 | 11.50 | 1.06 | 0.14 | 10.43 | 3.80 | 3.52 |
| HR 5740 | S | 5956 | 29 | 9 | 0.65 | | | 1.30 | 4.50 | 1.45 | 2.75 | 1.33 | 4.00 | 1.36 | 0.15 | 3.75 | 1.75 | 3.97 |
| HR 6105 | S | 5934 | 37 | 2 | 0.33 | 1.00 | 9.25 | 0.88 | 11.50 | 1.13 | 5.50 | 1.06 | 7.50 | 1.02 | 0.25 | 8.44 | 6.00 | 4.16 |
| HR 6106 | S | 6009 | 38 | 3 | 0.60 | 1.09 | 6.31 | 1.16 | 5.65 | 1.28 | 4.33 | 1.19 | 5.17 | 1.18 | 0.19 | 5.36 | 1.98 | 3.97 |
| HR 6269 | S | 5634 | 57 | 6 | 0.46 | 1.01 | 9.47 | 1.11 | 7.75 | 1.01 | 9.00 | 1.07 | 8.42 | 1.05 | 0.10 | 8.66 | 1.72 | 3.95 |
| HR 6301 | S | 4955 | 81 | 9 | 1.05 | 1.44 | 2.99 | 1.60 | 2.50 | 1.74 | 1.97 | 1.54 | 2.50 | 1.58 | 0.30 | 2.49 | 1.02 | 3.31 |
| HR 6372 | S | 5744 | 48 | 9 | 0.48 | 1.16 | 5.01 | 1.24 | 5.31 | 1.18 | 6.00 | 1.17 | 6.75 | 1.19 | 0.08 | 5.77 | 1.74 | 4.02 |
| HR 6465 | S | 5705 | 31 | 7 | 0.00 | 1.05 | 2.20 | 1.00 | 5.63 | 1.07 | 2.66 | 1.00 | 6.57 | 1.03 | 0.07 | 4.26 | 4.37 | 4.42 |
| HR 6516 | S | 5583 | 20 | 7 | 0.21 | | | 0.95 | 12.50 | | | 1.00 | 12.00 | 0.98 | 0.05 | 12.25 | 0.50 | 4.15 |
| HR 6669 | S | 6134 | 72 | 9 | 0.35 | 1.24 | 1.02 | 1.17 | 3.88 | 1.28 | 1.70 | 1.20 | 3.50 | 1.22 | 0.11 | 2.52 | 2.86 | 4.27 |
| HR 672 | S | 6079 | 57 | 7 | 0.45 | 1.02 | 7.94 | 0.94 | 9.00 | | | 1.03 | 7.50 | 1.00 | 0.09 | 8.15 | 1.50 | 4.07 |
| HR 6722 | S | 5539 | 64 | 7 | 0.74 | 1.24 | 3.98 | 1.14 | 5.58 | 1.34 | 4.00 | 1.34 | 3.46 | 1.27 | 0.20 | 4.26 | 2.12 | 3.72 |
| HR 6756 | S | 4907 | 3 | 2 | 0.86 | 1.27 | 5.78 | 1.30 | 4.75 | 1.18 | 7.00 | 1.27 | 5.25 | 1.26 | 0.12 | 5.69 | 2.25 | 3.39 |
| HR 6806 | S | 5042 | 40 | 9 | -0.46 | 0.84 | 5.36 | 0.83 | 3.88 | 0.84 | 2.88 | 0.82 | 5.29 | 0.83 | 0.02 | 4.35 | 2.48 | 4.58 |
| HR 6847 | S | 5791 | 24 | 9 | 0.13 | 1.09 | 2.51 | 1.00 | 8.00 | 1.13 | 3.00 | 1.03 | 7.67 | 1.06 | 0.13 | 5.29 | 5.49 | 4.33 |
| HR 6907 | S | 6311 | 33 | 9 | 0.67 | | | | | | | 1.32 | 3.25 | 1.32 | 0.00 | 3.25 | 0.00 | 4.04 |
| HR 6950 | S | 5337 | 76 | 9 | 0.45 | 0.99 | 10.22 | 0.96 | 10.50 | 1.10 | 8.00 | 1.01 | 9.36 | 1.02 | 0.14 | 9.52 | 2.50 | 3.85 |
| HR 7079 | S | 6319 | 39 | 9 | 0.35 | 1.16 | 2.86 | 1.12 | 3.75 | 1.23 | 1.60 | 1.23 | 1.46 | 1.19 | 0.11 | 2.42 | 2.29 | 4.31 |
| HR 7291 | S | 6173 | 45 | 8 | 0.29 | 1.15 | 3.16 | 1.23 | 1.36 | 1.23 | 1.55 | 1.22 | 1.67 | 1.21 | 0.08 | 1.94 | 1.80 | 4.34 |
| HR 7522 | S | 6026 | 29 | 7 | 0.70 | 1.22 | 4.50 | 1.23 | 4.67 | 1.18 | 5.00 | 1.25 | 4.38 | 1.22 | 0.07 | 4.63 | 0.63 | 3.89 |
| HR 7569 | S | 5774 | 59 | 7 | 0.33 | 1.03 | 8.52 | 1.03 | 9.00 | | | 0.94 | 11.38 | 1.00 | 0.09 | 9.63 | 2.86 | 4.10 |
| HR 761 | S | 6230 | 74 | 9 | 0.81 | 1.37 | 3.16 | 1.17 | 4.69 | 1.22 | 4.00 | 1.29 | 3.42 | 1.26 | 0.20 | 3.82 | 1.53 | 3.85 |
| HR 7637 | S | 5947 | 38 | 8 | 0.26 | 1.17 | 2.00 | 0.86 | 12.00 | 1.12 | 5.00 | 0.98 | 9.25 | 1.03 | 0.31 | 7.06 | 10.00 | 4.24 |
| HR 7793 | S | 6261 | 18 | 7 | 0.28 | 1.19 | 1.16 | 1.12 | 3.19 | 1.21 | 1.06 | 1.16 | 1.95 | 1.17 | 0.09 | 1.84 | 2.13 | 4.36 |
| HR 7855 | S | 6217 | 84 | 9 | 0.49 | 1.29 | 2.37 | 0.85 | 11.00 | 1.34 | 2.00 | 1.17 | 4.63 | 1.16 | 0.49 | 5.00 | 9.00 | 4.13 |
| HR 7907 | S | 6189 | 53 | 9 | 0.41 | 1.28 | 1.13 | 1.26 | 2.67 | 1.32 | 1.22 | 1.30 | 2.00 | 1.29 | 0.06 | 1.75 | 1.54 | 4.25 |
| HR 8133 | S | 5874 | 36 | 9 | 0.63 | 1.15 | 5.66 | 1.00 | 7.75 | 1.18 | 5.50 | 1.16 | 5.50 | 1.12 | 0.18 | 6.10 | 2.25 | 3.88 |
| HR 8148 | S | 5433 | 52 | 4 | -0.16 | 0.88 | 11.75 | 0.88 | 10.00 | 0.91 | 8.50 | | | 0.89 | 0.03 | 10.08 | 3.25 | 4.43 |
| HR 8170 | S | 5903 | 162 | 9 | 0.23 | | | 1.04 | 6.50 | 1.18 | 3.00 | 0.90 | 12.00 | 1.04 | 0.28 | 7.17 | 9.00 | 4.26 |

| Name | Type | | | | | | | | | | | | | | | | |
|---|---|---|---|---|---|---|---|---|---|---|---|---|---|---|---|---|---|
| HR 857 | S | 5201 | 61 | 12 | -0.40 | 0.90 | 0.79 | 0.88 | 0.95 | 0.87 | 2.56 | 0.88 | 1.39 | 0.88 | 0.03 | 1.42 | 1.76 | 4.59 |
| HR 8631 | S | 5453 | 11 | 3 | 0.72 | 1.18 | 4.50 | 1.06 | 6.94 | 1.27 | 4.00 | 1.30 | 3.54 | 1.20 | 0.24 | 4.74 | 3.40 | 3.69 |
| HR 8832 | S | 4883 | 56 | 9 | -0.55 | 0.81 | 5.36 | 0.82 | 2.74 | 0.79 | 6.06 | 0.80 | 4.34 | 0.81 | 0.03 | 4.62 | 3.31 | 4.60 |
| HR 8924 | S | 4784 | 97 | 9 | 0.91 | | | 1.13 | 8.25 | 1.27 | 6.00 | 1.10 | 9.25 | 1.17 | 0.17 | 7.83 | 3.25 | 3.26 |
| HR 8964 | S | 5821 | 37 | 3 | 0.03 | 1.02 | 4.50 | 1.06 | 2.67 | 1.04 | 3.17 | 1.06 | 2.59 | 1.05 | 0.04 | 3.23 | 1.90 | 4.43 |
| HR 9074 | S | 6227 | 60 | 2 | 0.27 | 1.18 | 1.42 | 1.16 | 2.00 | 1.17 | 1.68 | 1.16 | 1.99 | 1.17 | 0.02 | 1.77 | 0.58 | 4.36 |
| HR 9075 | S | 6112 | 51 | 3 | 0.19 | 1.14 | 1.46 | 1.07 | 3.40 | 1.14 | 1.71 | 1.10 | 2.50 | 1.11 | 0.07 | 2.27 | 1.94 | 4.39 |
| LP 837-53 | S | 3666 | 157 | 8 | -1.58 | | | | | | | | | | | | | 4.89 |
| NAME 23 H. Cam | S | 6215 | 31 | 7 | 0.24 | 1.18 | 0.92 | 1.12 | 2.33 | 1.17 | 1.42 | 1.12 | 2.42 | 1.15 | 0.06 | 1.77 | 1.50 | 4.38 |
| V* AK Lep | S | 4925 | 32 | 6 | -0.53 | 0.82 | 5.36 | 0.83 | 2.64 | 0.80 | 3.63 | 0.81 | 4.63 | 0.82 | 0.03 | 4.06 | 2.72 | 4.60 |
| V* AR Lac | S | 5342 | 68 | 9 | 0.76 | 1.27 | 4.09 | 1.05 | 8.00 | 1.16 | 5.00 | 1.24 | 3.63 | 1.18 | 0.22 | 5.18 | 4.38 | 3.61 |
| V* DE Boo | S | 5260 | 27 | 5 | -0.30 | 0.89 | 6.71 | 0.89 | 5.29 | 0.92 | 3.95 | 0.90 | 5.44 | 0.90 | 0.03 | 5.35 | 2.76 | 4.52 |
| V* HN Peg | S | 5972 | 32 | 7 | 0.04 | 0.92 | 7.94 | 1.05 | 2.10 | 0.97 | 4.50 | 0.94 | 5.75 | 0.97 | 0.13 | 5.07 | 5.85 | 4.44 |
| V* IL Aqr | S | 3935 | 59 | 2 | -2.42 | | | | | | | | | | | | | 5.98 |
| V* pi.01 UMa | S | 5921 | 44 | 9 | -0.02 | 0.92 | 7.13 | 0.84 | 9.00 | 0.97 | 3.50 | 0.94 | 4.44 | 0.92 | 0.13 | 6.02 | 5.50 | 4.46 |
| V* V2213 Oph | S | 6030 | 34 | 10 | 0.15 | 1.10 | 2.56 | 1.05 | 3.96 | 1.08 | 3.08 | 1.10 | 2.68 | 1.08 | 0.05 | 3.07 | 1.40 | 4.39 |
| V* V2215 Oph | S | 4463 | 25 | 10 | -0.84 | | | 0.72 | 0.84 | 0.72 | 3.99 | 0.71 | 1.28 | 0.72 | 0.01 | 2.03 | 3.15 | 4.68 |
| V* V2502 Oph | S | 6969 | 62 | 3 | 0.84 | 1.26 | 2.84 | 1.20 | 3.50 | | | 1.35 | 2.21 | 1.27 | 0.15 | 2.85 | 1.29 | 4.02 |
| V* V2689 Ori | S | 4073 | 89 | 11 | -1.15 | | | 0.53 | 3.94 | 0.55 | 4.44 | 0.59 | 5.29 | 0.56 | 0.06 | 4.56 | 1.35 | 4.72 |
| V* V376 Peg | S | 6066 | 42 | 9 | 0.25 | 1.17 | 1.80 | 1.06 | 5.30 | 1.22 | 1.31 | 1.14 | 3.47 | 1.15 | 0.16 | 2.97 | 3.99 | 4.33 |
| V* V450 And | S | 5653 | 33 | 9 | -0.08 | 0.97 | 5.10 | 0.96 | 4.54 | 0.94 | 6.10 | 1.01 | 3.20 | 0.97 | 0.07 | 4.73 | 2.90 | 4.46 |
| V* V819 Her | S | 5699 | 93 | 3 | 1.45 | 2.00 | 1.00 | 2.08 | 1.00 | 2.07 | 1.10 | 2.20 | 0.70 | 2.09 | 0.20 | 0.95 | 0.40 | 3.28 |
| Wolf 1008 | S | 5787 | 63 | 9 | 0.34 | 0.97 | 10.60 | 0.95 | 10.50 | 0.93 | 11.00 | 0.92 | 11.57 | 0.94 | 0.05 | 10.92 | 1.07 | 4.07 |
| HD 116442 | H | 5281 | 45 | 9 | -0.47 | | | 0.74 | 4.78 | 0.74 | 8.00 | 0.79 | 1.05 | 0.76 | 0.05 | 4.61 | 6.95 | 4.62 |
| HD 130087 | H | 5991 | 25 | 9 | 0.45 | 1.14 | 5.66 | | | 1.19 | 5.00 | 1.26 | 4.05 | 1.20 | 0.12 | 4.90 | 1.61 | 4.12 |
| HD 139323 | H | 5046 | 51 | 9 | -0.34 | | | 0.87 | 9.30 | 0.88 | 8.86 | | | 0.88 | 0.01 | 9.08 | 0.44 | 4.48 |
| HD 147044 | H | 5890 | 33 | 9 | 0.08 | 0.96 | 8.12 | 1.05 | 3.77 | 1.00 | 5.50 | 1.07 | 3.30 | 1.02 | 0.11 | 5.17 | 4.83 | 4.39 |
| HD 152792 | H | 5675 | 56 | 9 | 0.58 | | | 1.00 | 8.60 | 1.05 | 7.00 | 1.06 | 6.75 | 1.04 | 0.06 | 7.45 | 1.85 | 3.84 |
| HD 170778 | H | 5932 | 31 | 9 | 0.04 | 1.06 | 2.74 | 1.04 | 2.10 | 1.05 | 2.21 | 1.02 | 3.63 | 1.04 | 0.04 | 2.67 | 1.52 | 4.46 |
| HD 210640 | H | 6377 | 53 | 9 | 0.66 | | | 0.98 | 7.00 | 1.13 | 5.00 | 1.14 | 4.50 | 1.08 | 0.16 | 5.50 | 2.50 | 3.98 |
| HD 24213 | H | 6053 | 31 | 9 | 0.38 | 1.24 | 2.00 | 1.12 | 5.50 | | | 0.99 | 8.50 | 1.12 | 0.25 | 5.33 | 6.50 | 4.18 |
| HD 47127 | H | 5616 | 20 | 9 | 0.07 | 1.04 | 5.01 | 0.99 | 8.88 | 1.03 | 7.00 | 1.00 | 10.33 | 1.02 | 0.05 | 7.81 | 5.32 | 4.32 |
| HD 5372 | H | 5847 | 37 | 9 | 0.12 | 1.10 | 2.19 | 1.05 | 5.17 | 1.13 | 2.36 | 1.05 | 5.80 | 1.08 | 0.08 | 3.88 | 3.61 | 4.37 |
| HD 58781 | H | 5576 | 22 | 9 | -0.05 | 0.96 | 7.66 | 0.94 | 8.50 | 1.02 | 4.50 | 0.96 | 9.00 | 0.97 | 0.08 | 7.42 | 4.50 | 4.41 |
| HD 68017 | H | 5565 | 36 | 9 | -0.10 | 0.92 | 9.47 | 0.93 | 7.50 | | | | | 0.93 | 0.01 | 8.49 | 1.97 | 4.43 |
| HD 72760 | H | 5328 | 33 | 12 | -0.31 | 0.92 | 2.19 | 0.91 | 1.78 | 0.91 | 2.40 | 0.89 | 3.42 | 0.91 | 0.03 | 2.45 | 1.64 | 4.56 |
| HD 76752 | H | 5685 | 56 | 9 | 0.12 | 0.95 | 11.48 | 0.98 | 9.50 | 1.05 | 7.00 | 0.97 | 11.25 | 0.99 | 0.10 | 9.81 | 4.48 | 4.28 |

| Name | | T | | | | | | | | | | | | | | | |
|---|---|---|---|---|---|---|---|---|---|---|---|---|---|---|---|---|---|
| HD 76909 | H | 5598 | 53 | 9 | 0.24 | 0.98 | 11.75 | 0.97 | 11.83 | 1.13 | 6.50 | 1.01 | 11.50 | 1.02 | 0.16 | 10.40 | 5.33 | 4.15 |
| HD 8648 | H | 5711 | 35 | 9 | 0.26 | 1.03 | 8.27 | 1.04 | 9.00 | 0.98 | 10.50 | 1.01 | 10.67 | 1.02 | 0.06 | 9.61 | 2.39 | 4.16 |
| HD 98618 | H | 5735 | 38 | 9 | 0.08 | 1.07 | 2.84 | 1.00 | 7.20 | 1.00 | 7.33 | 1.00 | 8.60 | 1.02 | 0.07 | 6.49 | 5.76 | 4.35 |
| HD 9986 | H | 5791 | 37 | 9 | 0.04 | 1.05 | 3.47 | 1.03 | 4.10 | 1.08 | 2.94 | 1.05 | 3.98 | 1.05 | 0.05 | 3.62 | 1.16 | 4.42 |
| 1 Hya | E | 6359 | 35 | 10 | 0.51 | 1.13 | 5.01 | 0.95 | 8.00 | | | | | 1.04 | 0.18 | 6.51 | 2.99 | 4.11 |
| 101 Tau | E | 6465 | 58 | 10 | 0.42 | 1.26 | 1.13 | 1.30 | 0.92 | 1.29 | 0.98 | 1.27 | 0.93 | 1.28 | 0.04 | 0.99 | 0.21 | 4.31 |
| 14 Boo | E | 6180 | 22 | 5 | 0.76 | | | 1.28 | 4.00 | | | 1.35 | 3.25 | 1.32 | 0.07 | 3.63 | 0.75 | 3.91 |
| 15 Peg | E | 6452 | 42 | 9 | 0.57 | 1.13 | 5.01 | 0.91 | 8.50 | 1.18 | 4.00 | 1.12 | 4.50 | 1.09 | 0.27 | 5.50 | 4.50 | 4.09 |
| 22 Lyn | E | 6395 | 16 | 3 | 0.36 | 1.24 | 1.02 | 1.22 | 1.50 | 1.24 | 1.25 | 1.23 | 1.29 | 1.23 | 0.02 | 1.26 | 0.48 | 4.34 |
| 30 Ari B | E | 6257 | 49 | 6 | 0.30 | 1.19 | 1.61 | 1.13 | 3.20 | 1.19 | 1.83 | 1.16 | 2.38 | 1.17 | 0.06 | 2.26 | 1.59 | 4.34 |
| 34 Peg | E | 6258 | 38 | 7 | 0.77 | 1.33 | 3.22 | 1.29 | 3.85 | 1.25 | 4.00 | 1.38 | 2.96 | 1.31 | 0.13 | 3.51 | 1.04 | 3.92 |
| 36 Dra | E | 6522 | 30 | 3 | 0.64 | 1.20 | 3.98 | 1.13 | 5.00 | | | 1.21 | 3.44 | 1.18 | 0.08 | 4.14 | 1.56 | 4.07 |
| 38 Cet | E | 6480 | 17 | 7 | 0.88 | | | 1.22 | 3.94 | | | 1.38 | 2.58 | 1.30 | 0.16 | 3.26 | 1.35 | 3.87 |
| 4 Aqr | E | 6440 | 69 | 6 | 1.04 | 1.58 | 2.09 | 1.55 | 2.42 | 1.68 | 1.80 | 1.60 | 1.81 | 1.60 | 0.13 | 2.03 | 0.62 | 3.79 |
| 40 Leo | E | 6467 | 36 | 3 | 0.64 | 1.22 | 3.57 | | | 1.46 | 1.40 | | | 1.34 | 0.24 | 2.49 | 2.17 | 4.11 |
| 47 Ari | E | 6644 | 44 | 7 | 0.62 | | | 1.38 | 1.75 | 1.44 | 1.10 | 1.43 | 1.05 | 1.42 | 0.06 | 1.30 | 0.70 | 4.21 |
| 49 Peg | E | 6275 | 93 | 9 | 0.71 | 1.41 | 2.51 | 1.02 | 6.25 | 1.14 | 5.00 | 1.29 | 3.50 | 1.22 | 0.39 | 4.32 | 3.74 | 3.95 |
| 51 Ari | E | 5603 | 45 | 9 | -0.08 | 1.02 | 2.17 | 0.97 | 5.35 | 1.03 | 3.05 | 0.97 | 6.76 | 1.00 | 0.06 | 4.34 | 4.59 | 4.46 |
| 6 And | E | 6338 | 82 | 7 | 0.41 | 1.09 | 4.97 | 1.04 | 6.05 | 1.19 | 3.00 | 1.17 | 3.42 | 1.12 | 0.15 | 4.36 | 3.05 | 4.23 |
| 6 Cet | E | 6242 | 56 | 12 | 0.50 | 1.02 | 7.13 | 1.08 | 6.25 | | | | | 1.05 | 0.06 | 6.69 | 0.88 | 4.09 |
| 68 Eri | E | 6421 | 58 | 8 | 0.66 | | | 0.95 | 7.50 | 1.39 | 2.50 | 1.19 | 4.00 | 1.18 | 0.44 | 4.67 | 5.00 | 4.03 |
| 71 Ori | E | 6560 | 41 | 5 | 0.45 | 1.30 | 0.81 | 1.28 | 1.34 | 1.29 | 1.20 | 1.28 | 1.15 | 1.29 | 0.02 | 1.13 | 0.54 | 4.31 |
| 84 Cet | E | 6236 | 90 | 7 | 0.33 | 1.18 | 2.51 | 1.09 | 4.58 | 1.11 | 4.00 | 1.13 | 3.75 | 1.13 | 0.09 | 3.71 | 2.07 | 4.29 |
| 89 Leo | E | 6538 | 67 | 9 | 0.46 | 1.29 | 1.13 | 1.31 | 1.17 | 1.31 | 0.98 | 1.30 | 0.93 | 1.30 | 0.02 | 1.05 | 0.24 | 4.30 |
| b Her | E | 6000 | 66 | 9 | 0.28 | 1.12 | 5.01 | 0.98 | 8.25 | | | 0.90 | 10.83 | 1.00 | 0.22 | 8.03 | 5.82 | 4.22 |
| c Boo | E | 6560 | 47 | 7 | 0.50 | 1.31 | 1.58 | 1.30 | 1.69 | 1.33 | 1.30 | 1.32 | 1.38 | 1.32 | 0.03 | 1.49 | 0.39 | 4.27 |
| eta UMi | E | 6788 | 11 | 3 | 0.85 | 1.44 | 2.05 | 1.07 | 4.75 | 1.65 | 1.00 | 1.38 | 2.25 | 1.39 | 0.58 | 2.51 | 3.75 | 4.00 |
| iot Vir | E | 6217 | 31 | 12 | 0.97 | 1.49 | 2.25 | 1.25 | 3.50 | 1.64 | 2.15 | 1.45 | 2.38 | 1.46 | 0.39 | 2.57 | 1.35 | 3.75 |
| kap CrB | E | 4863 | 85 | 9 | 1.09 | 1.32 | 4.41 | 1.43 | 3.50 | 1.41 | 5.00 | 1.34 | 4.38 | 1.38 | 0.11 | 4.32 | 1.50 | 3.18 |
| phi Vir | E | 5551 | 73 | 12 | 1.14 | 1.71 | 1.58 | 1.49 | 2.06 | 1.74 | 1.80 | | | 1.65 | 0.25 | 1.82 | 0.48 | 3.44 |
| tau01 Hya | E | 6507 | 65 | 6 | 0.54 | 1.36 | 1.31 | 1.10 | 5.38 | 1.38 | 1.30 | 1.34 | 1.45 | 1.30 | 0.28 | 2.36 | 4.08 | 4.21 |
| tet Dra | E | 6208 | 48 | 5 | 0.95 | 1.53 | 2.03 | 1.55 | 2.70 | 1.59 | 2.30 | 1.48 | 2.33 | 1.54 | 0.11 | 2.34 | 0.67 | 3.79 |
| BD+01 2063 | E | 4978 | 16 | 8 | -0.60 | 0.75 | 5.36 | 0.67 | 9.75 | 0.74 | 3.66 | 0.73 | 4.64 | 0.72 | 0.08 | 5.85 | 6.09 | 4.63 |
| BD+12 4499 | E | 4653 | 84 | 8 | -0.73 | | | 0.77 | 0.75 | 0.74 | 2.93 | | | 0.76 | 0.03 | 1.84 | 2.18 | 4.66 |
| BD+17 4708 | E | 6180 | 52 | 12 | 0.30 | | | 0.82 | 12.25 | 0.82 | 12.00 | 0.82 | 12.00 | 0.82 | 0.00 | 12.08 | 0.25 | 4.16 |
| BD+23 465 | E | 5240 | 43 | 8 | -0.28 | 0.92 | 4.16 | 0.84 | 10.25 | 0.93 | 4.36 | 0.89 | 8.25 | 0.90 | 0.09 | 6.75 | 6.09 | 4.49 |
| BD+29 366 | E | 5777 | 22 | 9 | -0.11 | 0.89 | 6.34 | 0.80 | 10.00 | 0.84 | 9.00 | 0.81 | 10.00 | 0.84 | 0.09 | 8.83 | 3.66 | 4.46 |
| BD+33 99 | E | 4499 | 66 | 8 | -0.75 | 0.76 | 5.36 | 0.71 | 9.75 | | | 0.72 | 12.00 | 0.73 | 0.05 | 9.04 | 6.64 | 4.61 |

| Name | | Teff | | | [Fe/H] | | | | | | | | | | | | |
|---|---|---|---|---|---|---|---|---|---|---|---|---|---|---|---|---|---|
| BD+41 3306 | E | 5053 | 45 | 9 | -0.43 | | | 0.81 | 7.88 | | | | | 0.81 | 0.00 | 7.88 | 0.00 | 4.54 |
| BD+43 699 | E | 4802 | 47 | 9 | -0.71 | 0.72 | 5.36 | 0.65 | 6.28 | 0.70 | 3.66 | 0.68 | 5.29 | 0.69 | 0.07 | 5.15 | 2.63 | 4.66 |
| BD+46 1635 | E | 4215 | 58 | 9 | -0.92 | 0.70 | 5.36 | 0.70 | 4.01 | | | | | 0.70 | 0.00 | 4.69 | 1.35 | 4.65 |
| BD+52 2815 | E | 4203 | 54 | 8 | -0.96 | 0.67 | 5.36 | 0.66 | 2.07 | 0.65 | 11.50 | | | 0.66 | 0.02 | 6.31 | 9.43 | 4.66 |
| CCDM J20051-0418AB | E | 6370 | 37 | 3 | 0.58 | 1.17 | 4.47 | 0.94 | 7.88 | 1.13 | 5.00 | 1.19 | 4.00 | 1.11 | 0.25 | 5.34 | 3.88 | 4.07 |
| CCDM J21031+0132AB | E | 6423 | 48 | 3 | 1.07 | 1.64 | 1.77 | 1.40 | 2.71 | 1.63 | 2.14 | 1.58 | 1.83 | 1.56 | 0.24 | 2.11 | 0.95 | 3.74 |
| CCDM J22071+0034AB | E | 6059 | 33 | 3 | 0.66 | 1.14 | 5.40 | 1.04 | 6.91 | 1.21 | 4.80 | 1.17 | 5.10 | 1.14 | 0.17 | 5.55 | 2.11 | 3.91 |
| GJ 1067 | E | 4378 | 59 | 8 | -0.88 | 0.71 | 4.51 | 0.70 | 0.95 | 0.70 | 5.30 | 0.69 | 4.26 | 0.70 | 0.02 | 3.75 | 4.35 | 4.68 |
| GJ 697 | E | 4881 | 25 | 8 | -0.59 | | | 0.82 | 1.02 | 0.80 | 1.73 | 0.80 | 1.39 | 0.81 | 0.02 | 1.38 | 0.72 | 4.64 |
| HD 10086 | E | 5652 | 62 | 9 | -0.04 | 1.03 | 2.20 | 1.03 | 2.90 | 1.07 | 1.81 | 1.03 | 3.89 | 1.04 | 0.04 | 2.70 | 2.08 | 4.45 |
| HD 10145 | E | 5642 | 32 | 9 | 0.00 | 1.00 | 6.03 | 0.94 | 9.00 | 1.03 | 4.90 | 0.96 | 9.81 | 0.98 | 0.09 | 7.43 | 4.91 | 4.38 |
| HD 105631 | E | 5363 | 24 | 8 | -0.24 | 0.93 | 5.39 | 0.94 | 3.54 | 0.97 | 2.45 | 0.93 | 5.46 | 0.94 | 0.04 | 4.21 | 3.00 | 4.52 |
| HD 106116 | E | 5665 | 36 | 12 | 0.05 | 0.98 | 8.15 | 0.97 | 8.75 | 1.03 | 6.00 | 0.96 | 10.75 | 0.99 | 0.07 | 8.41 | 4.75 | 4.34 |
| HD 106691 | E | 6647 | 60 | 12 | 0.67 | 1.38 | 1.77 | 1.09 | 5.04 | 1.47 | 1.31 | 1.35 | 2.13 | 1.32 | 0.38 | 2.56 | 3.73 | 4.13 |
| HD 107611 | E | 6391 | 38 | 12 | 0.49 | 1.27 | 2.24 | 1.03 | 6.63 | 1.33 | 1.62 | 1.32 | 1.28 | 1.24 | 0.30 | 2.94 | 5.35 | 4.21 |
| HD 10853 | E | 4655 | 53 | 11 | -0.72 | 0.78 | 1.54 | 0.77 | 1.19 | 0.73 | 4.64 | 0.75 | 3.23 | 0.76 | 0.05 | 2.65 | 3.45 | 4.66 |
| HD 110463 | E | 4926 | 18 | 8 | -0.54 | 0.83 | 2.57 | 0.82 | 2.30 | 0.81 | 2.59 | 0.80 | 5.29 | 0.82 | 0.03 | 3.19 | 2.99 | 4.61 |
| HD 111069 | E | 5865 | 34 | 9 | 0.09 | 1.06 | 3.62 | 1.06 | 3.81 | 1.10 | 2.88 | 1.06 | 4.15 | 1.07 | 0.04 | 3.62 | 1.27 | 4.40 |
| HD 11373 | E | 4759 | 31 | 9 | -0.62 | 0.79 | 5.36 | 0.79 | 3.42 | 0.79 | 4.79 | 0.78 | 5.28 | 0.79 | 0.01 | 4.71 | 1.94 | 4.61 |
| HD 115274 | E | 6153 | 42 | 9 | 0.83 | 1.39 | 2.97 | 1.22 | 4.33 | 1.42 | 3.17 | 1.35 | 3.19 | 1.35 | 0.20 | 3.41 | 1.35 | 3.84 |
| HD 116956 | E | 5308 | 36 | 9 | -0.27 | 0.91 | 5.85 | 0.90 | 4.89 | 0.94 | 3.66 | 0.91 | 5.29 | 0.92 | 0.04 | 4.92 | 2.18 | 4.52 |
| HD 117635 | E | 5217 | 32 | 12 | 0.03 | | | | | | | | | | | | | 4.24 |
| HD 118096 | E | 4568 | 75 | 8 | -0.89 | | | | | 0.66 | 1.07 | | | 0.66 | 0.00 | 1.07 | 0.00 | 4.73 |
| HD 119332 | E | 5213 | 22 | 8 | -0.34 | 0.85 | 8.69 | 0.86 | 6.02 | 0.87 | 6.00 | 0.88 | 5.47 | 0.87 | 0.03 | 6.54 | 3.21 | 4.53 |
| HD 119802 | E | 4716 | 47 | 8 | -0.67 | 0.79 | 1.92 | 0.78 | 1.45 | 0.75 | 5.16 | 0.76 | 4.02 | 0.77 | 0.04 | 3.14 | 3.71 | 4.64 |
| HD 12051 | E | 5397 | 16 | 9 | -0.10 | | | 0.93 | 10.43 | 0.98 | 7.00 | | | 0.96 | 0.05 | 8.71 | 3.43 | 4.39 |
| HD 122120 | E | 4486 | 29 | 8 | -0.70 | | | | | | | | | | | | | 4.62 |
| HD 124292 | E | 5497 | 35 | 9 | -0.18 | 0.93 | 5.81 | 0.92 | 5.28 | 0.92 | 6.00 | 0.90 | 8.00 | 0.92 | 0.03 | 6.27 | 2.72 | 4.49 |
| HD 124642 | E | 4664 | 26 | 8 | -0.65 | 0.79 | 5.36 | 0.76 | 6.89 | 0.79 | 4.54 | 0.78 | 5.29 | 0.78 | 0.03 | 5.52 | 2.35 | 4.60 |
| HD 128429 | E | 6456 | 68 | 9 | 0.44 | 1.14 | 3.95 | 1.13 | 4.42 | 1.28 | 1.40 | 1.19 | 2.63 | 1.19 | 0.15 | 3.10 | 3.02 | 4.26 |
| HD 12846 | E | 5733 | 36 | 9 | -0.11 | 0.85 | 10.99 | 0.97 | 3.03 | 0.87 | 7.80 | 0.85 | 9.68 | 0.89 | 0.12 | 7.87 | 7.96 | 4.48 |
| HD 130307 | E | 5043 | 23 | 12 | -0.53 | | | | | 0.73 | 10.00 | 0.73 | 9.25 | 0.73 | 0.00 | 9.63 | 0.75 | 4.59 |
| HD 132142 | E | 5229 | 23 | 12 | -0.39 | 0.90 | 0.79 | 0.89 | 0.95 | 0.77 | 11.50 | | | 0.85 | 0.13 | 4.41 | 10.71 | 4.58 |
| HD 132254 | E | 6279 | 32 | 9 | 0.46 | 1.15 | 4.54 | 1.10 | 5.38 | 1.35 | 1.50 | 1.31 | 2.25 | 1.23 | 0.25 | 3.42 | 3.88 | 4.21 |
| HD 133002 | E | 5562 | 48 | 5 | 0.97 | 1.36 | 2.51 | 1.24 | 3.63 | 1.51 | 2.50 | 1.54 | 1.75 | 1.41 | 0.30 | 2.60 | 1.88 | 3.55 |

| ID | | Teff | | | | | | | | | | | | | | | | |
|---|---|---|---|---|---|---|---|---|---|---|---|---|---|---|---|---|---|---|
| HD 13403 | E | 5617 | 42 | 9 | 0.37 | | | 1.02 | 10.00 | 0.97 | 10.67 | 0.91 | 12.06 | 0.97 | 0.11 | 10.91 | 2.06 | 4.00 |
| HD 135204 | E | 5457 | 83 | 12 | -0.18 | 0.94 | 5.41 | 0.90 | 7.25 | 0.98 | 3.04 | 0.94 | 5.67 | 0.94 | 0.08 | 5.34 | 4.21 | 4.49 |
| HD 135599 | E | 5270 | 43 | 8 | -0.39 | | | | | 0.78 | 9.00 | 0.76 | 10.75 | 0.77 | 0.02 | 9.88 | 1.75 | 4.55 |
| HD 13579 | E | 5133 | 43 | 8 | -0.33 | 0.87 | 7.61 | 0.91 | 4.63 | 0.91 | 3.66 | 0.87 | 9.96 | 0.89 | 0.04 | 6.46 | 6.30 | 4.51 |
| HD 139777 | E | 5775 | 51 | 9 | -0.03 | 1.04 | 1.88 | 1.04 | 1.75 | 1.02 | 2.57 | 0.98 | 4.47 | 1.02 | 0.06 | 2.67 | 2.72 | 4.47 |
| HD 139813 | E | 5394 | 74 | 9 | -0.30 | | | 0.92 | 0.83 | 0.81 | 10.00 | 0.93 | 0.80 | 0.89 | 0.12 | 3.88 | 9.20 | 4.56 |
| HD 140283 | E | 5823 | 74 | 11 | 0.58 | 0.86 | 11.75 | | | | | | | 0.86 | 0.00 | 11.75 | 0.00 | 3.80 |
| HD 14348 | E | 6036 | 27 | 9 | 0.54 | 1.18 | 5.01 | 1.14 | 5.83 | 1.16 | 5.50 | 1.17 | 5.50 | 1.16 | 0.04 | 5.46 | 0.82 | 4.03 |
| HD 14374 | E | 5474 | 75 | 9 | -0.26 | 0.89 | 6.08 | 0.94 | 1.18 | 0.90 | 4.74 | 0.88 | 5.36 | 0.90 | 0.06 | 4.34 | 4.90 | 4.55 |
| HD 144287 | E | 5315 | 44 | 9 | -0.17 | 0.89 | 10.73 | | | 0.96 | 6.50 | | | 0.93 | 0.07 | 8.61 | 4.23 | 4.42 |
| HD 145435 | E | 6083 | 107 | 9 | 0.37 | | | 0.92 | 9.50 | | | | | 0.92 | 0.00 | 9.50 | 0.00 | 4.12 |
| HD 145729 | E | 6028 | 27 | 9 | 0.15 | 1.10 | 2.56 | 1.05 | 3.96 | 1.09 | 2.78 | 1.05 | 4.27 | 1.07 | 0.05 | 3.39 | 1.72 | 4.39 |
| HD 146946 | E | 5738 | 89 | 9 | 0.10 | | | 0.98 | 8.50 | | | | | 0.98 | 0.00 | 8.50 | 0.00 | 4.31 |
| HD 153525 | E | 4820 | 35 | 9 | -0.61 | 0.81 | 1.21 | 0.80 | 1.45 | 0.78 | 2.59 | 0.78 | 3.42 | 0.79 | 0.03 | 2.17 | 2.21 | 4.63 |
| HD 154931 | E | 5869 | 19 | 9 | 0.52 | 1.16 | 5.81 | 0.93 | 9.93 | 1.12 | 6.50 | 1.12 | 6.70 | 1.08 | 0.23 | 7.23 | 4.12 | 3.97 |
| HD 155712 | E | 4936 | 30 | 8 | -0.51 | 0.82 | 5.36 | 0.80 | 6.20 | 0.79 | 8.00 | 0.82 | 5.97 | 0.81 | 0.03 | 6.38 | 2.64 | 4.58 |
| HD 15632 | E | 5749 | 26 | 9 | -0.07 | 0.96 | 5.22 | 1.01 | 1.63 | 1.00 | 2.69 | 0.95 | 4.97 | 0.98 | 0.06 | 3.63 | 3.59 | 4.48 |
| HD 156985 | E | 4778 | 31 | 8 | -0.60 | 0.79 | 5.90 | 0.76 | 7.07 | 0.81 | 3.99 | 0.75 | 10.00 | 0.78 | 0.06 | 6.74 | 6.01 | 4.59 |
| HD 157089 | E | 5830 | 54 | 12 | 0.26 | | | | | 0.89 | 12.00 | | | 0.89 | 0.00 | 12.00 | 0.00 | 4.14 |
| HD 159062 | E | 5385 | 24 | 9 | -0.23 | 0.87 | 10.99 | 0.88 | 7.50 | | | | | 0.88 | 0.01 | 9.24 | 3.49 | 4.48 |
| HD 159482 | E | 5805 | 56 | 12 | 0.00 | 0.87 | 11.49 | 0.80 | 12.50 | 0.84 | 11.50 | 0.83 | 12.50 | 0.84 | 0.07 | 12.00 | 1.01 | 4.36 |
| HD 160964 | E | 4589 | 46 | 8 | -0.76 | 0.77 | 1.00 | 0.75 | 0.95 | 0.73 | 6.19 | 0.74 | 1.97 | 0.75 | 0.04 | 2.53 | 5.25 | 4.67 |
| HD 161098 | E | 5637 | 12 | 9 | -0.17 | 0.84 | 11.24 | 0.99 | 0.75 | 0.85 | 8.25 | 0.83 | 10.25 | 0.88 | 0.16 | 7.62 | 10.49 | 4.50 |
| HD 163183 | E | 5928 | 43 | 9 | 0.03 | 1.10 | 0.82 | 1.04 | 2.11 | 1.02 | 3.33 | 0.98 | 5.13 | 1.04 | 0.12 | 2.84 | 4.31 | 4.46 |
| HD 16397 | E | 5821 | 29 | 9 | 0.06 | 1.01 | 5.66 | 1.01 | 5.15 | 0.91 | 9.00 | | | 0.98 | 0.10 | 6.60 | 3.85 | 4.37 |
| HD 164651 | E | 5599 | 54 | 9 | -0.08 | 1.02 | 2.21 | 0.94 | 7.25 | 1.05 | 2.03 | 0.96 | 7.15 | 0.99 | 0.11 | 4.66 | 5.22 | 4.45 |
| HD 165173 | E | 5484 | 33 | 9 | -0.22 | 0.96 | 2.19 | 0.95 | 1.67 | 0.93 | 3.56 | 0.93 | 3.23 | 0.94 | 0.03 | 2.66 | 1.88 | 4.54 |
| HD 165401 | E | 5816 | 29 | 10 | -0.03 | 0.98 | 5.22 | 1.01 | 2.33 | 0.88 | 8.50 | 0.85 | 10.86 | 0.93 | 0.16 | 6.73 | 8.53 | 4.44 |
| HD 165476 | E | 5816 | 20 | 9 | 0.21 | 1.08 | 5.55 | 1.02 | 8.25 | 1.05 | 7.00 | 0.99 | 9.50 | 1.04 | 0.09 | 7.58 | 3.95 | 4.25 |
| HD 165590 | E | 5663 | 109 | 6 | 0.30 | 1.01 | 10.24 | 0.99 | 10.75 | 0.99 | 10.50 | 0.98 | 11.50 | 0.99 | 0.03 | 10.75 | 1.26 | 4.09 |
| HD 165670 | E | 6285 | 37 | 9 | 0.33 | 1.20 | 1.79 | 1.13 | 3.56 | 1.22 | 1.70 | 1.17 | 2.50 | 1.18 | 0.09 | 2.39 | 1.86 | 4.32 |
| HD 165672 | E | 5866 | 56 | 9 | 0.14 | 1.11 | 1.94 | 1.14 | 2.14 | 1.16 | 1.65 | 1.12 | 3.39 | 1.13 | 0.05 | 2.28 | 1.74 | 4.37 |
| HD 166183 | E | 6380 | 18 | 9 | 0.65 | | | 0.98 | 7.00 | 1.47 | 1.80 | 1.29 | 3.31 | 1.25 | 0.49 | 4.04 | 5.20 | 4.05 |
| HD 166435 | E | 5811 | 39 | 9 | -0.02 | 1.00 | 4.42 | 1.01 | 2.70 | 1.03 | 2.39 | 1.00 | 3.65 | 1.01 | 0.03 | 3.29 | 2.03 | 4.47 |
| HD 167278 | E | 6483 | 118 | 6 | 0.97 | 1.53 | 2.17 | 1.32 | 3.28 | 1.50 | 2.37 | 1.43 | 2.31 | 1.45 | 0.21 | 2.53 | 1.11 | 3.82 |
| HD 169822 | E | 5529 | 58 | 9 | -0.26 | 0.84 | 8.13 | 0.75 | 11.50 | 0.87 | 3.42 | 0.84 | 5.29 | 0.83 | 0.12 | 7.09 | 8.08 | 4.53 |
| HD 170008 | E | 4978 | 73 | 9 | 0.87 | 1.19 | 5.01 | | | 0.93 | 11.50 | | | 1.06 | 0.26 | 8.26 | 6.49 | 3.33 |
| HD 170291 | E | 6327 | 42 | 9 | 0.44 | 1.30 | 1.31 | 1.03 | 6.85 | 1.20 | 3.25 | 1.16 | 3.75 | 1.17 | 0.27 | 3.79 | 5.54 | 4.22 |

| Star | | Teff | | | | | | | | | | | | | | | | | |
|---|---|---|---|---|---|---|---|---|---|---|---|---|---|---|---|---|---|---|---|
| HD 170512 | E | 6152 |  | 1 | 0.38 | 1.25 | 1.02 | 1.28 | 1.54 | 1.30 | 1.15 | 1.27 | 2.19 | 1.28 | 0.05 | 1.48 | 1.17 | 4.27 |
| HD 170579 | E | 6417 | 59 | 9 | 0.47 | 1.26 | 2.20 | 1.06 | 6.06 | 1.20 | 3.20 | 1.21 | 2.75 | 1.18 | 0.20 | 3.55 | 3.86 | 4.22 |
| HD 171314 | E | 4530 | 28 | 9 | -0.72 | 0.76 | 5.36 | 0.73 | 8.53 | 0.73 | 12.00 |  |  | 0.74 | 0.03 | 8.63 | 6.64 | 4.60 |
| HD 171888 | E | 6095 | 28 | 9 | 0.61 | 1.25 | 4.21 | 1.01 | 7.35 | 1.31 | 4.00 | 1.28 | 3.97 | 1.21 | 0.30 | 4.88 | 3.38 | 4.00 |
| HD 171951 | E | 6094 | 23 | 9 | 0.51 | 1.10 | 6.39 | 0.92 | 9.65 | 1.06 | 6.50 | 1.04 | 6.90 | 1.03 | 0.18 | 7.36 | 3.26 | 4.03 |
| HD 171953 | E | 6480 | 77 | 9 | 0.97 | 1.52 | 2.20 | 1.32 | 3.21 | 1.48 | 2.46 | 1.43 | 2.33 | 1.44 | 0.20 | 2.55 | 1.02 | 3.82 |
| HD 172675 | E | 6303 | 36 | 9 | 0.26 |  |  | 1.18 | 1.27 | 1.10 | 2.75 | 1.05 | 3.80 | 1.11 | 0.13 | 2.61 | 2.53 | 4.37 |
| HD 172718 | E | 6132 | 37 | 9 | 0.65 | 1.25 | 4.05 | 1.11 | 6.00 | 1.19 | 5.00 | 1.14 | 5.17 | 1.17 | 0.14 | 5.05 | 1.95 | 3.95 |
| HD 172961 | E | 6571 | 75 | 9 | 0.42 | 1.25 | 1.53 | 1.24 | 1.67 | 1.20 | 1.98 | 1.18 | 2.18 | 1.22 | 0.07 | 1.84 | 0.65 | 4.32 |
| HD 173174 | E | 6009 | 48 | 5 | 0.60 | 1.20 | 4.83 | 1.18 | 5.55 | 1.26 | 4.40 | 1.18 | 5.39 | 1.21 | 0.08 | 5.04 | 1.15 | 3.98 |
| HD 173605 | E | 5722 | 48 | 9 | 0.39 | 0.94 | 11.49 | 0.93 | 11.17 | 0.99 | 9.67 | 0.96 | 10.64 | 0.96 | 0.06 | 10.74 | 1.82 | 4.01 |
| HD 173634 | E | 6505 | 46 | 9 | 1.08 | 1.60 | 1.89 | 1.44 | 2.61 | 1.64 | 1.90 | 1.65 | 1.61 | 1.58 | 0.21 | 2.00 | 1.00 | 3.76 |
| HD 17382 | E | 5245 | 28 | 9 | -0.27 | 0.91 | 4.83 | 0.87 | 9.78 | 0.93 | 4.68 | 0.87 | 11.50 | 0.90 | 0.06 | 7.70 | 6.82 | 4.49 |
| HD 174080 | E | 4676 | 19 | 9 | -0.64 | 0.79 | 5.36 | 0.76 | 7.09 | 0.79 | 4.54 | 0.78 | 5.29 | 0.78 | 0.03 | 5.57 | 2.56 | 4.60 |
| HD 174719 | E | 5647 | 25 | 9 | -0.16 | 0.84 | 11.24 | 0.99 | 0.83 | 0.88 | 7.00 | 0.83 | 10.50 | 0.89 | 0.16 | 7.39 | 10.40 | 4.50 |
| HD 175272 | E | 6638 | 46 | 3 | 0.82 | 1.31 | 2.84 | 1.12 | 4.56 | 1.58 | 1.20 | 1.38 | 2.38 | 1.35 | 0.46 | 2.74 | 3.36 | 3.98 |
| HD 175726 | E | 6069 | 49 | 9 | 0.07 | 0.97 | 5.66 | 0.89 | 7.25 | 1.01 | 2.75 | 0.99 | 3.54 | 0.97 | 0.12 | 4.80 | 4.50 | 4.43 |
| HD 175805 | E | 6318 | 61 | 9 | 1.08 | 1.67 | 1.75 | 1.59 | 2.28 | 1.66 | 2.11 | 1.55 | 1.92 | 1.62 | 0.12 | 2.02 | 0.53 | 3.72 |
| HD 175806 | E | 6171 | 34 | 9 | 1.11 | 1.60 | 1.89 | 1.38 | 2.66 | 1.71 | 2.00 | 1.56 | 1.75 | 1.56 | 0.33 | 2.08 | 0.91 | 3.63 |
| HD 176118 | E | 6665 | 68 | 9 | 0.77 | 1.40 | 2.03 | 1.09 | 4.94 | 1.55 | 1.00 | 1.36 | 2.35 | 1.35 | 0.46 | 2.58 | 3.94 | 4.04 |
| HD 176377 | E | 5868 | 28 | 9 | -0.03 |  |  | 0.99 | 3.39 | 0.92 | 6.00 | 0.89 | 8.00 | 0.93 | 0.10 | 5.80 | 4.61 | 4.46 |
| HD 177749 | E | 6403 | 33 | 3 | 0.77 | 1.31 | 3.44 | 1.28 | 3.63 | 1.36 | 3.00 | 1.39 | 2.68 | 1.34 | 0.11 | 3.18 | 0.95 | 3.97 |
| HD 177904 | E | 6902 | 50 | 3 | 1.09 | 1.59 | 1.71 | 1.35 | 2.58 | 1.68 | 1.45 | 1.62 | 1.60 | 1.56 | 0.33 | 1.83 | 1.13 | 3.84 |
| HD 178126 | E | 4541 | 55 | 9 | -0.79 | 0.74 | 5.36 | 0.74 | 1.45 | 0.70 | 8.50 | 0.72 | 5.29 | 0.73 | 0.04 | 5.15 | 7.05 | 4.66 |
| HD 180161 | E | 5400 | 6 | 8 | -0.25 | 0.93 | 3.23 | 0.94 | 2.58 | 0.93 | 2.84 | 0.92 | 4.63 | 0.93 | 0.02 | 3.32 | 2.06 | 4.53 |
| HD 180945 | E | 6415 | 36 | 9 | 0.66 |  |  | 1.01 | 6.50 | 1.40 | 2.50 | 1.28 | 3.13 | 1.23 | 0.39 | 4.04 | 4.00 | 4.04 |
| HD 181096 | E | 6270 | 67 | 5 | 0.75 | 1.29 | 3.56 | 1.16 | 5.05 | 1.25 | 4.00 | 1.28 | 3.63 | 1.25 | 0.13 | 4.06 | 1.49 | 3.92 |
| HD 181420 | E | 6606 | 58 | 9 | 0.62 | 1.45 | 0.63 | 1.10 | 4.75 | 1.43 | 1.20 | 1.43 | 1.00 | 1.35 | 0.35 | 1.90 | 4.12 | 4.18 |
| HD 181806 | E | 6404 | 60 | 9 | 0.81 | 1.35 | 3.09 | 1.17 | 4.50 | 1.36 | 3.00 | 1.42 | 2.56 | 1.33 | 0.25 | 3.29 | 1.94 | 3.92 |
| HD 182274 | E | 6307 | 83 | 9 | 0.30 | 1.15 | 2.60 | 1.14 | 2.58 | 1.06 | 4.00 | 1.04 | 4.75 | 1.10 | 0.11 | 3.48 | 2.17 | 4.32 |
| HD 182736 | E | 5237 | 37 | 9 | 0.70 |  |  | 1.13 | 6.08 |  |  | 1.37 | 3.38 | 1.25 | 0.24 | 4.73 | 2.71 | 3.66 |
| HD 182905 | E | 5376 | 55 | 9 | 0.44 | 1.01 | 9.87 | 0.99 | 10.14 | 1.07 | 8.60 | 1.05 | 9.08 | 1.03 | 0.08 | 9.42 | 1.54 | 3.88 |
| HD 183341 | E | 5952 | 47 | 9 | 0.25 | 1.17 | 2.00 | 1.01 | 7.58 | 1.10 | 5.00 | 1.05 | 7.33 | 1.08 | 0.16 | 5.48 | 5.59 | 4.27 |
| HD 183658 | E | 5828 | 52 | 9 | 0.01 | 0.99 | 5.25 | 1.06 | 1.73 | 1.03 | 2.90 | 1.00 | 4.46 | 1.02 | 0.07 | 3.58 | 3.52 | 4.45 |
| HD 183870 | E | 5015 | 19 | 9 | -0.51 |  |  | 0.84 | 1.02 | 0.84 | 1.26 | 0.83 | 1.17 | 0.84 | 0.01 | 1.15 | 0.24 | 4.62 |
| HD 184499 | E | 5807 | 11 | 9 | 0.29 |  |  |  |  | 0.89 | 12.00 |  |  | 0.89 | 0.00 | 12.00 | 0.00 | 4.10 |
| HD 184768 | E | 5635 | 23 | 9 | 0.09 | 1.05 | 5.01 | 0.94 | 11.50 | 0.98 | 10.00 | 0.98 | 11.00 | 0.99 | 0.11 | 9.38 | 6.49 | 4.29 |
| HD 184960 | E | 6290 | 78 | 9 | 0.42 | 1.20 | 2.82 | 1.00 | 7.33 | 1.32 | 1.28 |  |  | 1.17 | 0.32 | 3.81 | 6.05 | 4.23 |
| HD 185269 | E | 5987 | 57 | 3 | 0.65 | 1.32 | 3.16 | 1.29 | 4.42 | 1.42 | 3.00 | 1.30 | 4.13 | 1.33 | 0.13 | 3.68 | 1.42 | 3.97 |
| HD 185414 | E | 5806 | 75 | 9 | 0.00 | 1.03 | 2.65 | 1.01 | 3.38 | 1.03 | 2.90 | 1.01 | 4.11 | 1.02 | 0.02 | 3.26 | 1.46 | 4.45 |
| HD 186104 | E | 5759 | 17 | 9 | 0.11 | 1.08 | 2.84 | 1.01 | 7.20 | 1.06 | 5.00 | 1.00 | 8.75 | 1.04 | 0.08 | 5.95 | 5.91 | 4.33 |
| HD 186226 | E | 6371 | 35 | 3 | 0.80 | 1.31 | 3.44 | 1.27 | 3.75 | 1.40 | 2.75 | 1.43 | 2.52 | 1.35 | 0.16 | 3.11 | 1.23 | 3.93 |
| HD 186379 | E | 5923 | 41 | 9 | 0.48 | 0.97 | 9.44 | 0.96 | 9.38 | 0.97 | 8.50 | 1.00 | 8.30 | 0.98 | 0.04 | 8.90 | 1.14 | 3.98 |
| HD 186413 | E | 5918 | 17 | 9 | 0.32 | 0.99 | 9.59 | 0.94 | 10.17 | 1.10 | 6.00 | 1.04 | 8.25 | 1.02 | 0.16 | 8.50 | 4.17 | 4.16 |
| HD 18757 | E | 5674 | 32 | 9 | 0.05 | 0.96 | 10.00 | 0.95 | 9.25 | 0.98 | 8.00 | | | 0.96 | 0.03 | 9.08 | 2.00 | 4.33 |
| HD 18768 | E | 5815 | 36 | 9 | 0.60 | 0.99 | 7.94 | 1.02 | 7.90 | 1.05 | 7.00 | 1.05 | 6.58 | 1.03 | 0.06 | 7.36 | 1.36 | 3.85 |
| HD 187897 | E | 5905 | 49 | 9 | 0.16 | 1.14 | 0.92 | 1.07 | 4.85 | 1.12 | 3.47 | 1.05 | 5.82 | 1.10 | 0.09 | 3.76 | 4.90 | 4.35 |
| HD 188326 | E | 5342 | 23 | 9 | 0.44 | 1.01 | 9.74 | 0.97 | 10.50 | 1.01 | 9.50 | 1.00 | 9.83 | 1.00 | 0.04 | 9.89 | 1.00 | 3.85 |
| HD 189509 | E | 6368 | 74 | 9 | 0.34 | 1.23 | 0.92 | 1.18 | 2.17 | 1.23 | 1.25 | 1.20 | 1.63 | 1.21 | 0.05 | 1.49 | 1.25 | 4.34 |
| HD 189558 | E | 5773 | 18 | 3 | 0.46 | 0.91 | 11.24 | 0.92 | 10.25 | 0.91 | 10.50 | 0.88 | 10.92 | 0.91 | 0.04 | 10.73 | 0.99 | 3.93 |
| HD 19019 | E | 6113 | 60 | 9 | 0.15 | 1.03 | 4.94 | 1.11 | 1.39 | 1.09 | 2.30 | 1.00 | 5.10 | 1.06 | 0.11 | 3.43 | 3.71 | 4.40 |
| HD 190404 | E | 5088 | 84 | 9 | -0.51 | 0.77 | 5.98 | | | 0.75 | 8.00 | 0.75 | 5.50 | 0.76 | 0.02 | 6.49 | 2.50 | 4.60 |
| HD 190412 | E | 5388 | 97 | 9 | 0.05 | | | | | | | | | | | | | 4.29 |
| HD 190498 | E | 6415 | 29 | 9 | 0.80 | | | 1.16 | 4.38 | 1.36 | 3.00 | 1.39 | 2.75 | 1.30 | 0.23 | 3.38 | 1.63 | 3.93 |
| HD 191533 | E | 6254 | 62 | 9 | 0.87 | 1.29 | 3.16 | 1.27 | 3.86 | 1.56 | 2.25 | 1.41 | 2.70 | 1.38 | 0.29 | 2.99 | 1.61 | 3.84 |
| HD 191785 | E | 5213 | 13 | 8 | -0.33 | 0.90 | 5.13 | 0.84 | 8.50 | 0.92 | 3.59 | 0.82 | 12.25 | 0.87 | 0.10 | 7.37 | 8.66 | 4.52 |
| HD 19308 | E | 5767 | 32 | 9 | 0.23 | | | 1.05 | 8.00 | 1.04 | 8.00 | 0.99 | 10.50 | 1.03 | 0.06 | 8.83 | 2.50 | 4.21 |
| HD 193374 | E | 6522 | 60 | 9 | 0.89 | 1.45 | 2.23 | 1.35 | 2.97 | 1.43 | 2.63 | 1.48 | 2.21 | 1.43 | 0.13 | 2.51 | 0.76 | 3.91 |
| HD 194154 | E | 6456 | 53 | 9 | 0.41 | 1.22 | 1.89 | 1.22 | 1.67 | 1.24 | 1.61 | 1.22 | 1.70 | 1.23 | 0.02 | 1.72 | 0.28 | 4.30 |
| HD 19445 | E | 6052 | 50 | 9 | -0.10 | | | | | 0.75 | 11.50 | | | 0.75 | 0.00 | 11.50 | 0.00 | 4.49 |
| HD 194598 | E | 6126 | 88 | 12 | 0.11 | 0.87 | 10.73 | 0.90 | 6.75 | 0.89 | 8.00 | 0.89 | 8.50 | 0.89 | 0.03 | 8.49 | 3.98 | 4.37 |
| HD 195005 | E | 6149 | 26 | 9 | 0.18 | 1.12 | 2.18 | 1.12 | 1.55 | 1.14 | 1.46 | 1.10 | 2.56 | 1.12 | 0.04 | 1.94 | 1.10 | 4.41 |
| HD 195104 | E | 6226 | 36 | 9 | 0.20 | 1.02 | 5.01 | 1.14 | 1.38 | 1.04 | 3.50 | 1.02 | 4.21 | 1.06 | 0.12 | 3.53 | 3.64 | 4.38 |
| HD 195633 | E | 6024 | 37 | 9 | 0.50 | 0.93 | 9.98 | 0.94 | 9.45 | 1.02 | 7.50 | 1.00 | 7.69 | 0.97 | 0.09 | 8.66 | 2.48 | 3.99 |
| HD 196218 | E | 6204 | 43 | 9 | 0.39 | 1.05 | 6.49 | 0.94 | 9.33 | 1.15 | 4.00 | 1.08 | 5.67 | 1.06 | 0.21 | 6.37 | 5.33 | 4.19 |
| HD 198061 | E | 6379 | 20 | 9 | 0.72 | 1.21 | 3.98 | 1.09 | 5.58 | 1.34 | 3.17 | 1.31 | 3.18 | 1.24 | 0.25 | 3.98 | 2.42 | 3.98 |
| HD 199598 | E | 5918 | 93 | 9 | 0.15 | 1.14 | 0.81 | 1.04 | 5.50 | 1.18 | 1.24 | 1.07 | 4.86 | 1.11 | 0.14 | 3.10 | 4.69 | 4.37 |
| HD 20039 | E | 5331 | 24 | 8 | 0.63 | 1.03 | 8.04 | 1.12 | 6.46 | 1.03 | 7.50 | 1.02 | 7.34 | 1.05 | 0.10 | 7.33 | 1.58 | 3.68 |
| HD 200391 | E | 5709 | 80 | 9 | 0.43 | 0.97 | 10.28 | 0.95 | 10.33 | 0.98 | 9.50 | 0.98 | 9.88 | 0.97 | 0.03 | 10.00 | 0.83 | 3.97 |
| HD 200560 | E | 4894 | 32 | 7 | -0.46 | | | 0.80 | 12.00 | 0.84 | 8.00 | | | 0.82 | 0.04 | 10.00 | 4.00 | 4.52 |
| HD 200580 | E | 5870 | 79 | 9 | 0.43 | 0.93 | 10.99 | 0.95 | 9.75 | 0.93 | 10.00 | 0.91 | 10.55 | 0.93 | 0.04 | 10.32 | 1.24 | 4.00 |
| HD 201099 | E | 5947 | 15 | 9 | 0.23 | 0.92 | 11.48 | 0.96 | 9.00 | | | 0.90 | 11.50 | 0.93 | 0.06 | 10.66 | 2.50 | 4.22 |
| HD 20165 | E | 5137 | 43 | 9 | -0.42 | 0.88 | 2.19 | 0.86 | 2.07 | 0.87 | 1.73 | 0.84 | 4.43 | 0.86 | 0.04 | 2.61 | 2.70 | 4.58 |
| HD 201891 | E | 5998 | 31 | 9 | 0.04 | | | 0.86 | 8.50 | 0.84 | 10.00 | 0.86 | 9.50 | 0.85 | 0.02 | 9.33 | 1.50 | 4.39 |
| HD 202575 | E | 4737 | 37 | 12 | -0.67 | | | 0.79 | 1.11 | 0.76 | 3.28 | 0.77 | 2.22 | 0.77 | 0.03 | 2.20 | 2.18 | 4.65 |
| HD 203235 | E | 6242 | 80 | 3 | 0.49 | 1.24 | 3.05 | 1.15 | 4.95 | 1.29 | 2.60 | 1.18 | 4.35 | 1.22 | 0.14 | 3.74 | 2.35 | 4.16 |

| Star | Type | Teff | | | [Fe/H] | | | | | | | | | | | | | |
|---|---|---|---|---|---|---|---|---|---|---|---|---|---|---|---|---|---|---|
| HD 204426 | E | 5658 | 27 | 9 | 0.39 | 0.94 | 11.75 | 0.95 | 11.00 | 0.95 | 10.50 | 0.91 | 11.58 | 0.94 | 0.04 | 11.21 | 1.25 | 3.98 |
| HD 204734 | E | 5262 | 81 | 8 | -0.38 | | | 0.90 | 0.80 | 0.78 | 10.50 | 0.89 | 0.70 | 0.86 | 0.12 | 4.00 | 9.80 | 4.58 |
| HD 20512 | E | 5270 | 25 | 9 | 0.84 | 1.30 | 3.39 | 1.25 | 4.65 | 1.41 | 3.10 | 1.42 | 2.60 | 1.35 | 0.17 | 3.43 | 2.04 | 3.56 |
| HD 205434 | E | 4454 | 48 | 8 | -0.86 | | | | | 0.69 | 4.26 | | | 0.69 | 0.00 | 4.26 | 0.00 | 4.68 |
| HD 205702 | E | 6060 | 32 | 9 | 0.38 | 1.24 | 2.00 | 1.08 | 6.33 | 1.22 | 3.50 | 0.99 | 8.50 | 1.13 | 0.25 | 5.08 | 6.50 | 4.19 |
| HD 206374 | E | 5579 | 27 | 9 | -0.17 | 0.99 | 1.40 | 0.97 | 1.45 | 0.95 | 3.19 | 0.97 | 1.85 | 0.97 | 0.04 | 1.97 | 1.80 | 4.53 |
| HD 208038 | E | 4995 | 23 | 8 | -0.53 | | | 0.84 | 0.95 | 0.83 | 1.07 | 0.83 | 0.75 | 0.83 | 0.01 | 0.92 | 0.32 | 4.63 |
| HD 208313 | E | 5030 | 38 | 8 | -0.49 | 0.86 | 1.89 | 0.84 | 1.45 | 0.84 | 1.73 | 0.83 | 3.81 | 0.84 | 0.03 | 2.22 | 2.36 | 4.61 |
| HD 209472 | E | 6480 | 29 | 3 | 0.52 | 1.15 | 4.05 | 1.11 | 4.94 | 1.30 | 1.70 | 1.14 | 3.86 | 1.18 | 0.19 | 3.64 | 3.24 | 4.18 |
| HD 210752 | E | 6024 | 37 | 9 | 0.06 | 0.95 | 6.31 | 1.03 | 2.82 | 0.92 | 7.00 | 0.90 | 7.91 | 0.95 | 0.13 | 6.01 | 5.09 | 4.42 |
| HD 21197 | E | 4562 | 26 | 12 | -0.66 | | | 0.76 | 10.00 | | | | | 0.76 | 0.00 | 10.00 | 0.00 | 4.56 |
| HD 214683 | E | 4893 | 54 | 8 | -0.59 | | | 0.82 | 0.88 | | | | | 0.82 | 0.00 | 0.88 | 0.00 | 4.65 |
| HD 216259 | E | 5002 | 32 | 9 | -0.65 | 0.69 | 7.51 | 0.70 | 1.58 | 0.70 | 3.66 | | | 0.70 | 0.01 | 4.25 | 5.93 | 4.67 |
| HD 216520 | E | 5103 | 20 | 8 | -0.42 | 0.85 | 5.36 | 0.84 | 4.33 | 0.85 | 3.43 | 0.83 | 6.24 | 0.84 | 0.02 | 4.84 | 2.81 | 4.56 |
| HD 217813 | E | 5849 | 50 | 9 | 0.05 | 1.03 | 4.50 | 1.05 | 3.38 | 1.01 | 5.33 | 1.05 | 3.93 | 1.04 | 0.04 | 4.28 | 1.96 | 4.42 |
| HD 218059 | E | 6382 | 37 | 9 | 0.38 | 1.23 | 1.42 | 1.22 | 1.88 | 1.25 | 1.30 | 1.18 | 2.45 | 1.22 | 0.07 | 1.76 | 1.15 | 4.31 |
| HD 218209 | E | 5623 | 27 | 9 | -0.11 | 0.97 | 4.20 | 0.96 | 4.15 | 0.86 | 10.50 | | | 0.93 | 0.11 | 6.28 | 6.35 | 4.46 |
| HD 218566 | E | 4846 | 67 | 9 | -0.48 | | | 0.84 | 7.41 | 0.82 | 9.18 | | | 0.83 | 0.02 | 8.30 | 1.77 | 4.53 |
| HD 218687 | E | 5902 | 25 | 9 | 0.09 | 0.97 | 8.12 | 1.00 | 5.63 | 1.03 | 5.00 | 1.08 | 3.28 | 1.02 | 0.11 | 5.51 | 4.84 | 4.39 |
| HD 218868 | E | 5509 | 31 | 8 | -0.07 | 0.98 | 5.01 | 0.95 | 8.29 | 1.00 | 5.58 | 0.95 | 10.00 | 0.97 | 0.05 | 7.22 | 4.99 | 4.41 |
| HD 219396 | E | 5649 | 35 | 9 | 0.37 | 0.93 | 12.02 | 1.03 | 9.75 | 1.13 | 7.33 | 0.99 | 10.63 | 1.02 | 0.20 | 9.93 | 4.69 | 4.03 |
| HD 219420 | E | 6175 | 40 | 9 | 0.42 | 1.23 | 2.19 | 0.88 | 10.50 | 1.30 | 1.70 | 1.17 | 4.28 | 1.15 | 0.42 | 4.67 | 8.80 | 4.19 |
| HD 219538 | E | 5078 | 29 | 9 | -0.46 | 0.87 | 1.85 | 0.85 | 1.55 | 0.85 | 1.53 | 0.84 | 3.62 | 0.85 | 0.03 | 2.13 | 2.09 | 4.60 |
| HD 219623 | E | 6177 | 23 | 5 | 0.30 | 1.16 | 3.16 | 1.12 | 3.90 | 1.17 | 2.70 | 1.20 | 1.93 | 1.16 | 0.08 | 2.92 | 1.98 | 4.31 |
| HD 220140 | E | 5075 | 36 | 9 | -0.44 | 0.85 | 5.36 | 0.86 | 2.80 | 0.84 | 3.43 | 0.85 | 4.53 | 0.85 | 0.02 | 4.03 | 2.56 | 4.58 |
| HD 220182 | E | 5335 | 19 | 9 | -0.31 | 0.92 | 2.19 | 0.91 | 1.76 | 0.91 | 2.40 | 0.90 | 3.04 | 0.91 | 0.02 | 2.35 | 1.29 | 4.56 |
| HD 220221 | E | 4783 | 23 | 8 | -0.57 | 0.81 | 5.36 | 0.81 | 5.03 | 0.82 | 3.40 | 0.81 | 6.27 | 0.81 | 0.01 | 5.01 | 2.87 | 4.58 |
| HD 221354 | E | 5282 | 75 | 9 | -0.26 | 0.94 | 2.57 | 0.85 | 10.25 | 0.95 | 3.13 | 0.89 | 9.50 | 0.91 | 0.10 | 6.36 | 7.68 | 4.49 |
| HD 221585 | E | 5530 | 39 | 9 | 0.48 | 1.09 | 7.94 | 1.13 | 7.43 | 1.18 | 6.50 | 1.15 | 7.25 | 1.14 | 0.09 | 7.28 | 1.44 | 3.93 |
| HD 221851 | E | 5192 | 24 | 9 | -0.42 | | | 0.88 | 0.65 | 0.76 | 11.00 | | | 0.82 | 0.12 | 5.83 | 10.35 | 4.58 |
| HD 222155 | E | 5694 | 20 | 9 | 0.47 | 0.92 | 11.12 | 1.01 | 9.00 | 1.05 | 8.00 | 1.01 | 8.88 | 1.00 | 0.13 | 9.25 | 3.12 | 3.94 |
| HD 224465 | E | 5770 | 21 | 5 | 0.04 | 1.09 | 0.92 | 1.03 | 4.84 | 1.11 | 1.59 | 1.03 | 5.27 | 1.07 | 0.08 | 3.16 | 4.35 | 4.42 |
| HD 22879 | E | 5932 | 40 | 12 | 0.06 | 0.90 | 10.24 | 0.84 | 11.00 | | | 0.86 | 11.33 | 0.87 | 0.06 | 10.86 | 1.10 | 4.36 |
| HD 232781 | E | 4679 | 36 | 8 | -0.80 | | | 0.64 | 3.23 | 0.67 | 3.99 | 0.67 | 2.09 | 0.66 | 0.03 | 3.10 | 1.90 | 4.68 |
| HD 23439A | E | 5192 | 18 | 9 | -0.58 | 0.73 | 5.36 | 0.67 | 5.19 | 0.71 | 3.43 | 0.67 | 5.52 | 0.70 | 0.06 | 4.87 | 2.09 | 4.67 |
| HD 238087 | E | 4213 | 46 | 8 | -0.88 | | | | | | | | | | | | | 4.64 |
| HD 24040 | E | 5756 | 40 | 9 | 0.27 | | | 1.08 | 7.50 | | | 1.00 | 10.50 | 1.04 | 0.08 | 9.00 | 3.00 | 4.17 |
| HD 24238 | E | 5031 | 32 | 8 | -0.48 | 0.85 | 2.57 | 0.84 | 2.30 | | | | | 0.85 | 0.01 | 2.43 | 0.27 | 4.60 |

| | | | | | | | | | | | | | | | | | |
|---|---|---|---|---|---|---|---|---|---|---|---|---|---|---|---|---|---|
| HD 24409 | E | 5568 | 76 | 9 | 0.01 | | | 0.91 | 11.75 | 1.00 | 7.50 | | | 0.96 | 0.09 | 9.63 | 4.25 | 4.34 |
| HD 24451 | E | 4547 | 35 | 9 | -0.75 | 0.74 | 5.36 | 0.75 | 2.85 | 0.76 | 4.03 | 0.74 | 5.65 | 0.75 | 0.02 | 4.47 | 2.80 | 4.64 |
| HD 24496 | E | 5429 | 76 | 9 | -0.15 | 0.96 | 4.50 | 0.88 | 11.50 | 0.96 | 5.70 | 0.91 | 11.25 | 0.93 | 0.08 | 8.24 | 7.00 | 4.44 |
| HD 245 | E | 5805 | 70 | 6 | 0.09 | | | 1.00 | 6.50 | | | | | 1.00 | 0.00 | 6.50 | 0.00 | 4.35 |
| HD 24552 | E | 5916 | 35 | 9 | 0.06 | 1.02 | 4.60 | 1.06 | 2.74 | 1.06 | 2.29 | 1.05 | 3.13 | 1.05 | 0.04 | 3.19 | 2.31 | 4.43 |
| HD 25329 | E | 4855 | 37 | 9 | -0.83 | 0.61 | 3.02 | 0.56 | 9.50 | | | 0.58 | 5.29 | 0.58 | 0.05 | 5.94 | 6.48 | 4.73 |
| HD 25457 | E | 6268 | 65 | 6 | 0.31 | 1.19 | 1.79 | 1.12 | 3.56 | 1.19 | 2.00 | 1.19 | 1.75 | 1.17 | 0.07 | 2.28 | 1.81 | 4.33 |
| HD 25621 | E | 6307 | 27 | 3 | 0.86 | 1.43 | 2.37 | 1.28 | 3.81 | | | 1.43 | 2.63 | 1.38 | 0.15 | 2.94 | 1.44 | 3.86 |
| HD 25825 | E | 5976 | 38 | 12 | 0.11 | 1.06 | 3.40 | 1.08 | 2.51 | 1.08 | 2.29 | 1.07 | 3.03 | 1.07 | 0.02 | 2.81 | 1.11 | 4.41 |
| HD 26345 | E | 6676 | 68 | 12 | 0.54 | 1.35 | 1.13 | 1.36 | 1.15 | 1.37 | 0.90 | 1.36 | 0.90 | 1.36 | 0.02 | 1.02 | 0.25 | 4.28 |
| HD 26784 | E | 6261 | 41 | 12 | 0.38 | 1.19 | 2.51 | 1.25 | 1.93 | 1.30 | 1.16 | 1.29 | 1.28 | 1.26 | 0.11 | 1.72 | 1.35 | 4.29 |
| HD 27808 | E | 6230 | 23 | 12 | 0.35 | 1.18 | 2.51 | 1.21 | 2.60 | 1.29 | 1.16 | 1.23 | 2.06 | 1.23 | 0.11 | 2.08 | 1.44 | 4.30 |
| HD 28005 | E | 5726 | 55 | 7 | 0.15 | 1.03 | 7.23 | 1.03 | 7.67 | | | 1.01 | 9.33 | 1.02 | 0.02 | 8.08 | 2.11 | 4.28 |
| HD 28099 | E | 5738 | 30 | 12 | 0.02 | 1.05 | 2.90 | 1.05 | 3.30 | 1.10 | 1.87 | 1.03 | 4.47 | 1.06 | 0.07 | 3.13 | 2.60 | 4.42 |
| HD 28344 | E | 5921 | 46 | 12 | 0.12 | 0.97 | 8.12 | 1.12 | 2.27 | 1.13 | 2.32 | 1.08 | 3.67 | 1.08 | 0.16 | 4.10 | 5.86 | 4.39 |
| HD 283704 | E | 5524 | 63 | 8 | -0.16 | 0.96 | 4.16 | 1.01 | 2.09 | 1.01 | 2.08 | 1.00 | 3.51 | 1.00 | 0.05 | 2.96 | 2.07 | 4.51 |
| HD 283750 | E | 4405 | 71 | 6 | -0.64 | | | | | | | | | | | | | 4.51 |
| HD 284248 | E | 6157 | 73 | 7 | 0.11 | | | 0.78 | 12.00 | 0.77 | 12.00 | | | 0.78 | 0.01 | 12.00 | 0.00 | 4.32 |
| HD 284574 | E | 5370 | 57 | 8 | -0.20 | 0.92 | 5.18 | 0.88 | 7.13 | 0.95 | 4.28 | 0.91 | 6.40 | 0.92 | 0.07 | 5.75 | 2.84 | 4.47 |
| HD 284930 | E | 4667 | 59 | 7 | -0.70 | 0.75 | 5.36 | 0.73 | 5.60 | 0.76 | 4.99 | 0.73 | 5.29 | 0.74 | 0.03 | 5.31 | 0.60 | 4.63 |
| HD 285690 | E | 4975 | 75 | 12 | -0.36 | 0.82 | 10.99 | 0.82 | 10.75 | 0.87 | 7.05 | | | 0.84 | 0.05 | 9.59 | 3.94 | 4.45 |
| HD 28946 | E | 5366 | 28 | 9 | -0.34 | 0.81 | 8.13 | 0.90 | 1.77 | 0.81 | 7.50 | 0.82 | 4.93 | 0.84 | 0.09 | 5.58 | 6.36 | 4.57 |
| HD 28992 | E | 5865 | 56 | 12 | 0.08 | 1.04 | 4.49 | 1.05 | 3.75 | 1.09 | 2.95 | 1.05 | 4.18 | 1.06 | 0.05 | 3.84 | 1.54 | 4.40 |
| HD 29310 | E | 5841 | 81 | 9 | 0.31 | 1.01 | 8.92 | 1.03 | 8.38 | 1.02 | 8.50 | 0.96 | 10.71 | 1.01 | 0.07 | 9.13 | 2.34 | 4.14 |
| HD 29419 | E | 6058 | 73 | 9 | 0.21 | 1.11 | 3.57 | 1.17 | 1.80 | 1.19 | 1.56 | 1.17 | 1.85 | 1.16 | 0.08 | 2.19 | 2.01 | 4.37 |
| HD 30562 | E | 5894 | 25 | 5 | 0.46 | 1.22 | 3.98 | 1.17 | 5.88 | | | 1.16 | 6.33 | 1.18 | 0.06 | 5.40 | 2.35 | 4.08 |
| HD 3268 | E | 6232 | 55 | 9 | 0.51 | 1.20 | 3.95 | 1.03 | 7.17 | | | | | 1.12 | 0.17 | 5.56 | 3.22 | 4.10 |
| HD 32850 | E | 5226 | 38 | 9 | -0.36 | 0.89 | 2.57 | 0.88 | 2.52 | 0.89 | 1.73 | 0.87 | 4.22 | 0.88 | 0.02 | 2.76 | 2.49 | 4.56 |
| HD 330 | E | 5932 | 62 | 9 | 0.77 | 1.20 | 4.49 | 1.12 | 5.40 | 1.23 | 4.40 | 1.25 | 4.01 | 1.20 | 0.13 | 4.58 | 1.38 | 3.79 |
| HD 332518 | E | 4386 | 91 | 8 | -0.90 | | | | | 0.68 | 3.73 | | | 0.68 | 0.00 | 3.73 | 0.00 | 4.69 |
| HD 33564 | E | 6393 | 31 | 5 | 0.51 | 1.29 | 1.80 | 1.31 | 2.13 | 1.37 | 1.33 | | | 1.32 | 0.08 | 1.75 | 0.80 | 4.22 |
| HD 345957 | E | 5943 | 70 | 9 | 0.34 | 0.89 | 11.75 | | | 0.89 | 11.00 | 0.85 | 12.00 | 0.88 | 0.04 | 11.58 | 1.00 | 4.08 |
| HD 3628 | E | 5810 | 32 | 9 | 0.38 | 0.98 | 10.22 | 0.88 | 11.75 | 0.92 | 11.00 | 0.96 | 10.43 | 0.94 | 0.10 | 10.85 | 1.53 | 4.03 |
| HD 37008 | E | 5091 | 20 | 8 | -0.48 | | | 0.86 | 0.64 | 0.74 | 11.00 | | | 0.80 | 0.12 | 5.82 | 10.36 | 4.60 |
| HD 3765 | E | 4987 | 32 | 9 | -0.45 | 0.86 | 4.58 | 0.82 | 7.68 | 0.87 | 3.56 | 0.83 | 10.00 | 0.85 | 0.05 | 6.45 | 6.44 | 4.55 |
| HD 38230 | E | 5212 | 19 | 9 | -0.27 | 0.87 | 10.17 | | | 0.92 | 6.00 | | | 0.90 | 0.05 | 8.09 | 4.17 | 4.47 |
| HD 40512 | E | 6471 | 62 | 5 | 0.46 | 1.27 | 1.42 | 1.30 | 1.25 | 1.27 | 1.71 | 1.31 | 1.33 | 1.29 | 0.04 | 1.43 | 0.47 | 4.28 |
| HD 40616 | E | 5769 | 19 | 9 | 0.43 | 0.96 | 10.38 | 0.97 | 9.90 | 1.02 | 8.67 | 0.98 | 9.72 | 0.98 | 0.06 | 9.67 | 1.72 | 3.99 |
| HD 42250 | E | 5410 | 31 | 9 | -0.19 | 0.94 | 4.76 | 0.88 | 9.25 | 0.99 | 2.91 | 0.92 | 8.32 | 0.93 | 0.11 | 6.31 | 6.34 | 4.48 |
| HD 4256 | E | 4904 | 32 | 11 | -0.51 | 0.84 | 4.16 | 0.83 | 4.73 | 0.85 | 3.02 | 0.83 | 5.58 | 0.84 | 0.02 | 4.37 | 2.56 | 4.58 |
| HD 42618 | E | 5775 | 25 | 9 | -0.07 | 0.86 | 10.99 | 0.99 | 2.40 | 0.91 | 6.50 | 0.87 | 9.10 | 0.91 | 0.13 | 7.25 | 8.59 | 4.46 |
| HD 42983 | E | 4918 | 49 | 8 | 0.60 | 1.23 | 5.01 | 1.10 | 8.25 | 1.18 | 6.50 | 1.11 | 8.40 | 1.16 | 0.13 | 7.04 | 3.39 | 3.61 |
| HD 43318 | E | 6241 | 70 | 5 | 0.79 | 1.29 | 3.57 | 1.21 | 4.50 | | | 1.22 | 3.83 | 1.24 | 0.08 | 3.97 | 0.93 | 3.87 |
| HD 43856 | E | 6150 | 48 | 9 | 0.41 | 1.20 | 2.95 | 1.03 | 7.44 | 1.29 | 1.80 | 1.11 | 5.69 | 1.16 | 0.26 | 4.47 | 5.64 | 4.19 |
| HD 44966 | E | 6349 | 16 | 9 | 0.43 | | | 1.04 | 6.63 | 1.31 | 1.63 | 1.28 | 1.57 | 1.21 | 0.27 | 3.28 | 5.06 | 4.25 |
| HD 45282 | E | 5422 | 47 | 10 | 1.12 | 1.17 | 4.25 | 1.00 | 7.10 | 0.98 | 7.27 | 1.15 | 3.83 | 1.08 | 0.19 | 5.61 | 3.44 | 3.23 |
| HD 46090 | E | 5575 | 37 | 9 | -0.02 | 0.91 | 12.02 | 0.92 | 10.75 | 1.00 | 6.50 | | | 0.94 | 0.09 | 9.76 | 5.52 | 4.36 |
| HD 46301 | E | 6396 | 17 | 9 | 1.06 | 1.65 | 1.77 | 1.44 | 2.61 | 1.58 | 2.33 | 1.58 | 1.88 | 1.56 | 0.21 | 2.14 | 0.85 | 3.74 |
| HD 4635 | E | 5036 | 19 | 8 | -0.42 | 0.88 | 3.55 | 0.83 | 7.46 | 0.88 | 3.92 | 0.84 | 9.25 | 0.86 | 0.05 | 6.05 | 5.70 | 4.55 |
| HD 46780 | E | 5793 | 5 | 2 | 0.21 | 1.03 | 7.94 | 1.00 | 8.50 | 1.15 | 4.00 | 1.06 | 8.00 | 1.06 | 0.15 | 7.11 | 4.50 | 4.25 |
| HD 47157 | E | 5689 | 39 | 9 | 0.03 | 1.05 | 2.10 | 1.08 | 2.68 | 1.09 | 2.11 | 1.02 | 6.76 | 1.06 | 0.07 | 3.41 | 4.66 | 4.40 |
| HD 47309 | E | 5792 | 38 | 9 | 0.10 | 1.09 | 1.84 | 1.06 | 4.66 | 1.13 | 2.13 | 1.05 | 5.98 | 1.08 | 0.08 | 3.65 | 4.14 | 4.37 |
| HD 47752 | E | 4698 | 54 | 9 | -0.67 | 0.77 | 5.36 | 0.77 | 2.63 | 0.75 | 7.16 | 0.75 | 5.29 | 0.76 | 0.02 | 5.11 | 4.52 | 4.62 |
| HD 48565 | E | 6024 | 71 | 9 | 0.42 | 0.91 | 10.72 | 1.02 | 7.75 | 0.99 | 8.00 | 0.97 | 8.75 | 0.97 | 0.11 | 8.80 | 2.97 | 4.07 |
| HD 49385 | E | 6127 | 40 | 3 | 0.66 | 1.25 | 4.05 | 1.24 | 4.54 | 1.32 | 3.83 | 1.32 | 3.61 | 1.28 | 0.08 | 4.01 | 0.92 | 3.98 |
| HD 50206 | E | 6459 | 33 | 9 | 0.90 | 1.40 | 2.58 | 1.33 | 3.13 | 1.51 | 2.28 | 1.49 | 2.23 | 1.43 | 0.18 | 2.55 | 0.90 | 3.88 |
| HD 50867 | E | 6256 | 33 | 9 | 0.33 | 1.19 | 2.03 | 1.08 | 5.00 | 1.18 | 2.16 | 1.16 | 2.79 | 1.15 | 0.11 | 3.00 | 2.97 | 4.30 |
| HD 51219 | E | 5626 | 26 | 12 | -0.03 | 0.99 | 5.63 | 0.94 | 8.50 | 1.05 | 2.98 | 0.95 | 9.39 | 0.98 | 0.11 | 6.63 | 6.41 | 4.41 |
| HD 51419 | E | 5723 | 29 | 9 | -0.06 | 1.02 | 2.25 | 0.99 | 3.50 | 1.01 | 2.61 | 0.84 | 12.50 | 0.97 | 0.18 | 5.22 | 10.25 | 4.46 |
| HD 51866 | E | 4860 | 8 | 8 | -0.55 | 0.80 | 7.68 | 0.80 | 5.04 | 0.82 | 5.08 | 0.81 | 6.27 | 0.81 | 0.02 | 6.02 | 2.64 | 4.59 |
| HD 52634 | E | 5999 | 101 | 8 | 0.47 | 1.12 | 6.12 | 0.94 | 9.25 | 1.21 | 4.86 | 1.12 | 6.21 | 1.10 | 0.27 | 6.61 | 4.39 | 4.07 |
| HD 5294 | E | 5749 | 26 | 9 | -0.10 | 0.86 | 10.99 | 0.95 | 3.80 | 0.90 | 6.00 | 0.85 | 9.50 | 0.89 | 0.10 | 7.57 | 7.19 | 4.47 |
| HD 5351 | E | 4608 | 39 | 8 | -0.77 | | | | | 0.73 | 2.59 | 0.65 | 11.50 | 0.69 | 0.08 | 7.04 | 8.91 | 4.65 |
| HD 53927 | E | 4941 | 38 | 9 | -0.61 | 0.74 | 5.36 | 0.66 | 11.75 | 0.72 | 6.25 | 0.72 | 5.48 | 0.71 | 0.08 | 7.21 | 6.39 | 4.62 |
| HD 54371 | E | 5562 | 55 | 9 | -0.08 | 0.97 | 5.94 | 0.94 | 7.50 | 1.03 | 2.76 | 0.95 | 8.85 | 0.97 | 0.09 | 6.26 | 6.09 | 4.43 |
| HD 5600 | E | 6378 | 35 | 9 | 0.98 | 1.41 | 2.67 | 1.35 | 3.18 | 1.45 | 2.63 | 1.41 | 2.41 | 1.41 | 0.10 | 2.72 | 0.77 | 3.77 |
| HD 56303 | E | 5912 | 41 | 9 | 0.22 | 1.15 | 2.00 | 1.07 | 5.75 | 1.19 | 3.00 | 1.08 | 6.50 | 1.12 | 0.12 | 4.31 | 4.50 | 4.30 |
| HD 56515 | E | 5993 | 79 | 6 | 0.45 | 1.10 | 6.42 | 0.93 | 9.25 | 1.13 | 6.00 | 1.12 | 6.25 | 1.07 | 0.20 | 6.98 | 3.25 | 4.07 |
| HD 57707 | E | 5021 | 63 | 6 | 1.14 | | | 1.90 | 1.50 | 1.91 | 1.47 | 1.81 | 1.38 | 1.87 | 0.10 | 1.45 | 0.13 | 3.32 |
| HD 59090 | E | 6428 | 61 | 9 | 0.84 | 1.40 | 2.58 | 1.32 | 3.36 | 1.40 | 2.75 | 1.46 | 2.32 | 1.40 | 0.14 | 2.75 | 1.04 | 3.92 |
| HD 59747 | E | 5099 | 46 | 9 | -0.49 | | | | | 0.74 | 10.50 | 0.73 | 11.25 | 0.74 | 0.01 | 10.88 | 0.75 | 4.57 |
| HD 59984 | E | 6005 | 27 | 7 | 0.44 | 0.91 | 10.72 | 0.93 | 9.50 | 1.00 | 8.00 | 0.92 | 9.75 | 0.94 | 0.09 | 9.49 | 2.72 | 4.03 |
| HD 62161 | E | 6449 | 56 | 9 | 0.45 | 1.19 | 3.03 | 1.13 | 4.33 | 1.27 | 1.61 | 1.22 | 2.21 | 1.20 | 0.14 | 2.80 | 2.72 | 4.25 |
| HD 62323 | E | 6113 | 63 | 9 | 0.52 | 1.20 | 4.67 | 0.94 | 9.20 | | | 1.23 | 4.45 | 1.12 | 0.29 | 6.11 | 4.75 | 4.06 |
| HD 62613 | E | 5539 | 35 | 9 | -0.20 | 0.98 | 1.40 | 0.96 | 1.36 | 0.94 | 3.30 | 0.96 | 1.73 | 0.96 | 0.04 | 1.95 | 1.94 | 4.54 |
| HD 63433 | E | 5673 | 26 | 3 | -0.14 | 0.84 | 11.24 | 0.99 | 0.88 | 0.88 | 7.00 | 0.90 | 7.04 | 0.90 | 0.15 | 6.54 | 10.35 | 4.50 |
| HD 64021 | E | 6864 | 84 | 9 | 1.00 | 1.33 | 2.51 | 1.27 | 3.03 | 1.48 | 2.20 | 1.52 | 1.86 | 1.40 | 0.25 | 2.40 | 1.17 | 3.88 |
| HD 64090 | E | 5531 | 35 | 9 | -0.43 | | | 0.68 | 9.75 | | | 0.67 | 12.50 | 0.68 | 0.01 | 11.13 | 2.75 | 4.62 |
| HD 64468 | E | 4914 | 11 | 8 | -0.44 | | | | | 0.85 | 8.50 | | | 0.85 | 0.00 | 8.50 | 0.00 | 4.52 |
| HD 64606 | E | 5250 | 52 | 12 | -0.38 | | | | | | | | | | | | | 4.66 |
| HD 64815 | E | 5722 | 31 | 9 | 0.57 | 1.03 | 8.05 | 0.97 | 9.00 | 1.09 | 6.80 | 1.08 | 6.96 | 1.04 | 0.12 | 7.70 | 2.20 | 3.86 |
| HD 65583 | E | 5341 | 22 | 9 | -0.37 | 0.81 | 6.63 | 0.72 | 12.25 | | | 0.76 | 10.25 | 0.76 | 0.09 | 9.71 | 5.62 | 4.55 |
| HD 65874 | E | 5911 | 21 | 9 | 0.70 | 1.29 | 3.83 | 1.24 | 4.88 | 1.36 | 4.00 | 1.25 | 4.46 | 1.29 | 0.12 | 4.30 | 1.05 | 3.88 |
| HD 68168 | E | 5750 | 44 | 9 | 0.04 | 1.08 | 1.05 | 1.02 | 5.36 | 1.08 | 2.77 | 1.02 | 6.03 | 1.05 | 0.06 | 3.80 | 4.97 | 4.40 |
| HD 68284 | E | 5945 | 53 | 9 | 0.58 | 1.09 | 6.42 | 0.97 | 8.64 | 1.04 | 7.00 | 1.04 | 6.68 | 1.04 | 0.12 | 7.18 | 2.21 | 3.92 |
| HD 68380 | E | 6538 | 85 | 7 | 0.40 | 1.20 | 2.19 | 1.22 | 1.76 | 1.21 | 1.51 | 1.19 | 1.87 | 1.21 | 0.03 | 1.83 | 0.67 | 4.33 |
| HD 68638 | E | 5364 | 30 | 9 | -0.02 | | | | | | | | | | | | | 4.35 |
| HD 69611 | E | 5846 | 22 | 9 | 0.21 | | | 1.02 | 7.50 | | | | | 1.02 | 0.00 | 7.50 | 0.00 | 4.25 |
| HD 70088 | E | 5566 | 26 | 9 | -0.21 | 0.91 | 5.85 | 0.96 | 1.18 | 0.90 | 5.13 | 0.88 | 6.19 | 0.91 | 0.08 | 4.58 | 5.01 | 4.54 |
| HD 70516 | E | 5744 | 22 | 9 | 0.01 | 1.05 | 2.37 | 0.99 | 5.44 | 1.05 | 3.45 | 1.03 | 3.91 | 1.03 | 0.06 | 3.79 | 3.07 | 4.42 |
| HD 70923 | E | 6021 | 51 | 9 | 0.35 | 1.21 | 2.00 | 1.19 | 4.25 | | | 1.09 | 6.50 | 1.16 | 0.12 | 4.25 | 4.50 | 4.22 |
| HD 71431 | E | 5887 | 25 | 9 | 0.62 | 1.21 | 4.62 | 1.01 | 7.75 | 1.29 | 4.60 | 1.19 | 5.38 | 1.18 | 0.28 | 5.59 | 3.15 | 3.91 |
| HD 71595 | E | 6682 | 48 | 9 | 0.77 | 1.35 | 2.41 | 1.07 | 5.17 | 1.39 | 2.40 | 1.33 | 2.53 | 1.29 | 0.32 | 3.13 | 2.77 | 4.02 |
| HD 71640 | E | 6098 | 61 | 9 | 0.25 | 1.05 | 5.55 | 1.08 | 4.71 | 1.11 | 3.83 | 1.06 | 5.17 | 1.08 | 0.06 | 4.82 | 1.72 | 4.31 |
| HD 73393 | E | 5703 | 34 | 9 | -0.06 | 1.01 | 2.57 | 1.01 | 2.33 | 1.00 | 2.84 | 1.01 | 3.21 | 1.01 | 0.01 | 2.74 | 0.89 | 4.47 |
| HD 74011 | E | 5751 | 57 | 9 | 0.30 | | | 1.02 | 9.50 | | | 0.89 | 12.50 | 0.96 | 0.13 | 11.00 | 3.00 | 4.10 |
| HD 75318 | E | 5422 | 44 | 9 | -0.12 | 0.95 | 6.31 | 0.87 | 12.25 | 0.90 | 11.00 | | | 0.91 | 0.08 | 9.85 | 5.94 | 4.40 |
| HD 75767 | E | 5782 | 59 | 9 | 0.06 | 1.09 | 1.05 | 1.02 | 5.25 | 1.11 | 2.17 | 1.03 | 5.84 | 1.06 | 0.09 | 3.58 | 4.79 | 4.40 |
| HD 7590 | E | 5979 | 26 | 9 | 0.01 | 0.94 | 5.66 | 0.86 | 8.00 | 0.99 | 2.75 | 0.97 | 3.28 | 0.94 | 0.13 | 4.92 | 5.25 | 4.45 |
| HD 75935 | E | 5451 | 26 | 8 | -0.20 | 0.94 | 4.31 | 0.91 | 5.44 | 0.97 | 3.04 | 0.93 | 5.30 | 0.94 | 0.06 | 4.52 | 2.40 | 4.50 |
| HD 76932 | E | 5966 | 18 | 9 | 0.24 | | | | | | | 0.86 | 12.50 | 0.86 | 0.00 | 12.50 | 0.00 | 4.18 |
| HD 77407 | E | 5989 | 48 | 3 | 0.06 | 0.93 | 7.94 | 1.06 | 2.00 | 1.08 | 1.83 | 1.02 | 3.60 | 1.02 | 0.15 | 3.84 | 6.12 | 4.44 |
| HD 7924 | E | 5172 | 23 | 8 | -0.42 | | | 0.88 | 0.88 | 0.86 | 3.06 | 0.87 | 0.96 | 0.87 | 0.02 | 1.64 | 2.18 | 4.60 |
| HD 82443 | E | 5315 | 35 | 8 | -0.34 | | | 0.91 | 0.95 | 0.86 | 5.19 | 0.90 | 1.62 | 0.89 | 0.05 | 2.58 | 4.24 | 4.58 |
| HD 84937 | E | 6383 | 73 | 9 | 0.29 | | | | | | | | | | | | | 4.48 |
| HD 87883 | E | 4940 | 35 | 8 | -0.48 | 0.85 | 5.36 | 0.82 | 7.62 | 0.86 | 3.13 | 0.82 | 10.00 | 0.84 | 0.04 | 6.53 | 6.87 | 4.56 |
| HD 88725 | E | 5753 | 15 | 9 | -0.07 | 0.86 | 11.49 | 1.01 | 1.36 | 0.83 | 11.50 | | | 0.90 | 0.18 | 8.12 | 10.14 | 4.45 |
| HD 89269 | E | 5632 | 18 | 9 | -0.11 | 0.97 | 4.20 | 0.96 | 3.88 | 0.95 | 4.90 | 0.96 | 4.56 | 0.96 | 0.02 | 4.38 | 1.03 | 4.48 |
| HD 90875 | E | 4567 | 26 | 8 | -0.65 | | | 0.77 | 10.00 | | | | | 0.77 | 0.00 | 10.00 | 0.00 | 4.56 |
| HD 93215 | E | 5786 | | 1 | 0.08 | 1.09 | 1.21 | 1.13 | 1.58 | 1.12 | 1.25 | 1.09 | 3.65 | 1.11 | 0.04 | 1.92 | 2.44 | 4.40 |
| HD 9407 | E | 5652 | 35 | 8 | -0.04 | 1.03 | 2.20 | 0.99 | 5.28 | 1.04 | 2.94 | 0.98 | 6.39 | 1.01 | 0.06 | 4.20 | 4.19 | 4.44 |
| HD 94765 | E | 5033 | 30 | 8 | -0.46 | 0.84 | 5.36 | 0.85 | 3.11 | 0.83 | 4.69 | 0.84 | 4.79 | 0.84 | 0.02 | 4.49 | 2.25 | 4.58 |
| HD 97503 | E | 4451 | 41 | 10 | -0.81 | 0.72 | 5.36 | 0.73 | 2.53 | 0.72 | 9.00 | 0.72 | 5.79 | 0.72 | 0.01 | 5.67 | 6.47 | 4.65 |
| HD 97658 | E | 5157 | 29 | 8 | -0.46 | 0.75 | 11.49 | | | 0.76 | 10.00 | 0.75 | 9.75 | 0.75 | 0.01 | 10.41 | 1.74 | 4.57 |

| Name | Sp | T | Sig | N | logL/Ls | | | | | | | | | | | | |
|---|---|---|---|---|---|---|---|---|---|---|---|---|---|---|---|---|---|
| HD 98630 | E | 6033 | 33 | 9 | 0.80 | 1.43 | 2.71 | 1.41 | 3.59 | 1.50 | 2.93 | 1.41 | 3.12 | 1.44 | 0.09 | 3.09 | 0.88 | 3.86 |
| HD 98800 | E | 4213 | 119 | 6 | -0.06 | | | | | | | | | | | | | 3.82 |
| HD 99747 | E | 6676 | 42 | 9 | 0.60 | 1.20 | 3.53 | 1.15 | 4.38 | | | 1.18 | 3.25 | 1.18 | 0.05 | 3.72 | 1.13 | 4.15 |
| HR 1687 | E | 6540 | 41 | 5 | 0.69 | 1.46 | 1.00 | | | 1.53 | 0.80 | | | 1.50 | 0.07 | 0.90 | 0.20 | 4.13 |
| HR 3144 | E | 6064 | 56 | 9 | 1.04 | | | 1.56 | 2.60 | 1.72 | 1.95 | 1.69 | 1.75 | 1.66 | 0.16 | 2.10 | 0.85 | 3.70 |
| HR 4657 | E | 6316 | 96 | 9 | 0.16 | 1.06 | 1.94 | 0.91 | 6.50 | 0.98 | 4.06 | 1.01 | 2.61 | 0.99 | 0.15 | 3.78 | 4.56 | 4.42 |
| HR 4867 | E | 6293 | 70 | 5 | 0.33 | 1.18 | 1.75 | 1.12 | 3.67 | 1.19 | 1.85 | 1.17 | 2.16 | 1.17 | 0.07 | 2.36 | 1.91 | 4.32 |
| HR 5307 | E | 6461 | 61 | 11 | 0.63 | | | | | 1.47 | 1.03 | | | 1.47 | 0.00 | 1.03 | 0.00 | 4.16 |
| HR 7438 | E | 6726 | 99 | 9 | 0.54 | 1.36 | 0.81 | 1.28 | 1.97 | 1.34 | 1.13 | 1.33 | 1.15 | 1.33 | 0.08 | 1.27 | 1.16 | 4.28 |
| HR 784 | E | 6245 | 34 | 5 | 0.27 | 1.19 | 1.05 | 1.12 | 2.80 | 1.20 | 0.98 | 1.17 | 1.47 | 1.17 | 0.08 | 1.58 | 1.82 | 4.37 |
| HR 7955 | E | 6205 | 44 | 5 | 0.97 | 1.49 | 2.25 | 1.56 | 2.60 | 1.66 | 1.97 | 1.50 | 2.25 | 1.55 | 0.17 | 2.27 | 0.63 | 3.78 |
| V* BW Ari | E | 5186 | 13 | 8 | -0.39 | 0.87 | 4.94 | 0.88 | 2.17 | 0.88 | 2.27 | 0.87 | 4.23 | 0.88 | 0.01 | 3.40 | 2.77 | 4.58 |
| V* BZ Cet | E | 5014 | 65 | 9 | -0.40 | 0.83 | 10.48 | 0.83 | 10.75 | 0.89 | 4.58 | | | 0.85 | 0.06 | 8.60 | 6.17 | 4.52 |
| V* DI Cam | E | 6337 | 26 | 9 | 0.80 | 1.28 | 3.57 | 1.15 | 4.63 | 1.34 | 3.25 | 1.23 | 3.58 | 1.25 | 0.19 | 3.76 | 1.38 | 3.89 |
| V* EI Eri | E | 5494 | 62 | 9 | 0.54 | 0.98 | 9.53 | 1.02 | 8.45 | 1.09 | 7.17 | 1.04 | 7.42 | 1.03 | 0.11 | 8.14 | 2.36 | 3.82 |
| V* GM Com | E | 6709 | 67 | 12 | 0.52 | 1.32 | 1.23 | 1.31 | 1.27 | 1.31 | 1.28 | 1.31 | 1.25 | 1.31 | 0.01 | 1.26 | 0.05 | 4.29 |
| V* KX Cnc | E | 6008 | 26 | 3 | 0.45 | 1.17 | 5.01 | 0.87 | 11.00 | 1.19 | 5.00 | 1.19 | 5.10 | 1.11 | 0.32 | 6.53 | 6.00 | 4.09 |
| V* MV UMa | E | 4665 | 96 | 9 | -0.61 | | | | | | | | | | | | | 4.62 |
| V* NX Aqr | E | 5656 | 64 | 9 | -0.15 | 0.84 | 11.49 | 0.99 | 0.88 | 0.88 | 7.00 | 0.88 | 8.05 | 0.90 | 0.15 | 6.86 | 10.61 | 4.50 |
| V* V1309 Tau | E | 5791 | 37 | 12 | -0.02 | 1.00 | 4.34 | 1.04 | 1.78 | 1.04 | 1.73 | 1.04 | 2.23 | 1.03 | 0.04 | 2.52 | 2.61 | 4.47 |
| V* V1386 Ori | E | 5294 | 12 | 9 | -0.36 | | | 0.90 | 0.65 | 0.78 | 10.50 | | | 0.84 | 0.12 | 5.58 | 9.85 | 4.56 |
| V* V1709 Aql | E | 6913 | 83 | 3 | 0.77 | 1.47 | 1.26 | 1.20 | 3.75 | 1.57 | 0.90 | 1.52 | 0.96 | 1.44 | 0.37 | 1.72 | 2.85 | 4.13 |
| V* V401 Hya | E | 5836 | 45 | 9 | -0.01 | 0.99 | 4.97 | 1.06 | 1.26 | 1.06 | 1.26 | 1.04 | 1.50 | 1.04 | 0.07 | 2.25 | 3.72 | 4.48 |
| V* V457 Vul | E | 5433 | 27 | 9 | -0.20 | 0.94 | 5.64 | 0.90 | 7.00 | 0.98 | 2.79 | 0.94 | 5.37 | 0.94 | 0.08 | 5.20 | 4.21 | 4.50 |
| V* V566 Oph | E | 6358 | 60 | 5 | 0.59 | 1.18 | 4.36 | 1.02 | 6.90 | | | 1.12 | 4.83 | 1.11 | 0.16 | 5.36 | 2.54 | 4.05 |

| Notes: | Column | Unit | Description |
|---|---|---|---|
| | Sp | N/A | Source for spectroscopic material |
| | T | K | Effective Temperature |
| | Sig | N/A | Standard deviation of the effective temperature |
| | N | N/A | Number of colors used in the effective temperature determination |
| | logL/Ls | Solar | Luminosity in logarithmic solar units |
| B1 | Mass | Solar | Mass in solar units determined from the Bertelli et al. (1994) isochrones |
| | Age | Gyr | Age in gigayears determined from the Bertelli et al. (1994) isochrones |
| D | Mass | Solar | Mass in solar units determined from the Dotter et al. (2008) isochrones |
| | Age | Gyr | Age in gigayears determined from the Dotter et al. (2008) isochrones |
| Y | Mass | Solar | Mass in solar units determined from the Demarque et al. (2004) isochrones |
| | Age | Gyr | Age in gigayears determined from the Demarque et al. (2004) isochrones |
| B2 | Mass | Solar | Mass in solar units determined from the BaSTI Team (2016) isochrones |
| | Age | Gyr | Age in gigayears determined from the BaSTI Team (2016) isochrones |
| | <Mass> | Solar | Average mass in solar masses |
| | Range | Solar | Range in mass determination |
| | <Age> | Gyr | Average age in gigayears |
| | Range | Gyr | Range in age determination |
| | log g | cm s-2 | Surface acceleration computed from average mass, temperature, and luminosity. |

Table 3
Parameter and Iron Data

| Primary | Sp | Mass-Derived Gravity | | | | | | | | | Ionization Balance Gravity | | | | | | | | |
|---|---|---|---|---|---|---|---|---|---|---|---|---|---|---|---|---|---|---|---|
| | | T (K) | G (cm s$^{-2}$) | V$_t$ (km s$^{-1}$) | Fe I (log ε) | σ | N | Fe II (log ε) | σ | N | T (K) | G (cm s$^{-2}$) | V$_t$ (km s$^{-1}$) | Fe I (log ε) | σ | N | Fe II (log ε) | σ | N |
| 10 CVn | S | 5987 | 4.44 | 1.08 | 7.01 | 0.06 | 313 | 6.95 | 0.04 | 16 | 5987 | 4.57 | 0.84 | 7.02 | 0.07 | 313 | 7.02 | 0.04 | 16 |
| 10 Tau | S | 6013 | 4.05 | 1.48 | 7.41 | 0.06 | 354 | 7.38 | 0.06 | 26 | 6013 | 4.11 | 1.44 | 7.41 | 0.06 | 354 | 7.41 | 0.06 | 26 |
| 107 Psc | S | 5259 | 4.58 | 0.56 | 7.48 | 0.07 | 234 | 7.58 | 0.15 | 17 | 5259 | 4.38 | 0.80 | 7.45 | 0.07 | 234 | 7.46 | 0.14 | 17 |
| 109 Psc | S | 5604 | 3.94 | 1.22 | 7.59 | 0.06 | 401 | 7.57 | 0.06 | 19 | 5604 | 3.98 | 1.18 | 7.59 | 0.06 | 401 | 7.59 | 0.06 | 19 |
| 11 Aql | S | 6144 | 3.57 | 2.83 | 7.50 | 0.15 | 195 | 7.43 | 0.11 | 17 | 6144 | 3.74 | 2.78 | 7.50 | 0.15 | 195 | 7.50 | 0.12 | 17 |
| 11 Aqr | S | 5929 | 4.24 | 1.40 | 7.73 | 0.06 | 387 | 7.71 | 0.07 | 28 | 5929 | 4.30 | 1.33 | 7.74 | 0.06 | 387 | 7.74 | 0.07 | 28 |
| 11 LMi | S | 5498 | 4.43 | 1.32 | 7.74 | 0.07 | 400 | 7.84 | 0.13 | 24 | 5498 | 4.19 | 1.59 | 7.70 | 0.08 | 400 | 7.69 | 0.13 | 24 |
| 110 Her | S | 6457 | 3.94 | 2.37 | 7.56 | 0.13 | 252 | 7.52 | 0.10 | 24 | 6457 | 4.04 | 2.33 | 7.56 | 0.13 | 252 | 7.56 | 0.10 | 24 |
| 111 Tau | S | 6184 | 4.38 | 2.04 | 7.57 | 0.11 | 277 | 7.51 | 0.08 | 19 | 6184 | 4.51 | 1.91 | 7.58 | 0.11 | 277 | 7.58 | 0.08 | 19 |
| 111 Tau B | S | 4576 | 4.66 | 0.80 | 7.65 | 0.14 | 213 | 8.30 | 0.47 | 15 | 4576 | 3.66 | 1.40 | 7.21 | 0.15 | 213 | 7.36 | 0.49 | 15 |
| 112 Psc | S | 6031 | 4.03 | 1.68 | 7.75 | 0.07 | 397 | 7.69 | 0.07 | 22 | 6031 | 4.17 | 1.58 | 7.76 | 0.07 | 397 | 7.76 | 0.08 | 22 |
| 12 Oph | S | 5262 | 4.57 | 1.02 | 7.50 | 0.07 | 369 | 7.54 | 0.06 | 12 | 5262 | 4.47 | 1.32 | 7.46 | 0.07 | 369 | 7.46 | 0.06 | 12 |
| 13 Cet | S | 6080 | 4.07 | 0.70 | 7.40 | 0.16 | 255 | 7.00 | 0.05 | 9 | 6080 | 4.46 | 0.20 | 7.42 | 0.15 | 255 | 7.20 | 0.08 | 9 |
| 13 Ori | S | 5800 | 4.07 | 1.22 | 7.30 | 0.05 | 345 | 7.23 | 0.04 | 20 | 5800 | 4.23 | 1.04 | 7.31 | 0.05 | 345 | 7.32 | 0.05 | 20 |
| 13 Tri | S | 5957 | 3.95 | 1.55 | 7.31 | 0.05 | 336 | 7.23 | 0.06 | 21 | 5957 | 4.12 | 1.42 | 7.32 | 0.05 | 336 | 7.32 | 0.05 | 21 |
| 14 Cet | S | 6512 | 3.83 | 1.80 | 7.35 | 0.10 | 326 | 7.26 | 0.05 | 22 | 6512 | 4.05 | 1.72 | 7.35 | 0.09 | 326 | 7.36 | 0.06 | 22 |
| 14 Her | S | 5248 | 4.41 | 0.93 | 7.94 | 0.07 | 299 | 8.10 | 0.05 | 9 | 5248 | 4.13 | 1.12 | 7.86 | 0.07 | 299 | 7.82 | 0.04 | 9 |
| 15 LMi | S | 5916 | 4.06 | 1.42 | 7.56 | 0.06 | 365 | 7.51 | 0.06 | 23 | 5916 | 4.17 | 1.32 | 7.57 | 0.06 | 365 | 7.57 | 0.06 | 23 |
| 15 Sge | S | 5946 | 4.40 | 1.21 | 7.52 | 0.06 | 373 | 7.48 | 0.04 | 18 | 5946 | 4.47 | 1.11 | 7.52 | 0.06 | 373 | 7.52 | 0.03 | 18 |
| 16 Cyg A | S | 5800 | 4.28 | 1.13 | 7.57 | 0.05 | 368 | 7.53 | 0.06 | 26 | 5800 | 4.34 | 1.02 | 7.58 | 0.05 | 368 | 7.58 | 0.06 | 26 |
| 16 Cyg B | S | 5753 | 4.34 | 0.93 | 7.56 | 0.05 | 372 | 7.58 | 0.06 | 24 | 5753 | 4.31 | 1.00 | 7.56 | 0.05 | 372 | 7.56 | 0.06 | 24 |
| 17 Crt A | S | 6240 | 4.17 | 1.80 | 7.50 | 0.07 | 372 | 7.45 | 0.08 | 24 | 6240 | 4.29 | 1.71 | 7.50 | 0.07 | 372 | 7.51 | 0.08 | 24 |
| 17 Crt B | S | 6269 | 4.20 | 1.76 | 7.52 | 0.07 | 374 | 7.43 | 0.06 | 21 | 6269 | 4.38 | 1.62 | 7.53 | 0.07 | 374 | 7.52 | 0.06 | 21 |
| 17 Vir | S | 6146 | 4.33 | 1.60 | 7.60 | 0.07 | 340 | 7.61 | 0.06 | 22 | 6146 | 4.32 | 1.61 | 7.60 | 0.07 | 340 | 7.60 | 0.06 | 22 |
| 18 Cam | S | 5958 | 3.93 | 1.57 | 7.50 | 0.06 | 352 | 7.47 | 0.05 | 24 | 5958 | 3.99 | 1.53 | 7.50 | 0.05 | 352 | 7.50 | 0.05 | 24 |
| 18 Cet | S | 5861 | 3.99 | 1.29 | 7.28 | 0.05 | 347 | 7.21 | 0.05 | 21 | 5861 | 4.15 | 1.13 | 7.29 | 0.05 | 347 | 7.29 | 0.06 | 21 |
| 18 Sco | S | 5791 | 4.42 | 1.23 | 7.48 | 0.05 | 389 | 7.51 | 0.06 | 17 | 5791 | 4.37 | 1.30 | 7.48 | 0.05 | 389 | 7.48 | 0.07 | 17 |
| 20 LMi | S | 5771 | 4.32 | 1.33 | 7.67 | 0.06 | 399 | 7.68 | 0.07 | 19 | 5771 | 4.29 | 1.37 | 7.66 | 0.06 | 399 | 7.66 | 0.07 | 19 |
| 21 Eri | S | 5150 | 3.66 | 1.31 | 7.51 | 0.08 | 359 | 7.43 | 0.12 | 18 | 5150 | 3.82 | 1.12 | 7.55 | 0.08 | 359 | 7.55 | 0.11 | 18 |
| 23 Lib | S | 5717 | 4.26 | 1.38 | 7.71 | 0.07 | 400 | 7.74 | 0.07 | 20 | 5717 | 4.19 | 1.46 | 7.70 | 0.07 | 400 | 7.70 | 0.07 | 20 |
| 24 LMi | S | 5760 | 4.03 | 1.35 | 7.47 | 0.06 | 359 | 7.46 | 0.07 | 20 | 5760 | 4.04 | 1.34 | 7.47 | 0.06 | 359 | 7.46 | 0.07 | 20 |

| Name | | Teff | log g | vt | [Fe/H]a | σ | N | [Fe/H]b | σ | N | Teff | log g | vt | [Fe/H]a | σ | N | [Fe/H]b | σ | N |
|---|---|---|---|---|---|---|---|---|---|---|---|---|---|---|---|---|---|---|---|
| 26 Dra | S | 5925 | 4.37 | 1.39 | 7.45 | 0.06 | 381 | 7.48 | 0.06 | 19 | 5925 | 4.30 | 1.46 | 7.44 | 0.06 | 381 | 7.44 | 0.06 | 19 |
| 33 Sex | S | 5124 | 3.72 | 1.24 | 7.47 | 0.07 | 358 | 7.53 | 0.08 | 19 | 5124 | 3.55 | 1.38 | 7.43 | 0.08 | 358 | 7.43 | 0.09 | 19 |
| 35 Leo | S | 5736 | 3.93 | 1.39 | 7.50 | 0.05 | 412 | 7.47 | 0.06 | 23 | 5736 | 3.99 | 1.34 | 7.50 | 0.05 | 412 | 7.50 | 0.06 | 23 |
| 36 And | S | 4809 | 3.19 | 1.50 | 7.58 | 0.12 | 350 | 7.61 | 0.15 | 19 | 4809 | 3.10 | 1.60 | 7.54 | 0.12 | 350 | 7.54 | 0.15 | 19 |
| 36 Oph A | S | 5103 | 4.64 | 0.92 | 7.23 | 0.06 | 235 | 7.39 | 0.13 | 16 | 5103 | 4.34 | 1.29 | 7.18 | 0.06 | 235 | 7.21 | 0.13 | 16 |
| 36 Oph B | S | 5199 | 4.62 | 1.01 | 7.24 | 0.05 | 223 | 7.31 | 0.15 | 16 | 5199 | 4.48 | 1.16 | 7.22 | 0.05 | 223 | 7.22 | 0.14 | 16 |
| 36 UMa | S | 6173 | 4.40 | 1.33 | 7.38 | 0.05 | 359 | 7.39 | 0.05 | 27 | 6173 | 4.39 | 1.34 | 7.38 | 0.05 | 359 | 7.39 | 0.05 | 27 |
| 37 Gem | S | 5932 | 4.40 | 1.02 | 7.32 | 0.05 | 337 | 7.31 | 0.04 | 19 | 5932 | 4.42 | 0.99 | 7.32 | 0.05 | 337 | 7.32 | 0.04 | 19 |
| 38 LMi | S | 6090 | 3.71 | 2.66 | 7.82 | 0.12 | 266 | 7.72 | 0.11 | 21 | 6090 | 3.93 | 2.58 | 7.82 | 0.12 | 266 | 7.82 | 0.11 | 21 |
| 39 Gem | S | 6112 | 3.81 | 1.78 | 7.06 | 0.07 | 299 | 7.01 | 0.05 | 19 | 6112 | 3.94 | 1.71 | 7.06 | 0.07 | 299 | 7.06 | 0.05 | 19 |
| 39 Leo | S | 6187 | 4.29 | 1.55 | 7.11 | 0.06 | 297 | 7.12 | 0.05 | 21 | 6187 | 4.27 | 1.57 | 7.11 | 0.06 | 297 | 7.11 | 0.05 | 21 |
| 39 Ser | S | 5830 | 4.45 | 1.02 | 7.04 | 0.05 | 303 | 7.02 | 0.03 | 17 | 5830 | 4.50 | 0.92 | 7.05 | 0.05 | 303 | 7.05 | 0.03 | 17 |
| 39 Tau | S | 5836 | 4.44 | 1.33 | 7.50 | 0.05 | 369 | 7.51 | 0.03 | 21 | 5836 | 4.42 | 1.35 | 7.50 | 0.05 | 369 | 7.50 | 0.03 | 21 |
| 4 Equ | S | 6086 | 3.76 | 1.74 | 7.46 | 0.07 | 345 | 7.37 | 0.05 | 23 | 6086 | 3.95 | 1.64 | 7.47 | 0.07 | 345 | 7.47 | 0.06 | 23 |
| 42 Cap | S | 5706 | 3.65 | 1.62 | 7.45 | 0.11 | 351 | 7.34 | 0.11 | 20 | 5706 | 3.90 | 1.46 | 7.47 | 0.11 | 351 | 7.47 | 0.11 | 20 |
| 44 And | S | 5951 | 3.57 | 1.96 | 7.50 | 0.10 | 315 | 7.39 | 0.09 | 20 | 5951 | 3.80 | 1.87 | 7.50 | 0.10 | 315 | 7.50 | 0.08 | 20 |
| 47 UMa | S | 5960 | 4.34 | 1.20 | 7.54 | 0.05 | 406 | 7.48 | 0.05 | 24 | 5960 | 4.46 | 1.02 | 7.55 | 0.05 | 406 | 7.55 | 0.05 | 24 |
| 49 Lib | S | 6297 | 3.91 | 2.04 | 7.48 | 0.09 | 339 | 7.34 | 0.05 | 17 | 6297 | 4.23 | 1.87 | 7.49 | 0.08 | 339 | 7.49 | 0.05 | 17 |
| 49 Per | S | 4984 | 3.45 | 1.06 | 7.57 | 0.07 | 282 | 7.57 | 0.06 | 11 | 4984 | 3.45 | 1.06 | 7.57 | 0.07 | 282 | 7.57 | 0.06 | 11 |
| 5 Ser | S | 6134 | 3.95 | 1.69 | 7.46 | 0.06 | 334 | 7.43 | 0.05 | 22 | 6134 | 4.00 | 1.66 | 7.46 | 0.05 | 334 | 7.46 | 0.05 | 22 |
| 50 Per | S | 6313 | 4.33 | 2.21 | 7.73 | 0.12 | 254 | 7.65 | 0.10 | 19 | 6313 | 4.50 | 2.10 | 7.73 | 0.12 | 254 | 7.74 | 0.10 | 19 |
| 51 Boo Bn | S | 5821 | 4.43 | 1.11 | 7.60 | 0.08 | 347 | 7.71 | 0.10 | 21 | 5821 | 4.23 | 1.24 | 7.58 | 0.09 | 347 | 7.55 | 0.09 | 21 |
| 51 Boo Bs | S | 5990 | 4.33 | 1.77 | 7.58 | 0.10 | 350 | 7.57 | 0.06 | 22 | 5990 | 4.36 | 1.70 | 7.58 | 0.09 | 350 | 7.54 | 0.06 | 22 |
| 51 Peg | S | 5799 | 4.34 | 1.28 | 7.67 | 0.07 | 363 | 7.69 | 0.08 | 25 | 5799 | 4.30 | 1.33 | 7.66 | 0.07 | 363 | 7.66 | 0.09 | 25 |
| 54 Cnc | S | 5824 | 3.94 | 1.46 | 7.59 | 0.06 | 369 | 7.55 | 0.07 | 25 | 5824 | 4.03 | 1.45 | 7.60 | 0.06 | 369 | 7.64 | 0.08 | 25 |
| 54 Psc | S | 5289 | 4.52 | 0.72 | 7.70 | 0.07 | 244 | 7.82 | 0.16 | 19 | 5289 | 4.26 | 0.95 | 7.66 | 0.07 | 244 | 7.67 | 0.17 | 19 |
| 55 Vir | S | 5054 | 3.29 | 1.42 | 7.12 | 0.07 | 369 | 7.17 | 0.08 | 19 | 5054 | 3.15 | 1.49 | 7.10 | 0.07 | 369 | 7.10 | 0.08 | 19 |
| 58 Eri | S | 5839 | 4.47 | 1.29 | 7.48 | 0.06 | 377 | 7.46 | 0.04 | 20 | 5839 | 4.51 | 1.23 | 7.49 | 0.06 | 377 | 7.49 | 0.04 | 20 |
| 59 Eri | S | 6275 | 3.90 | 1.99 | 7.61 | 0.08 | 334 | 7.53 | 0.08 | 24 | 6275 | 4.09 | 1.91 | 7.62 | 0.08 | 334 | 7.61 | 0.08 | 24 |
| 61 Cyg A | S | 4481 | 4.67 | 0.15 | 7.33 | 0.12 | 273 | 7.41 | 0.21 | 3 | 4481 | 4.28 | 1.48 | 7.13 | 0.13 | 273 | 7.00 | 0.19 | 3 |
| 61 Cyg B | S | 4171 | 4.70 | 0.15 | 7.39 | 0.20 | 222 | 7.52 | 0.03 | 2 | 4171 | 4.37 | 0.90 | 7.29 | 0.20 | 222 | 7.27 | 0.02 | 2 |
| 61 Psc | S | 6273 | 3.92 | 2.11 | 7.65 | 0.10 | 337 | 7.65 | 0.08 | 28 | 6273 | 3.92 | 2.11 | 7.65 | 0.10 | 337 | 7.65 | 0.08 | 28 |
| 61 UMa | S | 5507 | 4.54 | 1.08 | 7.41 | 0.05 | 400 | 7.48 | 0.06 | 17 | 5507 | 4.41 | 1.15 | 7.38 | 0.05 | 400 | 7.37 | 0.07 | 17 |
| 61 Vir | S | 5578 | 4.44 | 0.92 | 7.46 | 0.05 | 335 | 7.49 | 0.06 | 18 | 5578 | 4.39 | 1.04 | 7.45 | 0.05 | 335 | 7.45 | 0.06 | 18 |
| 63 Eri | S | 5422 | 3.33 | 1.69 | 7.30 | 0.08 | 365 | 7.12 | 0.06 | 18 | 5422 | 3.72 | 1.46 | 7.33 | 0.08 | 365 | 7.33 | 0.05 | 18 |
| 64 Aql | S | 4736 | 3.14 | 1.17 | 7.47 | 0.10 | 218 | 7.60 | 0.12 | 15 | 4736 | 3.00 | 1.19 | 7.44 | 0.10 | 218 | 7.51 | 0.12 | 15 |
| 66 Cet | S | 6102 | 3.83 | 1.68 | 7.51 | 0.06 | 350 | 7.33 | 0.06 | 21 | 6102 | 4.22 | 1.45 | 7.53 | 0.06 | 350 | 7.52 | 0.06 | 21 |
| 66 Cet B | S | 5722 | 4.22 | 1.31 | 7.56 | 0.07 | 355 | 7.45 | 0.06 | 21 | 5722 | 4.44 | 1.18 | 7.59 | 0.07 | 355 | 7.62 | 0.05 | 21 |

| Star | | Teff | logg | vt | [Fe/H] | σ | N | [Fe/H]II | σ | N | Teff | logg | vt | [Fe/H] | σ | N | [Fe/H]II | σ | N |
|---|---|---|---|---|---|---|---|---|---|---|---|---|---|---|---|---|---|---|---|
| 70 Oph A | S | 5244 | 4.47 | 0.91 | 7.48 | 0.05 | 269 | 7.62 | 0.11 | 16 | 5244 | 4.15 | 1.12 | 7.44 | 0.06 | 269 | 7.44 | 0.12 | 16 |
| 70 Vir | S | 5538 | 3.90 | 1.27 | 7.39 | 0.06 | 418 | 7.37 | 0.07 | 25 | 5538 | 3.93 | 1.25 | 7.39 | 0.05 | 418 | 7.39 | 0.07 | 25 |
| 79 Cet | S | 5765 | 4.03 | 1.40 | 7.60 | 0.07 | 377 | 7.59 | 0.07 | 26 | 5765 | 4.06 | 1.38 | 7.60 | 0.07 | 377 | 7.60 | 0.07 | 26 |
| 83 Leo | S | 5380 | 4.42 | 1.04 | 7.76 | 0.07 | 268 | 7.90 | 0.11 | 22 | 5380 | 4.09 | 1.20 | 7.72 | 0.07 | 268 | 7.72 | 0.12 | 22 |
| 83 Leo B | S | 4973 | 4.58 | 0.75 | 7.79 | 0.09 | 281 | 7.84 | 0.12 | 10 | 4973 | 4.49 | 0.98 | 7.75 | 0.08 | 281 | 7.76 | 0.11 | 10 |
| 84 Her | S | 5810 | 3.77 | 1.76 | 7.71 | 0.07 | 356 | 7.68 | 0.09 | 24 | 5810 | 3.83 | 1.73 | 7.71 | 0.07 | 356 | 7.71 | 0.09 | 24 |
| 85 Peg | S | 5454 | 4.54 | 0.58 | 6.72 | 0.06 | 270 | 6.89 | 0.14 | 27 | 5454 | 4.22 | 1.36 | 6.67 | 0.06 | 270 | 6.68 | 0.14 | 27 |
| 88 Leo | S | 6030 | 4.39 | 1.28 | 7.47 | 0.05 | 345 | 7.45 | 0.05 | 21 | 6030 | 4.43 | 1.24 | 7.47 | 0.05 | 345 | 7.48 | 0.05 | 21 |
| 9 Cet | S | 5822 | 4.46 | 1.62 | 7.69 | 0.07 | 354 | 7.74 | 0.09 | 23 | 5822 | 4.35 | 1.72 | 7.68 | 0.08 | 354 | 7.68 | 0.09 | 23 |
| 9 Com | S | 6239 | 3.91 | 2.06 | 7.68 | 0.08 | 329 | 7.61 | 0.09 | 24 | 6239 | 4.06 | 1.99 | 7.68 | 0.08 | 329 | 7.69 | 0.09 | 24 |
| 94 Aqr | S | 5379 | 3.82 | 1.40 | 7.54 | 0.08 | 348 | 7.57 | 0.10 | 19 | 5379 | 3.76 | 1.45 | 7.53 | 0.08 | 348 | 7.53 | 0.10 | 19 |
| 94 Aqr B | S | 5219 | 4.35 | 1.29 | 7.66 | 0.08 | 319 | 7.65 | 0.17 | 18 | 5219 | 4.37 | 1.26 | 7.66 | 0.08 | 319 | 7.66 | 0.17 | 18 |
| 94 Cet | S | 6064 | 4.04 | 1.82 | 7.66 | 0.08 | 356 | 7.66 | 0.06 | 26 | 6064 | 4.02 | 1.83 | 7.66 | 0.08 | 356 | 7.65 | 0.06 | 26 |
| alf Aql | S | 7377 | 3.95 | 5.68 | 7.94 | 0.26 | 21 | 8.02 | 0.13 | 2 | 7377 | 3.75 | 5.65 | 7.95 | 0.26 | 21 | 7.96 | 0.13 | 2 |
| alf Cep | S | 7217 | 3.69 | 3.82 | 7.94 | 0.20 | 27 | 7.08 | 0.11 | 4 | 7217 | 4.08 | 3.84 | 7.92 | 0.20 | 27 | 7.20 | 0.11 | 4 |
| alf CMi | S | 6654 | 3.95 | 2.25 | 7.44 | 0.07 | 327 | 7.39 | 0.06 | 24 | 6654 | 4.09 | 2.20 | 7.44 | 0.07 | 327 | 7.44 | 0.06 | 24 |
| alf Com A | S | 6391 | 4.09 | 2.29 | 7.39 | 0.13 | 231 | 7.32 | 0.11 | 17 | 6391 | 4.24 | 2.21 | 7.39 | 0.13 | 231 | 7.39 | 0.10 | 17 |
| alf Crv | S | 7019 | 4.27 | 2.77 | 7.44 | 0.22 | 198 | 7.35 | 0.11 | 18 | 7019 | 4.50 | 2.67 | 7.44 | 0.22 | 198 | 7.44 | 0.11 | 18 |
| alf For A | S | 6195 | 3.95 | 1.77 | 7.26 | 0.07 | 297 | 7.18 | 0.04 | 19 | 6195 | 4.12 | 1.67 | 7.26 | 0.07 | 297 | 7.26 | 0.04 | 19 |
| alf PsA | S | 7671 | 4.03 | 3.90 | 7.76 | 0.19 | 30 | 7.89 | 0.18 | 8 | 7671 | 3.67 | 4.20 | 7.67 | 0.19 | 30 | 7.86 | 0.17 | 8 |
| b Aql | S | 5466 | 4.10 | 1.45 | 7.76 | 0.11 | 321 | 7.86 | 0.08 | 20 | 5466 | 3.88 | 1.61 | 7.74 | 0.11 | 321 | 7.73 | 0.08 | 20 |
| b01 Cyg | S | 5090 | 3.69 | 1.38 | 7.44 | 0.08 | 361 | 7.54 | 0.12 | 19 | 5090 | 3.44 | 1.57 | 7.39 | 0.10 | 361 | 7.38 | 0.13 | 19 |
| bet Aql | S | 5144 | 3.58 | 1.22 | 7.28 | 0.07 | 380 | 7.24 | 0.07 | 18 | 5144 | 3.66 | 1.13 | 7.30 | 0.07 | 380 | 7.30 | 0.07 | 18 |
| bet Com | S | 6022 | 4.40 | 1.31 | 7.53 | 0.05 | 330 | 7.49 | 0.05 | 22 | 6022 | 4.47 | 1.22 | 7.53 | 0.05 | 330 | 7.54 | 0.05 | 22 |
| bet CVn | S | 5865 | 4.40 | 1.04 | 7.25 | 0.05 | 371 | 7.27 | 0.05 | 18 | 5865 | 4.36 | 1.09 | 7.25 | 0.05 | 371 | 7.25 | 0.06 | 18 |
| bet Vir | S | 6159 | 4.08 | 1.60 | 7.61 | 0.06 | 375 | 7.57 | 0.05 | 28 | 6159 | 4.18 | 1.54 | 7.62 | 0.06 | 375 | 7.62 | 0.05 | 28 |
| c Eri | S | 7146 | 4.23 | 4.98 | 7.70 | 0.35 | 28 | 7.88 | 0.00 | 1 | 7146 | 3.75 | 5.12 | 7.70 | 0.35 | 28 | 7.72 | 0.00 | 1 |
| c UMa | S | 5995 | 4.12 | 1.48 | 7.52 | 0.05 | 357 | 7.46 | 0.07 | 23 | 5995 | 4.24 | 1.38 | 7.53 | 0.05 | 357 | 7.53 | 0.06 | 23 |
| chi Cnc | S | 6274 | 4.25 | 1.63 | 7.19 | 0.07 | 311 | 7.19 | 0.07 | 20 | 6274 | 4.26 | 1.62 | 7.19 | 0.07 | 311 | 7.20 | 0.07 | 20 |
| chi Dra | S | 6083 | 4.20 | 1.12 | 6.82 | 0.11 | 275 | 6.89 | 0.07 | 20 | 6083 | 4.04 | 1.26 | 6.81 | 0.11 | 275 | 6.80 | 0.07 | 20 |
| chi Her | S | 5890 | 3.96 | 1.44 | 6.97 | 0.06 | 304 | 6.95 | 0.04 | 17 | 5890 | 4.02 | 1.39 | 6.98 | 0.06 | 304 | 6.98 | 0.04 | 17 |
| chi01 Ori | S | 5983 | 4.44 | 1.83 | 7.49 | 0.08 | 334 | 7.44 | 0.07 | 19 | 5983 | 4.55 | 1.70 | 7.50 | 0.08 | 334 | 7.50 | 0.07 | 19 |
| del Cap | S | 7021 | 4.00 | 4.40 | 8.21 | 0.22 | 20 | 8.09 | 0.26 | 2 | 7021 | 4.32 | 4.81 | 8.19 | 0.22 | 20 | 8.19 | 0.25 | 2 |
| del Eri | S | 5076 | 3.77 | 1.19 | 7.55 | 0.08 | 365 | 7.58 | 0.11 | 20 | 5076 | 3.71 | 1.26 | 7.54 | 0.08 | 365 | 7.53 | 0.11 | 20 |
| del Tri | S | 5796 | 4.37 | 1.22 | 7.00 | 0.12 | 322 | 7.00 | 0.06 | 19 | 5796 | 4.37 | 1.22 | 7.00 | 0.12 | 322 | 7.00 | 0.06 | 19 |
| e Vir | S | 5999 | 4.17 | 1.75 | 7.60 | 0.08 | 356 | 7.63 | 0.06 | 20 | 5999 | 4.11 | 1.77 | 7.59 | 0.07 | 356 | 7.55 | 0.07 | 20 |
| eps Eri | S | 5123 | 4.57 | 0.90 | 7.42 | 0.07 | 249 | 7.53 | 0.17 | 15 | 5123 | 4.32 | 1.15 | 7.38 | 0.07 | 249 | 7.38 | 0.16 | 15 |
| eps For | S | 5129 | 3.57 | 1.04 | 6.89 | 0.06 | 340 | 6.87 | 0.06 | 17 | 5129 | 3.61 | 1.00 | 6.89 | 0.06 | 340 | 6.90 | 0.06 | 17 |

| Name | | Teff | log g | vt | [M/H] | σ | N | [M/H] | σ | N | Teff | log g | vt | [M/H] | σ | N | [M/H] | σ | N |
|---|---|---|---|---|---|---|---|---|---|---|---|---|---|---|---|---|---|---|---|
| eta Ari   | S | 6485 | 3.98 | 1.84 | 7.33 | 0.08 | 295 | 7.26 | 0.06 | 23 | 6485 | 4.16 | 1.80 | 7.33 | 0.08 | 295 | 7.32 | 0.06 | 23 |
| eta Boo   | S | 6050 | 3.75 | 2.38 | 7.76 | 0.11 | 277 | 7.70 | 0.11 | 23 | 6050 | 3.89 | 2.33 | 7.76 | 0.11 | 277 | 7.76 | 0.11 | 23 |
| eta Cas   | S | 5937 | 4.37 | 1.14 | 7.22 | 0.06 | 350 | 7.21 | 0.04 | 21 | 5937 | 4.40 | 1.10 | 7.22 | 0.05 | 350 | 7.22 | 0.04 | 21 |
| eta Cep   | S | 5057 | 3.42 | 1.33 | 7.35 | 0.08 | 374 | 7.39 | 0.10 | 19 | 5057 | 3.31 | 1.35 | 7.32 | 0.08 | 374 | 7.28 | 0.10 | 19 |
| eta CrB A | S | 6060 | 4.45 | 1.40 | 7.44 | 0.05 | 335 | 7.43 | 0.02 | 19 | 6060 | 4.47 | 1.38 | 7.45 | 0.05 | 335 | 7.45 | 0.02 | 19 |
| eta CrB B | S | 5948 | 4.51 | 1.18 | 7.42 | 0.06 | 343 | 7.55 | 0.06 | 22 | 5948 | 4.27 | 1.47 | 7.40 | 0.07 | 343 | 7.40 | 0.06 | 22 |
| eta Ser   | S | 4985 | 3.15 | 1.37 | 7.30 | 0.08 | 374 | 7.29 | 0.05 | 16 | 4985 | 3.18 | 1.34 | 7.31 | 0.08 | 374 | 7.31 | 0.05 | 16 |
| gam Cep   | S | 4850 | 3.21 | 1.50 | 7.61 | 0.11 | 368 | 7.72 | 0.19 | 23 | 4850 | 3.00 | 1.56 | 7.54 | 0.11 | 368 | 7.52 | 0.20 | 23 |
| gam Lep   | S | 6352 | 4.31 | 1.70 | 7.45 | 0.08 | 330 | 7.40 | 0.05 | 22 | 6352 | 4.43 | 1.61 | 7.45 | 0.07 | 330 | 7.46 | 0.05 | 22 |
| gam Ser   | S | 6286 | 4.13 | 1.86 | 7.35 | 0.09 | 317 | 7.30 | 0.05 | 16 | 6286 | 4.25 | 1.77 | 7.36 | 0.09 | 317 | 7.36 | 0.05 | 16 |
| gam Vir A | S | 6922 | 4.27 | 2.83 | 7.54 | 0.20 | 171 | 7.59 | 0.18 | 17 | 6922 | 4.16 | 2.95 | 7.53 | 0.20 | 171 | 7.52 | 0.18 | 17 |
| gam01 Del | S | 6194 | 3.72 | 1.91 | 7.51 | 0.07 | 347 | 7.46 | 0.07 | 29 | 6194 | 3.85 | 1.86 | 7.52 | 0.07 | 347 | 7.52 | 0.07 | 29 |
| i Boo A   | S | 5878 | 4.38 | 3.80 | 6.38 | 0.10 | 10  | 7.93 | 0.19 | 2  | 5878 | 4.38 | 3.80 | 6.38 | 0.10 | 10  | 7.93 | 0.19 | 2  |
| i Boo B   | S | 5240 | 4.50 | 0.96 | 6.94 | 0.06 | 283 | 7.47 | 0.12 | 34 | 5240 | 3.56 | 1.53 | 6.85 | 0.07 | 283 | 6.94 | 0.11 | 34 |
| iot Peg   | S | 6504 | 4.19 | 2.02 | 7.37 | 0.13 | 296 | 7.31 | 0.08 | 22 | 6504 | 4.34 | 1.92 | 7.38 | 0.13 | 296 | 7.38 | 0.08 | 22 |
| iot Per   | S | 5995 | 4.15 | 1.49 | 7.59 | 0.06 | 369 | 7.55 | 0.05 | 24 | 5995 | 4.22 | 1.38 | 7.59 | 0.05 | 369 | 7.56 | 0.05 | 24 |
| iot Psc   | S | 6177 | 4.08 | 1.60 | 7.36 | 0.07 | 334 | 7.34 | 0.06 | 23 | 6177 | 4.14 | 1.56 | 7.37 | 0.07 | 334 | 7.37 | 0.06 | 23 |
| kap Del   | S | 5633 | 3.66 | 1.60 | 7.45 | 0.07 | 360 | 7.44 | 0.07 | 25 | 5633 | 3.68 | 1.58 | 7.45 | 0.06 | 360 | 7.45 | 0.07 | 25 |
| kap For   | S | 5867 | 3.96 | 1.43 | 7.40 | 0.07 | 366 | 7.39 | 0.04 | 20 | 5867 | 3.98 | 1.42 | 7.41 | 0.07 | 366 | 7.40 | 0.04 | 20 |
| kap01 Cet | S | 5730 | 4.50 | 1.47 | 7.53 | 0.06 | 377 | 7.49 | 0.04 | 15 | 5730 | 4.58 | 1.34 | 7.54 | 0.06 | 377 | 7.54 | 0.04 | 15 |
| ksi Boo A | S | 5480 | 4.53 | 1.38 | 7.29 | 0.05 | 371 | 7.33 | 0.07 | 20 | 5480 | 4.44 | 1.53 | 7.27 | 0.06 | 371 | 7.27 | 0.07 | 20 |
| ksi Boo B | S | 4767 | 5.00 | 0.15 | 7.45 | 0.15 | 310 | 7.39 | 0.29 | 7  | 4767 | 4.10 | 2.35 | 7.16 | 0.18 | 310 | 6.72 | 0.34 | 7  |
| ksi Peg   | S | 6250 | 4.01 | 1.77 | 7.28 | 0.08 | 313 | 7.21 | 0.04 | 20 | 6250 | 4.17 | 1.66 | 7.29 | 0.08 | 313 | 7.29 | 0.04 | 20 |
| ksi UMa B | S | 5667 | 4.44 | 0.77 | 6.79 | 0.10 | 264 | 6.93 | 0.11 | 13 | 5667 | 4.24 | 1.02 | 6.76 | 0.10 | 264 | 6.78 | 0.11 | 13 |
| lam Aur   | S | 5931 | 4.28 | 1.28 | 7.56 | 0.05 | 365 | 7.52 | 0.05 | 22 | 5931 | 4.37 | 1.16 | 7.57 | 0.05 | 365 | 7.58 | 0.05 | 22 |
| lam Ser   | S | 5908 | 4.13 | 1.45 | 7.44 | 0.06 | 382 | 7.40 | 0.04 | 17 | 5908 | 4.21 | 1.38 | 7.44 | 0.05 | 382 | 7.44 | 0.04 | 17 |
| m Per     | S | 6704 | 3.83 | 2.15 | 7.59 | 0.22 | 6   | 7.35 | 0.10 | 2  | 6704 | 4.22 | 0.15 | 7.63 | 0.20 | 6   | 7.52 | 0.12 | 2  |
| m Tau     | S | 5631 | 4.03 | 1.11 | 7.24 | 0.05 | 346 | 7.28 | 0.04 | 18 | 5631 | 3.92 | 1.22 | 7.23 | 0.06 | 346 | 7.22 | 0.05 | 18 |
| mu. Cas   | S | 5434 | 4.57 | 0.53 | 6.72 | 0.06 | 288 | 6.69 | 0.05 | 12 | 5434 | 4.64 | 0.10 | 6.73 | 0.06 | 288 | 6.73 | 0.05 | 12 |
| mu. Her   | S | 5560 | 3.99 | 1.36 | 7.70 | 0.08 | 404 | 7.73 | 0.09 | 20 | 5560 | 3.91 | 1.43 | 7.69 | 0.08 | 404 | 7.68 | 0.09 | 20 |
| mu.01 Cyg | S | 6354 | 3.93 | 1.96 | 7.32 | 0.09 | 298 | 7.26 | 0.08 | 21 | 6354 | 4.08 | 1.88 | 7.32 | 0.09 | 298 | 7.32 | 0.08 | 21 |
| mu.02 Cnc | S | 5857 | 3.97 | 1.55 | 7.62 | 0.07 | 361 | 7.56 | 0.07 | 22 | 5857 | 4.10 | 1.46 | 7.63 | 0.07 | 361 | 7.63 | 0.07 | 22 |
| mu.02 Cyg | S | 5998 | 4.33 | 1.27 | 7.23 | 0.06 | 330 | 7.25 | 0.04 | 19 | 5998 | 4.28 | 1.33 | 7.22 | 0.06 | 330 | 7.22 | 0.04 | 19 |
| ome Leo   | S | 5883 | 3.79 | 1.60 | 7.48 | 0.06 | 380 | 7.41 | 0.06 | 24 | 5883 | 3.93 | 1.52 | 7.48 | 0.06 | 380 | 7.48 | 0.06 | 24 |
| ome Sgr   | S | 5455 | 3.61 | 1.42 | 7.42 | 0.07 | 378 | 7.40 | 0.07 | 27 | 5455 | 3.65 | 1.39 | 7.42 | 0.07 | 378 | 7.42 | 0.07 | 27 |
| omi Aql   | S | 6173 | 4.20 | 1.53 | 7.64 | 0.06 | 365 | 7.58 | 0.06 | 27 | 6173 | 4.32 | 1.43 | 7.65 | 0.06 | 365 | 7.65 | 0.06 | 27 |
| omi02 Eri | S | 5202 | 4.59 | 0.33 | 7.22 | 0.07 | 345 | 7.25 | 0.09 | 14 | 5202 | 4.54 | 0.69 | 7.20 | 0.07 | 345 | 7.20 | 0.09 | 14 |
| phi02 Cet | S | 6218 | 4.39 | 1.32 | 7.38 | 0.06 | 345 | 7.41 | 0.06 | 22 | 6218 | 4.33 | 1.38 | 7.38 | 0.07 | 345 | 7.38 | 0.07 | 22 |

| Name | | Teff | log g | vt | A(Fe I) | σ | # | A(Fe II) | σ | # | Teff | log g | vt | A(Fe I) | σ | # | A(Fe II) | σ | # |
|---|---|---|---|---|---|---|---|---|---|---|---|---|---|---|---|---|---|---|---|
| pi.03 Ori | S | 6509 | 4.31 | 2.22 | 7.64 | 0.13 | 266 | 7.61 | 0.11 | 20 | 6509 | 4.38 | 2.19 | 7.64 | 0.13 | 266 | 7.64 | 0.11 | 20 |
| psi Cap | S | 6633 | 4.24 | 3.36 | 7.67 | 0.21 | 100 | 7.80 | 0.40 | 10 | 6633 | 3.90 | 3.48 | 7.67 | 0.22 | 100 | 7.67 | 0.40 | 10 |
| psi Cnc | S | 5305 | 3.53 | 1.51 | 7.38 | 0.09 | 374 | 7.35 | 0.08 | 18 | 5305 | 3.60 | 1.46 | 7.39 | 0.08 | 374 | 7.39 | 0.08 | 18 |
| psi Ser | S | 5635 | 4.45 | 1.07 | 7.48 | 0.05 | 391 | 7.53 | 0.05 | 16 | 5635 | 4.36 | 1.23 | 7.47 | 0.06 | 391 | 7.47 | 0.05 | 16 |
| psi01 Dra A | S | 6573 | 3.97 | 2.55 | 7.47 | 0.10 | 278 | 7.39 | 0.08 | 22 | 6573 | 4.16 | 2.47 | 7.47 | 0.10 | 278 | 7.47 | 0.08 | 22 |
| psi01 Dra B | S | 6237 | 4.31 | 1.60 | 7.52 | 0.07 | 338 | 7.46 | 0.05 | 21 | 6237 | 4.43 | 1.49 | 7.52 | 0.06 | 338 | 7.53 | 0.05 | 21 |
| psi05 Aur | S | 6128 | 4.34 | 1.47 | 7.59 | 0.06 | 378 | 7.57 | 0.05 | 21 | 6128 | 4.39 | 1.38 | 7.60 | 0.05 | 378 | 7.56 | 0.05 | 21 |
| rho Cap | S | 6856 | 3.95 | 4.61 | 7.33 | 0.37 | 121 | 7.14 | 0.39 | 5 | 6856 | 4.34 | 4.64 | 7.31 | 0.37 | 121 | 7.27 | 0.39 | 5 |
| rho CrB | S | 5850 | 4.17 | 1.11 | 7.27 | 0.05 | 374 | 7.21 | 0.05 | 22 | 5850 | 4.29 | 0.96 | 7.28 | 0.05 | 374 | 7.28 | 0.05 | 22 |
| rho01 Cnc | S | 5248 | 4.43 | 1.09 | 7.80 | 0.09 | 242 | 7.92 | 0.14 | 18 | 5248 | 4.14 | 1.25 | 7.76 | 0.09 | 242 | 7.75 | 0.15 | 18 |
| sig Boo | S | 6781 | 4.29 | 2.13 | 7.11 | 0.10 | 257 | 7.10 | 0.05 | 19 | 6781 | 4.33 | 2.11 | 7.12 | 0.10 | 257 | 7.12 | 0.05 | 19 |
| sig CrB A | S | 5923 | 4.12 | 1.52 | 7.47 | 0.29 | 141 | 7.43 | 0.39 | 8 | 5923 | 4.20 | 1.50 | 7.47 | 0.29 | 141 | 7.47 | 0.39 | 8 |
| sig CrB B | S | 5992 | 4.47 | 1.26 | 7.47 | 0.05 | 357 | 7.36 | 0.05 | 22 | 5992 | 4.70 | 0.89 | 7.50 | 0.05 | 357 | 7.50 | 0.05 | 22 |
| sig Dra | S | 5338 | 4.57 | 0.96 | 7.26 | 0.06 | 356 | 7.28 | 0.09 | 17 | 5338 | 4.53 | 1.08 | 7.25 | 0.06 | 356 | 7.25 | 0.09 | 17 |
| tau Boo | S | 6447 | 4.26 | 2.38 | 7.80 | 0.12 | 309 | 7.74 | 0.10 | 26 | 6447 | 4.38 | 2.31 | 7.80 | 0.12 | 309 | 7.80 | 0.10 | 26 |
| tau Cet | S | 5403 | 4.53 | 0.41 | 7.02 | 0.05 | 344 | 6.99 | 0.06 | 14 | 5403 | 4.60 | 0.10 | 7.02 | 0.05 | 344 | 7.03 | 0.06 | 14 |
| tau01 Eri | S | 6395 | 4.29 | 2.62 | 7.64 | 0.15 | 167 | 7.66 | 0.13 | 16 | 6395 | 4.24 | 2.64 | 7.64 | 0.15 | 167 | 7.64 | 0.13 | 16 |
| tet Boo | S | 6294 | 4.07 | 3.04 | 7.49 | 0.17 | 160 | 7.51 | 0.17 | 10 | 6294 | 4.02 | 3.06 | 7.49 | 0.17 | 160 | 7.49 | 0.17 | 10 |
| tet Per | S | 6310 | 4.32 | 1.64 | 7.55 | 0.07 | 304 | 7.52 | 0.06 | 21 | 6310 | 4.38 | 1.58 | 7.55 | 0.07 | 304 | 7.55 | 0.06 | 21 |
| tet UMa | S | 6371 | 3.80 | 2.02 | 7.35 | 0.09 | 326 | 7.27 | 0.05 | 22 | 6371 | 4.00 | 1.93 | 7.35 | 0.08 | 326 | 7.35 | 0.06 | 22 |
| ups And | S | 6269 | 4.12 | 1.87 | 7.67 | 0.09 | 334 | 7.57 | 0.07 | 25 | 6269 | 4.33 | 1.74 | 7.68 | 0.08 | 334 | 7.68 | 0.07 | 25 |
| w Her | S | 5780 | 4.33 | 1.09 | 7.10 | 0.05 | 348 | 7.06 | 0.05 | 15 | 5780 | 4.42 | 0.95 | 7.11 | 0.05 | 348 | 7.11 | 0.05 | 15 |
| zet Her | S | 5759 | 3.69 | 1.69 | 7.50 | 0.07 | 397 | 7.45 | 0.05 | 20 | 5759 | 3.79 | 1.63 | 7.50 | 0.07 | 397 | 7.50 | 0.05 | 20 |
| BD+04 701A | S | 6056 | 4.50 | 1.45 | 7.48 | 0.18 | 113 | 7.86 | 0.13 | 9 | 6056 | 3.58 | 1.77 | 7.47 | 0.18 | 113 | 7.43 | 0.12 | 9 |
| BD+04 701B | S | 5642 | 4.50 | 0.58 | 7.09 | 0.08 | 128 | 7.70 | 0.01 | 3 | 5642 | 3.50 | 1.03 | 7.04 | 0.08 | 128 | 7.13 | 0.02 | 3 |
| BD+18 2776 | S | 3644 | 4.46 | 0.67 | 7.89 | 0.58 | 176 | 10.90 | 0.96 | 11 | 3644 | 3.46 | 1.33 | 7.31 | 0.60 | 176 | 9.73 | 0.96 | 11 |
| BD+27 4120 | S | 3899 | 4.80 | 0.35 | 7.65 | 0.29 | 194 | 8.37 | 0.00 | 1 | 3899 | 4.80 | 0.30 | 7.66 | 0.29 | 194 | 8.37 | 0.00 | 1 |
| BD+29 2963 | S | 5569 | 4.15 | 1.07 | 7.09 | 0.09 | 332 | 7.23 | 0.06 | 19 | 5569 | 3.84 | 1.32 | 7.06 | 0.10 | 332 | 7.03 | 0.07 | 19 |
| BD+30 2512 | S | 4313 | 4.68 | 0.15 | 7.68 | 0.16 | 308 | 7.63 | 0.17 | 2 | 4313 | 4.87 | 0.10 | 7.70 | 0.17 | 308 | 7.70 | 0.17 | 2 |
| BD+33 529 | S | 3896 | 4.51 | 0.15 | 6.83 | 0.14 | 149 | 7.75 | 0.30 | 2 | 3896 | 3.51 | 1.80 | 6.14 | 0.18 | 149 | 6.34 | 0.34 | 2 |
| BD+61 195 | S | 3799 | 4.84 | 0.15 | 7.93 | 0.27 | 104 | 10.80 | 0.00 | 1 | 3799 | 4.84 | 0.10 | 7.93 | 0.27 | 104 | 10.80 | 0.00 | 1 |
| BD+67 1468A | S | 6680 | 4.13 | 2.89 | 7.80 | 0.16 | 192 | 7.75 | 0.10 | 18 | 6680 | 4.26 | 2.94 | 7.79 | 0.15 | 192 | 7.77 | 0.10 | 18 |
| BD+67 1468B | S | 6642 | 4.15 | 2.78 | 7.77 | 0.14 | 249 | 7.69 | 0.09 | 19 | 6642 | 4.34 | 2.70 | 7.77 | 0.14 | 249 | 7.77 | 0.09 | 19 |
| BD-04 782 | S | 4342 | 4.68 | 0.15 | 7.72 | 0.15 | 276 | 7.64 | 0.00 | 1 | 4342 | 4.68 | 0.10 | 7.72 | 0.15 | 276 | 7.64 | 0.00 | 1 |
| BD-10 3166 | S | 5314 | 4.66 | 0.89 | 7.90 | 0.09 | 232 | 8.16 | 0.14 | 19 | 5314 | 4.05 | 1.24 | 7.78 | 0.09 | 232 | 7.72 | 0.16 | 19 |
| CCDM J14534+1542AB | S | 6135 | 4.11 | 1.82 | 7.72 | 0.10 | 341 | 7.65 | 0.05 | 21 | 6135 | 4.25 | 1.73 | 7.72 | 0.10 | 341 | 7.72 | 0.05 | 21 |
| GJ 282 C | S | 3866 | 4.76 | 0.95 | 7.84 | 0.42 | 126 | 8.51 | 0.00 | 1 | 3866 | 4.76 | 0.95 | 7.84 | 0.42 | 126 | 8.51 | 0.00 | 1 |
| GJ 528 A | S | 4471 | 4.42 | 0.53 | 7.54 | 0.11 | 197 | 7.72 | 0.08 | 5 | 4471 | 4.03 | 1.21 | 7.42 | 0.10 | 197 | 7.44 | 0.08 | 5 |

| Star | | Teff | logg | vt | AFe | sigma | N | AFe2 | sigma2 | N2 | Teff | logg | vt | AFe | sigma | N | AFe2 | sigma2 | N2 |
|---|---|---|---|---|---|---|---|---|---|---|---|---|---|---|---|---|---|---|---|
| GJ 528 B | S | 4384 | 4.59 | 0.15 | 7.66 | 0.13 | 352 | 7.67 | 0.09 | 2 | 4384 | 4.54 | 0.10 | 7.65 | 0.13 | 352 | 7.66 | 0.09 | 2 |
| HD 101177 | S | 5964 | 4.43 | 1.10 | 7.30 | 0.06 | 353 | 7.28 | 0.05 | 18 | 5964 | 4.47 | 1.04 | 7.31 | 0.06 | 353 | 7.31 | 0.05 | 18 |
| HD 101563 | S | 5868 | 3.89 | 1.57 | 7.43 | 0.07 | 378 | 7.38 | 0.06 | 21 | 5868 | 4.01 | 1.49 | 7.44 | 0.07 | 378 | 7.44 | 0.06 | 21 |
| HD 101959 | S | 6095 | 4.42 | 1.37 | 7.38 | 0.06 | 329 | 7.36 | 0.05 | 23 | 6095 | 4.46 | 1.33 | 7.38 | 0.06 | 329 | 7.38 | 0.05 | 23 |
| HD 102158 | S | 5781 | 4.31 | 0.86 | 7.05 | 0.06 | 316 | 7.02 | 0.03 | 15 | 5781 | 4.37 | 0.73 | 7.06 | 0.06 | 316 | 7.06 | 0.03 | 15 |
| HD 10307 | S | 5976 | 4.36 | 1.20 | 7.56 | 0.05 | 341 | 7.46 | 0.05 | 20 | 5976 | 4.54 | 0.88 | 7.58 | 0.06 | 341 | 7.58 | 0.05 | 20 |
| HD 103095 | S | 5178 | 4.72 | 1.30 | 6.22 | 0.07 | 281 | 6.28 | 0.10 | 12 | 5178 | 4.59 | 1.52 | 6.20 | 0.07 | 281 | 6.20 | 0.10 | 12 |
| HD 103932 | S | 4585 | 4.58 | 0.60 | 7.77 | 0.14 | 215 | 8.46 | 0.44 | 18 | 4585 | 3.58 | 1.38 | 7.39 | 0.14 | 215 | 7.65 | 0.47 | 18 |
| HD 104304 | S | 5555 | 4.42 | 0.97 | 7.77 | 0.07 | 274 | 7.84 | 0.11 | 23 | 5555 | 4.26 | 1.07 | 7.75 | 0.07 | 274 | 7.75 | 0.12 | 23 |
| HD 10436 | S | 4393 | 4.72 | 0.15 | 7.33 | 0.13 | 271 | 7.32 | 0.09 | 3 | 4393 | 4.76 | 0.10 | 7.27 | 0.14 | 271 | 7.25 | 0.08 | 3 |
| HD 106252 | S | 5890 | 4.37 | 1.17 | 7.39 | 0.05 | 331 | 7.36 | 0.05 | 18 | 5890 | 4.43 | 1.09 | 7.39 | 0.05 | 331 | 7.39 | 0.05 | 18 |
| HD 106640 | S | 6003 | 4.30 | 1.34 | 7.26 | 0.06 | 340 | 7.22 | 0.06 | 21 | 6003 | 4.40 | 1.23 | 7.26 | 0.06 | 340 | 7.27 | 0.06 | 21 |
| HD 108799 | S | 5878 | 4.33 | 1.46 | 7.31 | 0.08 | 325 | 7.38 | 0.06 | 19 | 5878 | 4.17 | 1.62 | 7.30 | 0.08 | 325 | 7.30 | 0.06 | 19 |
| HD 108874 | S | 5555 | 4.32 | 1.26 | 7.65 | 0.07 | 365 | 7.66 | 0.08 | 16 | 5555 | 4.31 | 1.27 | 7.65 | 0.07 | 365 | 7.65 | 0.08 | 16 |
| HD 108954 | S | 6024 | 4.43 | 1.24 | 7.39 | 0.05 | 332 | 7.42 | 0.04 | 22 | 6024 | 4.35 | 1.32 | 7.38 | 0.05 | 332 | 7.38 | 0.04 | 22 |
| HD 109057 | S | 6060 | 4.32 | 2.05 | 7.51 | 0.12 | 303 | 7.50 | 0.11 | 22 | 6060 | 4.36 | 2.02 | 7.52 | 0.12 | 303 | 7.52 | 0.10 | 22 |
| HD 11007 | S | 5994 | 4.01 | 1.47 | 7.27 | 0.06 | 342 | 7.22 | 0.05 | 24 | 5994 | 4.13 | 1.38 | 7.28 | 0.05 | 342 | 7.28 | 0.05 | 24 |
| HD 110315 | S | 4505 | 4.61 | 0.15 | 7.38 | 0.13 | 335 | 7.59 | 0.23 | 10 | 4505 | 4.09 | 1.60 | 7.19 | 0.15 | 335 | 7.19 | 0.21 | 10 |
| HD 11038 | S | 6044 | 4.36 | 1.12 | 7.21 | 0.05 | 131 | 7.13 | 0.03 | 9 | 6044 | 4.52 | 0.94 | 7.22 | 0.05 | 131 | 7.22 | 0.03 | 9 |
| HD 110745 | S | 6127 | 4.22 | 1.59 | 7.52 | 0.07 | 355 | 7.39 | 0.07 | 20 | 6127 | 4.49 | 1.33 | 7.53 | 0.07 | 355 | 7.54 | 0.07 | 20 |
| HD 110833 | S | 4972 | 4.56 | 1.36 | 7.61 | 0.07 | 191 | 7.86 | 0.16 | 13 | 4972 | 3.82 | 1.82 | 7.42 | 0.07 | 191 | 7.32 | 0.14 | 13 |
| HD 110869 | S | 5783 | 4.41 | 1.15 | 7.58 | 0.06 | 350 | 7.56 | 0.06 | 20 | 5783 | 4.44 | 1.10 | 7.59 | 0.06 | 350 | 7.58 | 0.06 | 20 |
| HD 111513 | S | 5822 | 4.33 | 1.14 | 7.58 | 0.06 | 371 | 7.58 | 0.06 | 20 | 5822 | 4.33 | 1.14 | 7.58 | 0.06 | 371 | 7.58 | 0.06 | 20 |
| HD 111540 | S | 5729 | 4.37 | 1.25 | 7.57 | 0.07 | 387 | 7.79 | 0.06 | 28 | 5729 | 3.83 | 1.68 | 7.52 | 0.09 | 387 | 7.43 | 0.10 | 28 |
| HD 111799 | S | 5842 | 4.38 | 1.33 | 7.59 | 0.07 | 356 | 7.71 | 0.05 | 23 | 5842 | 4.09 | 1.54 | 7.56 | 0.08 | 356 | 7.51 | 0.05 | 23 |
| HD 112068 | S | 5745 | 4.26 | 1.09 | 6.96 | 0.06 | 303 | 7.00 | 0.04 | 16 | 5745 | 4.17 | 1.19 | 6.95 | 0.06 | 303 | 6.95 | 0.04 | 16 |
| HD 112257 | S | 5659 | 4.31 | 1.10 | 7.42 | 0.06 | 371 | 7.44 | 0.04 | 17 | 5659 | 4.26 | 1.17 | 7.41 | 0.06 | 371 | 7.41 | 0.04 | 17 |
| HD 112758 | S | 5190 | 4.58 | 0.49 | 7.03 | 0.08 | 271 | 7.08 | 0.16 | 14 | 5190 | 4.45 | 0.70 | 7.01 | 0.08 | 271 | 7.01 | 0.15 | 14 |
| HD 113470 | S | 5839 | 4.27 | 1.32 | 7.26 | 0.07 | 332 | 7.26 | 0.05 | 20 | 5839 | 4.27 | 1.32 | 7.26 | 0.07 | 332 | 7.26 | 0.05 | 20 |
| HD 113713 | S | 6316 | 3.99 | 1.86 | 7.09 | 0.08 | 290 | 7.07 | 0.09 | 21 | 6316 | 4.05 | 1.86 | 7.08 | 0.08 | 290 | 7.08 | 0.09 | 21 |
| HD 114762 | S | 5921 | 4.27 | 1.26 | 6.74 | 0.07 | 295 | 6.75 | 0.03 | 16 | 5921 | 4.24 | 1.29 | 6.74 | 0.07 | 295 | 6.74 | 0.03 | 16 |
| HD 114783 | S | 5118 | 4.55 | 1.01 | 7.58 | 0.06 | 205 | 7.63 | 0.04 | 7 | 5118 | 4.45 | 1.00 | 7.53 | 0.06 | 205 | 7.48 | 0.04 | 7 |
| HD 11507 | S | 4011 | 4.72 | 0.65 | 7.88 | 0.28 | 216 | 8.03 | 0.00 | 1 | 4011 | 4.72 | 0.65 | 7.88 | 0.28 | 216 | 8.03 | 0.00 | 1 |
| HD 115404A | S | 5030 | 4.60 | 0.96 | 7.29 | 0.06 | 238 | 7.41 | 0.13 | 11 | 5030 | 4.36 | 1.27 | 7.25 | 0.06 | 238 | 7.26 | 0.13 | 11 |
| HD 115404B | S | 3903 | 4.90 | 0.35 | 7.57 | 0.28 | 147 | 8.52 | 0.00 | 1 | 3903 | 4.90 | 0.30 | 7.58 | 0.28 | 147 | 8.53 | 0.00 | 1 |
| HD 115953 | S | 3751 | 4.36 | 1.15 | 7.66 | 0.33 | 60 | 9.55 | 0.00 | 1 | 3751 | 4.36 | 1.15 | 7.66 | 0.33 | 60 | 9.55 | 0.00 | 1 |
| HD 117043 | S | 5558 | 4.36 | 1.20 | 7.63 | 0.06 | 361 | 7.71 | 0.09 | 19 | 5558 | 4.19 | 1.26 | 7.60 | 0.06 | 361 | 7.56 | 0.10 | 19 |
| HD 117845 | S | 5856 | 4.46 | 1.07 | 7.18 | 0.07 | 327 | 7.27 | 0.04 | 19 | 5856 | 4.29 | 1.29 | 7.17 | 0.07 | 327 | 7.17 | 0.04 | 19 |

| Star | | Teff | log g | vt | [Fe/H]₁ | σ₁ | N₁ | [Fe/H]₂ | σ₂ | N | Teff | log g | vt | [Fe/H]₁ | σ₁ | N₁ | [Fe/H]₂ | σ₂ | N |
|---|---|---|---|---|---|---|---|---|---|---|---|---|---|---|---|---|---|---|---|
| HD 118203 | S | 5768 | 3.92 | 1.68 | 7.62 | 0.08 | 353 | 7.63 | 0.07 | 23 | 5768 | 3.90 | 1.69 | 7.62 | 0.08 | 353 | 7.62 | 0.07 | 23 |
| HD 120066 | S | 5903 | 4.11 | 1.43 | 7.54 | 0.06 | 347 | 7.47 | 0.05 | 20 | 5903 | 4.27 | 1.28 | 7.55 | 0.06 | 347 | 7.56 | 0.06 | 20 |
| HD 120467 | S | 4369 | 4.60 | 0.35 | 7.92 | 0.14 | 281 | 7.98 | 0.02 | 2 | 4369 | 4.26 | 1.75 | 7.71 | 0.16 | 281 | 7.68 | 0.05 | 2 |
| HD 120690 | S | 5550 | 4.39 | 0.84 | 7.42 | 0.07 | 265 | 7.56 | 0.07 | 21 | 5550 | 4.07 | 1.03 | 7.40 | 0.07 | 265 | 7.40 | 0.07 | 21 |
| HD 120730 | S | 5300 | 4.49 | 1.16 | 7.46 | 0.10 | 351 | 7.51 | 0.09 | 16 | 5300 | 4.36 | 1.39 | 7.42 | 0.10 | 351 | 7.42 | 0.09 | 16 |
| HD 122064 | S | 4811 | 4.57 | 0.15 | 7.79 | 0.13 | 421 | 7.86 | 0.19 | 17 | 4811 | 4.43 | 1.10 | 7.69 | 0.12 | 421 | 7.70 | 0.18 | 17 |
| HD 122303 | S | 4067 | 5.00 | 0.25 | 7.47 | 0.35 | 136 | 9.11 | 0.00 | 1 | 4067 | 5.00 | 0.10 | 7.48 | 0.35 | 136 | 9.12 | 0.00 | 1 |
| HD 122742 | S | 5459 | 4.39 | 0.92 | 7.44 | 0.06 | 366 | 7.51 | 0.05 | 15 | 5459 | 4.27 | 1.17 | 7.41 | 0.07 | 366 | 7.43 | 0.06 | 15 |
| HD 122967 | S | 7008 | 4.16 | 4.32 | 8.10 | 0.39 | 36 | 8.13 | 0.20 | 3 | 7008 | 4.07 | 4.32 | 8.10 | 0.39 | 36 | 8.10 | 0.20 | 3 |
| HD 123A | S | 5823 | 4.43 | 1.26 | 7.56 | 0.06 | 356 | 7.56 | 0.08 | 23 | 5823 | 4.42 | 1.27 | 7.56 | 0.06 | 356 | 7.56 | 0.08 | 23 |
| HD 123B | S | 5230 | 4.47 | 0.23 | 7.56 | 0.07 | 347 | 7.74 | 0.08 | 17 | 5230 | 4.13 | 1.27 | 7.44 | 0.08 | 347 | 7.43 | 0.09 | 17 |
| HD 124553 | S | 6060 | 4.00 | 1.65 | 7.65 | 0.07 | 358 | 7.59 | 0.07 | 27 | 6060 | 4.13 | 1.57 | 7.65 | 0.06 | 358 | 7.65 | 0.08 | 27 |
| HD 125040 | S | 6357 | 4.28 | 3.24 | 7.48 | 0.14 | 91 | 7.50 | 0.18 | 7 | 6357 | 4.23 | 3.27 | 7.48 | 0.14 | 91 | 7.48 | 0.18 | 7 |
| HD 125184 | S | 5637 | 4.06 | 1.37 | 7.72 | 0.07 | 344 | 7.75 | 0.08 | 27 | 5637 | 4.00 | 1.42 | 7.71 | 0.07 | 344 | 7.71 | 0.09 | 27 |
| HD 126053 | S | 5727 | 4.46 | 0.94 | 7.14 | 0.05 | 330 | 7.11 | 0.03 | 15 | 5727 | 4.53 | 0.78 | 7.15 | 0.05 | 330 | 7.15 | 0.03 | 15 |
| HD 12661 | S | 5651 | 4.35 | 1.06 | 7.81 | 0.07 | 248 | 7.92 | 0.11 | 24 | 5651 | 4.11 | 1.17 | 7.80 | 0.07 | 248 | 7.79 | 0.12 | 24 |
| HD 127334 | S | 5643 | 4.25 | 1.32 | 7.64 | 0.06 | 369 | 7.67 | 0.06 | 22 | 5643 | 4.18 | 1.39 | 7.63 | 0.07 | 369 | 7.63 | 0.06 | 22 |
| HD 128165 | S | 4793 | 4.59 | 0.76 | 7.47 | 0.09 | 256 | 8.09 | 0.43 | 15 | 4793 | 3.63 | 1.38 | 7.16 | 0.10 | 256 | 7.31 | 0.43 | 15 |
| HD 128311 | S | 4903 | 4.58 | 0.77 | 7.58 | 0.06 | 176 | 7.73 | 0.14 | 12 | 4903 | 4.27 | 1.15 | 7.51 | 0.06 | 176 | 7.53 | 0.14 | 12 |
| HD 129132 | S | 6609 | 3.41 | 2.80 | 7.31 | 0.17 | 189 | 7.23 | 0.15 | 18 | 6609 | 3.63 | 2.77 | 7.31 | 0.17 | 189 | 7.31 | 0.15 | 18 |
| HD 129290 | S | 5837 | 4.19 | 1.26 | 7.34 | 0.07 | 344 | 7.34 | 0.05 | 21 | 5837 | 4.20 | 1.18 | 7.34 | 0.07 | 344 | 7.32 | 0.04 | 21 |
| HD 129829 | S | 6085 | 4.41 | 1.37 | 7.24 | 0.06 | 318 | 7.27 | 0.08 | 23 | 6085 | 4.33 | 1.46 | 7.24 | 0.06 | 318 | 7.23 | 0.09 | 23 |
| HD 130322 | S | 5422 | 4.54 | 0.88 | 7.51 | 0.06 | 365 | 7.55 | 0.11 | 19 | 5422 | 4.47 | 1.06 | 7.49 | 0.06 | 365 | 7.50 | 0.11 | 19 |
| HD 13043 | S | 5877 | 4.15 | 1.33 | 7.54 | 0.06 | 357 | 7.51 | 0.07 | 24 | 5877 | 4.20 | 1.27 | 7.55 | 0.05 | 357 | 7.55 | 0.07 | 24 |
| HD 130948 | S | 5983 | 4.43 | 1.50 | 7.48 | 0.07 | 353 | 7.50 | 0.06 | 20 | 5983 | 4.38 | 1.56 | 7.47 | 0.07 | 353 | 7.47 | 0.06 | 20 |
| HD 131976 | S | 4077 | 5.00 | 0.15 | 7.42 | 0.27 | 120 | 7.98 | 0.00 | 1 | 4077 | 5.00 | 0.10 | 7.42 | 0.27 | 120 | 7.98 | 0.00 | 1 |
| HD 131977 | S | 4625 | 4.59 | 1.31 | 7.59 | 0.09 | 164 | 7.91 | 0.20 | 8 | 4625 | 3.83 | 1.77 | 7.30 | 0.09 | 164 | 7.25 | 0.18 | 8 |
| HD 132375 | S | 6336 | 4.18 | 1.90 | 7.56 | 0.07 | 329 | 7.50 | 0.06 | 22 | 6336 | 4.32 | 1.81 | 7.57 | 0.06 | 329 | 7.57 | 0.06 | 22 |
| HD 133161 | S | 5946 | 4.31 | 1.51 | 7.63 | 0.07 | 361 | 7.70 | 0.06 | 23 | 5946 | 4.16 | 1.63 | 7.62 | 0.08 | 361 | 7.61 | 0.06 | 23 |
| HD 135101 | S | 5637 | 4.24 | 0.87 | 7.54 | 0.05 | 415 | 7.60 | 0.06 | 21 | 5637 | 4.16 | 1.02 | 7.52 | 0.05 | 415 | 7.54 | 0.05 | 21 |
| HD 135101B | S | 5529 | 4.07 | 1.27 | 7.49 | 0.06 | 410 | 7.35 | 0.08 | 20 | 5529 | 4.38 | 0.72 | 7.55 | 0.06 | 410 | 7.56 | 0.07 | 20 |
| HD 135145 | S | 5852 | 4.05 | 1.47 | 7.34 | 0.07 | 352 | 7.31 | 0.07 | 21 | 5852 | 4.11 | 1.39 | 7.34 | 0.06 | 352 | 7.31 | 0.06 | 21 |
| HD 136064 | S | 6144 | 3.98 | 1.77 | 7.48 | 0.06 | 337 | 7.44 | 0.05 | 23 | 6144 | 4.06 | 1.72 | 7.48 | 0.06 | 337 | 7.48 | 0.04 | 23 |
| HD 136118 | S | 6143 | 4.08 | 1.66 | 7.41 | 0.07 | 329 | 7.40 | 0.06 | 21 | 6143 | 4.09 | 1.65 | 7.41 | 0.07 | 329 | 7.41 | 0.06 | 21 |
| HD 136231 | S | 5881 | 4.17 | 1.27 | 7.16 | 0.08 | 180 | 7.19 | 0.05 | 11 | 5881 | 4.12 | 1.30 | 7.16 | 0.08 | 180 | 7.16 | 0.05 | 11 |
| HD 1388 | S | 5884 | 4.31 | 1.20 | 7.44 | 0.06 | 365 | 7.44 | 0.06 | 19 | 5884 | 4.30 | 1.21 | 7.44 | 0.06 | 365 | 7.44 | 0.06 | 19 |
| HD 140913 | S | 5946 | 4.46 | 1.70 | 7.52 | 0.08 | 331 | 7.53 | 0.07 | 22 | 5946 | 4.42 | 1.74 | 7.51 | 0.08 | 331 | 7.51 | 0.08 | 22 |
| HD 141715 | S | 5836 | 4.25 | 2.04 | 7.37 | 0.10 | 329 | 7.30 | 0.06 | 15 | 5836 | 4.42 | 1.89 | 7.38 | 0.09 | 329 | 7.39 | 0.05 | 15 |

| Star | | Teff | log g | vt | [X/H] | σ | N | [X/H] | σ | N | Teff | log g | vt | [X/H] | σ | N | [X/H] | σ | N |
|---|---|---|---|---|---|---|---|---|---|---|---|---|---|---|---|---|---|---|---|
| HD 141937 | S | 5885 | 4.44 | 1.26 | 7.55 | 0.06 | 384 | 7.57 | 0.04 | 20 | 5885 | 4.41 | 1.30 | 7.55 | 0.06 | 384 | 7.55 | 0.05 | 20 |
| HD 14412 | S | 5492 | 4.62 | 0.48 | 7.11 | 0.05 | 262 | 7.06 | 0.07 | 12 | 5492 | 4.73 | 0.10 | 7.12 | 0.06 | 262 | 7.12 | 0.08 | 12 |
| HD 144579 | S | 5308 | 4.67 | 0.15 | 6.84 | 0.08 | 427 | 6.90 | 0.10 | 20 | 5308 | 4.54 | 1.08 | 6.81 | 0.07 | 427 | 6.83 | 0.09 | 20 |
| HD 144585 | S | 5850 | 4.22 | 1.41 | 7.75 | 0.06 | 366 | 7.73 | 0.06 | 26 | 5850 | 4.26 | 1.37 | 7.76 | 0.06 | 366 | 7.76 | 0.06 | 26 |
| HD 145148 | S | 4923 | 3.68 | 1.30 | 7.62 | 0.10 | 350 | 7.74 | 0.13 | 20 | 4923 | 3.29 | 1.51 | 7.49 | 0.11 | 350 | 7.41 | 0.15 | 20 |
| HD 1461 | S | 5765 | 4.38 | 1.26 | 7.65 | 0.06 | 370 | 7.63 | 0.06 | 22 | 5765 | 4.41 | 1.22 | 7.65 | 0.06 | 370 | 7.65 | 0.07 | 22 |
| HD 14624 | S | 5599 | 4.07 | 1.31 | 7.58 | 0.08 | 356 | 7.62 | 0.06 | 20 | 5599 | 3.99 | 1.26 | 7.57 | 0.07 | 356 | 7.53 | 0.07 | 20 |
| HD 147681 | S | 6139 | 4.44 | 1.97 | 7.58 | 0.12 | 291 | 7.58 | 0.08 | 17 | 6139 | 4.44 | 1.94 | 7.58 | 0.11 | 291 | 7.54 | 0.08 | 17 |
| HD 149026 | S | 6003 | 4.13 | 1.73 | 7.70 | 0.07 | 346 | 7.73 | 0.07 | 24 | 6003 | 4.08 | 1.76 | 7.70 | 0.07 | 346 | 7.70 | 0.07 | 24 |
| HD 149143 | S | 5768 | 4.07 | 1.50 | 7.67 | 0.07 | 355 | 7.77 | 0.07 | 26 | 5768 | 3.85 | 1.64 | 7.66 | 0.07 | 355 | 7.65 | 0.08 | 26 |
| HD 15069 | S | 5722 | 4.11 | 1.22 | 7.48 | 0.06 | 351 | 7.52 | 0.05 | 21 | 5722 | 4.02 | 1.32 | 7.47 | 0.07 | 351 | 7.47 | 0.05 | 21 |
| HD 150706 | S | 5900 | 4.46 | 1.34 | 7.39 | 0.06 | 346 | 7.41 | 0.07 | 21 | 5900 | 4.43 | 1.37 | 7.39 | 0.06 | 346 | 7.39 | 0.07 | 21 |
| HD 150933 | S | 6087 | 4.28 | 1.37 | 7.59 | 0.06 | 358 | 7.62 | 0.06 | 24 | 6087 | 4.22 | 1.38 | 7.59 | 0.06 | 358 | 7.55 | 0.06 | 24 |
| HD 151288 | S | 4181 | 4.70 | 0.15 | 7.85 | 0.21 | 274 | 7.84 | 0.03 | 2 | 4181 | 4.71 | 0.10 | 7.85 | 0.21 | 274 | 7.85 | 0.03 | 2 |
| HD 151426 | S | 5711 | 4.28 | 1.07 | 7.37 | 0.06 | 347 | 7.40 | 0.06 | 19 | 5711 | 4.22 | 1.15 | 7.36 | 0.06 | 347 | 7.36 | 0.06 | 19 |
| HD 151450 | S | 6108 | 4.38 | 1.31 | 7.45 | 0.06 | 339 | 7.43 | 0.07 | 20 | 6108 | 4.43 | 1.26 | 7.45 | 0.06 | 339 | 7.45 | 0.07 | 20 |
| HD 15189 | S | 6051 | 4.47 | 0.93 | 7.48 | 0.08 | 202 | 7.57 | 0.06 | 12 | 6051 | 4.30 | 1.17 | 7.46 | 0.08 | 202 | 7.47 | 0.06 | 12 |
| HD 152391 | S | 5431 | 4.51 | 1.32 | 7.42 | 0.05 | 356 | 7.53 | 0.12 | 20 | 5431 | 4.26 | 1.54 | 7.37 | 0.06 | 356 | 7.34 | 0.12 | 20 |
| HD 154160 | S | 5380 | 3.90 | 1.36 | 7.71 | 0.08 | 356 | 7.91 | 0.11 | 25 | 5380 | 3.41 | 1.62 | 7.66 | 0.10 | 356 | 7.63 | 0.11 | 25 |
| HD 154363 | S | 4373 | 4.66 | 0.15 | 7.15 | 0.15 | 286 | 7.45 | 0.24 | 6 | 4373 | 3.66 | 2.18 | 6.83 | 0.20 | 286 | 6.71 | 0.23 | 6 |
| HD 154578 | S | 6294 | 4.14 | 1.91 | 7.26 | 0.10 | 277 | 7.23 | 0.06 | 17 | 6294 | 4.21 | 1.86 | 7.26 | 0.10 | 277 | 7.26 | 0.07 | 17 |
| HD 156062 | S | 5995 | 4.37 | 1.22 | 7.45 | 0.06 | 341 | 7.47 | 0.07 | 21 | 5995 | 4.33 | 1.27 | 7.45 | 0.06 | 341 | 7.45 | 0.07 | 21 |
| HD 156826 | S | 5155 | 3.52 | 1.19 | 7.22 | 0.06 | 370 | 7.22 | 0.07 | 20 | 5155 | 3.53 | 1.18 | 7.22 | 0.06 | 370 | 7.22 | 0.07 | 20 |
| HD 156846 | S | 6069 | 3.94 | 1.74 | 7.64 | 0.07 | 351 | 7.59 | 0.06 | 26 | 6069 | 4.05 | 1.68 | 7.65 | 0.07 | 351 | 7.65 | 0.06 | 26 |
| HD 156968 | S | 5948 | 4.37 | 1.18 | 7.38 | 0.05 | 338 | 7.39 | 0.05 | 17 | 5948 | 4.35 | 1.20 | 7.38 | 0.05 | 338 | 7.38 | 0.05 | 17 |
| HD 157881 | S | 4161 | 4.67 | 0.15 | 7.72 | 0.17 | 239 | 7.76 | 0.03 | 2 | 4161 | 4.56 | 0.10 | 7.71 | 0.17 | 239 | 7.71 | 0.03 | 2 |
| HD 158633 | S | 5313 | 4.57 | 0.55 | 7.06 | 0.06 | 317 | 7.05 | 0.06 | 14 | 5313 | 4.59 | 0.38 | 7.07 | 0.06 | 317 | 7.07 | 0.06 | 14 |
| HD 159222 | S | 5788 | 4.39 | 1.12 | 7.58 | 0.05 | 356 | 7.62 | 0.05 | 22 | 5788 | 4.31 | 1.25 | 7.56 | 0.06 | 356 | 7.57 | 0.05 | 22 |
| HD 160346 | S | 4864 | 4.52 | 0.29 | 7.50 | 0.10 | 376 | 7.58 | 0.13 | 18 | 4864 | 4.35 | 1.10 | 7.42 | 0.09 | 376 | 7.42 | 0.12 | 18 |
| HD 160933 | S | 5829 | 3.81 | 1.55 | 7.16 | 0.06 | 331 | 7.14 | 0.04 | 18 | 5829 | 3.85 | 1.53 | 7.16 | 0.06 | 331 | 7.16 | 0.04 | 18 |
| HD 16160 | S | 4886 | 4.63 | 0.15 | 7.43 | 0.09 | 322 | 7.37 | 0.19 | 15 | 4886 | 4.77 | 0.10 | 7.43 | 0.10 | 322 | 7.43 | 0.18 | 15 |
| HD 163492 | S | 5834 | 4.37 | 1.27 | 7.57 | 0.07 | 356 | 7.56 | 0.05 | 20 | 5834 | 4.39 | 1.24 | 7.57 | 0.07 | 356 | 7.57 | 0.05 | 20 |
| HD 163840 | S | 5760 | 4.10 | 1.26 | 7.48 | 0.08 | 355 | 7.50 | 0.06 | 20 | 5760 | 4.07 | 1.29 | 7.48 | 0.08 | 355 | 7.48 | 0.06 | 20 |
| HD 164595 | S | 5728 | 4.42 | 1.00 | 7.38 | 0.06 | 356 | 7.42 | 0.04 | 17 | 5728 | 4.35 | 1.11 | 7.37 | 0.06 | 356 | 7.37 | 0.05 | 17 |
| HD 166 | S | 5481 | 4.52 | 1.27 | 7.55 | 0.06 | 373 | 7.61 | 0.09 | 20 | 5481 | 4.40 | 1.45 | 7.53 | 0.07 | 373 | 7.53 | 0.09 | 20 |
| HD 166601 | S | 6289 | 4.01 | 1.89 | 7.39 | 0.08 | 312 | 7.38 | 0.06 | 20 | 6289 | 4.04 | 1.87 | 7.40 | 0.08 | 312 | 7.40 | 0.06 | 20 |
| HD 167588 | S | 5909 | 3.91 | 1.58 | 7.10 | 0.06 | 311 | 7.09 | 0.03 | 18 | 5909 | 3.93 | 1.57 | 7.10 | 0.06 | 311 | 7.10 | 0.03 | 18 |
| HD 167665 | S | 6179 | 4.25 | 1.55 | 7.33 | 0.06 | 331 | 7.33 | 0.07 | 20 | 6179 | 4.26 | 1.55 | 7.33 | 0.06 | 331 | 7.31 | 0.07 | 20 |

| | | | | | | | | | | | | | | | | |
|---|---|---|---|---|---|---|---|---|---|---|---|---|---|---|---|---|
| HD 168443 | S | 5565 | 4.04 | 1.19 | 7.51 | 0.06 | 370 | 7.52 | 0.07 | 24 | 5565 | 4.01 | 1.22 | 7.50 | 0.07 | 370 | 7.50 | 0.07 | 24 |
| HD 168746 | S | 5576 | 4.33 | 0.96 | 7.39 | 0.06 | 361 | 7.41 | 0.07 | 22 | 5576 | 4.29 | 1.03 | 7.38 | 0.06 | 361 | 7.38 | 0.07 | 22 |
| HD 170657 | S | 5133 | 4.59 | 0.89 | 7.34 | 0.08 | 253 | 7.46 | 0.20 | 16 | 5133 | 4.32 | 1.05 | 7.28 | 0.07 | 253 | 7.25 | 0.20 | 16 |
| HD 171706 | S | 5935 | 4.09 | 1.35 | 7.40 | 0.06 | 344 | 7.38 | 0.04 | 21 | 5935 | 4.13 | 1.32 | 7.40 | 0.06 | 344 | 7.40 | 0.05 | 21 |
| HD 172051 | S | 5669 | 4.49 | 0.95 | 7.24 | 0.06 | 355 | 7.21 | 0.06 | 19 | 5669 | 4.57 | 0.75 | 7.26 | 0.06 | 355 | 7.26 | 0.06 | 19 |
| HD 173818 | S | 4245 | 4.68 | 0.15 | 7.43 | 0.20 | 311 | 7.38 | 0.06 | 2 | 4245 | 4.79 | 0.10 | 7.45 | 0.21 | 311 | 7.45 | 0.06 | 2 |
| HD 175225 | S | 5281 | 3.76 | 1.46 | 7.63 | 0.09 | 362 | 7.72 | 0.13 | 24 | 5281 | 3.55 | 1.62 | 7.59 | 0.10 | 362 | 7.59 | 0.13 | 24 |
| HD 175290 | S | 6359 | 4.02 | 1.84 | 7.17 | 0.11 | 275 | 7.19 | 0.05 | 21 | 6359 | 3.95 | 1.88 | 7.16 | 0.12 | 275 | 7.16 | 0.05 | 21 |
| HD 176051 | S | 5980 | 4.34 | 1.44 | 7.39 | 0.06 | 321 | 7.29 | 0.05 | 17 | 5980 | 4.54 | 1.21 | 7.41 | 0.06 | 321 | 7.41 | 0.03 | 17 |
| HD 176367 | S | 6087 | 4.47 | 2.58 | 7.58 | 0.15 | 218 | 7.56 | 0.05 | 12 | 6087 | 4.52 | 2.53 | 7.58 | 0.15 | 218 | 7.54 | 0.06 | 12 |
| HD 177830 | S | 4813 | 3.49 | 1.30 | 7.83 | 0.11 | 276 | 8.01 | 0.10 | 14 | 4813 | 3.00 | 1.55 | 7.75 | 0.12 | 276 | 7.76 | 0.09 | 14 |
| HD 178428 | S | 5646 | 4.21 | 1.16 | 7.58 | 0.06 | 356 | 7.60 | 0.07 | 21 | 5646 | 4.18 | 1.20 | 7.58 | 0.06 | 356 | 7.58 | 0.07 | 21 |
| HD 178911 | S | 5825 | 3.86 | 1.44 | 7.47 | 0.09 | 380 | 7.40 | 0.09 | 23 | 5825 | 4.01 | 1.32 | 7.48 | 0.09 | 380 | 7.48 | 0.09 | 23 |
| HD 178911B | S | 5602 | 4.37 | 1.29 | 7.69 | 0.07 | 364 | 7.72 | 0.12 | 23 | 5602 | 4.31 | 1.36 | 7.68 | 0.07 | 364 | 7.68 | 0.13 | 23 |
| HD 179957 | S | 5771 | 4.38 | 1.09 | 7.49 | 0.05 | 386 | 7.46 | 0.05 | 24 | 5771 | 4.44 | 0.98 | 7.50 | 0.05 | 386 | 7.50 | 0.05 | 24 |
| HD 179958 | S | 5807 | 4.33 | 1.13 | 7.53 | 0.05 | 385 | 7.49 | 0.04 | 26 | 5807 | 4.41 | 0.99 | 7.55 | 0.05 | 385 | 7.55 | 0.05 | 26 |
| HD 181655 | S | 5673 | 4.17 | 1.43 | 7.48 | 0.06 | 348 | 7.39 | 0.08 | 20 | 5673 | 4.37 | 1.21 | 7.50 | 0.06 | 348 | 7.50 | 0.07 | 20 |
| HD 182488 | S | 5362 | 4.45 | 0.88 | 7.64 | 0.06 | 242 | 7.76 | 0.11 | 17 | 5362 | 4.19 | 1.04 | 7.61 | 0.07 | 242 | 7.61 | 0.11 | 17 |
| HD 183263 | S | 5911 | 4.28 | 1.46 | 7.72 | 0.07 | 362 | 7.71 | 0.07 | 22 | 5911 | 4.30 | 1.44 | 7.72 | 0.07 | 362 | 7.72 | 0.07 | 22 |
| HD 184151 | S | 6430 | 3.77 | 2.21 | 7.21 | 0.13 | 260 | 7.15 | 0.12 | 22 | 6430 | 3.93 | 2.15 | 7.21 | 0.13 | 260 | 7.21 | 0.12 | 22 |
| HD 184152 | S | 5578 | 4.37 | 0.97 | 7.22 | 0.08 | 345 | 7.21 | 0.12 | 23 | 5578 | 4.41 | 0.88 | 7.23 | 0.08 | 345 | 7.23 | 0.11 | 23 |
| HD 18445 | S | 4874 | 4.36 | 1.13 | 7.36 | 0.07 | 185 | 7.53 | 0.19 | 10 | 4874 | 3.87 | 1.44 | 7.22 | 0.08 | 185 | 7.17 | 0.17 | 10 |
| HD 184489 | S | 4133 | 4.74 | 0.15 | 7.31 | 0.20 | 245 | 7.54 | 0.17 | 2 | 4133 | 4.23 | 0.10 | 7.16 | 0.19 | 245 | 7.15 | 0.19 | 2 |
| HD 184509 | S | 6069 | 4.33 | 1.30 | 7.35 | 0.06 | 332 | 7.33 | 0.05 | 21 | 6069 | 4.39 | 1.23 | 7.36 | 0.06 | 332 | 7.36 | 0.05 | 21 |
| HD 18455 | S | 5105 | 4.44 | 1.62 | 7.38 | 0.08 | 270 | 7.48 | 0.10 | 10 | 5105 | 4.19 | 1.84 | 7.31 | 0.09 | 270 | 7.27 | 0.09 | 10 |
| HD 184700 | S | 5745 | 4.23 | 1.02 | 7.31 | 0.06 | 363 | 7.25 | 0.06 | 19 | 5745 | 4.33 | 0.84 | 7.32 | 0.06 | 363 | 7.32 | 0.06 | 19 |
| HD 187123 | S | 5796 | 4.30 | 1.25 | 7.55 | 0.06 | 373 | 7.52 | 0.06 | 25 | 5796 | 4.38 | 1.12 | 7.56 | 0.06 | 373 | 7.56 | 0.06 | 25 |
| HD 188015 | S | 5667 | 4.26 | 1.35 | 7.68 | 0.08 | 367 | 7.68 | 0.08 | 24 | 5667 | 4.25 | 1.36 | 7.68 | 0.08 | 367 | 7.68 | 0.08 | 24 |
| HD 188088 | S | 4818 | 4.20 | 1.45 | 7.59 | 0.30 | 390 | 7.63 | 0.18 | 11 | 4818 | 4.12 | 1.58 | 7.56 | 0.30 | 390 | 7.57 | 0.17 | 11 |
| HD 188169 | S | 6509 | 4.25 | 1.91 | 7.37 | 0.11 | 287 | 7.40 | 0.07 | 22 | 6509 | 4.19 | 1.94 | 7.37 | 0.11 | 287 | 7.37 | 0.07 | 22 |
| HD 189712 | S | 6363 | 3.85 | 2.21 | 6.94 | 0.10 | 255 | 6.95 | 0.06 | 22 | 6363 | 3.82 | 2.22 | 6.94 | 0.10 | 255 | 6.94 | 0.06 | 22 |
| HD 189733 | S | 5044 | 4.60 | 0.83 | 7.48 | 0.05 | 171 | 7.57 | 0.11 | 12 | 5044 | 4.42 | 1.05 | 7.45 | 0.04 | 171 | 7.45 | 0.11 | 12 |
| HD 190007 | S | 4596 | 4.57 | 0.88 | 7.73 | 0.15 | 209 | 8.36 | 0.43 | 14 | 4596 | 3.57 | 1.40 | 7.30 | 0.15 | 209 | 7.41 | 0.43 | 14 |
| HD 190228 | S | 5264 | 3.71 | 1.16 | 7.19 | 0.06 | 382 | 7.21 | 0.05 | 19 | 5264 | 3.65 | 1.22 | 7.18 | 0.06 | 382 | 7.18 | 0.06 | 19 |
| HD 190360 | S | 5564 | 4.31 | 1.24 | 7.67 | 0.07 | 391 | 7.74 | 0.10 | 25 | 5564 | 4.14 | 1.45 | 7.63 | 0.08 | 391 | 7.63 | 0.11 | 25 |
| HD 190771 | S | 5751 | 4.44 | 1.29 | 7.58 | 0.06 | 351 | 7.65 | 0.08 | 23 | 5751 | 4.28 | 1.48 | 7.56 | 0.06 | 351 | 7.56 | 0.08 | 23 |
| HD 192145 | S | 6069 | 3.98 | 1.59 | 7.16 | 0.07 | 312 | 7.18 | 0.04 | 20 | 6069 | 3.93 | 1.62 | 7.15 | 0.07 | 312 | 7.15 | 0.04 | 20 |
| HD 192263 | S | 4974 | 4.61 | 0.93 | 7.50 | 0.09 | 233 | 7.74 | 0.22 | 17 | 4974 | 4.06 | 1.30 | 7.36 | 0.09 | 233 | 7.35 | 0.22 | 17 |

| Star | | Teff | log g | vt | logε(Fe I) | σ | # | logε(Fe II) | σ | # | Teff | log g | vt | logε(Fe I) | σ | # | logε(Fe II) | σ | # |
|---|---|---|---|---|---|---|---|---|---|---|---|---|---|---|---|---|---|---|---|
| HD 192310 | S | 5044 | 4.51 | 0.25 | 7.52 | 0.08 | 347 | 7.66 | 0.14 | 16 | 5044 | 4.24 | 1.21 | 7.41 | 0.09 | 347 | 7.44 | 0.12 | 16 |
| HD 193555 | S | 6133 | 3.72 | 2.83 | 7.73 | 0.16 | 197 | 7.64 | 0.08 | 14 | 6133 | 3.94 | 2.75 | 7.74 | 0.15 | 197 | 7.74 | 0.09 | 14 |
| HD 193664 | S | 5945 | 4.44 | 1.08 | 7.40 | 0.06 | 361 | 7.37 | 0.04 | 21 | 5945 | 4.50 | 0.99 | 7.40 | 0.06 | 361 | 7.41 | 0.04 | 21 |
| HD 19383 | S | 6396 | 4.32 | 2.14 | 7.61 | 0.12 | 290 | 7.65 | 0.09 | 23 | 6396 | 4.23 | 2.19 | 7.60 | 0.12 | 290 | 7.61 | 0.09 | 23 |
| HD 195019 | S | 5660 | 4.05 | 1.32 | 7.42 | 0.06 | 367 | 7.50 | 0.07 | 23 | 5660 | 3.86 | 1.49 | 7.41 | 0.07 | 367 | 7.40 | 0.08 | 23 |
| HD 195564 | S | 5619 | 3.95 | 1.31 | 7.48 | 0.06 | 362 | 7.47 | 0.04 | 19 | 5619 | 3.98 | 1.27 | 7.49 | 0.06 | 362 | 7.49 | 0.04 | 19 |
| HD 19617 | S | 5682 | 4.43 | 1.17 | 7.63 | 0.08 | 362 | 7.70 | 0.07 | 20 | 5682 | 4.31 | 1.13 | 7.62 | 0.08 | 362 | 7.59 | 0.07 | 20 |
| HD 196761 | S | 5483 | 4.53 | 0.82 | 7.21 | 0.06 | 352 | 7.23 | 0.07 | 17 | 5483 | 4.49 | 0.94 | 7.20 | 0.06 | 352 | 7.20 | 0.07 | 17 |
| HD 197076 | S | 5844 | 4.46 | 1.09 | 7.38 | 0.05 | 347 | 7.37 | 0.06 | 19 | 5844 | 4.50 | 1.03 | 7.39 | 0.05 | 347 | 7.39 | 0.05 | 19 |
| HD 198387 | S | 5056 | 3.51 | 1.21 | 7.28 | 0.06 | 362 | 7.25 | 0.08 | 18 | 5056 | 3.59 | 1.12 | 7.30 | 0.06 | 362 | 7.30 | 0.08 | 18 |
| HD 198483 | S | 5986 | 4.30 | 1.66 | 7.64 | 0.08 | 372 | 7.65 | 0.08 | 23 | 5986 | 4.29 | 1.67 | 7.64 | 0.08 | 372 | 7.65 | 0.08 | 23 |
| HD 198802 | S | 5736 | 3.81 | 1.46 | 7.43 | 0.06 | 360 | 7.36 | 0.05 | 20 | 5736 | 3.95 | 1.36 | 7.44 | 0.06 | 360 | 7.44 | 0.05 | 20 |
| HD 199604 | S | 5887 | 4.33 | 1.04 | 6.88 | 0.08 | 289 | 6.88 | 0.06 | 17 | 5887 | 4.35 | 1.01 | 6.89 | 0.08 | 289 | 6.89 | 0.06 | 17 |
| HD 200779 | S | 4389 | 4.61 | 0.15 | 7.72 | 0.14 | 284 | 7.57 | 0.11 | 3 | 4389 | 4.99 | 0.10 | 7.73 | 0.16 | 284 | 7.71 | 0.03 | 3 |
| HD 201456 | S | 6201 | 4.20 | 1.55 | 7.55 | 0.07 | 342 | 7.58 | 0.07 | 26 | 6201 | 4.13 | 1.61 | 7.55 | 0.07 | 342 | 7.55 | 0.08 | 26 |
| HD 201496 | S | 5944 | 4.37 | 1.19 | 7.42 | 0.06 | 339 | 7.43 | 0.04 | 18 | 5944 | 4.35 | 1.21 | 7.42 | 0.06 | 339 | 7.41 | 0.04 | 18 |
| HD 202206 | S | 5693 | 4.40 | 1.27 | 7.75 | 0.07 | 388 | 7.81 | 0.11 | 26 | 5693 | 4.27 | 1.43 | 7.73 | 0.08 | 388 | 7.72 | 0.11 | 26 |
| HD 202282 | S | 5796 | 4.30 | 1.31 | 7.51 | 0.06 | 365 | 7.44 | 0.07 | 19 | 5796 | 4.44 | 1.11 | 7.53 | 0.06 | 365 | 7.53 | 0.08 | 19 |
| HD 20339 | S | 5918 | 4.34 | 1.28 | 7.45 | 0.06 | 348 | 7.46 | 0.05 | 23 | 5918 | 4.32 | 1.30 | 7.45 | 0.06 | 348 | 7.45 | 0.05 | 23 |
| HD 20367 | S | 6050 | 4.36 | 1.48 | 7.50 | 0.06 | 351 | 7.46 | 0.04 | 21 | 6050 | 4.43 | 1.41 | 7.50 | 0.06 | 351 | 7.50 | 0.04 | 21 |
| HD 204153 | S | 6920 | 4.20 | 2.15 | 7.15 | 0.21 | 7 | 6.77 | 0.00 | 1 | 6920 | 4.20 | 1.16 | 7.30 | 0.15 | 7 | 6.79 | 0.00 | 1 |
| HD 204485 | S | 7125 | 4.13 | 2.62 | 7.64 | 0.13 | 316 | 7.60 | 0.08 | 30 | 7125 | 4.24 | 2.60 | 7.64 | 0.13 | 316 | 7.64 | 0.08 | 30 |
| HD 205027 | S | 5752 | 4.31 | 0.89 | 7.11 | 0.06 | 268 | 7.20 | 0.04 | 17 | 5752 | 4.11 | 1.02 | 7.11 | 0.06 | 268 | 7.10 | 0.04 | 17 |
| HD 205700 | S | 6656 | 4.11 | 1.98 | 7.32 | 0.10 | 297 | 7.32 | 0.05 | 23 | 6656 | 4.12 | 2.02 | 7.31 | 0.10 | 297 | 7.31 | 0.05 | 23 |
| HD 206282 | S | 6345 | 4.01 | 2.06 | 7.68 | 0.09 | 329 | 7.64 | 0.10 | 26 | 6345 | 4.11 | 2.01 | 7.69 | 0.09 | 329 | 7.69 | 0.09 | 26 |
| HD 207858 | S | 6290 | 3.97 | 2.14 | 7.59 | 0.10 | 327 | 7.49 | 0.09 | 24 | 6290 | 4.21 | 2.03 | 7.60 | 0.10 | 327 | 7.59 | 0.08 | 24 |
| HD 208801 | S | 4918 | 3.59 | 1.10 | 7.57 | 0.10 | 358 | 7.62 | 0.17 | 19 | 4918 | 3.47 | 1.27 | 7.53 | 0.10 | 358 | 7.52 | 0.18 | 19 |
| HD 21019 | S | 5514 | 3.79 | 1.21 | 7.04 | 0.05 | 336 | 6.98 | 0.03 | 17 | 5514 | 3.92 | 1.08 | 7.05 | 0.05 | 336 | 7.05 | 0.02 | 17 |
| HD 210277 | S | 5533 | 4.36 | 1.12 | 7.67 | 0.07 | 368 | 7.69 | 0.09 | 21 | 5533 | 4.32 | 1.20 | 7.66 | 0.07 | 368 | 7.66 | 0.09 | 21 |
| HD 210460 | S | 5529 | 3.52 | 1.39 | 7.17 | 0.06 | 347 | 7.12 | 0.04 | 19 | 5529 | 3.65 | 1.32 | 7.18 | 0.06 | 347 | 7.18 | 0.04 | 19 |
| HD 210483 | S | 5842 | 4.10 | 1.34 | 7.34 | 0.06 | 364 | 7.37 | 0.04 | 19 | 5842 | 4.04 | 1.36 | 7.34 | 0.06 | 364 | 7.31 | 0.04 | 19 |
| HD 210855 | S | 6255 | 3.78 | 2.20 | 7.63 | 0.09 | 313 | 7.55 | 0.08 | 23 | 6255 | 3.98 | 2.13 | 7.64 | 0.09 | 313 | 7.64 | 0.08 | 23 |
| HD 211476 | S | 5829 | 4.35 | 1.04 | 7.30 | 0.06 | 338 | 7.28 | 0.04 | 18 | 5829 | 4.38 | 0.99 | 7.30 | 0.06 | 338 | 7.30 | 0.04 | 18 |
| HD 211575 | S | 6589 | 4.23 | 2.62 | 7.68 | 0.14 | 233 | 7.62 | 0.09 | 18 | 6589 | 4.38 | 2.55 | 7.69 | 0.14 | 233 | 7.69 | 0.09 | 18 |
| HD 213338 | S | 5558 | 4.51 | 0.84 | 7.54 | 0.06 | 349 | 7.62 | 0.08 | 18 | 5558 | 4.40 | 1.14 | 7.51 | 0.07 | 349 | 7.53 | 0.08 | 18 |
| HD 214385 | S | 5711 | 4.47 | 0.74 | 7.18 | 0.07 | 268 | 7.22 | 0.06 | 16 | 5711 | 4.37 | 0.84 | 7.17 | 0.07 | 268 | 7.17 | 0.06 | 16 |
| HD 214749 | S | 4531 | 4.63 | 0.80 | 7.57 | 0.15 | 187 | 8.21 | 0.47 | 12 | 4531 | 3.63 | 1.48 | 7.20 | 0.16 | 187 | 7.36 | 0.47 | 12 |
| HD 215243 | S | 6393 | 4.12 | 1.94 | 7.53 | 0.07 | 322 | 7.46 | 0.07 | 25 | 6393 | 4.27 | 1.85 | 7.53 | 0.07 | 322 | 7.53 | 0.07 | 25 |

| | | | | | | | | | | | | | | | | | |
|---|---|---|---|---|---|---|---|---|---|---|---|---|---|---|---|---|---|
| HD 21531 | S | 4231 | 4.67 | 0.15 | 7.89 | 0.15 | 247 | 8.04 | 0.07 | 3 | 4231 | 4.28 | 1.70 | 7.68 | 0.17 | 247 | 7.69 | 0.06 | 3 |
| HD 215625 | S | 6228 | 4.39 | 1.42 | 7.55 | 0.08 | 350 | 7.52 | 0.06 | 22 | 6228 | 4.45 | 1.35 | 7.56 | 0.08 | 350 | 7.56 | 0.06 | 22 |
| HD 216133 | S | 3973 | 4.92 | 0.15 | 7.58 | 0.25 | 180 | 8.32 | 0.00 | 1 | 3973 | 4.92 | 0.10 | 7.58 | 0.25 | 180 | 8.32 | 0.00 | 1 |
| HD 216770 | S | 5411 | 4.48 | 0.95 | 7.85 | 0.08 | 221 | 8.03 | 0.15 | 18 | 5411 | 4.06 | 1.16 | 7.77 | 0.08 | 221 | 7.72 | 0.17 | 18 |
| HD 217107 | S | 5541 | 4.29 | 1.19 | 7.75 | 0.08 | 368 | 7.85 | 0.08 | 25 | 5541 | 4.10 | 1.41 | 7.72 | 0.08 | 368 | 7.72 | 0.09 | 25 |
| HD 217357 | S | 4192 | 4.82 | 0.15 | 7.61 | 0.20 | 235 | 7.72 | 0.03 | 2 | 4192 | 3.82 | 2.17 | 7.23 | 0.23 | 235 | 6.91 | 0.01 | 2 |
| HD 217577 | S | 5762 | 4.11 | 1.23 | 7.31 | 0.06 | 366 | 7.19 | 0.06 | 22 | 5762 | 4.35 | 0.88 | 7.34 | 0.06 | 366 | 7.34 | 0.06 | 22 |
| HD 217877 | S | 6015 | 4.35 | 1.29 | 7.39 | 0.05 | 338 | 7.38 | 0.04 | 21 | 6015 | 4.37 | 1.27 | 7.39 | 0.05 | 338 | 7.39 | 0.04 | 21 |
| HD 217958 | S | 5771 | 4.23 | 1.38 | 7.70 | 0.07 | 364 | 7.82 | 0.08 | 25 | 5771 | 3.94 | 1.64 | 7.67 | 0.09 | 364 | 7.65 | 0.10 | 25 |
| HD 218101 | S | 5217 | 3.81 | 1.26 | 7.54 | 0.08 | 361 | 7.62 | 0.11 | 23 | 5217 | 3.61 | 1.43 | 7.51 | 0.09 | 361 | 7.49 | 0.13 | 23 |
| HD 219428 | S | 5930 | 4.35 | 1.43 | 7.55 | 0.08 | 370 | 7.56 | 0.06 | 24 | 5930 | 4.32 | 1.45 | 7.55 | 0.08 | 370 | 7.55 | 0.06 | 24 |
| HD 220008 | S | 5653 | 3.82 | 1.35 | 7.26 | 0.06 | 339 | 7.23 | 0.04 | 19 | 5653 | 3.89 | 1.29 | 7.26 | 0.05 | 339 | 7.26 | 0.04 | 19 |
| HD 220689 | S | 5921 | 4.37 | 1.17 | 7.45 | 0.06 | 339 | 7.44 | 0.05 | 18 | 5921 | 4.40 | 1.12 | 7.46 | 0.06 | 339 | 7.46 | 0.05 | 18 |
| HD 22072 | S | 4974 | 3.48 | 1.13 | 7.15 | 0.07 | 371 | 7.22 | 0.07 | 17 | 4974 | 3.31 | 1.30 | 7.11 | 0.08 | 371 | 7.11 | 0.07 | 17 |
| HD 221356 | S | 6137 | 4.39 | 1.39 | 7.25 | 0.05 | 306 | 7.18 | 0.04 | 20 | 6137 | 4.53 | 1.23 | 7.26 | 0.05 | 306 | 7.26 | 0.04 | 20 |
| HD 221445 | S | 6230 | 3.77 | 1.87 | 7.32 | 0.08 | 328 | 7.16 | 0.07 | 22 | 6230 | 4.14 | 1.65 | 7.34 | 0.08 | 328 | 7.34 | 0.08 | 22 |
| HD 221503 | S | 4312 | 4.66 | 0.15 | 7.75 | 0.18 | 305 | 7.57 | 0.02 | 2 | 4312 | 4.99 | 0.10 | 7.78 | 0.19 | 305 | 7.70 | 0.02 | 2 |
| HD 222582 | S | 5796 | 4.37 | 1.09 | 7.49 | 0.06 | 368 | 7.48 | 0.08 | 25 | 5796 | 4.40 | 1.03 | 7.50 | 0.06 | 368 | 7.50 | 0.08 | 25 |
| HD 222645 | S | 6211 | 4.33 | 1.47 | 7.38 | 0.07 | 331 | 7.36 | 0.06 | 23 | 6211 | 4.38 | 1.42 | 7.39 | 0.07 | 331 | 7.39 | 0.06 | 23 |
| HD 22292 | S | 6483 | 4.31 | 2.64 | 7.51 | 0.17 | 199 | 7.51 | 0.15 | 15 | 6483 | 4.31 | 2.64 | 7.51 | 0.17 | 199 | 7.51 | 0.15 | 15 |
| HD 223084 | S | 5958 | 4.33 | 1.19 | 7.28 | 0.07 | 339 | 7.34 | 0.05 | 20 | 5958 | 4.21 | 1.34 | 7.27 | 0.07 | 339 | 7.27 | 0.05 | 20 |
| HD 223110 | S | 6516 | 3.96 | 2.60 | 7.61 | 0.15 | 262 | 7.54 | 0.11 | 22 | 6516 | 4.11 | 2.56 | 7.61 | 0.15 | 262 | 7.61 | 0.11 | 22 |
| HD 223238 | S | 5889 | 4.30 | 1.20 | 7.52 | 0.06 | 346 | 7.49 | 0.05 | 22 | 5889 | 4.37 | 1.10 | 7.53 | 0.06 | 346 | 7.53 | 0.05 | 22 |
| HD 22455 | S | 5886 | 4.37 | 1.22 | 7.51 | 0.07 | 356 | 7.49 | 0.06 | 23 | 5886 | 4.41 | 1.16 | 7.52 | 0.07 | 356 | 7.52 | 0.06 | 23 |
| HD 22468A | S | 4736 | 3.35 | 1.78 | 7.31 | 0.21 | 39 | 7.33 | 0.15 | 7 | 4736 | 3.29 | 1.82 | 7.32 | 0.21 | 39 | 7.36 | 0.15 | 7 |
| HD 22468B | S | 4631 | 4.47 | 0.15 | 7.56 | 0.12 | 373 | 7.81 | 0.13 | 15 | 4631 | 3.91 | 1.89 | 7.32 | 0.13 | 373 | 7.35 | 0.11 | 15 |
| HD 225239 | S | 5647 | 3.76 | 1.30 | 7.00 | 0.06 | 333 | 6.96 | 0.03 | 19 | 5647 | 3.85 | 1.22 | 7.01 | 0.06 | 333 | 7.01 | 0.03 | 19 |
| HD 231701 | S | 6240 | 4.17 | 1.65 | 7.49 | 0.08 | 342 | 7.36 | 0.07 | 22 | 6240 | 4.46 | 1.40 | 7.51 | 0.08 | 342 | 7.51 | 0.07 | 22 |
| HD 232979 | S | 3893 | 4.62 | 0.15 | 7.71 | 0.20 | 212 | 8.22 | 0.10 | 2 | 3893 | 3.85 | 1.00 | 7.35 | 0.20 | 212 | 7.40 | 0.14 | 2 |
| HD 23349 | S | 6018 | 4.26 | 1.42 | 7.59 | 0.07 | 356 | 7.58 | 0.06 | 27 | 6018 | 4.28 | 1.33 | 7.59 | 0.07 | 356 | 7.56 | 0.06 | 27 |
| HD 23356 | S | 4990 | 4.63 | 1.36 | 7.38 | 0.07 | 205 | 7.53 | 0.14 | 15 | 4990 | 4.22 | 1.64 | 7.27 | 0.08 | 205 | 7.23 | 0.15 | 15 |
| HD 234078 | S | 4157 | 4.68 | 0.15 | 7.61 | 0.15 | 225 | 7.82 | 0.05 | 2 | 4157 | 3.88 | 1.49 | 7.36 | 0.16 | 225 | 7.32 | 0.10 | 2 |
| HD 23453 | S | 3748 | 4.37 | 0.25 | 7.92 | 0.25 | 187 | 8.52 | 0.06 | 2 | 3748 | 3.71 | 1.74 | 7.54 | 0.24 | 187 | 7.56 | 0.10 | 2 |
| HD 23476 | S | 5662 | 4.48 | 0.81 | 7.09 | 0.06 | 329 | 7.08 | 0.10 | 17 | 5662 | 4.50 | 0.75 | 7.10 | 0.06 | 329 | 7.10 | 0.10 | 17 |
| HD 23596 | S | 6008 | 4.15 | 1.50 | 7.69 | 0.06 | 386 | 7.66 | 0.06 | 22 | 6008 | 4.22 | 1.45 | 7.70 | 0.06 | 386 | 7.70 | 0.06 | 22 |
| HD 239928 | S | 5899 | 4.43 | 1.08 | 7.52 | 0.06 | 349 | 7.53 | 0.06 | 21 | 5899 | 4.41 | 1.12 | 7.52 | 0.06 | 349 | 7.52 | 0.06 | 21 |
| HD 2475 | S | 6012 | 4.19 | 1.66 | 7.56 | 0.07 | 352 | 7.40 | 0.07 | 22 | 6012 | 4.52 | 1.31 | 7.58 | 0.07 | 352 | 7.59 | 0.07 | 22 |
| HD 2582 | S | 5705 | 4.19 | 1.41 | 7.60 | 0.08 | 363 | 7.63 | 0.09 | 22 | 5705 | 4.10 | 1.38 | 7.59 | 0.08 | 363 | 7.54 | 0.10 | 22 |

| Star | | Teff | log g | vt | [Fe/H]I | σ | N | [Fe/H]II | σ | N | Teff | log g | vt | [Fe/H]I | σ | N | [Fe/H]II | σ | N |
|---|---|---|---|---|---|---|---|---|---|---|---|---|---|---|---|---|---|---|---|
| HD 2589 | S | 5154 | 3.64 | 1.15 | 7.41 | 0.07 | 321 | 7.38 | 0.09 | 15 | 5154 | 3.72 | 1.05 | 7.43 | 0.07 | 321 | 7.44 | 0.09 | 15 |
| HD 2638 | S | 5120 | 4.54 | 1.33 | 7.66 | 0.07 | 63 | 7.89 | 0.06 | 6 | 5120 | 3.86 | 1.72 | 7.55 | 0.07 | 63 | 7.47 | 0.06 | 6 |
| HD 26505 | S | 5878 | 4.11 | 1.39 | 7.50 | 0.06 | 365 | 7.41 | 0.07 | 26 | 5878 | 4.31 | 1.16 | 7.52 | 0.06 | 365 | 7.52 | 0.07 | 26 |
| HD 26913 | S | 5661 | 4.52 | 1.55 | 7.45 | 0.06 | 401 | 7.46 | 0.07 | 18 | 5661 | 4.51 | 1.57 | 7.45 | 0.06 | 401 | 7.45 | 0.07 | 18 |
| HD 26923 | S | 5989 | 4.43 | 1.20 | 7.44 | 0.05 | 393 | 7.42 | 0.04 | 20 | 5989 | 4.47 | 1.15 | 7.44 | 0.05 | 393 | 7.44 | 0.04 | 20 |
| HD 2730 | S | 6192 | 3.99 | 1.71 | 7.37 | 0.09 | 315 | 7.28 | 0.07 | 21 | 6192 | 4.18 | 1.59 | 7.38 | 0.09 | 315 | 7.37 | 0.07 | 21 |
| HD 27530 | S | 5926 | 4.38 | 1.49 | 7.65 | 0.08 | 360 | 7.67 | 0.09 | 22 | 5926 | 4.36 | 1.50 | 7.65 | 0.08 | 360 | 7.65 | 0.09 | 22 |
| HD 28185 | S | 5658 | 4.33 | 1.27 | 7.69 | 0.07 | 375 | 7.66 | 0.09 | 21 | 5658 | 4.38 | 1.19 | 7.70 | 0.07 | 375 | 7.70 | 0.09 | 21 |
| HD 28343 | S | 4152 | 4.65 | 0.35 | 7.89 | 0.19 | 220 | 7.91 | 0.08 | 2 | 4152 | 4.59 | 0.40 | 7.88 | 0.19 | 220 | 7.88 | 0.09 | 2 |
| HD 285660 | S | 6055 | 4.27 | 1.65 | 7.47 | 0.09 | 325 | 7.52 | 0.07 | 19 | 6055 | 4.16 | 1.74 | 7.46 | 0.09 | 325 | 7.46 | 0.07 | 19 |
| HD 28571 | S | 5736 | 4.22 | 1.07 | 7.24 | 0.07 | 352 | 7.18 | 0.06 | 19 | 5736 | 4.34 | 0.87 | 7.26 | 0.07 | 352 | 7.25 | 0.06 | 19 |
| HD 28635 | S | 6140 | 4.35 | 1.54 | 7.57 | 0.07 | 353 | 7.52 | 0.05 | 22 | 6140 | 4.44 | 1.44 | 7.58 | 0.07 | 353 | 7.58 | 0.05 | 22 |
| HD 29587 | S | 5682 | 4.48 | 0.75 | 6.94 | 0.05 | 310 | 6.95 | 0.05 | 16 | 5682 | 4.47 | 0.77 | 6.94 | 0.05 | 310 | 6.94 | 0.05 | 16 |
| HD 29645 | S | 6002 | 4.02 | 1.63 | 7.58 | 0.07 | 354 | 7.51 | 0.05 | 21 | 6002 | 4.15 | 1.53 | 7.58 | 0.07 | 354 | 7.58 | 0.05 | 21 |
| HD 29697 | S | 4421 | 4.64 | 0.97 | 7.68 | 0.12 | 119 | 8.18 | 0.26 | 4 | 4421 | 3.72 | 1.50 | 7.28 | 0.11 | 119 | 7.29 | 0.23 | 4 |
| HD 31527 | S | 5908 | 4.36 | 1.08 | 7.30 | 0.06 | 340 | 7.25 | 0.05 | 17 | 5908 | 4.46 | 0.92 | 7.31 | 0.06 | 340 | 7.31 | 0.06 | 17 |
| HD 31949 | S | 6213 | 4.30 | 1.78 | 7.41 | 0.08 | 312 | 7.39 | 0.07 | 21 | 6213 | 4.35 | 1.74 | 7.41 | 0.08 | 312 | 7.41 | 0.08 | 21 |
| HD 32147 | S | 4790 | 4.57 | 0.15 | 7.86 | 0.13 | 399 | 7.95 | 0.22 | 14 | 4790 | 4.37 | 1.39 | 7.77 | 0.13 | 399 | 7.82 | 0.20 | 14 |
| HD 32715 | S | 6615 | 4.25 | 2.45 | 7.62 | 0.15 | 59 | 7.82 | 0.09 | 4 | 6615 | 3.78 | 2.53 | 7.62 | 0.15 | 59 | 7.62 | 0.10 | 4 |
| HD 332612 | S | 6124 | 4.10 | 1.74 | 7.47 | 0.08 | 337 | 7.44 | 0.04 | 20 | 6124 | 4.15 | 1.70 | 7.47 | 0.08 | 337 | 7.47 | 0.04 | 20 |
| HD 334372 | S | 5765 | 4.01 | 1.32 | 7.39 | 0.06 | 362 | 7.41 | 0.06 | 22 | 5765 | 3.97 | 1.35 | 7.39 | 0.06 | 362 | 7.39 | 0.06 | 22 |
| HD 33636 | S | 5953 | 4.45 | 1.11 | 7.38 | 0.06 | 365 | 7.41 | 0.04 | 19 | 5953 | 4.40 | 1.17 | 7.38 | 0.06 | 365 | 7.38 | 0.04 | 19 |
| HD 33866 | S | 5625 | 4.33 | 1.12 | 7.34 | 0.06 | 362 | 7.50 | 0.06 | 22 | 5625 | 4.01 | 1.41 | 7.30 | 0.07 | 362 | 7.28 | 0.08 | 22 |
| HD 34721 | S | 6004 | 4.14 | 1.45 | 7.40 | 0.06 | 350 | 7.36 | 0.05 | 19 | 6004 | 4.21 | 1.39 | 7.40 | 0.06 | 350 | 7.40 | 0.05 | 19 |
| HD 3556 | S | 6019 | 4.42 | 1.43 | 7.60 | 0.07 | 355 | 7.57 | 0.06 | 21 | 6019 | 4.47 | 1.38 | 7.60 | 0.07 | 355 | 7.60 | 0.06 | 21 |
| HD 35961 | S | 5740 | 4.37 | 0.90 | 7.24 | 0.06 | 273 | 7.33 | 0.03 | 15 | 5740 | 4.16 | 1.04 | 7.23 | 0.06 | 273 | 7.23 | 0.03 | 15 |
| HD 36003 | S | 4536 | 4.60 | 0.73 | 7.53 | 0.17 | 219 | 8.29 | 0.53 | 12 | 4536 | 3.60 | 1.45 | 7.14 | 0.17 | 219 | 7.44 | 0.54 | 12 |
| HD 37124 | S | 5561 | 4.39 | 0.70 | 7.06 | 0.05 | 347 | 7.04 | 0.03 | 15 | 5561 | 4.43 | 0.55 | 7.07 | 0.06 | 347 | 7.07 | 0.04 | 15 |
| HD 37605 | S | 5318 | 4.47 | 0.69 | 7.78 | 0.06 | 201 | 7.90 | 0.05 | 11 | 5318 | 4.24 | 0.88 | 7.74 | 0.06 | 201 | 7.76 | 0.05 | 11 |
| HD 3795 | S | 5456 | 3.89 | 1.02 | 6.93 | 0.06 | 324 | 6.85 | 0.05 | 17 | 5456 | 4.07 | 0.75 | 6.96 | 0.06 | 324 | 6.96 | 0.05 | 17 |
| HD 38529 | S | 5450 | 3.72 | 1.56 | 7.72 | 0.09 | 382 | 7.78 | 0.10 | 23 | 5450 | 3.59 | 1.63 | 7.71 | 0.09 | 382 | 7.71 | 0.10 | 23 |
| HD 38700 | S | 6049 | 4.45 | 1.57 | 7.45 | 0.09 | 350 | 7.48 | 0.06 | 20 | 6049 | 4.37 | 1.65 | 7.44 | 0.09 | 350 | 7.44 | 0.06 | 20 |
| HD 38858 | S | 5798 | 4.48 | 1.01 | 7.28 | 0.05 | 343 | 7.25 | 0.04 | 18 | 5798 | 4.55 | 0.87 | 7.29 | 0.05 | 343 | 7.29 | 0.04 | 18 |
| HD 38A | S | 4065 | 4.71 | 0.15 | 7.66 | 0.24 | 223 | 7.80 | 0.23 | 2 | 4065 | 4.43 | 0.10 | 7.61 | 0.23 | 223 | 7.61 | 0.25 | 2 |
| HD 38B | S | 4028 | 4.72 | 0.15 | 7.63 | 0.24 | 202 | 7.92 | 0.12 | 2 | 4028 | 4.12 | 0.40 | 7.42 | 0.22 | 202 | 7.41 | 0.14 | 2 |
| HD 39881 | S | 5719 | 4.27 | 1.07 | 7.33 | 0.05 | 356 | 7.33 | 0.06 | 18 | 5719 | 4.28 | 0.91 | 7.34 | 0.05 | 356 | 7.31 | 0.06 | 18 |
| HD 400 | S | 6240 | 4.13 | 1.64 | 7.28 | 0.07 | 323 | 7.23 | 0.06 | 22 | 6240 | 4.26 | 1.53 | 7.29 | 0.07 | 323 | 7.29 | 0.06 | 22 |
| HD 40979 | S | 6150 | 4.35 | 1.63 | 7.67 | 0.08 | 368 | 7.64 | 0.05 | 21 | 6150 | 4.41 | 1.58 | 7.67 | 0.08 | 368 | 7.68 | 0.05 | 21 |

| | | | | | | | | | | | | | | | | |
|---|---|---|---|---|---|---|---|---|---|---|---|---|---|---|---|---|
| HD 41330 | S | 5876 | 4.13 | 1.28 | 7.30 | 0.06 | 360 | 7.26 | 0.06 | 21 | 5876 | 4.21 | 1.19 | 7.31 | 0.06 | 360 | 7.31 | 0.06 | 21 |
| HD 41708 | S | 5867 | 4.47 | 1.03 | 7.52 | 0.06 | 360 | 7.53 | 0.05 | 18 | 5867 | 4.45 | 1.06 | 7.52 | 0.06 | 360 | 7.52 | 0.05 | 18 |
| HD 4203 | S | 5471 | 4.10 | 1.21 | 7.80 | 0.08 | 370 | 7.95 | 0.10 | 24 | 5471 | 3.71 | 1.54 | 7.74 | 0.09 | 370 | 7.71 | 0.12 | 24 |
| HD 4208 | S | 5674 | 4.47 | 0.94 | 7.24 | 0.07 | 355 | 7.25 | 0.07 | 20 | 5674 | 4.46 | 0.96 | 7.24 | 0.07 | 355 | 7.25 | 0.07 | 20 |
| HD 43162 | S | 5651 | 4.50 | 1.51 | 7.49 | 0.06 | 344 | 7.49 | 0.07 | 17 | 5651 | 4.50 | 1.51 | 7.49 | 0.06 | 344 | 7.49 | 0.07 | 17 |
| HD 43587 | S | 5859 | 4.27 | 1.24 | 7.40 | 0.05 | 351 | 7.44 | 0.03 | 21 | 5859 | 4.19 | 1.32 | 7.39 | 0.05 | 351 | 7.39 | 0.03 | 21 |
| HD 43745 | S | 6087 | 3.92 | 1.67 | 7.54 | 0.06 | 341 | 7.46 | 0.04 | 24 | 6087 | 4.09 | 1.55 | 7.55 | 0.06 | 341 | 7.55 | 0.04 | 24 |
| HD 45067 | S | 6049 | 3.95 | 1.70 | 7.41 | 0.07 | 335 | 7.35 | 0.04 | 20 | 6049 | 4.08 | 1.62 | 7.41 | 0.06 | 335 | 7.41 | 0.04 | 20 |
| HD 45088 | S | 4778 | 4.37 | 0.15 | 7.26 | 0.16 | 387 | 7.45 | 0.06 | 8 | 4778 | 3.89 | 1.52 | 7.07 | 0.19 | 387 | 7.07 | 0.10 | 8 |
| HD 45184 | S | 5852 | 4.41 | 1.22 | 7.51 | 0.06 | 360 | 7.52 | 0.05 | 22 | 5852 | 4.37 | 1.27 | 7.50 | 0.06 | 360 | 7.50 | 0.05 | 22 |
| HD 45205 | S | 5921 | 4.15 | 1.31 | 6.65 | 0.09 | 251 | 6.69 | 0.07 | 16 | 5921 | 4.06 | 1.39 | 6.65 | 0.09 | 251 | 6.65 | 0.07 | 16 |
| HD 45350 | S | 5567 | 4.22 | 1.33 | 7.69 | 0.07 | 362 | 7.76 | 0.10 | 23 | 5567 | 4.06 | 1.49 | 7.67 | 0.08 | 362 | 7.66 | 0.11 | 23 |
| HD 45588 | S | 6214 | 4.24 | 1.70 | 7.51 | 0.07 | 334 | 7.46 | 0.06 | 21 | 6214 | 4.36 | 1.60 | 7.52 | 0.07 | 334 | 7.52 | 0.06 | 21 |
| HD 45759 | S | 6132 | 4.37 | 2.26 | 7.59 | 0.11 | 277 | 7.62 | 0.09 | 19 | 6132 | 4.29 | 2.31 | 7.58 | 0.11 | 277 | 7.54 | 0.09 | 19 |
| HD 4628 | S | 5044 | 4.61 | 0.35 | 7.25 | 0.07 | 293 | 7.41 | 0.28 | 20 | 5044 | 4.23 | 0.91 | 7.18 | 0.07 | 293 | 7.19 | 0.28 | 20 |
| HD 46375 | S | 5265 | 4.33 | 1.24 | 7.68 | 0.08 | 364 | 7.67 | 0.10 | 13 | 5265 | 4.35 | 1.21 | 7.68 | 0.08 | 364 | 7.68 | 0.10 | 13 |
| HD 48938 | S | 6055 | 4.33 | 1.27 | 7.09 | 0.06 | 302 | 7.08 | 0.06 | 17 | 6055 | 4.37 | 1.23 | 7.10 | 0.06 | 302 | 7.10 | 0.06 | 17 |
| HD 49674 | S | 5621 | 4.39 | 1.37 | 7.70 | 0.08 | 372 | 7.70 | 0.08 | 21 | 5621 | 4.40 | 1.36 | 7.70 | 0.07 | 372 | 7.70 | 0.08 | 21 |
| HD 50281 | S | 4708 | 4.64 | 0.15 | 7.56 | 0.11 | 400 | 7.50 | 0.14 | 9 | 4708 | 4.76 | 0.10 | 7.56 | 0.12 | 400 | 7.55 | 0.15 | 9 |
| HD 50554 | S | 6019 | 4.41 | 1.21 | 7.46 | 0.05 | 356 | 7.46 | 0.05 | 20 | 6019 | 4.42 | 1.20 | 7.46 | 0.05 | 356 | 7.46 | 0.05 | 20 |
| HD 50806 | S | 5640 | 4.06 | 1.16 | 7.52 | 0.06 | 361 | 7.50 | 0.08 | 18 | 5640 | 4.09 | 1.12 | 7.52 | 0.06 | 361 | 7.52 | 0.08 | 18 |
| HD 52265 | S | 6086 | 4.29 | 1.51 | 7.65 | 0.07 | 373 | 7.66 | 0.06 | 23 | 6086 | 4.27 | 1.53 | 7.65 | 0.07 | 373 | 7.65 | 0.06 | 23 |
| HD 52698 | S | 5155 | 4.55 | 0.98 | 7.66 | 0.07 | 244 | 7.85 | 0.11 | 15 | 5155 | 4.12 | 1.23 | 7.56 | 0.08 | 244 | 7.50 | 0.12 | 15 |
| HD 52711 | S | 5992 | 4.41 | 1.26 | 7.41 | 0.06 | 367 | 7.35 | 0.03 | 17 | 5992 | 4.52 | 1.12 | 7.42 | 0.06 | 367 | 7.42 | 0.03 | 17 |
| HD 5494 | S | 6044 | 4.03 | 1.67 | 7.39 | 0.13 | 287 | 7.49 | 0.09 | 22 | 6044 | 3.80 | 1.81 | 7.38 | 0.13 | 287 | 7.38 | 0.10 | 22 |
| HD 55054 | S | 6219 | 4.14 | 1.70 | 7.42 | 0.08 | 260 | 7.32 | 0.05 | 15 | 6219 | 4.38 | 1.51 | 7.44 | 0.07 | 260 | 7.44 | 0.05 | 15 |
| HD 55575 | S | 5937 | 4.32 | 1.26 | 7.15 | 0.06 | 322 | 7.15 | 0.05 | 21 | 5937 | 4.33 | 1.25 | 7.15 | 0.06 | 322 | 7.16 | 0.05 | 21 |
| HD 55693 | S | 5854 | 4.32 | 1.32 | 7.73 | 0.06 | 383 | 7.73 | 0.07 | 23 | 5854 | 4.32 | 1.32 | 7.73 | 0.06 | 383 | 7.73 | 0.07 | 23 |
| HD 57006 | S | 6264 | 3.77 | 2.10 | 7.51 | 0.07 | 336 | 7.45 | 0.07 | 24 | 6264 | 3.91 | 2.05 | 7.51 | 0.07 | 336 | 7.51 | 0.07 | 24 |
| HD 603 | S | 5967 | 4.37 | 1.15 | 7.31 | 0.08 | 341 | 7.25 | 0.06 | 19 | 5967 | 4.50 | 0.83 | 7.32 | 0.08 | 341 | 7.30 | 0.07 | 19 |
| HD 6064 | S | 6363 | 3.78 | 2.03 | 7.60 | 0.10 | 332 | 7.50 | 0.08 | 22 | 6363 | 4.03 | 1.93 | 7.61 | 0.09 | 332 | 7.61 | 0.08 | 22 |
| HD 61606 | S | 4932 | 4.61 | 1.00 | 7.47 | 0.08 | 238 | 7.61 | 0.16 | 13 | 4932 | 4.29 | 1.31 | 7.40 | 0.09 | 238 | 7.41 | 0.16 | 13 |
| HD 61632 | S | 5632 | 4.02 | 1.09 | 6.93 | 0.06 | 309 | 6.93 | 0.04 | 17 | 5632 | 4.03 | 1.08 | 6.93 | 0.06 | 309 | 6.93 | 0.04 | 17 |
| HD 63598 | S | 5828 | 4.38 | 0.93 | 6.63 | 0.08 | 251 | 6.72 | 0.05 | 14 | 5828 | 4.15 | 1.19 | 6.61 | 0.08 | 251 | 6.59 | 0.05 | 14 |
| HD 63754 | S | 6040 | 3.92 | 1.78 | 7.61 | 0.07 | 356 | 7.61 | 0.06 | 24 | 6040 | 3.90 | 1.78 | 7.61 | 0.07 | 356 | 7.60 | 0.06 | 24 |
| HD 68988 | S | 5878 | 4.39 | 1.37 | 7.76 | 0.07 | 378 | 7.81 | 0.06 | 25 | 5878 | 4.27 | 1.50 | 7.74 | 0.07 | 378 | 7.74 | 0.06 | 25 |
| HD 69830 | S | 5443 | 4.53 | 0.92 | 7.45 | 0.06 | 379 | 7.48 | 0.07 | 18 | 5443 | 4.46 | 1.10 | 7.42 | 0.06 | 379 | 7.43 | 0.06 | 18 |
| HD 70889 | S | 6036 | 4.43 | 1.25 | 7.57 | 0.05 | 350 | 7.55 | 0.05 | 21 | 6036 | 4.47 | 1.20 | 7.57 | 0.05 | 350 | 7.57 | 0.05 | 21 |

| | | | | | | | | | | | | | | | | |
|---|---|---|---|---|---|---|---|---|---|---|---|---|---|---|---|---|
| HD 7091 | S | 6109 | 4.29 | 1.16 | 7.21 | 0.09 | 179 | 7.35 | 0.13 | 16 | 6109 | 3.93 | 1.46 | 7.19 | 0.09 | 179 | 7.18 | 0.14 | 16 |
| HD 71148 | S | 5835 | 4.39 | 1.17 | 7.49 | 0.06 | 357 | 7.47 | 0.05 | 17 | 5835 | 4.43 | 1.10 | 7.50 | 0.06 | 357 | 7.50 | 0.05 | 17 |
| HD 71881 | S | 5863 | 4.30 | 1.14 | 7.43 | 0.05 | 344 | 7.44 | 0.05 | 20 | 5863 | 4.28 | 1.16 | 7.43 | 0.05 | 344 | 7.43 | 0.05 | 20 |
| HD 7230 | S | 6637 | 4.17 | 3.73 | 7.75 | 0.23 | 34 | 8.18 | 0.53 | 4 | 6637 | 3.17 | 3.94 | 7.75 | 0.23 | 34 | 7.78 | 0.52 | 4 |
| HD 72659 | S | 5902 | 4.12 | 1.31 | 7.43 | 0.06 | 365 | 7.41 | 0.03 | 18 | 5902 | 4.15 | 1.28 | 7.43 | 0.06 | 365 | 7.43 | 0.03 | 18 |
| HD 72946 | S | 5674 | 4.49 | 1.23 | 7.57 | 0.06 | 414 | 7.53 | 0.07 | 18 | 5674 | 4.58 | 1.07 | 7.59 | 0.06 | 414 | 7.59 | 0.07 | 18 |
| HD 7352 | S | 6023 | 4.31 | 1.30 | 7.55 | 0.06 | 373 | 7.55 | 0.05 | 20 | 6023 | 4.32 | 1.29 | 7.55 | 0.06 | 373 | 7.55 | 0.05 | 20 |
| HD 73596 | S | 6744 | 3.41 | 2.68 | 7.49 | 0.26 | 159 | 7.30 | 0.12 | 13 | 6744 | 3.80 | 2.64 | 7.48 | 0.26 | 159 | 7.44 | 0.13 | 13 |
| HD 73668 | S | 5922 | 4.37 | 1.24 | 7.45 | 0.05 | 345 | 7.46 | 0.04 | 23 | 5922 | 4.36 | 1.26 | 7.45 | 0.05 | 345 | 7.45 | 0.04 | 23 |
| HD 73752 | S | 5654 | 3.98 | 1.51 | 7.68 | 0.11 | 317 | 7.65 | 0.12 | 21 | 5654 | 4.03 | 1.47 | 7.68 | 0.11 | 317 | 7.68 | 0.12 | 21 |
| HD 7397 | S | 6029 | 4.15 | 1.37 | 7.36 | 0.08 | 197 | 7.19 | 0.05 | 10 | 6029 | 4.51 | 1.04 | 7.38 | 0.08 | 197 | 7.38 | 0.07 | 10 |
| HD 74156 | S | 6005 | 4.06 | 1.62 | 7.52 | 0.07 | 368 | 7.43 | 0.07 | 23 | 6005 | 4.27 | 1.44 | 7.54 | 0.06 | 368 | 7.54 | 0.07 | 23 |
| HD 7514 | S | 5682 | 4.24 | 1.22 | 7.31 | 0.08 | 356 | 7.28 | 0.07 | 20 | 5682 | 4.30 | 1.01 | 7.32 | 0.07 | 356 | 7.29 | 0.07 | 20 |
| HD 75488 | S | 5996 | 4.18 | 1.36 | 7.04 | 0.06 | 303 | 7.02 | 0.05 | 19 | 5996 | 4.24 | 1.29 | 7.05 | 0.06 | 303 | 7.05 | 0.05 | 19 |
| HD 76151 | S | 5780 | 4.43 | 1.10 | 7.57 | 0.05 | 387 | 7.55 | 0.06 | 15 | 5780 | 4.46 | 1.04 | 7.57 | 0.05 | 387 | 7.57 | 0.06 | 15 |
| HD 78366 | S | 5981 | 4.42 | 1.41 | 7.50 | 0.06 | 355 | 7.47 | 0.05 | 20 | 5981 | 4.47 | 1.35 | 7.50 | 0.06 | 355 | 7.50 | 0.05 | 20 |
| HD 79210 | S | 3973 | 4.69 | 0.15 | 7.64 | 0.19 | 207 | 8.05 | 0.00 | 1 | 3973 | 4.69 | 0.10 | 7.64 | 0.19 | 207 | 8.05 | 0.00 | 1 |
| HD 79211 | S | 3763 | 4.49 | 0.15 | 7.81 | 0.22 | 186 | 8.65 | 0.00 | 1 | 3763 | 4.49 | 0.20 | 7.81 | 0.22 | 186 | 8.64 | 0.00 | 1 |
| HD 80372 | S | 6039 | 4.36 | 1.43 | 7.49 | 0.07 | 340 | 7.59 | 0.07 | 22 | 6039 | 4.13 | 1.64 | 7.48 | 0.08 | 340 | 7.47 | 0.07 | 22 |
| HD 80606 | S | 5467 | 3.52 | 1.50 | 7.70 | 0.09 | 247 | 7.43 | 0.11 | 18 | 5467 | 3.92 | 1.40 | 7.72 | 0.09 | 247 | 7.63 | 0.11 | 18 |
| HD 80607 | S | 5535 | 3.35 | 1.45 | 7.76 | 0.08 | 242 | 7.39 | 0.16 | 25 | 5535 | 3.75 | 1.38 | 7.78 | 0.08 | 242 | 7.58 | 0.15 | 25 |
| HD 81040 | S | 5753 | 4.48 | 1.17 | 7.38 | 0.06 | 345 | 7.39 | 0.04 | 17 | 5753 | 4.46 | 1.20 | 7.38 | 0.06 | 345 | 7.38 | 0.04 | 17 |
| HD 8173 | S | 6009 | 4.40 | 1.14 | 7.47 | 0.07 | 347 | 7.45 | 0.07 | 21 | 6009 | 4.44 | 1.09 | 7.47 | 0.07 | 347 | 7.47 | 0.07 | 21 |
| HD 81809 | S | 5666 | 3.72 | 1.27 | 7.14 | 0.07 | 336 | 7.04 | 0.05 | 18 | 5666 | 3.94 | 1.11 | 7.15 | 0.07 | 336 | 7.15 | 0.05 | 18 |
| HD 82106 | S | 4826 | 4.63 | 0.86 | 7.51 | 0.10 | 246 | 8.01 | 0.40 | 15 | 4826 | 3.83 | 1.42 | 7.26 | 0.11 | 246 | 7.39 | 0.39 | 15 |
| HD 82943 | S | 5919 | 4.35 | 1.36 | 7.67 | 0.06 | 386 | 7.69 | 0.05 | 24 | 5919 | 4.29 | 1.41 | 7.66 | 0.06 | 386 | 7.66 | 0.05 | 24 |
| HD 84117 | S | 6239 | 4.34 | 1.54 | 7.45 | 0.07 | 358 | 7.40 | 0.07 | 20 | 6239 | 4.45 | 1.43 | 7.46 | 0.07 | 358 | 7.46 | 0.07 | 20 |
| HD 84703 | S | 6125 | 4.14 | 1.60 | 7.50 | 0.06 | 344 | 7.43 | 0.07 | 23 | 6125 | 4.28 | 1.49 | 7.51 | 0.06 | 344 | 7.51 | 0.06 | 23 |
| HD 85725 | S | 5892 | 3.72 | 1.84 | 7.59 | 0.07 | 361 | 7.57 | 0.07 | 23 | 5892 | 3.77 | 1.83 | 7.59 | 0.07 | 361 | 7.55 | 0.07 | 23 |
| HD 8574 | S | 6045 | 4.19 | 1.43 | 7.45 | 0.06 | 360 | 7.46 | 0.05 | 23 | 6045 | 4.19 | 1.43 | 7.45 | 0.06 | 360 | 7.46 | 0.05 | 23 |
| HD 8673 | S | 6409 | 4.25 | 3.13 | 7.63 | 0.16 | 164 | 7.70 | 0.19 | 17 | 6409 | 4.07 | 3.20 | 7.63 | 0.16 | 164 | 7.62 | 0.19 | 17 |
| HD 87097 | S | 5993 | 4.40 | 1.88 | 7.50 | 0.09 | 331 | 7.52 | 0.09 | 20 | 5993 | 4.36 | 1.92 | 7.50 | 0.10 | 331 | 7.49 | 0.09 | 20 |
| HD 88133 | S | 5371 | 3.82 | 1.41 | 7.73 | 0.09 | 347 | 7.76 | 0.12 | 22 | 5371 | 3.75 | 1.46 | 7.72 | 0.09 | 347 | 7.72 | 0.12 | 22 |
| HD 88230 | S | 4215 | 4.70 | 0.15 | 7.68 | 0.19 | 251 | 7.78 | 0.01 | 2 | 4215 | 4.38 | 1.40 | 7.52 | 0.19 | 251 | 7.50 | 0.03 | 2 |
| HD 88371 | S | 5673 | 4.27 | 1.08 | 7.17 | 0.07 | 341 | 7.13 | 0.04 | 17 | 5673 | 4.35 | 0.96 | 7.18 | 0.07 | 341 | 7.18 | 0.03 | 17 |
| HD 88595 | S | 6322 | 4.12 | 1.91 | 7.54 | 0.07 | 342 | 7.49 | 0.05 | 20 | 6322 | 4.22 | 1.85 | 7.54 | 0.07 | 342 | 7.54 | 0.05 | 20 |
| HD 88737 | S | 6129 | 3.73 | 2.20 | 7.69 | 0.11 | 323 | 7.60 | 0.09 | 22 | 6129 | 3.92 | 2.13 | 7.69 | 0.10 | 323 | 7.69 | 0.09 | 22 |
| HD 89319 | S | 4942 | 3.35 | 1.61 | 7.70 | 0.12 | 339 | 7.84 | 0.26 | 22 | 4942 | 3.01 | 1.74 | 7.64 | 0.13 | 339 | 7.64 | 0.27 | 22 |

| | | | | | | | | | | | | | | | | |
|---|---|---|---|---|---|---|---|---|---|---|---|---|---|---|---|---|
| HD 89744 | S | 6207 | 3.92 | 2.11 | 7.65 | 0.09 | 318 | 7.60 | 0.08 | 23 | 6207 | 4.03 | 2.06 | 7.65 | 0.08 | 318 | 7.65 | 0.08 | 23 |
| HD 90508 | S | 5757 | 4.34 | 1.06 | 7.14 | 0.05 | 326 | 7.18 | 0.03 | 16 | 5757 | 4.25 | 1.17 | 7.13 | 0.05 | 326 | 7.13 | 0.04 | 16 |
| HD 9224  | S | 5848 | 4.16 | 1.24 | 7.46 | 0.06 | 356 | 7.43 | 0.05 | 22 | 5848 | 4.23 | 1.16 | 7.47 | 0.06 | 356 | 7.47 | 0.05 | 22 |
| HD 92788 | S | 5710 | 4.32 | 1.20 | 7.73 | 0.07 | 349 | 7.73 | 0.11 | 24 | 5710 | 4.32 | 1.20 | 7.73 | 0.07 | 349 | 7.73 | 0.11 | 24 |
| HD 9369  | S | 7237 | 4.09 | 3.50 | 8.19 | 0.10 | 23  | 8.27 | 0.18 | 3  | 7237 | 3.91 | 4.20 | 8.13 | 0.09 | 23  | 8.13 | 0.16 | 3 |
| HD 94132 | S | 4988 | 3.44 | 1.50 | 7.62 | 0.11 | 356 | 7.61 | 0.09 | 17 | 4988 | 3.47 | 1.50 | 7.63 | 0.11 | 356 | 7.63 | 0.09 | 17 |
| HD 94915 | S | 5985 | 4.41 | 1.33 | 7.38 | 0.06 | 328 | 7.34 | 0.04 | 18 | 5985 | 4.48 | 1.24 | 7.38 | 0.06 | 328 | 7.38 | 0.04 | 18 |
| HD 95650 | S | 3655 | 4.56 | 0.25 | 7.99 | 0.36 | 181 | 8.93 | 0.00 | 1  | 3655 | 4.56 | 0.10 | 8.00 | 0.36 | 181 | 8.93 | 0.00 | 1 |
| HD 96276 | S | 6040 | 4.34 | 1.38 | 7.41 | 0.07 | 341 | 7.42 | 0.05 | 22 | 6040 | 4.31 | 1.40 | 7.41 | 0.07 | 341 | 7.41 | 0.05 | 22 |
| HD 97100 | S | 5008 | 5.00 | 0.15 | 7.24 | 0.13 | 447 | 8.36 | 0.08 | 30 | 5008 | 4.63 | 1.17 | 7.16 | 0.12 | 447 | 8.12 | 0.07 | 30 |
| HD 97101 | S | 4189 | 4.67 | 0.70 | 8.10 | 0.29 | 125 | 9.86 | 1.63 | 2  | 4189 | 4.67 | 0.70 | 8.10 | 0.29 | 125 | 9.86 | 1.63 | 2 |
| HD 97334 | S | 5906 | 4.44 | 1.57 | 7.56 | 0.06 | 380 | 7.54 | 0.07 | 20 | 5906 | 4.48 | 1.52 | 7.56 | 0.06 | 380 | 7.56 | 0.06 | 20 |
| HD 975   | S | 6405 | 4.27 | 1.66 | 7.43 | 0.10 | 294 | 7.48 | 0.07 | 22 | 6405 | 4.16 | 1.74 | 7.42 | 0.10 | 294 | 7.42 | 0.07 | 22 |
| HD 97584 | S | 4732 | 4.62 | 0.91 | 7.46 | 0.11 | 223 | 7.89 | 0.46 | 16 | 4732 | 3.70 | 1.50 | 7.18 | 0.12 | 223 | 7.21 | 0.47 | 16 |
| HD 98388 | S | 6338 | 4.34 | 2.03 | 7.63 | 0.10 | 303 | 7.60 | 0.07 | 21 | 6338 | 4.41 | 1.98 | 7.64 | 0.10 | 303 | 7.64 | 0.07 | 21 |
| HD 98712 | S | 4300 | 4.72 | 0.15 | 7.59 | 0.16 | 237 | 7.80 | 0.00 | 1  | 4300 | 4.72 | 0.10 | 7.59 | 0.17 | 237 | 7.80 | 0.00 | 1 |
| HD 98823 | S | 6414 | 3.59 | 2.87 | 7.63 | 0.25 | 88  | 7.67 | 0.39 | 6  | 6414 | 3.49 | 2.89 | 7.63 | 0.25 | 88  | 7.63 | 0.39 | 6 |
| HR 1179  | S | 6292 | 4.17 | 1.77 | 7.35 | 0.08 | 300 | 7.28 | 0.05 | 18 | 6292 | 4.33 | 1.66 | 7.35 | 0.08 | 300 | 7.36 | 0.05 | 18 |
| HR 1232  | S | 4905 | 3.26 | 1.50 | 7.52 | 0.11 | 353 | 7.56 | 0.13 | 20 | 4905 | 3.18 | 1.52 | 7.51 | 0.11 | 353 | 7.51 | 0.13 | 20 |
| HR 159   | S | 5501 | 4.28 | 1.08 | 7.26 | 0.06 | 350 | 7.19 | 0.05 | 17 | 5501 | 4.42 | 0.81 | 7.29 | 0.06 | 350 | 7.28 | 0.05 | 17 |
| HR 1665  | S | 5935 | 3.92 | 1.63 | 7.48 | 0.06 | 347 | 7.43 | 0.05 | 21 | 5935 | 4.04 | 1.55 | 7.49 | 0.06 | 347 | 7.49 | 0.05 | 21 |
| HR 1685  | S | 4976 | 3.48 | 1.20 | 7.55 | 0.09 | 356 | 7.57 | 0.14 | 20 | 4976 | 3.41 | 1.32 | 7.51 | 0.09 | 356 | 7.51 | 0.14 | 20 |
| HR 1925  | S | 5243 | 4.53 | 1.00 | 7.54 | 0.06 | 238 | 7.61 | 0.08 | 13 | 5243 | 4.37 | 1.14 | 7.52 | 0.06 | 238 | 7.52 | 0.08 | 13 |
| HR 1980  | S | 6095 | 4.38 | 1.39 | 7.49 | 0.06 | 345 | 7.47 | 0.04 | 20 | 6095 | 4.44 | 1.32 | 7.50 | 0.05 | 345 | 7.50 | 0.04 | 20 |
| HR 2208  | S | 5762 | 4.50 | 1.42 | 7.45 | 0.06 | 356 | 7.45 | 0.06 | 19 | 5762 | 4.50 | 1.42 | 7.45 | 0.06 | 356 | 7.45 | 0.06 | 19 |
| HR 244   | S | 6201 | 4.09 | 1.58 | 7.62 | 0.07 | 368 | 7.58 | 0.07 | 29 | 6201 | 4.19 | 1.51 | 7.63 | 0.07 | 368 | 7.63 | 0.07 | 29 |
| HR 2692  | S | 4995 | 3.47 | 1.09 | 7.08 | 0.08 | 365 | 7.08 | 0.05 | 16 | 4995 | 3.47 | 0.97 | 7.08 | 0.07 | 365 | 7.04 | 0.06 | 16 |
| HR 2866  | S | 6358 | 4.25 | 1.76 | 7.40 | 0.07 | 301 | 7.36 | 0.06 | 18 | 6358 | 4.34 | 1.69 | 7.41 | 0.07 | 301 | 7.41 | 0.06 | 18 |
| HR 3193  | S | 5989 | 3.91 | 1.68 | 7.50 | 0.07 | 352 | 7.42 | 0.07 | 23 | 5989 | 4.07 | 1.58 | 7.50 | 0.06 | 352 | 7.50 | 0.07 | 23 |
| HR 3271  | S | 5975 | 3.96 | 1.62 | 7.56 | 0.06 | 364 | 7.51 | 0.05 | 25 | 5975 | 4.07 | 1.55 | 7.57 | 0.06 | 364 | 7.57 | 0.05 | 25 |
| HR 3395  | S | 6222 | 4.32 | 1.52 | 7.52 | 0.05 | 396 | 7.52 | 0.06 | 24 | 6222 | 4.32 | 1.52 | 7.52 | 0.05 | 396 | 7.52 | 0.06 | 24 |
| HR 357   | S | 6491 | 3.72 | 2.76 | 7.52 | 0.31 | 110 | 7.61 | 0.32 | 10 | 6491 | 3.49 | 2.91 | 7.51 | 0.31 | 110 | 7.50 | 0.33 | 10 |
| HR 3762  | S | 5153 | 3.40 | 1.31 | 7.10 | 0.06 | 366 | 7.16 | 0.08 | 20 | 5153 | 3.26 | 1.36 | 7.08 | 0.06 | 366 | 7.04 | 0.09 | 20 |
| HR 3901  | S | 6081 | 4.01 | 1.66 | 7.57 | 0.06 | 369 | 7.49 | 0.06 | 20 | 6081 | 4.19 | 1.53 | 7.58 | 0.06 | 369 | 7.58 | 0.05 | 20 |
| HR 4051  | S | 5978 | 4.18 | 1.44 | 7.45 | 0.06 | 339 | 7.45 | 0.06 | 22 | 5978 | 4.18 | 1.44 | 7.45 | 0.06 | 339 | 7.45 | 0.06 | 22 |
| HR 407   | S | 6520 | 3.72 | 3.15 | 7.41 | 0.25 | 131 | 7.46 | 0.33 | 17 | 6520 | 3.58 | 3.19 | 7.41 | 0.25 | 131 | 7.41 | 0.33 | 17 |
| HR 4285  | S | 5916 | 3.76 | 1.66 | 7.17 | 0.06 | 316 | 7.15 | 0.04 | 22 | 5916 | 3.81 | 1.63 | 7.18 | 0.06 | 316 | 7.18 | 0.04 | 22 |
| HR 448   | S | 5861 | 3.99 | 1.51 | 7.66 | 0.06 | 374 | 7.64 | 0.08 | 26 | 5861 | 4.04 | 1.48 | 7.66 | 0.06 | 374 | 7.67 | 0.08 | 26 |

| | | | | | | | | | | | | | | | | |
|---|---|---|---|---|---|---|---|---|---|---|---|---|---|---|---|---|
| HR 4864 | S | 5615 | 4.49 | 1.30 | 7.50 | 0.06 | 389 | 7.51 | 0.08 | 14 | 5615 | 4.46 | 1.34 | 7.50 | 0.06 | 389 | 7.49 | 0.08 | 14 |
| HR 495  | S | 4661 | 3.08 | 1.39 | 7.79 | 0.13 | 164 | 8.13 | 0.19 | 20 | 4661 | 3.00 | 1.39 | 7.78 | 0.13 | 164 | 8.08 | 0.19 | 20 |
| HR 511  | S | 5422 | 4.55 | 0.91 | 7.53 | 0.06 | 370 | 7.51 | 0.09 | 16 | 5422 | 4.58 | 0.81 | 7.54 | 0.06 | 370 | 7.54 | 0.09 | 16 |
| HR 5258 | S | 6417 | 3.71 | 2.57 | 7.58 | 0.16 | 242 | 7.42 | 0.14 | 18 | 6417 | 4.10 | 2.51 | 7.58 | 0.15 | 242 | 7.55 | 0.14 | 18 |
| HR 5317 | S | 6437 | 3.74 | 2.93 | 7.59 | 0.16 | 180 | 7.51 | 0.07 | 13 | 6437 | 3.95 | 2.96 | 7.59 | 0.16 | 180 | 7.56 | 0.07 | 13 |
| HR 5335 | S | 4862 | 3.29 | 1.68 | 7.76 | 0.14 | 334 | 7.80 | 0.20 | 18 | 4862 | 3.20 | 1.72 | 7.74 | 0.15 | 334 | 7.74 | 0.20 | 18 |
| HR 5387 | S | 6753 | 4.29 | 2.17 | 7.46 | 0.09 | 312 | 7.43 | 0.06 | 20 | 6753 | 4.36 | 2.13 | 7.46 | 0.08 | 312 | 7.46 | 0.06 | 20 |
| HR 5504 | S | 5990 | 4.01 | 1.78 | 7.65 | 0.07 | 340 | 7.51 | 0.07 | 22 | 5990 | 4.32 | 1.58 | 7.66 | 0.06 | 340 | 7.67 | 0.07 | 22 |
| HR 5630 | S | 6186 | 4.30 | 1.46 | 7.52 | 0.06 | 340 | 7.49 | 0.06 | 23 | 6186 | 4.36 | 1.40 | 7.53 | 0.05 | 340 | 7.53 | 0.06 | 23 |
| HR 5706 | S | 4785 | 3.52 | 1.02 | 7.86 | 0.10 | 248 | 8.06 | 0.12 | 11 | 4785 | 3.05 | 1.30 | 7.72 | 0.10 | 248 | 7.71 | 0.12 | 11 |
| HR 5740 | S | 5956 | 3.97 | 1.87 | 7.78 | 0.09 | 354 | 7.73 | 0.08 | 24 | 5956 | 4.07 | 1.80 | 7.78 | 0.08 | 354 | 7.78 | 0.08 | 24 |
| HR 6105 | S | 5934 | 4.16 | 1.38 | 7.57 | 0.06 | 355 | 7.53 | 0.06 | 25 | 5934 | 4.24 | 1.29 | 7.57 | 0.06 | 355 | 7.58 | 0.06 | 25 |
| HR 6106 | S | 6009 | 3.97 | 1.68 | 7.60 | 0.07 | 354 | 7.57 | 0.06 | 24 | 6009 | 4.04 | 1.64 | 7.61 | 0.07 | 354 | 7.61 | 0.06 | 24 |
| HR 6269 | S | 5634 | 3.95 | 1.34 | 7.57 | 0.06 | 361 | 7.56 | 0.07 | 22 | 5634 | 3.97 | 1.32 | 7.57 | 0.06 | 361 | 7.57 | 0.07 | 22 |
| HR 6301 | S | 4955 | 3.31 | 1.39 | 7.46 | 0.08 | 350 | 7.53 | 0.11 | 20 | 4955 | 3.14 | 1.48 | 7.43 | 0.09 | 350 | 7.43 | 0.11 | 20 |
| HR 6372 | S | 5744 | 4.02 | 1.49 | 7.73 | 0.07 | 369 | 7.72 | 0.08 | 25 | 5744 | 4.03 | 1.48 | 7.73 | 0.07 | 369 | 7.73 | 0.08 | 25 |
| HR 6465 | S | 5719 | 4.43 | 1.06 | 7.51 | 0.05 | 356 | 7.50 | 0.06 | 21 | 5719 | 4.44 | 1.04 | 7.51 | 0.05 | 356 | 7.51 | 0.06 | 21 |
| HR 6516 | S | 5583 | 4.15 | 1.69 | 7.49 | 0.08 | 341 | 7.43 | 0.11 | 19 | 5583 | 4.28 | 1.57 | 7.50 | 0.07 | 341 | 7.50 | 0.10 | 19 |
| HR 6669 | S | 6134 | 4.27 | 1.45 | 7.46 | 0.05 | 340 | 7.46 | 0.05 | 22 | 6134 | 4.27 | 1.45 | 7.46 | 0.05 | 340 | 7.46 | 0.05 | 22 |
| HR 672  | S | 6079 | 4.07 | 1.56 | 7.61 | 0.06 | 358 | 7.53 | 0.06 | 22 | 6079 | 4.24 | 1.46 | 7.62 | 0.06 | 358 | 7.62 | 0.05 | 22 |
| HR 6722 | S | 5539 | 3.72 | 1.54 | 7.50 | 0.07 | 361 | 7.49 | 0.10 | 22 | 5539 | 3.75 | 1.51 | 7.51 | 0.07 | 361 | 7.51 | 0.11 | 22 |
| HR 6756 | S | 4907 | 3.39 | 1.30 | 7.49 | 0.09 | 353 | 7.59 | 0.17 | 21 | 4907 | 3.14 | 1.47 | 7.44 | 0.10 | 353 | 7.42 | 0.17 | 21 |
| HR 6806 | S | 5042 | 4.58 | 0.98 | 7.30 | 0.05 | 163 | 7.60 | 0.06 | 6  | 5042 | 4.12 | 1.39 | 7.18 | 0.05 | 163 | 7.25 | 0.04 | 6  |
| HR 6847 | S | 5791 | 4.33 | 1.09 | 7.47 | 0.05 | 356 | 7.50 | 0.08 | 22 | 5791 | 4.29 | 1.14 | 7.47 | 0.05 | 356 | 7.47 | 0.07 | 22 |
| HR 6907 | S | 6311 | 4.04 | 1.76 | 7.61 | 0.08 | 355 | 7.57 | 0.06 | 25 | 6311 | 4.12 | 1.71 | 7.61 | 0.08 | 355 | 7.61 | 0.06 | 25 |
| HR 6950 | S | 5337 | 3.85 | 1.28 | 7.50 | 0.07 | 364 | 7.59 | 0.11 | 20 | 5337 | 3.64 | 1.50 | 7.46 | 0.09 | 364 | 7.45 | 0.12 | 20 |
| HR 7079 | S | 6319 | 4.31 | 1.73 | 7.41 | 0.06 | 314 | 7.40 | 0.06 | 23 | 6319 | 4.33 | 1.71 | 7.41 | 0.06 | 314 | 7.41 | 0.06 | 23 |
| HR 7291 | S | 6173 | 4.34 | 1.51 | 7.66 | 0.08 | 359 | 7.70 | 0.06 | 26 | 6173 | 4.27 | 1.57 | 7.66 | 0.08 | 359 | 7.66 | 0.06 | 26 |
| HR 7522 | S | 6026 | 3.89 | 1.86 | 7.60 | 0.07 | 355 | 7.56 | 0.07 | 23 | 6026 | 3.97 | 1.82 | 7.60 | 0.07 | 355 | 7.60 | 0.07 | 23 |
| HR 7569 | S | 5774 | 4.10 | 1.26 | 7.35 | 0.06 | 339 | 7.29 | 0.06 | 18 | 5774 | 4.23 | 1.12 | 7.36 | 0.06 | 339 | 7.37 | 0.06 | 18 |
| HR 761  | S | 6230 | 3.85 | 1.86 | 7.26 | 0.07 | 321 | 7.21 | 0.06 | 22 | 6230 | 3.95 | 1.81 | 7.26 | 0.07 | 321 | 7.26 | 0.06 | 22 |
| HR 7637 | S | 5947 | 4.24 | 1.29 | 7.42 | 0.09 | 347 | 7.34 | 0.06 | 21 | 5947 | 4.43 | 1.07 | 7.44 | 0.09 | 347 | 7.44 | 0.06 | 21 |
| HR 7793 | S | 6261 | 4.36 | 1.55 | 7.41 | 0.07 | 309 | 7.40 | 0.07 | 22 | 6261 | 4.37 | 1.54 | 7.41 | 0.07 | 309 | 7.41 | 0.07 | 22 |
| HR 7855 | S | 6217 | 4.13 | 1.69 | 7.50 | 0.06 | 339 | 7.43 | 0.06 | 21 | 6217 | 4.29 | 1.57 | 7.51 | 0.06 | 339 | 7.51 | 0.06 | 21 |
| HR 7907 | S | 6189 | 4.25 | 1.74 | 7.66 | 0.07 | 340 | 7.65 | 0.06 | 23 | 6189 | 4.27 | 1.73 | 7.66 | 0.07 | 340 | 7.66 | 0.06 | 23 |
| HR 8133 | S | 5874 | 3.88 | 1.45 | 7.47 | 0.06 | 346 | 7.32 | 0.04 | 21 | 5874 | 4.20 | 1.21 | 7.49 | 0.05 | 346 | 7.49 | 0.05 | 21 |
| HR 8148 | S | 5433 | 4.43 | 0.39 | 7.21 | 0.07 | 267 | 7.29 | 0.06 | 15 | 5433 | 4.26 | 0.63 | 7.20 | 0.07 | 267 | 7.20 | 0.06 | 15 |
| HR 8170 | S | 5903 | 4.26 | 1.81 | 7.35 | 0.15 | 250 | 7.44 | 0.12 | 17 | 5903 | 4.06 | 1.96 | 7.34 | 0.16 | 250 | 7.31 | 0.12 | 17 |

| Name | | T | log g | v | A1 | σ1 | N1 | A2 | σ2 | N2 | T | log g | v | A1 | σ1 | N1 | A2 | σ2 | N2 |
|---|---|---|---|---|---|---|---|---|---|---|---|---|---|---|---|---|---|---|---|
| HR 857 | S | 5201 | 4.59 | 1.56 | 7.54 | 0.08 | 297 | 7.64 | 0.16 | 18 | 5201 | 4.32 | 1.91 | 7.48 | 0.09 | 297 | 7.47 | 0.16 | 18 |
| HR 8631 | S | 5453 | 3.69 | 1.29 | 7.29 | 0.07 | 359 | 7.15 | 0.05 | 19 | 5453 | 3.99 | 0.97 | 7.33 | 0.06 | 359 | 7.33 | 0.07 | 19 |
| HR 8832 | S | 4883 | 4.60 | 1.27 | 7.58 | 0.07 | 166 | 7.77 | 0.20 | 11 | 4883 | 4.10 | 1.59 | 7.42 | 0.07 | 166 | 7.36 | 0.19 | 11 |
| HR 8924 | S | 4784 | 3.26 | 1.76 | 7.81 | 0.16 | 320 | 7.88 | 0.22 | 17 | 4784 | 3.09 | 1.83 | 7.76 | 0.16 | 320 | 7.76 | 0.22 | 17 |
| HR 8964 | S | 5821 | 4.43 | 1.43 | 7.57 | 0.06 | 357 | 7.52 | 0.05 | 20 | 5821 | 4.53 | 1.30 | 7.58 | 0.06 | 357 | 7.58 | 0.05 | 20 |
| HR 9074 | S | 6227 | 4.36 | 1.54 | 7.50 | 0.07 | 327 | 7.47 | 0.05 | 19 | 6227 | 4.43 | 1.47 | 7.50 | 0.07 | 327 | 7.51 | 0.05 | 19 |
| HR 9075 | S | 6112 | 4.39 | 1.37 | 7.53 | 0.06 | 339 | 7.50 | 0.06 | 22 | 6112 | 4.46 | 1.29 | 7.54 | 0.06 | 339 | 7.54 | 0.06 | 22 |
| LP 837-53 | S | 3666 | 4.89 | 1.25 | 7.82 | 0.20 | 22 | 11.39 | 0.00 | 1 | 3666 | 4.89 | 1.31 | 7.81 | 0.20 | 22 | 11.39 | 0.00 | 1 |
| 23 H. Cam | S | 6215 | 4.38 | 1.38 | 7.41 | 0.06 | 355 | 7.39 | 0.04 | 22 | 6215 | 4.43 | 1.33 | 7.41 | 0.06 | 355 | 7.41 | 0.04 | 22 |
| V* AK Lep | S | 4925 | 4.60 | 1.70 | 7.42 | 0.06 | 152 | 7.50 | 0.08 | 10 | 4925 | 4.40 | 1.91 | 7.38 | 0.06 | 152 | 7.38 | 0.07 | 10 |
| V* AR Lac | S | 5342 | 3.61 | 1.75 | 8.48 | 0.41 | 21 | 8.91 | 0.37 | 2 | 5342 | 3.00 | 2.32 | 8.27 | 0.41 | 21 | 8.50 | 0.32 | 2 |
| V* DE Boo | S | 5260 | 4.52 | 1.34 | 7.52 | 0.07 | 366 | 7.63 | 0.12 | 13 | 5260 | 4.25 | 1.67 | 7.47 | 0.08 | 366 | 7.46 | 0.12 | 13 |
| V* HN Peg | S | 5972 | 4.44 | 1.80 | 7.47 | 0.09 | 324 | 7.46 | 0.08 | 20 | 5972 | 4.45 | 1.79 | 7.47 | 0.09 | 324 | 7.47 | 0.08 | 20 |
| V* IL Aqr | S | 3935 | 5.00 | 1.75 | 6.49 | 0.43 | 9 | 10.51 | 0.00 | 1 | 3935 | 5.00 | 2.17 | 6.45 | 0.44 | 9 | 10.45 | 0.00 | 1 |
| V* pi.01 UMa | S | 5921 | 4.46 | 1.98 | 7.46 | 0.09 | 344 | 7.42 | 0.09 | 18 | 5921 | 4.55 | 1.89 | 7.47 | 0.09 | 344 | 7.47 | 0.09 | 18 |
| V* V2213 Oph | S | 6030 | 4.39 | 1.48 | 7.50 | 0.06 | 338 | 7.48 | 0.07 | 23 | 6030 | 4.43 | 1.44 | 7.50 | 0.06 | 338 | 7.50 | 0.07 | 23 |
| V* V2215 Oph | S | 4463 | 4.68 | 0.15 | 7.37 | 0.12 | 301 | 7.57 | 0.16 | 7 | 4463 | 4.14 | 1.60 | 7.19 | 0.14 | 301 | 7.18 | 0.14 | 7 |
| V* V2502 Oph | S | 6969 | 4.02 | 2.59 | 7.39 | 0.12 | 264 | 7.43 | 0.07 | 25 | 6969 | 3.91 | 2.61 | 7.39 | 0.12 | 264 | 7.39 | 0.07 | 25 |
| V* V2689 Ori | S | 4073 | 4.72 | 0.15 | 7.92 | 0.30 | 264 | 7.83 | 0.04 | 2 | 4073 | 4.88 | 0.10 | 7.95 | 0.31 | 264 | 7.95 | 0.04 | 2 |
| V* V376 Peg | S | 6066 | 4.33 | 1.33 | 7.50 | 0.07 | 360 | 7.48 | 0.06 | 25 | 6066 | 4.38 | 1.27 | 7.50 | 0.07 | 360 | 7.50 | 0.06 | 25 |
| V* V450 And | S | 5653 | 4.46 | 1.29 | 7.44 | 0.06 | 362 | 7.47 | 0.05 | 17 | 5653 | 4.38 | 1.41 | 7.42 | 0.07 | 362 | 7.42 | 0.06 | 17 |
| V* V819 Her | S | 5699 | 3.28 | 0.58 | 7.59 | 0.06 | 185 | 6.86 | 0.05 | 11 | 5699 | 3.68 | 0.44 | 7.59 | 0.06 | 185 | 7.05 | 0.06 | 11 |
| Wolf 1008 | S | 5787 | 4.07 | 1.39 | 7.12 | 0.06 | 313 | 7.02 | 0.06 | 18 | 5787 | 4.31 | 1.15 | 7.14 | 0.06 | 313 | 7.14 | 0.06 | 18 |
| 10 Tau | H | 6013 | 4.05 | 1.61 | 7.42 | 0.07 | 566 | 7.40 | 0.12 | 73 | 6013 | 4.10 | 1.58 | 7.42 | 0.07 | 566 | 7.42 | 0.12 | 73 |
| 11 LMi | H | 5498 | 4.43 | 1.54 | 7.79 | 0.09 | 572 | 7.96 | 0.21 | 70 | 5498 | 4.00 | 1.89 | 7.73 | 0.11 | 572 | 7.71 | 0.22 | 70 |
| 15 LMi | H | 5916 | 4.06 | 1.49 | 7.64 | 0.06 | 638 | 7.64 | 0.09 | 85 | 5916 | 4.06 | 1.49 | 7.64 | 0.06 | 638 | 7.64 | 0.09 | 85 |
| 16 Cyg B | H | 5753 | 4.34 | 1.22 | 7.58 | 0.06 | 613 | 7.63 | 0.14 | 75 | 5753 | 4.23 | 1.36 | 7.56 | 0.06 | 613 | 7.56 | 0.15 | 75 |
| 37 Gem | H | 5932 | 4.40 | 1.20 | 7.37 | 0.06 | 600 | 7.41 | 0.11 | 76 | 5932 | 4.30 | 1.30 | 7.36 | 0.06 | 600 | 7.36 | 0.12 | 76 |
| 88 Leo | H | 6030 | 4.39 | 1.25 | 7.54 | 0.06 | 616 | 7.56 | 0.11 | 84 | 6030 | 4.35 | 1.29 | 7.54 | 0.06 | 616 | 7.54 | 0.11 | 84 |
| b Aql | H | 5466 | 4.10 | 1.12 | 7.84 | 0.07 | 375 | 8.02 | 0.13 | 57 | 5466 | 3.69 | 1.29 | 7.81 | 0.07 | 375 | 7.80 | 0.14 | 57 |
| bet CVn | H | 5865 | 4.40 | 1.07 | 7.29 | 0.05 | 582 | 7.35 | 0.10 | 70 | 5865 | 4.29 | 1.21 | 7.28 | 0.06 | 582 | 7.29 | 0.10 | 70 |
| iot Psc | H | 6177 | 4.08 | 1.62 | 7.38 | 0.08 | 561 | 7.39 | 0.14 | 78 | 6177 | 4.06 | 1.64 | 7.38 | 0.08 | 561 | 7.38 | 0.14 | 78 |
| sig Dra | H | 5338 | 4.57 | 0.94 | 7.38 | 0.08 | 420 | 7.53 | 0.21 | 43 | 5338 | 4.20 | 1.15 | 7.32 | 0.08 | 420 | 7.29 | 0.22 | 43 |
| HD 10086 | H | 5658 | 4.46 | 1.16 | 7.66 | 0.09 | 477 | 7.79 | 0.10 | 49 | 5658 | 4.17 | 1.34 | 7.64 | 0.09 | 477 | 7.64 | 0.10 | 49 |
| HD 101177 | H | 5964 | 4.43 | 1.17 | 7.37 | 0.06 | 590 | 7.40 | 0.12 | 77 | 5964 | 4.35 | 1.26 | 7.36 | 0.06 | 590 | 7.36 | 0.12 | 77 |
| HD 102158 | H | 5781 | 4.31 | 1.09 | 7.09 | 0.06 | 551 | 7.17 | 0.14 | 70 | 5781 | 4.13 | 1.23 | 7.08 | 0.06 | 551 | 7.05 | 0.15 | 70 |
| HD 112257 | H | 5659 | 4.31 | 1.13 | 7.48 | 0.07 | 609 | 7.59 | 0.13 | 68 | 5659 | 4.09 | 1.40 | 7.45 | 0.08 | 609 | 7.45 | 0.14 | 68 |
| HD 116442 | H | 5281 | 4.62 | 0.88 | 7.19 | 0.06 | 404 | 7.26 | 0.10 | 33 | 5281 | 4.48 | 1.02 | 7.17 | 0.05 | 404 | 7.18 | 0.10 | 33 |
| HD 12051 | H | 5400 | 4.39 | 1.68 | 7.71 | 0.07 | 301 | 7.82 | 0.17 | 41 | 5400 | 4.12 | 1.77 | 7.68 | 0.07 | 301 | 7.68 | 0.17 | 41 |
| HD 130087 | H | 5991 | 4.12 | 1.55 | 7.75 | 0.08 | 602 | 7.60 | 0.14 | 72 | 5991 | 4.42 | 1.29 | 7.77 | 0.08 | 602 | 7.77 | 0.13 | 72 |
| HD 13043 | H | 5877 | 4.15 | 1.43 | 7.59 | 0.06 | 632 | 7.63 | 0.12 | 85 | 5877 | 4.04 | 1.47 | 7.58 | 0.06 | 632 | 7.54 | 0.12 | 85 |
| HD 130948 | H | 5983 | 4.43 | 1.47 | 7.50 | 0.07 | 583 | 7.53 | 0.10 | 73 | 5983 | 4.37 | 1.53 | 7.50 | 0.07 | 583 | 7.50 | 0.10 | 73 |
| HD 135101 | H | 5637 | 4.24 | 1.15 | 7.56 | 0.07 | 602 | 7.69 | 0.13 | 75 | 5637 | 3.98 | 1.45 | 7.52 | 0.08 | 602 | 7.53 | 0.14 | 75 |
| HD 139323 | H | 5046 | 4.48 | 0.79 | 7.90 | 0.10 | 258 | 8.00 | 0.13 | 15 | 5046 | 4.24 | 0.95 | 7.85 | 0.09 | 258 | 7.85 | 0.13 | 15 |
| HD 144579 | H | 5308 | 4.67 | 0.73 | 6.92 | 0.08 | 386 | 6.98 | 0.10 | 18 | 5308 | 4.57 | 0.86 | 6.91 | 0.07 | 386 | 6.92 | 0.09 | 18 |
| HD 147044 | H | 5890 | 4.39 | 1.43 | 7.45 | 0.09 | 570 | 7.49 | 0.15 | 72 | 5890 | 4.31 | 1.51 | 7.44 | 0.09 | 570 | 7.44 | 0.15 | 72 |
| HD 152792 | H | 5675 | 3.84 | 1.33 | 7.15 | 0.06 | 570 | 7.18 | 0.10 | 71 | 5675 | 3.78 | 1.36 | 7.15 | 0.06 | 570 | 7.15 | 0.11 | 71 |
| HD 159222 | H | 5788 | 4.39 | 1.36 | 7.63 | 0.07 | 621 | 7.75 | 0.12 | 78 | 5788 | 4.11 | 1.62 | 7.60 | 0.09 | 621 | 7.59 | 0.13 | 78 |
| HD 170778 | H | 5932 | 4.46 | 1.64 | 7.47 | 0.08 | 560 | 7.51 | 0.12 | 66 | 5932 | 4.38 | 1.72 | 7.47 | 0.08 | 560 | 7.47 | 0.12 | 66 |
| HD 182488 | H | 5362 | 4.45 | 1.29 | 7.67 | 0.09 | 559 | 7.90 | 0.22 | 62 | 5362 | 3.88 | 1.74 | 7.55 | 0.11 | 559 | 7.49 | 0.25 | 62 |
| HD 183341 | H | 5952 | 4.27 | 1.47 | 7.57 | 0.08 | 584 | 7.56 | 0.11 | 73 | 5952 | 4.28 | 1.38 | 7.57 | 0.07 | 584 | 7.53 | 0.11 | 73 |
| HD 193664 | H | 5945 | 4.44 | 1.25 | 7.41 | 0.05 | 572 | 7.40 | 0.08 | 67 | 5945 | 4.45 | 1.24 | 7.41 | 0.05 | 572 | 7.41 | 0.08 | 67 |
| HD 197076 | H | 5844 | 4.46 | 1.19 | 7.43 | 0.05 | 593 | 7.47 | 0.10 | 72 | 5844 | 4.37 | 1.30 | 7.42 | 0.05 | 593 | 7.42 | 0.11 | 72 |
| HD 210460 | H | 5529 | 3.52 | 1.45 | 7.22 | 0.06 | 611 | 7.24 | 0.13 | 77 | 5529 | 3.49 | 1.46 | 7.22 | 0.06 | 611 | 7.22 | 0.13 | 77 |
| HD 210640 | H | 6377 | 3.98 | 1.52 | 7.72 | 0.11 | 456 | 7.27 | 0.15 | 61 | 6377 | 4.37 | 1.43 | 7.72 | 0.11 | 456 | 7.43 | 0.15 | 61 |
| HD 223238 | H | 5889 | 4.30 | 1.35 | 7.56 | 0.07 | 592 | 7.54 | 0.12 | 67 | 5889 | 4.35 | 1.29 | 7.57 | 0.07 | 592 | 7.57 | 0.12 | 67 |
| HD 24213 | H | 6053 | 4.18 | 1.53 | 7.59 | 0.07 | 618 | 7.60 | 0.10 | 79 | 6053 | 4.16 | 1.52 | 7.59 | 0.06 | 618 | 7.56 | 0.11 | 79 |
| HD 28005 | H | 5727 | 4.27 | 1.38 | 7.78 | 0.09 | 593 | 7.89 | 0.15 | 68 | 5727 | 4.03 | 1.61 | 7.75 | 0.10 | 593 | 7.74 | 0.16 | 68 |
| HD 38858 | H | 5798 | 4.48 | 1.19 | 7.34 | 0.05 | 602 | 7.40 | 0.14 | 73 | 5798 | 4.37 | 1.24 | 7.33 | 0.05 | 602 | 7.31 | 0.15 | 73 |
| HD 39881 | H | 5719 | 4.27 | 1.08 | 7.41 | 0.06 | 616 | 7.52 | 0.13 | 76 | 5719 | 4.06 | 1.31 | 7.39 | 0.07 | 616 | 7.39 | 0.14 | 76 |
| HD 47127 | H | 5616 | 4.32 | 1.12 | 7.65 | 0.04 | 445 | 7.77 | 0.12 | 66 | 5616 | 4.03 | 1.26 | 7.63 | 0.04 | 445 | 7.63 | 0.12 | 66 |
| HD 5372 | H | 5847 | 4.37 | 1.44 | 7.73 | 0.09 | 599 | 7.81 | 0.12 | 72 | 5847 | 4.18 | 1.63 | 7.70 | 0.10 | 599 | 7.70 | 0.12 | 72 |
| HD 58781 | H | 5576 | 4.41 | 1.20 | 7.64 | 0.07 | 620 | 7.82 | 0.18 | 78 | 5576 | 4.04 | 1.51 | 7.57 | 0.09 | 620 | 7.54 | 0.19 | 78 |
| HD 64090 | H | 5528 | 4.62 | 2.34 | 5.78 | 0.19 | 154 | 5.83 | 0.42 | 17 | 5528 | 4.48 | 2.46 | 5.77 | 0.19 | 154 | 5.77 | 0.42 | 17 |
| HD 65583 | H | 5342 | 4.55 | 0.90 | 6.84 | 0.05 | 337 | 6.90 | 0.06 | 28 | 5342 | 4.43 | 0.95 | 6.82 | 0.05 | 337 | 6.79 | 0.06 | 28 |
| HD 68017 | H | 5565 | 4.43 | 0.66 | 7.14 | 0.07 | 463 | 7.33 | 0.21 | 60 | 5565 | 4.04 | 0.93 | 7.11 | 0.07 | 463 | 7.10 | 0.22 | 60 |
| HD 71148 | H | 5835 | 4.39 | 1.17 | 7.53 | 0.05 | 624 | 7.59 | 0.11 | 78 | 5835 | 4.27 | 1.32 | 7.52 | 0.05 | 624 | 7.52 | 0.12 | 78 |
| HD 72760 | H | 5328 | 4.56 | 1.41 | 7.53 | 0.08 | 550 | 7.73 | 0.20 | 57 | 5328 | 4.08 | 1.87 | 7.44 | 0.10 | 550 | 7.40 | 0.22 | 57 |
| HD 76752 | H | 5685 | 4.28 | 1.19 | 7.53 | 0.05 | 472 | 7.60 | 0.09 | 68 | 5685 | 4.13 | 1.35 | 7.51 | 0.05 | 472 | 7.51 | 0.09 | 68 |
| HD 76909 | H | 5598 | 4.15 | 1.50 | 7.79 | 0.09 | 583 | 7.90 | 0.19 | 74 | 5598 | 3.90 | 1.70 | 7.75 | 0.10 | 583 | 7.75 | 0.20 | 74 |
| HD 8648 | H | 5711 | 4.16 | 1.46 | 7.62 | 0.07 | 612 | 7.70 | 0.13 | 78 | 5711 | 3.98 | 1.59 | 7.61 | 0.08 | 612 | 7.60 | 0.14 | 78 |
| HD 89269 | H | 5635 | 4.49 | 1.07 | 7.36 | 0.05 | 626 | 7.51 | 0.14 | 73 | 5635 | 4.23 | 1.33 | 7.33 | 0.06 | 626 | 7.32 | 0.16 | 73 |
| HD 98618 | H | 5735 | 4.35 | 1.13 | 7.52 | 0.06 | 619 | 7.64 | 0.14 | 82 | 5735 | 4.11 | 1.42 | 7.48 | 0.07 | 619 | 7.49 | 0.15 | 82 |
| HD 9986 | H | 5791 | 4.42 | 1.32 | 7.56 | 0.07 | 571 | 7.60 | 0.13 | 69 | 5791 | 4.33 | 1.44 | 7.54 | 0.07 | 571 | 7.54 | 0.14 | 69 |
| HR 2208 | H | 5762 | 4.50 | 1.50 | 7.48 | 0.07 | 585 | 7.52 | 0.13 | 69 | 5762 | 4.42 | 1.59 | 7.47 | 0.07 | 585 | 7.47 | 0.13 | 69 |
| HR 511 | H | 5422 | 4.55 | 0.76 | 7.64 | 0.05 | 294 | 7.79 | 0.14 | 51 | 5422 | 4.26 | 0.96 | 7.59 | 0.05 | 294 | 7.55 | 0.15 | 51 |
| HR 6465 | H | 5719 | 4.43 | 1.16 | 7.56 | 0.06 | 602 | 7.64 | 0.13 | 69 | 5719 | 4.30 | 1.35 | 7.54 | 0.06 | 602 | 7.55 | 0.14 | 69 |
| HR 6847 | H | 5791 | 4.33 | 1.21 | 7.52 | 0.05 | 620 | 7.59 | 0.11 | 76 | 5791 | 4.17 | 1.38 | 7.50 | 0.06 | 620 | 7.50 | 0.12 | 76 |
| HR 8964 | H | 5821 | 4.43 | 1.50 | 7.63 | 0.07 | 639 | 7.71 | 0.12 | 82 | 5821 | 4.25 | 1.65 | 7.61 | 0.08 | 639 | 7.61 | 0.13 | 82 |
| V* BZ Cet | H | 5014 | 4.52 | 1.11 | 7.76 | 0.09 | 416 | 8.04 | 0.18 | 35 | 5014 | 3.77 | 1.56 | 7.56 | 0.10 | 416 | 7.50 | 0.19 | 35 |
| 1 Hya | E | 6359 | 4.11 | 2.69 | 7.54 | 0.35 | 86 | 7.68 | 0.37 | 15 | 6359 | 3.76 | 2.83 | 7.53 | 0.35 | 86 | 7.53 | 0.38 | 15 |
| 101 Tau | E | 6465 | 4.31 | 2.66 | 7.69 | 0.25 | 155 | 7.79 | 0.23 | 13 | 6465 | 4.10 | 2.75 | 7.69 | 0.25 | 155 | 7.69 | 0.22 | 13 |
| 14 Boo | E | 6180 | 3.91 | 1.81 | 7.58 | 0.09 | 501 | 7.58 | 0.15 | 76 | 6180 | 3.91 | 1.84 | 7.57 | 0.09 | 501 | 7.54 | 0.14 | 76 |
| 15 Peg | E | 6452 | 4.09 | 1.95 | 6.96 | 0.09 | 373 | 6.92 | 0.13 | 66 | 6452 | 4.18 | 1.91 | 6.96 | 0.09 | 373 | 6.96 | 0.13 | 66 |
| 22 Lyn | E | 6395 | 4.34 | 1.65 | 7.27 | 0.09 | 418 | 7.29 | 0.16 | 69 | 6395 | 4.30 | 1.68 | 7.27 | 0.09 | 418 | 7.27 | 0.16 | 69 |
| 30 Ari B | E | 6257 | 4.34 | 3.36 | 7.69 | 0.33 | 95 | 7.88 | 0.40 | 13 | 6257 | 3.87 | 3.51 | 7.68 | 0.33 | 95 | 7.68 | 0.40 | 13 |
| 34 Peg | E | 6258 | 3.92 | 1.94 | 7.52 | 0.09 | 461 | 7.46 | 0.13 | 67 | 6258 | 4.07 | 1.87 | 7.52 | 0.09 | 461 | 7.53 | 0.13 | 67 |
| 36 Dra | E | 6522 | 4.07 | 1.98 | 7.17 | 0.11 | 380 | 7.13 | 0.19 | 71 | 6522 | 4.17 | 1.94 | 7.17 | 0.11 | 380 | 7.17 | 0.18 | 71 |
| 38 Cet | E | 6480 | 3.87 | 1.89 | 7.30 | 0.10 | 426 | 7.29 | 0.16 | 74 | 6480 | 3.90 | 1.88 | 7.31 | 0.10 | 426 | 7.31 | 0.16 | 74 |
| 4 Aqr | E | 6440 | 3.79 | 3.48 | 7.48 | 0.28 | 198 | 7.33 | 0.18 | 21 | 6440 | 4.14 | 3.38 | 7.48 | 0.28 | 198 | 7.48 | 0.19 | 21 |
| 40 Leo | E | 6467 | 4.11 | 2.57 | 7.59 | 0.16 | 331 | 7.57 | 0.15 | 45 | 6467 | 4.17 | 2.55 | 7.60 | 0.16 | 331 | 7.59 | 0.15 | 45 |
| 47 Ari | E | 6644 | 4.21 | 3.01 | 7.72 | 0.18 | 223 | 7.79 | 0.19 | 33 | 6644 | 4.05 | 3.07 | 7.71 | 0.18 | 223 | 7.71 | 0.19 | 33 |
| 49 Peg | E | 6275 | 3.95 | 1.68 | 7.29 | 0.10 | 473 | 7.27 | 0.16 | 80 | 6275 | 4.01 | 1.65 | 7.30 | 0.10 | 473 | 7.29 | 0.16 | 80 |
| 51 Ari | E | 5603 | 4.46 | 0.85 | 7.64 | 0.08 | 381 | 7.88 | 0.19 | 55 | 5603 | 3.91 | 1.07 | 7.61 | 0.08 | 381 | 7.61 | 0.20 | 55 |
| 6 And | E | 6338 | 4.23 | 2.43 | 7.34 | 0.16 | 296 | 7.41 | 0.16 | 43 | 6338 | 4.08 | 2.55 | 7.33 | 0.16 | 296 | 7.31 | 0.16 | 43 |
| 6 Cet | E | 6242 | 4.09 | 1.53 | 7.17 | 0.10 | 454 | 7.14 | 0.16 | 78 | 6242 | 4.17 | 1.48 | 7.18 | 0.10 | 454 | 7.18 | 0.16 | 78 |
| 68 Eri | E | 6421 | 4.03 | 2.01 | 7.19 | 0.10 | 389 | 7.20 | 0.17 | 69 | 6421 | 4.02 | 2.01 | 7.19 | 0.10 | 389 | 7.20 | 0.17 | 69 |
| 71 Ori | E | 6560 | 4.31 | 1.59 | 7.60 | 0.10 | 475 | 7.62 | 0.17 | 85 | 6560 | 4.28 | 1.61 | 7.60 | 0.10 | 475 | 7.60 | 0.17 | 85 |
| 84 Cet | E | 6236 | 4.29 | 2.64 | 7.50 | 0.24 | 161 | 7.48 | 0.22 | 15 | 6236 | 4.33 | 2.62 | 7.50 | 0.24 | 161 | 7.50 | 0.22 | 15 |
| 89 Leo | E | 6538 | 4.30 | 2.22 | 7.63 | 0.12 | 374 | 7.58 | 0.14 | 54 | 6538 | 4.41 | 2.17 | 7.63 | 0.12 | 374 | 7.63 | 0.14 | 54 |
| b Her | E | 6000 | 4.22 | 1.26 | 6.90 | 0.08 | 415 | 6.93 | 0.13 | 71 | 6000 | 4.14 | 1.33 | 6.89 | 0.08 | 415 | 6.89 | 0.14 | 71 |
| c Boo | E | 6560 | 4.27 | 3.22 | 7.64 | 0.30 | 107 | 7.70 | 0.28 | 17 | 6560 | 4.13 | 3.28 | 7.63 | 0.30 | 107 | 7.64 | 0.28 | 17 |
| eta UMi | E | 6788 | 4.00 | 3.75 | 7.50 | 0.26 | 14 | 7.71 | 0.00 | 1 | 6788 | 3.27 | 5.09 | 7.42 | 0.26 | 14 | 7.40 | 0.00 | 1 |
| iot Vir | E | 6217 | 3.75 | 2.41 | 7.41 | 0.14 | 345 | 7.41 | 0.16 | 53 | 6217 | 3.75 | 2.41 | 7.41 | 0.14 | 345 | 7.41 | 0.16 | 53 |
| kap CrB | E | 4863 | 3.18 | 1.25 | 7.67 | 0.12 | 282 | 7.91 | 0.29 | 44 | 4863 | 3.00 | 1.26 | 7.64 | 0.12 | 282 | 7.81 | 0.29 | 44 |
| phi Vir | E | 5551 | 3.44 | 2.33 | 7.43 | 0.15 | 351 | 7.60 | 0.20 | 41 | 5551 | 3.06 | 2.44 | 7.41 | 0.16 | 351 | 7.42 | 0.20 | 41 |
| tau01 Hya | E | 6507 | 4.21 | 2.49 | 7.63 | 0.23 | 167 | 7.77 | 0.24 | 24 | 6507 | 3.90 | 2.62 | 7.63 | 0.24 | 167 | 7.63 | 0.23 | 24 |
| tet Dra | E | 6208 | 3.79 | 3.03 | 7.66 | 0.23 | 175 | 7.71 | 0.17 | 15 | 6208 | 3.66 | 3.06 | 7.65 | 0.23 | 175 | 7.66 | 0.17 | 15 |
| BD+01 2063 | E | 4978 | 4.63 | 1.50 | 7.17 | 0.08 | 254 | 7.36 | 0.19 | 26 | 4978 | 4.16 | 1.80 | 7.06 | 0.08 | 254 | 7.02 | 0.18 | 26 |
| BD+12 4499 | E | 4653 | 4.66 | 0.69 | 7.66 | 0.15 | 275 | 8.24 | 0.38 | 30 | 4653 | 3.66 | 1.23 | 7.24 | 0.17 | 275 | 7.34 | 0.43 | 30 |
| BD+17 4708 | E | 6180 | 4.16 | 1.54 | 5.99 | 0.19 | 181 | 5.94 | 0.18 | 38 | 6180 | 4.28 | 1.49 | 5.99 | 0.19 | 181 | 5.99 | 0.18 | 38 |
| BD+23 465 | E | 5240 | 4.49 | 1.22 | 7.68 | 0.10 | 410 | 7.78 | 0.13 | 33 | 5240 | 4.24 | 1.41 | 7.63 | 0.10 | 410 | 7.63 | 0.13 | 33 |
| BD+29 366 | E | 5777 | 4.46 | 1.34 | 6.60 | 0.11 | 399 | 6.67 | 0.21 | 58 | 5777 | 4.32 | 1.45 | 6.59 | 0.11 | 399 | 6.57 | 0.22 | 58 |
| BD+33 99 | E | 4499 | 4.61 | 0.50 | 7.62 | 0.20 | 262 | 8.34 | 0.52 | 26 | 4499 | 3.61 | 1.41 | 7.30 | 0.21 | 262 | 7.62 | 0.53 | 26 |

| Name | | Teff | logg | vt | [Fe/H]ₐ | σ | n | [Fe/H]ᵦ | σ | n | Teff | logg | vt | [Fe/H]ₐ | σ | n | [Fe/H]ᵦ | σ | n |
|---|---|---|---|---|---|---|---|---|---|---|---|---|---|---|---|---|---|---|---|
| BD+41 3306 | E | 5053 | 4.54 | 0.15 | 7.02 | 0.10 | 453 | 7.14 | 0.14 | 36 | 5053 | 4.26 | 1.20 | 6.94 | 0.09 | 453 | 6.90 | 0.15 | 36 |
| BD+43 699 | E | 4802 | 4.66 | 0.15 | 7.23 | 0.10 | 428 | 7.44 | 0.21 | 33 | 4802 | 4.18 | 1.76 | 7.04 | 0.12 | 428 | 7.02 | 0.20 | 33 |
| BD+46 1635 | E | 4215 | 4.65 | 1.36 | 7.90 | 0.15 | 114 | 8.39 | 0.16 | 6 | 4215 | 3.65 | 2.02 | 7.45 | 0.14 | 114 | 7.49 | 0.14 | 6 |
| BD+52 2815 | E | 4203 | 4.66 | 0.15 | 7.60 | 0.18 | 302 | 8.04 | 0.22 | 7 | 4203 | 3.66 | 2.26 | 7.18 | 0.21 | 302 | 7.22 | 0.23 | 7 |
| CCDM J20051-0418AB | E | 6370 | 4.07 | 1.78 | 7.30 | 0.14 | 432 | 7.18 | 0.22 | 70 | 6370 | 4.35 | 1.61 | 7.31 | 0.13 | 432 | 7.29 | 0.21 | 70 |
| CCDM J21031+0132AB | E | 6423 | 3.74 | 2.48 | 7.47 | 0.18 | 343 | 7.37 | 0.24 | 53 | 6423 | 3.98 | 2.40 | 7.47 | 0.18 | 343 | 7.47 | 0.25 | 53 |
| CCDM J22071+0034AB | E | 6059 | 3.91 | 1.47 | 7.35 | 0.09 | 501 | 7.26 | 0.17 | 70 | 6059 | 4.12 | 1.33 | 7.37 | 0.09 | 501 | 7.37 | 0.17 | 70 |
| GJ 1067 | E | 4378 | 4.68 | 0.57 | 7.72 | 0.22 | 242 | 8.94 | 0.76 | 25 | 4378 | 3.68 | 1.37 | 7.32 | 0.22 | 242 | 8.09 | 0.76 | 25 |
| GJ 697 | E | 4881 | 4.64 | 1.03 | 7.44 | 0.13 | 320 | 7.90 | 0.35 | 39 | 4881 | 3.64 | 1.54 | 7.21 | 0.15 | 320 | 7.26 | 0.37 | 39 |
| HD 10086 | E | 5658 | 4.46 | 1.04 | 7.57 | 0.08 | 376 | 7.72 | 0.18 | 60 | 5658 | 4.12 | 1.20 | 7.55 | 0.08 | 376 | 7.55 | 0.19 | 60 |
| HD 10145 | E | 5642 | 4.38 | 0.82 | 7.52 | 0.07 | 396 | 7.67 | 0.19 | 53 | 5642 | 4.04 | 1.00 | 7.50 | 0.08 | 396 | 7.50 | 0.20 | 53 |
| HD 105631 | E | 5363 | 4.52 | 1.01 | 7.67 | 0.09 | 347 | 7.96 | 0.24 | 45 | 5363 | 3.84 | 1.26 | 7.61 | 0.09 | 347 | 7.60 | 0.24 | 45 |
| HD 106116 | E | 5665 | 4.34 | 1.14 | 7.65 | 0.07 | 544 | 7.79 | 0.22 | 80 | 5665 | 4.01 | 1.40 | 7.60 | 0.08 | 544 | 7.55 | 0.24 | 80 |
| HD 106691 | E | 6647 | 4.13 | 3.08 | 7.58 | 0.22 | 146 | 7.45 | 0.16 | 19 | 6647 | 4.41 | 3.02 | 7.58 | 0.22 | 146 | 7.55 | 0.17 | 19 |
| HD 107611 | E | 6391 | 4.21 | 2.54 | 7.45 | 0.18 | 284 | 7.36 | 0.16 | 40 | 6391 | 4.40 | 2.43 | 7.46 | 0.18 | 284 | 7.46 | 0.16 | 40 |
| HD 10853 | E | 4655 | 4.66 | 0.48 | 7.51 | 0.15 | 289 | 8.10 | 0.44 | 31 | 4655 | 3.66 | 1.35 | 7.18 | 0.15 | 289 | 7.33 | 0.46 | 31 |
| HD 110463 | E | 4926 | 4.61 | 1.03 | 7.51 | 0.12 | 322 | 7.82 | 0.33 | 35 | 4926 | 3.78 | 1.47 | 7.35 | 0.12 | 322 | 7.34 | 0.33 | 35 |
| HD 111069 | E | 5865 | 4.40 | 1.14 | 7.72 | 0.08 | 533 | 7.84 | 0.20 | 81 | 5865 | 4.17 | 1.41 | 7.69 | 0.09 | 533 | 7.69 | 0.22 | 81 |
| HD 11373 | E | 4759 | 4.61 | 0.85 | 7.67 | 0.13 | 483 | 7.74 | 0.12 | 17 | 4759 | 4.47 | 1.35 | 7.59 | 0.13 | 483 | 7.59 | 0.12 | 17 |
| HD 115274 | E | 6153 | 3.84 | 1.27 | 7.44 | 0.17 | 466 | 7.46 | 0.20 | 70 | 6153 | 3.78 | 1.31 | 7.43 | 0.17 | 466 | 7.43 | 0.20 | 70 |
| HD 116956 | E | 5308 | 4.52 | 1.26 | 7.65 | 0.09 | 373 | 7.74 | 0.10 | 29 | 5308 | 4.33 | 1.43 | 7.62 | 0.09 | 373 | 7.62 | 0.10 | 29 |
| HD 117635 | E | 5217 | 4.24 | 0.80 | 7.13 | 0.12 | 371 | 7.23 | 0.26 | 43 | 5217 | 4.00 | 0.94 | 7.10 | 0.12 | 371 | 7.10 | 0.26 | 43 |
| HD 118096 | E | 4568 | 4.73 | 0.15 | 7.15 | 0.11 | 363 | 7.31 | 0.19 | 17 | 4568 | 4.35 | 1.60 | 7.00 | 0.12 | 363 | 6.99 | 0.19 | 17 |
| HD 119332 | E | 5213 | 4.53 | 1.04 | 7.44 | 0.06 | 249 | 7.49 | 0.11 | 26 | 5213 | 4.41 | 1.13 | 7.42 | 0.06 | 249 | 7.42 | 0.11 | 26 |
| HD 119802 | E | 4716 | 4.64 | 1.19 | 7.64 | 0.10 | 216 | 7.92 | 0.23 | 13 | 4716 | 3.96 | 1.58 | 7.41 | 0.10 | 216 | 7.37 | 0.23 | 13 |
| HD 12051 | E | 5397 | 4.39 | 0.91 | 7.73 | 0.08 | 347 | 7.96 | 0.21 | 51 | 5397 | 3.86 | 1.11 | 7.68 | 0.09 | 347 | 7.68 | 0.21 | 51 |
| HD 122120 | E | 4486 | 4.62 | 0.49 | 7.79 | 0.19 | 249 | 8.60 | 0.59 | 28 | 4486 | 3.62 | 1.30 | 7.40 | 0.19 | 249 | 7.78 | 0.62 | 28 |
| HD 124292 | E | 5497 | 4.49 | 0.89 | 7.42 | 0.05 | 332 | 7.50 | 0.11 | 30 | 5497 | 4.32 | 1.02 | 7.40 | 0.05 | 332 | 7.40 | 0.11 | 30 |
| HD 124642 | E | 4664 | 4.60 | 1.26 | 7.59 | 0.10 | 200 | 7.90 | 0.23 | 17 | 4664 | 3.98 | 1.78 | 7.43 | 0.10 | 200 | 7.48 | 0.23 | 17 |
| HD 128429 | E | 6456 | 4.26 | 2.26 | 7.49 | 0.16 | 321 | 7.44 | 0.15 | 43 | 6456 | 4.35 | 2.21 | 7.49 | 0.16 | 321 | 7.49 | 0.15 | 43 |
| HD 12846 | E | 5733 | 4.48 | 0.66 | 7.30 | 0.07 | 412 | 7.41 | 0.16 | 56 | 5733 | 4.28 | 0.83 | 7.29 | 0.07 | 412 | 7.30 | 0.16 | 56 |
| HD 130307 | E | 5043 | 4.59 | 0.66 | 7.40 | 0.09 | 300 | 7.50 | 0.12 | 23 | 5043 | 4.37 | 0.81 | 7.33 | 0.09 | 300 | 7.31 | 0.11 | 23 |
| HD 132142 | E | 5229 | 4.58 | 0.47 | 7.13 | 0.09 | 384 | 7.28 | 0.28 | 48 | 5229 | 4.25 | 0.85 | 7.09 | 0.09 | 384 | 7.10 | 0.28 | 48 |
| HD 132254 | E | 6279 | 4.21 | 1.75 | 7.59 | 0.10 | 474 | 7.59 | 0.17 | 70 | 6279 | 4.22 | 1.76 | 7.59 | 0.10 | 474 | 7.56 | 0.17 | 70 |
| HD 133002 | E | 5562 | 3.55 | 1.29 | 7.15 | 0.07 | 517 | 7.20 | 0.18 | 73 | 5562 | 3.43 | 1.35 | 7.14 | 0.08 | 517 | 7.14 | 0.18 | 73 |
| HD 13403 | E | 5617 | 4.00 | 0.98 | 7.18 | 0.07 | 512 | 7.27 | 0.18 | 75 | 5617 | 3.80 | 1.17 | 7.16 | 0.07 | 512 | 7.16 | 0.19 | 75 |
| HD 135204 | E | 5457 | 4.49 | 0.93 | 7.47 | 0.09 | 543 | 7.41 | 0.11 | 42 | 5457 | 4.58 | 0.54 | 7.51 | 0.10 | 543 | 7.52 | 0.11 | 42 |
| HD 135599 | E | 5270 | 4.55 | 1.03 | 7.50 | 0.09 | 341 | 7.71 | 0.24 | 42 | 5270 | 4.04 | 1.28 | 7.44 | 0.09 | 341 | 7.44 | 0.25 | 42 |

| | | | | | | | | | | | | | | | | |
|---|---|---|---|---|---|---|---|---|---|---|---|---|---|---|---|---|
| HD 13579 | E | 5133 | 4.51 | 0.55 | 7.88 | 0.10 | 270 | 7.96 | 0.12 | 24 | 5133 | 4.33 | 0.75 | 7.84 | 0.09 | 270 | 7.84 | 0.12 | 24 |
| HD 139777 | E | 5775 | 4.47 | 1.52 | 7.49 | 0.09 | 508 | 7.56 | 0.15 | 63 | 5775 | 4.32 | 1.68 | 7.48 | 0.09 | 508 | 7.47 | 0.16 | 63 |
| HD 139813 | E | 5394 | 4.56 | 1.45 | 7.51 | 0.10 | 504 | 7.69 | 0.22 | 57 | 5394 | 4.16 | 1.91 | 7.44 | 0.12 | 504 | 7.44 | 0.24 | 57 |
| HD 140283 | E | 5823 | 3.80 | 2.24 | 5.15 | 0.14 | 96 | 5.23 | 0.18 | 26 | 5823 | 3.59 | 2.27 | 5.15 | 0.14 | 96 | 5.15 | 0.18 | 26 |
| HD 14348 | E | 6036 | 4.03 | 1.62 | 7.65 | 0.09 | 504 | 7.70 | 0.13 | 76 | 6036 | 3.94 | 1.67 | 7.65 | 0.09 | 504 | 7.65 | 0.13 | 76 |
| HD 14374 | E | 5474 | 4.55 | 0.71 | 7.56 | 0.07 | 313 | 7.57 | 0.13 | 28 | 5474 | 4.52 | 0.74 | 7.55 | 0.07 | 313 | 7.56 | 0.13 | 28 |
| HD 144287 | E | 5315 | 4.42 | 0.88 | 7.40 | 0.08 | 315 | 7.55 | 0.13 | 31 | 5315 | 4.09 | 1.09 | 7.36 | 0.08 | 315 | 7.36 | 0.13 | 31 |
| HD 145435 | E | 6083 | 4.12 | 1.42 | 7.49 | 0.08 | 519 | 7.53 | 0.17 | 86 | 6083 | 4.03 | 1.49 | 7.48 | 0.09 | 519 | 7.48 | 0.17 | 86 |
| HD 145729 | E | 6028 | 4.39 | 1.09 | 7.44 | 0.09 | 506 | 7.51 | 0.16 | 79 | 6028 | 4.23 | 1.26 | 7.42 | 0.09 | 506 | 7.42 | 0.16 | 79 |
| HD 146946 | E | 5738 | 4.31 | 0.96 | 7.18 | 0.08 | 496 | 7.33 | 0.17 | 77 | 5738 | 4.07 | 1.19 | 7.15 | 0.09 | 496 | 7.16 | 0.18 | 77 |
| HD 153525 | E | 4820 | 4.63 | 1.75 | 7.37 | 0.07 | 171 | 7.58 | 0.21 | 23 | 4820 | 4.10 | 2.08 | 7.22 | 0.07 | 171 | 7.18 | 0.22 | 23 |
| HD 154931 | E | 5869 | 3.97 | 1.42 | 7.37 | 0.08 | 508 | 7.43 | 0.15 | 78 | 5869 | 3.84 | 1.49 | 7.36 | 0.08 | 508 | 7.37 | 0.16 | 78 |
| HD 155712 | E | 4936 | 4.58 | 0.24 | 7.48 | 0.13 | 488 | 7.90 | 0.35 | 56 | 4936 | 3.64 | 1.91 | 7.18 | 0.18 | 488 | 7.25 | 0.40 | 56 |
| HD 15632 | E | 5749 | 4.48 | 0.81 | 7.60 | 0.09 | 393 | 7.77 | 0.17 | 58 | 5749 | 4.09 | 1.00 | 7.57 | 0.09 | 393 | 7.52 | 0.17 | 58 |
| HD 156985 | E | 4778 | 4.59 | 0.51 | 7.47 | 0.13 | 327 | 8.05 | 0.43 | 36 | 4778 | 3.59 | 1.20 | 7.18 | 0.14 | 327 | 7.31 | 0.44 | 36 |
| HD 157089 | E | 5830 | 4.14 | 1.17 | 6.92 | 0.08 | 444 | 7.02 | 0.15 | 72 | 5830 | 3.94 | 1.36 | 6.91 | 0.08 | 444 | 6.91 | 0.15 | 72 |
| HD 159062 | E | 5385 | 4.48 | 0.27 | 7.19 | 0.09 | 516 | 7.34 | 0.20 | 54 | 5385 | 4.16 | 1.18 | 7.10 | 0.10 | 516 | 7.09 | 0.23 | 54 |
| HD 159482 | E | 5805 | 4.36 | 0.86 | 6.75 | 0.09 | 418 | 6.83 | 0.15 | 65 | 5805 | 4.23 | 1.07 | 6.74 | 0.10 | 418 | 6.76 | 0.16 | 65 |
| HD 160964 | E | 4589 | 4.67 | 0.78 | 7.45 | 0.18 | 299 | 8.11 | 0.45 | 34 | 4589 | 3.67 | 1.38 | 7.05 | 0.18 | 299 | 7.26 | 0.49 | 34 |
| HD 161098 | E | 5637 | 4.50 | 0.69 | 7.33 | 0.08 | 395 | 7.43 | 0.20 | 51 | 5637 | 4.27 | 0.83 | 7.30 | 0.08 | 395 | 7.28 | 0.21 | 51 |
| HD 163183 | E | 5928 | 4.46 | 1.70 | 7.43 | 0.10 | 463 | 7.48 | 0.17 | 58 | 5928 | 4.36 | 1.80 | 7.42 | 0.10 | 463 | 7.42 | 0.18 | 58 |
| HD 16397 | E | 5821 | 4.37 | 0.92 | 7.00 | 0.07 | 463 | 7.05 | 0.11 | 68 | 5821 | 4.28 | 1.05 | 6.99 | 0.07 | 463 | 7.00 | 0.12 | 68 |
| HD 164651 | E | 5599 | 4.45 | 0.76 | 7.52 | 0.09 | 379 | 7.62 | 0.17 | 54 | 5599 | 4.23 | 0.91 | 7.51 | 0.09 | 379 | 7.50 | 0.17 | 54 |
| HD 165173 | E | 5484 | 4.54 | 0.74 | 7.59 | 0.09 | 376 | 7.83 | 0.20 | 49 | 5484 | 4.01 | 0.99 | 7.53 | 0.09 | 376 | 7.49 | 0.21 | 49 |
| HD 165401 | E | 5816 | 4.44 | 0.97 | 7.09 | 0.08 | 470 | 7.17 | 0.17 | 69 | 5816 | 4.31 | 1.06 | 7.08 | 0.08 | 470 | 7.07 | 0.18 | 69 |
| HD 165476 | E | 5816 | 4.25 | 1.02 | 7.46 | 0.07 | 525 | 7.58 | 0.16 | 81 | 5816 | 4.03 | 1.27 | 7.43 | 0.08 | 525 | 7.44 | 0.18 | 81 |
| HD 165590 | E | 5663 | 4.09 | 1.56 | 7.29 | 0.36 | 104 | 7.29 | 0.60 | 11 | 5663 | 4.08 | 1.56 | 7.29 | 0.36 | 104 | 7.29 | 0.60 | 11 |
| HD 165670 | E | 6285 | 4.32 | 1.88 | 7.47 | 0.13 | 408 | 7.57 | 0.15 | 58 | 6285 | 4.11 | 2.02 | 7.46 | 0.13 | 408 | 7.46 | 0.16 | 58 |
| HD 165672 | E | 5866 | 4.37 | 1.26 | 7.66 | 0.09 | 528 | 7.83 | 0.20 | 82 | 5866 | 3.97 | 1.60 | 7.62 | 0.11 | 528 | 7.61 | 0.22 | 82 |
| HD 166183 | E | 6380 | 4.05 | 1.92 | 7.45 | 0.11 | 423 | 7.44 | 0.16 | 68 | 6380 | 4.07 | 1.91 | 7.45 | 0.11 | 423 | 7.45 | 0.16 | 68 |
| HD 166435 | E | 5811 | 4.47 | 1.75 | 7.47 | 0.10 | 490 | 7.49 | 0.11 | 53 | 5811 | 4.43 | 1.79 | 7.47 | 0.10 | 490 | 7.47 | 0.11 | 53 |
| HD 167278 | E | 6483 | 3.82 | 2.70 | 7.25 | 0.20 | 291 | 7.08 | 0.24 | 41 | 6483 | 4.21 | 2.54 | 7.26 | 0.20 | 291 | 7.25 | 0.23 | 41 |
| HD 169822 | E | 5529 | 4.53 | 1.13 | 7.31 | 0.07 | 406 | 7.38 | 0.11 | 47 | 5529 | 4.44 | 1.14 | 7.29 | 0.07 | 406 | 7.29 | 0.11 | 47 |
| HD 170008 | E | 4978 | 3.33 | 0.90 | 7.15 | 0.10 | 386 | 7.50 | 0.21 | 51 | 4978 | 3.00 | 0.96 | 7.12 | 0.10 | 386 | 7.33 | 0.21 | 51 |
| HD 170291 | E | 6327 | 4.22 | 1.90 | 7.55 | 0.12 | 432 | 7.51 | 0.14 | 65 | 6327 | 4.30 | 1.85 | 7.55 | 0.12 | 432 | 7.55 | 0.14 | 65 |
| HD 170512 | E | 6152 | 4.27 | 1.58 | 7.68 | 0.10 | 502 | 7.67 | 0.13 | 71 | 6152 | 4.29 | 1.56 | 7.68 | 0.09 | 502 | 7.69 | 0.13 | 71 |
| HD 170579 | E | 6417 | 4.22 | 2.03 | 7.25 | 0.15 | 358 | 7.26 | 0.17 | 58 | 6417 | 4.19 | 2.05 | 7.25 | 0.15 | 358 | 7.24 | 0.17 | 58 |
| HD 171314 | E | 4530 | 4.60 | 0.69 | 7.75 | 0.19 | 262 | 8.42 | 0.60 | 28 | 4530 | 3.60 | 1.30 | 7.33 | 0.20 | 262 | 7.54 | 0.61 | 28 |

| | | | | | | | | | | | | | | | | |
|---|---|---|---|---|---|---|---|---|---|---|---|---|---|---|---|---|
| HD 171888 | E | 6095 | 4.00 | 1.56 | 7.54 | 0.10 | 501 | 7.54 | 0.15 | 79 | 6095 | 4.00 | 1.56 | 7.54 | 0.10 | 501 | 7.54 | 0.15 | 79 |
| HD 171951 | E | 6094 | 4.03 | 1.39 | 7.21 | 0.09 | 479 | 7.23 | 0.19 | 77 | 6094 | 3.98 | 1.43 | 7.21 | 0.09 | 479 | 7.21 | 0.19 | 77 |
| HD 171953 | E | 6480 | 3.82 | 2.28 | 7.77 | 0.32 | 44 | 7.65 | 0.48 | 5 | 6480 | 4.13 | 2.26 | 7.77 | 0.32 | 44 | 7.77 | 0.47 | 5 |
| HD 172675 | E | 6303 | 4.37 | 1.66 | 7.42 | 0.11 | 453 | 7.40 | 0.16 | 63 | 6303 | 4.42 | 1.62 | 7.43 | 0.11 | 453 | 7.43 | 0.16 | 63 |
| HD 172718 | E | 6132 | 3.95 | 1.58 | 7.38 | 0.09 | 476 | 7.39 | 0.17 | 78 | 6132 | 3.93 | 1.60 | 7.38 | 0.09 | 476 | 7.38 | 0.17 | 78 |
| HD 172961 | E | 6571 | 4.32 | 2.21 | 7.43 | 0.16 | 326 | 7.36 | 0.14 | 47 | 6571 | 4.48 | 2.11 | 7.44 | 0.16 | 326 | 7.44 | 0.14 | 47 |
| HD 173174 | E | 6009 | 3.98 | 1.70 | 7.67 | 0.11 | 510 | 7.59 | 0.16 | 80 | 6009 | 4.14 | 1.60 | 7.68 | 0.10 | 510 | 7.68 | 0.16 | 80 |
| HD 173605 | E | 5722 | 4.01 | 2.40 | 7.50 | 0.16 | 257 | 7.73 | 0.25 | 38 | 5722 | 3.50 | 2.60 | 7.48 | 0.17 | 257 | 7.48 | 0.25 | 38 |
| HD 173634 | E | 6505 | 3.76 | 2.94 | 7.60 | 0.29 | 132 | 7.46 | 0.14 | 15 | 6505 | 4.08 | 2.85 | 7.61 | 0.29 | 132 | 7.61 | 0.14 | 15 |
| HD 17382 | E | 5245 | 4.49 | 1.06 | 7.54 | 0.10 | 343 | 7.76 | 0.21 | 42 | 5245 | 3.96 | 1.29 | 7.48 | 0.10 | 343 | 7.47 | 0.21 | 42 |
| HD 174080 | E | 4676 | 4.60 | 1.12 | 7.67 | 0.11 | 203 | 7.95 | 0.23 | 15 | 4676 | 3.87 | 1.48 | 7.43 | 0.10 | 203 | 7.39 | 0.23 | 15 |
| HD 174719 | E | 5647 | 4.50 | 0.88 | 7.34 | 0.09 | 398 | 7.51 | 0.26 | 52 | 5647 | 4.12 | 1.07 | 7.31 | 0.09 | 398 | 7.28 | 0.26 | 52 |
| HD 175272 | E | 6638 | 3.98 | 2.93 | 7.60 | 0.19 | 248 | 7.55 | 0.17 | 37 | 6638 | 4.11 | 2.90 | 7.60 | 0.19 | 248 | 7.60 | 0.17 | 37 |
| HD 175726 | E | 6069 | 4.43 | 2.00 | 7.43 | 0.12 | 393 | 7.43 | 0.14 | 50 | 6069 | 4.44 | 1.99 | 7.43 | 0.12 | 393 | 7.43 | 0.14 | 50 |
| HD 175805 | E | 6318 | 3.72 | 3.42 | 7.86 | 0.24 | 136 | 7.86 | 0.46 | 14 | 6318 | 3.72 | 3.33 | 7.86 | 0.25 | 136 | 7.91 | 0.47 | 14 |
| HD 175806 | E | 6171 | 3.63 | 1.99 | 7.45 | 0.11 | 487 | 7.37 | 0.17 | 83 | 6171 | 3.82 | 1.92 | 7.45 | 0.10 | 487 | 7.45 | 0.17 | 83 |
| HD 176118 | E | 6665 | 4.04 | 2.14 | 7.69 | 0.13 | 449 | 7.69 | 0.18 | 75 | 6665 | 4.02 | 2.15 | 7.68 | 0.13 | 449 | 7.68 | 0.18 | 75 |
| HD 176377 | E | 5868 | 4.46 | 0.99 | 7.24 | 0.07 | 480 | 7.34 | 0.17 | 75 | 5868 | 4.29 | 1.23 | 7.22 | 0.08 | 480 | 7.24 | 0.18 | 75 |
| HD 177749 | E | 6403 | 3.97 | 1.92 | 7.48 | 0.10 | 488 | 7.46 | 0.16 | 86 | 6403 | 4.01 | 1.90 | 7.48 | 0.10 | 488 | 7.48 | 0.16 | 86 |
| HD 177904 | E | 6902 | 3.84 | 3.31 | 7.64 | 0.22 | 171 | 7.48 | 0.16 | 24 | 6902 | 4.23 | 3.22 | 7.64 | 0.22 | 171 | 7.64 | 0.17 | 24 |
| HD 178126 | E | 4541 | 4.66 | 0.15 | 7.13 | 0.13 | 369 | 7.26 | 0.24 | 17 | 4541 | 4.38 | 0.90 | 7.07 | 0.13 | 369 | 7.07 | 0.24 | 17 |
| HD 180161 | E | 5400 | 4.53 | 1.34 | 7.62 | 0.09 | 515 | 7.90 | 0.22 | 64 | 5400 | 3.87 | 1.82 | 7.49 | 0.12 | 515 | 7.40 | 0.26 | 64 |
| HD 180945 | E | 6415 | 4.04 | 2.36 | 7.56 | 0.15 | 338 | 7.54 | 0.16 | 54 | 6415 | 4.09 | 2.35 | 7.56 | 0.15 | 338 | 7.56 | 0.16 | 54 |
| HD 181096 | E | 6270 | 3.92 | 1.85 | 7.23 | 0.11 | 440 | 7.25 | 0.16 | 78 | 6270 | 3.88 | 1.87 | 7.23 | 0.11 | 440 | 7.23 | 0.16 | 78 |
| HD 181420 | E | 6606 | 4.18 | 2.72 | 7.60 | 0.18 | 278 | 7.56 | 0.14 | 39 | 6606 | 4.26 | 2.69 | 7.60 | 0.18 | 278 | 7.60 | 0.14 | 39 |
| HD 181806 | E | 6404 | 3.92 | 1.82 | 7.53 | 0.12 | 496 | 7.47 | 0.18 | 83 | 6404 | 4.06 | 1.75 | 7.54 | 0.11 | 496 | 7.54 | 0.18 | 83 |
| HD 182274 | E | 6307 | 4.32 | 1.47 | 7.28 | 0.12 | 427 | 7.31 | 0.15 | 70 | 6307 | 4.24 | 1.54 | 7.27 | 0.12 | 427 | 7.27 | 0.16 | 70 |
| HD 182736 | E | 5237 | 3.66 | 1.08 | 7.34 | 0.08 | 541 | 7.49 | 0.18 | 70 | 5237 | 3.30 | 1.38 | 7.28 | 0.09 | 541 | 7.28 | 0.20 | 70 |
| HD 182905 | E | 5376 | 3.88 | 1.18 | 7.57 | 0.09 | 543 | 7.77 | 0.23 | 69 | 5376 | 3.48 | 1.51 | 7.50 | 0.11 | 543 | 7.47 | 0.24 | 69 |
| HD 183341 | E | 5952 | 4.27 | 1.22 | 7.58 | 0.08 | 532 | 7.64 | 0.18 | 83 | 5952 | 4.12 | 1.30 | 7.56 | 0.09 | 532 | 7.52 | 0.19 | 83 |
| HD 183658 | E | 5828 | 4.45 | 0.87 | 7.62 | 0.09 | 386 | 7.75 | 0.17 | 55 | 5828 | 4.15 | 1.00 | 7.61 | 0.09 | 386 | 7.60 | 0.18 | 55 |
| HD 183870 | E | 5015 | 4.62 | 0.51 | 7.55 | 0.09 | 252 | 7.62 | 0.10 | 18 | 5015 | 4.47 | 0.73 | 7.52 | 0.08 | 252 | 7.52 | 0.10 | 18 |
| HD 184499 | E | 5807 | 4.10 | 1.11 | 6.98 | 0.08 | 465 | 7.02 | 0.18 | 77 | 5807 | 4.01 | 1.20 | 6.97 | 0.08 | 465 | 6.97 | 0.18 | 77 |
| HD 184768 | E | 5635 | 4.29 | 0.85 | 7.45 | 0.08 | 391 | 7.60 | 0.17 | 52 | 5635 | 3.94 | 1.01 | 7.43 | 0.08 | 391 | 7.43 | 0.17 | 52 |
| HD 184960 | E | 6290 | 4.23 | 1.63 | 7.45 | 0.10 | 453 | 7.45 | 0.16 | 70 | 6290 | 4.22 | 1.64 | 7.45 | 0.10 | 453 | 7.45 | 0.16 | 70 |
| HD 185269 | E | 5987 | 3.97 | 1.59 | 7.63 | 0.09 | 514 | 7.65 | 0.12 | 72 | 5987 | 3.94 | 1.61 | 7.63 | 0.09 | 514 | 7.63 | 0.12 | 72 |
| HD 185414 | E | 5806 | 4.45 | 0.73 | 7.40 | 0.07 | 412 | 7.52 | 0.14 | 57 | 5806 | 4.16 | 0.91 | 7.38 | 0.07 | 412 | 7.38 | 0.15 | 57 |
| HD 186104 | E | 5759 | 4.33 | 1.09 | 7.61 | 0.08 | 538 | 7.75 | 0.21 | 79 | 5759 | 4.05 | 1.30 | 7.57 | 0.09 | 538 | 7.54 | 0.23 | 79 |

| | | | | | | | | | | | | | | | | | |
|---|---|---|---|---|---|---|---|---|---|---|---|---|---|---|---|---|---|
| HD 186226 | E | 6371 | 3.93 | 2.33 | 7.69 | 0.14 | 382 | 7.66 | 0.16 | 55 | 6371 | 4.00 | 2.31 | 7.69 | 0.14 | 382 | 7.69 | 0.16 | 55 |
| HD 186379 | E | 5923 | 3.98 | 1.33 | 7.15 | 0.08 | 487 | 7.18 | 0.15 | 75 | 5923 | 3.92 | 1.38 | 7.15 | 0.08 | 487 | 7.14 | 0.16 | 75 |
| HD 186413 | E | 5918 | 4.16 | 1.29 | 7.50 | 0.09 | 517 | 7.58 | 0.15 | 79 | 5918 | 3.98 | 1.45 | 7.48 | 0.09 | 517 | 7.48 | 0.16 | 79 |
| HD 18757 | E | 5674 | 4.33 | 0.70 | 7.23 | 0.06 | 413 | 7.31 | 0.15 | 59 | 5674 | 4.16 | 0.83 | 7.22 | 0.06 | 413 | 7.22 | 0.16 | 59 |
| HD 18768 | E | 5815 | 3.85 | 1.37 | 6.94 | 0.07 | 457 | 6.97 | 0.13 | 71 | 5815 | 3.79 | 1.41 | 6.94 | 0.07 | 457 | 6.94 | 0.13 | 71 |
| HD 187897 | E | 5905 | 4.35 | 1.39 | 7.61 | 0.09 | 525 | 7.69 | 0.17 | 77 | 5905 | 4.17 | 1.55 | 7.59 | 0.11 | 525 | 7.59 | 0.18 | 77 |
| HD 188326 | E | 5342 | 3.85 | 1.00 | 7.41 | 0.09 | 543 | 7.59 | 0.23 | 73 | 5342 | 3.45 | 1.32 | 7.34 | 0.10 | 543 | 7.31 | 0.25 | 73 |
| HD 189509 | E | 6368 | 4.34 | 2.82 | 7.55 | 0.20 | 164 | 7.61 | 0.17 | 23 | 6368 | 4.23 | 2.89 | 7.55 | 0.20 | 164 | 7.55 | 0.17 | 23 |
| HD 189558 | E | 5773 | 3.93 | 1.40 | 6.39 | 0.11 | 352 | 6.41 | 0.16 | 57 | 5773 | 3.89 | 1.43 | 6.39 | 0.12 | 352 | 6.39 | 0.16 | 57 |
| HD 19019 | E | 6113 | 4.40 | 1.18 | 7.43 | 0.07 | 497 | 7.46 | 0.16 | 79 | 6113 | 4.34 | 1.25 | 7.43 | 0.08 | 497 | 7.43 | 0.17 | 79 |
| HD 190404 | E | 5088 | 4.60 | 0.54 | 6.93 | 0.08 | 431 | 7.01 | 0.14 | 38 | 5088 | 4.44 | 1.20 | 6.86 | 0.08 | 431 | 6.85 | 0.15 | 38 |
| HD 190412 | E | 5388 | 4.29 | 0.97 | 7.16 | 0.09 | 307 | 7.39 | 0.14 | 41 | 5388 | 3.79 | 1.24 | 7.12 | 0.10 | 307 | 7.12 | 0.15 | 41 |
| HD 190498 | E | 6415 | 3.93 | 2.69 | 7.67 | 0.17 | 224 | 7.71 | 0.22 | 36 | 6415 | 3.84 | 2.72 | 7.67 | 0.17 | 224 | 7.67 | 0.22 | 36 |
| HD 191533 | E | 6254 | 3.84 | 1.98 | 7.47 | 0.10 | 450 | 7.40 | 0.16 | 68 | 6254 | 3.99 | 1.91 | 7.47 | 0.09 | 450 | 7.47 | 0.16 | 68 |
| HD 191785 | E | 5213 | 4.52 | 0.68 | 7.43 | 0.09 | 500 | 7.50 | 0.12 | 36 | 5213 | 4.41 | 1.06 | 7.38 | 0.08 | 500 | 7.38 | 0.12 | 36 |
| HD 19308 | E | 5767 | 4.21 | 1.31 | 7.60 | 0.09 | 538 | 7.74 | 0.19 | 79 | 5767 | 3.88 | 1.53 | 7.56 | 0.10 | 538 | 7.51 | 0.21 | 79 |
| HD 193374 | E | 6522 | 3.91 | 2.73 | 7.62 | 0.18 | 286 | 7.55 | 0.14 | 39 | 6522 | 4.09 | 2.68 | 7.63 | 0.17 | 286 | 7.63 | 0.14 | 39 |
| HD 194154 | E | 6456 | 4.30 | 2.93 | 7.57 | 0.23 | 123 | 7.74 | 0.33 | 17 | 6456 | 3.92 | 3.17 | 7.56 | 0.24 | 123 | 7.52 | 0.34 | 17 |
| HD 19445 | E | 6052 | 4.49 | 1.97 | 5.56 | 0.11 | 127 | 5.48 | 0.09 | 23 | 6052 | 4.71 | 1.86 | 5.57 | 0.11 | 127 | 5.57 | 0.09 | 23 |
| HD 194598 | E | 6126 | 4.37 | 1.77 | 6.41 | 0.10 | 284 | 6.36 | 0.14 | 52 | 6126 | 4.49 | 1.66 | 6.41 | 0.10 | 284 | 6.41 | 0.14 | 52 |
| HD 195005 | E | 6149 | 4.41 | 1.33 | 7.50 | 0.08 | 496 | 7.53 | 0.14 | 72 | 6149 | 4.35 | 1.39 | 7.50 | 0.09 | 496 | 7.50 | 0.14 | 72 |
| HD 195104 | E | 6226 | 4.38 | 1.36 | 7.40 | 0.09 | 478 | 7.42 | 0.17 | 77 | 6226 | 4.32 | 1.41 | 7.39 | 0.09 | 478 | 7.39 | 0.17 | 77 |
| HD 195633 | E | 6024 | 3.99 | 1.59 | 6.87 | 0.10 | 415 | 6.91 | 0.16 | 71 | 6024 | 3.89 | 1.65 | 6.87 | 0.10 | 415 | 6.87 | 0.17 | 71 |
| HD 196218 | E | 6204 | 4.19 | 1.46 | 7.36 | 0.11 | 481 | 7.44 | 0.21 | 82 | 6204 | 4.01 | 1.59 | 7.35 | 0.12 | 481 | 7.35 | 0.21 | 82 |
| HD 198061 | E | 6379 | 3.98 | 3.53 | 7.47 | 0.21 | 174 | 7.56 | 0.13 | 23 | 6379 | 3.75 | 3.60 | 7.46 | 0.21 | 174 | 7.46 | 0.13 | 23 |
| HD 199598 | E | 5918 | 4.37 | 1.04 | 7.51 | 0.07 | 524 | 7.65 | 0.14 | 90 | 5918 | 4.12 | 1.34 | 7.48 | 0.08 | 524 | 7.50 | 0.16 | 90 |
| HD 20039 | E | 5331 | 3.68 | 0.40 | 6.99 | 0.12 | 430 | 7.06 | 0.26 | 55 | 5331 | 3.53 | 0.51 | 6.98 | 0.12 | 430 | 6.99 | 0.26 | 55 |
| HD 200391 | E | 5709 | 3.97 | 0.40 | 7.65 | 0.10 | 3 | 7.66 | 0.00 | 1 | 5709 | 3.94 | 0.10 | 7.69 | 0.10 | 3 | 7.78 | 0.00 | 1 |
| HD 200560 | E | 4894 | 4.52 | 1.13 | 7.62 | 0.11 | 286 | 8.00 | 0.30 | 33 | 4894 | 3.53 | 1.55 | 7.44 | 0.12 | 286 | 7.42 | 0.31 | 33 |
| HD 200580 | E | 5870 | 4.00 | 1.36 | 6.90 | 0.11 | 446 | 6.81 | 0.12 | 63 | 5870 | 4.18 | 1.19 | 6.92 | 0.10 | 446 | 6.91 | 0.13 | 63 |
| HD 201099 | E | 5947 | 4.22 | 1.16 | 7.06 | 0.09 | 458 | 7.16 | 0.16 | 76 | 5947 | 4.00 | 1.38 | 7.04 | 0.09 | 458 | 7.04 | 0.17 | 76 |
| HD 20165 | E | 5137 | 4.58 | 0.15 | 7.55 | 0.10 | 466 | 7.62 | 0.14 | 37 | 5137 | 4.41 | 1.10 | 7.45 | 0.08 | 466 | 7.43 | 0.13 | 37 |
| HD 201891 | E | 5998 | 4.39 | 1.39 | 6.46 | 0.09 | 326 | 6.44 | 0.12 | 52 | 5998 | 4.45 | 1.32 | 6.47 | 0.09 | 326 | 6.47 | 0.12 | 52 |
| HD 202575 | E | 4737 | 4.65 | 1.15 | 7.48 | 0.14 | 296 | 7.95 | 0.38 | 29 | 4737 | 3.65 | 1.67 | 7.22 | 0.15 | 296 | 7.28 | 0.39 | 29 |
| HD 203235 | E | 6242 | 4.16 | 1.62 | 7.64 | 0.10 | 514 | 7.60 | 0.18 | 88 | 6242 | 4.26 | 1.55 | 7.65 | 0.09 | 514 | 7.65 | 0.17 | 88 |
| HD 204426 | E | 5658 | 3.98 | 1.00 | 7.11 | 0.08 | 508 | 7.22 | 0.18 | 73 | 5658 | 3.75 | 1.17 | 7.08 | 0.09 | 508 | 7.06 | 0.19 | 73 |
| HD 204734 | E | 5262 | 4.58 | 1.15 | 7.56 | 0.10 | 319 | 7.77 | 0.24 | 38 | 5262 | 4.05 | 1.41 | 7.49 | 0.10 | 319 | 7.49 | 0.26 | 38 |
| HD 20512 | E | 5270 | 3.56 | 1.31 | 7.34 | 0.09 | 549 | 7.46 | 0.22 | 77 | 5270 | 3.30 | 1.43 | 7.31 | 0.10 | 549 | 7.28 | 0.23 | 77 |

| Star | | Teff | logg | vt | [Fe/H]I | σI | nI | [Fe/H]II | σII | nII | Teff | logg | vt | [Fe/H]I | σI | nI | [Fe/H]II | σII | nII |
|---|---|---|---|---|---|---|---|---|---|---|---|---|---|---|---|---|---|---|---|
| HD 205434 | E | 4454 | 4.68 | 0.60 | 7.53 | 0.14 | 177 | 7.85 | 0.20 | 12 | 4454 | 4.02 | 1.31 | 7.30 | 0.13 | 177 | 7.34 | 0.20 | 12 |
| HD 205702 | E | 6060 | 4.19 | 1.46 | 7.57 | 0.09 | 511 | 7.61 | 0.16 | 80 | 6060 | 4.09 | 1.52 | 7.56 | 0.09 | 511 | 7.52 | 0.17 | 80 |
| HD 206374 | E | 5579 | 4.53 | 0.83 | 7.43 | 0.06 | 344 | 7.47 | 0.05 | 30 | 5579 | 4.45 | 0.89 | 7.42 | 0.06 | 344 | 7.42 | 0.05 | 30 |
| HD 208038 | E | 4995 | 4.63 | 0.53 | 7.49 | 0.08 | 249 | 7.76 | 0.19 | 31 | 4995 | 4.14 | 0.97 | 7.37 | 0.08 | 249 | 7.40 | 0.19 | 31 |
| HD 208313 | E | 5030 | 4.61 | 0.52 | 7.52 | 0.10 | 505 | 7.73 | 0.20 | 46 | 5030 | 4.17 | 1.60 | 7.36 | 0.11 | 505 | 7.36 | 0.22 | 46 |
| HD 209472 | E | 6480 | 4.18 | 2.16 | 7.38 | 0.16 | 313 | 7.41 | 0.14 | 49 | 6480 | 4.12 | 2.19 | 7.38 | 0.16 | 313 | 7.38 | 0.14 | 49 |
| HD 210752 | E | 6024 | 4.42 | 1.09 | 6.94 | 0.09 | 436 | 6.99 | 0.17 | 71 | 6024 | 4.32 | 1.21 | 6.93 | 0.09 | 436 | 6.93 | 0.18 | 71 |
| HD 21197 | E | 4562 | 4.56 | 0.82 | 7.87 | 0.19 | 241 | 8.70 | 0.61 | 28 | 4562 | 3.56 | 1.33 | 7.43 | 0.20 | 241 | 7.76 | 0.66 | 28 |
| HD 214683 | E | 4893 | 4.65 | 0.56 | 7.26 | 0.12 | 339 | 7.67 | 0.36 | 39 | 4893 | 3.86 | 1.21 | 7.08 | 0.13 | 339 | 7.14 | 0.36 | 39 |
| HD 216259 | E | 5002 | 4.67 | 0.95 | 6.87 | 0.09 | 429 | 6.89 | 0.12 | 31 | 5002 | 4.64 | 1.08 | 6.85 | 0.09 | 429 | 6.86 | 0.12 | 31 |
| HD 216520 | E | 5103 | 4.56 | 1.11 | 7.32 | 0.07 | 260 | 7.52 | 0.17 | 28 | 5103 | 4.06 | 1.42 | 7.21 | 0.07 | 260 | 7.17 | 0.18 | 28 |
| HD 217813 | E | 5849 | 4.42 | 1.24 | 7.50 | 0.08 | 521 | 7.63 | 0.19 | 77 | 5849 | 4.18 | 1.52 | 7.47 | 0.09 | 521 | 7.48 | 0.21 | 77 |
| HD 218059 | E | 6382 | 4.31 | 1.52 | 7.21 | 0.09 | 433 | 7.18 | 0.14 | 67 | 6382 | 4.36 | 1.48 | 7.21 | 0.09 | 433 | 7.21 | 0.14 | 67 |
| HD 218209 | E | 5623 | 4.46 | 0.56 | 7.05 | 0.07 | 399 | 7.16 | 0.16 | 60 | 5623 | 4.26 | 0.77 | 7.04 | 0.07 | 399 | 7.06 | 0.16 | 60 |
| HD 218566 | E | 4846 | 4.53 | 0.95 | 7.84 | 0.14 | 258 | 8.31 | 0.38 | 30 | 4846 | 3.53 | 1.37 | 7.58 | 0.15 | 258 | 7.61 | 0.39 | 30 |
| HD 218687 | E | 5902 | 4.39 | 1.83 | 7.42 | 0.11 | 442 | 7.47 | 0.15 | 47 | 5902 | 4.29 | 1.92 | 7.41 | 0.12 | 442 | 7.41 | 0.15 | 47 |
| HD 218868 | E | 5509 | 4.41 | 1.03 | 7.71 | 0.08 | 354 | 7.89 | 0.21 | 52 | 5509 | 3.99 | 1.18 | 7.68 | 0.08 | 354 | 7.68 | 0.21 | 52 |
| HD 219396 | E | 5649 | 4.03 | 1.06 | 7.41 | 0.08 | 530 | 7.57 | 0.19 | 78 | 5649 | 3.69 | 1.33 | 7.37 | 0.08 | 530 | 7.34 | 0.21 | 78 |
| HD 219420 | E | 6175 | 4.19 | 1.49 | 7.49 | 0.09 | 492 | 7.44 | 0.13 | 74 | 6175 | 4.30 | 1.39 | 7.50 | 0.09 | 492 | 7.50 | 0.13 | 74 |
| HD 219538 | E | 5078 | 4.60 | 0.15 | 7.58 | 0.13 | 549 | 7.75 | 0.20 | 42 | 5078 | 4.19 | 1.60 | 7.44 | 0.13 | 549 | 7.45 | 0.20 | 42 |
| HD 219623 | E | 6177 | 4.31 | 1.43 | 7.57 | 0.07 | 508 | 7.57 | 0.15 | 78 | 6177 | 4.30 | 1.44 | 7.57 | 0.08 | 508 | 7.57 | 0.15 | 78 |
| HD 220140 | E | 5075 | 4.58 | 2.77 | 7.55 | 0.15 | 271 | 7.90 | 0.26 | 25 | 5075 | 3.70 | 3.23 | 7.41 | 0.17 | 271 | 7.40 | 0.27 | 25 |
| HD 220182 | E | 5335 | 4.56 | 1.16 | 7.54 | 0.09 | 328 | 7.78 | 0.22 | 41 | 5335 | 4.00 | 1.41 | 7.49 | 0.09 | 328 | 7.48 | 0.23 | 41 |
| HD 220221 | E | 4783 | 4.58 | 1.16 | 7.69 | 0.11 | 234 | 7.93 | 0.20 | 18 | 4783 | 3.97 | 1.47 | 7.50 | 0.11 | 234 | 7.45 | 0.20 | 18 |
| HD 221354 | E | 5282 | 4.49 | 0.89 | 7.62 | 0.07 | 287 | 7.67 | 0.14 | 26 | 5282 | 4.37 | 0.98 | 7.60 | 0.06 | 287 | 7.60 | 0.14 | 26 |
| HD 221585 | E | 5530 | 3.93 | 1.37 | 7.75 | 0.09 | 548 | 7.90 | 0.20 | 78 | 5530 | 3.56 | 1.63 | 7.70 | 0.11 | 548 | 7.69 | 0.21 | 78 |
| HD 221851 | E | 5192 | 4.58 | 1.00 | 7.43 | 0.07 | 306 | 7.56 | 0.17 | 33 | 5192 | 4.27 | 1.23 | 7.38 | 0.07 | 306 | 7.38 | 0.18 | 33 |
| HD 222155 | E | 5694 | 3.94 | 1.23 | 7.34 | 0.07 | 526 | 7.47 | 0.18 | 83 | 5694 | 3.66 | 1.41 | 7.32 | 0.08 | 526 | 7.28 | 0.19 | 83 |
| HD 224465 | E | 5770 | 4.42 | 0.98 | 7.58 | 0.07 | 394 | 7.68 | 0.17 | 64 | 5770 | 4.19 | 1.10 | 7.57 | 0.07 | 394 | 7.57 | 0.18 | 64 |
| HD 22879 | E | 5932 | 4.36 | 1.22 | 6.65 | 0.08 | 383 | 6.65 | 0.13 | 60 | 5932 | 4.37 | 1.21 | 6.65 | 0.08 | 383 | 6.65 | 0.13 | 60 |
| HD 232781 | E | 4679 | 4.68 | 0.73 | 7.16 | 0.16 | 326 | 7.76 | 0.52 | 34 | 4679 | 3.72 | 1.35 | 6.84 | 0.17 | 326 | 6.99 | 0.53 | 34 |
| HD 23439A | E | 5192 | 4.67 | 0.15 | 6.57 | 0.09 | 435 | 6.62 | 0.12 | 37 | 5192 | 4.55 | 0.72 | 6.55 | 0.09 | 435 | 6.53 | 0.11 | 37 |
| HD 238087 | E | 4213 | 4.64 | 0.98 | 7.77 | 0.25 | 210 | 9.14 | 0.96 | 17 | 4213 | 3.64 | 1.60 | 7.29 | 0.27 | 210 | 8.20 | 0.98 | 17 |
| HD 24040 | E | 5756 | 4.17 | 1.26 | 7.67 | 0.08 | 544 | 7.78 | 0.15 | 83 | 5756 | 3.90 | 1.49 | 7.63 | 0.09 | 544 | 7.63 | 0.17 | 83 |
| HD 24238 | E | 5031 | 4.60 | 0.15 | 7.06 | 0.08 | 411 | 7.19 | 0.18 | 42 | 5031 | 4.31 | 1.40 | 6.96 | 0.08 | 411 | 6.92 | 0.19 | 42 |
| HD 24409 | E | 5568 | 4.34 | 0.84 | 7.35 | 0.06 | 309 | 7.55 | 0.10 | 48 | 5568 | 3.88 | 1.06 | 7.31 | 0.06 | 309 | 7.28 | 0.10 | 48 |
| HD 24451 | E | 4547 | 4.64 | 0.82 | 7.67 | 0.16 | 268 | 8.33 | 0.41 | 28 | 4547 | 3.64 | 1.30 | 7.23 | 0.17 | 268 | 7.41 | 0.43 | 28 |
| HD 24496 | E | 5429 | 4.44 | 0.97 | 7.41 | 0.05 | 270 | 7.58 | 0.11 | 34 | 5429 | 4.05 | 1.18 | 7.37 | 0.06 | 270 | 7.37 | 0.12 | 34 |

| Star | | Teff | log g | vt | [Fe/H]a | σ | Na | [Fe/H]b | σ | Nb | Teff | log g | vt | [Fe/H]a | σ | Na | [Fe/H]b | σ | Nb |
|---|---|---|---|---|---|---|---|---|---|---|---|---|---|---|---|---|---|---|---|
| HD 245 | E | 5805 | 4.35 | 0.66 | 6.96 | 0.08 | 382 | 7.09 | 0.15 | 58 | 5805 | 4.09 | 0.85 | 6.95 | 0.09 | 382 | 6.96 | 0.15 | 58 |
| HD 24552 | E | 5916 | 4.43 | 1.09 | 7.53 | 0.08 | 527 | 7.61 | 0.19 | 76 | 5916 | 4.28 | 1.30 | 7.51 | 0.09 | 527 | 7.51 | 0.20 | 76 |
| HD 25329 | E | 4855 | 4.73 | 1.63 | 5.89 | 0.06 | 195 | 6.16 | 0.16 | 22 | 4855 | 3.93 | 2.32 | 5.74 | 0.09 | 195 | 5.70 | 0.17 | 22 |
| HD 25457 | E | 6268 | 4.33 | 2.41 | 7.58 | 0.14 | 314 | 7.61 | 0.15 | 39 | 6268 | 4.25 | 2.49 | 7.57 | 0.14 | 314 | 7.53 | 0.14 | 39 |
| HD 25621 | E | 6307 | 3.86 | 2.54 | 7.57 | 0.14 | 326 | 7.56 | 0.17 | 44 | 6307 | 3.88 | 2.53 | 7.57 | 0.14 | 326 | 7.57 | 0.17 | 44 |
| HD 25825 | E | 5976 | 4.41 | 1.24 | 7.58 | 0.12 | 506 | 7.67 | 0.17 | 73 | 5976 | 4.21 | 1.37 | 7.56 | 0.12 | 506 | 7.52 | 0.18 | 73 |
| HD 26345 | E | 6676 | 4.28 | 3.02 | 7.67 | 0.20 | 193 | 7.64 | 0.18 | 32 | 6676 | 4.34 | 2.99 | 7.67 | 0.20 | 193 | 7.67 | 0.18 | 32 |
| HD 26784 | E | 6261 | 4.29 | 2.33 | 7.70 | 0.15 | 333 | 7.78 | 0.18 | 43 | 6261 | 4.11 | 2.43 | 7.69 | 0.16 | 333 | 7.69 | 0.18 | 43 |
| HD 27808 | E | 6230 | 4.30 | 2.03 | 7.63 | 0.13 | 409 | 7.66 | 0.15 | 55 | 6230 | 4.24 | 2.07 | 7.63 | 0.13 | 409 | 7.63 | 0.15 | 55 |
| HD 28005 | E | 5727 | 4.27 | 1.22 | 7.76 | 0.09 | 546 | 7.93 | 0.18 | 83 | 5727 | 3.92 | 1.56 | 7.71 | 0.11 | 546 | 7.71 | 0.20 | 83 |
| HD 28099 | E | 5738 | 4.42 | 1.56 | 7.62 | 0.10 | 524 | 7.77 | 0.17 | 68 | 5738 | 4.09 | 1.81 | 7.59 | 0.11 | 524 | 7.58 | 0.18 | 68 |
| HD 28344 | E | 5921 | 4.39 | 1.82 | 7.62 | 0.10 | 494 | 7.67 | 0.11 | 62 | 5921 | 4.27 | 1.91 | 7.61 | 0.10 | 494 | 7.61 | 0.11 | 62 |
| HD 283704 | E | 5524 | 4.51 | 1.30 | 7.68 | 0.10 | 518 | 7.87 | 0.23 | 68 | 5524 | 4.05 | 1.63 | 7.61 | 0.11 | 518 | 7.55 | 0.26 | 68 |
| HD 283750 | E | 4405 | 4.51 | 1.99 | 7.96 | 0.17 | 148 | 8.84 | 0.35 | 18 | 4405 | 3.51 | 2.29 | 7.62 | 0.18 | 148 | 8.08 | 0.34 | 18 |
| HD 284248 | E | 6157 | 4.32 | 2.00 | 5.88 | 0.12 | 161 | 5.88 | 0.13 | 39 | 6157 | 4.32 | 2.00 | 5.88 | 0.12 | 161 | 5.88 | 0.13 | 39 |
| HD 284574 | E | 5370 | 4.47 | 0.98 | 7.75 | 0.08 | 351 | 7.91 | 0.12 | 36 | 5370 | 4.12 | 1.26 | 7.70 | 0.08 | 351 | 7.70 | 0.13 | 36 |
| HD 284930 | E | 4667 | 4.63 | 1.15 | 7.74 | 0.14 | 255 | 8.40 | 0.40 | 32 | 4667 | 3.63 | 1.67 | 7.44 | 0.15 | 255 | 7.68 | 0.42 | 32 |
| HD 285690 | E | 4975 | 4.45 | 0.78 | 7.67 | 0.10 | 252 | 7.74 | 0.17 | 16 | 4975 | 4.30 | 1.02 | 7.68 | 0.10 | 252 | 7.73 | 0.17 | 16 |
| HD 28946 | E | 5366 | 4.57 | 0.78 | 7.40 | 0.09 | 376 | 7.64 | 0.23 | 47 | 5366 | 4.07 | 1.07 | 7.34 | 0.09 | 376 | 7.32 | 0.23 | 47 |
| HD 28992 | E | 5865 | 4.40 | 1.54 | 7.57 | 0.09 | 505 | 7.66 | 0.16 | 72 | 5865 | 4.20 | 1.74 | 7.55 | 0.09 | 505 | 7.55 | 0.17 | 72 |
| HD 29310 | E | 5841 | 4.14 | 1.98 | 7.54 | 0.12 | 495 | 7.57 | 0.17 | 68 | 5841 | 4.08 | 2.02 | 7.54 | 0.12 | 495 | 7.54 | 0.18 | 68 |
| HD 29419 | E | 6058 | 4.37 | 1.51 | 7.63 | 0.08 | 522 | 7.69 | 0.14 | 83 | 6058 | 4.24 | 1.61 | 7.62 | 0.08 | 522 | 7.62 | 0.15 | 83 |
| HD 30562 | E | 5894 | 4.08 | 1.54 | 7.70 | 0.08 | 529 | 7.76 | 0.14 | 82 | 5894 | 3.94 | 1.63 | 7.68 | 0.09 | 529 | 7.68 | 0.15 | 82 |
| HD 3268 | E | 6232 | 4.10 | 1.41 | 7.39 | 0.10 | 491 | 7.45 | 0.17 | 80 | 6232 | 3.97 | 1.48 | 7.39 | 0.10 | 491 | 7.38 | 0.18 | 80 |
| HD 32850 | E | 5226 | 4.56 | 1.20 | 7.31 | 0.06 | 283 | 7.48 | 0.10 | 24 | 5226 | 4.17 | 1.43 | 7.24 | 0.06 | 283 | 7.20 | 0.10 | 24 |
| HD 330 | E | 5932 | 3.79 | 1.44 | 7.37 | 0.09 | 515 | 7.40 | 0.18 | 83 | 5932 | 3.73 | 1.47 | 7.37 | 0.09 | 515 | 7.37 | 0.19 | 83 |
| HD 332518 | E | 4386 | 4.69 | 0.15 | 7.47 | 0.15 | 349 | 7.59 | 0.15 | 8 | 4386 | 4.43 | 1.10 | 7.38 | 0.14 | 349 | 7.37 | 0.14 | 8 |
| HD 33564 | E | 6393 | 4.22 | 2.21 | 7.61 | 0.12 | 384 | 7.61 | 0.13 | 58 | 6393 | 4.21 | 2.21 | 7.61 | 0.12 | 384 | 7.61 | 0.13 | 58 |
| HD 345957 | E | 5943 | 4.08 | 1.55 | 6.16 | 0.14 | 268 | 6.18 | 0.15 | 46 | 5943 | 4.03 | 1.58 | 6.16 | 0.14 | 268 | 6.16 | 0.15 | 46 |
| HD 3628 | E | 5810 | 4.03 | 1.21 | 7.39 | 0.09 | 533 | 7.41 | 0.18 | 81 | 5810 | 3.98 | 1.25 | 7.39 | 0.09 | 533 | 7.39 | 0.18 | 81 |
| HD 37008 | E | 5091 | 4.60 | 0.62 | 7.13 | 0.09 | 473 | 7.15 | 0.13 | 31 | 5091 | 4.56 | 0.84 | 7.11 | 0.08 | 473 | 7.10 | 0.13 | 31 |
| HD 3765 | E | 4987 | 4.55 | 0.89 | 7.69 | 0.08 | 229 | 7.90 | 0.21 | 24 | 4987 | 4.10 | 1.16 | 7.55 | 0.07 | 229 | 7.52 | 0.21 | 24 |
| HD 38230 | E | 5212 | 4.47 | 0.36 | 7.53 | 0.09 | 472 | 7.53 | 0.12 | 40 | 5212 | 4.47 | 0.37 | 7.53 | 0.09 | 472 | 7.53 | 0.12 | 40 |
| HD 40512 | E | 6471 | 4.28 | 1.80 | 7.95 | 0.32 | 28 | 8.53 | 0.52 | 3 | 6471 | 4.28 | 1.81 | 7.95 | 0.32 | 28 | 8.53 | 0.51 | 3 |
| HD 40616 | E | 5769 | 3.99 | 1.17 | 7.23 | 0.08 | 514 | 7.31 | 0.16 | 78 | 5769 | 3.80 | 1.33 | 7.21 | 0.08 | 514 | 7.21 | 0.16 | 78 |
| HD 42250 | E | 5410 | 4.48 | 0.71 | 7.59 | 0.10 | 367 | 7.83 | 0.23 | 45 | 5410 | 3.94 | 0.97 | 7.51 | 0.10 | 367 | 7.47 | 0.25 | 45 |
| HD 4256 | E | 4904 | 4.58 | 0.79 | 7.82 | 0.10 | 203 | 7.96 | 0.23 | 17 | 4904 | 4.31 | 1.07 | 7.75 | 0.09 | 203 | 7.77 | 0.23 | 17 |
| HD 42618 | E | 5775 | 4.46 | 0.79 | 7.44 | 0.07 | 402 | 7.56 | 0.15 | 58 | 5775 | 4.19 | 0.95 | 7.42 | 0.07 | 402 | 7.42 | 0.15 | 58 |

| | | | | | | | | | | | | | | | | | |
|---|---|---|---|---|---|---|---|---|---|---|---|---|---|---|---|---|---|
| HD 42983 | E | 4918 | 3.61 | 1.07 | 7.62 | 0.13 | 314 | 7.80 | 0.30 | 40 | 4918 | 3.17 | 1.16 | 7.55 | 0.13 | 314 | 7.55 | 0.31 | 40 |
| HD 43318 | E | 6241 | 3.87 | 1.70 | 7.34 | 0.10 | 474 | 7.33 | 0.16 | 80 | 6241 | 3.89 | 1.73 | 7.33 | 0.10 | 474 | 7.31 | 0.15 | 80 |
| HD 43856 | E | 6150 | 4.19 | 1.31 | 7.31 | 0.08 | 478 | 7.36 | 0.15 | 77 | 6150 | 4.09 | 1.40 | 7.31 | 0.08 | 478 | 7.30 | 0.15 | 77 |
| HD 44966 | E | 6349 | 4.25 | 1.86 | 7.60 | 0.12 | 429 | 7.59 | 0.15 | 61 | 6349 | 4.29 | 1.84 | 7.61 | 0.12 | 429 | 7.61 | 0.15 | 61 |
| HD 45282 | E | 5422 | 3.23 | 1.54 | 6.12 | 0.11 | 341 | 6.07 | 0.15 | 53 | 5422 | 3.36 | 1.50 | 6.12 | 0.11 | 341 | 6.13 | 0.15 | 53 |
| HD 46090 | E | 5575 | 4.36 | 1.04 | 7.40 | 0.09 | 387 | 7.54 | 0.13 | 45 | 5575 | 4.21 | 1.15 | 7.37 | 0.09 | 387 | 7.41 | 0.13 | 45 |
| HD 46301 | E | 6396 | 3.74 | 3.57 | 7.77 | 0.32 | 71 | 7.84 | 0.82 | 10 | 6396 | 3.58 | 3.60 | 7.77 | 0.32 | 71 | 7.77 | 0.82 | 10 |
| HD 4635 | E | 5036 | 4.55 | 0.81 | 7.67 | 0.11 | 324 | 8.00 | 0.28 | 37 | 5036 | 3.71 | 1.17 | 7.48 | 0.12 | 324 | 7.43 | 0.30 | 37 |
| HD 46780 | E | 5793 | 4.25 | 1.95 | 7.56 | 0.15 | 466 | 7.55 | 0.17 | 49 | 5793 | 4.25 | 1.96 | 7.55 | 0.15 | 466 | 7.55 | 0.17 | 49 |
| HD 47157 | E | 5689 | 4.40 | 0.42 | 8.02 | 0.07 | 318 | 8.14 | 0.10 | 49 | 5689 | 4.17 | 1.05 | 7.92 | 0.06 | 318 | 7.92 | 0.09 | 49 |
| HD 47309 | E | 5792 | 4.37 | 1.01 | 7.55 | 0.08 | 385 | 7.67 | 0.16 | 58 | 5792 | 4.11 | 1.13 | 7.54 | 0.08 | 385 | 7.54 | 0.16 | 58 |
| HD 47752 | E | 4698 | 4.62 | 0.71 | 7.48 | 0.14 | 300 | 8.09 | 0.49 | 35 | 4698 | 3.62 | 1.33 | 7.14 | 0.15 | 300 | 7.31 | 0.53 | 35 |
| HD 48565 | E | 6024 | 4.07 | 1.33 | 6.83 | 0.11 | 405 | 6.94 | 0.15 | 67 | 6024 | 3.82 | 1.51 | 6.81 | 0.11 | 405 | 6.80 | 0.15 | 67 |
| HD 49385 | E | 6127 | 3.98 | 1.66 | 7.62 | 0.10 | 526 | 7.65 | 0.17 | 86 | 6127 | 3.91 | 1.70 | 7.62 | 0.10 | 526 | 7.62 | 0.17 | 86 |
| HD 50206 | E | 6459 | 3.88 | 2.12 | 7.61 | 0.11 | 490 | 7.56 | 0.16 | 80 | 6459 | 4.00 | 2.08 | 7.61 | 0.10 | 490 | 7.61 | 0.16 | 80 |
| HD 50867 | E | 6256 | 4.30 | 1.52 | 7.43 | 0.10 | 461 | 7.41 | 0.15 | 72 | 6256 | 4.34 | 1.49 | 7.43 | 0.10 | 461 | 7.43 | 0.15 | 72 |
| HD 51219 | E | 5626 | 4.41 | 0.81 | 7.55 | 0.09 | 381 | 7.74 | 0.22 | 54 | 5626 | 3.98 | 1.04 | 7.53 | 0.09 | 381 | 7.53 | 0.22 | 54 |
| HD 51419 | E | 5723 | 4.46 | 0.64 | 7.17 | 0.07 | 408 | 7.26 | 0.12 | 56 | 5723 | 4.28 | 0.79 | 7.16 | 0.07 | 408 | 7.16 | 0.12 | 56 |
| HD 51866 | E | 4860 | 4.59 | 0.31 | 7.70 | 0.10 | 255 | 7.86 | 0.12 | 15 | 4860 | 4.22 | 0.82 | 7.61 | 0.09 | 255 | 7.61 | 0.12 | 15 |
| HD 52634 | E | 5999 | 4.07 | 2.77 | 7.45 | 0.20 | 241 | 7.43 | 0.21 | 38 | 5999 | 4.11 | 2.75 | 7.45 | 0.20 | 241 | 7.45 | 0.21 | 38 |
| HD 5294 | E | 5749 | 4.47 | 0.82 | 7.42 | 0.08 | 406 | 7.55 | 0.13 | 56 | 5749 | 4.17 | 0.99 | 7.41 | 0.08 | 406 | 7.41 | 0.14 | 56 |
| HD 5351 | E | 4608 | 4.65 | 0.15 | 7.11 | 0.10 | 322 | 7.57 | 0.19 | 21 | 4608 | 3.65 | 2.10 | 6.79 | 0.15 | 322 | 6.84 | 0.19 | 21 |
| HD 53927 | E | 4941 | 4.62 | 0.21 | 7.26 | 0.12 | 499 | 7.67 | 0.34 | 54 | 4941 | 3.65 | 2.02 | 6.98 | 0.17 | 499 | 6.98 | 0.39 | 54 |
| HD 54371 | E | 5562 | 4.43 | 1.36 | 7.54 | 0.09 | 518 | 7.69 | 0.14 | 59 | 5562 | 4.13 | 1.72 | 7.49 | 0.10 | 518 | 7.50 | 0.16 | 59 |
| HD 5600 | E | 6378 | 3.77 | 2.28 | 7.25 | 0.14 | 394 | 7.23 | 0.19 | 63 | 6378 | 3.81 | 2.27 | 7.25 | 0.13 | 394 | 7.24 | 0.19 | 63 |
| HD 56303 | E | 5912 | 4.30 | 1.19 | 7.66 | 0.09 | 536 | 7.76 | 0.16 | 81 | 5912 | 4.08 | 1.41 | 7.63 | 0.10 | 536 | 7.63 | 0.17 | 81 |
| HD 56515 | E | 5993 | 4.07 | 1.41 | 7.54 | 0.09 | 508 | 7.48 | 0.17 | 85 | 5993 | 4.21 | 1.29 | 7.55 | 0.09 | 508 | 7.55 | 0.17 | 85 |
| HD 57707 | E | 5021 | 3.32 | 1.40 | 7.57 | 0.13 | 532 | 7.66 | 0.26 | 55 | 5021 | 3.12 | 1.55 | 7.52 | 0.14 | 532 | 7.52 | 0.26 | 55 |
| HD 59090 | E | 6428 | 3.92 | 2.43 | 7.68 | 0.15 | 345 | 7.57 | 0.17 | 52 | 6428 | 4.16 | 2.35 | 7.68 | 0.15 | 345 | 7.68 | 0.17 | 52 |
| HD 59747 | E | 5099 | 4.57 | 0.43 | 7.62 | 0.08 | 243 | 7.80 | 0.15 | 21 | 5099 | 4.21 | 0.85 | 7.55 | 0.07 | 243 | 7.57 | 0.15 | 21 |
| HD 59984 | E | 6005 | 4.03 | 1.38 | 6.77 | 0.08 | 398 | 6.76 | 0.12 | 68 | 6005 | 4.06 | 1.35 | 6.77 | 0.08 | 398 | 6.77 | 0.12 | 68 |
| HD 62161 | E | 6449 | 4.25 | 3.42 | 7.64 | 0.31 | 120 | 7.52 | 0.25 | 14 | 6449 | 4.54 | 3.29 | 7.65 | 0.31 | 120 | 7.65 | 0.25 | 14 |
| HD 62323 | E | 6113 | 4.06 | 1.60 | 7.50 | 0.10 | 500 | 7.56 | 0.19 | 85 | 6113 | 3.93 | 1.69 | 7.49 | 0.10 | 500 | 7.49 | 0.20 | 85 |
| HD 62613 | E | 5539 | 4.54 | 0.55 | 7.48 | 0.07 | 403 | 7.68 | 0.20 | 50 | 5539 | 4.15 | 0.87 | 7.45 | 0.07 | 403 | 7.47 | 0.20 | 50 |
| HD 63433 | E | 5673 | 4.50 | 1.53 | 7.49 | 0.10 | 495 | 7.59 | 0.16 | 60 | 5673 | 4.26 | 1.79 | 7.46 | 0.11 | 495 | 7.45 | 0.18 | 60 |
| HD 64021 | E | 6864 | 3.88 | 3.80 | 7.77 | 0.47 | 58 | 7.85 | 0.49 | 10 | 6864 | 3.70 | 3.95 | 7.76 | 0.47 | 58 | 7.74 | 0.48 | 10 |
| HD 64090 | E | 5528 | 4.62 | 2.03 | 5.82 | 0.12 | 242 | 5.88 | 0.26 | 31 | 5528 | 4.48 | 2.15 | 5.81 | 0.12 | 242 | 5.81 | 0.26 | 31 |
| HD 64468 | E | 4914 | 4.52 | 0.85 | 7.71 | 0.13 | 301 | 8.13 | 0.35 | 35 | 4914 | 3.52 | 1.27 | 7.44 | 0.14 | 301 | 7.39 | 0.38 | 35 |

| | | | | | | | | | | | | | | | | | |
|---|---|---|---|---|---|---|---|---|---|---|---|---|---|---|---|---|---|
| HD 64606 | E | 5250 | 4.66 | 0.15 | 6.69 | 0.08 | 393 | 6.96 | 0.14 | 35 | 5250 | 4.11 | 1.57 | 6.57 | 0.09 | 393 | 6.59 | 0.12 | 35 |
| HD 64815 | E | 5722 | 3.86 | 1.19 | 7.11 | 0.07 | 495 | 7.21 | 0.17 | 74 | 5722 | 3.64 | 1.34 | 7.09 | 0.08 | 495 | 7.09 | 0.18 | 74 |
| HD 65583 | E | 5342 | 4.55 | 0.41 | 6.85 | 0.09 | 401 | 6.99 | 0.22 | 47 | 5342 | 4.25 | 0.79 | 6.82 | 0.09 | 401 | 6.83 | 0.23 | 47 |
| HD 65874 | E | 5911 | 3.88 | 1.49 | 7.54 | 0.08 | 535 | 7.57 | 0.18 | 87 | 5911 | 3.82 | 1.53 | 7.54 | 0.08 | 535 | 7.54 | 0.19 | 87 |
| HD 68168 | E | 5750 | 4.40 | 0.97 | 7.65 | 0.08 | 385 | 7.75 | 0.14 | 58 | 5750 | 4.17 | 1.07 | 7.64 | 0.08 | 385 | 7.64 | 0.14 | 58 |
| HD 68284 | E | 5945 | 3.92 | 1.40 | 6.98 | 0.09 | 446 | 7.04 | 0.17 | 73 | 5945 | 3.78 | 1.49 | 6.97 | 0.10 | 446 | 6.97 | 0.17 | 73 |
| HD 68380 | E | 6538 | 4.33 | 1.86 | 7.43 | 0.12 | 402 | 7.43 | 0.19 | 63 | 6538 | 4.34 | 1.85 | 7.43 | 0.12 | 402 | 7.43 | 0.19 | 63 |
| HD 68638 | E | 5364 | 4.35 | 1.48 | 7.24 | 0.10 | 498 | 7.30 | 0.21 | 50 | 5364 | 4.24 | 1.63 | 7.22 | 0.11 | 498 | 7.23 | 0.22 | 50 |
| HD 69611 | E | 5846 | 4.25 | 1.06 | 6.92 | 0.08 | 435 | 6.97 | 0.14 | 67 | 5846 | 4.16 | 1.18 | 6.91 | 0.08 | 435 | 6.91 | 0.14 | 67 |
| HD 70088 | E | 5566 | 4.54 | 0.99 | 7.45 | 0.08 | 363 | 7.69 | 0.16 | 50 | 5566 | 3.98 | 1.24 | 7.42 | 0.09 | 363 | 7.41 | 0.16 | 50 |
| HD 70516 | E | 5744 | 4.42 | 2.16 | 7.56 | 0.13 | 393 | 7.62 | 0.15 | 44 | 5744 | 4.28 | 2.27 | 7.55 | 0.13 | 393 | 7.55 | 0.15 | 44 |
| HD 70923 | E | 6021 | 4.22 | 1.33 | 7.60 | 0.08 | 529 | 7.66 | 0.17 | 81 | 6021 | 4.11 | 1.41 | 7.60 | 0.08 | 529 | 7.60 | 0.17 | 81 |
| HD 71431 | E | 5887 | 3.91 | 1.42 | 7.47 | 0.09 | 524 | 7.50 | 0.17 | 83 | 5887 | 3.85 | 1.46 | 7.47 | 0.09 | 524 | 7.47 | 0.17 | 83 |
| HD 71595 | E | 6682 | 4.02 | 1.95 | 7.45 | 0.12 | 419 | 7.43 | 0.19 | 74 | 6682 | 4.06 | 1.94 | 7.45 | 0.12 | 419 | 7.45 | 0.19 | 74 |
| HD 71640 | E | 6098 | 4.31 | 1.29 | 7.32 | 0.09 | 485 | 7.44 | 0.18 | 82 | 6098 | 4.06 | 1.49 | 7.30 | 0.10 | 485 | 7.28 | 0.19 | 82 |
| HD 73393 | E | 5703 | 4.47 | 0.88 | 7.56 | 0.07 | 386 | 7.69 | 0.16 | 59 | 5703 | 4.16 | 1.05 | 7.54 | 0.08 | 386 | 7.54 | 0.17 | 59 |
| HD 74011 | E | 5751 | 4.10 | 1.11 | 6.87 | 0.08 | 451 | 6.97 | 0.13 | 71 | 5751 | 3.88 | 1.31 | 6.85 | 0.08 | 451 | 6.85 | 0.14 | 71 |
| HD 75318 | E | 5422 | 4.40 | 0.78 | 7.36 | 0.08 | 385 | 7.57 | 0.24 | 51 | 5422 | 3.90 | 1.04 | 7.30 | 0.09 | 385 | 7.27 | 0.25 | 51 |
| HD 75767 | E | 5782 | 4.40 | 0.91 | 7.44 | 0.07 | 401 | 7.59 | 0.16 | 58 | 5782 | 4.07 | 1.05 | 7.43 | 0.08 | 401 | 7.43 | 0.17 | 58 |
| HD 7590 | E | 5979 | 4.45 | 1.38 | 7.45 | 0.09 | 479 | 7.47 | 0.14 | 65 | 5979 | 4.40 | 1.43 | 7.44 | 0.09 | 479 | 7.44 | 0.14 | 65 |
| HD 75935 | E | 5451 | 4.50 | 1.45 | 7.50 | 0.09 | 508 | 7.66 | 0.25 | 61 | 5451 | 4.12 | 1.84 | 7.44 | 0.11 | 508 | 7.44 | 0.28 | 61 |
| HD 76932 | E | 5966 | 4.18 | 1.26 | 6.66 | 0.11 | 374 | 6.65 | 0.15 | 58 | 5966 | 4.19 | 1.25 | 6.66 | 0.11 | 374 | 6.66 | 0.15 | 58 |
| HD 77407 | E | 5989 | 4.44 | 1.70 | 7.57 | 0.10 | 509 | 7.58 | 0.12 | 70 | 5989 | 4.40 | 1.67 | 7.56 | 0.10 | 509 | 7.52 | 0.11 | 70 |
| HD 7924 | E | 5172 | 4.60 | 0.83 | 7.32 | 0.09 | 349 | 7.46 | 0.12 | 22 | 5172 | 4.32 | 1.10 | 7.28 | 0.08 | 349 | 7.29 | 0.11 | 22 |
| HD 82443 | E | 5315 | 4.58 | 1.60 | 7.47 | 0.10 | 483 | 7.66 | 0.19 | 46 | 5315 | 4.15 | 1.98 | 7.38 | 0.11 | 483 | 7.35 | 0.20 | 46 |
| HD 84937 | E | 6383 | 4.48 | 2.76 | 5.45 | 0.15 | 85 | 5.48 | 0.14 | 25 | 6383 | 4.38 | 2.77 | 5.45 | 0.15 | 85 | 5.45 | 0.14 | 25 |
| HD 87883 | E | 4940 | 4.56 | 0.70 | 7.70 | 0.13 | 312 | 8.21 | 0.38 | 41 | 4940 | 3.56 | 1.16 | 7.41 | 0.14 | 312 | 7.43 | 0.43 | 41 |
| HD 88725 | E | 5753 | 4.45 | 0.52 | 6.94 | 0.08 | 378 | 7.02 | 0.15 | 60 | 5753 | 4.29 | 0.69 | 6.93 | 0.08 | 378 | 6.94 | 0.15 | 60 |
| HD 89269 | E | 5635 | 4.49 | 0.63 | 7.37 | 0.07 | 409 | 7.53 | 0.16 | 55 | 5635 | 4.16 | 0.84 | 7.34 | 0.07 | 409 | 7.32 | 0.16 | 55 |
| HD 90875 | E | 4567 | 4.56 | 0.64 | 7.92 | 0.14 | 189 | 8.26 | 0.23 | 10 | 4567 | 3.91 | 1.18 | 7.67 | 0.13 | 189 | 7.69 | 0.22 | 10 |
| HD 93215 | E | 5786 | 4.40 | 1.23 | 7.73 | 0.08 | 541 | 7.86 | 0.17 | 75 | 5786 | 4.12 | 1.50 | 7.69 | 0.09 | 541 | 7.69 | 0.18 | 75 |
| HD 9407 | E | 5652 | 4.44 | 0.69 | 7.55 | 0.07 | 399 | 7.71 | 0.17 | 55 | 5652 | 4.12 | 0.89 | 7.53 | 0.07 | 399 | 7.54 | 0.17 | 55 |
| HD 94765 | E | 5033 | 4.58 | 1.00 | 7.55 | 0.10 | 312 | 7.88 | 0.29 | 37 | 5033 | 3.73 | 1.41 | 7.41 | 0.11 | 312 | 7.40 | 0.29 | 37 |
| HD 97503 | E | 4451 | 4.65 | 0.54 | 7.62 | 0.20 | 249 | 8.39 | 0.60 | 25 | 4451 | 3.65 | 1.42 | 7.29 | 0.21 | 249 | 7.68 | 0.62 | 25 |
| HD 97658 | E | 5157 | 4.57 | 0.90 | 7.20 | 0.07 | 301 | 7.29 | 0.09 | 28 | 5157 | 4.39 | 1.08 | 7.17 | 0.06 | 301 | 7.18 | 0.09 | 28 |
| HD 98630 | E | 6033 | 3.86 | 1.83 | 7.72 | 0.10 | 510 | 7.71 | 0.13 | 73 | 6033 | 3.88 | 1.82 | 7.72 | 0.10 | 510 | 7.72 | 0.13 | 73 |
| HD 98800 | E | 4213 | 3.82 | 1.43 | 7.53 | 0.25 | 191 | 8.47 | 0.57 | 25 | 4213 | 3.00 | 1.64 | 7.26 | 0.26 | 191 | 7.89 | 0.59 | 25 |
| HD 99747 | E | 6676 | 4.15 | 2.22 | 7.00 | 0.13 | 312 | 6.99 | 0.18 | 60 | 6676 | 4.17 | 2.21 | 7.00 | 0.13 | 312 | 7.00 | 0.18 | 60 |

| Name | | Teff | logg | vt | [A] | σ | n | [B] | σ | n | Teff | logg | vt | [A] | σ | n | [B] | σ | n |
|---|---|---|---|---|---|---|---|---|---|---|---|---|---|---|---|---|---|---|---|
| HR 1687 | E | 6540 | 4.13 | 2.53 | 7.75 | 0.14 | 351 | 7.72 | 0.16 | 48 | 6540 | 4.20 | 2.51 | 7.75 | 0.14 | 351 | 7.75 | 0.16 | 48 |
| HR 3144 | E | 6064 | 3.70 | 2.49 | 7.69 | 0.14 | 376 | 7.69 | 0.16 | 50 | 6064 | 3.69 | 2.49 | 7.69 | 0.14 | 376 | 7.69 | 0.16 | 50 |
| HR 4657 | E | 6316 | 4.42 | 1.50 | 6.83 | 0.12 | 347 | 6.84 | 0.19 | 61 | 6316 | 4.41 | 1.50 | 6.82 | 0.12 | 347 | 6.82 | 0.19 | 61 |
| HR 4867 | E | 6293 | 4.32 | 3.19 | 7.64 | 0.47 | 63 | 7.52 | 0.25 | 15 | 6293 | 4.58 | 3.05 | 7.64 | 0.47 | 63 | 7.64 | 0.25 | 15 |
| HR 5307 | E | 6461 | 4.16 | 1.72 | 7.90 | 0.20 | 87 | 7.76 | 0.15 | 5 | 6461 | 4.52 | 1.57 | 7.91 | 0.20 | 87 | 7.94 | 0.15 | 5 |
| HR 7438 | E | 6726 | 4.28 | 5.02 | 7.70 | 0.26 | 34 | 7.66 | 0.66 | 8 | 6726 | 4.38 | 5.01 | 7.70 | 0.26 | 34 | 7.70 | 0.66 | 8 |
| HR 784 | E | 6245 | 4.37 | 1.40 | 7.51 | 0.10 | 479 | 7.54 | 0.16 | 69 | 6245 | 4.30 | 1.46 | 7.51 | 0.10 | 479 | 7.51 | 0.16 | 69 |
| HR 7955 | E | 6205 | 3.78 | 1.89 | 7.66 | 0.09 | 487 | 7.52 | 0.13 | 71 | 6205 | 4.09 | 1.76 | 7.67 | 0.09 | 487 | 7.67 | 0.12 | 71 |
| V* BW Ari | E | 5186 | 4.58 | 1.40 | 7.65 | 0.11 | 495 | 7.91 | 0.22 | 45 | 5186 | 3.98 | 1.87 | 7.47 | 0.13 | 495 | 7.40 | 0.25 | 45 |
| V* BZ Cet | E | 5014 | 4.52 | 1.60 | 7.65 | 0.13 | 485 | 7.99 | 0.29 | 53 | 5014 | 3.52 | 2.23 | 7.38 | 0.16 | 485 | 7.25 | 0.34 | 53 |
| V* DI Cam | E | 6337 | 3.89 | 2.28 | 7.35 | 0.27 | 259 | 7.41 | 0.22 | 39 | 6337 | 3.76 | 2.32 | 7.35 | 0.27 | 259 | 7.35 | 0.22 | 39 |
| V* EI Eri | E | 5494 | 3.82 | 2.67 | 7.42 | 0.25 | 37 | 7.34 | 0.24 | 5 | 5494 | 4.02 | 2.58 | 7.44 | 0.25 | 37 | 7.44 | 0.23 | 5 |
| V* GM Com | E | 6709 | 4.29 | 2.86 | 7.42 | 0.15 | 279 | 7.36 | 0.17 | 41 | 6709 | 4.43 | 2.80 | 7.43 | 0.15 | 279 | 7.42 | 0.17 | 41 |
| V* KX Cnc | E | 6008 | 4.09 | 1.60 | 7.54 | 0.10 | 506 | 7.42 | 0.15 | 73 | 6008 | 4.34 | 1.38 | 7.56 | 0.09 | 506 | 7.56 | 0.14 | 73 |
| V* MV UMa | E | 4665 | 4.62 | 0.57 | 7.28 | 0.16 | 325 | 7.81 | 0.43 | 39 | 4665 | 3.62 | 1.37 | 6.97 | 0.17 | 325 | 7.07 | 0.43 | 39 |
| V* NX Aqr | E | 5656 | 4.50 | 1.21 | 7.47 | 0.09 | 526 | 7.63 | 0.19 | 68 | 5656 | 4.17 | 1.56 | 7.42 | 0.10 | 526 | 7.40 | 0.21 | 68 |
| V* V1309 Tau | E | 5791 | 4.47 | 1.51 | 7.62 | 0.10 | 516 | 7.79 | 0.16 | 70 | 5791 | 4.09 | 1.82 | 7.59 | 0.11 | 516 | 7.58 | 0.18 | 70 |
| V* V1386 Ori | E | 5294 | 4.56 | 1.14 | 7.54 | 0.09 | 332 | 7.78 | 0.27 | 44 | 5294 | 3.97 | 1.41 | 7.48 | 0.09 | 332 | 7.47 | 0.27 | 44 |
| V* V1709 Aql | E | 6913 | 4.13 | 3.36 | 7.66 | 0.31 | 125 | 7.68 | 0.26 | 16 | 6913 | 4.08 | 3.38 | 7.66 | 0.31 | 125 | 7.66 | 0.26 | 16 |
| V* V401 Hya | E | 5836 | 4.48 | 1.42 | 7.60 | 0.09 | 523 | 7.69 | 0.17 | 74 | 5836 | 4.28 | 1.53 | 7.58 | 0.10 | 523 | 7.54 | 0.19 | 74 |
| V* V457 Vul | E | 5433 | 4.50 | 1.16 | 7.53 | 0.07 | 322 | 7.53 | 0.07 | 22 | 5433 | 4.49 | 1.17 | 7.53 | 0.07 | 322 | 7.53 | 0.07 | 22 |
| V* V566 Oph | E | 6358 | 4.05 | 2.80 | 6.66 | 0.53 | 14 | 7.43 | 0.05 | 2 | 6358 | 3.05 | 2.50 | 6.64 | 0.53 | 14 | 7.11 | 0.05 | 2 |

| Column Count | Column | Unit | Description |
| --- | --- | --- | --- |
| 2 | Sp | | Source for spectroscopic material |
| 3 | T | K | Effective Temperature |
| 4 | G | cm s$^{-2}$ | Logarithm of the surface acceleration (gravity) computed from average mass, temperature, and luminosity. |
| 5 | $V_t$ | km s$^{-1}$ | Microturbulent velocity |
| 6 | Fe I | log ε | Total iron abundance computed from neutral iron lines.  The solar iron abundance is 7.47 |
| 7 | σ | | Standard deviation of the neutral iron line abundances |
| 8 | N | | Number of neutral iron lines used |
| 9 | Fe II | log ε | Total iron abundance computed from first ionization stage iron lines.  The solar iron abundance is 7.47 |
| 10 | σ | | Standard deviation of the first ionization stage iron line abundances |
| 11 | N | | Number of first ionization stage iron lines used |
| 12 | T | K | Effective Temperature |
| 13 | G | cm s$^{-2}$ | Logarithm of the surface acceleration (gravity) computed from ionization balance |
| 14 | $V_t$ | km s$^{-1}$ | Microturbulent velocity |
| 15 | Fe I | log ε | Total iron abundance computed from neutral iron lines.  The solar iron abundance is 7.47 |
| 16 | σ | | Standard deviation of the neutral iron line abundances |
| 17 | N | | Number of neutral iron lines used |
| 18 | Fe II | log ε | Total iron abundance computed from first ionization stage iron lines.  The solar iron abundance is 7.47 |
| 19 | σ | | Standard deviation of the first ionization stage iron line abundances |
| 20 | N | | Number of first ionization stage iron lines used |

Columns 2 - 11 comprise the mass-derived gravity results
Columns 12 - 20 comprise the ionization balance gravity results
The effective temperature is the same in both cases

Table 4
Z > 10 Abundances for Mass-Derived Gravities

| Primary | Sp | T | G | V$_t$ | Na | Mg | Al | Si | S | Ca | Sc | Ti | V | Cr | Mn | Fe | Co | Ni | Cu | Zn | Sr | Y | Zr | Ba | La | Ce | Nd | Sm | Eu |
|---|---|---|---|---|---|---|---|---|---|---|---|---|---|---|---|---|---|---|---|---|---|---|---|---|---|---|---|---|---|
| 10 CVn | S | 5987 | 4.44 | 1.08 | -0.51 | -0.33 | -0.39 | -0.39 | -0.23 | -0.35 | -0.26 | -0.33 | -0.45 | -0.43 | -0.61 | -0.46 | -0.35 | -0.46 | -0.55 | -0.59 | 0.19 | -0.50 | | -0.49 | | 0.27 | -0.33 | -0.20 | |
| 10 Tau | S | 6013 | 4.05 | 1.48 | 0.04 | 0.06 | 0.02 | 0.00 | 0.03 | 0.02 | 0.05 | -0.01 | -0.07 | -0.03 | -0.11 | -0.06 | 0.01 | -0.06 | -0.11 | -0.02 | 0.23 | 0.01 | 0.03 | -0.08 | -0.08 | 0.05 | 0.09 | -0.11 | 0.13 |
| 107 Psc | S | 5259 | 4.58 | 0.56 | 0.08 | 0.06 | 0.13 | 0.04 | 0.48 | 0.07 | 0.14 | 0.18 | 0.24 | 0.13 | 0.16 | 0.02 | 0.09 | 0.11 | 0.25 | -0.03 | 0.30 | 0.23 | -0.07 | -0.04 | 0.29 | 0.28 | 0.56 | 1.23 | 0.30 |
| 109 Psc | S | 5604 | 3.94 | 1.22 | 0.24 | 0.31 | 0.29 | 0.19 | 0.30 | 0.21 | 0.17 | 0.16 | 0.11 | 0.18 | 0.21 | 0.12 | 0.12 | 0.15 | 0.30 | 0.06 | 0.39 | 0.10 | 0.14 | 0.05 | 0.30 | 0.24 | 0.20 | 0.46 | 0.43 |
| 11 Aql | S | 6144 | 3.57 | 2.83 | 0.35 | -0.18 | 0.27 | 0.14 | 0.14 | 0.17 | 0.07 | 0.15 | 0.04 | 0.12 | -0.04 | 0.02 | 0.17 | 0.03 | -0.35 | | | 0.09 | 0.76 | -0.02 | | 0.20 | 0.00 | | 0.11 |
| 11 Aqr | S | 5929 | 4.24 | 1.40 | 0.53 | 0.33 | 0.35 | 0.32 | 0.37 | 0.32 | 0.38 | 0.31 | 0.32 | 0.33 | 0.38 | 0.26 | 0.34 | 0.34 | 0.34 | 0.28 | 0.53 | 0.35 | 0.37 | 0.09 | 0.44 | 0.40 | 0.52 | 0.50 | 0.62 |
| 11 LMi | S | 5498 | 4.43 | 1.32 | 0.52 | 0.41 | 0.45 | 0.36 | 0.61 | 0.36 | 0.32 | 0.34 | 0.39 | 0.37 | 0.48 | 0.28 | 0.35 | 0.36 | 0.40 | 0.29 | 0.43 | 0.37 | 0.16 | 0.14 | 0.61 | 0.46 | 0.55 | 1.46 | 0.63 |
| 110 Her | S | 6457 | 3.94 | 2.37 | 0.26 | -0.15 | 0.25 | 0.15 | 0.21 | 0.22 | 0.12 | 0.20 | 0.33 | 0.09 | 0.04 | 0.09 | 0.40 | 0.08 | -0.30 | 0.43 | | 0.18 | 0.75 | 0.18 | 0.22 | 0.60 | 0.20 | 0.67 | 0.42 |
| 111 Tau | S | 6184 | 4.38 | 2.04 | 0.26 | 0.17 | 0.17 | 0.14 | 0.17 | 0.19 | 0.24 | 0.19 | 0.27 | 0.16 | 0.01 | 0.09 | 0.33 | 0.09 | -0.32 | 0.33 | | 0.26 | 0.63 | 0.28 | 0.27 | 0.39 | 0.27 | 0.56 | |
| 111 Tau B | S | 4576 | 4.66 | 0.80 | 0.36 | 0.28 | 0.08 | 0.37 | 1.13 | 0.39 | 0.20 | 0.16 | 0.28 | 0.27 | 0.21 | 0.22 | 0.31 | 0.21 | 0.24 | 0.59 | 0.69 | 0.19 | -0.20 | 0.19 | 0.83 | 1.46 | 1.37 | 2.01 | 0.42 |
| 112 Psc | S | 6031 | 4.03 | 1.68 | 0.68 | 0.47 | 0.47 | 0.35 | 0.38 | 0.36 | 0.31 | 0.34 | 0.35 | 0.35 | 0.41 | 0.28 | 0.38 | 0.37 | 0.24 | 0.18 | 0.72 | 0.34 | 0.68 | 0.14 | 0.20 | 0.22 | 0.42 | 0.64 | 0.45 |
| 12 Oph | S | 5262 | 4.57 | 1.02 | 0.08 | 0.20 | 0.21 | 0.11 | 0.49 | 0.10 | 0.07 | 0.14 | 0.14 | 0.12 | 0.16 | 0.03 | 0.10 | 0.07 | 0.16 | 0.06 | 0.20 | 0.14 | -0.04 | 0.07 | 0.22 | 0.29 | 0.48 | 1.37 | 0.42 |
| 13 Cet | S | 6080 | 4.07 | 0.70 | -0.06 | 0.08 | 0.04 | 0.07 | 0.28 | 0.01 | 0.04 | 0.09 | 0.26 | 0.00 | -0.11 | -0.08 | 0.06 | -0.10 | -0.82 | -0.20 | | -0.20 | 0.46 | 0.05 | -0.24 | 0.09 | 0.60 | -0.53 | 0.05 |
| 13 Ori | S | 5800 | 4.07 | 1.22 | -0.17 | 0.04 | 0.08 | -0.04 | -0.09 | 0.03 | 0.00 | 0.02 | -0.07 | -0.14 | -0.30 | -0.17 | -0.07 | -0.17 | -0.21 | -0.24 | 0.27 | -0.11 | -0.20 | -0.16 | -0.09 | 0.21 | 0.14 | 0.06 | 0.26 |
| 13 Tri | S | 5957 | 3.95 | 1.55 | -0.05 | -0.09 | -0.04 | -0.08 | -0.10 | -0.06 | -0.10 | -0.11 | -0.12 | -0.13 | -0.17 | -0.17 | -0.10 | -0.16 | -0.27 | -0.31 | 0.40 | -0.14 | -0.20 | -0.16 | -0.09 | -0.02 | -0.04 | -0.22 | 0.02 |
| 14 Cet | S | 6512 | 3.83 | 1.80 | -0.10 | -0.03 | -0.02 | -0.08 | -0.17 | 0.00 | 0.00 | -0.07 | 0.07 | -0.10 | -0.22 | -0.13 | 0.21 | -0.18 | -0.35 | -0.27 | 0.64 | 0.01 | 0.30 | 0.27 | 0.29 | 0.16 | 0.01 | -0.65 | 0.13 |
| 14 Her | S | 5248 | 4.41 | 0.93 | 0.77 | 0.65 | 0.65 | 0.57 | 0.90 | 0.52 | 0.62 | 0.56 | 0.60 | 0.53 | 0.71 | 0.48 | 0.61 | 0.63 | 0.81 | 0.88 | 0.17 | 0.69 | 0.20 | 0.40 | 0.95 | 0.83 | 0.83 | 1.58 | 0.83 |
| 15 LMi | S | 5916 | 4.06 | 1.42 | 0.25 | 0.23 | 0.26 | 0.15 | 0.17 | 0.19 | 0.19 | 0.15 | 0.13 | 0.13 | 0.10 | 0.09 | 0.15 | 0.12 | 0.15 | 0.00 | 0.50 | 0.13 | 0.14 | 0.04 | 0.30 | 0.13 | 0.23 | 0.41 | 0.51 |
| 15 Sge | S | 5946 | 4.40 | 1.21 | 0.13 | 0.07 | 0.13 | 0.09 | 0.16 | 0.11 | 0.11 | 0.08 | 0.06 | 0.10 | 0.07 | 0.04 | 0.05 | 0.04 | -0.08 | -0.03 | 0.49 | 0.17 | 0.07 | 0.07 | | 0.48 | 0.26 | 0.29 | 0.44 |
| 16 Cyg A | S | 5800 | 4.28 | 1.13 | 0.16 | 0.23 | 0.23 | 0.13 | 0.19 | 0.17 | 0.18 | 0.17 | 0.16 | 0.18 | 0.14 | 0.10 | 0.17 | 0.13 | 0.19 | 0.04 | 0.48 | 0.24 | 0.13 | 0.01 | 0.30 | 0.42 | 0.27 | 0.31 | 0.49 |
| 16 Cyg B | S | 5753 | 4.34 | 0.93 | 0.19 | 0.14 | 0.24 | 0.13 | 0.26 | 0.17 | 0.20 | 0.17 | 0.11 | 0.15 | 0.18 | 0.09 | 0.14 | 0.13 | 0.19 | 0.09 | 0.66 | 0.28 | 0.10 | 0.09 | 0.37 | 0.14 | 0.45 | 0.17 | 0.43 |
| 17 Crt A | S | 6240 | 4.17 | 1.80 | 0.14 | 0.15 | 0.06 | 0.09 | 0.05 | 0.10 | 0.17 | 0.19 | 0.10 | 0.14 | -0.01 | 0.03 | 0.21 | 0.04 | -0.15 | -0.01 | | 0.15 | 0.47 | 0.15 | 0.15 | 0.36 | 0.26 | 0.30 | 0.18 |
| 17 Crt B | S | 6269 | 4.20 | 1.76 | 0.11 | 0.11 | 0.06 | 0.09 | 0.08 | 0.11 | 0.17 | 0.17 | 0.12 | 0.08 | 0.00 | 0.04 | 0.18 | 0.06 | -0.14 | -0.03 | | 0.15 | 0.25 | 0.12 | 0.31 | 0.34 | 0.27 | 0.18 | 0.19 |
| 17 Vir | S | 6146 | 4.33 | 1.60 | 0.22 | 0.25 | 0.18 | 0.18 | 0.16 | 0.23 | 0.22 | 0.21 | 0.14 | 0.19 | 0.15 | 0.13 | 0.22 | 0.15 | -0.14 | 0.15 | | 0.29 | 0.43 | 0.26 | 0.15 | 0.54 | 0.43 | 0.22 | 0.31 |
| 18 Cam | S | 5958 | 3.93 | 1.57 | 0.12 | 0.12 | 0.12 | 0.10 | 0.13 | 0.15 | 0.19 | 0.08 | 0.03 | 0.06 | 0.01 | 0.03 | 0.07 | 0.03 | -0.07 | -0.08 | 0.58 | 0.11 | 0.15 | 0.08 | 0.12 | 0.20 | 0.11 | 0.17 | 0.16 |
| 18 Cet | S | 5861 | 3.99 | 1.29 | -0.18 | -0.06 | -0.02 | -0.12 | -0.15 | -0.09 | -0.09 | -0.11 | -0.16 | -0.17 | -0.27 | -0.20 | -0.13 | -0.20 | -0.25 | -0.21 | | -0.05 | -0.10 | -0.19 | 0.14 | 0.08 | -0.03 | -0.17 | 0.17 |
| 18 Sco | S | 5791 | 4.42 | 1.23 | 0.09 | 0.12 | 0.15 | 0.06 | 0.16 | 0.12 | 0.09 | 0.04 | 0.05 | 0.08 | 0.03 | 0.01 | 0.05 | 0.03 | -0.02 | -0.07 | 0.43 | 0.19 | -0.04 | 0.04 | 0.42 | 0.40 | 0.22 | 0.45 | 0.44 |
| 20 LMi | S | 5771 | 4.32 | 1.33 | 0.41 | 0.34 | 0.34 | 0.26 | 0.27 | 0.27 | 0.30 | 0.25 | 0.28 | 0.26 | 0.29 | 0.20 | 0.28 | 0.27 | 0.11 | 0.14 | 0.52 | 0.30 | 0.22 | 0.09 | 0.49 | 0.54 | 0.43 | 0.66 | 0.59 |
| 21 Eri | S | 5150 | 3.66 | 1.31 | 0.16 | 0.18 | 0.22 | 0.10 | 0.37 | 0.23 | 0.08 | 0.22 | 0.29 | 0.15 | 0.27 | 0.03 | 0.13 | 0.09 | 0.41 | 0.07 | 0.18 | 0.16 | -0.08 | -0.01 | 0.36 | 0.27 | 0.35 | 0.83 | 0.19 |
| 23 Lib | S | 5717 | 4.26 | 1.38 | 0.57 | 0.36 | 0.43 | 0.33 | 0.49 | 0.29 | 0.32 | 0.30 | 0.31 | 0.30 | 0.39 | 0.24 | 0.35 | 0.33 | 0.63 | 0.30 | 0.44 | 0.37 | 0.28 | 0.10 | 0.63 | 0.40 | 0.49 | 0.82 | 0.54 |
| 24 LMi | S | 5760 | 4.03 | 1.35 | 0.13 | 0.10 | 0.12 | 0.08 | 0.09 | 0.08 | 0.11 | 0.04 | 0.04 | 0.05 | -0.02 | -0.01 | 0.06 | 0.02 | 0.02 | 0.04 | 0.47 | 0.04 | 0.04 | -0.05 | 0.13 | 0.19 | 0.09 | 0.14 | 0.37 |
| 26 Dra | S | 5925 | 4.37 | 1.39 | -0.01 | -0.03 | 0.11 | 0.04 | 0.14 | 0.10 | 0.10 | 0.02 | 0.04 | 0.01 | -0.05 | -0.02 | -0.03 | -0.04 | -0.32 | -0.08 | 0.52 | 0.18 | 0.13 | 0.15 | 0.20 | 0.27 | 0.25 | 0.24 | 0.41 |
| 33 Sex | S | 5124 | 3.72 | 1.24 | 0.11 | 0.10 | 0.18 | 0.10 | 0.36 | 0.18 | 0.04 | 0.09 | 0.10 | 0.08 | 0.16 | 0.00 | 0.09 | 0.05 | 0.17 | 0.07 | 0.39 | 0.03 | -0.14 | 0.01 | 0.37 | 0.63 | 0.29 | 1.25 | 0.17 |
| 35 Leo | S | 5736 | 3.93 | 1.39 | -0.01 | 0.17 | 0.22 | 0.10 | 0.15 | 0.12 | 0.12 | 0.08 | 0.05 | 0.08 | 0.05 | 0.03 | 0.11 | 0.07 | 0.07 | -0.03 | 0.42 | 0.15 | 0.01 | -0.02 | 0.15 | 0.16 | 0.17 | 0.19 | 0.35 |
| 36 And | S | 4809 | 3.19 | 1.50 | 0.49 | 0.39 | 0.29 | 0.36 | 0.92 | 0.23 | 0.16 | 0.22 | 0.33 | 0.24 | 0.36 | 0.11 | 0.22 | 0.23 | 0.49 | 1.18 | 0.16 | 0.04 | -0.17 | -0.13 | 0.59 | 0.22 | 0.34 | 1.14 | 0.24 |

| Name | | T | logg | [Fe/H] | | | | | | | | | | | | | | | | | | | | | | | | |
|---|---|---|---|---|---|---|---|---|---|---|---|---|---|---|---|---|---|---|---|---|---|---|---|---|---|---|---|---|
| 36 Oph A | S | 5103 | 4.64 | 0.92 | -0.21 | -0.11 | -0.16 | -0.15 | 0.30 | -0.17 | -0.20 | -0.11 | -0.11 | -0.12 | -0.17 | -0.23 | -0.16 | -0.20 | -0.14 | -0.27 | 0.30 | -0.17 | -0.15 | -0.08 | 0.13 | 0.63 | 0.51 | 0.41 | -0.08 |
| 36 Oph B | S | 5199 | 4.62 | 1.01 | -0.18 | -0.04 | -0.12 | -0.19 | 0.32 | -0.08 | -0.16 | 0.00 | 0.01 | -0.09 | -0.14 | -0.22 | -0.14 | -0.21 | -0.18 | -0.29 | 0.14 | -0.05 | -0.13 | -0.10 | 0.11 | 0.65 | 0.45 | 0.62 | 0.14 |
| 36 UMa | S | 6173 | 4.40 | 1.33 | -0.11 | -0.15 | -0.09 | -0.08 | 0.02 | -0.01 | -0.01 | -0.05 | -0.08 | -0.06 | -0.16 | -0.09 | -0.04 | -0.13 | -0.29 | -0.25 | 0.55 | 0.03 | 0.27 | 0.14 | 0.22 | 0.24 | 0.26 | -0.13 | 0.32 |
| 37 Gem | S | 5932 | 4.40 | 1.02 | -0.10 | -0.18 | -0.04 | -0.10 | -0.10 | -0.05 | -0.06 | -0.06 | -0.10 | -0.11 | -0.18 | -0.15 | -0.06 | -0.15 | -0.19 | -0.19 | 0.32 | 0.03 | 0.06 | -0.06 | 0.22 | 0.04 | 0.17 | 0.22 | 0.30 |
| 38 LMi | S | 6090 | 3.71 | 2.66 | 0.81 | 0.50 | 0.57 | 0.47 | 0.60 | 0.32 | 0.50 | 0.44 | 0.48 | 0.41 | 0.50 | 0.34 | 0.53 | 0.44 | 0.16 | 0.50 | | 0.38 | 0.67 | -0.11 | 0.32 | 0.28 | 0.25 | 0.68 | 0.52 |
| 39 Gem | S | 6112 | 3.81 | 1.78 | -0.27 | -0.24 | -0.49 | -0.25 | -0.17 | -0.25 | -0.30 | -0.28 | -0.32 | -0.39 | -0.50 | -0.42 | -0.27 | -0.42 | -0.56 | -0.34 | 0.35 | -0.46 | -0.30 | -0.25 | | -0.01 | -0.27 | -0.22 | -0.14 |
| 39 Leo | S | 6187 | 4.29 | 1.55 | -0.29 | -0.24 | -0.36 | -0.24 | -0.29 | -0.22 | -0.30 | -0.24 | -0.27 | -0.30 | -0.43 | -0.36 | -0.14 | -0.35 | -0.53 | -0.38 | 0.45 | -0.40 | -0.26 | -0.26 | 0.11 | -0.33 | -0.12 | -0.15 | 0.20 |
| 39 Ser | S | 5830 | 4.45 | 1.02 | -0.41 | -0.31 | -0.28 | -0.32 | -0.13 | -0.31 | -0.30 | -0.33 | -0.33 | -0.39 | -0.54 | -0.43 | -0.34 | -0.45 | -0.51 | -0.44 | | -0.39 | 0.12 | -0.43 | | | -0.20 | -0.10 | 0.27 |
| 39 Tau | S | 5836 | 4.44 | 1.33 | 0.00 | 0.10 | 0.09 | 0.08 | 0.19 | 0.15 | 0.08 | 0.07 | 0.04 | 0.12 | 0.04 | 0.03 | 0.03 | 0.02 | -0.13 | -0.03 | 0.50 | 0.19 | 0.17 | 0.17 | 0.29 | 0.30 | 0.31 | 0.48 | 0.45 |
| 4 Equ | S | 6086 | 3.76 | 1.74 | 0.19 | 0.03 | 0.09 | 0.08 | 0.16 | 0.10 | 0.06 | -0.01 | 0.03 | 0.03 | -0.03 | -0.02 | 0.02 | 0.00 | -0.15 | -0.15 | 0.65 | 0.06 | 0.10 | 0.01 | 0.05 | 0.16 | 0.02 | 0.05 | 0.06 |
| 42 Cap | S | 5706 | 3.65 | 1.62 | 0.16 | 0.04 | 0.06 | 0.03 | 0.14 | 0.07 | 0.02 | 0.08 | 0.02 | 0.10 | 0.08 | -0.03 | 0.04 | 0.03 | -0.03 | -0.18 | 0.51 | 0.12 | 0.02 | -0.03 | -0.07 | 0.30 | 0.10 | 0.22 | 0.21 |
| 44 And | S | 5951 | 3.57 | 1.96 | 0.09 | 0.12 | 0.18 | 0.12 | 0.06 | 0.19 | 0.09 | 0.12 | 0.10 | 0.12 | -0.03 | 0.02 | 0.02 | 0.01 | -0.15 | 0.08 | | 0.14 | 0.20 | 0.27 | -0.03 | 0.22 | 0.19 | 0.00 | 0.14 |
| 47 UMa | S | 5960 | 4.34 | 1.20 | 0.18 | 0.17 | 0.15 | 0.07 | 0.10 | 0.12 | 0.15 | 0.12 | 0.13 | 0.10 | 0.14 | 0.07 | 0.14 | 0.10 | 0.16 | -0.06 | 0.53 | 0.18 | 0.12 | 0.00 | 0.28 | 0.34 | 0.27 | 0.20 | 0.43 |
| 49 Lib | S | 6297 | 3.91 | 2.04 | 0.47 | 0.34 | 0.26 | 0.17 | 0.11 | 0.16 | 0.19 | 0.19 | 0.20 | 0.04 | -0.04 | 0.00 | 0.21 | 0.07 | -0.17 | 0.16 | | 0.08 | 0.25 | -0.16 | 0.02 | 0.32 | -0.07 | -0.17 | 0.14 |
| 49 Per | S | 4984 | 3.45 | 1.06 | 0.24 | 0.21 | 0.31 | 0.26 | 0.70 | 0.31 | 0.12 | 0.22 | 0.27 | 0.21 | 0.31 | 0.10 | 0.14 | 0.21 | 0.25 | 0.35 | 0.18 | 0.12 | -0.14 | 0.15 | 0.28 | 0.23 | 0.32 | 0.90 | 0.29 |
| 5 Ser | S | 6134 | 3.95 | 1.69 | 0.13 | 0.11 | 0.10 | 0.05 | 0.08 | 0.09 | 0.12 | 0.04 | 0.01 | 0.04 | -0.05 | -0.02 | 0.07 | -0.01 | -0.23 | -0.03 | 0.62 | 0.13 | 0.08 | -0.04 | -0.02 | 0.29 | 0.06 | -0.11 | 0.12 |
| 50 Per | S | 6313 | 4.33 | 2.21 | 0.58 | 0.77 | 0.48 | 0.37 | 0.41 | 0.37 | 0.28 | 0.41 | 0.41 | 0.32 | 0.23 | 0.25 | 0.46 | 0.24 | -0.12 | 0.60 | | 0.61 | 0.52 | 0.29 | 0.60 | | 0.53 | 0.74 | 0.43 |
| 51 Boo Bn | S | 5821 | 4.43 | 1.11 | 0.21 | 0.22 | 0.19 | 0.20 | 0.27 | 0.27 | 0.26 | 0.14 | 0.22 | 0.19 | 0.25 | 0.14 | 0.18 | 0.14 | 0.04 | 0.18 | 0.41 | 0.33 | 0.26 | 0.32 | 0.29 | 0.21 | 0.26 | 0.51 | 0.69 |
| 51 Boo Bs | S | 5990 | 4.33 | 1.77 | 0.19 | 0.11 | 0.18 | 0.17 | 0.25 | 0.16 | 0.22 | 0.21 | 0.16 | 0.24 | | 0.11 | | 0.16 | | 0.11 | 0.53 | 0.31 | 0.30 | 0.09 | 0.33 | 0.21 | 0.46 | 0.49 | 0.56 |
| 51 Peg | S | 5799 | 4.34 | 1.28 | 0.39 | 0.24 | 0.33 | 0.24 | 0.31 | 0.26 | 0.30 | 0.26 | 0.28 | 0.28 | 0.33 | 0.20 | 0.30 | 0.27 | 0.21 | 0.13 | 0.58 | 0.32 | 0.27 | 0.13 | 0.78 | 0.23 | 0.34 | 0.71 | 0.52 |
| 54 Cnc | S | 5824 | 3.94 | 1.46 | 0.33 | 0.09 | 0.21 | 0.18 | 0.24 | 0.20 | 0.23 | 0.14 | 0.18 | 0.20 | 0.19 | 0.12 | 0.20 | 0.19 | 0.24 | 0.12 | 0.54 | 0.22 | 0.23 | 0.01 | -0.10 | 0.13 | 0.14 | 0.27 | 0.43 |
| 54 Psc | S | 5289 | 4.52 | 0.72 | 0.42 | 0.36 | 0.33 | 0.27 | 0.58 | 0.28 | 0.33 | 0.39 | 0.46 | 0.33 | 0.43 | 0.24 | 0.33 | 0.34 | 0.59 | 0.44 | 0.41 | 0.31 | 0.09 | 0.21 | 0.82 | 0.81 | 0.88 | 2.10 | 0.62 |
| 55 Vir | S | 5054 | 3.29 | 1.42 | -0.21 | -0.18 | -0.09 | -0.16 | 0.06 | -0.18 | -0.28 | -0.23 | -0.26 | -0.26 | -0.31 | -0.35 | -0.25 | -0.31 | -0.15 | -0.15 | 0.00 | -0.32 | -0.53 | -0.26 | -0.05 | -0.07 | -0.07 | 0.12 | -0.07 |
| 58 Eri | S | 5839 | 4.47 | 1.29 | 0.00 | 0.00 | 0.07 | 0.05 | 0.12 | 0.11 | 0.05 | 0.06 | 0.08 | 0.08 | 0.01 | 0.01 | 0.06 | -0.02 | -0.14 | -0.06 | 0.49 | 0.25 | 0.14 | 0.21 | 0.44 | 0.59 | 0.40 | 0.29 | 0.48 |
| 59 Eri | S | 6275 | 3.90 | 1.99 | 0.25 | 0.14 | 0.17 | 0.18 | 0.13 | 0.24 | 0.30 | 0.26 | 0.23 | 0.19 | 0.08 | 0.13 | 0.23 | 0.14 | -0.05 | 0.11 | | 0.29 | 0.43 | 0.34 | 0.44 | 0.39 | 0.31 | -0.14 | 0.28 |
| 61 Cyg A | S | 4481 | 4.67 | 0.15 | 0.30 | -0.04 | -0.02 | 0.26 | 1.24 | 0.26 | 0.11 | 0.04 | 0.23 | 0.14 | 0.01 | -0.14 | 0.27 | 0.08 | 0.20 | | 0.52 | 0.19 | -0.67 | -0.22 | 0.65 | 1.38 | 1.70 | 2.19 | 0.25 |
| 61 Cyg B | S | 4171 | 4.70 | 0.15 | 0.28 | 0.31 | 0.12 | 0.95 | 2.16 | 0.31 | 0.29 | 0.08 | 0.32 | 0.41 | 0.02 | -0.08 | 0.65 | 0.34 | 0.54 | 1.30 | 0.96 | 0.33 | -0.64 | -0.31 | 1.24 | 1.71 | 2.09 | 2.46 | 1.02 |
| 61 Psc | S | 6273 | 3.92 | 2.11 | 0.47 | 0.26 | 0.27 | 0.26 | 0.35 | 0.29 | 0.32 | 0.28 | 0.32 | 0.27 | 0.17 | 0.18 | 0.36 | 0.22 | -0.04 | 0.15 | | 0.38 | 0.43 | 0.28 | 0.20 | 0.45 | 0.29 | 0.04 | 0.48 |
| 61 UMa | S | 5507 | 4.54 | 1.08 | -0.11 | -0.02 | 0.03 | 0.00 | 0.31 | 0.03 | -0.03 | 0.00 | 0.00 | 0.04 | 0.02 | -0.06 | -0.01 | -0.05 | -0.21 | -0.10 | 0.30 | 0.15 | 0.03 | 0.08 | 0.41 | 0.31 | 0.34 | 0.66 | 0.24 |
| 61 Vir | S | 5578 | 4.44 | 0.92 | -0.01 | 0.12 | 0.14 | 0.05 | 0.17 | 0.09 | 0.08 | 0.08 | 0.04 | 0.07 | 0.06 | -0.01 | 0.05 | 0.02 | 0.09 | -0.03 | | 0.21 | -0.03 | -0.06 | 0.33 | 0.52 | 0.32 | 0.89 | 0.49 |
| 63 Eri | S | 5422 | 3.33 | 1.69 | -0.09 | -0.07 | 0.03 | -0.14 | -0.10 | -0.05 | -0.23 | -0.06 | -0.07 | -0.09 | -0.17 | -0.18 | -0.08 | -0.20 | -0.28 | -0.25 | 0.38 | -0.13 | -0.21 | 0.06 | -0.03 | -0.01 | -0.05 | -0.12 | 0.12 |
| 64 Aql | S | 4736 | 3.14 | 1.17 | 0.16 | 0.26 | 0.35 | 0.28 | 0.78 | 0.15 | 0.07 | 0.16 | 0.31 | 0.14 | 0.18 | 0.01 | 0.12 | 0.15 | 0.64 | 0.53 | 0.06 | 0.00 | -0.33 | -0.12 | 0.43 | 0.17 | 0.44 | 0.90 | 0.14 |
| 66 Cet | S | 6102 | 3.83 | 1.68 | 0.24 | 0.11 | 0.15 | 0.08 | 0.08 | 0.14 | 0.09 | 0.07 | 0.06 | 0.09 | 0.04 | 0.03 | 0.12 | 0.07 | -0.05 | 0.00 | 0.48 | 0.03 | 0.19 | -0.06 | 0.19 | -0.05 | 0.06 | -0.11 | 0.06 |
| 66 Cet B | S | 5722 | 4.22 | 1.31 | 0.23 | 0.20 | 0.25 | 0.13 | 0.12 | 0.18 | 0.09 | 0.18 | 0.14 | 0.20 | 0.17 | 0.08 | 0.18 | 0.13 | 0.07 | 0.06 | 0.63 | 0.11 | 0.20 | -0.13 | 0.20 | 0.33 | 0.28 | 0.48 | 0.31 |
| 70 Oph A | S | 5244 | 4.47 | 0.91 | 0.08 | 0.08 | 0.08 | 0.11 | 0.39 | 0.06 | 0.07 | 0.10 | 0.09 | 0.11 | 0.12 | 0.02 | 0.06 | 0.08 | 0.16 | -0.06 | 0.16 | 0.13 | -0.07 | 0.06 | 0.34 | 0.63 | 0.63 | 0.91 | 0.33 |
| 70 Vir | S | 5538 | 3.90 | 1.27 | -0.07 | 0.05 | 0.05 | -0.01 | 0.07 | 0.03 | -0.01 | 0.01 | -0.01 | -0.03 | -0.07 | -0.09 | -0.02 | -0.10 | 0.03 | -0.23 | 0.19 | -0.10 | -0.10 | -0.12 | 0.07 | 0.36 | 0.22 | -0.06 | 0.25 |
| 79 Cet | S | 5765 | 4.03 | 1.40 | 0.15 | 0.26 | 0.22 | 0.14 | 0.27 | 0.21 | 0.15 | 0.17 | 0.16 | 0.19 | 0.16 | 0.13 | 0.15 | 0.15 | -0.01 | 0.11 | 0.41 | 0.15 | 0.20 | 0.06 | 0.31 | 0.31 | 0.34 | 0.09 | 0.47 |
| 83 Leo | S | 5380 | 4.42 | 1.04 | 0.59 | 0.48 | 0.48 | 0.40 | 0.81 | 0.34 | 0.39 | 0.32 | 0.37 | 0.34 | 0.48 | 0.30 | 0.37 | 0.39 | 0.48 | 0.51 | 0.35 | 0.45 | 0.11 | 0.15 | 0.56 | 0.51 | 0.60 | 1.89 | 0.65 |

| Name | | T | logg | vt | | | | | | | | | | | | | | | | | | | | | | |
|---|---|---|---|---|---|---|---|---|---|---|---|---|---|---|---|---|---|---|---|---|---|---|---|---|---|---|
| 83 Leo B | S | 4973 | 4.58 | 0.75 | 0.69 | 0.51 | 0.56 | 0.43 | 0.97 | 0.44 | 0.47 | 0.50 | 0.78 | 0.45 | 0.51 | 0.32 | 0.48 | 0.47 | 0.58 | 1.27 | 0.41 | 0.47 | 0.07 | 0.16 | 0.67 | 0.79 | 1.21 | 2.18 | 0.85 |
| 84 Her | S | 5810 | 3.77 | 1.76 | 0.63 | 0.36 | 0.35 | 0.33 | 0.37 | 0.32 | 0.33 | 0.30 | 0.29 | 0.32 | 0.30 | 0.23 | 0.29 | 0.31 | -0.07 | 0.27 | 0.81 | 0.30 | 0.45 | 0.19 | 0.31 | 0.34 | 0.39 | 0.39 | 0.39 |
| 85 Peg | S | 5454 | 4.54 | 0.58 | -0.63 | -0.45 | -0.38 | -0.43 | -0.02 | -0.44 | -0.47 | -0.38 | -0.47 | -0.64 | -0.95 | -0.73 | -0.53 | -0.69 | -0.81 | -0.40 | | -0.63 | -0.17 | -0.75 | -0.06 | 0.62 | -0.29 | -0.15 | |
| 88 Leo | S | 6030 | 4.39 | 1.28 | 0.03 | 0.05 | 0.08 | 0.03 | 0.11 | 0.10 | 0.07 | 0.06 | 0.03 | 0.06 | -0.02 | 0.00 | 0.05 | 0.00 | -0.28 | -0.08 | 0.67 | 0.16 | 0.11 | 0.14 | 0.04 | 0.10 | 0.28 | 0.15 | 0.45 |
| 9 Cet | S | 5822 | 4.46 | 1.62 | 0.35 | 0.24 | 0.30 | 0.28 | 0.39 | 0.33 | 0.32 | 0.30 | 0.31 | 0.31 | 0.28 | 0.22 | 0.28 | 0.24 | 0.10 | 0.12 | 0.64 | 0.33 | 0.46 | 0.33 | 0.51 | 0.54 | 0.52 | 0.60 | 0.45 |
| 9 Com | S | 6239 | 3.91 | 2.06 | 0.55 | 0.47 | 0.36 | 0.29 | 0.33 | 0.32 | 0.33 | 0.28 | 0.24 | 0.27 | 0.26 | 0.21 | 0.30 | 0.25 | -0.08 | 0.30 | | 0.25 | 0.50 | 0.16 | -0.07 | 0.55 | 0.25 | 0.31 | 0.34 |
| 94 Aqr | S | 5379 | 3.82 | 1.40 | 0.32 | 0.20 | 0.30 | 0.24 | 0.34 | 0.21 | 0.13 | 0.14 | 0.08 | 0.20 | 0.23 | 0.07 | 0.15 | 0.15 | 0.22 | 0.11 | 0.39 | 0.25 | -0.03 | -0.14 | 0.29 | 0.46 | 0.19 | 1.03 | 0.40 |
| 94 Aqr B | S | 5219 | 4.35 | 1.29 | 0.49 | 0.35 | 0.39 | 0.24 | 0.58 | 0.32 | 0.28 | 0.34 | 0.42 | 0.31 | 0.38 | 0.19 | 0.33 | 0.29 | 0.53 | 0.55 | 0.30 | 0.30 | 0.03 | -0.06 | 0.59 | 1.04 | 0.80 | 1.42 | 0.44 |
| 94 Cet | S | 6064 | 4.04 | 1.82 | 0.49 | 0.33 | 0.28 | 0.28 | 0.35 | 0.29 | 0.40 | 0.28 | 0.28 | 0.29 | 0.20 | 0.19 | 0.25 | 0.25 | 0.02 | 0.20 | | 0.30 | 0.52 | 0.13 | 0.00 | 0.44 | 0.38 | 0.52 | 0.37 |
| alf Aql | S | 7377 | 3.95 | 5.68 | | | 0.36 | | -0.49 | -1.28 | 1.19 | 1.75 | 0.64 | 1.27 | 0.48 | 1.47 | 1.19 | | | | | 1.11 | 1.50 | -0.62 | | 0.64 | 1.46 | | 0.99 |
| alf Cep | S | 7217 | 3.69 | 3.82 | 0.29 | -0.11 | | 0.51 | | -0.12 | 0.54 | 1.12 | 1.32 | 1.37 | 0.43 | 0.36 | 1.37 | 0.83 | | | | 1.71 | | | | 1.47 | 1.63 | 0.28 | 0.66 |
| alf CMi | S | 6654 | 3.95 | 2.25 | 0.15 | -0.04 | 0.05 | 0.06 | 0.02 | 0.01 | 0.13 | 0.06 | 0.01 | 0.02 | 0.00 | -0.04 | 0.24 | -0.01 | -0.11 | -0.08 | 0.74 | 0.09 | 0.17 | 0.03 | | 0.14 | 0.08 | 0.02 | 0.11 |
| alf Com A | S | 6391 | 4.09 | 2.29 | 0.03 | 0.21 | | 0.00 | -0.03 | 0.07 | -0.28 | 0.06 | 0.14 | -0.06 | -0.22 | -0.09 | 0.34 | -0.13 | -0.54 | 0.63 | | 0.02 | 0.98 | -0.01 | 0.28 | 0.21 | 0.15 | | 0.87 |
| alf Crv | S | 7019 | 4.27 | 2.77 | 0.08 | -0.12 | 0.25 | 0.08 | -0.02 | -0.04 | 0.02 | 0.13 | 0.54 | 0.08 | -0.12 | -0.04 | 0.52 | -0.01 | -0.54 | 0.04 | | 0.53 | 0.72 | -0.03 | | 0.31 | | | 1.38 |
| alf For A | S | 6195 | 3.95 | 1.77 | -0.11 | -0.22 | -0.15 | -0.12 | -0.15 | -0.08 | -0.10 | -0.14 | -0.07 | -0.19 | -0.23 | -0.22 | -0.02 | -0.23 | -0.40 | -0.09 | | -0.27 | 0.30 | -0.08 | -0.15 | -0.08 | -0.03 | -0.55 | |
| alf PsA | S | 7671 | 4.03 | 3.90 | 1.75 | 0.86 | | 0.11 | 0.17 | 0.06 | 1.43 | 1.66 | 1.51 | 1.09 | 0.34 | 0.31 | 1.56 | 1.01 | | | | 0.84 | | -0.08 | 1.34 | 0.71 | | 2.12 | |
| b Aql | S | 5466 | 4.10 | 1.45 | 0.73 | 0.47 | 0.51 | 0.45 | 0.81 | 0.33 | 0.43 | 0.37 | 0.30 | 0.41 | 0.53 | 0.30 | 0.41 | 0.40 | 0.81 | 0.43 | | 0.45 | 0.20 | 0.08 | 0.67 | 0.72 | 0.46 | 1.31 | 0.69 |
| b01 Cyg | S | 5090 | 3.69 | 1.38 | 0.14 | 0.17 | 0.22 | 0.17 | 0.52 | 0.12 | -0.02 | 0.04 | 0.06 | 0.06 | 0.03 | -0.03 | 0.03 | 0.01 | 0.17 | 0.22 | 0.14 | -0.07 | -0.17 | -0.08 | 0.30 | 0.17 | 0.24 | 0.80 | 0.21 |
| bet Aql | S | 5144 | 3.58 | 1.22 | -0.17 | -0.02 | -0.03 | -0.06 | 0.15 | -0.04 | -0.17 | -0.08 | -0.08 | -0.11 | -0.10 | -0.19 | -0.12 | -0.15 | -0.05 | -0.01 | 0.15 | -0.16 | -0.36 | -0.15 | -0.01 | -0.05 | 0.05 | 0.53 | 0.01 |
| bet Com | S | 6022 | 4.40 | 1.31 | 0.05 | 0.11 | 0.05 | 0.08 | 0.20 | 0.15 | 0.10 | 0.09 | 0.04 | 0.10 | 0.06 | 0.06 | 0.04 | 0.03 | -0.10 | -0.10 | | 0.22 | 0.13 | 0.16 | 0.29 | 0.31 | 0.36 | 0.70 | 0.31 |
| bet CVn | S | 5865 | 4.40 | 1.04 | -0.18 | -0.10 | -0.10 | -0.13 | -0.13 | -0.13 | -0.11 | -0.14 | -0.19 | -0.17 | -0.26 | -0.22 | -0.17 | -0.20 | -0.24 | -0.30 | 0.45 | -0.11 | -0.11 | -0.15 | 0.21 | 0.03 | 0.14 | 0.11 | 0.16 |
| bet Vir | S | 6159 | 4.08 | 1.60 | 0.34 | 0.21 | 0.20 | 0.19 | 0.18 | 0.23 | 0.27 | 0.19 | 0.18 | 0.21 | 0.17 | 0.14 | 0.21 | 0.17 | 0.06 | 0.04 | 0.74 | 0.27 | 0.41 | 0.17 | 0.30 | 0.26 | 0.23 | -0.06 | 0.30 |
| c Eri | S | 7146 | 4.23 | 4.98 | | | 0.44 | 0.20 | | 0.63 | 0.58 | 0.88 | 1.19 | 0.56 | 0.58 | 0.24 | 1.25 | 0.56 | | | | 0.87 | 2.09 | | 1.20 | | 0.65 | | 2.00 |
| c UMa | S | 5995 | 4.12 | 1.48 | 0.23 | 0.05 | 0.13 | 0.12 | 0.14 | 0.13 | 0.17 | 0.09 | 0.10 | 0.10 | 0.05 | 0.05 | 0.14 | 0.09 | 0.03 | -0.03 | 0.43 | 0.09 | 0.10 | 0.00 | 0.17 | 0.27 | 0.23 | 0.03 | 0.30 |
| chi Cnc | S | 6274 | 4.25 | 1.63 | -0.19 | -0.27 | -0.16 | -0.16 | -0.15 | -0.15 | -0.11 | -0.16 | -0.06 | -0.25 | -0.36 | -0.28 | -0.14 | -0.30 | -0.45 | -0.21 | | -0.21 | -0.22 | -0.12 | -0.01 | 0.19 | -0.14 | 0.06 | 0.01 |
| chi Dra | S | 6083 | 4.20 | 1.12 | -0.53 | -0.53 | -0.78 | -0.51 | -0.40 | -0.51 | -0.42 | -0.42 | -0.34 | -0.51 | -0.60 | -0.64 | 0.04 | -0.62 | -0.96 | -0.39 | | -0.36 | 0.06 | -0.62 | | 0.32 | -0.22 | 0.18 | |
| chi Her | S | 5890 | 3.96 | 1.44 | -0.49 | -0.35 | -0.33 | -0.37 | -0.44 | -0.33 | -0.42 | -0.34 | -0.40 | -0.45 | -0.61 | -0.50 | -0.40 | -0.50 | -0.60 | -0.47 | | -0.59 | -0.36 | -0.48 | | -0.60 | -0.34 | -0.54 | -0.01 |
| chi01 Ori | S | 5983 | 4.44 | 1.83 | 0.04 | 0.09 | 0.11 | 0.05 | 0.11 | 0.11 | 0.01 | 0.11 | 0.11 | 0.09 | -0.01 | 0.02 | 0.16 | -0.01 | -0.25 | 0.04 | | 0.18 | 0.14 | 0.18 | 0.60 | 0.43 | 0.34 | 0.16 | 0.36 |
| del Cap | S | 7021 | 4.00 | 4.40 | -0.62 | 1.55 | | 1.12 | -0.60 | -0.98 | 0.94 | 1.25 | 1.90 | 0.97 | -0.15 | 0.73 | 1.53 | 0.72 | | | | 0.94 | | | 1.56 | | | 1.27 | |
| del Eri | S | 5076 | 3.77 | 1.19 | 0.24 | 0.25 | 0.33 | 0.22 | 0.62 | 0.19 | 0.13 | 0.25 | 0.30 | 0.20 | 0.32 | 0.08 | 0.21 | 0.16 | 0.38 | 0.31 | 0.14 | 0.17 | -0.17 | -0.05 | 0.41 | 0.53 | 0.43 | 1.27 | 0.25 |
| del Tri | S | 5796 | 4.37 | 1.22 | -0.40 | -0.41 | -0.36 | -0.36 | -0.26 | -0.34 | -0.27 | -0.22 | -0.22 | -0.26 | -0.28 | -0.47 | -0.26 | -0.45 | -0.51 | -0.10 | | -0.18 | 0.09 | -0.52 | 0.06 | 0.68 | 0.06 | 0.00 | -0.02 |
| e Vir | S | 5999 | 4.17 | 1.75 | 0.32 | 0.19 | 0.24 | 0.21 | 0.24 | 0.25 | 0.29 | 0.20 | 0.20 | 0.21 | 0.09 | 0.13 | 0.20 | 0.16 | -0.08 | 0.10 | 0.64 | 0.27 | 0.46 | 0.22 | 0.33 | 0.49 | 0.42 | 0.14 | 0.38 |
| eps Eri | S | 5123 | 4.57 | 0.90 | 0.01 | 0.12 | 0.03 | -0.03 | 0.35 | 0.08 | 0.01 | 0.08 | 0.09 | 0.08 | 0.10 | -0.04 | -0.02 | -0.01 | 0.11 | -0.07 | 0.27 | 0.00 | -0.14 | 0.07 | 0.36 | 0.52 | 0.74 | 1.34 | 0.18 |
| eps For | S | 5129 | 3.57 | 1.04 | -0.43 | -0.22 | -0.26 | -0.29 | -0.10 | -0.23 | -0.41 | -0.23 | -0.34 | -0.48 | -0.75 | -0.58 | -0.42 | -0.52 | -0.50 | -0.47 | -0.11 | -0.44 | -0.50 | -0.60 | -0.62 | -0.31 | -0.27 | 0.05 | -0.32 |
| eta Ari | S | 6485 | 3.98 | 1.84 | -0.12 | 0.01 | -0.08 | -0.08 | -0.09 | 0.00 | -0.08 | -0.02 | -0.01 | -0.05 | -0.21 | -0.14 | 0.09 | -0.18 | -0.37 | -0.44 | | -0.01 | -0.16 | 0.23 | 0.59 | 0.17 | 0.11 | 0.23 | 0.22 |
| eta Boo | S | 6050 | 3.75 | 2.38 | 0.88 | 0.48 | 0.53 | 0.42 | 0.56 | 0.38 | 0.43 | 0.36 | 0.35 | 0.35 | 0.45 | 0.28 | 0.41 | 0.39 | 0.25 | 0.55 | | 0.44 | 0.50 | 0.12 | 0.29 | | 0.14 | 0.30 | 0.48 |
| eta Cas | S | 5937 | 4.37 | 1.14 | -0.19 | -0.11 | -0.29 | -0.18 | -0.16 | -0.13 | -0.22 | -0.19 | -0.21 | -0.18 | -0.32 | -0.25 | -0.15 | -0.24 | -0.36 | -0.30 | | -0.15 | 0.00 | -0.22 | 0.04 | -0.20 | -0.01 | 0.05 | 0.16 |
| eta Cep | S | 5057 | 3.42 | 1.33 | 0.01 | 0.02 | 0.05 | 0.02 | 0.30 | 0.02 | -0.06 | -0.02 | 0.02 | -0.02 | -0.02 | -0.12 | -0.03 | -0.09 | 0.16 | -0.01 | 0.20 | -0.10 | -0.35 | -0.08 | 0.25 | 0.16 | 0.18 | 0.40 | 0.04 |

| Name | | T | log g | ξ | | | | | | | | | | | | | | | | | | | | | | | | |
|---|---|---|---|---|---|---|---|---|---|---|---|---|---|---|---|---|---|---|---|---|---|---|---|---|---|---|---|---|
| eta CrB A | S | 6060 | 4.45 | 1.40 | 0.00 | 0.04 | 0.01 | 0.04 | -0.02 | 0.10 | 0.05 | 0.03 | 0.03 | 0.01 | -0.08 | -0.03 | 0.00 | -0.03 | -0.32 | -0.13 | 0.56 | 0.09 | 0.21 | 0.16 | 0.18 | 0.34 | 0.25 | 0.36 | 0.21 |
| eta CrB B | S | 5948 | 4.51 | 1.18 | -0.04 | 0.03 | -0.06 | 0.02 | 0.20 | 0.06 | 0.07 | -0.01 | -0.03 | 0.01 | -0.09 | -0.04 | -0.06 | -0.07 | -0.21 | -0.09 | 0.53 | 0.17 | 0.17 | 0.22 | 0.14 | 0.23 | 0.24 | 0.11 | 0.20 |
| eta Ser | S | 4985 | 3.15 | 1.37 | -0.06 | -0.15 | 0.03 | -0.04 | 0.21 | 0.00 | -0.16 | -0.07 | -0.04 | -0.07 | -0.08 | -0.17 | -0.13 | -0.14 | 0.01 | -0.15 | 0.23 | -0.04 | -0.29 | -0.04 | 0.10 | -0.04 | 0.09 | 0.18 | 0.02 |
| gam Cep | S | 4850 | 3.21 | 1.50 | 0.37 | 0.37 | 0.30 | 0.30 | 0.76 | 0.29 | 0.20 | 0.26 | 0.42 | 0.28 | 0.41 | 0.15 | 0.26 | 0.25 | 0.15 | 0.99 | 0.25 | 0.13 | -0.12 | 0.08 | 0.66 | 0.36 | 0.51 | 1.09 | 0.29 |
| gam Lep | S | 6352 | 4.31 | 1.70 | 0.03 | -0.04 | -0.02 | 0.05 | 0.05 | 0.09 | 0.12 | 0.12 | 0.15 | 0.04 | -0.11 | -0.02 | 0.25 | -0.04 | -0.26 | -0.05 | | 0.10 | 0.23 | 0.18 | | 0.29 | 0.20 | 0.34 | 0.28 |
| gam Ser | S | 6286 | 4.13 | 1.86 | -0.06 | 0.02 | -0.04 | -0.05 | 0.00 | -0.01 | 0.04 | -0.03 | -0.01 | -0.03 | -0.16 | -0.12 | 0.14 | -0.13 | -0.41 | -0.21 | | 0.06 | 0.60 | 0.01 | 0.05 | 0.20 | 0.04 | -0.04 | 0.59 |
| gam Vir A | S | 6922 | 4.27 | 2.83 | 0.30 | -0.30 | 0.34 | 0.20 | 0.26 | 0.10 | 0.05 | 0.36 | 0.67 | 0.16 | 0.22 | 0.08 | 0.91 | 0.08 | -0.46 | | | 0.30 | 0.18 | 0.29 | | | | | 0.97 |
| gam01 Del | S | 6194 | 3.72 | 1.91 | 0.18 | 0.12 | 0.12 | 0.16 | 0.15 | 0.15 | 0.11 | 0.11 | 0.10 | 0.13 | 0.03 | 0.04 | 0.15 | 0.07 | -0.15 | -0.18 | 0.83 | 0.13 | 0.29 | 0.18 | 0.03 | 0.14 | 0.07 | 0.05 | 0.08 |
| i Boo A | S | 5878 | 4.38 | 3.80 | | | 1.22 | | | -0.72 | -0.23 | -0.27 | -0.22 | -0.26 | | -0.83 | 0.29 | -0.44 | | | | -0.45 | 1.45 | | 0.78 | 0.25 | -0.39 | 0.67 | |
| i Boo B | S | 5240 | 4.50 | 0.96 | -0.58 | -0.36 | -0.40 | -0.13 | 0.26 | -0.57 | -0.36 | -0.63 | -0.75 | -0.55 | -0.76 | -0.48 | -0.61 | -0.55 | -0.78 | -0.02 | -0.25 | -0.33 | -0.59 | -0.37 | 0.04 | 0.17 | -0.04 | -0.06 | -0.03 |
| iot Peg | S | 6504 | 4.19 | 2.02 | -0.10 | -0.09 | -0.15 | -0.06 | -0.05 | -0.05 | 0.14 | 0.07 | 0.19 | -0.03 | -0.08 | -0.10 | 0.37 | -0.11 | -0.59 | -0.23 | | 0.06 | 0.47 | -0.13 | 0.40 | 0.44 | 0.11 | 0.10 | 0.15 |
| iot Per | S | 5995 | 4.15 | 1.49 | 0.28 | 0.19 | 0.23 | 0.17 | 0.24 | 0.18 | 0.26 | 0.17 | 0.15 | 0.16 | 0.17 | 0.11 | 0.21 | 0.18 | 0.08 | 0.11 | 0.66 | 0.19 | 0.28 | 0.07 | 0.19 | 0.47 | 0.35 | 0.45 | 0.33 |
| iot Psc | S | 6177 | 4.08 | 1.60 | -0.07 | 0.02 | -0.04 | -0.04 | -0.02 | 0.03 | 0.10 | -0.04 | -0.06 | -0.06 | -0.17 | -0.11 | -0.01 | -0.13 | -0.24 | -0.24 | | 0.14 | 0.23 | 0.12 | 0.23 | 0.02 | 0.10 | 0.20 | 0.32 |
| kap Del | S | 5633 | 3.66 | 1.60 | 0.09 | 0.14 | 0.15 | 0.09 | 0.25 | 0.09 | 0.05 | 0.01 | 0.00 | 0.03 | -0.01 | -0.03 | 0.03 | -0.01 | -0.09 | -0.21 | 0.51 | 0.01 | 0.02 | 0.11 | 0.11 | 0.18 | 0.14 | 0.00 | 0.28 |
| kap For | S | 5867 | 3.96 | 1.43 | 0.10 | 0.07 | 0.09 | 0.03 | 0.07 | -0.01 | 0.07 | 0.00 | -0.02 | 0.01 | -0.05 | -0.07 | 0.00 | -0.02 | -0.06 | -0.08 | 0.54 | 0.05 | -0.01 | -0.09 | | -0.04 | -0.01 | -0.10 | 0.38 |
| kap01 Cet | S | 5730 | 4.50 | 1.47 | 0.04 | 0.14 | 0.19 | 0.10 | 0.26 | 0.14 | 0.10 | 0.10 | 0.12 | 0.13 | 0.10 | 0.06 | 0.05 | 0.05 | -0.05 | 0.01 | 0.42 | 0.20 | 0.26 | 0.18 | 0.27 | 0.40 | 0.44 | 0.86 | 0.28 |
| ksi Boo A | S | 5480 | 4.53 | 1.38 | -0.20 | -0.08 | -0.10 | -0.10 | 0.17 | -0.08 | -0.14 | -0.10 | -0.14 | -0.09 | -0.18 | -0.18 | -0.20 | -0.21 | -0.34 | -0.33 | 0.19 | -0.13 | -0.06 | 0.08 | 0.21 | 0.23 | 0.31 | 0.12 | 0.24 |
| ksi Boo B | S | 4767 | 5.00 | 0.15 | 0.07 | 0.12 | 0.15 | 0.19 | 1.00 | 0.56 | 0.36 | 0.41 | 0.60 | 0.38 | 0.23 | -0.02 | 0.33 | 0.14 | 0.21 | 0.71 | 0.89 | 0.61 | 0.31 | 0.24 | 1.55 | 1.65 | 1.91 | 2.25 | 0.46 |
| ksi Peg | S | 6250 | 4.01 | 1.77 | -0.06 | -0.13 | -0.15 | -0.12 | -0.11 | -0.08 | -0.04 | -0.08 | -0.05 | -0.15 | -0.24 | -0.19 | 0.04 | -0.18 | -0.30 | -0.26 | | -0.16 | -0.35 | -0.02 | | 0.29 | -0.08 | -0.03 | 0.15 |
| ksi UMa B | S | 5667 | 4.44 | 0.77 | -0.75 | -0.47 | -0.47 | -0.41 | -0.08 | -0.51 | -0.40 | -0.50 | -0.62 | -0.45 | -0.49 | -0.68 | -0.23 | -0.68 | -0.88 | -0.38 | | -0.48 | -0.35 | -0.30 | 0.24 | 0.41 | -0.21 | -0.09 | 0.10 |
| lam Aur | S | 5931 | 4.28 | 1.28 | 0.22 | 0.20 | 0.20 | 0.13 | 0.22 | 0.17 | 0.19 | 0.15 | 0.12 | 0.14 | 0.13 | 0.09 | 0.17 | 0.13 | 0.14 | 0.10 | 0.37 | 0.23 | 0.11 | 0.06 | 0.42 | 0.54 | 0.27 | 0.17 | 0.46 |
| lam Ser | S | 5908 | 4.13 | 1.45 | 0.07 | 0.06 | 0.17 | 0.03 | -0.05 | 0.06 | 0.16 | 0.03 | 0.02 | 0.01 | -0.06 | -0.04 | 0.06 | 0.00 | -0.07 | -0.16 | 0.27 | 0.08 | -0.04 | -0.06 | 0.29 | 0.05 | 0.21 | 0.16 | 0.32 |
| m Per | S | 6704 | 3.83 | 2.15 | -1.13 | 1.33 | | | | -0.60 | 1.74 | 1.46 | 1.64 | 1.79 | | 0.06 | | -0.05 | | | | 0.90 | | | | | | | 0.82 |
| m Tau | S | 5631 | 4.03 | 1.11 | -0.14 | -0.07 | -0.04 | -0.08 | 0.02 | -0.07 | -0.11 | -0.10 | -0.21 | -0.17 | -0.33 | -0.23 | -0.17 | -0.23 | -0.20 | -0.11 | 0.27 | -0.03 | -0.16 | -0.15 | 0.02 | 0.00 | 0.04 | 0.19 | 0.19 |
| mu. Cas | S | 5434 | 4.57 | 0.53 | -0.67 | -0.51 | -0.49 | -0.52 | 0.02 | -0.45 | -0.52 | -0.37 | -0.51 | -0.66 | -0.89 | -0.75 | -0.50 | -0.69 | -0.79 | -0.57 | -0.15 | -0.51 | -0.06 | -0.77 | | 0.19 | -0.22 | 0.09 | |
| mu. Her | S | 5560 | 3.99 | 1.36 | 0.43 | 0.39 | 0.46 | 0.30 | 0.41 | 0.29 | 0.30 | 0.29 | 0.28 | 0.29 | 0.35 | 0.23 | 0.28 | 0.32 | 0.05 | 0.31 | 0.48 | 0.28 | 0.16 | -0.02 | 0.52 | 0.42 | 0.38 | 0.79 | 0.56 |
| mu.01 Cyg | S | 6354 | 3.93 | 1.96 | -0.06 | -0.25 | 0.05 | -0.05 | -0.02 | 0.01 | -0.01 | -0.08 | 0.03 | -0.07 | -0.28 | -0.16 | 0.02 | -0.17 | -0.54 | -0.18 | | 0.06 | 0.34 | 0.11 | 0.25 | 0.15 | 0.04 | 0.20 | 0.11 |
| mu.02 Cnc | S | 5857 | 3.97 | 1.55 | 0.38 | 0.24 | 0.29 | 0.22 | 0.28 | 0.24 | 0.30 | 0.21 | 0.18 | 0.24 | 0.21 | 0.14 | 0.24 | 0.21 | 0.24 | 0.20 | 0.52 | 0.19 | 0.35 | 0.15 | 0.19 | 0.34 | 0.24 | 0.33 | 0.54 |
| mu.02 Cyg | S | 5998 | 4.33 | 1.27 | -0.20 | -0.21 | -0.19 | -0.14 | -0.04 | -0.10 | -0.16 | -0.19 | -0.21 | -0.20 | -0.31 | -0.24 | -0.19 | -0.26 | -0.51 | -0.29 | 0.45 | -0.13 | -0.06 | -0.07 | 0.44 | 0.23 | -0.01 | -0.16 | |
| ome Leo | S | 5883 | 3.79 | 1.60 | 0.17 | 0.08 | 0.13 | 0.10 | 0.16 | 0.09 | 0.09 | 0.02 | 0.02 | 0.06 | 0.00 | 0.00 | 0.07 | 0.01 | -0.04 | -0.10 | 0.53 | 0.03 | 0.13 | 0.00 | 0.18 | 0.14 | 0.17 | 0.08 | 0.15 |
| ome Sgr | S | 5455 | 3.61 | 1.42 | 0.00 | 0.00 | 0.06 | 0.04 | 0.12 | 0.07 | -0.01 | -0.02 | -0.02 | 0.01 | -0.01 | -0.05 | -0.05 | -0.06 | -0.17 | -0.13 | 0.52 | 0.07 | 0.05 | 0.02 | 0.09 | 0.02 | 0.10 | -0.03 | 0.30 |
| omi Aql | S | 6173 | 4.20 | 1.53 | 0.27 | 0.24 | 0.21 | 0.19 | 0.14 | 0.25 | 0.37 | 0.23 | 0.22 | 0.22 | 0.21 | 0.16 | 0.24 | 0.19 | 0.06 | -0.05 | 0.83 | 0.42 | 0.28 | 0.17 | 0.07 | 0.37 | 0.35 | 0.26 | 0.37 |
| omi02 Eri | S | 5202 | 4.59 | 0.33 | -0.11 | -0.02 | 0.07 | -0.06 | 0.33 | -0.07 | 0.00 | 0.10 | 0.13 | -0.07 | -0.22 | -0.25 | -0.03 | -0.14 | 0.11 | 0.08 | 0.22 | -0.08 | -0.26 | -0.33 | 0.72 | 0.59 | 0.69 | 1.30 | 0.17 |
| phi02 Cet | S | 6218 | 4.39 | 1.32 | -0.13 | -0.15 | -0.10 | -0.06 | 0.00 | -0.02 | -0.01 | -0.02 | -0.04 | -0.03 | -0.12 | -0.09 | 0.07 | -0.12 | -0.33 | -0.10 | 0.62 | 0.17 | 0.19 | 0.18 | | 0.39 | 0.17 | 0.06 | 0.60 |
| pi.03 Ori | S | 6509 | 4.31 | 2.22 | 0.43 | 0.17 | 0.19 | 0.19 | 0.23 | 0.30 | 0.12 | 0.25 | 0.37 | 0.20 | 0.18 | 0.17 | 0.54 | 0.18 | -0.14 | 0.45 | | 0.57 | 0.66 | 0.25 | 0.80 | 0.54 | 0.62 | 0.78 | 0.44 |
| psi Cap | S | 6633 | 4.24 | 3.36 | 0.45 | | 0.05 | 0.31 | 0.32 | 0.18 | 0.21 | 0.63 | 0.74 | 0.39 | 0.42 | 0.22 | 0.75 | 0.37 | -0.11 | | | 0.55 | 0.67 | -0.17 | 0.78 | 1.02 | 1.64 | | 0.52 |
| psi Cnc | S | 5305 | 3.53 | 1.51 | 0.04 | -0.03 | 0.14 | 0.02 | 0.19 | 0.06 | -0.06 | 0.00 | 0.03 | 0.00 | -0.01 | -0.10 | -0.01 | -0.10 | 0.00 | -0.02 | 0.18 | -0.11 | -0.15 | -0.19 | 0.04 | 0.07 | 0.08 | 0.25 | 0.18 |
| psi Ser | S | 5635 | 4.45 | 1.07 | 0.06 | 0.09 | 0.15 | 0.07 | 0.15 | 0.13 | 0.10 | 0.06 | 0.06 | 0.08 | 0.07 | 0.01 | 0.04 | 0.02 | -0.02 | -0.12 | 0.37 | 0.19 | 0.13 | 0.06 | 0.28 | 0.31 | 0.26 | 0.58 | 0.49 |

| Name | | Teff | logg | | | | | | | | | | | | | | | | | | | | | | | | | |
|---|---|---|---|---|---|---|---|---|---|---|---|---|---|---|---|---|---|---|---|---|---|---|---|---|---|---|---|---|
| psi01 Dra A | S | 6573 | 3.97 | 2.55 | 0.16 | 0.24 | 0.05 | 0.06 | 0.12 | 0.20 | 0.14 | 0.22 | 0.33 | 0.10 | 0.00 | -0.01 | 0.33 | 0.05 | -0.32 | 0.01 | | 0.12 | 0.86 | 0.06 | 0.30 | 0.27 | 0.14 | 0.78 | 0.30 |
| psi01 Dra B | S | 6237 | 4.31 | 1.60 | 0.14 | 0.10 | 0.11 | 0.08 | 0.14 | 0.17 | 0.17 | 0.14 | 0.10 | 0.11 | 0.03 | 0.04 | 0.12 | 0.06 | -0.31 | 0.06 | 0.70 | 0.17 | 0.26 | 0.11 | 0.16 | 0.43 | 0.17 | -0.11 | 0.21 |
| psi05 Aur | S | 6128 | 4.34 | 1.47 | 0.20 | 0.11 | 0.17 | 0.15 | 0.28 | 0.21 | 0.26 | 0.20 | 0.14 | 0.17 | 0.17 | 0.12 | 0.20 | 0.14 | 0.01 | 0.08 | 0.84 | 0.29 | 0.17 | 0.24 | 0.23 | 0.35 | 0.25 | 0.56 | 0.29 |
| rho Cap | S | 6856 | 3.95 | 4.61 | 0.40 | -0.50 | -0.04 | 0.03 | 0.50 | 0.30 | -0.37 | 0.45 | 1.00 | 0.27 | 0.09 | -0.15 | 0.63 | 0.04 | -0.70 | | | 0.17 | 1.69 | | 1.59 | 0.17 | 0.70 | | 0.50 |
| rho CrB | S | 5850 | 4.17 | 1.11 | -0.16 | -0.01 | -0.02 | -0.10 | -0.08 | -0.03 | -0.10 | -0.04 | -0.11 | -0.16 | -0.22 | -0.20 | -0.12 | -0.20 | -0.24 | -0.20 | | -0.18 | -0.05 | -0.21 | 0.31 | 0.08 | 0.09 | 0.36 | 0.28 |
| rho01 Cnc | S | 5248 | 4.43 | 1.09 | 0.66 | 0.47 | 0.56 | 0.48 | 0.77 | 0.37 | 0.46 | 0.43 | 0.55 | 0.41 | 0.62 | 0.34 | 0.49 | 0.49 | 0.67 | 0.36 | 0.28 | 0.46 | 0.13 | 0.09 | 0.61 | 0.70 | 0.70 | 2.16 | 0.73 |
| sig Boo | S | 6781 | 4.29 | 2.13 | -0.25 | -0.20 | -0.26 | -0.23 | -0.25 | -0.23 | -0.21 | -0.21 | 0.18 | -0.34 | -0.42 | -0.36 | 0.34 | -0.39 | -0.60 | -0.24 | | -0.18 | -0.18 | -0.10 | | 0.35 | -0.08 | | 0.25 |
| sig CrB A | S | 5923 | 4.12 | 1.52 | 0.29 | -0.44 | 0.42 | 0.12 | 0.43 | 0.31 | 0.38 | 0.60 | 0.48 | 0.69 | 0.55 | -0.01 | 0.83 | 0.13 | 1.10 | | 1.43 | 0.85 | 0.85 | 0.24 | | 0.45 | 1.53 | 2.69 | 0.70 |
| sig CrB B | S | 5992 | 4.47 | 1.26 | -0.05 | 0.01 | 0.04 | 0.02 | -0.05 | 0.10 | 0.02 | 0.06 | 0.03 | 0.08 | 0.02 | -0.01 | 0.04 | -0.03 | -0.29 | -0.14 | 0.50 | 0.20 | 0.19 | 0.09 | 0.35 | 0.47 | 0.31 | 0.57 | 0.52 |
| sig Dra | S | 5338 | 4.57 | 0.96 | -0.16 | -0.09 | -0.07 | -0.16 | 0.36 | -0.10 | -0.16 | -0.06 | -0.04 | -0.11 | -0.14 | -0.21 | -0.08 | -0.22 | -0.12 | -0.10 | 0.23 | -0.06 | -0.24 | -0.27 | 0.35 | 0.27 | 0.32 | 0.74 | 0.11 |
| tau Boo | S | 6447 | 4.26 | 2.38 | 0.54 | 0.45 | 0.48 | 0.39 | 0.43 | 0.40 | 0.52 | 0.50 | 0.46 | 0.41 | 0.42 | 0.32 | 0.43 | 0.39 | 0.02 | 0.17 | | 0.53 | 1.06 | 0.32 | 0.91 | 0.72 | 0.56 | 0.91 | 0.48 |
| tau Cet | S | 5403 | 4.53 | 0.41 | -0.44 | -0.13 | -0.16 | -0.27 | 0.09 | -0.26 | -0.26 | -0.17 | -0.24 | -0.37 | -0.54 | -0.45 | -0.25 | -0.41 | -0.36 | -0.33 | -0.11 | -0.23 | -0.42 | -0.45 | 0.09 | 0.30 | 0.10 | 0.47 | 0.03 |
| tau01 Eri | S | 6395 | 4.29 | 2.62 | 0.50 | -0.16 | 0.35 | 0.31 | 0.47 | 0.30 | 0.28 | 0.38 | 0.57 | 0.21 | 0.16 | 0.17 | 0.56 | 0.30 | -0.35 | | | 0.48 | 0.24 | -0.12 | 0.52 | | 0.60 | | 1.71 |
| tet Boo | S | 6294 | 4.07 | 3.04 | 0.33 | 0.18 | | 0.17 | 0.28 | 0.06 | 0.11 | 0.29 | 0.24 | 0.05 | 0.02 | 0.03 | 0.30 | 0.12 | -0.49 | | | 0.17 | 0.51 | -0.26 | | | 0.49 | 0.83 | | |
| tet Per | S | 6310 | 4.32 | 1.64 | 0.13 | 0.21 | 0.07 | 0.09 | 0.19 | 0.18 | 0.17 | 0.16 | 0.20 | 0.21 | 0.01 | 0.08 | 0.23 | 0.06 | -0.13 | -0.03 | | 0.20 | 0.33 | 0.17 | 0.31 | 0.37 | 0.36 | 0.07 | 0.25 |
| tet UMa | S | 6371 | 3.80 | 2.02 | -0.09 | -0.12 | -0.09 | -0.06 | -0.08 | 0.00 | 0.00 | 0.01 | 0.00 | -0.03 | -0.19 | -0.13 | 0.10 | -0.13 | -0.36 | -0.37 | | 0.05 | 0.09 | 0.21 | 0.11 | 0.19 | 0.13 | 0.10 | 0.07 |
| ups And | S | 6269 | 4.12 | 1.87 | 0.48 | 0.27 | 0.28 | 0.23 | 0.27 | 0.31 | 0.34 | 0.33 | 0.32 | 0.28 | 0.22 | 0.19 | 0.30 | 0.24 | 0.06 | 0.13 | | 0.28 | 0.40 | 0.20 | 0.18 | 0.52 | 0.35 | 0.25 | 0.44 |
| w Her | S | 5780 | 4.33 | 1.09 | -0.24 | -0.04 | 0.02 | -0.14 | -0.06 | -0.11 | -0.16 | -0.06 | -0.21 | -0.30 | -0.49 | -0.37 | -0.16 | -0.30 | -0.41 | -0.16 | 0.09 | -0.28 | -0.10 | -0.43 | 0.11 | 0.11 | -0.05 | -0.04 | 0.26 |
| zet Her | S | 5759 | 3.69 | 1.69 | 0.22 | 0.16 | 0.25 | 0.13 | 0.19 | 0.14 | 0.10 | 0.06 | 0.04 | 0.07 | 0.06 | 0.02 | 0.07 | 0.07 | 0.02 | -0.06 | 0.50 | 0.05 | 0.06 | -0.01 | 0.02 | 0.17 | 0.11 | 0.12 | -0.07 |
| BD+04 701A | S | 6056 | 4.50 | 1.45 | 0.21 | 0.19 | 0.15 | 0.24 | 1.01 | 0.18 | 0.32 | 0.28 | 0.57 | 0.30 | -0.10 | 0.04 | 0.41 | 0.12 | -0.68 | | | 0.67 | 0.66 | 0.62 | 0.81 | | 0.73 | 0.12 | 1.52 |
| BD+04 701B | S | 5642 | 4.50 | 0.58 | -0.34 | -0.28 | -0.07 | 0.01 | 0.71 | -0.05 | 0.22 | -0.23 | -0.19 | -0.14 | -0.53 | -0.37 | -0.32 | -0.27 | -0.61 | 0.01 | | 0.23 | 0.14 | 0.36 | | 0.47 | 0.14 | 0.12 | |
| BD+18 2776 | S | 3644 | 4.46 | 0.67 | -0.13 | 0.61 | 0.20 | 1.70 | 3.47 | 1.05 | 0.48 | -0.01 | 0.01 | 0.51 | -0.29 | 0.59 | 0.72 | 0.71 | 0.84 | 2.60 | 0.23 | 0.14 | -1.07 | -0.68 | 1.35 | 2.13 | 1.81 | 2.04 | -0.77 |
| BD+27 4120 | S | 3899 | 4.80 | 0.35 | 0.29 | 0.72 | 0.69 | 1.90 | 3.41 | 1.52 | 0.96 | 0.30 | 0.66 | 0.82 | 0.40 | 0.19 | 1.21 | 0.95 | 0.56 | 2.81 | 1.29 | 0.95 | -0.41 | 0.12 | 2.40 | 2.52 | 2.43 | 2.45 | 1.93 |
| BD+29 2963 | S | 5569 | 4.15 | 1.07 | -0.30 | -0.14 | -0.10 | -0.19 | 0.06 | -0.21 | -0.17 | -0.21 | -0.29 | -0.25 | -0.50 | -0.37 | -0.28 | -0.32 | -0.34 | -0.19 | 0.04 | -0.25 | -0.21 | -0.39 | 0.28 | -0.04 | 0.18 | 0.43 | 0.25 |
| BD+30 2512 | S | 4313 | 4.68 | 0.15 | 0.46 | 0.29 | 0.22 | 0.63 | 1.66 | 0.62 | 0.43 | 0.28 | 0.47 | 0.43 | 0.30 | 0.21 | 0.43 | 0.38 | 0.43 | | 0.59 | 0.39 | 0.09 | 0.16 | 1.01 | 1.58 | 1.91 | 2.35 | 1.02 |
| BD+33 529 | S | 3896 | 4.51 | 0.15 | -0.67 | -0.41 | -0.27 | 1.23 | 2.60 | 0.06 | 0.04 | -0.33 | -0.25 | 0.17 | -0.50 | -0.63 | 0.31 | 0.09 | -0.21 | 1.71 | 0.67 | -0.14 | -0.73 | -0.47 | 1.00 | 1.55 | 1.85 | 2.19 | -2.21 |
| BD+61 195 | S | 3799 | 4.84 | 0.15 | 0.36 | 0.52 | 0.79 | 2.41 | 3.83 | 1.82 | 1.07 | 0.41 | 0.74 | 1.05 | 0.63 | 0.49 | 1.34 | 1.10 | 0.06 | 3.50 | 0.80 | 1.30 | -0.50 | 0.01 | 2.02 | 2.12 | 2.53 | | 2.48 |
| BD+67 1468A | S | 6680 | 4.13 | 2.89 | 0.69 | 0.12 | 0.42 | 0.44 | 0.48 | 0.43 | 0.51 | 0.51 | 0.61 | 0.44 | 0.48 | 0.33 | 0.73 | 0.41 | 0.00 | -0.16 | | 0.49 | 0.60 | 0.04 | 0.81 | | 0.70 | 1.09 | 0.57 |
| BD+67 1468B | S | 6642 | 4.15 | 2.78 | 0.60 | 0.82 | 0.44 | 0.43 | 0.46 | 0.43 | 0.43 | 0.40 | 0.50 | 0.33 | 0.43 | 0.29 | 0.56 | 0.40 | 0.05 | 0.34 | | 0.42 | 1.00 | 0.09 | | 0.40 | 0.34 | | 0.51 |
| BD-04 782 | S | 4342 | 4.68 | 0.15 | 0.52 | 0.47 | 0.22 | 0.64 | 1.46 | 0.41 | 0.37 | 0.23 | 0.41 | 0.44 | 0.39 | 0.25 | 0.55 | 0.43 | 0.63 | 1.14 | 0.95 | 0.50 | -0.36 | 0.16 | 1.05 | 1.60 | 1.80 | 2.33 | 0.60 |
| BD-10 3166 | S | 5314 | 4.66 | 0.89 | 0.61 | 0.43 | 0.53 | 0.52 | 0.92 | 0.39 | 0.69 | 0.53 | 0.61 | 0.52 | 0.65 | 0.45 | 0.58 | 0.61 | 0.77 | 0.59 | 0.36 | 0.58 | 0.27 | 0.40 | 0.95 | 0.98 | 0.90 | 1.91 | 0.70 |
| CCDM J14534+1542AB | S | 6135 | 4.11 | 1.82 | 0.65 | 0.55 | 0.40 | 0.33 | 0.48 | 0.33 | 0.39 | 0.32 | 0.31 | 0.31 | 0.34 | 0.24 | 0.33 | 0.31 | 0.21 | 0.40 | | 0.25 | 0.48 | 0.08 | 0.32 | 0.24 | 0.30 | 0.27 | 0.45 |
| GJ 282 C | S | 3866 | 4.76 | 0.95 | -1.68 | 0.54 | 0.63 | 1.95 | 3.48 | 0.46 | 0.71 | 0.24 | 0.15 | 0.78 | -0.01 | 0.38 | 1.24 | 1.14 | | | 1.30 | 0.37 | 0.21 | -1.08 | 1.22 | 2.88 | 2.08 | | 0.33 |
| GJ 528 A | S | 4471 | 4.42 | 0.53 | 0.24 | 0.20 | 0.09 | 0.30 | 1.10 | 0.13 | 0.05 | -0.02 | 0.14 | 0.13 | 0.14 | 0.07 | 0.13 | 0.18 | 0.19 | 0.24 | 0.09 | -0.06 | -0.47 | -0.06 | 0.43 | 1.77 | 1.37 | 2.16 | 0.25 |
| GJ 528 B | S | 4384 | 4.59 | 0.15 | 0.33 | 0.49 | 0.17 | 0.47 | 1.44 | 0.49 | 0.27 | 0.22 | 0.42 | 0.36 | 0.25 | 0.19 | 0.37 | 0.31 | 0.46 | | 0.33 | 0.28 | -0.10 | 0.16 | 0.84 | 1.55 | 2.01 | 2.29 | 0.33 |
| HD 101177 | S | 5964 | 4.43 | 1.10 | -0.15 | -0.08 | -0.08 | -0.14 | -0.07 | -0.04 | -0.08 | -0.09 | -0.17 | -0.11 | -0.17 | -0.17 | -0.09 | -0.18 | -0.29 | -0.26 | 0.26 | 0.02 | 0.19 | -0.07 | 0.06 | 0.06 | 0.17 | -0.04 | 0.34 |
| HD 101563 | S | 5868 | 3.89 | 1.57 | 0.15 | 0.14 | 0.17 | 0.05 | 0.02 | 0.06 | 0.08 | 0.03 | -0.01 | 0.04 | 0.00 | -0.04 | 0.07 | 0.00 | 0.07 | 0.02 | 0.84 | 0.42 | 0.43 | 0.47 | 0.28 | 0.47 | 0.34 | 0.16 | 0.01 |

| Star | | Teff | logg | vt | [Fe/H] | [Na/Fe] | [Mg/Fe] | [Al/Fe] | [Si/Fe] | [Ca/Fe] | [Sc/Fe] | [Ti/Fe] | [V/Fe] | [Cr/Fe] | [Mn/Fe] | [Co/Fe] | [Ni/Fe] | [Cu/Fe] | [Zn/Fe] | [Y/Fe] | [Zr/Fe] | [Ba/Fe] | [La/Fe] | [Ce/Fe] | [Nd/Fe] | [Sm/Fe] | [Eu/Fe] |
|---|---|---|---|---|---|---|---|---|---|---|---|---|---|---|---|---|---|---|---|---|---|---|---|---|---|---|---|
| HD 101959 | S | 6095 | 4.42 | 1.37 | -0.11 | -0.12 | -0.08 | -0.07 | 0.01 | -0.03 | -0.01 | -0.01 | -0.06 | -0.01 | -0.16 | -0.09 | -0.06 | -0.11 | -0.32 | -0.08 | 0.44 | 0.13 | 0.09 | 0.06 | 0.22 | 0.20 | 0.26 | 0.04 | 0.33 |
| HD 102158 | S | 5781 | 4.31 | 0.86 | -0.33 | -0.09 | -0.08 | -0.20 | -0.21 | -0.16 | -0.20 | -0.12 | -0.26 | -0.38 | -0.62 | -0.42 | -0.26 | -0.36 | -0.49 | -0.28 | 0.08 | -0.34 | -0.20 | -0.42 | -0.10 | | -0.04 | -0.20 | 0.25 |
| HD 10307 | S | 5976 | 4.36 | 1.20 | 0.19 | 0.08 | 0.20 | 0.08 | 0.07 | 0.13 | 0.19 | 0.14 | 0.14 | 0.13 | 0.13 | 0.08 | 0.18 | 0.13 | -0.13 | -0.08 | 0.52 | 0.19 | 0.07 | 0.04 | 0.22 | 0.35 | 0.34 | 0.25 | 0.45 |
| HD 103095 | S | 5178 | 4.72 | 1.30 | -1.48 | -1.13 | -0.85 | -1.02 | | -0.96 | -1.11 | -0.87 | -1.03 | -1.07 | -1.45 | -1.24 | -0.77 | -1.22 | -1.49 | | -0.98 | -0.14 | -1.25 | | 0.61 | -0.05 | -0.01 | | |
| HD 103932 | S | 4585 | 4.58 | 0.60 | 0.62 | 0.37 | 0.39 | 0.55 | 1.21 | 0.47 | 0.40 | 0.32 | 0.67 | 0.42 | 0.37 | 0.35 | 0.42 | 0.45 | | 0.57 | 0.44 | 0.28 | -0.23 | 0.08 | 1.00 | 1.34 | 1.81 | 1.98 | 1.17 |
| HD 104304 | S | 5555 | 4.42 | 0.97 | 0.59 | 0.38 | 0.41 | 0.35 | 0.55 | 0.37 | 0.35 | 0.37 | 0.40 | 0.37 | 0.52 | 0.30 | 0.42 | 0.41 | 0.30 | 0.48 | 0.46 | 0.45 | 0.19 | 0.19 | 0.47 | 0.60 | 0.64 | 1.40 | 0.76 |
| HD 10436 | S | 4393 | 4.72 | 0.15 | 0.22 | 0.06 | -0.08 | 0.37 | 1.25 | 0.09 | 0.11 | 0.01 | 0.13 | 0.16 | -0.01 | -0.14 | 0.29 | 0.06 | 0.25 | | 0.76 | 0.12 | -0.57 | -0.09 | 1.09 | 1.57 | 1.69 | 2.12 | 0.32 |
| HD 106252 | S | 5890 | 4.37 | 1.17 | -0.05 | 0.06 | 0.04 | -0.04 | 0.01 | -0.02 | 0.03 | -0.04 | -0.09 | -0.05 | -0.12 | -0.09 | -0.04 | -0.07 | -0.27 | -0.24 | 0.37 | -0.06 | 0.13 | -0.06 | 0.19 | 0.30 | 0.25 | 0.38 | 0.33 |
| HD 106640 | S | 6003 | 4.30 | 1.34 | -0.15 | -0.14 | -0.10 | -0.14 | -0.20 | -0.13 | -0.10 | -0.13 | -0.16 | -0.20 | -0.26 | -0.22 | -0.07 | -0.23 | -0.30 | -0.23 | 0.26 | -0.09 | -0.17 | -0.13 | 0.14 | 0.03 | -0.04 | -0.23 | 0.14 |
| HD 108799 | S | 5878 | 4.33 | 1.46 | -0.09 | -0.07 | -0.04 | -0.05 | 0.14 | 0.00 | -0.07 | -0.12 | -0.15 | -0.08 | -0.21 | -0.15 | -0.15 | -0.17 | -0.60 | -0.23 | 0.34 | 0.05 | -0.13 | 0.15 | | 0.30 | 0.10 | 0.30 | 0.31 |
| HD 108874 | S | 5555 | 4.32 | 1.26 | 0.31 | 0.31 | 0.29 | 0.23 | 0.32 | 0.27 | 0.22 | 0.26 | 0.26 | 0.26 | 0.31 | 0.18 | 0.23 | 0.25 | 0.39 | 0.36 | 0.51 | 0.29 | 0.14 | 0.03 | 0.34 | 0.50 | 0.60 | 1.06 | 0.55 |
| HD 108954 | S | 6024 | 4.43 | 1.24 | -0.04 | -0.16 | -0.06 | -0.02 | 0.01 | 0.03 | 0.06 | -0.02 | -0.05 | -0.03 | -0.13 | -0.08 | -0.03 | -0.11 | -0.34 | -0.08 | 0.42 | 0.08 | 0.09 | 0.14 | 0.11 | 0.50 | 0.21 | 0.13 | 0.13 |
| HD 109057 | S | 6060 | 4.32 | 2.05 | 0.14 | 0.13 | 0.21 | 0.17 | 0.15 | 0.10 | 0.21 | 0.24 | 0.08 | 0.18 | 0.03 | 0.04 | 0.21 | 0.03 | -0.39 | 0.12 | 0.59 | 0.20 | 0.69 | 0.28 | | 0.39 | 0.37 | 0.52 | 0.21 |
| HD 11007 | S | 5994 | 4.01 | 1.47 | -0.13 | -0.16 | -0.10 | -0.12 | -0.10 | -0.10 | -0.09 | -0.11 | -0.12 | -0.18 | -0.26 | -0.20 | -0.12 | -0.21 | -0.30 | -0.09 | | -0.21 | -0.22 | -0.17 | -0.05 | -0.04 | -0.13 | -0.10 | 0.00 |
| HD 110315 | S | 4505 | 4.61 | 0.15 | 0.38 | 0.22 | 0.15 | 0.28 | 1.19 | 0.32 | 0.27 | 0.16 | 0.33 | 0.13 | -0.01 | -0.08 | 0.25 | 0.10 | 0.39 | 0.53 | 0.34 | 0.12 | -0.68 | -0.31 | 0.65 | 1.29 | 1.46 | 2.07 | 0.30 |
| HD 11038 | S | 6044 | 4.36 | 1.12 | -0.32 | -0.24 | -0.32 | -0.21 | -0.13 | -0.15 | -0.08 | -0.12 | -0.11 | -0.11 | -0.45 | -0.27 | -0.12 | -0.30 | | -0.35 | | -0.20 | 0.33 | -0.05 | | | 0.19 | 0.29 |
| HD 110745 | S | 6127 | 4.22 | 1.59 | 0.17 | 0.03 | 0.08 | 0.08 | 0.08 | 0.14 | 0.07 | 0.08 | 0.14 | 0.10 | 0.06 | 0.04 | 0.10 | 0.06 | -0.10 | 0.06 | 0.54 | 0.14 | 0.31 | -0.01 | 0.31 | 0.19 | 0.11 | 0.03 | 0.34 |
| HD 110833 | S | 4972 | 4.56 | 1.36 | 0.20 | 0.20 | 0.15 | 0.25 | 0.69 | 0.14 | 0.22 | 0.14 | 0.21 | 0.21 | 0.25 | 0.16 | 0.24 | 0.21 | 0.11 | 0.33 | 0.20 | 0.14 | -0.05 | -0.03 | 0.87 | 0.73 | 0.76 | 1.92 | 0.24 |
| HD 110869 | S | 5783 | 4.41 | 1.15 | 0.15 | 0.22 | 0.19 | 0.14 | 0.25 | 0.20 | 0.17 | 0.17 | 0.15 | 0.18 | 0.13 | 0.11 | 0.14 | 0.14 | 0.14 | 0.06 | 0.47 | 0.26 | 0.04 | 0.15 | 0.21 | 0.42 | 0.38 | 0.64 | 0.54 |
| HD 111513 | S | 5822 | 4.33 | 1.14 | 0.23 | 0.12 | 0.22 | 0.16 | 0.23 | 0.17 | 0.21 | 0.15 | 0.15 | 0.15 | 0.17 | 0.11 | 0.16 | 0.16 | 0.13 | 0.16 | 0.40 | 0.18 | 0.15 | 0.00 | 0.20 | 0.20 | 0.20 | 0.49 | 0.47 |
| HD 111540 | S | 5729 | 4.37 | 1.25 | 0.24 | 0.17 | 0.25 | 0.25 | 0.46 | 0.18 | 0.29 | 0.12 | 0.10 | 0.19 | 0.17 | 0.12 | 0.13 | 0.18 | 0.09 | 0.31 | 0.43 | 0.29 | 0.18 | 0.09 | 0.42 | 0.43 | 0.44 | 0.60 | 0.52 |
| HD 111799 | S | 5842 | 4.38 | 1.33 | 0.38 | 0.22 | 0.26 | 0.22 | 0.42 | 0.20 | 0.31 | 0.18 | 0.18 | 0.21 | 0.20 | 0.13 | 0.21 | 0.21 | 0.18 | 0.27 | 0.52 | 0.26 | 0.25 | 0.09 | 0.31 | 0.42 | 0.36 | 0.39 | 0.50 |
| HD 112068 | S | 5745 | 4.26 | 1.09 | -0.42 | -0.11 | -0.22 | -0.21 | -0.21 | -0.25 | -0.28 | -0.24 | -0.35 | -0.45 | -0.69 | -0.51 | -0.32 | -0.47 | -0.57 | -0.26 | | -0.36 | | -0.48 | | | -0.18 | -0.07 | -0.05 |
| HD 112257 | S | 5659 | 4.31 | 1.10 | 0.05 | 0.13 | 0.16 | 0.07 | 0.20 | 0.08 | 0.13 | 0.07 | 0.03 | 0.01 | -0.09 | -0.05 | 0.06 | -0.02 | -0.05 | 0.12 | 0.12 | 0.08 | 0.10 | -0.02 | 0.13 | 0.15 | 0.28 | 0.56 | 0.34 |
| HD 112758 | S | 5190 | 4.58 | 0.49 | -0.36 | -0.06 | -0.16 | -0.21 | 0.27 | -0.27 | -0.20 | -0.10 | -0.09 | -0.28 | -0.48 | -0.44 | -0.27 | -0.37 | -0.20 | -0.26 | -0.31 | -0.36 | -0.59 | -0.61 | -0.03 | 0.56 | 0.40 | 0.55 | -0.06 |
| HD 113470 | S | 5839 | 4.27 | 1.32 | -0.13 | -0.04 | -0.21 | -0.12 | 0.10 | -0.13 | -0.17 | -0.17 | -0.23 | -0.16 | -0.17 | -0.21 | -0.13 | -0.21 | -0.29 | -0.19 | | -0.16 | 0.04 | -0.18 | 0.47 | 0.02 | 0.04 | -0.15 | 0.26 |
| HD 113713 | S | 6316 | 3.99 | 1.86 | -0.29 | -0.29 | -0.36 | -0.24 | -0.25 | -0.21 | -0.18 | -0.26 | -0.19 | -0.30 | -0.41 | -0.38 | -0.07 | -0.38 | -0.54 | -0.26 | 0.47 | -0.35 | -0.29 | -0.29 | | 0.00 | -0.20 | 0.01 | 0.13 |
| HD 114762 | S | 5921 | 4.27 | 1.26 | -0.64 | -0.40 | -0.55 | -0.47 | -0.43 | -0.47 | -0.55 | -0.49 | -0.56 | -0.66 | -0.90 | -0.73 | -0.50 | -0.71 | -0.84 | -0.71 | | -0.70 | | -0.76 | | 0.08 | -0.45 | | |
| HD 114783 | S | 5118 | 4.55 | 1.01 | 0.32 | 0.26 | 0.26 | 0.22 | 0.58 | 0.19 | 0.26 | 0.25 | 0.33 | 0.19 | 0.33 | 0.11 | 0.27 | 0.23 | 0.44 | 0.48 | 0.22 | 0.27 | -0.05 | 0.01 | 0.91 | 0.57 | 0.75 | 1.88 | 0.41 |
| HD 11507 | S | 4011 | 4.72 | 0.65 | 0.64 | 0.73 | 0.57 | 1.51 | 3.10 | 1.43 | 0.81 | 0.44 | 0.81 | 0.74 | 0.45 | 0.41 | 0.98 | 0.81 | | | 1.25 | 0.74 | 0.06 | 0.14 | 1.43 | 2.29 | 2.40 | 2.48 | 1.86 |
| HD 115404A | S | 5030 | 4.60 | 0.96 | -0.05 | -0.06 | -0.03 | -0.10 | 0.50 | -0.07 | -0.09 | -0.01 | 0.02 | -0.04 | -0.07 | -0.17 | -0.11 | -0.15 | -0.05 | -0.22 | 0.02 | -0.17 | -0.22 | -0.05 | 0.36 | 0.54 | 0.83 | 0.90 | -0.07 |
| HD 115404B | S | 3903 | 4.90 | 0.35 | 0.00 | 0.71 | 0.88 | 1.83 | 3.08 | 1.56 | 1.21 | 0.35 | 0.81 | 0.88 | 0.43 | 0.11 | 1.26 | 1.04 | | 3.63 | 1.10 | 0.66 | -0.59 | 0.22 | 1.86 | 2.67 | 2.30 | | 1.36 |
| HD 115953 | S | 3751 | 4.36 | 1.15 | 0.24 | 0.41 | 0.58 | 1.96 | 3.19 | 0.79 | 0.72 | -0.03 | 0.07 | 0.77 | 0.15 | 0.22 | 1.06 | 0.85 | | 3.02 | 1.29 | 0.45 | -1.09 | -0.22 | 1.32 | 2.17 | 2.16 | | 2.08 |
| HD 117043 | S | 5558 | 4.36 | 1.20 | 0.38 | 0.21 | 0.30 | 0.23 | 0.46 | 0.23 | 0.23 | 0.22 | 0.26 | 0.25 | 0.30 | 0.17 | 0.24 | 0.23 | 0.63 | 0.39 | 0.30 | 0.28 | 0.17 | 0.05 | 0.51 | 0.47 | 0.46 | 1.02 | 0.55 |
| HD 117845 | S | 5856 | 4.46 | 1.07 | -0.29 | -0.21 | -0.29 | -0.18 | -0.01 | -0.13 | -0.21 | -0.23 | -0.21 | -0.19 | -0.45 | -0.29 | -0.26 | -0.35 | -0.50 | | 0.24 | 0.00 | 0.00 | 0.08 | 0.30 | 0.18 | 0.06 | 0.36 | 0.04 |
| HD 118203 | S | 5768 | 3.92 | 1.68 | 0.51 | 0.19 | 0.29 | 0.24 | 0.31 | 0.25 | 0.27 | 0.22 | 0.21 | 0.25 | 0.24 | 0.16 | 0.28 | 0.22 | 0.20 | 0.15 | 0.56 | 0.21 | 0.22 | 0.13 | 0.04 | 0.36 | 0.29 | 0.42 | 0.32 |
| HD 120066 | S | 5903 | 4.11 | 1.43 | 0.09 | 0.19 | 0.29 | 0.14 | 0.08 | 0.27 | 0.23 | 0.22 | 0.18 | 0.12 | -0.01 | 0.07 | 0.17 | 0.11 | -0.08 | -0.15 | 0.49 | 0.13 | 0.21 | 0.01 | 0.16 | 0.31 | 0.30 | 0.40 | 0.43 |
| HD 120467 | S | 4369 | 4.60 | 0.35 | 0.59 | 0.74 | 0.57 | 0.80 | 1.90 | 0.85 | 0.82 | 0.49 | 0.84 | 0.70 | 0.56 | 0.45 | 0.82 | 0.65 | 0.99 | 0.99 | 0.91 | 0.80 | -0.22 | 0.31 | 1.50 | 1.87 | 2.05 | 2.65 | |

| Star | | Teff | log g | | | | | | | | | | | | | | | | | | | | | | | | | | |
|---|---|---|---|---|---|---|---|---|---|---|---|---|---|---|---|---|---|---|---|---|---|---|---|---|---|---|---|---|---|
| HD 120690 | S | 5550 | 4.39 | 0.84 | 0.05 | 0.06 | 0.04 | 0.04 | 0.22 | 0.07 | 0.01 | 0.00 | -0.05 | 0.05 | -0.06 | -0.04 | 0.00 | 0.00 | -0.21 | -0.16 | 0.23 | 0.10 | -0.07 | 0.01 | 0.21 | 0.32 | 0.20 | 0.69 | 0.34 |
| HD 120730 | S | 5300 | 4.49 | 1.16 | 0.00 | 0.12 | 0.20 | 0.06 | 0.00 | 0.04 | 0.09 | 0.11 | 0.12 | 0.15 | 0.09 | -0.01 | 0.09 | 0.02 | -0.05 | 0.16 | 0.09 | 0.10 | 0.11 | -0.08 | 0.33 | 0.82 | 0.62 | 0.88 | 0.29 |
| HD 122064 | S | 4811 | 4.57 | 0.15 | 0.56 | 0.52 | 0.47 | 0.41 | 1.03 | 0.36 | 0.49 | 0.49 | 0.73 | 0.43 | 0.47 | 0.33 | 0.44 | 0.45 | 0.47 | 0.85 | 0.49 | 0.39 | -0.11 | 0.20 | 0.95 | 1.39 | 1.48 | 1.59 | 1.11 |
| HD 122303 | S | 4067 | 5.00 | 0.25 | 0.09 | 0.00 | 0.73 | 1.84 | 2.83 | 1.29 | 0.97 | 0.46 | 0.76 | 0.73 | 0.92 | 0.01 | 1.30 | 1.05 | | 2.98 | | 1.34 | -0.28 | 0.04 | 2.59 | 2.35 | 2.61 | 2.91 | 1.11 |
| HD 122742 | S | 5459 | 4.39 | 0.92 | 0.05 | 0.05 | 0.05 | 0.06 | 0.32 | 0.07 | 0.05 | 0.03 | 0.00 | 0.03 | 0.06 | -0.03 | 0.00 | 0.03 | 0.11 | 0.14 | 0.12 | 0.08 | -0.12 | -0.06 | 0.36 | 0.19 | 0.35 | 0.85 | 0.41 |
| HD 122967 | S | 7008 | 4.16 | 4.32 | | 1.44 | | 0.13 | 1.51 | 1.05 | 2.34 | 0.99 | 1.94 | 0.97 | | 0.63 | 1.21 | 0.71 | | | 0.40 | 2.20 | | 1.00 | 1.66 | 0.48 | | 1.63 |
| HD 123A | S | 5823 | 4.43 | 1.26 | 0.09 | 0.18 | 0.10 | 0.09 | 0.25 | 0.15 | 0.14 | 0.14 | 0.11 | 0.17 | 0.08 | 0.09 | 0.11 | 0.06 | -0.09 | -0.09 | 0.51 | 0.30 | 0.19 | 0.28 | 0.37 | 0.50 | 0.42 | 0.47 | 0.48 |
| HD 123B | S | 5230 | 4.47 | 0.23 | 0.03 | 0.02 | 0.09 | 0.17 | 0.58 | 0.14 | 0.10 | 0.14 | 0.08 | 0.17 | 0.20 | 0.09 | 0.06 | 0.13 | 0.20 | 0.04 | 0.32 | 0.20 | -0.03 | 0.21 | 0.33 | 0.50 | 0.50 | 1.07 | 0.49 |
| HD 124553 | S | 6060 | 4.00 | 1.65 | 0.40 | 0.26 | 0.24 | 0.23 | 0.19 | 0.28 | 0.32 | 0.22 | 0.21 | 0.23 | 0.21 | 0.17 | 0.26 | 0.22 | 0.06 | 0.12 | 0.67 | 0.28 | 0.25 | 0.13 | 0.18 | 0.45 | 0.29 | 0.24 | 0.53 |
| HD 125040 | S | 6357 | 4.28 | 3.24 | 0.29 | | 0.30 | 0.21 | 0.36 | 0.06 | 0.32 | 0.55 | 0.63 | 0.49 | 0.04 | 0.02 | 0.62 | 0.25 | | 0.07 | | 0.52 | | -0.07 | 0.58 | 0.94 | 0.29 | 1.17 | |
| HD 125184 | S | 5637 | 4.06 | 1.37 | 0.39 | 0.41 | 0.37 | 0.30 | 0.44 | 0.31 | 0.30 | 0.29 | 0.28 | 0.32 | 0.39 | 0.25 | 0.32 | 0.29 | 0.41 | 0.31 | | 0.40 | 0.25 | 0.13 | 0.47 | 0.53 | 0.42 | 0.81 | 0.58 |
| HD 126053 | S | 5727 | 4.46 | 0.94 | -0.34 | -0.28 | -0.12 | -0.24 | -0.26 | -0.22 | -0.22 | -0.20 | -0.28 | -0.29 | -0.46 | -0.33 | -0.22 | -0.33 | -0.31 | -0.28 | | -0.23 | -0.32 | -0.29 | | -0.37 | -0.01 | 0.17 | 0.19 |
| HD 12661 | S | 5651 | 4.35 | 1.06 | 0.64 | 0.45 | 0.45 | 0.40 | 0.65 | 0.42 | 0.47 | 0.42 | 0.45 | 0.41 | 0.57 | 0.35 | 0.46 | 0.47 | 0.74 | 0.40 | 0.65 | 0.45 | 0.34 | 0.18 | 0.68 | 0.45 | 0.52 | 0.58 | 0.64 |
| HD 127334 | S | 5643 | 4.25 | 1.32 | 0.36 | 0.30 | 0.32 | 0.25 | 0.47 | 0.28 | 0.23 | 0.19 | 0.21 | 0.22 | 0.24 | 0.17 | 0.22 | 0.23 | 0.48 | 0.12 | 0.38 | 0.22 | 0.15 | 0.07 | 0.54 | 0.52 | 0.48 | 0.78 | 0.53 |
| HD 128165 | S | 4793 | 4.59 | 0.76 | 0.28 | 0.09 | 0.16 | 0.11 | 1.07 | 0.14 | 0.10 | 0.13 | 0.27 | 0.12 | 0.21 | 0.03 | 0.12 | 0.13 | | -0.08 | 0.09 | 0.09 | -0.21 | -0.06 | 0.25 | 1.16 | 1.17 | 1.50 | |
| HD 128311 | S | 4903 | 4.58 | 0.77 | 0.21 | 0.14 | 0.17 | 0.23 | 0.73 | 0.27 | 0.17 | 0.24 | 0.35 | 0.29 | 0.31 | 0.12 | 0.17 | 0.21 | 0.45 | 1.03 | 0.16 | 0.28 | 0.00 | 0.20 | 0.56 | 0.66 | 0.99 | 1.63 | 0.21 |
| HD 129132 | S | 6609 | 3.41 | 2.80 | -0.11 | -0.06 | -0.17 | -0.10 | -0.16 | 0.01 | -0.14 | 0.09 | 0.51 | -0.12 | -0.15 | -0.17 | 0.30 | -0.14 | -0.28 | | -0.18 | 0.11 | -0.02 | 0.28 | | -0.08 | 1.34 | 0.37 | |
| HD 129290 | S | 5837 | 4.19 | 1.26 | -0.06 | 0.02 | 0.03 | -0.05 | -0.07 | -0.01 | -0.01 | -0.05 | -0.10 | -0.09 | -0.14 | -0.13 | -0.08 | -0.12 | -0.12 | -0.14 | 0.33 | -0.02 | -0.02 | -0.16 | | 0.25 | -0.03 | 0.07 | 0.18 |
| HD 129829 | S | 6085 | 4.41 | 1.37 | -0.19 | -0.08 | -0.36 | -0.15 | -0.14 | -0.14 | -0.01 | -0.16 | -0.25 | -0.18 | -0.34 | -0.23 | -0.10 | -0.26 | -0.41 | -0.26 | 0.29 | -0.06 | -0.12 | -0.08 | | 0.20 | -0.03 | 0.06 | 0.24 |
| HD 130322 | S | 5422 | 4.54 | 0.88 | 0.09 | 0.12 | 0.14 | 0.09 | 0.37 | 0.15 | 0.07 | 0.16 | 0.18 | 0.13 | 0.18 | 0.04 | 0.09 | 0.09 | 0.17 | -0.03 | 0.27 | 0.09 | -0.03 | 0.07 | 0.49 | 0.33 | 0.51 | 1.09 | 0.41 |
| HD 13043 | S | 5877 | 4.15 | 1.33 | 0.22 | 0.16 | 0.20 | 0.14 | 0.12 | 0.15 | 0.17 | 0.11 | 0.13 | 0.12 | 0.10 | 0.07 | 0.14 | 0.10 | 0.00 | 0.06 | 0.51 | 0.16 | 0.25 | -0.05 | 0.21 | 0.18 | 0.25 | 0.38 | 0.42 |
| HD 130948 | S | 5983 | 4.43 | 1.50 | -0.01 | -0.07 | 0.10 | 0.05 | 0.08 | 0.14 | 0.07 | 0.07 | 0.01 | 0.07 | -0.05 | 0.01 | 0.06 | -0.04 | -0.33 | -0.02 | 0.50 | 0.21 | 0.23 | 0.26 | 0.47 | 0.18 | 0.35 | 0.18 | 0.24 |
| HD 131976 | S | 4077 | 5.00 | 0.15 | 0.26 | -0.11 | | 1.64 | 3.07 | 1.42 | 0.85 | 0.30 | 0.67 | 0.74 | 0.30 | -0.05 | 1.08 | 1.03 | 0.66 | | | 1.01 | -0.40 | 0.02 | 2.36 | 2.80 | 2.35 | | 1.07 |
| HD 131977 | S | 4625 | 4.59 | 1.31 | 0.31 | 0.18 | 0.18 | 0.33 | 1.04 | 0.31 | 0.12 | 0.09 | 0.30 | 0.25 | 0.23 | 0.14 | 0.32 | 0.28 | 0.25 | | 0.49 | 0.13 | -0.35 | -0.10 | 0.84 | 1.32 | 1.22 | 1.92 | 0.85 |
| HD 132375 | S | 6336 | 4.18 | 1.90 | 0.25 | 0.16 | 0.11 | 0.15 | 0.19 | 0.18 | 0.21 | 0.20 | 0.17 | 0.16 | 0.10 | 0.09 | 0.18 | 0.12 | -0.14 | 0.01 | | 0.22 | 0.49 | 0.17 | 0.11 | 0.22 | 0.24 | -0.26 | 0.33 |
| HD 133161 | S | 5946 | 4.31 | 1.51 | 0.41 | 0.26 | 0.25 | 0.28 | 0.43 | 0.22 | 0.31 | 0.21 | 0.23 | 0.25 | 0.25 | 0.16 | 0.24 | 0.24 | 0.15 | 0.19 | 0.66 | 0.28 | 0.32 | 0.04 | 0.34 | 0.71 | 0.18 | 0.36 | 0.50 |
| HD 135101 | S | 5637 | 4.24 | 0.87 | 0.15 | 0.28 | 0.35 | 0.16 | 0.33 | 0.20 | 0.23 | 0.21 | 0.14 | 0.13 | 0.03 | 0.07 | 0.15 | 0.14 | 0.30 | 0.17 | 0.29 | 0.26 | 0.16 | 0.11 | 0.32 | 0.30 | 0.36 | 0.67 | 0.50 |
| HD 135101B | S | 5529 | 4.07 | 1.27 | 0.17 | 0.27 | 0.36 | 0.12 | 0.29 | 0.15 | 0.06 | 0.18 | 0.16 | 0.10 | 0.06 | 0.01 | 0.13 | 0.05 | 0.11 | 0.00 | 0.24 | 0.19 | 0.00 | -0.23 | 0.13 | 0.50 | 0.18 | 0.87 | 0.37 |
| HD 135145 | S | 5852 | 4.05 | 1.47 | -0.05 | 0.01 | -0.01 | -0.06 | -0.09 | -0.04 | -0.04 | -0.07 | -0.10 | -0.08 | -0.16 | -0.14 | -0.08 | -0.14 | -0.22 | -0.15 | 0.26 | 0.01 | 0.02 | -0.24 | 0.05 | -0.03 | 0.06 | -0.02 | 0.35 |
| HD 136064 | S | 6144 | 3.98 | 1.77 | 0.15 | 0.08 | 0.06 | 0.09 | 0.11 | 0.12 | 0.22 | 0.08 | 0.04 | 0.07 | 0.01 | 0.00 | 0.13 | 0.02 | -0.13 | -0.01 | 0.75 | 0.08 | 0.10 | 0.09 | 0.14 | 0.20 | 0.15 | -0.02 | 0.18 |
| HD 136118 | S | 6143 | 4.08 | 1.66 | 0.03 | 0.09 | 0.02 | 0.04 | 0.09 | 0.06 | 0.11 | 0.04 | 0.00 | 0.04 | -0.11 | -0.06 | 0.06 | -0.08 | -0.19 | -0.12 | | 0.09 | 0.08 | 0.10 | | 0.25 | 0.04 | 0.12 | 0.16 |
| HD 136231 | S | 5881 | 4.17 | 1.27 | -0.31 | -0.20 | -0.11 | -0.24 | -0.19 | -0.18 | -0.07 | -0.13 | -0.18 | -0.31 | -0.45 | -0.31 | -0.07 | -0.35 | | -0.17 | | -0.21 | | -0.26 | | | -0.23 | | -0.10 |
| HD 1388 | S | 5884 | 4.31 | 1.20 | 0.00 | -0.06 | -0.03 | 0.02 | 0.10 | 0.02 | 0.09 | 0.04 | -0.02 | 0.04 | -0.07 | -0.03 | 0.00 | -0.02 | -0.16 | -0.07 | 0.33 | 0.13 | 0.05 | 0.04 | 0.26 | 0.26 | 0.28 | 0.12 | 0.40 |
| HD 140913 | S | 5946 | 4.46 | 1.70 | 0.13 | 0.13 | 0.09 | 0.13 | 0.16 | 0.14 | 0.15 | 0.13 | 0.13 | 0.16 | 0.06 | 0.05 | 0.14 | 0.05 | -0.19 | -0.13 | | 0.21 | 0.23 | 0.18 | 0.50 | 0.34 | 0.33 | 0.59 | 0.39 |
| HD 141715 | S | 5836 | 4.25 | 2.04 | 0.02 | 0.07 | 0.05 | -0.03 | 0.09 | 0.02 | -0.02 | 0.01 | 0.02 | -0.01 | -0.11 | -0.10 | -0.04 | -0.14 | -0.39 | -0.11 | | 0.06 | 0.03 | -0.02 | | 0.11 | 0.28 | 0.52 | 0.36 |
| HD 141937 | S | 5885 | 4.44 | 1.26 | 0.19 | 0.04 | 0.18 | 0.15 | 0.22 | 0.16 | 0.15 | 0.12 | 0.14 | 0.16 | 0.10 | 0.08 | 0.15 | 0.11 | 0.06 | 0.13 | 0.40 | 0.17 | 0.17 | 0.10 | 0.37 | 0.20 | 0.38 | 0.45 | 0.54 |
| HD 14412 | S | 5492 | 4.62 | 0.48 | -0.34 | -0.26 | -0.28 | -0.29 | -0.11 | -0.21 | -0.24 | -0.17 | -0.19 | -0.29 | -0.36 | -0.36 | -0.26 | -0.32 | -0.83 | -0.48 | -0.04 | -0.09 | -0.49 | -0.17 | | 0.59 | 0.34 | 0.46 | 0.09 |
| HD 144579 | S | 5308 | 4.67 | 0.15 | -0.44 | -0.30 | -0.33 | -0.38 | 0.20 | -0.43 | -0.44 | -0.28 | -0.37 | -0.54 | -0.78 | -0.63 | -0.44 | -0.58 | -0.53 | -0.36 | | -0.53 | -0.34 | -0.65 | | 0.75 | 0.06 | 0.37 | -0.34 |

| Star | | Teff | logg | | | | | | | | | | | | | | | | | | | | | | | | |
|---|---|---|---|---|---|---|---|---|---|---|---|---|---|---|---|---|---|---|---|---|---|---|---|---|---|---|---|
| HD 144585 | S | 5850 | 4.22 | 1.41 | 0.48 | 0.43 | 0.42 | 0.33 | 0.38 | 0.35 | 0.37 | 0.33 | 0.37 | 0.34 | 0.36 | 0.28 | 0.36 | 0.34 | 0.15 | 0.27 | 0.70 | 0.32 | 0.35 | 0.17 | 0.52 | 0.30 | 0.38 | 0.53 | 0.60 |
| HD 145148 | S | 4923 | 3.68 | 1.30 | 0.29 | 0.32 | 0.35 | 0.29 | 0.68 | 0.21 | 0.21 | 0.28 | 0.38 | 0.26 | 0.32 | 0.15 | 0.26 | 0.23 | | 0.71 | 0.28 | 0.22 | -0.11 | 0.06 | 0.68 | 0.70 | 0.65 | 1.25 | 0.30 |
| HD 1461 | S | 5765 | 4.38 | 1.26 | 0.41 | 0.23 | 0.32 | 0.24 | 0.30 | 0.24 | 0.28 | 0.26 | 0.26 | 0.26 | 0.33 | 0.18 | 0.29 | 0.26 | 0.25 | 0.22 | 0.47 | 0.27 | 0.31 | -0.02 | 0.37 | 0.37 | 0.33 | 0.74 | 0.53 |
| HD 14624 | S | 5599 | 4.07 | 1.31 | 0.26 | 0.26 | 0.31 | 0.22 | 0.41 | 0.21 | 0.28 | 0.25 | 0.18 | 0.21 | 0.15 | 0.11 | 0.22 | 0.19 | 0.15 | 0.33 | 0.44 | 0.20 | 0.11 | 0.08 | 0.29 | 0.39 | 0.37 | 0.82 | 0.51 |
| HD 147681 | S | 6139 | 4.44 | 1.97 | 0.09 | 0.21 | 0.11 | 0.18 | 0.35 | 0.22 | 0.23 | 0.23 | 0.29 | 0.23 | 0.06 | 0.11 | 0.33 | 0.10 | -0.38 | 0.25 | | 0.37 | 0.64 | 0.52 | | 0.24 | 0.59 | 0.39 | 0.68 |
| HD 149026 | S | 6003 | 4.13 | 1.73 | 0.54 | 0.40 | 0.34 | 0.33 | 0.42 | 0.35 | 0.37 | 0.28 | 0.27 | 0.32 | 0.28 | 0.23 | 0.31 | 0.29 | 0.10 | 0.28 | | 0.37 | 0.43 | 0.08 | 0.30 | 0.24 | 0.39 | 0.58 | 0.34 |
| HD 149143 | S | 5768 | 4.07 | 1.50 | 0.48 | 0.35 | 0.34 | 0.32 | 0.40 | 0.29 | 0.35 | 0.26 | 0.23 | 0.27 | 0.28 | 0.21 | 0.25 | 0.28 | 0.30 | 0.36 | 0.56 | 0.29 | 0.36 | 0.17 | 0.39 | 0.40 | 0.41 | 0.49 | 0.53 |
| HD 15069 | S | 5722 | 4.11 | 1.22 | -0.04 | 0.04 | 0.15 | 0.06 | 0.11 | 0.16 | 0.11 | 0.07 | 0.04 | 0.08 | -0.01 | 0.01 | -0.02 | 0.01 | -0.02 | -0.04 | 0.37 | 0.12 | 0.11 | 0.08 | 0.21 | 0.38 | 0.25 | 0.28 | 0.42 |
| HD 150706 | S | 5900 | 4.46 | 1.34 | -0.09 | 0.01 | -0.02 | -0.05 | -0.02 | 0.01 | -0.03 | -0.02 | -0.08 | 0.02 | -0.07 | -0.08 | -0.03 | -0.11 | -0.26 | -0.18 | 0.41 | 0.10 | 0.12 | 0.18 | 0.46 | 0.34 | 0.24 | 0.31 | 0.21 |
| HD 150933 | S | 6087 | 4.28 | 1.37 | 0.29 | 0.17 | 0.21 | 0.17 | 0.24 | 0.19 | 0.27 | 0.16 | 0.16 | 0.19 | 0.14 | 0.12 | 0.19 | 0.18 | -0.29 | 0.17 | 0.66 | 0.23 | 0.33 | 0.15 | 0.43 | 0.83 | 0.14 | 0.31 | 0.55 |
| HD 151288 | S | 4181 | 4.70 | 0.15 | 0.61 | 0.83 | 0.45 | 0.84 | 2.31 | 0.95 | 0.76 | 0.45 | 0.68 | 0.67 | 0.57 | 0.38 | 0.80 | 0.67 | 0.86 | 1.56 | 0.92 | 0.73 | -0.08 | 0.16 | 1.32 | 1.94 | 2.36 | 2.89 | 0.83 |
| HD 151426 | S | 5711 | 4.28 | 1.07 | -0.06 | 0.11 | 0.06 | -0.01 | 0.14 | 0.04 | -0.01 | 0.02 | -0.01 | -0.02 | -0.15 | -0.10 | -0.01 | -0.09 | -0.03 | -0.06 | 0.22 | 0.03 | -0.04 | -0.08 | 0.16 | 0.19 | 0.26 | 0.38 | 0.31 |
| HD 151450 | S | 6108 | 4.38 | 1.31 | 0.01 | -0.04 | -0.04 | 0.00 | 0.03 | 0.08 | 0.03 | 0.03 | 0.06 | 0.03 | -0.08 | -0.02 | 0.05 | -0.02 | -0.10 | -0.03 | 0.62 | 0.11 | 0.30 | 0.09 | 0.39 | 0.28 | 0.02 | -0.15 | 0.49 |
| HD 15189 | S | 6051 | 4.47 | 0.93 | 0.03 | 0.07 | 0.03 | 0.04 | 0.18 | 0.08 | 0.21 | 0.09 | 0.06 | 0.17 | -0.07 | 0.02 | -0.03 | 0.02 | | 0.04 | | 0.25 | 0.62 | 0.25 | 0.24 | 0.58 | 0.41 | 1.20 | 0.43 |
| HD 152391 | S | 5431 | 4.51 | 1.32 | -0.06 | 0.05 | 0.00 | 0.03 | 0.28 | 0.03 | 0.01 | -0.01 | 0.01 | 0.05 | -0.03 | -0.05 | 0.01 | -0.04 | -0.08 | -0.02 | 0.33 | 0.14 | -0.13 | 0.11 | 0.39 | 0.21 | 0.37 | 0.76 | 0.27 |
| HD 154160 | S | 5380 | 3.90 | 1.36 | 0.59 | 0.43 | 0.42 | 0.44 | 0.76 | 0.31 | 0.33 | 0.26 | 0.24 | 0.34 | 0.45 | 0.25 | 0.32 | 0.35 | 0.26 | 0.42 | 0.47 | 0.39 | 0.09 | 0.17 | 0.55 | 0.45 | 0.46 | 0.98 | 0.56 |
| HD 154363 | S | 4373 | 4.66 | 0.15 | 0.17 | 0.12 | 0.00 | 0.20 | 1.37 | 0.22 | -0.07 | -0.04 | 0.04 | -0.03 | -0.32 | -0.32 | 0.10 | -0.04 | | 0.54 | 0.49 | 0.04 | -0.59 | -0.57 | 0.86 | 0.61 | 1.66 | 1.83 | |
| HD 154578 | S | 6294 | 4.14 | 1.91 | -0.13 | -0.06 | -0.18 | -0.13 | 0.02 | -0.06 | 0.06 | -0.06 | -0.08 | -0.10 | -0.33 | -0.21 | 0.13 | -0.20 | -0.37 | -0.09 | | 0.01 | 0.09 | -0.09 | -0.08 | 0.39 | 0.11 | -0.10 | 0.24 |
| HD 156062 | S | 5995 | 4.37 | 1.22 | 0.07 | 0.06 | 0.03 | 0.01 | 0.12 | 0.02 | 0.01 | 0.02 | 0.02 | 0.04 | -0.02 | -0.02 | 0.02 | -0.02 | -0.16 | -0.01 | 0.57 | 0.18 | 0.23 | 0.03 | 0.44 | 0.16 | 0.17 | 0.47 | 0.67 |
| HD 156826 | S | 5155 | 3.52 | 1.19 | -0.18 | -0.20 | -0.06 | -0.14 | 0.07 | -0.04 | -0.19 | -0.12 | -0.15 | -0.17 | -0.23 | -0.25 | -0.21 | -0.22 | -0.43 | -0.16 | 0.11 | -0.24 | -0.38 | -0.20 | -0.08 | 0.05 | 0.05 | 0.26 | -0.09 |
| HD 156846 | S | 6069 | 3.94 | 1.74 | 0.35 | 0.23 | 0.31 | 0.23 | 0.23 | 0.28 | 0.29 | 0.22 | 0.24 | 0.23 | 0.20 | 0.17 | 0.25 | 0.21 | 0.08 | 0.08 | 0.73 | 0.25 | 0.36 | 0.20 | 0.27 | 0.44 | 0.22 | 0.44 | 0.35 |
| HD 156968 | S | 5948 | 4.37 | 1.18 | -0.02 | -0.03 | -0.06 | -0.04 | -0.14 | 0.01 | 0.04 | -0.02 | -0.04 | -0.02 | -0.16 | -0.09 | -0.01 | -0.09 | -0.21 | -0.21 | 0.36 | 0.09 | 0.07 | 0.02 | 0.51 | 1.15 | 0.14 | -0.02 | 0.24 |
| HD 157881 | S | 4161 | 4.67 | 0.15 | 0.55 | 0.66 | 0.45 | 0.93 | 2.42 | 0.86 | 0.60 | 0.32 | 0.57 | 0.63 | 0.45 | 0.25 | 0.83 | 0.57 | 0.93 | | 1.07 | 0.46 | -0.32 | 0.03 | 1.01 | 1.64 | 2.22 | 2.60 | 0.67 |
| HD 158633 | S | 5313 | 4.57 | 0.55 | -0.37 | -0.27 | -0.27 | -0.32 | 0.04 | -0.28 | -0.31 | -0.23 | -0.27 | -0.32 | -0.41 | -0.41 | -0.30 | -0.38 | -0.31 | -0.42 | 0.00 | -0.20 | -0.51 | -0.39 | 0.11 | 0.42 | 0.20 | 0.37 | -0.04 |
| HD 159222 | S | 5788 | 4.39 | 1.12 | 0.22 | 0.19 | 0.23 | 0.17 | 0.33 | 0.18 | 0.22 | 0.13 | 0.11 | 0.18 | 0.16 | 0.11 | 0.12 | 0.13 | 0.11 | 0.11 | 0.53 | 0.23 | 0.24 | 0.14 | 0.36 | 0.20 | 0.31 | 0.38 | 0.42 |
| HD 160346 | S | 4864 | 4.52 | 0.29 | 0.06 | 0.10 | 0.03 | 0.14 | 0.74 | 0.09 | 0.07 | 0.13 | 0.30 | 0.16 | 0.19 | 0.04 | 0.13 | 0.13 | 0.39 | -0.03 | 0.29 | 0.19 | -0.41 | -0.01 | 0.56 | 0.97 | 1.09 | 1.81 | 0.41 |
| HD 160933 | S | 5829 | 3.81 | 1.55 | -0.17 | -0.17 | -0.22 | -0.18 | -0.16 | -0.23 | -0.23 | -0.25 | -0.27 | -0.27 | -0.31 | -0.31 | -0.24 | -0.31 | -0.42 | -0.27 | 0.30 | -0.28 | -0.37 | -0.18 | -0.22 | -0.17 | -0.18 | -0.27 | -0.19 |
| HD 16160 | S | 4886 | 4.63 | 0.15 | 0.16 | 0.19 | 0.24 | 0.14 | 0.80 | 0.12 | 0.23 | 0.28 | 0.54 | 0.14 | 0.07 | -0.04 | 0.18 | 0.11 | | 0.15 | 0.22 | 0.24 | -0.27 | -0.10 | 0.58 | 1.00 | 1.27 | 1.48 | 0.47 |
| HD 163492 | S | 5834 | 4.37 | 1.27 | 0.37 | 0.23 | 0.20 | 0.15 | 0.25 | 0.16 | 0.18 | 0.16 | 0.12 | 0.19 | 0.12 | 0.10 | 0.19 | 0.16 | 0.15 | 0.12 | 0.69 | 0.24 | 0.09 | 0.02 | 0.20 | 0.24 | 0.39 | 0.34 | 0.44 |
| HD 163840 | S | 5760 | 4.10 | 1.26 | 0.25 | 0.19 | 0.16 | 0.14 | 0.32 | 0.16 | 0.16 | 0.07 | 0.04 | 0.15 | 0.03 | 0.01 | 0.12 | 0.11 | -0.09 | 0.06 | 0.44 | 0.20 | 0.17 | -0.12 | 0.18 | 0.35 | 0.24 | 0.43 | 0.34 |
| HD 164595 | S | 5728 | 4.42 | 1.00 | -0.02 | -0.08 | 0.02 | -0.02 | 0.11 | -0.01 | 0.07 | 0.00 | -0.06 | -0.03 | -0.09 | -0.09 | -0.03 | -0.07 | -0.05 | 0.03 | 0.06 | 0.09 | -0.17 | -0.01 | 0.22 | 0.10 | 0.26 | 0.44 | 0.29 |
| HD 166 | S | 5481 | 4.52 | 1.27 | 0.07 | 0.17 | 0.04 | 0.13 | 0.36 | 0.20 | 0.09 | 0.13 | 0.15 | 0.20 | 0.11 | 0.08 | 0.09 | 0.09 | -0.01 | -0.07 | 0.54 | 0.14 | 0.09 | 0.14 | 0.39 | 0.30 | 0.59 | 0.33 | 0.45 |
| HD 166601 | S | 6289 | 4.01 | 1.89 | -0.01 | -0.03 | -0.02 | -0.01 | 0.06 | 0.02 | 0.15 | 0.05 | -0.04 | 0.02 | -0.17 | -0.08 | 0.23 | -0.11 | -0.26 | -0.18 | | -0.03 | 0.40 | 0.16 | 0.14 | 0.22 | 0.08 | -0.25 | 0.17 |
| HD 167588 | S | 5909 | 3.91 | 1.58 | -0.25 | -0.25 | -0.24 | -0.24 | -0.17 | -0.26 | -0.23 | -0.26 | -0.28 | -0.31 | -0.42 | -0.37 | -0.24 | -0.35 | -0.57 | -0.25 | 0.22 | -0.43 | -0.25 | -0.32 | -0.02 | 0.07 | -0.16 | -0.27 | -0.06 |
| HD 167665 | S | 6179 | 4.25 | 1.55 | 0.02 | -0.06 | -0.10 | -0.08 | 0.00 | -0.02 | 0.02 | -0.03 | -0.04 | -0.08 | -0.19 | -0.14 | -0.02 | -0.14 | -0.34 | -0.18 | 0.33 | -0.03 | 0.06 | -0.05 | -0.13 | 0.15 | 0.02 | -0.12 | 0.18 |
| HD 168443 | S | 5565 | 4.04 | 1.19 | 0.14 | 0.23 | 0.26 | 0.15 | 0.26 | 0.17 | 0.18 | 0.18 | 0.15 | 0.10 | 0.09 | 0.04 | 0.13 | 0.08 | 0.30 | 0.07 | 0.23 | 0.05 | 0.04 | 0.00 | 0.27 | 0.26 | 0.33 | 0.42 | 0.43 |
| HD 168746 | S | 5576 | 4.33 | 0.96 | -0.02 | 0.06 | 0.22 | 0.07 | 0.15 | 0.08 | 0.09 | 0.12 | 0.11 | -0.02 | -0.14 | -0.08 | 0.06 | -0.01 | 0.09 | 0.07 | 0.12 | 0.06 | -0.04 | -0.18 | 0.37 | 0.09 | 0.22 | 0.20 | 0.33 |
| HD 170657 | S | 5133 | 4.59 | 0.89 | -0.09 | -0.20 | -0.02 | -0.08 | 0.37 | -0.02 | -0.07 | 0.01 | 0.08 | -0.02 | -0.03 | -0.12 | -0.02 | -0.08 | -0.04 | -0.05 | 0.20 | 0.08 | -0.20 | 0.00 | 0.22 | 0.36 | 0.65 | 0.93 | 0.06 |

| Star | | Teff | log g | ξ | [NaI/Fe] | [MgI/Fe] | [AlI/Fe] | [SiI/Fe] | [CaI/Fe] | [ScII/Fe] | [TiI/Fe] | [VI/Fe] | [CrI/Fe] | [MnI/Fe] | [CoI/Fe] | [NiI/Fe] | [CuI/Fe] | [ZnI/Fe] | [SrI/Fe] | [YII/Fe] | [ZrII/Fe] | [BaII/Fe] | [LaII/Fe] | [CeII/Fe] | [NdII/Fe] | [EuII/Fe] |
|---|---|---|---|---|---|---|---|---|---|---|---|---|---|---|---|---|---|---|---|---|---|---|---|---|---|---|
| HD 171706 | S | 5935 | 4.09 | 1.35 | -0.03 | 0.11 | 0.07 | 0.01 | 0.12 | 0.06 | 0.12 | 0.02 | 0.00 | 0.00 | -0.12 | -0.07 | 0.01 | -0.04 | -0.16 | -0.09 | 0.40 | 0.05 | 0.02 | -0.05 | 0.05 | 0.28 | 0.23 | 0.26 | 0.34 |
| HD 172051 | S | 5669 | 4.49 | 0.95 | -0.27 | -0.20 | -0.13 | -0.18 | -0.07 | -0.10 | -0.11 | -0.12 | -0.15 | -0.17 | -0.22 | -0.23 | -0.16 | -0.23 | -0.45 | -0.21 | | -0.04 | -0.29 | -0.11 | 0.14 | 0.28 | 0.26 | 0.26 | 0.15 |
| HD 173818 | S | 4245 | 4.68 | 0.15 | 0.26 | 0.39 | 0.03 | 0.66 | 1.98 | 0.43 | 0.19 | 0.06 | 0.20 | 0.29 | -0.01 | -0.04 | 0.43 | 0.26 | 0.21 | | 0.57 | 0.21 | -0.51 | -0.14 | 0.80 | 1.32 | 1.64 | 2.26 | 0.31 |
| HD 175225 | S | 5281 | 3.76 | 1.46 | 0.32 | 0.32 | 0.35 | 0.32 | 0.59 | 0.25 | 0.24 | 0.21 | 0.24 | 0.27 | 0.34 | 0.17 | 0.26 | 0.27 | 0.51 | 0.15 | 0.22 | 0.29 | 0.03 | -0.04 | 0.55 | 0.32 | 0.34 | 0.99 | 0.44 |
| HD 175290 | S | 6359 | 4.02 | 1.84 | -0.20 | -0.09 | -0.10 | -0.16 | -0.05 | -0.12 | -0.02 | -0.13 | 0.04 | -0.20 | -0.43 | -0.30 | 0.09 | -0.35 | -0.51 | -0.37 | | -0.22 | 0.46 | 0.02 | 0.14 | 0.36 | -0.01 | -0.26 | -0.11 |
| HD 176051 | S | 5980 | 4.34 | 1.44 | -0.04 | 0.02 | 0.05 | -0.06 | -0.01 | 0.11 | 0.04 | 0.07 | 0.20 | 0.03 | -0.09 | -0.09 | 0.04 | -0.10 | -0.26 | -0.19 | 0.48 | 0.06 | 0.08 | -0.06 | 0.17 | 0.31 | 0.19 | 0.17 | 0.39 |
| HD 176367 | S | 6087 | 4.47 | 2.58 | 0.39 | 0.04 | 0.13 | 0.21 | 0.27 | 0.25 | 0.13 | 0.21 | 0.38 | 0.20 | 0.16 | 0.11 | 0.49 | 0.13 | -0.36 | | | 0.29 | 0.28 | 0.04 | 0.80 | 0.38 | 0.85 | 0.31 | 0.95 |
| HD 177830 | S | 4813 | 3.49 | 1.30 | 0.75 | 0.66 | 0.60 | 0.56 | 1.25 | 0.51 | 0.54 | 0.52 | 0.85 | 0.51 | 0.67 | 0.37 | 0.55 | 0.56 | 0.78 | 0.14 | 0.39 | 0.40 | 0.12 | 0.14 | 1.10 | 0.65 | 0.75 | 1.01 | 0.73 |
| HD 178428 | S | 5646 | 4.21 | 1.16 | 0.19 | 0.29 | 0.35 | 0.20 | 0.34 | 0.22 | 0.17 | 0.20 | 0.17 | 0.18 | 0.17 | 0.11 | 0.18 | 0.15 | 0.22 | 0.13 | 0.31 | 0.21 | 0.06 | 0.05 | 0.30 | 0.48 | 0.30 | 0.79 | 0.42 |
| HD 178911 | S | 5825 | 3.86 | 1.44 | 0.30 | 0.26 | 0.14 | 0.15 | 0.21 | 0.12 | 0.09 | 0.04 | 0.00 | 0.11 | 0.06 | 0.00 | 0.16 | 0.07 | -0.20 | 0.04 | 0.81 | 0.21 | -0.02 | -0.37 | 0.12 | 0.12 | 0.08 | 0.34 | 0.43 |
| HD 178911B | S | 5602 | 4.37 | 1.29 | 0.47 | 0.33 | 0.38 | 0.27 | 0.40 | 0.30 | 0.28 | 0.31 | 0.34 | 0.32 | 0.39 | 0.23 | 0.33 | 0.31 | 0.39 | 0.18 | 0.45 | 0.44 | 0.36 | 0.05 | 0.48 | 0.47 | 0.45 | 1.16 | 0.59 |
| HD 179957 | S | 5771 | 4.38 | 1.09 | 0.05 | 0.14 | 0.18 | 0.06 | 0.11 | 0.10 | 0.11 | 0.12 | 0.12 | 0.10 | 0.07 | 0.02 | 0.11 | 0.05 | 0.05 | -0.03 | 0.50 | 0.11 | -0.02 | -0.04 | 0.26 | 0.17 | 0.18 | 0.62 | 0.37 |
| HD 179958 | S | 5807 | 4.33 | 1.13 | 0.11 | 0.11 | 0.20 | 0.10 | 0.10 | 0.17 | 0.16 | 0.15 | 0.12 | 0.12 | 0.11 | 0.06 | 0.13 | 0.09 | 0.05 | -0.05 | 0.41 | 0.18 | 0.02 | 0.01 | 0.25 | 0.35 | 0.27 | 0.33 | 0.44 |
| HD 181655 | S | 5673 | 4.17 | 1.43 | 0.05 | 0.12 | 0.08 | 0.03 | 0.14 | 0.06 | -0.05 | 0.04 | 0.03 | 0.09 | 0.06 | 0.00 | 0.01 | 0.00 | -0.31 | -0.11 | 0.31 | 0.10 | 0.03 | -0.09 | 0.22 | 0.29 | 0.35 | 0.35 | 0.34 |
| HD 182488 | S | 5362 | 4.45 | 0.88 | 0.28 | 0.32 | 0.29 | 0.24 | 0.52 | 0.21 | 0.23 | 0.25 | 0.28 | 0.25 | 0.34 | 0.18 | 0.24 | 0.29 | 0.43 | 0.39 | 0.26 | 0.33 | 0.02 | 0.10 | 0.66 | 0.55 | 0.61 | 1.41 | 0.51 |
| HD 183263 | S | 5911 | 4.28 | 1.46 | 0.44 | 0.38 | 0.39 | 0.30 | 0.37 | 0.30 | 0.38 | 0.29 | 0.32 | 0.31 | 0.38 | 0.25 | 0.36 | 0.34 | 0.37 | 0.25 | 0.59 | 0.35 | 0.41 | 0.12 | 0.56 | 0.45 | 0.50 | 0.71 | 0.66 |
| HD 184151 | S | 6430 | 3.77 | 2.21 | -0.25 | -0.05 | -0.22 | -0.14 | -0.17 | -0.11 | -0.10 | -0.11 | 0.04 | -0.17 | -0.41 | -0.27 | 0.25 | -0.27 | -0.46 | -0.47 | | -0.23 | -0.04 | 0.07 | 0.08 | -0.08 | 0.01 | 0.45 | 0.02 |
| HD 184152 | S | 5578 | 4.37 | 0.97 | -0.16 | 0.07 | 0.12 | -0.09 | 0.11 | -0.05 | -0.11 | 0.04 | -0.04 | -0.14 | -0.28 | -0.25 | -0.08 | -0.18 | -0.04 | -0.05 | 0.03 | -0.08 | -0.14 | -0.36 | | 0.38 | 0.15 | 0.60 | 0.23 |
| HD 18445 | S | 4874 | 4.36 | 1.13 | 0.26 | 0.10 | 0.08 | 0.06 | 0.78 | 0.13 | 0.01 | 0.09 | 0.15 | 0.07 | 0.03 | -0.10 | 0.10 | 0.05 | 0.18 | -0.07 | | 0.22 | -0.32 | -0.03 | 0.34 | 0.39 | 0.93 | 1.95 | 0.05 |
| HD 184489 | S | 4133 | 4.74 | 0.15 | 0.25 | 0.24 | 0.06 | 0.91 | 2.29 | 0.62 | 0.27 | 0.04 | 0.18 | 0.41 | 0.01 | -0.16 | 0.60 | 0.26 | 0.26 | 1.53 | 0.73 | 0.17 | -0.52 | -0.09 | 1.24 | 1.83 | 1.74 | 2.33 | 0.50 |
| HD 184509 | S | 6069 | 4.33 | 1.30 | -0.10 | -0.08 | -0.20 | -0.07 | -0.07 | -0.05 | -0.04 | -0.05 | -0.09 | -0.08 | -0.19 | -0.12 | -0.06 | -0.13 | -0.31 | -0.12 | 0.36 | 0.06 | 0.20 | -0.04 | 0.17 | 0.22 | -0.03 | 0.31 | 0.20 |
| HD 18455 | S | 5105 | 4.44 | 1.62 | 0.07 | 0.08 | 0.02 | 0.03 | 0.61 | 0.01 | -0.07 | -0.01 | 0.01 | 0.04 | -0.01 | -0.08 | 0.05 | -0.04 | 0.12 | -0.17 | | 0.03 | -0.18 | -0.24 | 0.24 | 0.50 | 0.68 | 1.13 | 0.37 |
| HD 184700 | S | 5745 | 4.23 | 1.02 | -0.05 | 0.07 | 0.08 | 0.00 | -0.12 | 0.03 | 0.04 | 0.05 | -0.04 | -0.09 | -0.23 | -0.17 | -0.02 | -0.13 | -0.35 | -0.02 | 0.32 | -0.09 | -0.02 | -0.22 | 0.06 | 0.21 | 0.09 | 0.53 | 0.26 |
| HD 187123 | S | 5796 | 4.30 | 1.25 | 0.22 | 0.18 | 0.08 | 0.14 | 0.14 | 0.15 | 0.18 | 0.13 | 0.12 | 0.14 | 0.15 | 0.08 | 0.13 | 0.11 | 0.15 | -0.08 | 0.55 | 0.19 | 0.08 | -0.08 | 0.43 | 0.20 | 0.31 | 0.11 | 0.46 |
| HD 188015 | S | 5667 | 4.26 | 1.35 | 0.54 | 0.38 | 0.41 | 0.32 | 0.43 | 0.30 | 0.30 | 0.27 | 0.31 | 0.29 | 0.37 | 0.21 | 0.33 | 0.32 | 0.42 | 0.41 | 0.31 | 0.25 | 0.23 | 0.03 | 0.40 | 0.33 | 0.44 | 0.86 | 0.57 |
| HD 188088 | S | 4818 | 4.20 | 1.45 | 0.36 | 0.47 | 0.19 | 0.40 | 1.07 | 0.02 | 0.12 | 0.15 | 0.15 | 0.37 | 0.46 | 0.12 | 0.34 | 0.33 | | 0.52 | 0.26 | 0.35 | 0.20 | -0.71 | 0.84 | 1.22 | 1.05 | 1.64 | 0.80 |
| HD 188169 | S | 6509 | 4.25 | 1.91 | -0.02 | -0.06 | -0.02 | 0.01 | 0.08 | 0.03 | 0.13 | 0.04 | 0.10 | 0.01 | -0.25 | -0.10 | 0.23 | -0.11 | -0.33 | -0.22 | | 0.17 | 0.24 | 0.19 | 0.11 | 0.42 | 0.04 | 0.25 | |
| HD 189712 | S | 6363 | 3.85 | 2.21 | -0.41 | -0.31 | -0.22 | -0.35 | -0.29 | -0.36 | -0.30 | -0.36 | -0.34 | -0.40 | -0.55 | -0.53 | -0.01 | -0.52 | -0.71 | -0.45 | | -0.47 | -0.31 | -0.24 | | -0.16 | -0.23 | -0.81 | -0.27 |
| HD 189733 | S | 5044 | 4.60 | 0.83 | 0.16 | 0.15 | 0.09 | 0.10 | 0.60 | 0.23 | 0.09 | 0.16 | 0.24 | 0.16 | 0.18 | 0.02 | 0.10 | 0.11 | 0.59 | 0.10 | 0.22 | 0.19 | -0.15 | 0.06 | 0.45 | 0.62 | 0.84 | 1.71 | 0.09 |
| HD 190007 | S | 4596 | 4.57 | 0.88 | 0.60 | 0.44 | 0.30 | 0.43 | 1.17 | 0.45 | 0.33 | 0.29 | 0.61 | 0.41 | 0.35 | 0.30 | 0.41 | 0.38 | 0.03 | 0.33 | 0.49 | 0.31 | -0.44 | 0.08 | 0.58 | 1.31 | 1.44 | 1.90 | 0.53 |
| HD 190228 | S | 5264 | 3.71 | 1.16 | -0.20 | -0.14 | -0.13 | -0.16 | 0.04 | -0.14 | -0.17 | -0.18 | -0.19 | -0.22 | -0.30 | -0.28 | -0.20 | -0.25 | -0.20 | -0.32 | 0.05 | -0.22 | -0.35 | -0.17 | -0.15 | 0.22 | 0.11 | 0.13 | -0.02 |
| HD 190360 | S | 5564 | 4.31 | 1.24 | 0.36 | 0.38 | 0.41 | 0.30 | 0.50 | 0.29 | 0.33 | 0.31 | 0.32 | 0.28 | 0.32 | 0.20 | 0.34 | 0.28 | 0.35 | 0.29 | 0.40 | 0.38 | 0.13 | 0.12 | 0.52 | 0.49 | 0.51 | 1.24 | 0.62 |
| HD 190771 | S | 5751 | 4.44 | 1.29 | 0.14 | 0.17 | 0.16 | 0.17 | 0.30 | 0.21 | 0.15 | 0.13 | 0.14 | 0.19 | 0.15 | 0.11 | 0.08 | 0.13 | -0.04 | 0.23 | 0.52 | 0.16 | 0.19 | 0.20 | 0.27 | 0.50 | 0.18 | 0.45 | 0.68 |
| HD 192145 | S | 6069 | 3.98 | 1.59 | -0.21 | -0.31 | -0.21 | -0.18 | -0.11 | -0.18 | -0.15 | -0.22 | -0.21 | -0.27 | -0.37 | -0.31 | -0.12 | -0.31 | -0.53 | -0.28 | | -0.24 | -0.27 | -0.17 | | 0.09 | -0.18 | -0.27 | 0.19 |
| HD 192263 | S | 4974 | 4.61 | 0.93 | 0.01 | 0.10 | 0.12 | 0.14 | 0.79 | 0.10 | 0.08 | 0.12 | 0.24 | 0.16 | 0.12 | 0.04 | 0.12 | 0.10 | 0.29 | -0.16 | 0.32 | 0.01 | -0.09 | 0.04 | 0.64 | 0.97 | 0.96 | 1.12 | 0.25 |
| HD 192310 | S | 5044 | 4.51 | 0.25 | 0.21 | 0.16 | 0.15 | 0.16 | 0.62 | 0.13 | 0.20 | 0.19 | 0.30 | 0.16 | 0.22 | 0.06 | 0.12 | 0.16 | 0.38 | 0.32 | 0.13 | 0.25 | -0.22 | 0.02 | 0.83 | 0.90 | 0.84 | 1.61 | 0.22 |
| HD 193555 | S | 6133 | 3.72 | 2.83 | 0.38 | 0.02 | 0.68 | 0.45 | 0.57 | 0.34 | 0.42 | 0.41 | 0.47 | 0.42 | 0.37 | 0.26 | 0.48 | 0.41 | -0.03 | -0.24 | | 0.47 | 0.96 | 0.03 | | 0.53 | | 0.47 | |
| HD 193664 | S | 5945 | 4.44 | 1.08 | -0.17 | -0.04 | -0.03 | -0.05 | 0.08 | 0.03 | -0.05 | -0.03 | -0.01 | -0.05 | -0.11 | -0.07 | -0.03 | -0.09 | -0.20 | -0.17 | 0.54 | 0.12 | 0.09 | 0.06 | 0.21 | 0.38 | 0.30 | 0.34 | 0.34 |

| | | | | | | | | | | | | | | | | | | | | | | | | | | | |
|---|---|---|---|---|---|---|---|---|---|---|---|---|---|---|---|---|---|---|---|---|---|---|---|---|---|---|---|
| HD 19383 | S | 6396 | 4.32 | 2.14 | 0.23 | 0.20 | 0.28 | 0.21 | 0.39 | 0.25 | 0.31 | 0.29 | 0.45 | 0.18 | 0.13 | 0.14 | 0.23 | 0.17 | -0.24 | 0.25 | | 0.34 | 0.14 | 0.21 | 0.97 | 0.49 | 0.38 | 0.62 | 0.50 |
| HD 195019 | S | 5660 | 4.05 | 1.32 | 0.00 | 0.00 | 0.02 | 0.06 | 0.13 | 0.07 | 0.07 | -0.03 | -0.05 | 0.03 | -0.03 | -0.04 | -0.04 | -0.05 | -0.17 | -0.11 | 0.36 | 0.04 | 0.03 | -0.07 | 0.12 | 0.27 | 0.08 | 0.01 | 0.35 |
| HD 195564 | S | 5619 | 3.95 | 1.31 | 0.05 | 0.16 | 0.20 | 0.09 | 0.15 | 0.14 | 0.08 | 0.06 | 0.03 | 0.09 | 0.02 | 0.01 | 0.04 | 0.04 | -0.08 | 0.02 | 0.36 | 0.01 | -0.02 | -0.09 | 0.14 | 0.28 | 0.18 | 0.38 | 0.34 |
| HD 19617 | S | 5682 | 4.43 | 1.17 | 0.27 | 0.17 | 0.30 | 0.23 | 0.33 | 0.22 | 0.32 | 0.24 | 0.22 | 0.24 | 0.22 | 0.16 | 0.20 | 0.22 | 0.22 | -0.03 | 0.45 | 0.29 | 0.23 | 0.16 | 0.76 | 0.46 | 0.47 | 0.69 | 0.51 |
| HD 196761 | S | 5483 | 4.53 | 0.82 | -0.21 | -0.07 | -0.15 | -0.19 | 0.01 | -0.18 | -0.14 | -0.13 | -0.15 | -0.18 | -0.24 | -0.26 | -0.18 | -0.23 | -0.17 | -0.27 | | -0.18 | 0.00 | -0.19 | 0.15 | 0.01 | 0.15 | 0.54 | 0.10 |
| HD 197076 | S | 5844 | 4.46 | 1.09 | -0.11 | -0.06 | -0.03 | -0.04 | -0.13 | 0.04 | 0.03 | -0.03 | -0.05 | -0.04 | -0.13 | -0.09 | -0.04 | -0.10 | -0.23 | -0.09 | | 0.18 | 0.08 | 0.10 | 0.14 | 0.50 | 0.18 | 0.35 | 0.39 |
| HD 198387 | S | 5056 | 3.51 | 1.21 | -0.09 | -0.02 | 0.04 | -0.05 | 0.29 | -0.05 | -0.14 | -0.06 | -0.06 | -0.10 | -0.08 | -0.19 | -0.12 | -0.14 | 0.06 | -0.03 | 0.03 | -0.14 | -0.42 | -0.26 | 0.10 | 0.16 | 0.04 | 0.48 | 0.01 |
| HD 198483 | S | 5986 | 4.30 | 1.66 | 0.42 | 0.16 | 0.29 | 0.28 | 0.33 | 0.26 | 0.27 | 0.25 | 0.21 | 0.27 | 0.27 | 0.17 | 0.25 | 0.21 | 0.12 | 0.02 | 0.63 | 0.29 | 0.31 | 0.09 | 0.26 | 0.23 | 0.34 | 0.36 | 0.44 |
| HD 198802 | S | 5736 | 3.81 | 1.46 | 0.06 | 0.02 | 0.06 | 0.04 | 0.04 | 0.04 | 0.07 | -0.01 | -0.01 | 0.02 | -0.04 | -0.04 | 0.00 | -0.01 | -0.07 | -0.10 | 0.46 | 0.07 | -0.03 | -0.07 | 0.11 | 0.15 | 0.08 | 0.03 | 0.23 |
| HD 199604 | S | 5887 | 4.33 | 1.04 | -0.54 | -0.37 | -0.47 | -0.35 | -0.37 | -0.34 | -0.35 | -0.30 | -0.38 | -0.55 | -0.78 | -0.59 | -0.38 | -0.56 | -0.73 | -0.36 | | -0.57 | | -0.58 | | -0.48 | -0.25 | | |
| HD 200779 | S | 4389 | 4.61 | 0.15 | 0.53 | 0.60 | 0.35 | 0.58 | 1.62 | 0.66 | 0.45 | 0.34 | 0.59 | 0.48 | 0.32 | 0.25 | 0.55 | 0.39 | 0.61 | | | 0.67 | 0.53 | -0.39 | 0.11 | 1.05 | 1.08 | 1.93 | 2.29 | 0.51 |
| HD 201456 | S | 6201 | 4.20 | 1.55 | 0.20 | 0.21 | 0.04 | 0.15 | 0.21 | 0.18 | 0.21 | 0.15 | 0.18 | 0.19 | 0.06 | 0.09 | 0.24 | 0.09 | -0.13 | 0.10 | 0.65 | 0.22 | 0.42 | 0.18 | 0.15 | 0.27 | 0.35 | -0.21 | 0.15 |
| HD 201496 | S | 5944 | 4.37 | 1.19 | 0.00 | -0.01 | 0.00 | -0.01 | -0.06 | 0.03 | 0.10 | 0.02 | 0.01 | 0.01 | -0.05 | -0.05 | 0.01 | -0.02 | -0.04 | 0.03 | 0.33 | 0.01 | 0.17 | -0.02 | 0.14 | -0.03 | 0.09 | 0.13 | 0.56 |
| HD 202206 | S | 5693 | 4.40 | 1.27 | 0.45 | 0.37 | 0.33 | 0.31 | 0.53 | 0.33 | 0.36 | 0.32 | 0.35 | 0.37 | 0.42 | 0.28 | 0.34 | 0.35 | 0.41 | 0.24 | 0.54 | 0.44 | 0.27 | 0.14 | 0.60 | 0.60 | 0.49 | 1.24 | 0.77 |
| HD 202282 | S | 5796 | 4.30 | 1.31 | 0.16 | 0.15 | 0.09 | 0.07 | 0.14 | 0.10 | 0.10 | 0.09 | 0.08 | 0.14 | 0.08 | 0.04 | 0.10 | 0.07 | -0.04 | 0.01 | 0.61 | 0.14 | 0.13 | -0.04 | 0.18 | 0.12 | 0.25 | 0.51 | 0.38 |
| HD 20339 | S | 5918 | 4.34 | 1.28 | 0.02 | -0.01 | 0.11 | 0.02 | 0.16 | 0.11 | 0.08 | 0.06 | 0.04 | 0.04 | -0.02 | -0.02 | 0.08 | -0.01 | -0.08 | -0.06 | 0.34 | 0.09 | 0.21 | 0.01 | -0.18 | 0.49 | 0.18 | 0.27 | 0.41 |
| HD 20367 | S | 6050 | 4.36 | 1.48 | 0.08 | 0.05 | 0.10 | 0.08 | 0.18 | 0.12 | 0.03 | 0.06 | 0.05 | 0.11 | 0.02 | 0.03 | 0.07 | 0.00 | -0.17 | -0.11 | 0.58 | 0.22 | 0.13 | 0.29 | 0.36 | 0.23 | 0.18 | 0.24 | 0.40 |
| HD 204153 | S | 6920 | 4.20 | 2.15 | | | | 0.86 | | | 0.85 | -0.72 | 1.89 | 1.04 | 1.10 | 1.06 | -0.37 | 1.74 | 1.01 | | | 1.24 | 2.33 | | | | | | |
| HD 204485 | S | 7125 | 4.13 | 2.62 | 0.26 | -0.01 | 0.11 | 0.21 | 0.16 | 0.26 | 0.27 | 0.29 | 0.52 | 0.32 | 0.26 | 0.16 | 0.71 | 0.18 | -0.07 | -0.26 | | 0.35 | 0.75 | 0.07 | 0.94 | 0.47 | 0.46 | -0.03 | 0.43 |
| HD 205027 | S | 5752 | 4.31 | 0.89 | -0.27 | -0.03 | -0.16 | -0.14 | -0.06 | -0.16 | -0.12 | -0.12 | -0.23 | -0.28 | -0.51 | -0.35 | -0.22 | -0.32 | -0.53 | -0.17 | 0.09 | -0.19 | -0.01 | -0.25 | | -0.14 | -0.01 | 0.24 | 0.24 |
| HD 205700 | S | 6656 | 4.11 | 1.98 | -0.08 | -0.22 | -0.28 | -0.09 | -0.07 | -0.07 | 0.05 | 0.01 | 0.14 | -0.11 | -0.14 | -0.15 | 0.34 | -0.18 | -0.43 | -0.32 | | 0.03 | -0.09 | 0.12 | | 0.27 | 0.11 | -0.09 | |
| HD 206282 | S | 6345 | 4.01 | 2.06 | 0.47 | 0.34 | 0.27 | 0.26 | 0.26 | 0.30 | 0.36 | 0.30 | 0.33 | 0.31 | 0.21 | 0.21 | 0.33 | 0.25 | 0.09 | 0.17 | | 0.34 | 0.57 | 0.32 | 0.45 | 0.48 | 0.34 | 0.57 | 0.41 |
| HD 207858 | S | 6290 | 3.97 | 2.14 | 0.28 | 0.20 | 0.25 | 0.20 | 0.17 | 0.20 | 0.31 | 0.22 | 0.20 | 0.22 | 0.04 | 0.11 | 0.37 | 0.15 | -0.03 | 0.16 | | 0.27 | 0.79 | -0.01 | 0.11 | 0.23 | 0.14 | 0.34 | 0.25 |
| HD 208801 | S | 4918 | 3.59 | 1.10 | 0.32 | 0.29 | 0.39 | 0.25 | 0.72 | 0.28 | 0.26 | 0.37 | 0.47 | 0.25 | 0.34 | 0.11 | 0.26 | 0.21 | 0.09 | 0.38 | 0.25 | 0.17 | -0.06 | -0.22 | 0.54 | 0.43 | 0.55 | 1.54 | 0.18 |
| HD 21019 | S | 5514 | 3.79 | 1.21 | -0.37 | -0.29 | -0.24 | -0.31 | -0.19 | -0.29 | -0.35 | -0.31 | -0.35 | -0.39 | -0.51 | -0.43 | -0.32 | -0.44 | -0.44 | -0.45 | -0.09 | -0.36 | -0.44 | -0.41 | -0.30 | -0.01 | -0.19 | -0.08 | -0.13 |
| HD 210277 | S | 5533 | 4.36 | 1.12 | 0.34 | 0.39 | 0.43 | 0.28 | 0.51 | 0.29 | 0.33 | 0.34 | 0.31 | 0.29 | 0.31 | 0.20 | 0.32 | 0.30 | 0.53 | 0.35 | 0.33 | 0.34 | 0.14 | 0.06 | 0.39 | 0.41 | 0.55 | 1.34 | 0.67 |
| HD 210460 | S | 5529 | 3.52 | 1.39 | -0.19 | -0.15 | -0.14 | -0.20 | -0.11 | -0.15 | -0.26 | -0.23 | -0.25 | -0.23 | -0.37 | -0.30 | -0.25 | -0.32 | -0.64 | -0.39 | 0.29 | -0.22 | -0.34 | 0.03 | -0.17 | -0.08 | -0.08 | -0.21 | -0.02 |
| HD 210483 | S | 5842 | 4.10 | 1.34 | -0.10 | 0.04 | 0.03 | -0.04 | 0.04 | 0.00 | 0.06 | -0.05 | -0.05 | -0.06 | -0.19 | -0.13 | -0.10 | -0.13 | -0.18 | -0.02 | 0.15 | -0.01 | -0.12 | -0.14 | 0.12 | 0.18 | 0.03 | -0.01 | 0.28 |
| HD 210855 | S | 6255 | 3.78 | 2.20 | 0.49 | 0.27 | 0.31 | 0.26 | 0.29 | 0.24 | 0.32 | 0.25 | 0.27 | 0.23 | 0.17 | 0.16 | 0.31 | 0.20 | 0.04 | 0.19 | | 0.28 | 0.46 | 0.19 | 0.11 | 0.37 | 0.07 | 0.45 | 0.34 |
| HD 211476 | S | 5829 | 4.35 | 1.04 | -0.14 | -0.11 | -0.15 | -0.10 | -0.09 | -0.08 | -0.06 | -0.10 | -0.14 | -0.12 | -0.20 | -0.17 | -0.15 | -0.17 | -0.20 | -0.21 | 0.23 | 0.01 | 0.02 | -0.12 | 0.36 | -0.03 | 0.08 | 0.00 | 0.35 |
| HD 211575 | S | 6589 | 4.23 | 2.62 | 0.37 | 0.22 | 0.34 | 0.32 | 0.34 | 0.39 | 0.28 | 0.38 | 0.60 | 0.33 | 0.29 | 0.21 | 0.77 | 0.29 | -0.11 | 0.54 | | 0.58 | 0.58 | 0.19 | 0.89 | | 0.47 | | 0.42 |
| HD 213338 | S | 5558 | 4.51 | 0.84 | 0.10 | 0.17 | 0.13 | 0.11 | 0.20 | 0.17 | 0.07 | 0.13 | 0.08 | 0.16 | 0.10 | 0.08 | 0.08 | 0.09 | 0.12 | -0.04 | 0.43 | 0.21 | 0.09 | 0.10 | 0.43 | 0.40 | 0.46 | 1.11 | 0.40 |
| HD 214385 | S | 5711 | 4.47 | 0.74 | -0.28 | -0.16 | -0.29 | -0.23 | -0.06 | -0.14 | -0.14 | -0.18 | -0.21 | -0.24 | -0.33 | -0.29 | -0.22 | -0.27 | -0.30 | -0.17 | 0.14 | -0.09 | 0.03 | -0.27 | 0.27 | 0.15 | 0.04 | 0.15 | 0.09 |
| HD 214749 | S | 4531 | 4.63 | 0.80 | 0.25 | 0.29 | 0.05 | 0.32 | 1.11 | 0.27 | 0.12 | 0.13 | 0.33 | 0.24 | 0.17 | 0.14 | 0.23 | 0.20 | | 0.18 | 0.60 | 0.27 | -0.43 | 0.09 | 0.55 | 0.70 | 1.30 | 1.89 | 0.19 |
| HD 215243 | S | 6393 | 4.12 | 1.94 | 0.18 | 0.11 | 0.10 | 0.11 | 0.13 | 0.13 | 0.14 | 0.20 | 0.19 | 0.16 | 0.01 | 0.05 | 0.23 | 0.07 | -0.19 | -0.13 | | 0.21 | 0.47 | 0.11 | -0.04 | 0.19 | 0.13 | 0.09 | 0.33 |
| HD 21531 | S | 4231 | 4.67 | 0.15 | 0.63 | 0.22 | 0.47 | 0.98 | 1.99 | 0.77 | 0.73 | 0.49 | 0.78 | 0.74 | 0.63 | 0.42 | 0.80 | 0.69 | 1.13 | | | 0.80 | 0.89 | 0.26 | 0.33 | 1.61 | 2.11 | 2.14 | 2.85 | 0.74 |
| HD 215625 | S | 6228 | 4.39 | 1.42 | 0.21 | 0.15 | 0.08 | 0.11 | 0.07 | 0.17 | 0.22 | 0.16 | 0.10 | 0.15 | 0.07 | 0.08 | 0.16 | 0.09 | -0.10 | 0.06 | 0.62 | 0.31 | 0.37 | 0.17 | -0.01 | 0.35 | 0.38 | 0.31 | 0.11 |
| HD 216133 | S | 3973 | 4.92 | 0.15 | 0.21 | 0.54 | 0.50 | 1.60 | 3.42 | 1.33 | 0.78 | 0.26 | 0.42 | 0.76 | 0.20 | 0.11 | 0.96 | 0.81 | | | 2.67 | 0.90 | 0.57 | -0.01 | 0.09 | 1.60 | 2.35 | 2.29 | 2.08 | 1.67 |

| Star | | Teff | logg | | | | | | | | | | | | | | | | | | | | | | | | | | |
|---|---|---|---|---|---|---|---|---|---|---|---|---|---|---|---|---|---|---|---|---|---|---|---|---|---|---|---|---|---|
| HD 216770 | S | 5411 | 4.48 | 0.95 | 0.51 | 0.50 | 0.45 | 0.45 | 0.61 | 0.35 | 0.54 | 0.48 | 0.52 | 0.45 | 0.63 | 0.39 | 0.50 | 0.54 | 0.63 | 0.68 | 0.61 | 0.57 | 0.22 | 0.32 | 0.71 | 0.78 | 0.83 | 1.94 | 0.86 |
| HD 217107 | S | 5541 | 4.29 | 1.19 | 0.51 | 0.45 | 0.44 | 0.37 | 0.55 | 0.37 | 0.38 | 0.33 | 0.33 | 0.35 | 0.47 | 0.29 | 0.37 | 0.37 | 0.47 | 0.39 | 0.47 | 0.39 | 0.17 | 0.10 | 0.71 | 0.49 | 0.57 | 1.04 | 0.68 |
| HD 217357 | S | 4192 | 4.82 | 0.15 | 0.36 | 0.22 | 0.27 | 0.94 | 2.22 | 0.52 | 0.54 | 0.26 | 0.37 | 0.55 | 0.29 | 0.14 | 0.73 | 0.51 | 0.75 | | 0.93 | 0.32 | -0.32 | 0.11 | 1.39 | 1.94 | 2.19 | 2.49 | 1.18 |
| HD 217577 | S | 5762 | 4.11 | 1.23 | -0.21 | -0.06 | -0.04 | -0.14 | -0.03 | -0.03 | -0.10 | -0.10 | -0.14 | -0.12 | -0.21 | -0.17 | -0.13 | -0.20 | -0.55 | -0.11 | 0.16 | -0.05 | -0.14 | -0.25 | -0.13 | 0.13 | 0.07 | 0.05 | 0.47 |
| HD 217877 | S | 6015 | 4.35 | 1.29 | 0.02 | -0.02 | 0.02 | -0.03 | 0.02 | 0.01 | 0.10 | -0.01 | -0.04 | 0.02 | -0.05 | -0.08 | -0.01 | -0.07 | -0.13 | -0.22 | 0.40 | 0.14 | 0.08 | 0.01 | 0.11 | 0.19 | 0.20 | 0.46 | 0.32 |
| HD 217958 | S | 5771 | 4.23 | 1.38 | 0.53 | 0.26 | 0.32 | 0.36 | 0.57 | 0.28 | 0.45 | 0.30 | 0.31 | 0.32 | 0.33 | 0.24 | 0.35 | 0.35 | 0.48 | 0.52 | 0.46 | 0.32 | 0.55 | 0.16 | 0.47 | 0.36 | 0.52 | 0.57 | 0.69 |
| HD 218101 | S | 5217 | 3.81 | 1.26 | 0.20 | 0.19 | 0.26 | 0.18 | 0.47 | 0.16 | 0.11 | 0.15 | 0.20 | 0.18 | 0.29 | 0.08 | 0.17 | 0.16 | 0.28 | 0.00 | 0.24 | 0.20 | -0.06 | -0.05 | 0.35 | 0.29 | 0.32 | 0.98 | 0.34 |
| HD 219428 | S | 5930 | 4.35 | 1.43 | 0.19 | 0.04 | 0.14 | 0.14 | 0.21 | 0.18 | 0.19 | 0.11 | 0.14 | 0.18 | 0.12 | 0.08 | 0.14 | 0.12 | -0.01 | 0.11 | 0.46 | 0.19 | 0.50 | 0.06 | 0.34 | 0.33 | 0.20 | 0.23 | 0.16 |
| HD 220008 | S | 5653 | 3.82 | 1.35 | -0.15 | -0.18 | -0.04 | -0.11 | 0.01 | -0.11 | -0.12 | -0.16 | -0.18 | -0.17 | -0.25 | -0.22 | -0.18 | -0.21 | -0.23 | -0.28 | | -0.25 | -0.22 | -0.24 | -0.24 | -0.13 | 0.05 | 0.01 | 0.14 |
| HD 220689 | S | 5921 | 4.37 | 1.17 | 0.08 | 0.08 | 0.09 | 0.02 | -0.07 | 0.05 | 0.10 | 0.03 | 0.00 | 0.02 | 0.01 | -0.02 | 0.04 | 0.01 | -0.07 | -0.04 | 0.52 | 0.07 | -0.08 | -0.01 | 0.32 | 0.47 | 0.26 | 0.38 | 0.60 |
| HD 22072 | S | 4974 | 3.48 | 1.13 | -0.24 | -0.05 | 0.01 | -0.07 | 0.29 | -0.10 | -0.18 | -0.10 | -0.11 | -0.25 | -0.34 | -0.31 | -0.20 | -0.26 | -0.02 | -0.07 | 0.03 | -0.20 | -0.44 | -0.28 | -0.03 | 0.15 | 0.07 | 0.41 | -0.01 |
| HD 221356 | S | 6137 | 4.39 | 1.39 | -0.25 | -0.12 | -0.20 | -0.16 | -0.25 | -0.06 | -0.11 | -0.09 | -0.10 | -0.19 | -0.28 | -0.23 | -0.09 | -0.23 | -0.40 | -0.23 | | -0.09 | -0.06 | -0.12 | 0.39 | -0.20 | 0.05 | -0.15 | 0.25 |
| HD 221445 | S | 6230 | 3.77 | 1.87 | -0.08 | -0.04 | -0.04 | -0.07 | -0.11 | -0.02 | -0.06 | -0.09 | -0.04 | -0.15 | -0.25 | -0.16 | -0.05 | -0.16 | -0.34 | -0.42 | | -0.13 | -0.04 | -0.13 | 0.17 | -0.10 | -0.01 | -0.04 | -0.02 |
| HD 221503 | S | 4312 | 4.66 | 0.15 | 0.49 | 0.34 | 0.21 | 0.66 | 1.66 | 0.64 | 0.52 | 0.32 | 0.50 | 0.51 | 0.29 | 0.28 | 0.52 | 0.44 | | 1.05 | 0.92 | 0.50 | -0.28 | 0.21 | 1.06 | 1.63 | 2.16 | 2.21 | 0.58 |
| HD 222582 | S | 5796 | 4.37 | 1.09 | 0.08 | -0.03 | 0.11 | 0.07 | 0.12 | 0.10 | 0.18 | 0.11 | 0.10 | 0.08 | 0.10 | 0.02 | 0.12 | 0.05 | 0.13 | 0.06 | 0.55 | 0.14 | 0.22 | -0.10 | 0.16 | 0.41 | 0.29 | 0.01 | 0.58 |
| HD 222645 | S | 6211 | 4.33 | 1.47 | -0.13 | -0.12 | -0.21 | -0.02 | -0.07 | 0.04 | -0.01 | 0.01 | -0.04 | -0.03 | -0.17 | -0.09 | 0.02 | -0.11 | -0.21 | -0.11 | 0.64 | 0.01 | 0.11 | 0.12 | 0.18 | 0.46 | 0.18 | 0.07 | 0.12 |
| HD 22292 | S | 6483 | 4.31 | 2.64 | 0.27 | -0.14 | 0.22 | 0.22 | 0.47 | 0.14 | 0.24 | 0.28 | 0.23 | 0.19 | -0.10 | 0.04 | 0.68 | 0.17 | -0.43 | | 0.27 | 0.74 | 0.27 | 0.67 | | 0.70 | 0.95 | 0.32 |
| HD 223084 | S | 5958 | 4.33 | 1.19 | -0.20 | -0.16 | -0.15 | -0.09 | 0.06 | -0.08 | -0.14 | -0.17 | -0.18 | -0.15 | -0.31 | -0.19 | -0.16 | -0.24 | -0.43 | -0.33 | 0.63 | 0.00 | 0.11 | 0.05 | | 0.19 | 0.03 | 0.20 | 0.10 |
| HD 223110 | S | 6516 | 3.96 | 2.60 | 0.28 | 0.05 | 0.33 | 0.20 | 0.20 | 0.25 | 0.17 | 0.24 | 0.32 | 0.20 | 0.13 | 0.13 | 0.43 | 0.21 | -0.21 | 0.46 | | 0.31 | 0.91 | 0.21 | 0.58 | 0.57 | 0.19 | | 0.66 |
| HD 223238 | S | 5889 | 4.30 | 1.20 | 0.18 | 0.13 | 0.19 | 0.08 | 0.11 | 0.12 | 0.16 | 0.09 | 0.11 | 0.11 | 0.06 | 0.05 | 0.12 | 0.08 | 0.04 | 0.04 | 0.44 | 0.09 | 0.03 | -0.02 | -0.04 | 0.31 | 0.33 | 0.38 | 0.46 |
| HD 22455 | S | 5886 | 4.37 | 1.22 | 0.15 | 0.05 | 0.12 | 0.08 | 0.18 | 0.14 | 0.10 | 0.12 | 0.09 | 0.16 | 0.09 | 0.04 | 0.12 | 0.10 | 0.04 | 0.13 | 0.50 | 0.20 | 0.05 | -0.05 | 0.30 | 0.52 | 0.20 | 0.44 | 0.35 |
| HD 22468A | S | 4736 | 3.35 | 1.78 | -0.15 | -0.08 | 0.25 | 0.43 | 0.30 | 0.59 | 0.11 | 0.09 | 0.39 | 0.06 | -0.04 | -0.16 | -0.30 | 0.08 | | | 0.22 | 0.13 | 0.10 | | -0.02 | 0.60 | | 0.40 |
| HD 22468B | S | 4631 | 4.47 | 0.15 | -0.03 | 0.19 | 0.05 | 0.24 | 1.31 | -0.13 | 0.10 | 0.03 | 0.20 | 0.14 | 0.08 | 0.10 | 0.10 | 0.19 | 0.25 | 0.91 | 0.16 | 0.07 | -0.44 | -0.01 | 0.63 | 0.84 | 1.12 | 2.08 | |
| HD 225239 | S | 5647 | 3.76 | 1.30 | -0.41 | -0.32 | -0.24 | -0.33 | -0.33 | -0.32 | -0.39 | -0.39 | -0.47 | -0.43 | -0.58 | -0.47 | -0.35 | -0.47 | -0.57 | -0.44 | | -0.50 | | -0.58 | -0.18 | -0.39 | -0.31 | -0.33 | -0.10 |
| HD 231701 | S | 6240 | 4.17 | 1.65 | 0.11 | -0.05 | -0.04 | 0.06 | 0.14 | 0.10 | 0.17 | 0.07 | 0.13 | 0.08 | 0.11 | 0.01 | 0.18 | 0.05 | -0.07 | 0.01 | 0.44 | 0.08 | 0.27 | -0.02 | 0.22 | 0.37 | 0.15 | 0.18 | 0.27 |
| HD 232979 | S | 3893 | 4.62 | 0.15 | -0.15 | -0.05 | 0.25 | 1.47 | 3.14 | 1.14 | 0.56 | 0.22 | 0.37 | 0.67 | 0.24 | 0.24 | 0.93 | 0.75 | 0.92 | | 0.73 | 0.43 | -0.01 | 0.22 | 1.70 | 2.03 | 2.32 | 2.25 | 1.30 |
| HD 23349 | S | 6018 | 4.26 | 1.42 | 0.29 | 0.43 | 0.21 | 0.17 | 0.26 | 0.17 | 0.27 | 0.19 | 0.19 | 0.18 | 0.09 | 0.12 | 0.13 | 0.15 | 0.09 | 0.15 | 0.95 | 0.32 | 0.29 | 0.09 | 0.26 | 0.22 | 0.20 | 0.48 | 0.49 |
| HD 23356 | S | 4990 | 4.63 | 1.36 | 0.06 | -0.01 | 0.00 | -0.01 | 0.71 | 0.06 | -0.05 | 0.00 | 0.09 | 0.02 | 0.02 | -0.08 | -0.01 | -0.07 | 0.11 | -0.07 | 0.10 | -0.02 | -0.33 | -0.17 | 0.34 | 0.89 | 0.73 | 1.39 | 0.46 |
| HD 234078 | S | 4157 | 4.68 | 0.15 | 0.43 | 0.67 | 0.26 | 0.68 | 2.10 | 0.67 | 0.67 | 0.25 | 0.46 | 0.43 | 0.38 | 0.15 | 0.63 | 0.43 | 0.70 | 1.27 | 0.84 | 0.48 | -0.29 | 0.01 | 0.91 | 1.75 | 1.96 | 2.55 | 1.19 |
| HD 23453 | S | 3748 | 4.37 | 0.25 | 0.27 | 0.03 | 0.81 | 1.85 | 3.77 | 1.25 | 0.81 | 0.31 | 0.48 | 0.91 | 0.51 | 0.45 | 1.11 | 1.05 | 1.48 | 2.65 | 1.05 | 0.69 | -0.51 | -0.05 | 1.76 | 2.14 | 2.26 | | 1.06 |
| HD 23476 | S | 5662 | 4.48 | 0.81 | -0.29 | -0.10 | -0.09 | -0.17 | -0.24 | -0.08 | -0.10 | -0.07 | -0.16 | -0.29 | -0.53 | -0.38 | -0.17 | -0.31 | -0.26 | -0.21 | -0.02 | -0.28 | -0.22 | -0.40 | | -0.01 | 0.07 | 0.29 | 0.35 |
| HD 23596 | S | 6008 | 4.15 | 1.50 | 0.53 | 0.19 | 0.33 | 0.30 | 0.36 | 0.29 | 0.39 | 0.27 | 0.27 | 0.30 | 0.31 | 0.22 | 0.28 | 0.31 | 0.34 | 0.23 | 0.69 | 0.31 | 0.33 | 0.10 | 0.30 | 0.49 | 0.30 | 0.31 | 0.42 |
| HD 239928 | S | 5899 | 4.43 | 1.08 | 0.12 | 0.01 | 0.07 | 0.08 | 0.12 | 0.10 | 0.14 | 0.10 | 0.10 | 0.12 | 0.08 | 0.05 | 0.10 | 0.06 | 0.08 | 0.00 | | 0.26 | 0.10 | 0.07 | 0.17 | 0.19 | 0.24 | 0.39 | 0.36 |
| HD 2475 | S | 6012 | 4.19 | 1.66 | 0.11 | 0.15 | 0.13 | 0.10 | 0.11 | 0.19 | 0.07 | 0.13 | 0.10 | 0.13 | 0.12 | 0.08 | 0.12 | 0.07 | -0.22 | -0.06 | 0.51 | 0.24 | 0.16 | 0.16 | 0.38 | 0.34 | 0.18 | 0.14 | 0.31 |
| HD 2582 | S | 5705 | 4.19 | 1.41 | 0.25 | 0.18 | 0.27 | 0.24 | 0.37 | 0.18 | 0.20 | 0.18 | 0.20 | 0.25 | 0.18 | 0.13 | 0.25 | 0.21 | 0.26 | 0.18 | 0.29 | 0.24 | 0.21 | 0.00 | 0.21 | 0.37 | 0.37 | 0.59 | 0.68 |
| HD 2589 | S | 5154 | 3.64 | 1.15 | -0.01 | 0.14 | 0.21 | 0.06 | 0.37 | 0.11 | 0.03 | 0.11 | 0.11 | 0.03 | 0.03 | -0.06 | 0.03 | -0.01 | 0.30 | 0.05 | 0.07 | 0.02 | -0.19 | -0.08 | 0.25 | 0.25 | 0.30 | 1.92 | 0.26 |
| HD 2638 | S | 5120 | 4.54 | 1.33 | 0.35 | | 0.37 | 0.35 | 0.94 | 0.37 | 0.31 | 0.25 | 0.38 | 0.25 | 0.31 | 0.21 | 0.30 | 0.29 | 0.33 | | 0.34 | 0.33 | 0.02 | | 1.17 | 0.74 | 0.80 | 1.46 | 0.25 |
| HD 26505 | S | 5878 | 4.11 | 1.39 | 0.08 | -0.01 | 0.10 | 0.05 | 0.03 | 0.08 | 0.10 | 0.08 | 0.10 | 0.08 | 0.00 | 0.02 | 0.12 | 0.02 | 0.02 | 0.18 | 0.41 | 0.13 | -0.09 | -0.08 | 0.15 | 0.28 | 0.34 | 0.38 | 0.30 |

| Star | | Teff | log g | | | | | | | | | | | | | | | | | | | | | | | | | | |
|---|---|---|---|---|---|---|---|---|---|---|---|---|---|---|---|---|---|---|---|---|---|---|---|---|---|---|---|---|---|
| HD 26913 | S | 5661 | 4.52 | 1.55 | -0.03 | 0.01 | 0.07 | 0.02 | 0.08 | 0.07 | 0.06 | 0.05 | 0.04 | 0.09 | -0.06 | -0.02 | 0.00 | -0.05 | -0.25 | -0.10 | 0.36 | 0.10 | 0.03 | 0.19 | 0.32 | 0.33 | 0.43 | 0.53 | 0.25 |
| HD 26923 | S | 5989 | 4.43 | 1.20 | -0.10 | -0.06 | 0.00 | -0.02 | 0.03 | 0.06 | 0.05 | 0.00 | -0.02 | 0.03 | -0.08 | -0.03 | -0.03 | -0.09 | -0.27 | -0.18 | 0.60 | 0.22 | 0.19 | 0.22 | 0.31 | 0.44 | 0.25 | 0.20 | 0.38 |
| HD 2730 | S | 6192 | 3.99 | 1.71 | -0.05 | -0.13 | -0.02 | 0.01 | -0.15 | 0.02 | 0.05 | 0.02 | 0.07 | 0.02 | -0.17 | -0.11 | 0.06 | -0.09 | -0.29 | -0.05 | 0.66 | -0.05 | 0.30 | 0.05 | 0.06 | 0.12 | 0.10 | -0.01 | 0.29 |
| HD 27530 | S | 5926 | 4.38 | 1.49 | 0.39 | 0.20 | 0.22 | 0.24 | 0.35 | 0.23 | 0.32 | 0.24 | 0.19 | 0.30 | 0.27 | 0.19 | 0.24 | 0.25 | 0.13 | 0.16 | 0.65 | 0.33 | 0.20 | 0.01 | 0.54 | 0.33 | 0.46 | 0.43 | 0.73 |
| HD 28185 | S | 5658 | 4.33 | 1.27 | 0.52 | 0.32 | 0.35 | 0.28 | 0.52 | 0.33 | 0.33 | 0.29 | 0.30 | 0.29 | 0.38 | 0.22 | 0.34 | 0.30 | 0.38 | 0.31 | 0.40 | 0.39 | 0.23 | 0.00 | 0.36 | 0.50 | 0.45 | 0.92 | 0.55 |
| HD 28343 | S | 4152 | 4.65 | 0.35 | 0.72 | 0.98 | 0.61 | 1.06 | 2.20 | 0.95 | 0.82 | 0.53 | 0.91 | 0.81 | 0.63 | 0.42 | 0.97 | 0.80 | 1.30 | | 0.94 | 0.63 | 0.23 | 0.22 | 1.53 | 2.40 | 2.46 | 2.92 | 0.74 |
| HD 285660 | S | 6055 | 4.27 | 1.65 | 0.09 | 0.19 | 0.06 | 0.11 | -0.01 | 0.09 | 0.13 | 0.10 | 0.12 | 0.06 | -0.03 | 0.00 | 0.20 | 0.02 | -0.14 | 0.28 | | 0.25 | 0.26 | 0.11 | 0.31 | 0.64 | 0.29 | 0.15 | 0.45 |
| HD 28571 | S | 5736 | 4.22 | 1.07 | -0.09 | 0.11 | 0.06 | -0.05 | -0.14 | -0.04 | -0.03 | 0.02 | -0.03 | -0.15 | -0.31 | -0.23 | -0.04 | -0.16 | -0.10 | -0.04 | 0.17 | -0.09 | -0.05 | -0.34 | -0.06 | -0.01 | 0.19 | 0.39 | 0.28 |
| HD 28635 | S | 6140 | 4.35 | 1.54 | 0.22 | 0.08 | 0.15 | 0.15 | 0.18 | 0.21 | 0.17 | 0.14 | 0.17 | 0.17 | 0.10 | 0.10 | 0.15 | 0.12 | -0.03 | 0.03 | 0.54 | 0.29 | 0.27 | 0.08 | 0.58 | 0.47 | 0.20 | 0.27 | 0.60 |
| HD 29587 | S | 5682 | 4.48 | 0.75 | -0.48 | -0.18 | -0.22 | -0.30 | -0.16 | -0.33 | -0.33 | -0.28 | -0.39 | -0.48 | -0.61 | -0.53 | -0.33 | -0.48 | -0.54 | -0.36 | | -0.45 | | -0.53 | | -0.43 | -0.12 | -0.04 | 0.05 |
| HD 29645 | S | 6002 | 4.02 | 1.63 | 0.40 | 0.22 | 0.21 | 0.19 | 0.27 | 0.18 | 0.22 | 0.18 | 0.14 | 0.19 | 0.14 | 0.10 | 0.16 | 0.16 | 0.14 | 0.09 | 0.60 | 0.16 | 0.19 | -0.01 | 0.11 | 0.33 | 0.17 | 0.08 | 0.13 |
| HD 29697 | S | 4421 | 4.64 | 0.97 | -0.03 | 0.39 | 0.00 | 0.57 | 1.51 | 0.18 | 0.24 | 0.14 | 0.32 | 0.32 | 0.19 | 0.23 | 0.37 | 0.32 | 0.56 | 1.31 | | 0.30 | -0.29 | 0.53 | 0.90 | 0.82 | 1.60 | 1.57 | |
| HD 31527 | S | 5908 | 4.36 | 1.08 | -0.13 | -0.14 | -0.14 | -0.13 | -0.11 | -0.11 | -0.08 | -0.09 | -0.15 | -0.11 | -0.22 | -0.17 | -0.10 | -0.17 | -0.23 | -0.19 | 0.41 | -0.02 | -0.14 | -0.16 | 0.22 | 0.02 | 0.06 | -0.13 | 0.42 |
| HD 31949 | S | 6213 | 4.30 | 1.78 | -0.05 | 0.03 | -0.01 | 0.04 | 0.11 | 0.03 | -0.02 | -0.01 | 0.01 | 0.05 | -0.14 | -0.06 | 0.06 | -0.10 | -0.32 | -0.07 | | 0.15 | 0.23 | 0.17 | -0.07 | 0.41 | 0.15 | -0.13 | 0.18 |
| HD 32147 | S | 4790 | 4.57 | 0.15 | 0.66 | 0.28 | 0.63 | 0.46 | 1.24 | 0.39 | 0.58 | 0.57 | 0.85 | 0.51 | 0.50 | 0.39 | 0.57 | 0.51 | 0.52 | 0.64 | 0.54 | 0.48 | 0.07 | 0.23 | 0.67 | 1.57 | 1.55 | 1.53 | 0.37 |
| HD 32715 | S | 6615 | 4.25 | 2.45 | 0.52 | | | 0.38 | 1.26 | 0.33 | 0.30 | 0.80 | 0.86 | 0.53 | 0.68 | 0.17 | 0.91 | 0.29 | -0.31 | | 2.07 | 1.09 | 1.29 | 0.41 | | | 0.49 | 0.87 | |
| HD 332612 | S | 6124 | 4.10 | 1.74 | 0.02 | 0.11 | 0.08 | 0.07 | 0.04 | 0.15 | 0.09 | 0.14 | 0.06 | 0.07 | -0.06 | 0.00 | 0.04 | 0.01 | -0.24 | 0.07 | | 0.16 | 0.24 | 0.09 | | 0.34 | 0.22 | -0.01 | 0.05 |
| HD 334372 | S | 5765 | 4.01 | 1.32 | -0.11 | 0.05 | 0.12 | 0.01 | 0.07 | 0.06 | 0.07 | 0.03 | -0.03 | -0.01 | -0.15 | -0.08 | 0.00 | -0.09 | -0.11 | -0.01 | 0.38 | 0.06 | 0.09 | -0.12 | -0.03 | 0.40 | 0.07 | -0.13 | 0.29 |
| HD 33636 | S | 5953 | 4.45 | 1.11 | -0.12 | -0.08 | -0.06 | -0.04 | -0.02 | 0.03 | 0.06 | 0.01 | -0.07 | -0.05 | -0.10 | -0.09 | -0.03 | -0.11 | -0.22 | -0.03 | 0.39 | 0.06 | 0.08 | 0.11 | 0.27 | 0.55 | 0.24 | 0.11 | 0.37 |
| HD 33866 | S | 5625 | 4.33 | 1.12 | -0.08 | 0.00 | -0.06 | 0.01 | 0.24 | 0.01 | -0.06 | -0.12 | -0.11 | -0.05 | -0.17 | -0.12 | -0.08 | -0.10 | -0.23 | 0.07 | 0.27 | 0.06 | -0.05 | -0.09 | 0.40 | 0.12 | 0.13 | 0.41 | 0.43 |
| HD 34721 | S | 6004 | 4.14 | 1.45 | 0.02 | -0.07 | 0.01 | -0.01 | -0.01 | 0.05 | 0.07 | -0.02 | -0.04 | -0.03 | -0.11 | -0.08 | -0.03 | -0.07 | -0.16 | -0.12 | 0.36 | -0.06 | -0.06 | -0.06 | 0.04 | 0.16 | 0.12 | 0.25 | 0.20 |
| HD 3556 | S | 6019 | 4.42 | 1.43 | 0.11 | 0.21 | 0.20 | 0.16 | 0.13 | 0.21 | 0.27 | 0.21 | 0.19 | 0.20 | 0.14 | 0.13 | 0.21 | 0.14 | -0.20 | 0.07 | 0.59 | 0.28 | 0.36 | 0.27 | 0.22 | 0.70 | 0.44 | 0.64 | 0.49 |
| HD 35961 | S | 5740 | 4.37 | 0.90 | -0.22 | -0.07 | -0.10 | -0.13 | 0.10 | -0.11 | -0.08 | -0.18 | -0.18 | -0.16 | -0.29 | -0.23 | -0.17 | -0.22 | -0.26 | -0.08 | 0.27 | -0.08 | 0.16 | -0.11 | 0.13 | 0.25 | 0.04 | -0.03 | 0.18 |
| HD 36003 | S | 4536 | 4.60 | 0.73 | 0.30 | 0.17 | 0.19 | 0.34 | 0.95 | 0.42 | 0.19 | 0.17 | 0.34 | 0.22 | 0.08 | 0.10 | 0.26 | 0.12 | 0.24 | 0.59 | 0.37 | -0.05 | -0.58 | -0.20 | 0.75 | 1.31 | 1.43 | 2.07 | 0.32 |
| HD 37124 | S | 5561 | 4.39 | 0.70 | -0.29 | -0.07 | -0.04 | -0.20 | -0.16 | -0.15 | -0.14 | -0.11 | -0.19 | -0.34 | -0.49 | -0.41 | -0.21 | -0.35 | -0.37 | -0.23 | -0.03 | -0.27 | -0.48 | -0.50 | | 0.25 | -0.05 | 0.31 | 0.18 |
| HD 37605 | S | 5318 | 4.47 | 0.69 | 0.52 | 0.42 | 0.38 | 0.41 | 0.70 | 0.34 | 0.40 | 0.40 | 0.48 | 0.43 | 0.59 | 0.31 | 0.40 | 0.45 | 0.65 | 0.66 | 0.34 | 0.51 | 0.17 | 0.26 | 0.76 | 0.61 | 0.65 | 1.78 | 0.65 |
| HD 3795 | S | 5456 | 3.89 | 1.02 | -0.38 | -0.14 | -0.20 | -0.30 | -0.23 | -0.23 | -0.36 | -0.21 | -0.32 | -0.44 | -0.68 | -0.54 | -0.36 | -0.49 | -0.52 | -0.38 | -0.02 | -0.43 | -0.35 | -0.46 | -0.20 | 0.05 | -0.17 | -0.11 | 0.03 |
| HD 38529 | S | 5450 | 3.72 | 1.56 | 0.58 | 0.41 | 0.43 | 0.41 | 0.72 | 0.36 | 0.28 | 0.28 | 0.31 | 0.33 | 0.40 | 0.25 | 0.33 | 0.34 | 0.50 | 0.26 | 0.56 | 0.29 | 0.13 | 0.22 | 0.36 | 0.32 | 0.36 | 0.77 | 0.48 |
| HD 38700 | S | 6049 | 4.45 | 1.57 | -0.04 | -0.13 | 0.08 | 0.04 | 0.10 | 0.08 | 0.14 | 0.05 | 0.08 | 0.07 | -0.04 | -0.02 | 0.01 | -0.07 | -0.27 | -0.05 | 0.32 | 0.16 | 0.22 | 0.29 | 0.27 | 0.44 | 0.32 | 0.42 | 0.55 |
| HD 38858 | S | 5798 | 4.48 | 1.01 | -0.17 | -0.16 | -0.13 | -0.15 | -0.10 | -0.11 | -0.08 | -0.11 | -0.13 | -0.14 | -0.21 | -0.19 | -0.10 | -0.20 | -0.24 | -0.18 | 0.19 | -0.11 | -0.09 | -0.18 | | 0.19 | 0.17 | 0.15 | 0.18 |
| HD 38A | S | 4065 | 4.71 | 0.15 | 0.42 | -0.07 | 0.43 | 1.12 | 2.48 | 0.95 | 0.70 | 0.31 | 0.54 | 0.72 | 0.29 | 0.19 | 0.82 | 0.69 | 1.05 | | 1.18 | 0.27 | -0.15 | 0.01 | 1.68 | 2.15 | 2.36 | 2.25 | 2.17 |
| HD 38B | S | 4028 | 4.72 | 0.15 | 0.33 | -0.15 | 0.40 | 1.40 | 2.70 | 1.16 | 0.70 | 0.32 | 0.58 | 0.77 | 0.32 | 0.17 | 0.97 | 0.76 | 1.33 | 1.82 | 1.06 | 0.70 | -0.10 | 0.09 | 1.69 | 2.08 | 2.32 | 2.62 | 0.86 |
| HD 39881 | S | 5719 | 4.27 | 1.07 | -0.03 | 0.11 | 0.10 | 0.01 | 0.06 | 0.04 | 0.09 | 0.05 | -0.03 | -0.08 | -0.19 | -0.14 | -0.02 | -0.10 | 0.01 | -0.09 | 0.36 | 0.03 | -0.15 | -0.11 | -0.03 | 0.24 | 0.17 | 0.40 | 0.29 |
| HD 400 | S | 6240 | 4.13 | 1.64 | -0.12 | -0.09 | -0.11 | -0.11 | -0.04 | -0.07 | -0.11 | -0.10 | -0.08 | -0.15 | -0.26 | -0.19 | -0.05 | -0.21 | -0.46 | -0.24 | | -0.15 | -0.17 | -0.10 | -0.10 | 0.30 | -0.06 | 0.18 | 0.13 |
| HD 40979 | S | 6150 | 4.35 | 1.63 | 0.37 | 0.11 | 0.24 | 0.26 | 0.37 | 0.27 | 0.35 | 0.31 | 0.28 | 0.27 | 0.24 | 0.20 | 0.23 | 0.23 | 0.07 | 0.19 | 0.81 | 0.31 | 0.44 | 0.22 | | 0.49 | 0.31 | 0.15 | 0.47 |
| HD 41330 | S | 5876 | 4.13 | 1.28 | -0.16 | 0.03 | -0.10 | -0.09 | -0.06 | -0.04 | -0.07 | -0.08 | -0.13 | -0.13 | -0.19 | -0.17 | -0.10 | -0.19 | -0.27 | -0.16 | 0.26 | -0.13 | -0.16 | -0.19 | 0.19 | -0.04 | -0.03 | -0.03 | 0.15 |
| HD 41708 | S | 5867 | 4.47 | 1.03 | 0.06 | -0.05 | 0.10 | 0.08 | 0.14 | 0.10 | 0.15 | 0.08 | 0.04 | 0.14 | 0.02 | 0.05 | 0.09 | 0.07 | -0.07 | 0.01 | 0.44 | 0.26 | 0.33 | 0.14 | 0.01 | 0.59 | 0.38 | 0.33 | 0.40 |
| HD 4203 | S | 5471 | 4.10 | 1.21 | 0.63 | 0.40 | 0.51 | 0.46 | 0.65 | 0.38 | 0.47 | 0.38 | 0.35 | 0.41 | 0.52 | 0.34 | 0.39 | 0.44 | 0.60 | 0.48 | 0.62 | 0.41 | 0.24 | 0.21 | 0.43 | 0.56 | 0.55 | 1.16 | 0.64 |

| Star | | Teff | log g | | | | | | | | | | | | | | | | | | | | | | | | | |
|---|---|---|---|---|---|---|---|---|---|---|---|---|---|---|---|---|---|---|---|---|---|---|---|---|---|---|---|---|
| HD 4208 | S | 5674 | 4.47 | 0.94 | -0.22 | -0.11 | -0.08 | -0.15 | -0.11 | -0.10 | -0.11 | -0.08 | -0.17 | -0.15 | -0.22 | -0.23 | -0.11 | -0.21 | -0.16 | -0.18 | 0.36 | -0.10 | 0.05 | -0.27 | -0.08 | 0.16 | 0.23 | 0.49 | 0.38 |
| HD 43162 | S | 5651 | 4.50 | 1.51 | 0.04 | 0.10 | 0.03 | 0.05 | 0.27 | 0.14 | 0.00 | 0.10 | 0.07 | 0.10 | 0.02 | 0.02 | 0.06 | 0.00 | -0.15 | -0.15 | 0.61 | 0.13 | 0.18 | 0.19 | 0.38 | 0.38 | 0.46 | 0.54 | 0.42 |
| HD 43587 | S | 5859 | 4.27 | 1.24 | 0.01 | 0.01 | 0.03 | -0.01 | 0.09 | 0.02 | 0.06 | -0.03 | -0.06 | -0.02 | -0.07 | -0.07 | -0.01 | -0.04 | -0.12 | -0.11 | 0.44 | 0.08 | -0.07 | -0.08 | 0.12 | 0.38 | 0.06 | 0.19 | 0.32 |
| HD 43745 | S | 6087 | 3.92 | 1.67 | 0.21 | 0.12 | 0.13 | 0.12 | 0.17 | 0.11 | 0.15 | 0.11 | 0.13 | 0.14 | 0.07 | 0.06 | 0.14 | 0.09 | 0.05 | -0.04 | 0.44 | 0.17 | 0.29 | 0.01 | 0.13 | 0.22 | 0.07 | 0.17 | 0.39 |
| HD 45067 | S | 6049 | 3.95 | 1.70 | 0.05 | -0.03 | -0.01 | 0.00 | 0.04 | 0.07 | 0.08 | 0.01 | -0.05 | 0.00 | -0.11 | -0.07 | 0.03 | -0.07 | -0.23 | -0.20 | 0.52 | -0.04 | 0.08 | 0.01 | 0.09 | 0.09 | 0.09 | -0.19 | 0.03 |
| HD 45088 | S | 4778 | 4.37 | 0.15 | -0.36 | -0.34 | -0.22 | 0.03 | 0.50 | -0.21 | -0.18 | -0.09 | -0.31 | 0.00 | -0.14 | -0.21 | -0.04 | -0.04 | 0.05 | 0.41 | -0.10 | 0.11 | -0.40 | -0.19 | 0.66 | 0.73 | 0.92 | 1.59 | 1.01 |
| HD 45184 | S | 5852 | 4.41 | 1.22 | 0.07 | 0.10 | 0.17 | 0.09 | 0.18 | 0.12 | 0.13 | 0.09 | 0.11 | 0.11 | 0.07 | 0.04 | 0.09 | 0.05 | -0.04 | -0.11 | 0.36 | 0.12 | 0.07 | 0.07 | 0.13 | 0.42 | 0.30 | 0.38 | 0.44 |
| HD 45205 | S | 5921 | 4.15 | 1.31 | -0.71 | -0.40 | -0.43 | -0.53 | -0.05 | -0.55 | -0.46 | -0.52 | -0.41 | -0.68 | -0.88 | -0.82 | -0.51 | -0.79 | -0.97 | -0.52 | | -0.64 | 1.11 | -0.84 | | | -0.52 | -0.37 | -0.20 |
| HD 45350 | S | 5567 | 4.22 | 1.33 | 0.37 | 0.37 | 0.39 | 0.32 | 0.43 | 0.35 | 0.30 | 0.31 | 0.31 | 0.32 | 0.31 | 0.23 | 0.29 | 0.27 | 0.22 | 0.16 | 0.44 | 0.34 | 0.17 | 0.09 | 0.57 | 0.50 | 0.52 | 0.91 | 0.65 |
| HD 45588 | S | 6214 | 4.24 | 1.70 | 0.10 | 0.16 | 0.10 | 0.09 | 0.11 | 0.18 | 0.18 | 0.15 | 0.15 | 0.07 | -0.03 | 0.04 | 0.15 | 0.06 | -0.16 | -0.15 | 0.67 | 0.12 | 0.20 | 0.10 | 0.03 | 0.20 | 0.21 | -0.23 | 0.22 |
| HD 45759 | S | 6132 | 4.37 | 2.26 | 0.29 | 0.30 | 0.21 | 0.18 | 0.21 | 0.19 | 0.18 | 0.27 | 0.27 | 0.20 | 0.13 | 0.12 | 0.33 | 0.13 | -0.29 | 0.27 | | 0.36 | 0.45 | 0.21 | 0.67 | 0.45 | 0.39 | 0.53 | 0.54 |
| HD 4628 | S | 5044 | 4.61 | 0.35 | -0.14 | -0.08 | -0.04 | -0.13 | 0.43 | -0.11 | -0.11 | -0.01 | 0.10 | -0.06 | -0.17 | -0.21 | -0.10 | -0.13 | 0.04 | -0.21 | 0.10 | -0.02 | -0.16 | -0.25 | 0.36 | 1.15 | 0.84 | 1.28 | 0.07 |
| HD 46375 | S | 5265 | 4.33 | 1.24 | 0.46 | 0.41 | 0.39 | 0.33 | 0.68 | 0.30 | 0.32 | 0.34 | 0.42 | 0.31 | 0.48 | 0.21 | 0.34 | 0.33 | 0.53 | 0.38 | 0.24 | 0.40 | 0.11 | -0.01 | 0.77 | 0.63 | 0.73 | 1.58 | 0.52 |
| HD 48938 | S | 6055 | 4.33 | 1.27 | -0.34 | -0.24 | -0.27 | -0.28 | -0.40 | -0.27 | -0.20 | -0.25 | -0.27 | -0.35 | -0.52 | -0.38 | -0.29 | -0.38 | -0.50 | -0.31 | | -0.42 | -0.15 | -0.29 | | -0.16 | -0.10 | | 0.32 |
| HD 49674 | S | 5621 | 4.39 | 1.37 | 0.36 | 0.25 | 0.40 | 0.29 | 0.30 | 0.33 | 0.28 | 0.30 | 0.32 | 0.34 | 0.38 | 0.23 | 0.31 | 0.29 | 0.30 | 0.11 | 0.59 | 0.33 | 0.25 | 0.19 | 0.60 | 0.72 | 0.46 | 1.00 | 0.65 |
| HD 50281 | S | 4708 | 4.64 | 0.15 | 0.26 | 0.07 | 0.03 | 0.20 | 0.89 | 0.26 | 0.15 | 0.18 | 0.37 | 0.23 | 0.19 | 0.09 | 0.17 | 0.18 | 0.22 | 0.47 | 0.30 | 0.28 | -0.09 | 0.19 | 1.01 | 1.67 | 1.45 | 2.02 | 0.30 |
| HD 50554 | S | 6019 | 4.41 | 1.21 | 0.04 | -0.05 | 0.04 | 0.04 | 0.03 | 0.09 | 0.09 | 0.07 | 0.04 | 0.04 | -0.05 | -0.01 | 0.01 | -0.02 | -0.18 | -0.11 | 0.57 | 0.24 | 0.19 | 0.11 | 0.20 | 0.30 | 0.25 | 0.13 | 0.38 |
| HD 50806 | S | 5640 | 4.06 | 1.16 | 0.17 | 0.23 | 0.29 | 0.16 | 0.25 | 0.18 | 0.20 | 0.19 | 0.16 | 0.12 | 0.07 | 0.04 | 0.18 | 0.11 | 0.39 | 0.04 | 0.43 | 0.15 | 0.03 | -0.05 | 0.19 | 0.35 | 0.33 | 0.60 | 0.37 |
| HD 52265 | S | 6086 | 4.29 | 1.51 | 0.41 | 0.15 | 0.25 | 0.25 | 0.35 | 0.30 | 0.30 | 0.24 | 0.23 | 0.30 | 0.29 | 0.18 | 0.24 | 0.23 | 0.12 | 0.09 | 0.67 | 0.35 | 0.38 | 0.14 | 0.27 | 0.43 | 0.26 | 0.88 | 0.47 |
| HD 52698 | S | 5155 | 4.55 | 0.98 | 0.21 | 0.25 | 0.22 | 0.27 | 0.63 | 0.24 | 0.29 | 0.28 | 0.33 | 0.30 | 0.36 | 0.21 | 0.27 | 0.29 | 0.43 | 0.32 | 0.26 | 0.34 | 0.08 | 0.28 | 0.82 | 0.69 | 0.96 | 1.51 | 0.54 |
| HD 52711 | S | 5992 | 4.41 | 1.26 | 0.00 | -0.04 | 0.02 | -0.05 | -0.01 | 0.04 | 0.03 | 0.04 | -0.01 | -0.01 | -0.08 | -0.07 | 0.02 | -0.05 | -0.08 | -0.06 | 0.19 | 0.09 | 0.09 | -0.05 | 0.25 | 0.09 | 0.25 | -0.13 | 0.38 |
| HD 5494 | S | 6044 | 4.03 | 1.67 | 0.03 | 0.08 | -0.03 | 0.05 | 0.27 | 0.04 | 0.13 | 0.05 | 0.05 | 0.08 | -0.11 | -0.08 | 0.22 | -0.05 | -0.25 | 0.01 | | 0.20 | 0.56 | 0.10 | 0.19 | 0.23 | 0.20 | 1.01 | 0.30 |
| HD 55054 | S | 6219 | 4.14 | 1.70 | 0.20 | -0.09 | 0.27 | -0.01 | -0.09 | 0.01 | 0.07 | 0.06 | 0.03 | 0.07 | -0.09 | -0.05 | 0.08 | -0.02 | -0.15 | | 0.57 | 0.02 | 0.39 | 0.01 | 0.13 | 0.46 | 0.14 | -0.26 | 0.31 |
| HD 55575 | S | 5937 | 4.32 | 1.26 | -0.33 | -0.16 | -0.18 | -0.22 | -0.17 | -0.18 | -0.19 | -0.19 | -0.27 | -0.26 | -0.39 | -0.32 | -0.20 | -0.32 | -0.38 | -0.35 | | -0.34 | -0.29 | -0.36 | 0.07 | -0.31 | 0.01 | -0.10 | 0.17 |
| HD 55693 | S | 5854 | 4.32 | 1.32 | 0.49 | 0.25 | 0.36 | 0.31 | 0.38 | 0.32 | 0.34 | 0.29 | 0.31 | 0.32 | 0.33 | 0.26 | 0.32 | 0.33 | 0.38 | 0.22 | 0.57 | 0.37 | 0.46 | 0.06 | 0.66 | 0.31 | 0.45 | 0.61 | 0.62 |
| HD 57006 | S | 6264 | 3.77 | 2.10 | 0.13 | 0.07 | 0.08 | 0.09 | 0.09 | 0.17 | 0.13 | 0.14 | 0.12 | 0.10 | 0.02 | 0.04 | 0.21 | 0.02 | -0.15 | -0.04 | 0.94 | 0.10 | 0.18 | 0.22 | 0.18 | 0.29 | 0.13 | -0.25 | 0.07 |
| HD 603 | S | 5967 | 4.37 | 1.15 | -0.17 | -0.05 | -0.17 | -0.12 | -0.05 | -0.05 | -0.10 | -0.07 | 0.00 | -0.15 | -0.19 | -0.17 | -0.01 | -0.18 | -0.37 | -0.14 | 0.68 | 0.03 | 0.07 | 0.06 | 0.23 | 0.02 | 0.19 | 0.27 | 0.60 |
| HD 6064 | S | 6363 | 3.78 | 2.03 | 0.39 | -0.03 | 0.28 | 0.21 | 0.12 | 0.25 | 0.31 | 0.27 | 0.33 | 0.15 | 0.05 | 0.13 | 0.30 | 0.13 | -0.16 | 0.13 | | 0.39 | 0.40 | 0.26 | | 0.37 | 0.15 | 0.39 | 0.20 |
| HD 61606 | S | 4932 | 4.61 | 1.00 | 0.14 | -0.06 | 0.05 | 0.11 | 0.59 | 0.10 | 0.03 | 0.10 | 0.19 | 0.16 | 0.14 | 0.01 | 0.08 | 0.05 | 0.32 | 0.04 | 0.25 | 0.11 | -0.19 | 0.08 | 0.83 | 1.12 | 0.85 | 1.63 | |
| HD 61632 | S | 5632 | 4.02 | 1.09 | -0.41 | -0.21 | -0.23 | -0.27 | -0.06 | -0.26 | -0.28 | -0.24 | -0.34 | -0.48 | -0.67 | -0.54 | -0.37 | -0.49 | -0.58 | -0.34 | | -0.46 | 0.13 | -0.51 | | -0.38 | -0.27 | -0.35 | -0.02 |
| HD 63598 | S | 5828 | 4.38 | 0.93 | -0.76 | -0.41 | -0.62 | -0.52 | -0.15 | -0.49 | -0.61 | -0.51 | -0.65 | -0.68 | -1.18 | -0.84 | -0.50 | -0.80 | -1.01 | -0.57 | | -0.69 | -0.28 | -0.75 | 0.20 | -0.01 | -0.43 | 0.05 | -0.17 |
| HD 63754 | S | 6040 | 3.92 | 1.78 | 0.43 | 0.14 | 0.16 | 0.22 | 0.31 | 0.23 | 0.31 | 0.18 | 0.20 | 0.19 | 0.14 | 0.14 | 0.25 | 0.19 | 0.12 | 0.15 | 0.84 | 0.28 | 0.37 | 0.09 | 0.22 | 0.39 | 0.23 | 0.30 | 0.26 |
| HD 68988 | S | 5878 | 4.39 | 1.37 | 0.71 | 0.37 | 0.43 | 0.37 | 0.56 | 0.34 | 0.44 | 0.35 | 0.32 | 0.38 | 0.48 | 0.29 | 0.37 | 0.39 | 0.44 | 0.31 | 0.83 | 0.40 | 0.43 | 0.12 | 0.67 | 0.62 | 0.36 | 0.88 | 0.72 |
| HD 69830 | S | 5443 | 4.53 | 0.92 | 0.07 | 0.10 | 0.12 | 0.04 | 0.18 | 0.02 | 0.06 | 0.10 | 0.12 | 0.06 | 0.03 | -0.02 | 0.09 | 0.03 | 0.07 | -0.06 | 0.25 | 0.09 | -0.10 | -0.03 | 0.51 | 0.28 | 0.30 | 1.07 | 0.32 |
| HD 70889 | S | 6036 | 4.43 | 1.25 | 0.10 | 0.18 | 0.11 | 0.11 | 0.15 | 0.19 | 0.15 | 0.12 | 0.10 | 0.18 | 0.13 | 0.10 | 0.14 | 0.09 | -0.01 | 0.02 | 0.55 | 0.29 | 0.21 | 0.17 | 0.30 | 0.43 | 0.42 | 0.30 | 0.45 |
| HD 7091 | S | 6109 | 4.29 | 1.16 | -0.26 | -0.22 | -0.25 | -0.21 | -0.17 | -0.08 | -0.01 | -0.11 | -0.09 | -0.11 | -0.45 | -0.25 | 0.06 | -0.27 | | -0.18 | | -0.09 | 1.04 | 0.06 | 0.22 | | | 0.32 | 0.37 |
| HD 71148 | S | 5835 | 4.39 | 1.17 | 0.07 | 0.05 | 0.13 | 0.05 | 0.12 | 0.09 | 0.09 | 0.07 | 0.04 | 0.08 | 0.01 | 0.02 | 0.07 | 0.03 | -0.03 | 0.04 | 0.40 | 0.17 | 0.03 | 0.02 | 0.24 | 0.31 | 0.26 | 0.41 | 0.60 |
| HD 71881 | S | 5863 | 4.30 | 1.14 | 0.07 | 0.04 | 0.04 | 0.03 | 0.03 | 0.03 | 0.08 | 0.02 | -0.02 | 0.01 | -0.01 | -0.04 | 0.06 | 0.00 | -0.04 | 0.02 | 0.49 | 0.13 | -0.09 | -0.03 | 0.27 | 0.04 | 0.23 | 0.24 | 0.37 |

| Name | | Teff | logg | | | | | | | | | | | | | | | | | | | | | | | | | | |
|---|---|---|---|---|---|---|---|---|---|---|---|---|---|---|---|---|---|---|---|---|---|---|---|---|---|---|---|---|---|
| HD 7230 | S | 6637 | 4.17 | 3.73 | 0.32 | | | 0.36 | 0.36 | 0.49 | 0.82 | 0.79 | 1.29 | 1.09 | 0.26 | 0.32 | 1.35 | 0.69 | | | 1.31 | 2.21 | 0.01 | | | 1.48 | | | |
| HD 72659 | S | 5902 | 4.12 | 1.31 | 0.08 | 0.07 | 0.03 | 0.03 | 0.08 | 0.04 | 0.10 | 0.01 | 0.04 | 0.01 | -0.06 | -0.05 | 0.03 | -0.02 | -0.02 | -0.07 | 0.34 | 0.07 | 0.11 | -0.10 | 0.13 | 0.05 | 0.21 | 0.08 | 0.32 |
| HD 72946 | S | 5674 | 4.49 | 1.23 | 0.14 | 0.20 | 0.23 | 0.13 | 0.21 | 0.21 | 0.11 | 0.17 | 0.20 | 0.17 | 0.17 | 0.10 | 0.17 | 0.11 | 0.04 | 0.04 | 0.53 | 0.25 | 0.25 | 0.13 | 0.39 | 0.44 | 0.44 | 0.82 | 0.50 |
| HD 7352 | S | 6023 | 4.31 | 1.30 | 0.17 | 0.06 | 0.15 | 0.13 | 0.26 | 0.17 | 0.24 | 0.13 | 0.07 | 0.14 | 0.04 | 0.08 | 0.12 | 0.09 | 0.02 | 0.11 | 0.69 | 0.21 | 0.24 | 0.19 | 0.39 | 0.30 | 0.07 | 0.34 | 0.29 |
| HD 73596 | S | 6744 | 3.41 | 2.68 | 0.40 | -0.20 | 1.21 | 0.22 | 0.26 | 0.37 | 0.19 | 0.14 | 0.77 | 0.26 | 0.40 | 0.00 | 1.21 | 0.24 | -0.35 | | 0.01 | | -0.16 | | 0.66 | 1.55 | | 0.02 | | |
| HD 73668 | S | 5922 | 4.37 | 1.24 | 0.01 | 0.05 | 0.06 | 0.05 | 0.06 | 0.09 | 0.11 | 0.01 | -0.01 | 0.04 | -0.06 | -0.02 | -0.02 | -0.02 | -0.14 | -0.07 | 0.49 | 0.19 | 0.02 | 0.13 | -0.02 | 0.33 | 0.13 | 0.21 | 0.50 |
| HD 73752 | S | 5654 | 3.98 | 1.51 | 0.60 | 0.40 | 0.53 | 0.36 | 0.53 | 0.33 | 0.33 | 0.28 | 0.30 | 0.36 | 0.40 | 0.21 | 0.37 | 0.32 | 0.20 | 0.40 | 0.65 | 0.44 | 0.25 | -0.09 | 0.46 | 0.50 | 0.33 | 1.07 | 0.65 |
| HD 7397 | S | 6029 | 4.15 | 1.37 | -0.07 | -0.04 | -0.03 | -0.08 | -0.31 | -0.01 | -0.05 | 0.02 | -0.04 | 0.04 | -0.28 | -0.12 | 0.06 | -0.15 | | -0.18 | | 0.27 | 0.65 | -0.01 | 0.17 | | 0.17 | 0.79 | 0.44 |
| HD 74156 | S | 6005 | 4.06 | 1.62 | 0.26 | 0.05 | 0.16 | 0.13 | 0.20 | 0.16 | 0.11 | 0.11 | 0.07 | 0.12 | 0.09 | 0.05 | 0.11 | 0.10 | 0.03 | 0.07 | 0.75 | 0.15 | 0.11 | -0.07 | 0.02 | 0.02 | 0.14 | 0.07 | 0.23 |
| HD 7514 | S | 5682 | 4.24 | 1.22 | -0.12 | 0.13 | 0.04 | -0.04 | 0.00 | 0.00 | 0.02 | 0.02 | -0.05 | -0.07 | -0.13 | -0.17 | -0.03 | -0.13 | -0.12 | 0.02 | 0.28 | 0.09 | 0.27 | -0.44 | 0.39 | 0.47 | 0.26 | 0.50 | 0.49 |
| HD 75488 | S | 5996 | 4.18 | 1.36 | -0.41 | -0.34 | -0.45 | -0.32 | -0.30 | -0.31 | -0.33 | -0.35 | -0.32 | -0.40 | -0.55 | -0.43 | -0.33 | -0.44 | -0.54 | -0.43 | | -0.51 | | -0.45 | | -0.15 | -0.33 | -0.18 | -0.11 | |
| HD 76151 | S | 5780 | 4.43 | 1.10 | 0.22 | 0.18 | 0.27 | 0.14 | 0.29 | 0.17 | 0.14 | 0.15 | 0.15 | 0.15 | 0.16 | 0.10 | 0.13 | 0.14 | 0.17 | -0.02 | 0.26 | 0.27 | 0.19 | 0.08 | 0.21 | 0.34 | 0.30 | 0.57 | 0.37 |
| HD 78366 | S | 5981 | 4.42 | 1.41 | -0.01 | 0.03 | 0.06 | 0.05 | 0.10 | 0.14 | 0.10 | 0.09 | 0.06 | 0.09 | 0.01 | 0.03 | 0.03 | -0.01 | -0.17 | -0.10 | 0.52 | 0.17 | 0.14 | 0.22 | 0.27 | 0.30 | 0.40 | 0.31 | 0.18 |
| HD 79210 | S | 3973 | 4.69 | 0.15 | 0.33 | 0.58 | 0.37 | 1.23 | 2.47 | 1.18 | 0.64 | 0.27 | 0.45 | 0.63 | 0.26 | 0.17 | 0.92 | 0.67 | 0.90 | | 0.80 | 0.60 | -0.25 | 0.18 | 1.79 | 2.04 | 2.54 | 2.98 | 1.44 |
| HD 79211 | S | 3763 | 4.49 | 0.15 | 0.28 | 0.85 | 0.45 | 1.75 | 3.52 | 1.43 | 0.59 | 0.21 | 0.46 | 0.82 | 0.27 | 0.34 | 0.71 | 0.84 | 1.27 | | 0.71 | 0.39 | -0.63 | 0.01 | 1.76 | 2.13 | 2.38 | 2.74 | 1.46 |
| HD 80372 | S | 6039 | 4.36 | 1.43 | 0.19 | 0.04 | 0.04 | 0.10 | 0.21 | 0.14 | 0.18 | 0.12 | 0.07 | 0.12 | 0.02 | 0.03 | 0.11 | 0.06 | -0.04 | 0.10 | 0.53 | 0.15 | 0.21 | 0.11 | 0.73 | 0.53 | 0.20 | -0.05 | 0.16 |
| HD 80606 | S | 5467 | 3.52 | 1.50 | 0.77 | 0.50 | 0.44 | 0.34 | 0.42 | 0.45 | 0.19 | 0.27 | 0.28 | 0.31 | 0.46 | 0.21 | 0.28 | 0.31 | 0.41 | 0.19 | 0.33 | 0.18 | 0.16 | -0.29 | 0.17 | 0.29 | 0.10 | 0.78 | 0.58 |
| HD 80607 | S | 5535 | 3.35 | 1.45 | 0.98 | 0.56 | 0.51 | 0.33 | 0.27 | 0.52 | 0.13 | 0.38 | 0.46 | 0.41 | 0.61 | 0.26 | 0.38 | 0.37 | 0.66 | 0.06 | 0.53 | 0.21 | 0.13 | -0.30 | 0.04 | 0.30 | 0.14 | 0.70 | 0.62 |
| HD 81040 | S | 5753 | 4.48 | 1.17 | -0.13 | -0.10 | -0.06 | -0.05 | -0.05 | 0.02 | -0.02 | -0.03 | -0.09 | -0.02 | -0.08 | -0.09 | -0.08 | -0.12 | -0.19 | -0.19 | 0.53 | 0.10 | -0.22 | 0.07 | 0.22 | 0.39 | 0.35 | 0.38 | 0.27 |
| HD 8173 | S | 6009 | 4.40 | 1.14 | 0.04 | 0.04 | 0.00 | 0.00 | -0.07 | 0.06 | 0.08 | 0.06 | 0.08 | 0.04 | 0.00 | -0.01 | 0.10 | 0.01 | -0.07 | -0.08 | 0.42 | 0.20 | 0.24 | 0.12 | 0.17 | 0.48 | 0.25 | 0.19 | 0.56 |
| HD 81809 | S | 5666 | 3.72 | 1.27 | -0.16 | 0.00 | -0.01 | -0.08 | -0.19 | -0.03 | -0.11 | -0.09 | -0.18 | -0.28 | -0.43 | -0.34 | -0.18 | -0.28 | -0.43 | -0.28 | | -0.28 | -0.18 | -0.41 | -0.28 | -0.66 | -0.15 | -0.08 | 0.12 | |
| HD 82106 | S | 4826 | 4.63 | 0.86 | 0.19 | 0.10 | 0.11 | 0.15 | 0.88 | 0.22 | 0.13 | 0.16 | 0.31 | 0.17 | 0.23 | 0.07 | 0.09 | 0.12 | | 0.30 | 0.31 | 0.27 | -0.13 | 0.14 | 0.51 | 1.05 | 1.23 | 1.69 | 0.19 | |
| HD 82943 | S | 5919 | 4.35 | 1.36 | 0.38 | 0.24 | 0.26 | 0.29 | 0.38 | 0.27 | 0.32 | 0.23 | 0.22 | 0.27 | 0.26 | 0.20 | 0.23 | 0.25 | 0.21 | 0.27 | 0.66 | 0.33 | 0.31 | 0.14 | 0.19 | 0.21 | 0.28 | 0.52 | 0.61 |
| HD 84117 | S | 6239 | 4.34 | 1.54 | 0.00 | -0.01 | -0.02 | 0.02 | 0.01 | 0.03 | 0.14 | 0.04 | -0.01 | 0.02 | -0.05 | -0.02 | 0.08 | -0.02 | -0.18 | -0.15 | 0.69 | 0.07 | 0.17 | 0.02 | | -0.05 | 0.11 | 0.00 | 0.13 | |
| HD 84703 | S | 6125 | 4.14 | 1.60 | 0.16 | 0.05 | 0.09 | 0.10 | 0.03 | 0.10 | 0.15 | 0.08 | 0.06 | 0.08 | 0.02 | 0.02 | 0.08 | 0.04 | -0.06 | 0.12 | 0.86 | 0.15 | 0.09 | 0.03 | -0.11 | 0.27 | 0.23 | 0.17 | 0.59 |
| HD 85725 | S | 5892 | 3.72 | 1.84 | 0.31 | 0.18 | 0.20 | 0.20 | 0.19 | 0.24 | 0.23 | 0.19 | 0.14 | 0.19 | 0.12 | 0.12 | 0.18 | 0.15 | 0.01 | 0.08 | 0.76 | 0.22 | 0.42 | 0.17 | 0.31 | 0.38 | 0.25 | 0.32 | 0.30 |
| HD 8574 | S | 6045 | 4.19 | 1.43 | 0.12 | 0.07 | 0.09 | 0.05 | 0.15 | 0.06 | 0.12 | 0.06 | 0.04 | 0.02 | -0.07 | -0.02 | 0.07 | 0.00 | -0.16 | -0.02 | 0.40 | 0.13 | 0.16 | 0.03 | 0.05 | 0.33 | 0.18 | -0.04 | 0.17 |
| HD 8673 | S | 6409 | 4.25 | 3.13 | 0.91 | 0.08 | 0.55 | 0.38 | 0.34 | 0.25 | 0.43 | 0.43 | 0.50 | 0.47 | 0.14 | 0.17 | 0.52 | 0.28 | -0.16 | | 0.46 | 1.21 | -0.14 | | 0.34 | 0.79 | | | |
| HD 87097 | S | 5993 | 4.40 | 1.88 | 0.01 | 0.20 | 0.20 | 0.08 | 0.16 | 0.19 | 0.21 | 0.13 | 0.05 | 0.17 | -0.01 | 0.03 | 0.09 | -0.01 | -0.19 | -0.02 | | 0.23 | 0.33 | 0.26 | 0.28 | 0.58 | 0.48 | 0.65 | 0.42 |
| HD 88133 | S | 5371 | 3.82 | 1.41 | 0.50 | 0.36 | 0.49 | 0.35 | 0.51 | 0.36 | 0.27 | 0.32 | 0.36 | 0.36 | 0.45 | 0.26 | 0.35 | 0.38 | 0.67 | 0.37 | 0.42 | 0.28 | 0.11 | 0.12 | 0.64 | 0.49 | 0.43 | 1.06 | 0.52 |
| HD 88230 | S | 4215 | 4.70 | 0.15 | 0.56 | 0.55 | 0.36 | 0.75 | 2.12 | 0.87 | 0.72 | 0.36 | 0.61 | 0.62 | 0.39 | 0.21 | 0.59 | 0.62 | | 1.55 | 0.87 | 0.40 | -0.07 | -0.03 | 1.23 | 2.07 | 2.51 | 2.98 | 0.66 | |
| HD 88371 | S | 5673 | 4.27 | 1.08 | -0.19 | 0.03 | 0.07 | -0.07 | -0.14 | -0.04 | -0.06 | -0.01 | -0.08 | -0.18 | -0.38 | -0.31 | -0.09 | -0.23 | -0.23 | -0.17 | | -0.17 | 0.02 | -0.42 | | 0.15 | 0.04 | 0.15 | 0.17 |
| HD 88595 | S | 6322 | 4.12 | 1.91 | 0.24 | -0.05 | 0.15 | 0.15 | 0.16 | 0.16 | 0.23 | 0.13 | 0.10 | 0.14 | 0.06 | 0.06 | 0.19 | 0.09 | -0.06 | 0.02 | | 0.15 | 0.30 | 0.08 | | 0.54 | 0.25 | 0.06 | 0.13 |
| HD 88737 | S | 6129 | 3.73 | 2.20 | 0.56 | 0.33 | 0.37 | 0.30 | 0.36 | 0.32 | 0.26 | 0.31 | 0.22 | 0.31 | 0.24 | 0.21 | 0.27 | 0.25 | -0.03 | 0.08 | | 0.32 | 0.44 | 0.25 | 0.09 | 0.32 | 0.18 | 0.37 | 0.28 |
| HD 89319 | S | 4942 | 3.35 | 1.61 | 0.56 | 0.53 | 0.49 | 0.42 | 1.05 | 0.44 | 0.29 | 0.37 | 0.51 | 0.39 | 0.52 | 0.24 | 0.38 | 0.36 | 0.47 | 0.87 | 0.30 | 0.24 | 0.00 | 0.03 | 0.68 | 0.56 | 0.54 | 1.24 | 0.43 |
| HD 89744 | S | 6207 | 3.92 | 2.11 | 0.45 | 0.28 | 0.33 | 0.25 | 0.32 | 0.32 | 0.34 | 0.27 | 0.25 | 0.25 | 0.18 | 0.17 | 0.30 | 0.22 | -0.08 | 0.10 | | 0.34 | 0.44 | 0.28 | 0.34 | 0.45 | 0.29 | 0.39 | 0.36 |
| HD 90508 | S | 5757 | 4.34 | 1.06 | -0.34 | -0.07 | -0.13 | -0.17 | -0.13 | -0.19 | -0.13 | -0.18 | -0.30 | -0.30 | -0.41 | -0.33 | -0.24 | -0.33 | -0.37 | -0.20 | | -0.20 | -0.27 | -0.34 | | -0.09 | 0.38 | 0.09 | | |
| HD 9224 | S | 5848 | 4.16 | 1.24 | 0.08 | 0.11 | 0.10 | 0.05 | 0.12 | 0.07 | 0.08 | 0.05 | 0.02 | 0.06 | -0.03 | -0.01 | 0.07 | 0.03 | 0.05 | -0.09 | 0.43 | 0.04 | 0.07 | -0.01 | 0.41 | 0.05 | 0.11 | 0.24 | 0.43 |

| Name | | Teff | logg | | | | | | | | | | | | | | | | | | | | | | | | | |
|---|---|---|---|---|---|---|---|---|---|---|---|---|---|---|---|---|---|---|---|---|---|---|---|---|---|---|---|---|
| HD 92788 | S | 5710 | 4.32 | 1.20 | 0.51 | 0.40 | 0.36 | 0.33 | 0.45 | 0.36 | 0.34 | 0.33 | 0.31 | 0.32 | 0.42 | 0.26 | 0.34 | 0.36 | 0.30 | 0.25 | 0.48 | 0.45 | 0.29 | 0.10 | 0.41 | 0.46 | 0.40 | 0.90 | 0.67 |
| HD 9369 | S | 7237 | 4.09 | 3.50 | | | 1.11 | 1.50 | 0.90 | 1.16 | 1.29 | 1.40 | 1.22 | 0.38 | 0.73 | 1.36 | 1.04 | | | 2.70 | 2.35 | 1.26 | | 2.12 | | 3.57 | | |
| HD 94132 | S | 4988 | 3.44 | 1.50 | 0.40 | 0.35 | 0.31 | 0.28 | 0.61 | 0.28 | 0.25 | 0.31 | 0.43 | 0.29 | 0.40 | 0.15 | 0.30 | 0.26 | 0.47 | 0.67 | 0.22 | 0.16 | 0.00 | 0.04 | 0.71 | 0.46 | 0.54 | 0.94 | 0.40 |
| HD 94915 | S | 5985 | 4.41 | 1.33 | -0.02 | 0.00 | -0.09 | -0.05 | -0.03 | -0.02 | 0.03 | -0.03 | -0.04 | -0.04 | -0.17 | -0.10 | -0.04 | -0.10 | -0.19 | -0.01 | 0.41 | 0.01 | -0.03 | 0.03 | | 0.24 | 0.13 | | 0.43 |
| HD 95650 | S | 3655 | 4.56 | 0.25 | 0.17 | -0.07 | 0.81 | 2.15 | 3.44 | 2.05 | 0.99 | 0.39 | 0.64 | 1.04 | 0.42 | 0.53 | 1.08 | 1.17 | 1.68 | 3.96 | 0.93 | 0.84 | 0.00 | 0.11 | 1.98 | 2.18 | 2.68 | | 1.04 |
| HD 96276 | S | 6040 | 4.34 | 1.38 | 0.01 | 0.03 | 0.05 | -0.01 | 0.13 | 0.03 | 0.08 | -0.01 | 0.00 | 0.02 | -0.06 | -0.06 | 0.01 | -0.05 | -0.15 | -0.08 | 0.59 | 0.05 | 0.03 | 0.03 | 0.32 | 0.12 | 0.22 | 0.05 | 0.34 |
| HD 97100 | S | 5008 | 5.00 | 0.15 | -0.64 | -0.14 | -0.14 | 0.20 | 0.81 | -0.46 | 0.41 | -0.07 | -0.17 | -0.26 | -0.51 | -0.16 | 0.04 | -0.01 | 0.06 | 0.61 | 0.02 | 0.25 | -0.06 | 0.18 | 0.67 | 0.73 | 0.86 | 0.78 | 1.09 |
| HD 97101 | S | 4189 | 4.67 | 0.70 | 0.42 | 0.80 | 0.54 | 1.01 | 1.63 | 0.57 | 0.69 | 0.32 | 0.65 | 0.49 | 0.75 | 0.65 | 0.58 | 0.63 | | 2.54 | 0.67 | 0.21 | -0.22 | 0.18 | 1.45 | 1.58 | 1.51 | | 0.86 |
| HD 97334 | S | 5906 | 4.44 | 1.57 | 0.13 | -0.04 | 0.16 | 0.12 | 0.27 | 0.24 | 0.17 | 0.14 | 0.11 | 0.17 | 0.12 | 0.09 | 0.11 | 0.07 | -0.12 | 0.04 | 0.57 | 0.21 | 0.19 | 0.25 | 0.36 | 0.40 | 0.30 | 0.34 | 0.24 |
| HD 975 | S | 6405 | 4.27 | 1.66 | -0.04 | 0.04 | -0.05 | 0.04 | 0.02 | 0.08 | 0.13 | 0.06 | 0.17 | 0.03 | -0.19 | -0.04 | 0.25 | -0.05 | -0.45 | -0.02 | | 0.22 | 0.68 | 0.26 | 0.28 | 0.45 | 0.21 | 0.32 | 0.43 |
| HD 97584 | S | 4732 | 4.62 | 0.91 | 0.17 | 0.01 | 0.04 | 0.12 | 1.02 | 0.19 | 0.08 | 0.11 | 0.28 | 0.17 | 0.13 | 0.02 | 0.10 | 0.07 | 0.07 | 0.11 | 0.48 | 0.16 | -0.34 | 0.03 | 0.69 | 1.27 | 1.06 | 1.92 | 0.35 |
| HD 98388 | S | 6338 | 4.34 | 2.03 | 0.33 | 0.32 | 0.36 | 0.20 | 0.22 | 0.26 | 0.35 | 0.30 | 0.24 | 0.24 | 0.02 | 0.16 | 0.25 | 0.17 | -0.22 | -0.05 | | 0.30 | 0.59 | 0.30 | 0.24 | 0.49 | 0.47 | 0.42 | 0.33 |
| HD 98712 | S | 4300 | 4.72 | 0.15 | 0.34 | -0.11 | 0.15 | 0.77 | 1.99 | 0.69 | 0.36 | 0.27 | 0.54 | 0.50 | 0.28 | 0.12 | 0.61 | 0.38 | 0.62 | 1.30 | 0.99 | 0.59 | -0.11 | 0.28 | 1.60 | 1.78 | 2.29 | 2.86 | 0.59 |
| HD 98823 | S | 6414 | 3.59 | 2.87 | 0.88 | 0.34 | | 0.24 | 0.35 | 0.26 | 0.35 | 0.67 | 0.42 | 0.69 | 0.51 | 0.16 | 1.48 | 0.46 | -0.23 | | 1.90 | 0.83 | 1.92 | | 1.13 | 0.51 | | 0.11 |
| HR 1179 | S | 6292 | 4.17 | 1.77 | -0.05 | 0.09 | -0.21 | -0.06 | -0.04 | -0.01 | 0.07 | 0.01 | -0.03 | -0.07 | -0.26 | -0.13 | -0.02 | -0.12 | -0.33 | -0.13 | | -0.01 | 0.07 | 0.03 | -0.07 | 0.20 | 0.09 | 0.30 | 0.16 |
| HR 1232 | S | 4905 | 3.26 | 1.50 | 0.23 | 0.17 | 0.22 | 0.21 | 0.67 | 0.15 | 0.07 | 0.18 | 0.29 | 0.19 | 0.27 | 0.05 | 0.17 | 0.14 | 0.25 | 0.70 | 0.18 | 0.07 | -0.14 | -0.06 | 0.46 | 0.19 | 0.33 | 0.79 | 0.19 |
| HR 159 | S | 5501 | 4.28 | 1.08 | -0.13 | -0.05 | -0.01 | -0.14 | -0.09 | -0.10 | -0.17 | -0.10 | -0.12 | -0.14 | -0.18 | -0.21 | -0.13 | -0.19 | -0.16 | -0.22 | 0.18 | -0.10 | -0.31 | -0.32 | -0.02 | -0.06 | -0.03 | 0.42 | 0.06 |
| HR 1665 | S | 5935 | 3.92 | 1.63 | 0.07 | 0.07 | 0.17 | 0.07 | 0.09 | 0.14 | 0.12 | 0.08 | 0.04 | 0.06 | 0.01 | 0.01 | 0.10 | 0.02 | -0.08 | -0.03 | 0.61 | 0.08 | 0.04 | 0.10 | 0.24 | 0.18 | 0.16 | -0.03 | 0.26 |
| HR 1685 | S | 4976 | 3.48 | 1.20 | 0.22 | 0.27 | 0.35 | 0.19 | 0.61 | 0.29 | 0.15 | 0.30 | 0.42 | 0.21 | 0.20 | 0.08 | 0.18 | 0.16 | 0.42 | 0.46 | 0.08 | 0.06 | -0.06 | 0.00 | 0.46 | 0.31 | 0.51 | 0.97 | 0.36 |
| HR 1925 | S | 5243 | 4.53 | 1.00 | 0.10 | 0.17 | 0.12 | 0.14 | 0.55 | 0.17 | 0.09 | 0.15 | 0.18 | 0.18 | 0.21 | 0.08 | 0.11 | 0.12 | 0.04 | -0.12 | 0.18 | 0.28 | -0.09 | 0.20 | 0.64 | 0.47 | 0.57 | 1.32 | 0.35 |
| HR 1980 | S | 6095 | 4.38 | 1.39 | 0.10 | 0.05 | 0.04 | 0.06 | 0.10 | 0.10 | 0.13 | 0.06 | 0.07 | 0.08 | 0.02 | 0.02 | 0.10 | 0.03 | -0.09 | -0.05 | 0.59 | 0.16 | 0.20 | 0.07 | 0.48 | 0.06 | 0.31 | 0.05 | 0.41 |
| HR 2208 | S | 5762 | 4.50 | 1.42 | -0.02 | 0.12 | 0.01 | 0.02 | 0.09 | 0.10 | -0.01 | 0.02 | 0.05 | 0.07 | -0.02 | -0.02 | 0.01 | -0.05 | -0.26 | -0.15 | 0.24 | 0.10 | 0.02 | 0.12 | 0.39 | 0.54 | 0.38 | 0.47 | 0.57 |
| HR 244 | S | 6201 | 4.09 | 1.58 | 0.21 | 0.16 | 0.12 | 0.16 | 0.05 | 0.27 | 0.27 | 0.24 | 0.26 | 0.18 | 0.11 | 0.15 | 0.30 | 0.16 | 0.04 | -0.03 | 0.81 | 0.31 | 0.44 | 0.33 | 0.30 | 0.21 | 0.39 | 0.00 | 0.31 |
| HR 2692 | S | 4995 | 3.47 | 1.09 | -0.15 | -0.02 | -0.01 | -0.08 | 0.25 | -0.04 | -0.20 | -0.08 | -0.13 | -0.28 | -0.45 | -0.39 | -0.22 | -0.28 | -0.09 | -0.15 | -0.19 | -0.40 | -0.45 | -0.51 | -0.24 | 0.23 | -0.06 | 0.29 | -0.09 |
| HR 2866 | S | 6358 | 4.25 | 1.76 | 0.04 | 0.10 | -0.04 | -0.01 | 0.00 | 0.04 | 0.08 | 0.06 | 0.09 | 0.01 | -0.14 | -0.07 | 0.15 | -0.06 | -0.28 | -0.07 | | 0.09 | 0.29 | 0.10 | | 0.37 | 0.43 | -0.19 | 0.09 |
| HR 3193 | S | 5989 | 3.91 | 1.68 | 0.18 | 0.14 | 0.12 | 0.10 | 0.14 | 0.16 | 0.15 | 0.07 | 0.10 | 0.06 | 0.02 | 0.02 | 0.09 | 0.06 | -0.04 | -0.03 | 0.80 | 0.04 | 0.19 | 0.05 | 0.14 | 0.10 | 0.08 | 0.22 | 0.17 |
| HR 3271 | S | 5975 | 3.96 | 1.62 | 0.29 | 0.16 | 0.18 | 0.14 | 0.22 | 0.19 | 0.19 | 0.14 | 0.10 | 0.16 | 0.10 | 0.09 | 0.17 | 0.12 | 0.15 | 0.04 | 0.66 | 0.12 | 0.23 | 0.10 | 0.10 | 0.24 | 0.30 | 0.32 | 0.30 |
| HR 3395 | S | 6222 | 4.32 | 1.52 | 0.09 | 0.15 | 0.10 | 0.12 | 0.19 | 0.15 | 0.19 | 0.10 | 0.12 | 0.11 | -0.01 | 0.05 | 0.11 | 0.05 | -0.08 | 0.03 | 0.78 | 0.20 | 0.35 | 0.18 | 0.29 | 0.39 | 0.21 | 0.34 | 0.20 |
| HR 357 | S | 6491 | 3.72 | 2.76 | 0.18 | -0.19 | | 0.29 | 0.74 | 0.35 | 0.31 | 0.78 | 0.31 | 0.46 | 0.57 | 0.06 | 0.76 | 0.15 | 0.13 | 0.54 | 1.63 | 0.73 | 0.59 | 0.03 | 0.75 | 0.83 | 1.20 | | 1.72 |
| HR 3762 | S | 5153 | 3.40 | 1.31 | -0.27 | -0.23 | -0.15 | -0.21 | -0.07 | -0.21 | -0.30 | -0.25 | -0.24 | -0.29 | -0.40 | -0.37 | -0.29 | -0.35 | -0.31 | -0.18 | 0.07 | -0.28 | -0.41 | -0.22 | -0.14 | -0.10 | 0.00 | -0.02 | -0.03 |
| HR 3901 | S | 6081 | 4.01 | 1.66 | 0.31 | 0.12 | 0.21 | 0.16 | 0.21 | 0.18 | 0.23 | 0.17 | 0.14 | 0.19 | 0.13 | 0.10 | 0.19 | 0.13 | 0.11 | 0.01 | 0.84 | 0.13 | 0.36 | 0.09 | 0.20 | 0.23 | 0.14 | 0.02 | 0.30 |
| HR 4051 | S | 5978 | 4.18 | 1.44 | 0.18 | 0.12 | 0.01 | 0.08 | 0.20 | 0.10 | 0.12 | 0.04 | 0.03 | 0.07 | -0.03 | -0.02 | 0.02 | 0.02 | -0.08 | 0.00 | 0.39 | 0.07 | 0.12 | 0.00 | 0.10 | 0.27 | 0.15 | 0.22 | 0.13 |
| HR 407 | S | 6520 | 3.72 | 3.15 | 0.13 | | 0.29 | 0.14 | 0.57 | -0.03 | 0.27 | 0.52 | 0.74 | 0.15 | 0.26 | -0.05 | 0.48 | 0.19 | | | 0.17 | 1.15 | -0.18 | 0.80 | | 0.10 | 0.90 | | |
| HR 4285 | S | 5916 | 3.76 | 1.66 | -0.24 | -0.16 | -0.28 | -0.17 | -0.33 | -0.13 | -0.14 | -0.21 | -0.24 | -0.26 | -0.36 | -0.30 | -0.21 | -0.31 | -0.52 | -0.30 | 0.42 | -0.26 | -0.14 | -0.05 | -0.18 | 0.10 | 0.01 | -0.41 | -0.18 |
| HR 448 | S | 5861 | 3.99 | 1.51 | 0.41 | 0.32 | 0.30 | 0.25 | 0.22 | 0.29 | 0.27 | 0.27 | 0.23 | 0.28 | 0.27 | 0.19 | 0.29 | 0.27 | 0.29 | 0.24 | 0.58 | 0.27 | 0.23 | 0.16 | 0.30 | 0.39 | 0.31 | 0.36 | 0.33 |
| HR 4864 | S | 5615 | 4.49 | 1.30 | 0.06 | -0.04 | 0.15 | 0.07 | 0.23 | 0.15 | 0.05 | 0.10 | 0.09 | 0.12 | 0.07 | 0.03 | 0.05 | 0.01 | -0.02 | -0.20 | 0.39 | 0.19 | 0.28 | 0.20 | 0.39 | 0.49 | 0.33 | 0.50 | 0.26 |
| HR 495 | S | 4661 | 3.08 | 1.39 | 0.77 | 0.74 | 0.48 | 0.75 | 1.43 | 0.50 | 0.26 | 0.26 | 0.43 | 0.41 | 0.54 | 0.36 | 0.42 | 0.52 | 0.06 | 1.83 | 0.18 | 0.14 | -0.24 | 0.01 | 0.78 | 0.41 | 0.43 | 1.65 | 0.55 |
| HR 511 | S | 5422 | 4.55 | 0.91 | 0.05 | 0.16 | 0.11 | 0.08 | 0.28 | 0.15 | 0.08 | 0.18 | 0.20 | 0.16 | 0.17 | 0.06 | 0.10 | 0.07 | 0.15 | -0.02 | 0.32 | 0.18 | 0.09 | 0.08 | 0.43 | 0.32 | 0.50 | 0.98 | 0.53 |

| Name | | Teff | logg | | | | | | | | | | | | | | | | | | | | | | | | | |
|---|---|---|---|---|---|---|---|---|---|---|---|---|---|---|---|---|---|---|---|---|---|---|---|---|---|---|---|---|
| HR 5258 | S | 6417 | 3.71 | 2.57 | 0.33 | 0.65 | 0.49 | 0.17 | 0.23 | 0.11 | 0.22 | 0.35 | 0.24 | 0.23 | 0.18 | 0.10 | 0.47 | 0.17 | -0.18 | 0.42 | | 0.27 | 0.31 | 0.07 | 0.49 | 0.28 | 0.21 | |
| HR 5317 | S | 6437 | 3.74 | 2.93 | 0.67 | 0.09 | 0.52 | 0.23 | 0.40 | 0.27 | 0.10 | 0.31 | 0.53 | 0.24 | 0.12 | 0.12 | 0.68 | 0.17 | -0.33 | | | 0.18 | | -0.06 | | 0.47 | | 0.77 |
| HR 5335 | S | 4862 | 3.29 | 1.68 | 0.77 | 0.71 | 0.68 | 0.58 | 1.16 | 0.50 | 0.42 | 0.44 | 0.71 | 0.50 | 0.62 | 0.30 | 0.52 | 0.48 | 0.61 | 1.22 | 0.41 | 0.31 | 0.04 | -0.05 | 0.70 | 0.43 | 0.47 | 1.60 | 0.74 |
| HR 5387 | S | 6753 | 4.29 | 2.17 | 0.05 | 0.00 | 0.07 | 0.03 | 0.07 | 0.08 | 0.14 | 0.10 | 0.21 | 0.04 | -0.11 | -0.01 | 0.39 | -0.03 | -0.42 | 0.01 | | 0.19 | 0.56 | 0.22 | | 0.41 | 0.16 | 0.27 | 0.25 |
| HR 5504 | S | 5990 | 4.01 | 1.78 | 0.48 | 0.34 | 0.29 | 0.24 | 0.20 | 0.28 | 0.22 | 0.24 | 0.23 | 0.24 | 0.23 | 0.17 | 0.24 | 0.23 | 0.09 | 0.15 | | 0.18 | 0.21 | -0.06 | 0.22 | 0.33 | 0.19 | 0.62 | 0.32 |
| HR 5630 | S | 6186 | 4.30 | 1.46 | 0.13 | -0.01 | 0.05 | 0.11 | 0.13 | 0.15 | 0.17 | 0.09 | 0.04 | 0.12 | 0.04 | 0.05 | 0.12 | 0.04 | -0.10 | 0.06 | 0.55 | 0.18 | 0.24 | 0.13 | 0.33 | 0.49 | 0.06 | 0.34 | 0.63 |
| HR 5706 | S | 4785 | 3.52 | 1.02 | 0.67 | 0.66 | 0.66 | 0.57 | 1.25 | 0.60 | 0.52 | 0.59 | 0.86 | 0.52 | 0.64 | 0.40 | 0.54 | 0.55 | 0.81 | 2.38 | 0.39 | 0.41 | 0.10 | 0.37 | 0.97 | 0.67 | 0.76 | 1.46 | 0.64 |
| HR 5740 | S | 5956 | 3.97 | 1.87 | 0.74 | 0.44 | 0.47 | 0.41 | 0.48 | 0.41 | 0.45 | 0.39 | 0.38 | 0.39 | 0.41 | 0.30 | 0.42 | 0.40 | 0.35 | 0.26 | | 0.43 | 0.43 | 0.09 | 0.28 | 0.41 | 0.24 | 0.36 | 0.48 |
| HR 6105 | S | 5934 | 4.16 | 1.38 | 0.22 | 0.18 | 0.23 | 0.16 | 0.21 | 0.23 | 0.20 | 0.14 | 0.11 | 0.15 | 0.08 | 0.09 | 0.14 | 0.13 | -0.04 | 0.05 | 0.54 | 0.13 | 0.20 | 0.06 | 0.31 | 0.39 | 0.41 | 0.33 | 0.49 |
| HR 6106 | S | 6009 | 3.97 | 1.68 | 0.26 | 0.17 | 0.22 | 0.18 | 0.23 | 0.22 | 0.30 | 0.19 | 0.17 | 0.20 | 0.15 | 0.13 | 0.20 | 0.17 | -0.02 | 0.03 | 0.75 | 0.20 | 0.29 | 0.20 | 0.34 | 0.39 | 0.29 | 0.09 | 0.30 |
| HR 6269 | S | 5634 | 3.95 | 1.34 | 0.28 | 0.16 | 0.25 | 0.18 | 0.34 | 0.22 | 0.21 | 0.14 | 0.15 | 0.18 | 0.11 | 0.10 | 0.14 | 0.14 | 0.28 | 0.12 | 0.46 | 0.10 | 0.03 | -0.01 | 0.24 | 0.35 | 0.25 | 0.35 | 0.35 |
| HR 6301 | S | 4955 | 3.31 | 1.39 | 0.12 | 0.19 | 0.20 | 0.17 | 0.48 | 0.20 | -0.01 | 0.08 | 0.11 | 0.11 | 0.18 | -0.01 | 0.02 | 0.06 | 0.09 | 0.30 | 0.19 | 0.02 | -0.11 | 0.08 | 0.36 | 0.26 | 0.35 | 0.61 | 0.23 |
| HR 6372 | S | 5744 | 4.02 | 1.49 | 0.56 | 0.38 | 0.44 | 0.33 | 0.46 | 0.30 | 0.37 | 0.31 | 0.33 | 0.34 | 0.38 | 0.26 | 0.34 | 0.34 | 0.34 | 0.26 | 0.66 | 0.30 | 0.31 | 0.06 | 0.37 | 0.38 | 0.25 | 0.67 | 0.63 |
| HR 6465 | S | 5719 | 4.43 | 1.06 | 0.06 | 0.15 | 0.19 | 0.07 | 0.12 | 0.12 | 0.09 | 0.09 | 0.12 | 0.10 | 0.05 | 0.04 | 0.10 | 0.07 | 0.09 | 0.02 | 0.28 | 0.21 | 0.00 | 0.04 | 0.31 | 0.25 | 0.23 | 0.59 | 0.48 |
| HR 6516 | S | 5583 | 4.15 | 1.69 | 0.30 | 0.24 | 0.16 | 0.10 | 0.29 | 0.08 | 0.05 | 0.10 | 0.12 | 0.11 | 0.18 | 0.01 | 0.10 | 0.07 | 0.04 | 0.06 | 0.33 | 0.07 | -0.08 | -0.29 | 0.22 | 0.19 | 0.48 | 0.61 | 0.23 |
| HR 6669 | S | 6134 | 4.27 | 1.45 | 0.09 | 0.08 | 0.10 | 0.05 | 0.10 | 0.10 | 0.10 | 0.03 | 0.04 | 0.05 | 0.00 | -0.01 | 0.04 | 0.00 | -0.22 | -0.09 | 0.50 | 0.12 | 0.14 | 0.11 | 0.36 | 0.43 | 0.18 | 0.13 | 0.13 |
| HR 672 | S | 6079 | 4.07 | 1.56 | 0.31 | 0.27 | 0.25 | 0.20 | 0.24 | 0.24 | 0.23 | 0.19 | 0.18 | 0.21 | 0.16 | 0.13 | 0.21 | 0.17 | 0.14 | 0.04 | 0.58 | 0.25 | 0.22 | 0.00 | 0.12 | 0.20 | 0.23 | 0.36 | 0.50 |
| HR 6722 | S | 5539 | 3.72 | 1.54 | 0.20 | 0.11 | 0.20 | 0.13 | 0.18 | 0.17 | 0.06 | 0.09 | 0.07 | 0.12 | 0.13 | 0.03 | 0.09 | 0.07 | -0.09 | -0.07 | 0.47 | 0.13 | 0.12 | -0.03 | 0.14 | 0.12 | 0.11 | 0.27 | 0.36 |
| HR 6756 | S | 4907 | 3.39 | 1.30 | 0.24 | 0.17 | 0.26 | 0.20 | 0.59 | 0.16 | 0.03 | 0.15 | 0.27 | 0.14 | 0.28 | 0.03 | 0.11 | 0.12 | 0.33 | 0.41 | 0.06 | -0.07 | -0.18 | -0.12 | 0.34 | 0.18 | 0.31 | 0.81 | 0.14 |
| HR 6806 | S | 5042 | 4.58 | 0.98 | -0.06 | 0.02 | 0.07 | 0.03 | 0.64 | -0.05 | 0.00 | 0.06 | 0.13 | -0.06 | -0.08 | -0.16 | 0.01 | -0.07 | 0.53 | 0.01 | 0.18 | -0.01 | -0.47 | -0.25 | 0.33 | 0.80 | 0.72 | 0.88 | 0.03 |
| HR 6847 | S | 5791 | 4.33 | 1.09 | 0.04 | -0.03 | 0.09 | 0.05 | 0.13 | 0.08 | 0.13 | 0.04 | 0.02 | 0.06 | 0.03 | 0.01 | 0.03 | 0.05 | 0.05 | 0.05 | 0.24 | 0.08 | -0.04 | 0.01 | 0.18 | 0.61 | 0.24 | 0.38 | 0.46 |
| HR 6907 | S | 6311 | 4.04 | 1.76 | 0.24 | 0.16 | 0.20 | 0.19 | 0.14 | 0.20 | 0.25 | 0.19 | 0.21 | 0.20 | 0.14 | 0.14 | 0.23 | 0.14 | 0.01 | 0.00 | 0.94 | 0.26 | 0.48 | 0.22 | 0.20 | 0.24 | 0.32 | 0.30 | 0.33 |
| HR 6950 | S | 5337 | 3.85 | 1.28 | 0.10 | 0.13 | 0.20 | 0.14 | 0.33 | 0.12 | 0.08 | 0.08 | 0.06 | 0.12 | 0.14 | 0.04 | 0.07 | 0.07 | 0.33 | 0.18 | 0.26 | 0.08 | -0.14 | -0.11 | 0.25 | 0.20 | 0.34 | 0.68 | 0.27 |
| HR 7079 | S | 6319 | 4.31 | 1.73 | 0.02 | 0.02 | 0.02 | -0.01 | -0.02 | 0.05 | 0.05 | 0.03 | 0.02 | 0.02 | -0.08 | -0.06 | 0.13 | -0.07 | -0.36 | -0.06 | | 0.05 | 0.16 | 0.07 | | 0.32 | 0.14 | 0.12 | 0.21 |
| HR 7291 | S | 6173 | 4.34 | 1.51 | 0.30 | 0.27 | 0.27 | 0.22 | 0.38 | 0.30 | 0.28 | 0.28 | 0.25 | 0.25 | 0.19 | 0.20 | 0.22 | 0.21 | -0.09 | 0.32 | 0.60 | 0.35 | 0.49 | 0.28 | 0.43 | 0.61 | 0.44 | 0.19 | 0.38 |
| HR 7522 | S | 6026 | 3.89 | 1.86 | 0.40 | 0.11 | 0.20 | 0.19 | 0.30 | 0.20 | 0.25 | 0.19 | 0.17 | 0.20 | 0.18 | 0.13 | 0.21 | 0.18 | -0.05 | 0.14 | 0.80 | 0.19 | 0.34 | 0.07 | 0.18 | 0.43 | 0.36 | 0.13 | 0.32 |
| HR 7569 | S | 5774 | 4.10 | 1.26 | -0.06 | 0.01 | 0.11 | -0.01 | -0.05 | 0.02 | 0.04 | 0.05 | -0.02 | -0.07 | -0.19 | -0.12 | 0.00 | -0.09 | -0.08 | -0.04 | 0.45 | -0.06 | -0.12 | -0.21 | 0.00 | -0.01 | 0.28 | 0.33 | 0.24 |
| HR 761 | S | 6230 | 3.85 | 1.86 | -0.13 | -0.04 | -0.16 | -0.09 | -0.16 | -0.03 | -0.06 | -0.08 | -0.04 | -0.17 | -0.30 | -0.22 | -0.05 | -0.23 | -0.41 | -0.29 | 0.51 | -0.19 | -0.01 | 0.04 | | 0.08 | -0.05 | -0.19 | -0.02 |
| HR 7637 | S | 5947 | 4.24 | 1.29 | 0.01 | 0.15 | 0.04 | -0.04 | 0.04 | 0.10 | 0.00 | 0.04 | 0.11 | 0.05 | -0.05 | -0.06 | 0.01 | -0.07 | -0.31 | -0.19 | 0.30 | 0.09 | -0.12 | 0.09 | 0.61 | -0.02 | 0.24 | 0.23 | 0.09 |
| HR 7793 | S | 6261 | 4.36 | 1.55 | -0.04 | -0.05 | -0.07 | -0.01 | 0.05 | 0.02 | 0.01 | 0.01 | 0.04 | 0.00 | -0.14 | -0.06 | 0.12 | -0.08 | -0.43 | -0.06 | | 0.15 | 0.23 | 0.16 | 0.32 | 0.21 | 0.17 | 0.01 | 0.37 |
| HR 7855 | S | 6217 | 4.13 | 1.69 | 0.13 | 0.09 | 0.11 | 0.08 | 0.05 | 0.10 | 0.14 | 0.13 | 0.10 | 0.08 | 0.05 | 0.03 | 0.11 | 0.04 | -0.10 | -0.04 | 0.72 | 0.07 | 0.41 | 0.04 | 0.36 | 0.44 | 0.17 | 0.05 | 0.13 |
| HR 7907 | S | 6189 | 4.25 | 1.74 | 0.42 | 0.31 | 0.25 | 0.27 | 0.35 | 0.26 | 0.31 | 0.31 | 0.24 | 0.26 | 0.23 | 0.19 | 0.26 | 0.24 | 0.04 | 0.23 | | 0.36 | 0.34 | 0.18 | 0.53 | 0.54 | 0.28 | 0.49 | 0.54 |
| HR 8133 | S | 5874 | 3.88 | 1.45 | 0.14 | 0.11 | 0.13 | 0.07 | 0.06 | 0.13 | 0.03 | 0.04 | 0.04 | 0.05 | 0.01 | -0.01 | 0.03 | 0.01 | -0.05 | -0.15 | 0.50 | 0.01 | -0.04 | -0.14 | -0.05 | -0.01 | -0.07 | 0.00 | 0.22 |
| HR 8148 | S | 5433 | 4.43 | 0.39 | -0.24 | -0.09 | -0.09 | -0.13 | 0.14 | -0.15 | -0.17 | -0.13 | -0.18 | -0.16 | -0.30 | -0.25 | -0.17 | -0.22 | -0.10 | -0.41 | 0.36 | 0.05 | -0.21 | -0.27 | 0.68 | 0.27 | 0.22 | 0.82 | 0.18 |
| HR 8170 | S | 5903 | 4.26 | 1.81 | 0.01 | 0.32 | 0.33 | 0.06 | 0.35 | -0.02 | 0.04 | 0.02 | 0.02 | -0.05 | -0.23 | -0.11 | 0.16 | -0.09 | -0.68 | | | 0.05 | 0.40 | 0.13 | 0.65 | 0.24 | 0.25 | | 0.13 |
| HR 857 | S | 5201 | 4.59 | 1.56 | 0.25 | 0.15 | 0.16 | 0.16 | 0.43 | 0.20 | 0.10 | 0.20 | 0.23 | 0.22 | 0.27 | 0.08 | 0.15 | 0.12 | 0.14 | -0.11 | 0.32 | 0.29 | 0.02 | 0.12 | 0.49 | 0.50 | 0.66 | 1.16 | 0.43 |
| HR 8631 | S | 5453 | 3.69 | 1.29 | -0.09 | -0.07 | -0.02 | -0.13 | -0.06 | -0.04 | -0.20 | -0.06 | -0.02 | -0.11 | -0.19 | -0.19 | -0.09 | -0.18 | -0.21 | -0.42 | 0.18 | -0.14 | -0.17 | -0.26 | -0.10 | 0.12 | 0.05 | 0.22 | 0.02 |
| HR 8832 | S | 4883 | 4.60 | 1.27 | 0.37 | 0.26 | 0.25 | 0.26 | 0.92 | 0.16 | 0.28 | 0.21 | 0.39 | 0.26 | 0.29 | 0.13 | 0.31 | 0.24 | 0.22 | 0.24 | 0.24 | 0.19 | -0.21 | -0.13 | 0.64 | 1.65 | 1.28 | 2.14 | 0.60 |

| Name | Type | Col1 | Col2 | Col3 | Col4 | Col5 | Col6 | Col7 | Col8 | Col9 | Col10 | Col11 | Col12 | Col13 | Col14 | Col15 | Col16 | Col17 | Col18 | Col19 | Col20 | Col21 | Col22 | Col23 | Col24 | Col25 | Col26 | Col27 | Col28 |
|---|---|---|---|---|---|---|---|---|---|---|---|---|---|---|---|---|---|---|---|---|---|---|---|---|---|---|---|---|---|
| HR 8924 | S | 4784 | 3.26 | 1.76 | 1.00 | 0.69 | 0.67 | 0.62 | 1.34 | 0.49 | 0.47 | 0.47 | 0.75 | 0.54 | 0.69 | 0.34 | 0.62 | 0.53 | 0.78 | 1.60 | 0.44 | 0.45 | 0.14 | -0.07 | 0.96 | 0.81 | 0.64 | 1.73 | 0.79 |
| HR 8964 | S | 5821 | 4.43 | 1.43 | 0.10 | 0.09 | 0.20 | 0.12 | 0.20 | 0.20 | 0.08 | 0.15 | 0.16 | 0.18 | 0.12 | 0.09 | 0.12 | 0.08 | -0.04 | -0.06 | 0.56 | 0.27 | 0.24 | 0.18 | 0.35 | 0.59 | 0.44 | 0.63 | 0.43 |
| HR 9074 | S | 6227 | 4.36 | 1.54 | 0.08 | 0.12 | 0.02 | 0.07 | 0.12 | 0.13 | 0.08 | 0.13 | 0.15 | 0.10 | 0.02 | 0.03 | 0.07 | -0.01 | -0.35 | -0.01 | | 0.21 | 0.32 | 0.32 | 0.36 | 0.47 | 0.23 | 0.27 | 0.31 |
| HR 9075 | S | 6112 | 4.39 | 1.37 | 0.10 | 0.10 | 0.09 | 0.08 | 0.08 | 0.17 | 0.10 | 0.14 | 0.07 | 0.12 | 0.02 | 0.06 | 0.10 | 0.03 | -0.11 | -0.02 | 0.64 | 0.25 | 0.21 | 0.35 | 0.25 | 0.35 | 0.45 | 0.32 | 0.36 |
| LP 837-53 | S | 3666 | 4.89 | 1.25 | | 0.70 | | 3.15 | 4.62 | 1.20 | 1.20 | 0.48 | 0.45 | 0.99 | 1.43 | 0.50 | 1.63 | 1.50 | | 3.90 | | 0.69 | -0.45 | -0.05 | | 3.05 | 2.92 | | 2.24 |
| NAME 23 H. Cam | S | 6215 | 4.38 | 1.38 | -0.06 | 0.03 | -0.10 | 0.01 | 0.05 | 0.03 | 0.12 | 0.00 | 0.02 | 0.00 | -0.12 | -0.07 | 0.01 | -0.10 | -0.24 | -0.19 | 0.54 | 0.12 | 0.36 | 0.15 | 0.37 | 0.42 | 0.12 | 0.19 | 0.09 |
| V* AK Lep | S | 4925 | 4.60 | 1.70 | -0.03 | 0.12 | 0.07 | 0.12 | 0.71 | 0.05 | 0.02 | 0.03 | 0.10 | 0.08 | 0.05 | -0.04 | 0.07 | 0.00 | -0.10 | 0.06 | 0.13 | -0.04 | -0.29 | -0.16 | 0.53 | 0.81 | 0.87 | 1.41 | 0.44 |
| V* AR Lac | S | 5342 | 3.61 | 1.75 | -0.46 | | 1.13 | 0.73 | 0.78 | | | 1.39 | 1.11 | 1.66 | | 1.05 | 1.42 | 0.57 | | | | 0.73 | | | | 1.47 | | | |
| V* DE Boo | S | 5260 | 4.52 | 1.34 | 0.20 | 0.09 | 0.14 | 0.12 | 0.52 | 0.17 | 0.03 | 0.13 | 0.16 | 0.16 | 0.21 | 0.05 | 0.07 | 0.11 | -0.36 | 0.01 | 0.32 | 0.14 | -0.09 | 0.05 | 0.30 | 0.46 | 0.71 | 1.06 | 0.27 |
| V* HN Peg | S | 5972 | 4.44 | 1.80 | -0.04 | 0.07 | 0.15 | 0.06 | 0.22 | 0.08 | 0.14 | 0.10 | 0.10 | 0.11 | -0.06 | 0.00 | 0.05 | -0.03 | -0.22 | -0.07 | | 0.17 | 0.42 | 0.20 | -0.06 | 0.52 | 0.38 | 0.29 | 0.21 |
| V* IL Aqr | S | 3935 | 5.00 | 1.75 | -1.38 | 0.61 | 0.59 | 2.22 | 4.33 | 0.98 | 0.96 | 1.38 | 0.50 | 1.32 | 1.82 | -0.58 | 2.01 | 1.72 | | | 1.29 | 1.59 | | 1.32 | 2.92 | 2.48 | 2.17 | 3.22 |
| V* pi.01 UMa | S | 5921 | 4.46 | 1.98 | 0.06 | -0.06 | 0.08 | 0.02 | 0.10 | 0.09 | 0.00 | 0.13 | 0.05 | 0.07 | 0.04 | -0.01 | 0.03 | -0.04 | -0.27 | -0.16 | | 0.21 | 0.13 | 0.20 | 0.35 | 0.45 | 0.34 | 0.22 | 0.25 |
| V* V2213 Oph | S | 6030 | 4.39 | 1.48 | 0.05 | 0.11 | 0.05 | 0.07 | 0.13 | 0.11 | 0.07 | 0.07 | 0.05 | 0.11 | 0.05 | 0.02 | 0.04 | 0.00 | -0.35 | -0.05 | 0.47 | 0.24 | 0.22 | 0.11 | 0.18 | 0.45 | 0.24 | 0.42 | 0.19 |
| V* V2215 Oph | S | 4463 | 4.68 | 0.15 | -0.01 | 0.17 | -0.13 | 0.25 | 1.13 | 0.13 | -0.03 | -0.01 | 0.16 | 0.17 | 0.03 | -0.09 | 0.17 | 0.13 | 0.01 | 0.77 | 0.57 | 0.14 | -0.65 | -0.01 | 0.67 | 1.83 | 1.43 | 1.89 | 0.12 |
| V* V2502 Oph | S | 6969 | 4.02 | 2.59 | 0.00 | -0.30 | -0.16 | 0.04 | 0.07 | 0.07 | 0.14 | 0.01 | 0.39 | 0.00 | -0.10 | -0.08 | 0.32 | -0.11 | -0.38 | -0.14 | | 0.13 | 0.62 | 0.15 | | | -0.25 | 0.31 | 0.20 |
| V* V2689 Ori | S | 4073 | 4.72 | 0.15 | 0.40 | -0.07 | 0.42 | 1.29 | 2.76 | 0.96 | 0.69 | 0.42 | 0.61 | 0.77 | 0.48 | 0.45 | 1.00 | 0.82 | 1.10 | | 1.18 | 0.54 | -0.15 | 0.42 | 1.79 | 2.26 | 2.54 | 2.90 | 0.86 |
| V* V376 Peg | S | 6066 | 4.33 | 1.33 | 0.04 | 0.09 | 0.07 | 0.07 | 0.11 | 0.17 | 0.15 | 0.09 | 0.10 | 0.10 | -0.02 | 0.03 | 0.14 | 0.03 | -0.06 | -0.12 | 0.56 | 0.10 | 0.51 | 0.12 | 0.17 | 0.38 | 0.40 | -0.33 | 0.35 |
| V* V450 And | S | 5653 | 4.46 | 1.29 | -0.07 | -0.06 | 0.00 | 0.01 | 0.28 | 0.08 | 0.01 | 0.03 | 0.01 | 0.08 | -0.06 | -0.03 | -0.03 | -0.06 | -0.26 | -0.12 | 0.38 | 0.23 | 0.04 | 0.15 | 0.49 | 0.45 | 0.35 | 0.53 | 0.55 |
| V* V819 Her | S | 5699 | 3.28 | 0.58 | -0.06 | -0.05 | 0.22 | -0.15 | -0.04 | 0.10 | -0.05 | 0.28 | 0.43 | 0.21 | 0.17 | 0.08 | 0.25 | 0.05 | -0.38 | -0.46 | 0.56 | 0.11 | 0.31 | 0.09 | 0.21 | 0.60 | 0.46 | 0.25 | 0.40 |
| Wolf 1008 | S | 5787 | 4.07 | 1.39 | -0.20 | -0.28 | -0.17 | -0.28 | -0.18 | -0.18 | -0.24 | -0.15 | -0.04 | -0.26 | -0.36 | -0.36 | -0.20 | -0.38 | -0.50 | -0.46 | | -0.34 | -0.30 | -0.44 | | -0.07 | -0.12 | 0.01 | 0.20 |
| 10 Tau | H | 6013 | 4.05 | 1.61 | 0.06 | 0.11 | 0.07 | 0.03 | 0.01 | 0.04 | 0.05 | 0.01 | 0.01 | -0.03 | -0.03 | -0.06 | 0.01 | -0.05 | -0.11 | -0.07 | | 0.06 | 0.23 | -0.04 | -0.04 | 0.01 | 0.09 | -0.02 | 0.13 |
| 11 LMi | H | 5498 | 4.43 | 1.54 | 0.65 | 0.48 | 0.48 | 0.40 | 0.66 | 0.38 | 0.35 | 0.38 | 0.42 | 0.40 | 0.45 | 0.33 | 0.40 | 0.39 | 0.45 | 0.56 | 0.53 | 0.42 | 0.39 | 0.10 | 0.53 | 0.53 | 0.73 | 0.79 | 0.60 |
| 15 LMi | H | 5916 | 4.06 | 1.49 | 0.39 | 0.30 | 0.27 | 0.21 | 0.20 | 0.28 | 0.28 | 0.23 | 0.24 | 0.22 | 0.23 | 0.17 | 0.23 | 0.20 | 0.13 | 0.25 | -0.02 | 0.29 | 0.42 | 0.16 | 0.31 | 0.38 | 0.39 | 0.38 | 0.74 |
| 16 Cyg B | H | 5753 | 4.34 | 1.22 | 0.24 | 0.24 | 0.26 | 0.15 | 0.25 | 0.19 | 0.18 | 0.18 | 0.16 | 0.16 | 0.13 | 0.11 | 0.16 | 0.13 | 0.26 | 0.21 | 0.04 | 0.23 | 0.49 | -0.01 | 0.25 | 0.24 | 0.41 | 0.49 | 0.65 |
| 37 Gem | H | 5932 | 4.40 | 1.20 | -0.01 | -0.01 | -0.04 | -0.05 | 0.01 | -0.01 | -0.03 | -0.01 | -0.01 | -0.05 | -0.13 | -0.10 | -0.03 | -0.11 | -0.16 | -0.08 | 0.00 | 0.07 | 0.39 | 0.01 | 0.12 | 0.16 | 0.21 | 0.22 | 0.53 |
| 88 Leo | H | 6030 | 4.39 | 1.25 | 0.11 | 0.12 | 0.10 | 0.10 | 0.15 | 0.18 | 0.10 | 0.11 | 0.15 | 0.12 | 0.09 | 0.08 | 0.11 | 0.06 | -0.09 | 0.06 | 0.03 | 0.21 | 0.42 | 0.18 | 0.27 | 0.38 | 0.31 | 0.35 | 0.55 |
| b Aql | H | 5466 | 4.10 | 1.12 | 0.80 | 0.68 | 0.62 | 0.52 | 0.82 | 0.47 | 0.45 | 0.48 | 0.49 | 0.49 | 0.58 | 0.39 | 0.45 | 0.53 | 0.75 | 0.90 | 0.57 | 0.51 | 0.42 | 0.23 | 0.62 | 0.56 | 0.72 | 0.97 | 0.98 |
| bet CVn | H | 5865 | 4.40 | 1.07 | -0.12 | -0.08 | -0.08 | -0.11 | -0.06 | -0.08 | -0.08 | -0.10 | -0.12 | -0.13 | -0.19 | -0.17 | -0.15 | -0.16 | -0.20 | -0.15 | -0.32 | -0.02 | 0.33 | -0.09 | 0.04 | 0.10 | 0.07 | 0.14 | 0.45 |
| iot Psc | H | 6177 | 4.08 | 1.62 | 0.03 | 0.02 | -0.02 | 0.00 | 0.03 | 0.06 | 0.07 | 0.00 | -0.01 | -0.04 | -0.09 | -0.09 | -0.03 | -0.10 | -0.27 | -0.20 | -0.12 | 0.05 | 0.48 | 0.06 | 0.08 | 0.14 | 0.15 | 0.21 | 0.37 |
| sig Dra | H | 5338 | 4.57 | 0.94 | -0.01 | -0.01 | -0.03 | -0.06 | 0.09 | 0.01 | -0.06 | 0.05 | 0.06 | 0.00 | -0.07 | -0.08 | -0.05 | -0.07 | 0.05 | -0.21 | 0.35 | 0.13 | 0.16 | -0.01 | 0.65 | 0.27 | 0.61 | 0.40 | 0.21 |
| HD 10086 | H | 5658 | 4.46 | 1.16 | 0.29 | 0.33 | 0.13 | 0.19 | 0.38 | 0.30 | 0.28 | 0.26 | 0.24 | 0.29 | 0.23 | 0.21 | 0.18 | 0.23 | 0.11 | 0.17 | 0.49 | 0.26 | 0.60 | 0.38 | 0.55 | 0.61 | 0.65 | 0.63 | 0.35 |
| HD 101177 | H | 5964 | 4.43 | 1.17 | -0.09 | -0.07 | -0.09 | -0.08 | 0.00 | -0.02 | -0.05 | -0.03 | -0.05 | -0.07 | -0.11 | -0.10 | -0.04 | -0.12 | -0.22 | -0.12 | -0.18 | 0.12 | 0.36 | 0.03 | 0.14 | 0.19 | 0.20 | 0.22 | 0.19 |
| HD 102158 | H | 5781 | 4.31 | 1.09 | -0.25 | 0.02 | -0.06 | -0.14 | -0.10 | -0.11 | -0.14 | -0.06 | -0.15 | -0.31 | -0.49 | -0.37 | -0.18 | -0.32 | -0.39 | -0.18 | -0.38 | -0.15 | 0.14 | -0.39 | -0.04 | -0.09 | -0.05 | 0.11 | 0.38 |
| HD 112257 | H | 5659 | 4.31 | 1.13 | 0.10 | 0.25 | 0.23 | 0.11 | 0.23 | 0.13 | 0.09 | 0.14 | 0.08 | 0.07 | 0.02 | 0.02 | 0.07 | 0.05 | 0.13 | 0.24 | | 0.16 | 0.21 | -0.08 | 0.20 | 0.33 | 0.40 | 0.41 | 0.56 |
| HD 116442 | H | 5281 | 4.62 | 0.88 | -0.26 | -0.12 | -0.13 | -0.19 | 0.10 | -0.15 | -0.18 | -0.10 | -0.09 | -0.18 | -0.33 | -0.28 | -0.15 | -0.26 | -0.12 | -0.26 | | -0.03 | 0.00 | -0.39 | 0.20 | -0.04 | 0.40 | 0.40 | 0.13 |
| HD 12051 | H | 5400 | 4.39 | 1.68 | 0.36 | 0.37 | 0.35 | 0.30 | 0.60 | 0.27 | 0.22 | 0.23 | 0.28 | 0.26 | 0.22 | 0.25 | 0.29 | 0.22 | 0.22 | 0.36 | 0.20 | 0.34 | 0.28 | -0.10 | 0.56 | 0.38 | 0.62 | 0.73 | 0.57 |
| HD 130087 | H | 5991 | 4.12 | 1.55 | 0.50 | 0.51 | 0.44 | 0.28 | 0.21 | 0.36 | 0.35 | 0.43 | 0.46 | 0.32 | 0.35 | 0.26 | 0.40 | 0.32 | 0.32 | 0.32 | 0.12 | 0.32 | 0.52 | 0.04 | 0.34 | 0.44 | 0.49 | 0.48 | 0.92 |

| Star | | Teff | logg | | | | | | | | | | | | | | | | | | | | | | | | | | | |
|---|---|---|---|---|---|---|---|---|---|---|---|---|---|---|---|---|---|---|---|---|---|---|---|---|---|---|---|---|---|---|
| HD 13043 | H | 5877 | 4.15 | 1.43 | 0.32 | 0.22 | 0.23 | 0.18 | 0.18 | 0.18 | 0.22 | 0.17 | 0.19 | 0.16 | 0.16 | 0.12 | 0.18 | 0.16 | 0.18 | 0.24 | 0.50 | 0.27 | 0.35 | 0.06 | 0.24 | 0.36 | 0.38 | 0.36 | 0.75 |
| HD 130948 | H | 5983 | 4.43 | 1.47 | 0.07 | 0.07 | 0.00 | 0.07 | 0.13 | 0.12 | 0.00 | 0.09 | 0.06 | 0.10 | 0.03 | 0.04 | 0.07 | -0.01 | -0.23 | -0.16 | 0.01 | 0.16 | 0.49 | 0.29 | 0.27 | 0.37 | 0.36 | 0.32 | 0.63 |
| HD 135101 | H | 5637 | 4.24 | 1.15 | 0.20 | 0.35 | 0.33 | 0.18 | 0.31 | 0.23 | 0.23 | 0.25 | 0.19 | 0.15 | 0.07 | 0.10 | 0.17 | 0.14 | 0.31 | 0.36 | | 0.20 | 0.33 | 0.02 | 0.25 | 0.26 | 0.45 | 0.40 | 0.46 |
| HD 139323 | H | 5046 | 4.48 | 0.79 | 0.83 | 0.63 | 0.64 | 0.52 | 1.16 | 0.51 | 0.56 | 0.62 | 0.80 | 0.62 | 0.55 | 0.43 | 0.54 | 0.58 | 0.77 | 0.84 | 0.38 | 0.62 | 0.42 | 0.30 | 1.24 | 0.90 | 1.40 | 1.60 | 0.70 |
| HD 144579 | H | 5308 | 4.67 | 0.73 | -0.46 | -0.18 | -0.28 | -0.33 | 0.82 | -0.30 | -0.31 | -0.22 | -0.28 | -0.47 | -0.68 | -0.55 | -0.41 | -0.51 | -0.51 | -0.35 | -0.76 | -0.35 | 0.04 | -0.60 | -0.22 | 0.22 | 0.12 | 0.16 | 0.18 |
| HD 147044 | H | 5890 | 4.39 | 1.43 | 0.04 | 0.08 | 0.05 | 0.05 | 0.07 | 0.08 | 0.02 | 0.03 | 0.02 | 0.06 | 0.03 | -0.02 | 0.00 | -0.02 | -0.15 | -0.17 | 0.53 | 0.17 | 0.41 | -0.02 | 0.28 | 0.45 | 0.35 | 0.32 | 0.28 |
| HD 152792 | H | 5675 | 3.84 | 1.33 | -0.23 | -0.11 | -0.14 | -0.17 | -0.13 | -0.18 | -0.21 | -0.20 | -0.27 | -0.28 | -0.37 | -0.31 | -0.23 | -0.32 | -0.38 | -0.26 | | -0.33 | 0.07 | -0.33 | -0.20 | -0.11 | -0.13 | -0.08 | -0.12 |
| HD 159222 | H | 5788 | 4.39 | 1.36 | 0.31 | 0.23 | 0.22 | 0.23 | 0.32 | 0.25 | 0.26 | 0.20 | 0.20 | 0.23 | 0.24 | 0.17 | 0.19 | 0.20 | 0.03 | 0.29 | | 0.38 | 0.48 | 0.16 | 0.48 | 0.36 | 0.49 | 0.61 | 0.38 |
| HD 170778 | H | 5932 | 4.46 | 1.64 | 0.04 | 0.08 | -0.02 | 0.06 | 0.13 | 0.13 | 0.05 | 0.06 | 0.07 | 0.07 | 0.01 | 0.01 | 0.01 | -0.03 | -0.26 | -0.20 | | 0.18 | 0.62 | 0.26 | 0.19 | 0.42 | 0.37 | 0.30 | 0.26 |
| HD 182488 | H | 5362 | 4.45 | 1.29 | 0.47 | 0.40 | 0.32 | 0.29 | 0.63 | 0.25 | 0.28 | 0.27 | 0.29 | 0.27 | 0.33 | 0.22 | 0.28 | 0.29 | 0.39 | 0.50 | 0.25 | 0.30 | 0.42 | 0.02 | 0.60 | 0.47 | 0.78 | 0.84 | 0.52 |
| HD 183341 | H | 5952 | 4.27 | 1.47 | 0.16 | 0.11 | 0.11 | 0.12 | 0.15 | 0.17 | 0.22 | 0.13 | 0.14 | 0.10 | 0.10 | 0.09 | 0.18 | 0.11 | 0.03 | 0.14 | 0.77 | 0.22 | 0.51 | 0.11 | 0.29 | 0.18 | 0.33 | 0.36 | 0.52 |
| HD 193664 | H | 5945 | 4.44 | 1.25 | -0.02 | -0.02 | -0.05 | -0.03 | 0.07 | 0.04 | 0.00 | 0.01 | -0.01 | -0.04 | -0.07 | -0.06 | -0.03 | -0.09 | -0.17 | -0.14 | -0.06 | 0.13 | 0.09 | 0.09 | 0.17 | 0.26 | 0.22 | 0.16 | 0.39 |
| HD 197076 | H | 5844 | 4.46 | 1.19 | -0.06 | 0.04 | 0.01 | -0.01 | 0.10 | 0.05 | 0.09 | 0.02 | 0.00 | 0.00 | -0.05 | -0.04 | -0.02 | -0.06 | -0.17 | -0.08 | | 0.04 | 0.40 | 0.11 | 0.24 | 0.26 | 0.36 | 0.24 | 0.59 |
| HD 210460 | H | 5529 | 3.52 | 1.45 | -0.11 | -0.07 | -0.11 | -0.15 | -0.10 | -0.09 | -0.24 | -0.17 | -0.21 | -0.18 | -0.31 | -0.25 | -0.23 | -0.27 | -0.42 | -0.21 | -0.29 | -0.14 | 0.04 | 0.01 | -0.04 | 0.02 | -0.02 | -0.11 | 0.24 |
| HD 210640 | H | 6377 | 3.98 | 1.52 | 0.24 | 0.22 | 0.14 | 0.04 | -0.30 | 0.31 | 0.25 | 0.44 | 0.55 | 0.29 | 0.34 | 0.20 | 0.46 | 0.25 | 0.33 | 0.02 | | 0.32 | 0.63 | 0.50 | 0.44 | 0.48 | 0.50 | 0.45 | 0.27 |
| HD 223238 | H | 5889 | 4.30 | 1.35 | 0.22 | 0.19 | 0.20 | 0.13 | 0.14 | 0.18 | 0.13 | 0.15 | 0.14 | 0.14 | 0.18 | 0.09 | 0.14 | 0.12 | 0.20 | 0.18 | | 0.17 | 0.41 | 0.04 | 0.21 | 0.29 | 0.32 | 0.29 | 0.28 |
| HD 24213 | H | 6053 | 4.18 | 1.53 | 0.25 | 0.23 | 0.19 | 0.18 | 0.10 | 0.20 | 0.24 | 0.19 | 0.18 | 0.17 | 0.13 | 0.12 | 0.18 | 0.14 | -0.04 | 0.10 | 0.03 | 0.29 | 0.43 | 0.18 | 0.27 | 0.29 | 0.34 | 0.38 | 0.60 |
| HD 28005 | H | 5727 | 4.27 | 1.38 | 0.78 | 0.50 | 0.52 | 0.41 | 0.61 | 0.39 | 0.38 | 0.37 | 0.39 | 0.39 | 0.49 | 0.32 | 0.42 | 0.43 | 0.46 | 0.64 | | 0.39 | 0.50 | 0.11 | 0.56 | 0.45 | 0.55 | 0.66 | 0.51 |
| HD 38858 | H | 5798 | 4.48 | 1.19 | -0.08 | -0.06 | -0.08 | -0.10 | -0.04 | -0.04 | -0.08 | -0.05 | -0.03 | -0.08 | -0.13 | -0.12 | -0.07 | -0.15 | -0.25 | -0.13 | | 0.09 | 0.23 | -0.05 | 0.20 | 0.28 | 0.26 | 0.37 | 0.46 |
| HD 39881 | H | 5719 | 4.27 | 1.08 | 0.09 | 0.21 | 0.19 | 0.07 | 0.17 | 0.11 | 0.15 | 0.14 | 0.10 | 0.00 | -0.08 | -0.05 | 0.06 | -0.01 | 0.12 | 0.20 | -0.16 | 0.15 | 0.36 | -0.03 | 0.16 | 0.25 | 0.33 | 0.38 | 0.69 |
| HD 47127 | H | 5616 | 4.32 | 1.12 | 0.25 | 0.32 | 0.32 | 0.23 | 0.37 | 0.31 | 0.24 | 0.28 | 0.24 | 0.26 | 0.25 | 0.19 | 0.22 | 0.22 | 0.27 | 0.42 | | 0.35 | 0.36 | 0.15 | 0.50 | 0.51 | 0.53 | 0.63 | 0.68 |
| HD 5372 | H | 5847 | 4.37 | 1.44 | 0.52 | 0.42 | 0.38 | 0.35 | 0.45 | 0.33 | 0.32 | 0.32 | 0.35 | 0.36 | 0.37 | 0.27 | 0.34 | 0.34 | 0.17 | 0.48 | 0.56 | 0.42 | 0.67 | 0.04 | 0.45 | 0.51 | 0.62 | 0.78 | 0.58 |
| HD 58781 | H | 5576 | 4.41 | 1.20 | 0.36 | 0.34 | 0.27 | 0.24 | 0.45 | 0.22 | 0.22 | 0.23 | 0.22 | 0.24 | 0.21 | 0.19 | 0.23 | 0.23 | 0.36 | 0.41 | 0.32 | 0.32 | 0.35 | 0.08 | 0.48 | 0.42 | 0.60 | 0.62 | 0.73 |
| HD 64090 | H | 5528 | 4.62 | 2.34 | -0.72 | -1.27 | | -1.22 | | -1.48 | -1.37 | -1.38 | -1.07 | -1.55 | -2.01 | -1.69 | -0.86 | -1.66 | | -1.73 | -1.67 | -1.08 | -0.50 | -1.71 | -1.15 | 0.22 | 0.25 | -0.85 | | |
| HD 65583 | H | 5342 | 4.55 | 0.90 | -0.46 | -0.20 | -0.32 | -0.34 | 0.74 | -0.33 | -0.33 | -0.27 | -0.33 | -0.51 | -0.73 | -0.62 | -0.40 | -0.55 | -0.50 | -0.47 | | -0.33 | -0.01 | -0.62 | -0.21 | -0.23 | -0.01 | -0.08 | -0.09 |
| HD 68017 | H | 5565 | 4.43 | 0.66 | -0.21 | -0.01 | -0.03 | -0.14 | 0.02 | -0.09 | -0.08 | -0.04 | -0.10 | -0.22 | -0.39 | -0.31 | -0.17 | -0.26 | -0.24 | 0.02 | | -0.10 | 0.09 | -0.28 | 0.09 | -0.05 | 0.11 | 0.34 | 0.13 |
| HD 71148 | H | 5835 | 4.39 | 1.17 | 0.15 | 0.15 | 0.12 | 0.10 | 0.13 | 0.15 | 0.09 | 0.11 | 0.10 | 0.11 | 0.10 | 0.07 | 0.09 | 0.07 | 0.01 | 0.09 | -0.13 | 0.25 | 0.34 | 0.08 | 0.32 | 0.33 | 0.37 | 0.41 | 0.71 |
| HD 72760 | H | 5328 | 4.56 | 1.41 | 0.09 | 0.17 | 0.16 | 0.13 | 0.53 | 0.14 | 0.05 | 0.13 | 0.18 | 0.17 | 0.10 | 0.08 | 0.08 | 0.08 | -0.11 | -0.11 | 0.32 | 0.20 | 0.21 | 0.07 | 0.51 | 0.33 | 0.54 | 0.59 | 0.36 |
| HD 76752 | H | 5685 | 4.28 | 1.19 | 0.17 | 0.20 | 0.18 | 0.15 | 0.35 | 0.16 | 0.11 | 0.13 | 0.11 | 0.12 | 0.11 | 0.07 | 0.07 | 0.08 | 0.09 | 0.17 | 0.19 | 0.17 | 0.46 | 0.02 | 0.20 | 0.33 | 0.36 | 0.27 | 0.68 |
| HD 76909 | H | 5598 | 4.15 | 1.50 | 0.75 | 0.59 | 0.54 | 0.44 | 0.68 | 0.42 | 0.38 | 0.38 | 0.40 | 0.40 | 0.47 | 0.33 | 0.44 | 0.43 | 0.59 | 0.61 | 0.44 | 0.48 | 0.45 | 0.03 | 0.52 | 0.43 | 0.58 | 0.75 | 0.52 |
| HD 8648 | H | 5711 | 4.16 | 1.46 | 0.38 | 0.31 | 0.32 | 0.26 | 0.31 | 0.24 | 0.21 | 0.20 | 0.18 | 0.21 | 0.24 | 0.16 | 0.20 | 0.21 | 0.21 | 0.28 | | 0.32 | 0.38 | 0.03 | 0.30 | 0.34 | 0.44 | 0.39 | 0.60 |
| HD 89269 | H | 5635 | 4.49 | 1.07 | -0.03 | -0.02 | -0.01 | -0.05 | 0.12 | 0.00 | -0.06 | -0.02 | -0.03 | -0.05 | -0.13 | -0.09 | -0.03 | -0.10 | -0.11 | -0.04 | -0.15 | 0.19 | 0.27 | -0.05 | 0.25 | 0.29 | 0.32 | 0.35 | 0.49 |
| HD 98618 | H | 5735 | 4.35 | 1.13 | 0.13 | 0.16 | 0.14 | 0.11 | 0.29 | 0.14 | 0.11 | 0.10 | 0.07 | 0.09 | 0.06 | 0.06 | 0.08 | 0.08 | 0.01 | 0.11 | -0.03 | 0.19 | 0.50 | 0.00 | 0.24 | 0.36 | 0.33 | 0.35 | 0.24 |
| HD 9986 | H | 5791 | 4.42 | 1.32 | 0.17 | 0.13 | 0.18 | 0.14 | 0.25 | 0.16 | 0.10 | 0.12 | 0.12 | 0.15 | 0.11 | 0.09 | 0.12 | 0.09 | 0.05 | 0.15 | 0.54 | 0.25 | 0.43 | 0.02 | 0.17 | 0.20 | 0.37 | 0.32 | 0.34 |
| HR 2208 | H | 5762 | 4.50 | 1.50 | 0.06 | 0.03 | 0.03 | 0.03 | 0.16 | 0.12 | 0.00 | 0.06 | 0.07 | 0.08 | -0.01 | 0.02 | 0.04 | -0.01 | -0.16 | -0.16 | | 0.09 | 0.32 | 0.16 | 0.27 | 0.28 | 0.37 | 0.34 | 0.61 |
| HR 511 | H | 5422 | 4.55 | 0.76 | 0.26 | 0.24 | 0.21 | 0.17 | 0.46 | 0.30 | 0.17 | 0.33 | 0.30 | 0.29 | 0.25 | 0.20 | 0.19 | 0.24 | 0.33 | 0.32 | 0.20 | 0.43 | 0.47 | 0.39 | 0.64 | 0.55 | 0.79 | 0.80 | 0.54 |
| HR 6465 | H | 5719 | 4.43 | 1.16 | 0.14 | 0.21 | 0.22 | 0.12 | 0.29 | 0.16 | 0.16 | 0.17 | 0.17 | 0.16 | 0.15 | 0.10 | 0.14 | 0.12 | 0.16 | 0.16 | | 0.24 | 0.38 | 0.02 | 0.35 | 0.36 | 0.41 | 0.53 | 0.53 |
| HR 6847 | H | 5791 | 4.33 | 1.21 | 0.17 | 0.16 | 0.16 | 0.11 | 0.18 | 0.13 | 0.11 | 0.11 | 0.11 | 0.10 | 0.09 | 0.06 | 0.09 | 0.08 | 0.10 | 0.17 | 0.06 | 0.24 | 0.28 | -0.03 | 0.17 | 0.33 | 0.31 | 0.33 | 0.62 |

| Name | Type | T | logg | v | | | | | | | | | | | | | | | | | | | | | | | | | |
|---|---|---|---|---|---|---|---|---|---|---|---|---|---|---|---|---|---|---|---|---|---|---|---|---|---|---|---|---|---|
| HR 8964 | H | 5821 | 4.43 | 1.50 | 0.19 | 0.21 | 0.17 | 0.17 | 0.27 | 0.26 | 0.17 | 0.20 | 0.20 | 0.22 | 0.18 | 0.16 | 0.16 | 0.15 | -0.10 | 0.10 | 0.13 | 0.40 | 0.53 | 0.34 | 0.37 | 0.53 | 0.49 | 0.51 | 0.77 |
| V* BZ Cet | H | 5014 | 4.52 | 1.11 | 0.48 | 0.48 | 0.38 | 0.37 | 0.93 | 0.39 | 0.29 | 0.37 | 0.46 | 0.39 | 0.38 | 0.31 | 0.31 | 0.38 | 0.40 | 0.19 | 0.31 | 0.39 | 0.28 | 0.35 | 0.97 | 0.65 | 1.16 | 1.10 | 0.68 |
| 1 Hya | E | 6359 | 4.11 | 2.69 | 0.10 | | | 0.13 | 0.69 | 0.04 | 0.40 | 0.22 | 0.76 | 0.19 | 0.02 | 0.09 | 0.41 | 0.15 | | 0.20 | | 0.59 | 1.63 | -0.20 | 0.32 | 0.88 | 0.64 | 1.17 | 0.06 |
| 101 Tau | E | 6465 | 4.31 | 2.66 | 0.20 | 0.39 | 0.62 | 0.40 | 0.79 | 0.24 | 0.33 | 0.63 | 0.44 | 0.27 | 0.27 | 0.23 | 0.62 | 0.38 | | 0.17 | | 0.34 | 1.96 | -0.10 | | 0.40 | 1.70 | 1.04 | | |
| 14 Boo | E | 6180 | 3.91 | 1.81 | 0.34 | 0.26 | 0.15 | 0.18 | 0.15 | 0.18 | 0.24 | 0.14 | 0.24 | 0.17 | 0.16 | 0.11 | 0.26 | 0.14 | 0.05 | 0.02 | -0.03 | 0.22 | 0.18 | 0.15 | 0.22 | 0.29 | 0.20 | 0.15 | 0.53 |
| 15 Peg | E | 6452 | 4.09 | 1.95 | -0.36 | -0.27 | -0.45 | -0.33 | -0.39 | -0.36 | -0.36 | -0.38 | -0.43 | -0.50 | -0.64 | -0.52 | -0.12 | -0.49 | -0.69 | -0.61 | -0.58 | -0.54 | -0.37 | -0.31 | -0.06 | -0.21 | -0.31 | -0.11 | -0.42 |
| 22 Lyn | E | 6395 | 4.34 | 1.65 | -0.12 | -0.02 | -0.20 | -0.09 | -0.18 | -0.09 | -0.06 | -0.09 | -0.01 | -0.18 | -0.30 | -0.20 | 0.11 | -0.21 | -0.41 | -0.29 | -0.22 | -0.12 | -0.05 | 0.08 | 0.43 | 0.14 | 0.15 | 0.25 | 0.21 |
| 30 Ari B | E | 6257 | 4.34 | 3.36 | 0.35 | | 0.10 | 0.29 | 0.09 | 0.48 | 0.52 | 0.54 | 0.62 | 0.75 | 0.30 | 0.24 | 0.52 | 0.39 | | -0.28 | 1.10 | 0.51 | 2.11 | -0.08 | 1.10 | 1.60 | 2.38 | 1.05 | 1.41 |
| 34 Peg | E | 6258 | 3.92 | 1.94 | 0.15 | 0.15 | 0.14 | 0.12 | 0.10 | 0.16 | 0.13 | 0.12 | 0.11 | 0.10 | 0.05 | 0.04 | 0.25 | 0.03 | -0.15 | -0.04 | 0.01 | 0.12 | 0.27 | 0.17 | 0.24 | 0.27 | 0.24 | 0.10 | 0.53 |
| 36 Dra | E | 6522 | 4.07 | 1.98 | -0.20 | -0.14 | -0.30 | -0.18 | -0.29 | -0.19 | -0.22 | -0.17 | -0.22 | -0.26 | -0.42 | -0.31 | 0.03 | -0.31 | -0.45 | -0.47 | -0.18 | -0.22 | 0.05 | 0.01 | 0.12 | 0.03 | 0.06 | 0.03 | -0.08 |
| 38 Cet | E | 6480 | 3.87 | 1.89 | -0.07 | -0.02 | -0.11 | -0.08 | -0.14 | -0.03 | -0.04 | -0.08 | -0.20 | -0.16 | -0.34 | -0.17 | -0.12 | -0.20 | -0.45 | -0.38 | -0.09 | 0.01 | -0.06 | 0.30 | 0.17 | 0.08 | 0.01 | -0.03 | 0.07 |
| 4 Aqr | E | 6440 | 3.79 | 3.48 | 0.13 | -0.05 | 0.06 | 0.19 | 0.39 | -0.06 | 0.04 | 0.31 | 0.44 | 0.12 | 0.11 | -0.01 | 0.63 | 0.20 | -0.33 | -0.55 | | 0.41 | 0.41 | -0.72 | 0.50 | 0.00 | 0.22 | 0.99 | 0.48 | |
| 40 Leo | E | 6467 | 4.11 | 2.57 | 0.30 | 0.41 | 0.32 | 0.18 | 0.28 | 0.21 | 0.32 | 0.35 | 0.24 | 0.18 | 0.10 | 0.12 | 0.37 | 0.20 | -0.19 | -0.33 | 0.80 | 0.35 | 0.84 | 0.15 | 0.49 | 0.46 | 0.42 | 0.62 | 0.36 |
| 47 Ari | E | 6644 | 4.21 | 3.01 | 0.65 | 0.22 | | 0.38 | 0.45 | 0.31 | 0.54 | 0.49 | 0.43 | 0.31 | 0.38 | 0.26 | 0.55 | 0.32 | -0.15 | -0.34 | | 0.77 | 1.12 | -0.04 | | 0.40 | 0.53 | 1.02 | 0.26 | |
| 49 Peg | E | 6275 | 3.95 | 1.68 | -0.12 | -0.01 | -0.24 | -0.08 | -0.16 | -0.06 | -0.06 | -0.11 | -0.10 | -0.12 | -0.29 | -0.18 | 0.01 | -0.20 | -0.62 | -0.33 | -0.26 | -0.09 | 0.25 | 0.04 | 0.09 | 0.00 | -0.03 | -0.08 | -0.01 |
| 51 Ari | E | 5603 | 4.46 | 0.85 | 0.34 | 0.26 | 0.28 | 0.21 | 0.45 | 0.22 | 0.21 | 0.21 | 0.23 | 0.24 | 0.24 | 0.20 | 0.23 | 0.23 | 0.36 | 0.37 | 0.23 | 0.29 | 0.31 | 0.18 | 0.42 | 0.48 | 0.45 | 0.58 | 0.69 |
| 6 And | E | 6338 | 4.23 | 2.43 | -0.04 | 0.01 | -0.23 | 0.05 | 0.10 | 0.07 | -0.05 | 0.05 | 0.27 | -0.07 | -0.20 | -0.12 | 0.23 | -0.07 | -0.55 | -0.27 | | 0.14 | 0.54 | 0.15 | 0.24 | 0.32 | 0.24 | 0.53 | 0.57 | |
| 6 Cet | E | 6242 | 4.09 | 1.53 | -0.23 | -0.06 | -0.39 | -0.21 | -0.24 | -0.18 | -0.09 | -0.19 | -0.15 | -0.25 | -0.46 | -0.30 | 0.03 | -0.29 | -0.41 | -0.38 | -0.36 | -0.13 | 0.21 | -0.12 | 0.09 | -0.11 | -0.10 | -0.13 | 0.14 |
| 68 Eri | E | 6421 | 4.03 | 2.01 | -0.15 | -0.09 | -0.25 | -0.11 | -0.20 | -0.16 | -0.05 | -0.16 | -0.13 | -0.23 | -0.39 | -0.28 | 0.04 | -0.28 | -0.43 | -0.35 | | -0.10 | 0.02 | -0.04 | 0.53 | 0.08 | 0.05 | 0.03 | 0.00 | |
| 71 Ori | E | 6560 | 4.31 | 1.59 | 0.17 | 0.15 | 0.14 | 0.12 | 0.11 | 0.21 | 0.27 | 0.19 | 0.27 | 0.17 | 0.06 | 0.14 | 0.26 | 0.11 | -0.08 | -0.15 | 0.37 | 0.30 | 0.60 | 0.35 | 0.41 | 0.36 | 0.30 | 0.27 | 0.80 |
| 84 Cet | E | 6236 | 4.29 | 2.64 | 0.04 | -0.33 | 0.12 | 0.24 | 0.22 | 0.15 | 0.33 | 0.20 | 0.48 | 0.14 | 0.14 | 0.03 | 0.26 | 0.14 | -0.48 | -0.33 | | 0.40 | 1.94 | -0.16 | | -0.30 | 0.57 | | | |
| 89 Leo | E | 6538 | 4.30 | 2.22 | 0.26 | 0.20 | 0.12 | 0.20 | 0.29 | 0.21 | 0.41 | 0.25 | 0.26 | 0.22 | 0.12 | 0.15 | 0.14 | 0.19 | -0.14 | -0.13 | | 0.48 | 0.62 | 0.12 | 0.58 | 0.48 | 0.40 | 0.60 | 0.32 | |
| b Her | E | 6000 | 4.22 | 1.26 | -0.40 | -0.34 | -0.47 | -0.43 | -0.30 | -0.39 | -0.42 | -0.41 | -0.36 | -0.50 | -0.73 | -0.57 | -0.28 | -0.56 | -0.70 | -0.67 | -0.62 | -0.47 | -0.40 | -0.47 | -0.31 | -0.36 | -0.30 | -0.22 | -0.10 |
| c Boo | E | 6560 | 4.27 | 3.22 | 0.07 | | 0.22 | 0.26 | 0.59 | -0.01 | 0.42 | 0.66 | 1.10 | 0.50 | 0.20 | 0.18 | 0.68 | 0.27 | | -0.25 | | 1.80 | 1.99 | | | 0.76 | 1.10 | 1.01 | | |
| eta UMi | E | 6788 | 4.00 | 3.75 | | | 0.68 | | 0.38 | 1.05 | 0.72 | 1.28 | 0.93 | -0.15 | 0.05 | 1.68 | 0.62 | | | 1.02 | 1.51 | | | 0.37 | 1.42 | 1.37 | | | | |
| iot Vir | E | 6217 | 3.75 | 2.41 | 0.17 | 0.09 | 0.08 | 0.05 | 0.02 | 0.07 | 0.09 | 0.12 | 0.08 | -0.04 | -0.19 | -0.06 | 0.15 | -0.02 | -0.39 | -0.20 | | 0.33 | 0.47 | 0.12 | 0.59 | 0.33 | 0.21 | 0.27 | 0.19 | |
| kap CrB | E | 4863 | 3.18 | 1.25 | 0.48 | 0.36 | 0.34 | 0.36 | 0.72 | 0.34 | 0.23 | 0.28 | 0.44 | 0.30 | 0.29 | 0.23 | 0.25 | 0.33 | 0.56 | 0.33 | 0.16 | 0.28 | 0.17 | 0.25 | 0.49 | 0.62 | 0.55 | 0.50 | 0.32 |
| phi Vir | E | 5551 | 3.44 | 2.33 | 0.17 | 0.26 | 0.15 | 0.14 | 0.30 | 0.04 | -0.03 | 0.08 | -0.02 | 0.08 | -0.05 | -0.03 | 0.03 | 0.00 | -0.44 | -0.31 | | 0.12 | 0.38 | -0.04 | 0.44 | 0.20 | 0.15 | 0.28 | 0.15 | |
| tau01 Hya | E | 6507 | 4.21 | 2.49 | 0.17 | 0.46 | | 0.31 | 0.53 | 0.29 | 0.42 | 0.39 | 0.38 | 0.24 | 0.04 | 0.18 | 0.70 | 0.28 | -0.21 | -0.29 | | 0.77 | 1.52 | -0.13 | | 0.77 | 0.65 | 1.20 | | |
| tet Dra | E | 6208 | 3.79 | 3.03 | 0.42 | 0.52 | 0.25 | 0.38 | 0.32 | 0.23 | 0.38 | 0.41 | 0.35 | 0.29 | 0.20 | 0.19 | 0.48 | 0.28 | 0.10 | -0.45 | 0.75 | 0.44 | 1.56 | -0.21 | | 1.07 | 0.37 | 0.98 | 0.33 | |
| BD+01 2063 | E | 4978 | 4.63 | 1.50 | -0.32 | -0.32 | -0.26 | -0.17 | 0.49 | -0.21 | -0.25 | -0.23 | -0.20 | -0.15 | -0.35 | -0.28 | -0.14 | -0.26 | -0.19 | -0.45 | -0.10 | -0.05 | -0.09 | -0.20 | 0.21 | 0.41 | 0.75 | 0.26 | 0.36 |
| BD+12 4499 | E | 4653 | 4.66 | 0.69 | 0.32 | 0.28 | 0.16 | 0.30 | 0.97 | 0.25 | 0.31 | 0.22 | 0.38 | 0.29 | 0.15 | 0.25 | 0.35 | 0.30 | 0.28 | 0.18 | 0.43 | 0.33 | 0.14 | 0.20 | 0.77 | 1.10 | 1.26 | 1.13 | 0.88 |
| BD+17 4708 | E | 6180 | 4.16 | 1.54 | -1.73 | -1.21 | | -1.07 | | -1.14 | -1.52 | -1.15 | -0.65 | -1.15 | -1.90 | -1.49 | -0.46 | -1.45 | | -1.58 | -1.12 | -0.55 | -0.48 | -1.46 | -0.29 | -0.13 | 0.02 | -0.21 | -1.40 | |
| BD+23 465 | E | 5240 | 4.49 | 1.22 | 0.37 | 0.26 | 0.22 | 0.28 | 0.66 | 0.29 | 0.20 | 0.25 | 0.32 | 0.29 | 0.29 | 0.22 | 0.23 | 0.29 | 0.16 | 0.29 | 0.25 | 0.34 | 0.37 | 0.26 | 0.45 | 0.54 | 0.81 | 0.69 | 0.52 |
| BD+29 366 | E | 5777 | 4.46 | 1.34 | -0.64 | -0.54 | -0.84 | -0.63 | -0.19 | -0.71 | -0.61 | -0.60 | -0.52 | -0.80 | -1.10 | -0.86 | -0.45 | -0.81 | -0.98 | -0.87 | -0.91 | -0.81 | -0.41 | -0.96 | -0.35 | -0.37 | -0.46 | -0.04 | -0.77 |
| BD+33 99 | E | 4499 | 4.61 | 0.50 | 0.29 | 0.31 | 0.22 | 0.33 | 1.77 | 0.25 | 0.38 | 0.20 | 0.37 | 0.26 | 0.09 | 0.21 | 0.34 | 0.26 | 0.51 | 0.42 | 0.20 | 0.21 | -0.07 | -0.08 | 0.75 | 0.95 | 1.32 | 1.28 | 0.61 |
| BD+41 3306 | E | 5053 | 4.54 | 0.15 | -0.32 | -0.18 | -0.18 | -0.17 | 0.32 | -0.20 | -0.25 | -0.15 | -0.13 | -0.32 | -0.58 | -0.44 | -0.27 | -0.33 | -0.33 | -0.30 | -0.23 | -0.11 | -0.08 | -0.56 | 0.25 | -0.06 | 0.40 | 0.05 | 0.05 |
| BD+43 699 | E | 4802 | 4.66 | 0.15 | -0.15 | 0.01 | -0.18 | -0.08 | 0.71 | -0.07 | -0.13 | -0.07 | 0.06 | -0.12 | -0.30 | -0.22 | -0.11 | -0.16 | 0.18 | -0.41 | -0.07 | 0.02 | -0.10 | -0.19 | 0.27 | 0.73 | 0.87 | 0.44 | 0.36 |

| Name | | Teff | logg | | | | | | | | | | | | | | | | | | | | | | | | | |
|------|---|------|------|---|---|---|---|---|---|---|---|---|---|---|---|---|---|---|---|---|---|---|---|---|---|---|---|---|
| BD+46 1635 | E | 4215 | 4.65 | 1.36 | 0.55 | 0.76 | 0.47 | 0.80 | 1.98 | 0.70 | 0.61 | 0.33 | 0.61 | 0.50 | 0.41 | 0.45 | 0.72 | 0.64 | 0.93 | 1.26 | 0.96 | 0.45 | -0.01 | 0.08 | 1.14 | 1.37 | 1.65 | 1.32 | 0.66 |
| BD+52 2815 | E | 4203 | 4.66 | 0.15 | 0.15 | 0.49 | 0.20 | 0.67 | 2.21 | 0.47 | 0.51 | 0.27 | 0.35 | 0.32 | 0.19 | 0.14 | 0.57 | 0.44 | 0.83 | 0.63 | 0.75 | 0.41 | 0.13 | 0.11 | 1.24 | 1.23 | 1.72 | 1.34 | 0.88 |
| CCDM J20051-0418AB | E | 6370 | 4.07 | 1.78 | -0.13 | -0.07 | -0.28 | -0.10 | -0.24 | 0.01 | -0.11 | -0.08 | 0.06 | -0.14 | -0.28 | -0.18 | 0.10 | -0.17 | -0.40 | -0.46 | 0.03 | -0.04 | 0.17 | -0.09 | 0.35 | 0.09 | 0.02 | 0.01 | 0.15 |
| CCDM J21031+0132AB | E | 6423 | 3.74 | 2.48 | 0.06 | 0.21 | 0.21 | 0.13 | 0.26 | 0.13 | -0.16 | 0.07 | 0.28 | 0.05 | -0.10 | -0.02 | 0.37 | 0.04 | -0.19 | -0.05 | | -0.07 | 0.26 | -0.17 | 0.22 | 0.01 | -0.07 | 0.42 | -0.12 |
| CCDM J22071+0034AB | E | 6059 | 3.91 | 1.47 | -0.05 | -0.05 | -0.02 | -0.09 | -0.12 | -0.01 | -0.07 | -0.04 | -0.09 | -0.09 | -0.21 | -0.13 | 0.01 | -0.12 | -0.22 | -0.28 | 0.12 | 0.09 | 0.01 | -0.31 | 0.08 | 0.07 | 0.10 | 0.04 | 0.20 |
| GJ 1067 | E | 4378 | 4.68 | 0.57 | 0.03 | 0.39 | 0.12 | 0.51 | 1.58 | 0.44 | 0.38 | 0.22 | 0.39 | 0.35 | 0.21 | 0.37 | 0.41 | 0.34 | 0.55 | 0.41 | 0.93 | 0.36 | 0.29 | 0.26 | 0.94 | 1.07 | 1.53 | 1.32 | 0.74 |
| GJ 697 | E | 4881 | 4.64 | 1.03 | -0.02 | -0.12 | -0.10 | 0.09 | 0.76 | 0.01 | -0.01 | -0.01 | 0.06 | 0.04 | -0.05 | 0.02 | 0.04 | 0.02 | 0.25 | -0.16 | 0.01 | 0.10 | 0.11 | 0.11 | 0.50 | 0.65 | 0.92 | 0.61 | 0.71 |
| HD 10086 | E | 5658 | 4.46 | 1.04 | 0.20 | 0.13 | 0.17 | 0.14 | 0.19 | 0.19 | 0.06 | 0.12 | 0.15 | 0.18 | 0.12 | 0.12 | 0.13 | 0.11 | 0.03 | 0.15 | 0.11 | 0.16 | 0.19 | 0.17 | 0.28 | 0.42 | 0.38 | 0.42 | 0.66 |
| HD 10145 | E | 5642 | 4.38 | 0.82 | 0.09 | 0.16 | 0.09 | 0.09 | 0.18 | 0.12 | 0.06 | 0.14 | 0.10 | 0.10 | 0.04 | 0.07 | 0.05 | 0.04 | 0.07 | 0.15 | 0.05 | 0.15 | 0.13 | -0.02 | 0.40 | 0.31 | 0.43 | 0.27 | 0.51 |
| HD 105631 | E | 5363 | 4.52 | 1.01 | 0.31 | 0.30 | 0.32 | 0.23 | 0.60 | 0.27 | 0.26 | 0.24 | 0.28 | 0.29 | 0.28 | 0.24 | 0.25 | 0.26 | 0.29 | 0.31 | 0.60 | 0.35 | 0.35 | 0.25 | 0.49 | 0.71 | 0.65 | 0.68 | 0.74 |
| HD 106116 | E | 5665 | 4.34 | 1.14 | 0.35 | 0.30 | 0.28 | 0.22 | 0.38 | 0.23 | 0.24 | 0.21 | 0.26 | 0.24 | 0.22 | 0.20 | 0.20 | 0.23 | 0.16 | 0.23 | 0.20 | 0.30 | 0.26 | 0.12 | 0.50 | 0.42 | 0.37 | 0.63 | 0.71 |
| HD 106691 | E | 6647 | 4.13 | 3.08 | 0.13 | | -0.01 | 0.21 | 0.21 | 0.11 | 0.28 | 0.36 | 0.34 | 0.27 | -0.03 | 0.10 | 0.80 | 0.24 | -0.04 | -0.06 | | 0.53 | 1.96 | -0.18 | 0.87 | 0.90 | 0.91 | 0.59 | |
| HD 107611 | E | 6391 | 4.21 | 2.54 | 0.13 | -0.08 | 0.08 | 0.07 | 0.07 | 0.08 | 0.13 | 0.16 | 0.10 | -0.03 | -0.04 | -0.03 | 0.45 | 0.01 | -0.21 | -0.27 | | 0.27 | 0.57 | -0.08 | 0.86 | 0.27 | 0.33 | 0.50 | 1.30 |
| HD 10853 | E | 4655 | 4.66 | 0.48 | 0.10 | 0.10 | -0.05 | 0.17 | 0.93 | 0.16 | 0.07 | 0.07 | 0.20 | 0.14 | 0.00 | 0.09 | 0.17 | 0.09 | 0.23 | -0.02 | 0.22 | 0.20 | 0.33 | 0.11 | 0.59 | 0.80 | 1.24 | 0.75 | 0.62 |
| HD 110463 | E | 4926 | 4.61 | 1.03 | 0.12 | 0.16 | 0.07 | 0.10 | 0.65 | 0.08 | 0.05 | 0.10 | 0.19 | 0.15 | 0.01 | 0.07 | 0.07 | 0.06 | 0.29 | -0.11 | 0.26 | 0.18 | 0.23 | 0.14 | 0.48 | 0.58 | 0.96 | 0.53 | 0.75 |
| HD 111069 | E | 5865 | 4.40 | 1.14 | 0.48 | 0.36 | 0.31 | 0.25 | 0.35 | 0.29 | 0.35 | 0.30 | 0.35 | 0.33 | 0.31 | 0.27 | 0.31 | 0.31 | 0.31 | 0.20 | 0.33 | 0.38 | 0.39 | 0.21 | 0.68 | 0.46 | 0.47 | 0.61 | 0.73 |
| HD 11373 | E | 4759 | 4.61 | 0.85 | 0.34 | 0.31 | 0.18 | 0.32 | 0.91 | 0.15 | 0.23 | 0.24 | 0.42 | 0.30 | 0.17 | 0.20 | 0.31 | 0.29 | 0.27 | 0.37 | 0.41 | 0.30 | 0.12 | 0.15 | 0.77 | 1.24 | 1.07 | 1.06 | 0.95 |
| HD 115274 | E | 6153 | 3.84 | 1.27 | 0.13 | 0.03 | -0.16 | -0.01 | 0.14 | 0.00 | 0.16 | 0.11 | 0.21 | 0.21 | -0.01 | -0.03 | 0.34 | 0.04 | -0.19 | -0.11 | 0.43 | 0.25 | 0.60 | 0.00 | 0.44 | 0.18 | 0.27 | 0.34 | 0.32 |
| HD 116956 | E | 5308 | 4.52 | 1.26 | 0.26 | 0.21 | 0.14 | 0.22 | 0.54 | 0.28 | 0.15 | 0.17 | 0.21 | 0.27 | 0.18 | 0.19 | 0.15 | 0.22 | 0.15 | -0.02 | 0.15 | 0.39 | 0.25 | 0.35 | 0.56 | 0.64 | 0.81 | 0.81 | 1.17 |
| HD 117635 | E | 5217 | 4.24 | 0.80 | -0.07 | -0.10 | -0.06 | -0.16 | 0.28 | -0.05 | -0.22 | -0.05 | 0.01 | -0.19 | -0.37 | -0.33 | -0.22 | -0.28 | -0.10 | -0.32 | -0.08 | -0.15 | -0.15 | -0.41 | -0.10 | -0.04 | 0.43 | 0.24 | 0.35 |
| HD 118096 | E | 4568 | 4.73 | 0.15 | -0.33 | -0.16 | -0.22 | -0.08 | 0.97 | -0.11 | -0.13 | -0.12 | 0.00 | -0.15 | -0.30 | -0.31 | -0.04 | -0.18 | -0.22 | -0.41 | 0.05 | 0.13 | -0.19 | -0.18 | 0.51 | 0.71 | 1.14 | 0.75 | 0.83 |
| HD 119332 | E | 5213 | 4.53 | 1.04 | 0.10 | 0.03 | 0.04 | 0.06 | 0.58 | 0.01 | 0.05 | 0.04 | 0.16 | 0.08 | 0.04 | -0.02 | 0.04 | 0.03 | 0.05 | -0.06 | 0.12 | 0.16 | 0.30 | -0.03 | 0.18 | 0.55 | 0.56 | 0.49 | 0.56 |
| HD 119802 | E | 4716 | 4.64 | 1.19 | 0.35 | 0.29 | 0.14 | 0.30 | 0.93 | 0.18 | 0.23 | 0.17 | 0.33 | 0.25 | 0.16 | 0.18 | 0.29 | 0.26 | 0.22 | 0.15 | 0.32 | 0.40 | 0.04 | 0.09 | 0.68 | 0.63 | 1.09 | 1.12 | 0.78 |
| HD 12051 | E | 5397 | 4.39 | 0.91 | 0.38 | 0.36 | 0.34 | 0.29 | 0.71 | 0.28 | 0.28 | 0.27 | 0.38 | 0.32 | 0.34 | 0.29 | 0.30 | 0.31 | 0.43 | 0.49 | 0.23 | 0.34 | 0.36 | 0.20 | 0.58 | 0.81 | 0.78 | 0.78 | 0.80 |
| HD 122120 | E | 4486 | 4.62 | 0.49 | 0.47 | 0.42 | 0.29 | 0.46 | 1.32 | 0.45 | 0.41 | 0.36 | 0.51 | 0.41 | 0.25 | 0.40 | 0.47 | 0.46 | 0.57 | 0.32 | 0.87 | 0.42 | 0.13 | 0.14 | 0.89 | 1.28 | 1.58 | 1.29 | 0.75 |
| HD 124292 | E | 5497 | 4.49 | 0.89 | 0.06 | 0.09 | 0.09 | 0.02 | 0.16 | 0.01 | 0.01 | 0.04 | 0.07 | 0.03 | 0.01 | -0.05 | 0.02 | -0.02 | 0.12 | 0.06 | 0.35 | 0.06 | 0.03 | -0.09 | 0.39 | 0.29 | 0.39 | 0.28 | 0.52 |
| HD 124642 | E | 4664 | 4.60 | 1.26 | 0.29 | 0.18 | 0.08 | 0.21 | 1.01 | 0.09 | 0.12 | 0.08 | 0.24 | 0.22 | 0.10 | 0.14 | 0.21 | 0.16 | 0.23 | 0.07 | 0.48 | 0.28 | 0.20 | 0.04 | 0.49 | 0.61 | 0.94 | 0.94 | 0.62 |
| HD 128429 | E | 6456 | 4.26 | 2.26 | 0.33 | 0.36 | 0.23 | 0.12 | 0.12 | 0.22 | 0.29 | 0.25 | 0.42 | 0.07 | -0.09 | 0.01 | 0.39 | 0.11 | -0.12 | -0.42 | | 0.25 | 0.63 | 0.05 | 0.47 | 0.42 | 0.41 | 0.32 | |
| HD 12846 | E | 5733 | 4.48 | 0.66 | -0.12 | -0.04 | -0.06 | -0.11 | -0.04 | -0.08 | -0.08 | -0.06 | -0.09 | -0.11 | -0.19 | -0.16 | -0.10 | -0.16 | -0.20 | -0.09 | -0.25 | -0.03 | -0.01 | -0.12 | 0.23 | 0.02 | 0.21 | 0.11 | 0.44 |
| HD 130307 | E | 5043 | 4.59 | 0.66 | 0.01 | 0.00 | -0.05 | -0.08 | 0.67 | -0.01 | -0.01 | 0.06 | 0.11 | 0.05 | -0.08 | -0.06 | 0.00 | -0.03 | 0.03 | -0.17 | 0.10 | 0.21 | 0.07 | 0.07 | 0.49 | 0.66 | 0.85 | 0.50 | 0.45 |
| HD 132142 | E | 5229 | 4.58 | 0.47 | -0.19 | -0.08 | -0.01 | -0.16 | 0.42 | -0.14 | -0.14 | -0.04 | 0.01 | -0.21 | -0.39 | -0.33 | -0.15 | -0.29 | -0.20 | -0.13 | -0.06 | -0.18 | 0.10 | -0.47 | -0.04 | 0.09 | 0.25 | 0.15 | 0.20 |
| HD 132254 | E | 6279 | 4.21 | 1.75 | 0.15 | 0.25 | 0.15 | 0.15 | 0.08 | 0.26 | 0.27 | 0.22 | 0.23 | 0.19 | 0.10 | 0.12 | 0.19 | 0.14 | -0.06 | -0.07 | 0.06 | 0.29 | 0.24 | 0.20 | 0.44 | 0.40 | 0.38 | 0.24 | 0.56 |
| HD 133002 | E | 5562 | 3.55 | 1.29 | -0.11 | -0.17 | -0.11 | -0.20 | -0.25 | -0.18 | -0.25 | -0.22 | -0.28 | -0.28 | -0.45 | -0.32 | -0.24 | -0.33 | -0.40 | -0.25 | -0.11 | -0.27 | -0.26 | -0.20 | -0.20 | -0.14 | -0.15 | -0.32 | 0.43 |
| HD 13403 | E | 5617 | 4.00 | 0.98 | -0.16 | 0.06 | 0.03 | -0.09 | -0.09 | -0.08 | -0.04 | -0.05 | -0.15 | -0.22 | -0.41 | -0.28 | -0.16 | -0.24 | -0.23 | -0.07 | -0.16 | -0.25 | 0.12 | -0.30 | -0.05 | -0.01 | 0.01 | -0.02 | 0.37 |
| HD 135204 | E | 5457 | 4.49 | 0.93 | 0.13 | 0.18 | 0.23 | 0.07 | 0.28 | 0.10 | 0.13 | 0.19 | 0.24 | 0.05 | -0.01 | 0.00 | 0.13 | 0.00 | 0.01 | 0.19 | -0.01 | 0.12 | 0.13 | -0.06 | 0.28 | 0.35 | 0.44 | 0.45 | 0.60 |
| HD 135599 | E | 5270 | 4.55 | 1.03 | 0.09 | 0.07 | 0.05 | 0.06 | 0.26 | 0.15 | 0.00 | 0.11 | 0.18 | 0.14 | 0.04 | 0.05 | 0.05 | 0.03 | 0.03 | -0.04 | 0.32 | 0.28 | 0.20 | 0.23 | 0.49 | 0.50 | 0.62 | 0.59 | 0.51 |

| Star | | Teff | log g | | | | | | | | | | | | | | | | | | | | | | | | | | |
|---|---|---|---|---|---|---|---|---|---|---|---|---|---|---|---|---|---|---|---|---|---|---|---|---|---|---|---|---|---|
| HD 13579 | E | 5133 | 4.51 | 0.55 | 0.77 | 0.64 | 0.52 | 0.45 | 0.97 | 0.43 | 0.52 | 0.52 | 0.73 | 0.57 | 0.53 | 0.42 | 0.51 | 0.55 | 0.74 | 0.76 | 0.45 | 0.63 | 0.45 | 0.29 | 1.03 | 1.00 | 1.31 | 1.26 | 1.02 |
| HD 139777 | E | 5775 | 4.47 | 1.52 | 0.06 | 0.06 | 0.09 | 0.05 | 0.09 | 0.16 | 0.06 | 0.08 | 0.10 | 0.12 | -0.01 | 0.03 | 0.03 | 0.01 | -0.19 | -0.33 | 0.26 | 0.30 | 0.34 | 0.21 | 0.31 | 0.42 | 0.43 | 0.22 | 0.32 |
| HD 139813 | E | 5394 | 4.56 | 1.45 | 0.12 | 0.14 | 0.09 | 0.10 | 0.25 | 0.17 | 0.04 | 0.10 | 0.14 | 0.17 | 0.02 | 0.06 | 0.04 | 0.06 | -0.04 | -0.20 | 0.25 | 0.24 | 0.34 | 0.19 | 0.46 | 0.64 | 0.54 | 0.54 | 0.58 |
| HD 140283 | E | 5823 | 3.80 | 2.24 | -2.41 | -2.02 | | -1.51 | | -2.22 | | | -2.13 | -2.71 | -2.49 | -2.65 | -2.30 | -1.36 | -2.11 | | -2.29 | | -1.62 | 0.27 | -3.16 | | -0.45 | -0.85 | | |
| HD 14348 | E | 6036 | 4.03 | 1.62 | 0.47 | 0.41 | 0.26 | 0.26 | 0.30 | 0.24 | 0.32 | 0.23 | 0.26 | 0.24 | 0.24 | 0.19 | 0.26 | 0.26 | 0.11 | 0.12 | 0.62 | 0.29 | 0.38 | 0.15 | 0.32 | 0.40 | 0.29 | 0.33 | 0.69 |
| HD 14374 | E | 5474 | 4.55 | 0.71 | 0.15 | 0.10 | 0.13 | 0.09 | 0.46 | 0.18 | 0.09 | 0.19 | 0.20 | 0.19 | 0.17 | 0.09 | 0.13 | 0.11 | 0.06 | 0.17 | 0.38 | 0.25 | 0.30 | 0.20 | 0.62 | 0.64 | 0.73 | 0.56 | 0.78 |
| HD 144287 | E | 5315 | 4.42 | 0.88 | -0.01 | 0.05 | 0.09 | 0.03 | 0.40 | 0.00 | 0.03 | 0.04 | 0.02 | 0.02 | -0.11 | -0.06 | 0.01 | -0.06 | 0.02 | 0.09 | -0.15 | 0.07 | 0.12 | -0.13 | 0.45 | 0.43 | 0.47 | 0.47 | 0.71 |
| HD 145435 | E | 6083 | 4.12 | 1.42 | 0.15 | 0.10 | 0.00 | 0.10 | 0.18 | 0.12 | 0.07 | 0.01 | 0.05 | 0.05 | 0.00 | 0.02 | 0.06 | 0.01 | -0.18 | -0.08 | -0.10 | 0.17 | 0.24 | 0.08 | 0.18 | 0.24 | 0.11 | 0.15 | 0.61 |
| HD 145729 | E | 6028 | 4.39 | 1.09 | -0.04 | 0.01 | 0.05 | -0.04 | 0.10 | 0.05 | 0.09 | 0.01 | 0.05 | 0.06 | -0.04 | -0.02 | 0.08 | -0.06 | -0.15 | -0.04 | | 0.19 | 0.35 | 0.14 | 0.57 | 0.40 | 0.33 | 0.16 | 0.65 |
| HD 146946 | E | 5738 | 4.31 | 0.96 | -0.16 | -0.20 | -0.21 | -0.22 | -0.08 | -0.10 | -0.13 | -0.14 | -0.11 | -0.21 | -0.37 | -0.27 | -0.18 | -0.31 | -0.45 | -0.33 | -0.02 | -0.06 | 0.03 | -0.15 | 0.12 | 0.10 | 0.03 | 0.18 | 0.25 |
| HD 153525 | E | 4820 | 4.63 | 1.75 | -0.01 | -0.10 | -0.09 | 0.09 | 0.85 | 0.01 | -0.05 | -0.13 | -0.02 | 0.01 | -0.15 | -0.07 | 0.05 | -0.03 | -0.14 | -0.20 | -0.08 | 0.03 | -0.10 | -0.13 | 0.43 | 0.67 | 0.85 | 0.46 | 0.55 |
| HD 154931 | E | 5869 | 3.97 | 1.42 | 0.01 | 0.01 | -0.01 | 0.00 | -0.05 | -0.01 | 0.00 | -0.05 | -0.05 | -0.05 | -0.09 | -0.09 | -0.04 | -0.08 | -0.16 | -0.18 | 0.05 | -0.04 | 0.07 | -0.13 | 0.14 | 0.12 | 0.03 | 0.05 | 0.20 |
| HD 155712 | E | 4936 | 4.58 | 0.24 | 0.17 | 0.15 | 0.06 | 0.08 | 0.46 | 0.05 | 0.13 | 0.12 | 0.32 | 0.13 | -0.01 | 0.05 | 0.09 | 0.10 | 0.38 | 0.02 | 0.12 | 0.21 | 0.08 | 0.00 | 0.65 | 0.78 | 1.10 | 0.73 | 0.55 |
| HD 15632 | E | 5749 | 4.48 | 0.81 | 0.20 | 0.18 | 0.10 | 0.15 | 0.21 | 0.19 | 0.27 | 0.19 | 0.20 | 0.22 | 0.16 | 0.15 | 0.14 | 0.17 | 0.23 | 0.23 | 0.37 | 0.39 | 0.38 | 0.25 | 0.60 | 0.68 | 0.64 | 0.53 | 0.62 |
| HD 156985 | E | 4778 | 4.59 | 0.51 | 0.08 | 0.16 | 0.05 | 0.10 | 0.71 | 0.05 | 0.09 | 0.07 | 0.27 | 0.12 | -0.03 | 0.06 | 0.11 | 0.08 | 0.05 | 0.07 | -0.02 | 0.21 | 0.00 | -0.07 | 0.31 | 0.77 | 1.21 | 0.92 | 0.68 |
| HD 157089 | E | 5830 | 4.14 | 1.17 | -0.46 | -0.18 | -0.30 | -0.31 | -0.39 | -0.30 | -0.28 | -0.29 | -0.32 | -0.49 | -0.77 | -0.53 | -0.36 | -0.52 | -0.67 | -0.55 | -0.19 | -0.52 | 0.07 | -0.42 | -0.08 | -0.17 | -0.10 | -0.13 | 0.24 |
| HD 159062 | E | 5385 | 4.48 | 0.27 | -0.16 | -0.05 | -0.01 | -0.12 | 0.05 | -0.12 | -0.05 | 0.00 | -0.03 | -0.20 | -0.34 | -0.26 | -0.11 | -0.22 | -0.03 | -0.12 | 0.36 | 0.19 | 0.08 | 0.07 | 0.35 | 0.47 | 0.50 | 0.34 | 0.40 |
| HD 159482 | E | 5805 | 4.36 | 0.86 | -0.63 | -0.35 | -0.59 | -0.45 | -0.05 | -0.41 | -0.47 | -0.42 | -0.45 | -0.65 | -0.96 | -0.71 | -0.41 | -0.67 | -0.92 | -0.64 | -0.80 | -0.68 | -0.12 | -0.67 | -0.16 | -0.36 | -0.36 | -0.14 | -0.16 |
| HD 160964 | E | 4589 | 4.67 | 0.78 | 0.16 | 0.01 | -0.06 | 0.09 | 1.10 | 0.09 | 0.02 | 0.02 | 0.17 | 0.10 | -0.09 | 0.04 | 0.12 | 0.06 | 0.16 | 0.10 | 0.20 | 0.14 | 0.08 | -0.14 | 0.60 | 1.02 | 1.15 | 0.90 | 0.40 |
| HD 161098 | E | 5637 | 4.50 | 0.69 | -0.13 | -0.02 | -0.07 | -0.12 | 0.21 | -0.03 | -0.11 | -0.05 | -0.05 | -0.08 | -0.15 | -0.13 | -0.09 | -0.13 | -0.20 | -0.07 | -0.08 | 0.09 | 0.24 | -0.01 | 0.27 | 0.34 | 0.44 | 0.37 | 0.51 |
| HD 163183 | E | 5928 | 4.46 | 1.70 | -0.04 | 0.08 | 0.01 | 0.02 | 0.04 | 0.04 | -0.05 | 0.04 | 0.00 | 0.05 | -0.12 | -0.03 | 0.05 | -0.04 | -0.34 | -0.33 | -0.15 | 0.14 | 0.13 | 0.23 | 0.31 | 0.38 | 0.37 | 0.17 | 0.29 |
| HD 16397 | E | 5821 | 4.37 | 0.92 | -0.39 | -0.25 | -0.29 | -0.31 | -0.19 | -0.32 | -0.27 | -0.28 | -0.36 | -0.43 | -0.61 | -0.47 | -0.30 | -0.46 | -0.53 | -0.39 | -0.25 | -0.52 | -0.04 | -0.43 | 0.02 | -0.22 | -0.17 | -0.10 | 0.15 |
| HD 164651 | E | 5599 | 4.45 | 0.76 | 0.07 | 0.07 | 0.12 | 0.05 | 0.28 | 0.08 | 0.09 | 0.12 | 0.13 | 0.11 | 0.10 | 0.06 | 0.10 | 0.08 | 0.19 | 0.00 | -0.16 | 0.15 | 0.29 | 0.01 | 0.59 | 0.38 | 0.53 | 0.35 | 0.71 |
| HD 165173 | E | 5484 | 4.54 | 0.74 | 0.20 | 0.20 | 0.21 | 0.14 | 0.51 | 0.14 | 0.23 | 0.20 | 0.21 | 0.18 | 0.19 | 0.15 | 0.17 | 0.17 | 0.30 | 0.45 | 0.34 | 0.37 | 0.29 | 0.20 | 0.85 | 0.60 | 0.62 | 0.59 | 0.75 |
| HD 165401 | E | 5816 | 4.44 | 0.97 | -0.25 | -0.09 | -0.10 | -0.17 | -0.09 | -0.21 | -0.18 | -0.18 | -0.23 | -0.35 | -0.53 | -0.37 | -0.18 | -0.34 | -0.48 | -0.20 | -0.66 | -0.21 | 0.18 | -0.37 | 0.25 | -0.11 | -0.15 | 0.04 | 0.60 |
| HD 165476 | E | 5816 | 4.25 | 1.02 | 0.03 | 0.08 | 0.07 | 0.02 | 0.11 | 0.08 | 0.07 | 0.05 | 0.04 | 0.07 | -0.02 | 0.01 | 0.00 | -0.03 | -0.03 | 0.05 | 0.04 | 0.04 | 0.10 | 0.05 | 0.25 | 0.19 | 0.18 | 0.19 | 0.43 |
| HD 165590 | E | 5663 | 4.09 | 1.56 | -0.01 | -0.82 | 0.05 | -0.15 | -0.02 | -0.10 | -0.08 | -0.13 | 0.10 | -0.03 | -0.13 | -0.18 | 0.18 | -0.17 | | -0.95 | | 0.42 | 0.89 | | 1.34 | 0.41 | 0.23 | 0.75 | | |
| HD 165670 | E | 6285 | 4.32 | 1.88 | 0.09 | 0.21 | -0.08 | 0.09 | 0.08 | 0.09 | 0.25 | 0.16 | 0.29 | 0.08 | 0.02 | 0.01 | 0.33 | 0.02 | -0.24 | -0.01 | -0.13 | 0.17 | 0.94 | -0.01 | 0.08 | 0.46 | 0.40 | 0.39 | 0.68 |
| HD 165672 | E | 5866 | 4.37 | 1.26 | 0.30 | 0.31 | 0.21 | 0.23 | 0.34 | 0.27 | 0.29 | 0.22 | 0.27 | 0.26 | 0.27 | 0.21 | 0.25 | 0.22 | 0.09 | 0.25 | 0.30 | 0.41 | 0.41 | 0.18 | 0.42 | 0.58 | 0.45 | 0.61 | 0.97 |
| HD 166183 | E | 6380 | 4.05 | 1.92 | 0.10 | 0.06 | 0.12 | 0.07 | 0.01 | 0.08 | 0.12 | 0.08 | 0.14 | 0.00 | -0.11 | -0.02 | 0.26 | -0.03 | -0.22 | -0.28 | 0.33 | 0.21 | 0.44 | 0.13 | 0.17 | 0.29 | 0.14 | 0.26 | 0.40 |
| HD 166435 | E | 5811 | 4.47 | 1.75 | 0.08 | 0.00 | 0.03 | 0.03 | 0.19 | 0.12 | 0.02 | 0.07 | 0.06 | 0.07 | -0.05 | 0.00 | 0.05 | -0.06 | -0.31 | -0.28 | 0.03 | 0.18 | 0.28 | 0.28 | 0.43 | 0.34 | 0.35 | 0.30 | 0.43 |
| HD 167278 | E | 6483 | 3.82 | 2.70 | -0.07 | -0.01 | -0.09 | -0.04 | -0.18 | -0.12 | -0.25 | -0.01 | 0.15 | -0.07 | -0.22 | -0.24 | 0.31 | -0.18 | -0.28 | -0.55 | 1.03 | -0.03 | 0.19 | -0.47 | 0.28 | 0.39 | -0.03 | 0.08 | 0.66 |
| HD 169822 | E | 5529 | 4.53 | 1.13 | -0.07 | -0.14 | 0.03 | -0.04 | 0.00 | -0.11 | -0.06 | -0.08 | -0.03 | -0.07 | -0.22 | -0.16 | -0.08 | -0.13 | -0.24 | -0.12 | -0.39 | 0.01 | 0.11 | -0.23 | 0.15 | 0.14 | 0.26 | 0.16 | 0.36 |
| HD 170008 | E | 4978 | 3.33 | 0.90 | -0.21 | -0.23 | -0.23 | -0.12 | 0.15 | -0.13 | -0.27 | -0.27 | -0.31 | -0.25 | -0.35 | -0.28 | -0.31 | -0.25 | -0.30 | -0.09 | -0.21 | -0.21 | -0.29 | -0.13 | 0.01 | 0.10 | 0.21 | 0.01 | 0.18 |
| HD 170291 | E | 6327 | 4.22 | 1.90 | 0.21 | 0.22 | 0.14 | 0.14 | 0.26 | 0.14 | 0.19 | 0.18 | 0.23 | 0.16 | 0.10 | 0.07 | 0.29 | 0.09 | -0.05 | -0.01 | 0.45 | 0.18 | 0.50 | 0.02 | 0.55 | 0.23 | 0.24 | 0.40 | 0.48 |
| HD 170512 | E | 6152 | 4.27 | 1.58 | 0.42 | 0.35 | 0.25 | 0.27 | 0.38 | 0.33 | 0.30 | 0.28 | 0.26 | 0.29 | 0.27 | 0.21 | 0.30 | 0.28 | 0.12 | 0.20 | 0.18 | 0.31 | 0.66 | 0.11 | 0.44 | 0.55 | 0.46 | 0.43 | 0.55 |
| HD 170579 | E | 6417 | 4.22 | 2.03 | -0.13 | -0.05 | -0.23 | -0.08 | 0.08 | -0.16 | 0.00 | -0.04 | 0.08 | -0.11 | -0.33 | -0.22 | 0.24 | -0.17 | -0.38 | -0.17 | 0.53 | 0.23 | 0.78 | 0.49 | 0.90 | 0.52 | 0.46 | 0.41 | 0.40 |
| HD 171314 | E | 4530 | 4.60 | 0.69 | 0.36 | 0.38 | 0.28 | 0.36 | 1.07 | 0.43 | 0.47 | 0.26 | 0.44 | 0.36 | 0.22 | 0.34 | 0.42 | 0.34 | 0.49 | 0.30 | 0.46 | 0.28 | 0.24 | 0.06 | 1.06 | 0.96 | 1.44 | 1.14 | 0.97 |

| ID | Type | T | logg | [Fe/H] | | | | | | | | | | | | | | | | | | | | | | | | | |
|---|---|---|---|---|---|---|---|---|---|---|---|---|---|---|---|---|---|---|---|---|---|---|---|---|---|---|---|---|---|
| HD 171888 | E | 6095 | 4.00 | 1.56 | 0.22 | 0.10 | 0.08 | 0.11 | 0.16 | 0.14 | 0.10 | 0.13 | 0.14 | 0.15 | 0.08 | 0.07 | 0.14 | 0.09 | -0.02 | 0.01 | 0.25 | 0.25 | 0.31 | 0.10 | 0.15 | 0.23 | 0.18 | 0.17 | 0.46 |
| HD 171951 | E | 6094 | 4.03 | 1.39 | -0.18 | -0.09 | -0.36 | -0.18 | -0.21 | -0.14 | -0.03 | -0.17 | -0.20 | -0.22 | -0.32 | -0.26 | -0.01 | -0.27 | -0.42 | -0.39 | 0.93 | -0.17 | 0.43 | -0.13 | -0.09 | 0.05 | 0.11 | 0.01 | 0.14 |
| HD 171953 | E | 6480 | 3.82 | 2.28 | 0.95 | 0.77 | 0.60 | 0.26 | -0.22 | 0.71 | 0.40 | 0.88 | 0.73 | 0.55 | 0.81 | 0.29 | 0.45 | 0.79 | | -0.02 | | 1.49 | 0.43 | 0.53 | | 0.60 | 1.29 | 3.31 | 0.76 |
| HD 172675 | E | 6303 | 4.37 | 1.66 | -0.08 | -0.03 | -0.01 | 0.00 | 0.12 | 0.06 | 0.13 | 0.04 | 0.02 | 0.02 | -0.13 | -0.05 | 0.28 | -0.07 | -0.36 | -0.17 | -0.05 | 0.17 | 0.40 | 0.11 | 0.22 | 0.34 | 0.22 | 0.15 | 0.25 |
| HD 172718 | E | 6132 | 3.95 | 1.58 | 0.01 | 0.07 | -0.11 | -0.04 | -0.04 | 0.02 | 0.00 | -0.01 | -0.03 | -0.04 | -0.11 | -0.09 | 0.10 | -0.09 | -0.19 | -0.19 | 0.10 | 0.04 | 0.10 | 0.04 | 0.28 | 0.11 | 0.08 | 0.03 | 0.74 |
| HD 172961 | E | 6571 | 4.32 | 2.21 | 0.00 | 0.09 | -0.09 | 0.06 | -0.02 | 0.02 | 0.06 | 0.14 | 0.16 | -0.03 | 0.03 | -0.05 | 0.37 | -0.02 | -0.50 | -0.46 | | 0.37 | 0.14 | 0.23 | 0.13 | 0.31 | 0.34 | 0.35 | 0.78 |
| HD 173174 | E | 6009 | 3.98 | 1.70 | 0.30 | 0.27 | 0.31 | 0.20 | 0.14 | 0.31 | 0.19 | 0.21 | 0.26 | 0.25 | 0.23 | 0.19 | 0.28 | 0.21 | -0.02 | 0.00 | 0.40 | 0.30 | 0.79 | 0.12 | 0.29 | 0.32 | 0.37 | 0.36 | 0.42 |
| HD 173605 | E | 5722 | 4.01 | 2.40 | 0.18 | 0.33 | 0.39 | 0.24 | 0.77 | 0.15 | 0.28 | 0.17 | 0.14 | 0.21 | -0.01 | 0.06 | 0.33 | 0.11 | -0.27 | -0.06 | | 0.20 | 0.46 | -0.08 | 0.20 | 0.14 | 0.55 | 0.68 | 1.26 |
| HD 173634 | E | 6505 | 3.76 | 2.94 | 0.28 | | | 0.15 | 0.19 | 0.19 | 0.46 | 0.51 | 0.59 | 0.33 | 0.38 | 0.12 | 0.73 | 0.17 | -0.31 | -0.45 | | 0.53 | 2.05 | -0.12 | 0.62 | 0.34 | 0.45 | 0.85 | 0.79 |
| HD 17382 | E | 5245 | 4.49 | 1.06 | 0.17 | 0.11 | 0.08 | 0.11 | 0.35 | 0.15 | 0.04 | 0.10 | 0.16 | 0.18 | 0.11 | 0.09 | 0.06 | 0.09 | 0.00 | 0.13 | 0.25 | 0.20 | 0.17 | 0.17 | 0.54 | 0.45 | 0.55 | 0.60 | 0.68 |
| HD 174080 | E | 4676 | 4.60 | 1.12 | 0.46 | 0.33 | 0.26 | 0.44 | 1.20 | 0.37 | 0.24 | 0.24 | 0.44 | 0.32 | 0.20 | 0.22 | 0.37 | 0.32 | 0.31 | 0.21 | 0.29 | 0.25 | 0.12 | 0.11 | 0.55 | 0.80 | 1.27 | 0.75 | 0.76 |
| HD 174719 | E | 5647 | 4.50 | 0.88 | -0.17 | -0.07 | -0.10 | -0.11 | -0.02 | -0.05 | -0.03 | -0.03 | 0.06 | -0.02 | -0.17 | -0.11 | -0.03 | -0.13 | -0.22 | -0.15 | 0.14 | 0.10 | 0.43 | 0.07 | 0.43 | 0.64 | 0.43 | 0.37 | 0.40 |
| HD 175272 | E | 6638 | 3.98 | 2.93 | 0.28 | 0.11 | | 0.25 | 0.30 | 0.18 | 0.46 | 0.39 | 0.59 | 0.28 | 0.11 | 0.12 | 0.71 | 0.20 | -0.03 | -0.48 | | 0.35 | 1.08 | -0.36 | 0.30 | 0.38 | 0.51 | 1.06 | 0.74 |
| HD 175726 | E | 6069 | 4.43 | 2.00 | -0.07 | 0.05 | 0.21 | 0.04 | 0.04 | 0.14 | 0.10 | 0.09 | 0.01 | 0.06 | -0.10 | -0.04 | 0.19 | -0.03 | -0.21 | -0.37 | | 0.22 | 0.53 | 0.27 | 0.46 | 0.52 | 0.43 | 0.36 | 0.00 |
| HD 175805 | E | 6318 | 3.72 | 3.42 | | | 0.44 | 0.54 | 1.28 | 0.45 | 0.42 | 0.71 | 0.74 | 0.46 | 0.45 | 0.39 | 1.05 | 0.50 | 0.11 | -0.23 | 1.29 | 0.63 | 1.72 | -0.09 | 1.01 | 0.60 | 0.70 | 1.84 | |
| HD 175806 | E | 6171 | 3.63 | 1.99 | 0.18 | 0.18 | 0.04 | 0.07 | -0.03 | 0.11 | 0.05 | 0.06 | 0.12 | 0.05 | -0.08 | -0.04 | 0.06 | -0.02 | -0.34 | -0.20 | 0.22 | 0.10 | 0.18 | 0.05 | 0.09 | 0.10 | -0.05 | 0.00 | 0.17 |
| HD 176118 | E | 6665 | 4.04 | 2.14 | 0.38 | 0.40 | 0.17 | 0.29 | 0.29 | 0.30 | 0.38 | 0.32 | 0.49 | 0.30 | 0.13 | 0.22 | 0.62 | 0.25 | 0.02 | 0.01 | 0.20 | 0.34 | 0.91 | 0.32 | 0.42 | 0.42 | 0.33 | 0.38 | 0.52 |
| HD 176377 | E | 5868 | 4.46 | 0.99 | -0.18 | -0.12 | -0.13 | -0.16 | -0.14 | -0.13 | -0.15 | -0.16 | -0.22 | -0.19 | -0.28 | -0.22 | -0.18 | -0.26 | -0.40 | -0.23 | -0.20 | -0.09 | 0.00 | -0.02 | 0.06 | 0.15 | 0.11 | 0.15 | 0.35 |
| HD 177749 | E | 6403 | 3.97 | 1.92 | 0.12 | 0.10 | -0.13 | 0.08 | 0.09 | 0.19 | 0.14 | 0.10 | 0.15 | 0.09 | -0.06 | 0.00 | 0.15 | 0.01 | -0.13 | -0.12 | 0.44 | 0.26 | 0.40 | 0.17 | 0.21 | 0.28 | 0.26 | 0.12 | 0.24 |
| HD 177904 | E | 6902 | 3.84 | 3.31 | 0.20 | 0.75 | 0.55 | 0.23 | 0.07 | 0.29 | 0.21 | 0.50 | 1.01 | 0.26 | 0.19 | 0.15 | 0.72 | 0.32 | -0.43 | -0.44 | | 0.35 | 0.58 | -0.02 | 0.87 | 0.64 | 0.52 | 0.73 | 0.42 |
| HD 178126 | E | 4541 | 4.66 | 0.15 | -0.09 | 0.07 | -0.02 | 0.08 | 1.15 | -0.10 | 0.04 | 0.01 | 0.08 | -0.12 | -0.30 | -0.34 | 0.13 | -0.08 | 0.14 | 0.13 | -0.04 | -0.03 | -0.09 | -0.47 | 0.58 | 0.62 | 1.20 | 0.82 | 0.74 |
| HD 180161 | E | 5400 | 4.53 | 1.34 | 0.25 | 0.26 | 0.20 | 0.21 | 0.38 | 0.24 | 0.15 | 0.18 | 0.21 | 0.26 | 0.21 | 0.18 | 0.22 | 0.20 | 0.08 | 0.19 | 0.29 | 0.25 | 0.27 | 0.16 | 0.47 | 0.50 | 0.55 | 0.75 | 0.70 |
| HD 180945 | E | 6415 | 4.04 | 2.36 | 0.23 | 0.35 | 0.15 | 0.16 | 0.12 | 0.17 | 0.25 | 0.25 | 0.28 | 0.14 | -0.07 | 0.09 | 0.13 | 0.15 | -0.30 | -0.26 | | 0.19 | 0.52 | 0.22 | 0.68 | 0.37 | 0.37 | 0.41 | 0.67 |
| HD 181096 | E | 6270 | 3.92 | 1.85 | -0.14 | -0.11 | -0.08 | -0.11 | -0.12 | -0.08 | -0.06 | -0.11 | -0.12 | -0.18 | -0.32 | -0.24 | 0.13 | -0.25 | -0.41 | -0.37 | 0.11 | 0.02 | 0.09 | 0.01 | 0.12 | 0.07 | 0.03 | -0.05 | 0.17 |
| HD 181420 | E | 6606 | 4.18 | 2.72 | 0.28 | 0.10 | | 0.25 | 0.21 | 0.28 | 0.21 | 0.40 | 0.48 | 0.21 | 0.02 | 0.12 | 0.55 | 0.16 | 0.02 | -0.35 | | 0.38 | 1.09 | 0.13 | 0.91 | 0.39 | 0.39 | 0.81 | 0.55 |
| HD 181806 | E | 6404 | 3.92 | 1.82 | 0.23 | 0.16 | 0.24 | 0.07 | 0.08 | 0.21 | 0.12 | 0.17 | 0.29 | 0.16 | -0.01 | 0.05 | 0.31 | 0.06 | -0.21 | -0.21 | 0.44 | 0.30 | 0.40 | 0.10 | 0.27 | 0.18 | 0.09 | 0.19 | 0.19 |
| HD 182274 | E | 6307 | 4.32 | 1.47 | -0.09 | -0.12 | -0.05 | -0.09 | -0.11 | -0.09 | -0.07 | -0.10 | -0.06 | -0.12 | -0.29 | -0.19 | 0.13 | -0.22 | -0.30 | -0.18 | 0.84 | 0.61 | 0.70 | 0.69 | 0.74 | 0.55 | 0.40 | 0.20 | 0.72 |
| HD 182736 | E | 5237 | 3.66 | 1.08 | 0.02 | -0.03 | 0.00 | -0.03 | 0.32 | -0.03 | -0.11 | -0.06 | -0.09 | -0.05 | -0.09 | -0.11 | -0.11 | -0.11 | -0.20 | 0.03 | -0.05 | 0.06 | -0.06 | -0.07 | 0.19 | 0.21 | 0.25 | 0.23 | 0.40 |
| HD 182905 | E | 5376 | 3.88 | 1.18 | 0.26 | 0.26 | 0.26 | 0.19 | 0.38 | 0.20 | 0.14 | 0.15 | 0.20 | 0.20 | 0.18 | 0.12 | 0.15 | 0.15 | 0.30 | 0.21 | 0.18 | 0.20 | 0.12 | -0.01 | 0.46 | 0.57 | 0.41 | 0.29 | 0.38 |
| HD 183341 | E | 5952 | 4.27 | 1.22 | 0.16 | 0.15 | 0.15 | 0.11 | 0.04 | 0.19 | 0.18 | 0.13 | 0.14 | 0.17 | 0.16 | 0.11 | 0.15 | 0.12 | 0.02 | 0.06 | 0.19 | 0.24 | 0.35 | 0.17 | 0.46 | 0.47 | 0.44 | 0.39 | 0.38 |
| HD 183658 | E | 5828 | 4.45 | 0.87 | 0.25 | 0.22 | 0.16 | 0.12 | 0.34 | 0.21 | 0.29 | 0.22 | 0.24 | 0.23 | 0.19 | 0.16 | 0.20 | 0.20 | 0.24 | 0.30 | 0.26 | 0.37 | 0.54 | 0.22 | 0.65 | 0.50 | 0.52 | 0.46 | 0.77 |
| HD 183870 | E | 5015 | 4.62 | 0.51 | 0.11 | 0.02 | 0.06 | 0.10 | 0.54 | 0.20 | 0.14 | 0.18 | 0.27 | 0.20 | 0.06 | 0.09 | 0.15 | 0.11 | 0.25 | -0.11 | 0.43 | 0.32 | 0.23 | 0.28 | 0.68 | 0.89 | 1.14 | 1.10 | 0.53 |
| HD 184499 | E | 5807 | 4.10 | 1.11 | -0.35 | -0.13 | -0.18 | -0.26 | -0.35 | -0.23 | -0.24 | -0.22 | -0.29 | -0.44 | -0.68 | -0.49 | -0.28 | -0.44 | -0.62 | -0.37 | -0.62 | -0.49 | 0.08 | -0.52 | -0.15 | -0.24 | -0.13 | -0.26 | 0.30 |
| HD 184768 | E | 5635 | 4.29 | 0.85 | 0.09 | 0.22 | 0.19 | 0.10 | 0.24 | 0.08 | 0.14 | 0.14 | 0.08 | 0.05 | -0.03 | 0.00 | 0.05 | 0.04 | 0.24 | 0.23 | 0.35 | 0.18 | 0.26 | 0.00 | 0.18 | 0.34 | 0.40 | 0.32 | 0.55 |
| HD 184960 | E | 6290 | 4.23 | 1.63 | -0.02 | 0.16 | 0.04 | 0.06 | 0.08 | 0.14 | 0.08 | 0.12 | 0.03 | 0.04 | -0.16 | -0.02 | 0.09 | -0.05 | -0.34 | -0.26 | 0.37 | 0.16 | 0.38 | 0.23 | 0.57 | 0.34 | 0.12 | 0.14 | 1.05 |
| HD 185269 | E | 5987 | 3.97 | 1.59 | 0.36 | 0.25 | 0.21 | 0.22 | 0.26 | 0.26 | 0.27 | 0.27 | 0.18 | 0.25 | 0.16 | 0.17 | 0.24 | 0.19 | 0.15 | 0.26 | 0.25 | 0.31 | 0.44 | 0.28 | 0.39 | 0.39 | 0.44 | 0.28 | 0.69 |
| HD 185414 | E | 5806 | 4.45 | 0.73 | -0.05 | -0.01 | -0.01 | -0.01 | 0.05 | 0.04 | -0.07 | -0.02 | -0.05 | -0.02 | -0.12 | -0.06 | -0.09 | -0.08 | -0.20 | -0.08 | 0.13 | 0.22 | 0.17 | 0.20 | 0.24 | 0.42 | 0.26 | 0.23 | 0.61 |
| HD 186104 | E | 5759 | 4.33 | 1.09 | 0.27 | 0.22 | 0.20 | 0.17 | 0.22 | 0.21 | 0.19 | 0.18 | 0.20 | 0.20 | 0.19 | 0.16 | 0.17 | 0.17 | 0.17 | 0.06 | 0.40 | 0.31 | 0.45 | 0.20 | 0.27 | 0.53 | 0.41 | 0.36 | 0.70 |

| HD 186226 | E | 6371 | 3.93 | 2.33 | 0.45 | 0.36 | 0.47 | 0.29 | 0.32 | 0.33 | 0.40 | 0.31 | 0.32 | 0.27 | 0.25 | 0.21 | 0.49 | 0.30 | 0.31 | -0.02 | | 0.27 | 0.61 | 0.08 | 0.55 | 0.39 | 0.31 | 0.58 | 0.40 |
|---|---|---|---|---|---|---|---|---|---|---|---|---|---|---|---|---|---|---|---|---|---|---|---|---|---|---|---|---|---|
| HD 186379 | E | 5923 | 3.98 | 1.33 | -0.29 | -0.19 | -0.32 | -0.22 | -0.30 | -0.18 | -0.17 | -0.23 | -0.22 | -0.27 | -0.42 | -0.32 | -0.21 | -0.34 | -0.46 | -0.43 | -0.44 | | -0.24 | 0.06 | -0.25 | -0.05 | -0.11 | -0.16 | -0.19 | 0.18 |
| HD 186413 | E | 5918 | 4.16 | 1.29 | 0.08 | 0.20 | 0.12 | 0.08 | 0.09 | 0.15 | 0.08 | 0.08 | 0.12 | 0.11 | 0.06 | 0.04 | 0.12 | 0.06 | 0.00 | 0.08 | 0.36 | | 0.17 | 0.34 | -0.02 | 0.23 | 0.27 | 0.31 | 0.06 | 0.46 |
| HD 18757 | E | 5674 | 4.33 | 0.70 | -0.09 | 0.08 | 0.00 | -0.07 | -0.02 | -0.05 | 0.00 | 0.00 | -0.11 | -0.18 | -0.33 | -0.23 | -0.09 | -0.18 | -0.18 | 0.01 | -0.15 | | -0.13 | -0.06 | -0.25 | 0.16 | -0.06 | 0.10 | 0.09 | 0.40 |
| HD 18768 | E | 5815 | 3.85 | 1.37 | -0.44 | -0.37 | -0.53 | -0.37 | -0.38 | -0.40 | -0.40 | -0.41 | -0.39 | -0.46 | -0.69 | -0.52 | -0.37 | -0.52 | -0.63 | -0.53 | -0.17 | | -0.51 | -0.10 | -0.43 | -0.12 | -0.27 | -0.23 | -0.36 | 0.62 |
| HD 187897 | E | 5905 | 4.35 | 1.39 | 0.22 | 0.20 | 0.18 | 0.18 | 0.17 | 0.18 | 0.14 | 0.17 | 0.22 | 0.23 | 0.16 | 0.15 | 0.17 | 0.14 | 0.01 | 0.10 | 0.30 | | 0.33 | 0.40 | 0.24 | 0.43 | 0.51 | 0.36 | 0.58 | 0.53 |
| HD 188326 | E | 5342 | 3.85 | 1.00 | 0.10 | 0.15 | 0.19 | 0.07 | 0.28 | 0.09 | 0.03 | 0.07 | 0.05 | -0.02 | -0.06 | -0.04 | -0.02 | -0.02 | 0.13 | 0.23 | -0.01 | | 0.09 | -0.05 | -0.09 | 0.15 | 0.25 | 0.32 | 0.32 | 0.48 |
| HD 189509 | E | 6368 | 4.34 | 2.82 | 0.33 | 0.03 | 0.25 | 0.20 | 0.16 | 0.19 | 0.34 | 0.31 | 0.24 | 0.41 | 0.17 | 0.09 | 0.56 | 0.24 | -0.45 | -0.41 | | | 0.63 | 1.69 | | | 0.36 | 0.56 | 0.64 | 0.23 |
| HD 189558 | E | 5773 | 3.93 | 1.40 | -0.99 | -0.62 | -0.84 | -0.73 | -0.55 | -0.73 | -0.92 | -0.77 | -0.80 | -1.01 | -1.44 | -1.07 | -0.69 | -1.05 | -1.67 | -1.06 | -0.88 | | -0.79 | -0.20 | -0.79 | -0.59 | -0.58 | -0.53 | -0.36 | -0.36 |
| HD 19019 | E | 6113 | 4.40 | 1.18 | -0.06 | -0.04 | -0.13 | -0.03 | -0.04 | 0.06 | 0.06 | 0.00 | 0.02 | 0.01 | -0.08 | -0.03 | 0.03 | -0.08 | -0.31 | -0.16 | 0.23 | | 0.16 | 0.35 | 0.26 | 0.48 | 0.34 | 0.42 | 0.26 | 0.51 |
| HD 190404 | E | 5088 | 4.60 | 0.54 | -0.40 | -0.25 | -0.17 | -0.26 | 0.25 | -0.25 | -0.31 | -0.17 | -0.14 | -0.40 | -0.62 | -0.54 | -0.31 | -0.43 | -0.24 | -0.42 | -0.34 | | -0.39 | -0.23 | -0.65 | 0.07 | -0.06 | 0.31 | 0.05 | 0.05 |
| HD 190412 | E | 5388 | 4.29 | 0.97 | -0.22 | -0.29 | -0.18 | -0.16 | 0.07 | -0.21 | -0.16 | -0.24 | -0.31 | -0.19 | -0.33 | -0.28 | -0.18 | -0.26 | -0.38 | -0.27 | -0.22 | | -0.05 | 0.36 | -0.40 | 0.61 | 0.34 | 0.28 | 0.11 | 0.18 |
| HD 190498 | E | 6415 | 3.93 | 2.69 | 0.61 | 0.28 | 0.56 | 0.28 | 0.38 | 0.32 | 0.41 | 0.47 | 0.44 | 0.40 | 0.34 | 0.21 | 0.44 | 0.33 | -0.24 | -0.35 | | | 0.50 | 1.52 | -0.06 | | 0.75 | 0.46 | 0.66 | 0.40 |
| HD 191533 | E | 6254 | 3.84 | 1.98 | 0.04 | 0.10 | 0.03 | 0.08 | -0.03 | 0.08 | 0.13 | 0.05 | 0.08 | 0.04 | 0.01 | -0.01 | 0.13 | -0.01 | -0.20 | -0.20 | -0.03 | | 0.04 | 0.32 | 0.12 | 0.20 | 0.25 | 0.11 | 0.11 | 0.37 |
| HD 191785 | E | 5213 | 4.52 | 0.68 | 0.15 | 0.13 | 0.26 | 0.11 | 0.43 | 0.08 | 0.14 | 0.19 | 0.29 | 0.06 | -0.04 | -0.03 | 0.15 | 0.04 | 0.28 | 0.19 | 0.02 | | 0.09 | 0.03 | -0.17 | 0.36 | 0.39 | 0.66 | 0.40 | 0.53 |
| HD 19308 | E | 5767 | 4.21 | 1.31 | 0.37 | 0.27 | 0.23 | 0.22 | 0.35 | 0.21 | 0.23 | 0.17 | 0.20 | 0.21 | 0.21 | 0.15 | 0.26 | 0.18 | 0.24 | 0.25 | 0.31 | | 0.22 | 0.22 | 0.05 | 0.35 | 0.35 | 0.43 | 0.32 | 0.59 |
| HD 193374 | E | 6522 | 3.91 | 2.73 | 0.35 | 0.22 | 0.28 | 0.21 | 0.26 | 0.20 | 0.35 | 0.27 | 0.37 | 0.13 | 0.12 | 0.14 | 0.65 | 0.23 | 0.06 | -0.37 | | | 0.45 | 0.75 | 0.02 | 0.48 | 0.23 | 0.42 | 0.70 | |
| HD 194154 | E | 6456 | 4.30 | 2.93 | 0.09 | | | 0.31 | | 0.20 | 0.18 | 0.46 | 0.24 | 0.32 | 0.04 | 0.12 | 0.65 | 0.36 | -0.45 | -0.54 | | | 0.42 | 1.34 | -0.28 | | 0.48 | 1.67 | 0.99 | 0.46 |
| HD 19445 | E | 6052 | 4.49 | 1.97 | -1.92 | -1.44 | | -1.67 | | -1.63 | -1.75 | -1.63 | -1.17 | -2.01 | -2.45 | -1.92 | -1.88 | -1.91 | | -1.97 | | | -1.84 | -0.88 | -2.10 | | -0.09 | -0.46 | 0.01 | |
| HD 194598 | E | 6126 | 4.37 | 1.77 | -1.12 | -0.83 | | -0.89 | -0.52 | -0.85 | -1.01 | -0.84 | -0.93 | -1.10 | -1.39 | -1.07 | -0.72 | -1.10 | -1.60 | -1.20 | -1.18 | | -0.67 | -0.51 | -1.07 | -0.42 | -0.51 | -0.30 | -0.59 | -0.95 |
| HD 195005 | E | 6149 | 4.41 | 1.33 | 0.10 | 0.15 | -0.04 | 0.07 | 0.08 | 0.14 | 0.06 | 0.06 | 0.05 | 0.10 | 0.00 | 0.04 | 0.06 | 0.01 | -0.14 | -0.02 | 0.31 | | 0.24 | 0.42 | 0.23 | 0.36 | 0.38 | 0.36 | 0.24 | 0.39 |
| HD 195104 | E | 6226 | 4.38 | 1.36 | -0.10 | 0.01 | -0.11 | -0.04 | -0.01 | 0.01 | 0.14 | -0.01 | -0.03 | -0.03 | -0.21 | -0.07 | 0.11 | -0.11 | -0.36 | -0.20 | -0.10 | | 0.15 | 0.17 | 0.14 | 0.26 | 0.30 | 0.19 | 0.18 | 0.26 |
| HD 195633 | E | 6024 | 3.99 | 1.59 | -0.46 | -0.38 | -0.57 | -0.41 | -0.34 | -0.46 | -0.34 | -0.43 | -0.40 | -0.53 | -0.72 | -0.59 | -0.28 | -0.57 | -0.72 | -0.63 | -0.62 | | -0.43 | -0.37 | -0.47 | -0.04 | -0.39 | -0.32 | -0.29 | 0.00 |
| HD 196218 | E | 6204 | 4.19 | 1.46 | -0.06 | 0.00 | -0.23 | -0.04 | -0.02 | -0.01 | 0.10 | -0.03 | 0.05 | -0.02 | -0.21 | -0.10 | 0.22 | -0.12 | -0.34 | -0.17 | 0.09 | | 0.03 | 0.53 | 0.01 | 0.11 | 0.32 | 0.13 | 0.14 | 0.47 |
| HD 198061 | E | 6379 | 3.98 | 3.53 | 0.11 | 0.24 | 0.34 | 0.26 | 0.26 | 0.04 | 0.24 | 0.38 | 0.23 | 0.26 | 0.01 | 0.01 | 0.40 | 0.10 | -0.33 | -0.48 | | | 0.62 | 1.11 | -0.12 | 0.78 | 0.32 | 0.48 | 0.69 | 0.37 |
| HD 199598 | E | 5918 | 4.37 | 1.04 | 0.13 | 0.08 | 0.06 | 0.10 | 0.27 | 0.16 | 0.08 | 0.05 | 0.03 | 0.13 | 0.02 | 0.06 | 0.10 | 0.05 | -0.13 | -0.06 | 0.18 | | 0.20 | 0.27 | 0.14 | 0.21 | 0.28 | 0.22 | 0.14 | 0.73 |
| HD 20039 | E | 5331 | 3.68 | 0.40 | -0.54 | -0.39 | -0.53 | -0.45 | -0.22 | -0.34 | -0.43 | -0.28 | -0.34 | -0.38 | -0.70 | -0.47 | -0.44 | -0.51 | -0.66 | -0.75 | 0.01 | | -0.46 | 0.09 | -0.35 | -0.17 | 0.00 | -0.04 | 0.16 | 0.29 |
| HD 200391 | E | 5709 | 3.97 | 0.40 | 0.57 | 2.01 | | 0.72 | | 0.88 | | 0.77 | -0.22 | 1.51 | | 0.18 | 0.46 | 0.68 | | | | | | | 1.83 | 0.46 | | | | | |
| HD 200560 | E | 4894 | 4.52 | 1.13 | 0.25 | 0.12 | 0.18 | 0.27 | 1.06 | 0.19 | 0.14 | 0.16 | 0.27 | 0.26 | 0.11 | 0.19 | 0.19 | 0.21 | 0.11 | 0.03 | 0.27 | | 0.28 | 0.23 | 0.12 | 0.46 | 0.57 | 0.69 | 0.61 | 0.60 |
| HD 200580 | E | 5870 | 4.00 | 1.36 | -0.41 | -0.35 | -0.43 | -0.42 | -0.40 | -0.42 | -0.35 | -0.42 | -0.27 | -0.45 | -0.68 | -0.58 | -0.28 | -0.56 | -0.71 | -0.68 | -0.78 | | -0.59 | -0.04 | -0.60 | -0.34 | -0.40 | -0.39 | -0.17 | -0.32 |
| HD 201099 | E | 5947 | 4.22 | 1.16 | -0.40 | -0.32 | -0.42 | -0.31 | -0.17 | -0.31 | -0.22 | -0.29 | -0.29 | -0.37 | -0.55 | -0.40 | -0.23 | -0.43 | -0.60 | -0.43 | -0.52 | | -0.46 | -0.08 | -0.28 | -0.03 | -0.14 | -0.10 | -0.06 | 0.17 |
| HD 20165 | E | 5137 | 4.58 | 0.15 | 0.16 | 0.11 | 0.16 | 0.12 | 0.51 | 0.14 | 0.13 | 0.20 | 0.36 | 0.20 | 0.13 | 0.09 | 0.15 | 0.17 | 0.35 | 0.18 | 0.19 | | 0.22 | 0.28 | 0.10 | 0.62 | 0.98 | 0.85 | 0.91 | 0.67 |
| HD 201891 | E | 5998 | 4.39 | 1.39 | -0.93 | -0.58 | -0.73 | -0.71 | | -0.81 | -0.73 | -0.73 | -0.63 | -0.99 | -1.30 | -1.01 | -0.60 | -0.98 | -1.15 | -0.93 | -1.01 | | -1.06 | -0.66 | -1.02 | -0.78 | -0.65 | -0.49 | -0.54 | -0.93 |
| HD 202575 | E | 4737 | 4.65 | 1.15 | 0.05 | 0.03 | -0.03 | 0.17 | 1.08 | 0.12 | -0.03 | 0.02 | 0.11 | 0.10 | -0.08 | 0.05 | 0.13 | 0.04 | 0.07 | -0.01 | 0.24 | | 0.14 | 0.11 | 0.01 | 0.44 | 0.54 | 1.08 | 0.56 | 0.71 |
| HD 203235 | E | 6242 | 4.16 | 1.62 | 0.36 | 0.32 | 0.22 | 0.19 | 0.25 | 0.30 | 0.18 | 0.24 | 0.35 | 0.25 | 0.18 | 0.17 | 0.31 | 0.18 | -0.01 | -0.09 | 0.41 | | 0.27 | 0.61 | 0.14 | 0.32 | 0.31 | 0.22 | 0.27 | 0.35 |
| HD 204426 | E | 5658 | 3.98 | 1.00 | -0.22 | -0.07 | -0.12 | -0.15 | -0.12 | -0.15 | -0.15 | -0.16 | -0.23 | -0.33 | -0.52 | -0.35 | -0.23 | -0.33 | -0.41 | -0.10 | -0.20 | | -0.31 | -0.18 | -0.37 | -0.15 | -0.15 | -0.14 | -0.07 | 0.41 |
| HD 204734 | E | 5262 | 4.58 | 1.15 | 0.08 | 0.10 | 0.11 | 0.13 | 0.45 | 0.16 | 0.10 | 0.14 | 0.22 | 0.21 | 0.10 | 0.11 | 0.10 | 0.12 | 0.14 | -0.09 | 0.38 | | 0.38 | 0.16 | 0.16 | 0.48 | 0.56 | 0.68 | 0.61 | 0.45 |
| HD 20512 | E | 5270 | 3.56 | 1.31 | -0.02 | -0.06 | 0.02 | -0.05 | 0.11 | 0.03 | -0.15 | -0.08 | -0.07 | -0.04 | -0.10 | -0.11 | -0.12 | -0.12 | -0.10 | -0.18 | -0.15 | | -0.08 | -0.20 | -0.18 | -0.02 | 0.02 | 0.17 | 0.06 | 0.40 |

| | | | | | | | | | | | | | | | | | | | | | | | | | | | | | |
|---|---|---|---|---|---|---|---|---|---|---|---|---|---|---|---|---|---|---|---|---|---|---|---|---|---|---|---|---|---|
| HD 205434 | E | 4454 | 4.68 | 0.60 | 0.16 | 0.16 | -0.05 | 0.32 | 1.40 | 0.22 | 0.10 | 0.10 | 0.20 | 0.18 | 0.03 | 0.08 | 0.23 | 0.16 | 0.24 | 0.06 | 0.51 | 0.13 | 0.09 | 0.10 | 0.84 | 1.09 | 1.24 | 0.95 | 0.68 |
| HD 205702 | E | 6060 | 4.19 | 1.46 | 0.24 | 0.19 | 0.24 | 0.14 | 0.16 | 0.15 | 0.19 | 0.13 | 0.16 | 0.14 | 0.09 | 0.10 | 0.19 | 0.12 | 0.07 | 0.09 | -0.04 | 0.24 | 0.32 | 0.11 | 0.38 | 0.39 | 0.40 | 0.14 | 0.49 |
| HD 206374 | E | 5579 | 4.53 | 0.83 | 0.01 | 0.04 | -0.04 | 0.00 | 0.17 | 0.03 | -0.02 | 0.00 | 0.01 | 0.03 | -0.05 | -0.04 | -0.01 | -0.06 | -0.10 | 0.01 | 0.14 | 0.16 | 0.12 | 0.11 | 0.39 | 0.40 | 0.37 | 0.30 | 0.56 |
| HD 208038 | E | 4995 | 4.63 | 0.53 | 0.06 | 0.00 | -0.02 | 0.08 | 0.76 | 0.16 | 0.03 | 0.14 | 0.24 | 0.15 | 0.05 | 0.05 | 0.11 | 0.08 | 0.15 | 0.16 | 0.32 | 0.19 | 0.23 | 0.27 | 0.53 | 0.95 | 1.07 | 0.99 | 0.58 |
| HD 208313 | E | 5030 | 4.61 | 0.52 | 0.22 | 0.05 | 0.10 | 0.11 | 0.66 | 0.14 | 0.09 | 0.15 | 0.26 | 0.14 | 0.10 | 0.07 | 0.11 | 0.14 | 0.33 | 0.28 | 0.30 | 0.23 | 0.18 | 0.04 | 0.62 | 0.76 | 0.97 | 1.18 | 0.47 |
| HD 209472 | E | 6480 | 4.18 | 2.16 | 0.03 | 0.10 | -0.22 | 0.04 | 0.10 | 0.06 | -0.10 | 0.06 | 0.32 | -0.08 | -0.26 | -0.09 | 0.37 | -0.06 | -0.59 | -0.53 | | 0.16 | 0.67 | 0.20 | | 0.23 | 0.27 | 0.38 | 0.82 |
| HD 210752 | E | 6024 | 4.42 | 1.09 | -0.55 | -0.31 | -0.47 | -0.42 | -0.21 | -0.44 | -0.41 | -0.42 | -0.51 | -0.51 | -0.71 | -0.53 | -0.36 | -0.55 | -0.65 | -0.62 | -0.69 | -0.51 | 0.27 | -0.42 | -0.14 | -0.16 | -0.11 | 0.12 | 0.04 |
| HD 21197 | E | 4562 | 4.56 | 0.82 | 0.71 | 0.65 | 0.46 | 0.58 | 1.50 | 0.60 | 0.59 | 0.45 | 0.72 | 0.52 | 0.40 | 0.49 | 0.61 | 0.56 | 0.70 | 0.90 | 0.60 | 0.49 | 0.33 | 0.12 | 0.73 | 1.13 | 1.44 | 1.33 | |
| HD 214683 | E | 4893 | 4.65 | 0.56 | -0.16 | -0.09 | -0.12 | -0.12 | 0.52 | -0.09 | -0.08 | -0.04 | 0.06 | -0.08 | -0.26 | -0.16 | -0.04 | -0.17 | 0.02 | -0.10 | 0.04 | 0.04 | -0.12 | -0.03 | 0.44 | 0.65 | 0.83 | 0.43 | 0.43 |
| HD 216259 | E | 5002 | 4.67 | 0.95 | -0.50 | -0.38 | -0.27 | -0.36 | | -0.34 | -0.39 | -0.31 | -0.27 | -0.47 | -0.68 | -0.60 | -0.33 | -0.53 | -0.31 | -0.48 | -0.50 | -0.37 | -0.44 | -0.74 | -0.12 | 0.06 | 0.35 | 0.11 | -0.12 |
| HD 216520 | E | 5103 | 4.56 | 1.11 | -0.03 | -0.04 | -0.03 | -0.01 | 0.54 | -0.08 | -0.06 | -0.05 | 0.01 | -0.04 | -0.14 | -0.14 | 0.01 | -0.08 | -0.07 | -0.09 | -0.09 | 0.01 | -0.08 | -0.28 | 0.40 | 0.65 | 0.56 | 0.30 | 0.33 |
| HD 217813 | E | 5849 | 4.42 | 1.24 | -0.04 | 0.13 | 0.04 | 0.06 | 0.03 | 0.14 | 0.07 | 0.06 | 0.06 | 0.12 | 0.00 | 0.05 | 0.02 | 0.00 | -0.17 | -0.18 | 0.23 | 0.28 | 0.44 | 0.30 | 0.43 | 0.42 | 0.32 | 0.24 | 0.49 |
| HD 218059 | E | 6382 | 4.31 | 1.52 | -0.25 | -0.16 | -0.25 | -0.18 | -0.02 | -0.15 | -0.15 | -0.18 | -0.30 | -0.22 | -0.40 | -0.27 | -0.03 | -0.29 | -0.48 | -0.44 | -0.23 | -0.10 | 0.10 | 0.03 | 0.16 | 0.03 | -0.02 | -0.04 | 0.10 |
| HD 218209 | E | 5623 | 4.46 | 0.56 | -0.34 | -0.11 | -0.14 | -0.25 | -0.21 | -0.21 | -0.25 | -0.19 | -0.24 | -0.37 | -0.54 | -0.40 | -0.32 | -0.41 | -0.43 | -0.28 | -0.30 | -0.32 | -0.14 | -0.36 | 0.23 | -0.12 | -0.08 | -0.14 | 0.36 |
| HD 218566 | E | 4846 | 4.53 | 0.95 | 0.68 | 0.66 | 0.51 | 0.49 | 0.99 | 0.38 | 0.51 | 0.46 | 0.63 | 0.50 | 0.42 | 0.42 | 0.55 | 0.48 | 0.60 | 0.82 | 0.48 | 0.47 | 0.30 | 0.06 | 0.67 | 1.07 | 1.33 | 1.35 | 1.24 |
| HD 218687 | E | 5902 | 4.39 | 1.83 | -0.03 | 0.07 | 0.07 | 0.05 | 0.07 | 0.03 | 0.01 | 0.07 | 0.04 | 0.05 | -0.06 | -0.05 | 0.08 | -0.03 | -0.23 | -0.24 | -0.26 | 0.17 | 0.42 | 0.03 | 0.30 | 0.23 | 0.29 | 0.01 | 0.24 |
| HD 218868 | E | 5509 | 4.41 | 1.03 | 0.41 | 0.40 | 0.33 | 0.29 | 0.52 | 0.33 | 0.26 | 0.28 | 0.34 | 0.30 | 0.32 | 0.26 | 0.30 | 0.32 | 0.23 | 0.31 | 0.29 | 0.38 | 0.26 | 0.19 | 0.60 | 0.51 | 0.54 | 0.61 | 0.80 |
| HD 219396 | E | 5649 | 4.03 | 1.06 | 0.04 | 0.13 | 0.09 | 0.03 | 0.08 | 0.05 | 0.05 | 0.05 | 0.03 | 0.00 | -0.08 | -0.04 | -0.04 | -0.04 | -0.02 | 0.03 | -0.06 | 0.07 | 0.07 | 0.02 | 0.40 | 0.20 | 0.25 | 0.18 | 0.66 |
| HD 219420 | E | 6175 | 4.19 | 1.49 | 0.15 | 0.09 | -0.03 | 0.08 | 0.11 | 0.09 | 0.15 | 0.08 | 0.06 | 0.15 | 0.05 | 0.02 | 0.20 | 0.02 | -0.07 | -0.03 | 0.66 | 0.43 | 0.59 | 0.11 | 0.16 | 0.36 | 0.17 | 0.19 | 0.29 |
| HD 219538 | E | 5078 | 4.60 | 0.15 | 0.14 | 0.07 | 0.09 | 0.16 | 0.52 | 0.10 | 0.15 | 0.20 | 0.33 | 0.19 | 0.17 | 0.12 | 0.15 | 0.18 | 0.26 | 0.13 | 0.24 | 0.30 | 0.21 | 0.11 | 0.90 | 0.92 | 1.03 | 0.75 | 0.79 |
| HD 219623 | E | 6177 | 4.31 | 1.43 | 0.20 | 0.19 | 0.17 | 0.14 | 0.12 | 0.19 | 0.14 | 0.13 | 0.15 | 0.16 | 0.12 | 0.10 | 0.16 | 0.09 | -0.02 | -0.03 | 0.35 | 0.26 | 0.16 | 0.14 | 0.46 | 0.31 | 0.20 | 0.10 | 0.44 |
| HD 220140 | E | 5075 | 4.58 | 2.77 | 0.37 | 0.53 | 0.12 | 0.28 | 0.55 | 0.20 | 0.17 | 0.17 | 0.13 | 0.27 | 0.01 | 0.11 | 0.17 | 0.11 | | -0.08 | | 0.33 | -0.01 | 0.10 | 0.53 | 0.29 | 0.82 | 1.53 | |
| HD 220182 | E | 5335 | 4.56 | 1.16 | 0.11 | 0.12 | 0.12 | 0.11 | 0.30 | 0.16 | 0.04 | 0.12 | 0.15 | 0.17 | 0.09 | 0.10 | 0.09 | 0.08 | 0.02 | -0.06 | 0.30 | 0.27 | 0.25 | 0.19 | 0.56 | 0.54 | 0.67 | 0.51 | 0.65 |
| HD 220221 | E | 4783 | 4.58 | 1.16 | 0.51 | 0.40 | 0.28 | 0.34 | 1.19 | 0.43 | 0.30 | 0.24 | 0.43 | 0.33 | 0.23 | 0.24 | 0.37 | 0.33 | 0.34 | 0.51 | 0.37 | 0.37 | 0.07 | 0.04 | 0.65 | 0.62 | 1.22 | 1.27 | 0.96 |
| HD 221354 | E | 5282 | 4.49 | 0.89 | 0.34 | 0.36 | 0.36 | 0.22 | 0.49 | 0.24 | 0.27 | 0.30 | 0.42 | 0.25 | 0.20 | 0.15 | 0.31 | 0.24 | 0.40 | 0.43 | 0.20 | 0.31 | 0.28 | 0.10 | 0.58 | 0.71 | 0.84 | 0.83 | 0.74 |
| HD 221585 | E | 5530 | 3.93 | 1.37 | 0.54 | 0.41 | 0.43 | 0.36 | 0.57 | 0.33 | 0.34 | 0.30 | 0.34 | 0.36 | 0.38 | 0.30 | 0.34 | 0.34 | 0.38 | 0.58 | 0.25 | 0.41 | 0.29 | 0.19 | 0.55 | 0.55 | 0.47 | 0.61 | 0.70 |
| HD 221851 | E | 5192 | 4.58 | 1.00 | 0.05 | 0.02 | 0.06 | 0.04 | 0.57 | 0.05 | -0.01 | 0.04 | 0.13 | 0.05 | -0.02 | -0.03 | 0.04 | -0.01 | 0.07 | -0.01 | 0.15 | 0.26 | 0.08 | 0.04 | 0.39 | 0.46 | 0.54 | 0.43 | 0.52 |
| HD 222155 | E | 5694 | 3.94 | 1.23 | -0.05 | -0.02 | 0.03 | -0.04 | -0.01 | 0.00 | 0.03 | -0.06 | -0.09 | -0.08 | -0.19 | -0.11 | -0.06 | -0.13 | -0.19 | -0.11 | 0.16 | 0.10 | 0.09 | 0.14 | 0.19 | 0.22 | 0.19 | 0.08 | 0.51 |
| HD 224465 | E | 5770 | 4.42 | 0.98 | 0.25 | 0.18 | 0.19 | 0.12 | 0.27 | 0.16 | 0.16 | 0.13 | 0.16 | 0.16 | 0.17 | 0.12 | 0.14 | 0.15 | 0.17 | 0.14 | 0.22 | 0.25 | 0.32 | 0.06 | 0.33 | 0.35 | 0.42 | 0.24 | 0.66 |
| HD 22879 | E | 5932 | 4.36 | 1.22 | -0.74 | -0.47 | -0.61 | -0.50 | -0.18 | -0.54 | -0.65 | -0.52 | -0.62 | -0.80 | -1.12 | -0.82 | -0.48 | -0.78 | -1.01 | -0.70 | | -0.73 | -0.43 | -0.68 | -0.68 | -0.56 | -0.53 | -0.22 | -0.35 |
| HD 232781 | E | 4679 | 4.68 | 0.73 | -0.25 | -0.25 | -0.38 | -0.17 | 0.64 | -0.20 | -0.25 | -0.25 | -0.13 | -0.24 | -0.36 | -0.26 | -0.14 | -0.26 | -0.18 | -0.34 | -0.02 | -0.17 | -0.09 | -0.22 | 0.42 | 0.46 | 1.08 | 0.48 | 0.45 |
| HD 23439A | E | 5192 | 4.67 | 0.15 | -0.87 | -0.56 | -0.53 | -0.49 | | -0.61 | -0.70 | -0.54 | -0.59 | -0.77 | -1.15 | -0.90 | -0.64 | -0.86 | -0.93 | -0.76 | -0.25 | -0.36 | -0.11 | -0.56 | -0.09 | -0.25 | -0.13 | -0.13 | -0.26 |
| HD 238087 | E | 4213 | 4.64 | 0.98 | 0.15 | 0.51 | 0.15 | 0.67 | 1.84 | 0.51 | 0.33 | 0.15 | 0.29 | 0.39 | 0.17 | 0.40 | 0.53 | 0.42 | 0.70 | 0.53 | | 0.23 | 0.13 | 0.23 | 0.98 | 1.21 | 1.48 | 1.18 | 0.84 |
| HD 24040 | E | 5756 | 4.17 | 1.26 | 0.40 | 0.37 | 0.33 | 0.27 | 0.40 | 0.28 | 0.31 | 0.25 | 0.25 | 0.26 | 0.25 | 0.21 | 0.26 | 0.24 | 0.32 | 0.24 | 0.30 | 0.30 | 0.27 | 0.21 | 0.46 | 0.47 | 0.50 | 0.39 | 0.90 |
| HD 24238 | E | 5031 | 4.60 | 0.15 | -0.26 | -0.04 | -0.05 | -0.14 | 0.39 | -0.21 | -0.11 | -0.07 | -0.03 | -0.26 | -0.50 | -0.40 | -0.20 | -0.30 | -0.09 | -0.19 | -0.04 | -0.11 | -0.21 | -0.52 | 0.17 | 0.22 | 0.60 | 0.27 | 0.25 |
| HD 24409 | E | 5568 | 4.34 | 0.84 | 0.01 | -0.02 | -0.01 | 0.00 | 0.26 | 0.00 | -0.02 | -0.10 | -0.09 | -0.01 | -0.09 | -0.10 | -0.11 | -0.06 | -0.09 | -0.01 | -0.03 | 0.10 | 0.09 | -0.04 | 0.61 | 0.21 | 0.22 | 0.14 | 0.43 |
| HD 24451 | E | 4547 | 4.64 | 0.82 | 0.36 | 0.44 | 0.11 | 0.34 | 1.16 | 0.35 | 0.20 | 0.21 | 0.40 | 0.30 | 0.09 | 0.26 | 0.37 | 0.28 | 0.38 | 0.13 | 0.46 | 0.14 | 0.08 | 0.08 | 0.75 | 0.94 | 1.26 | 1.12 | 0.79 |
| HD 24496 | E | 5429 | 4.44 | 0.97 | 0.05 | 0.02 | 0.07 | 0.07 | 0.28 | 0.01 | -0.02 | -0.03 | -0.02 | -0.01 | -0.08 | -0.04 | 0.00 | -0.01 | -0.15 | -0.03 | 0.13 | 0.14 | 0.08 | 0.00 | 0.34 | 0.34 | 0.24 | 0.29 | 0.48 |

| HD 245 | E | 5805 | 4.35 | 0.66 | -0.42 | -0.20 | -0.23 | -0.28 | -0.21 | -0.32 | -0.23 | -0.27 | -0.36 | -0.44 | -0.69 | -0.50 | -0.29 | -0.47 | -0.60 | -0.35 | | -0.36 | -0.21 | -0.29 | 0.14 | -0.07 | -0.08 | 0.05 | 0.36 |
|---|---|---|---|---|---|---|---|---|---|---|---|---|---|---|---|---|---|---|---|---|---|---|---|---|---|---|---|---|---|
| HD 24552 | E | 5916 | 4.43 | 1.09 | 0.06 | 0.09 | 0.08 | 0.06 | 0.09 | 0.15 | 0.12 | 0.12 | 0.09 | 0.14 | 0.03 | 0.07 | 0.06 | 0.06 | -0.08 | 0.05 | 0.28 | 0.27 | 0.44 | 0.13 | 0.44 | 0.41 | 0.29 | 0.34 | 0.64 |
| HD 25329 | E | 4855 | 4.73 | 1.63 | -1.18 | -1.30 | -1.14 | -0.97 | | -1.21 | -1.24 | -1.23 | -1.30 | -1.33 | -1.84 | -1.55 | -0.86 | -1.35 | -1.65 | -1.45 | -1.12 | -1.02 | -0.50 | -1.22 | -0.43 | -0.04 | -0.19 | -0.20 | -1.10 |
| HD 25457 | E | 6268 | 4.33 | 2.41 | 0.35 | 0.36 | 0.20 | 0.19 | 0.13 | 0.21 | 0.22 | 0.27 | 0.16 | 0.24 | 0.13 | 0.11 | 0.29 | 0.11 | -0.31 | -0.38 | | 0.34 | 0.73 | 0.25 | 0.64 | 0.53 | 0.42 | 0.80 | 0.33 |
| HD 25621 | E | 6307 | 3.86 | 2.54 | 0.31 | 0.43 | | 0.18 | 0.15 | 0.22 | 0.26 | 0.28 | 0.16 | 0.17 | 0.06 | 0.10 | 0.30 | 0.17 | -0.20 | -0.26 | | 0.29 | 0.56 | 0.00 | 0.39 | 0.24 | 0.38 | 0.59 | |
| HD 25825 | E | 5976 | 4.41 | 1.24 | 0.14 | 0.17 | 0.13 | 0.17 | 0.22 | 0.19 | 0.20 | 0.14 | 0.21 | 0.24 | 0.13 | 0.12 | 0.28 | 0.13 | -0.16 | 0.26 | 0.36 | 0.49 | 0.61 | 0.18 | 0.34 | 0.75 | 0.69 | 0.63 | 0.59 |
| HD 26345 | E | 6676 | 4.28 | 3.02 | 0.32 | 0.35 | 0.59 | 0.27 | 0.44 | 0.34 | 0.22 | 0.45 | 0.40 | 0.20 | 0.25 | 0.19 | 0.58 | 0.37 | -0.23 | -0.20 | | 0.33 | 1.28 | -0.17 | 0.53 | 0.68 | 0.76 | 1.56 | 0.48 |
| HD 26784 | E | 6261 | 4.29 | 2.33 | 0.35 | 0.49 | 0.39 | 0.31 | 0.51 | 0.33 | 0.35 | 0.35 | 0.33 | 0.27 | 0.27 | 0.24 | 0.51 | 0.32 | -0.17 | 0.10 | | 0.43 | 0.68 | -0.07 | 0.64 | 0.48 | 0.40 | 0.62 | 0.01 |
| HD 27808 | E | 6230 | 4.30 | 2.03 | 0.24 | 0.36 | 0.25 | 0.22 | 0.22 | 0.26 | 0.31 | 0.26 | 0.25 | 0.26 | 0.25 | 0.16 | 0.35 | 0.21 | 0.04 | -0.11 | | 0.33 | 0.38 | 0.13 | 0.34 | 0.44 | 0.29 | 0.55 | 0.67 |
| HD 28005 | E | 5727 | 4.27 | 1.22 | 0.71 | 0.52 | 0.46 | 0.38 | 0.62 | 0.34 | 0.41 | 0.33 | 0.39 | 0.38 | 0.42 | 0.32 | 0.40 | 0.42 | 0.44 | 0.63 | 0.33 | 0.38 | 0.33 | 0.16 | 0.49 | 0.51 | 0.46 | 0.53 | 0.81 |
| HD 28099 | E | 5738 | 4.42 | 1.56 | 0.31 | 0.19 | 0.22 | 0.21 | 0.32 | 0.25 | 0.21 | 0.18 | 0.24 | 0.26 | 0.20 | 0.17 | 0.16 | 0.19 | 0.10 | 0.13 | 0.31 | 0.29 | 0.29 | 0.23 | 0.40 | 0.54 | 0.50 | 0.54 | 0.65 |
| HD 28344 | E | 5921 | 4.39 | 1.82 | 0.34 | 0.25 | 0.25 | 0.21 | 0.32 | 0.28 | 0.21 | 0.19 | 0.17 | 0.26 | 0.18 | 0.15 | 0.18 | 0.17 | -0.14 | -0.04 | -0.03 | 0.35 | 0.49 | 0.22 | 0.46 | 0.47 | 0.40 | 0.34 | 0.51 |
| HD 283704 | E | 5524 | 4.51 | 1.30 | 0.35 | 0.23 | 0.24 | 0.22 | 0.37 | 0.26 | 0.20 | 0.23 | 0.33 | 0.29 | 0.30 | 0.24 | 0.25 | 0.23 | 0.23 | 0.25 | 0.41 | 0.42 | 0.40 | 0.21 | 0.62 | 0.64 | 0.46 | 0.54 | 0.88 |
| HD 283750 | E | 4405 | 4.51 | 1.99 | 0.72 | 1.05 | 0.45 | 0.90 | 2.21 | 0.35 | 0.42 | 0.28 | 0.45 | 0.46 | 0.49 | 0.59 | 0.70 | 0.70 | 0.67 | 1.01 | 0.10 | 0.43 | -0.08 | 0.19 | 0.70 | 0.84 | 0.87 | 1.15 | 0.80 |
| HD 284248 | E | 6157 | 4.32 | 2.00 | -1.58 | -1.22 | | -1.08 | | -1.31 | -1.48 | -1.29 | -1.00 | -1.54 | -2.00 | -1.59 | -0.82 | -1.55 | | -1.66 | | -1.64 | -0.55 | -1.73 | -1.07 | -0.26 | -0.56 | 0.11 | -1.56 |
| HD 284574 | E | 5370 | 4.47 | 0.98 | 0.40 | 0.41 | 0.24 | 0.30 | 0.75 | 0.39 | 0.30 | 0.35 | 0.40 | 0.41 | 0.34 | 0.30 | 0.26 | 0.36 | 0.55 | 0.20 | 0.37 | 0.52 | 0.33 | 0.43 | 0.69 | 0.72 | 0.73 | 0.99 | 0.72 |
| HD 284930 | E | 4667 | 4.63 | 1.15 | 0.42 | 0.40 | 0.11 | 0.43 | 1.31 | 0.34 | 0.30 | 0.23 | 0.33 | 0.36 | 0.27 | 0.34 | 0.42 | 0.36 | 0.41 | 0.29 | 0.27 | 0.43 | 0.25 | 0.11 | 0.64 | 0.90 | 1.05 | 1.19 | 0.87 |
| HD 285690 | E | 4975 | 4.45 | 0.78 | 0.49 | 0.25 | 0.30 | 0.23 | 0.50 | 0.43 | 0.25 | 0.36 | 0.52 | 0.41 | 0.32 | 0.20 | 0.24 | 0.28 | 0.28 | -0.03 | 0.38 | 0.42 | 0.39 | 0.19 | 0.66 | 0.84 | 0.86 | 1.01 | 0.77 |
| HD 28946 | E | 5366 | 4.57 | 0.78 | -0.05 | -0.04 | 0.00 | -0.05 | 0.17 | 0.01 | -0.04 | 0.04 | 0.10 | 0.03 | -0.04 | -0.04 | 0.01 | -0.04 | 0.06 | -0.05 | 0.16 | 0.14 | 0.14 | 0.03 | 0.33 | 0.45 | 0.62 | 0.36 | 0.30 |
| HD 28992 | E | 5865 | 4.40 | 1.54 | 0.25 | 0.14 | 0.24 | 0.19 | 0.22 | 0.24 | 0.19 | 0.12 | 0.15 | 0.22 | 0.14 | 0.11 | 0.14 | 0.12 | -0.04 | 0.04 | 0.22 | 0.30 | 0.22 | 0.14 | 0.29 | 0.31 | 0.28 | 0.25 | 0.02 |
| HD 29310 | E | 5841 | 4.14 | 1.98 | 0.30 | 0.16 | 0.22 | 0.16 | 0.31 | 0.27 | 0.12 | 0.17 | 0.23 | 0.23 | 0.18 | 0.08 | 0.14 | 0.09 | -0.09 | -0.33 | -0.02 | 0.30 | 0.39 | 0.07 | 0.06 | 0.30 | 0.29 | 0.26 | 0.27 |
| HD 29419 | E | 6058 | 4.37 | 1.51 | 0.28 | 0.22 | 0.16 | 0.19 | 0.28 | 0.27 | 0.18 | 0.16 | 0.24 | 0.22 | 0.22 | 0.17 | 0.22 | 0.17 | 0.01 | 0.00 | 0.36 | 0.30 | 0.25 | 0.25 | 0.27 | 0.45 | 0.31 | 0.27 | 0.55 |
| HD 30562 | E | 5894 | 4.08 | 1.54 | 0.46 | 0.42 | 0.33 | 0.27 | 0.31 | 0.35 | 0.32 | 0.29 | 0.31 | 0.29 | 0.32 | 0.23 | 0.29 | 0.30 | 0.23 | 0.18 | 0.26 | 0.32 | 0.47 | 0.13 | 0.56 | 0.44 | 0.43 | 0.39 | 0.78 |
| HD 3268 | E | 6232 | 4.10 | 1.41 | 0.02 | -0.04 | -0.06 | -0.05 | -0.03 | 0.02 | 0.11 | 0.03 | 0.00 | 0.01 | -0.13 | -0.07 | 0.19 | -0.11 | -0.30 | -0.25 | 0.35 | 0.05 | 0.27 | 0.09 | 0.32 | 0.23 | 0.24 | 0.19 | 0.31 |
| HD 32850 | E | 5226 | 4.56 | 1.20 | -0.15 | -0.07 | -0.09 | -0.04 | 0.39 | -0.13 | -0.11 | -0.14 | -0.12 | -0.08 | -0.22 | -0.14 | -0.11 | -0.16 | -0.16 | -0.26 | -0.15 | 0.02 | -0.10 | -0.08 | 0.29 | 0.37 | 0.46 | 0.32 | 0.34 |
| HD 330 | E | 5932 | 3.79 | 1.44 | 0.02 | -0.02 | -0.02 | -0.04 | -0.11 | 0.00 | -0.02 | -0.04 | -0.08 | -0.05 | -0.17 | -0.10 | 0.01 | -0.13 | -0.28 | -0.21 | 0.24 | 0.03 | 0.01 | 0.08 | 0.07 | 0.09 | 0.09 | 0.00 | 0.15 |
| HD 332518 | E | 4386 | 4.69 | 0.15 | 0.06 | 0.13 | 0.05 | 0.36 | 1.67 | 0.28 | 0.19 | 0.10 | 0.25 | 0.19 | 0.03 | 0.00 | 0.27 | 0.19 | 0.49 | 0.44 | 0.00 | 0.28 | 0.05 | -0.12 | 1.05 | 0.94 | 1.55 | 1.02 | 0.73 |
| HD 33564 | E | 6393 | 4.22 | 2.21 | 0.27 | 0.21 | 0.19 | 0.19 | 0.32 | 0.27 | 0.35 | 0.26 | 0.25 | 0.20 | 0.10 | 0.14 | 0.37 | 0.18 | -0.15 | -0.13 | | 0.29 | 0.69 | 0.10 | | 0.35 | 0.35 | 0.58 | 0.58 |
| HD 345957 | E | 5943 | 4.08 | 1.55 | -1.38 | -0.89 | -0.93 | -0.94 | 0.21 | -1.01 | -1.12 | -0.98 | -1.00 | -1.28 | -1.78 | -1.30 | -0.53 | -1.34 | -1.35 | -1.26 | -1.26 | -0.91 | -0.35 | -1.17 | -0.37 | -0.46 | -0.33 | -0.25 | -1.23 |
| HD 3628 | E | 5810 | 4.03 | 1.21 | 0.06 | 0.12 | 0.03 | -0.02 | 0.02 | 0.03 | 0.04 | 0.05 | -0.01 | -0.04 | -0.19 | -0.07 | -0.02 | -0.09 | -0.19 | 0.02 | -0.26 | 0.02 | 0.11 | -0.12 | 0.05 | 0.12 | 0.29 | 0.07 | 0.47 |
| HD 37008 | E | 5091 | 4.60 | 0.62 | -0.20 | -0.10 | -0.06 | -0.16 | 0.37 | -0.13 | -0.13 | -0.04 | -0.02 | -0.22 | -0.44 | -0.34 | -0.14 | -0.25 | 0.03 | -0.23 | 0.06 | -0.18 | -0.23 | -0.46 | 0.08 | 0.23 | 0.58 | 0.19 | 0.20 |
| HD 3765 | E | 4987 | 4.55 | 0.89 | 0.39 | 0.32 | 0.34 | 0.34 | 0.86 | 0.23 | 0.36 | 0.30 | 0.48 | 0.31 | 0.28 | 0.24 | 0.34 | 0.33 | 0.48 | 0.59 | 0.15 | 0.29 | 0.24 | 0.11 | 0.72 | 0.79 | 1.11 | 1.09 | 0.78 |
| HD 38230 | E | 5212 | 4.47 | 0.36 | 0.18 | 0.23 | 0.27 | 0.12 | 0.48 | 0.12 | 0.22 | 0.25 | 0.35 | 0.15 | 0.08 | 0.06 | 0.15 | 0.13 | 0.33 | 0.37 | 0.12 | 0.23 | 0.08 | -0.03 | 0.50 | 0.57 | 0.79 | 0.66 | 0.57 |
| HD 40512 | E | 6471 | 4.28 | 1.80 | 0.80 | | 0.71 | | 0.75 | 0.86 | 1.00 | 2.10 | 1.21 | 0.71 | 0.54 | 0.63 | 0.67 | 0.11 | 0.31 | 1.83 | 2.23 | | 1.12 | | 1.68 | 2.06 | 3.25 | 1.08 |
| HD 40616 | E | 5769 | 3.99 | 1.17 | -0.28 | -0.09 | -0.11 | -0.19 | -0.08 | -0.15 | -0.15 | -0.20 | -0.29 | -0.20 | -0.33 | -0.23 | -0.23 | -0.29 | -0.36 | -0.35 | -0.15 | -0.22 | -0.11 | -0.28 | -0.06 | -0.02 | 0.00 | -0.19 | 0.31 |
| HD 42250 | E | 5410 | 4.48 | 0.71 | 0.24 | 0.37 | 0.25 | 0.16 | 0.30 | 0.19 | 0.28 | 0.27 | 0.32 | 0.20 | 0.16 | 0.14 | 0.23 | 0.19 | 0.30 | 0.42 | 0.27 | 0.38 | 0.13 | 0.20 | 0.76 | 0.69 | 0.77 | 0.70 | 0.75 |
| HD 4256 | E | 4904 | 4.58 | 0.79 | 0.67 | 0.64 | 0.52 | 0.49 | 1.11 | 0.41 | 0.45 | 0.46 | 0.67 | 0.48 | 0.45 | 0.36 | 0.52 | 0.48 | 0.41 | 0.73 | 0.42 | 0.47 | 0.33 | 0.17 | 0.90 | 1.14 | 1.51 | 1.17 | 0.99 |
| HD 42618 | E | 5775 | 4.46 | 0.79 | 0.01 | -0.03 | -0.04 | -0.02 | -0.03 | 0.04 | 0.04 | 0.02 | 0.00 | 0.02 | -0.05 | -0.02 | 0.00 | -0.02 | -0.03 | -0.01 | 0.10 | 0.15 | 0.15 | 0.09 | 0.35 | 0.30 | 0.30 | 0.20 | 0.67 |

| ID | Type | T | logg | v | col5 | col6 | col7 | col8 | col9 | col10 | col11 | col12 | col13 | col14 | col15 | col16 | col17 | col18 | col19 | col20 | col21 | col22 | col23 | col24 | col25 | col26 | col27 | col28 | col29 |
|---|---|---|---|---|---|---|---|---|---|---|---|---|---|---|---|---|---|---|---|---|---|---|---|---|---|---|---|---|---|
| HD 42983 | E | 4918 | 3.61 | 1.07 | 0.46 | 0.37 | 0.47 | 0.29 | 0.65 | 0.27 | 0.31 | 0.37 | 0.54 | 0.25 | 0.24 | 0.17 | 0.23 | 0.23 | 0.52 | 0.48 | 0.19 | 0.28 | 0.21 | -0.04 | 0.47 | 0.63 | 0.61 | 0.61 | 0.55 |
| HD 43318 | E | 6241 | 3.87 | 1.70 | -0.08 | -0.07 | -0.13 | -0.06 | -0.19 | -0.01 | 0.02 | -0.05 | -0.01 | -0.08 | -0.23 | -0.13 | 0.14 | -0.16 | -0.31 | -0.33 | 0.45 | 0.03 | 0.09 | 0.10 | 0.16 | 0.13 | 0.14 | 0.06 | 0.34 |
| HD 43856 | E | 6150 | 4.19 | 1.31 | -0.08 | -0.09 | -0.20 | -0.09 | -0.07 | -0.05 | 0.03 | -0.08 | -0.04 | -0.13 | -0.24 | -0.15 | -0.05 | -0.17 | -0.34 | -0.24 | 0.20 | -0.01 | -0.02 | 0.00 | 0.24 | 0.12 | 0.09 | 0.06 | 0.21 |
| HD 44966 | E | 6349 | 4.25 | 1.86 | 0.28 | 0.15 | 0.15 | 0.19 | 0.09 | 0.20 | 0.29 | 0.24 | 0.19 | 0.20 | 0.13 | 0.13 | 0.33 | 0.15 | -0.11 | -0.04 | 0.05 | 0.36 | 0.62 | 0.29 | 0.63 | 0.66 | 0.41 | 0.21 | 0.49 |
| HD 45282 | E | 5422 | 3.23 | 1.54 | -1.53 | -0.94 | | -1.14 | | -1.06 | -1.34 | -1.10 | -1.19 | -1.30 | -1.79 | -1.36 | -1.03 | -1.43 | -1.91 | -1.57 | -0.08 | -1.47 | -0.56 | -1.26 | -0.92 | -1.10 | -0.91 | -0.85 | -0.11 |
| HD 46090 | E | 5575 | 4.36 | 1.04 | 0.07 | 0.04 | 0.00 | 0.07 | 0.33 | -0.02 | 0.07 | -0.02 | -0.01 | 0.11 | -0.11 | -0.05 | 0.02 | 0.00 | -0.19 | 0.00 | 0.27 | 0.26 | 0.41 | -0.04 | 0.36 | 0.35 | 0.43 | 0.52 | 0.63 |
| HD 46301 | E | 6396 | 3.74 | 3.57 | | | -0.15 | 0.36 | 0.41 | 0.35 | 0.36 | 0.76 | 0.53 | 0.55 | 0.56 | 0.31 | 0.62 | 0.27 | -0.16 | -0.34 | 1.90 | 0.98 | 1.77 | 0.10 | | 0.51 | 1.19 | 1.36 | 1.44 |
| HD 4635 | E | 5036 | 4.55 | 0.81 | 0.36 | 0.29 | 0.21 | 0.28 | 0.84 | 0.15 | 0.25 | 0.21 | 0.39 | 0.28 | 0.27 | 0.23 | 0.30 | 0.29 | 0.37 | 0.52 | 0.24 | 0.33 | 0.41 | 0.21 | 0.71 | 0.68 | 1.13 | 1.10 | 0.72 |
| HD 46780 | E | 5793 | 4.25 | 1.95 | 0.24 | 0.18 | 0.21 | 0.10 | 0.43 | 0.24 | 0.07 | 0.20 | 0.26 | 0.26 | 0.13 | 0.09 | 0.09 | 0.08 | -0.14 | -0.23 | 0.21 | 0.30 | 0.47 | -0.04 | 0.24 | 0.42 | 0.27 | 0.27 | 0.47 |
| HD 47157 | E | 5689 | 4.40 | 0.42 | 0.91 | 0.61 | 0.58 | 0.50 | 0.83 | 0.57 | 0.70 | 0.64 | 0.67 | 0.66 | 0.66 | 0.57 | 0.57 | 0.70 | 1.04 | 0.87 | 0.82 | 0.80 | 0.65 | 0.55 | 0.82 | 1.10 | 1.09 | 1.38 | 1.14 |
| HD 47309 | E | 5792 | 4.37 | 1.01 | 0.21 | 0.18 | 0.16 | 0.13 | 0.25 | 0.14 | 0.19 | 0.12 | 0.14 | 0.14 | 0.09 | 0.10 | 0.11 | 0.11 | 0.01 | 0.13 | -0.07 | 0.17 | 0.24 | 0.09 | 0.33 | 0.32 | 0.29 | 0.33 | 0.45 |
| HD 47752 | E | 4698 | 4.62 | 0.71 | 0.08 | 0.08 | -0.02 | 0.06 | 0.87 | 0.09 | 0.05 | 0.07 | 0.21 | 0.10 | -0.01 | 0.07 | 0.14 | 0.10 | 0.09 | 0.03 | 0.26 | 0.22 | -0.11 | 0.00 | 0.50 | 0.71 | 1.17 | 0.98 | 0.23 |
| HD 48565 | E | 6024 | 4.07 | 1.33 | -0.56 | -0.47 | -0.75 | -0.43 | -0.36 | -0.48 | -0.39 | -0.46 | -0.45 | -0.51 | -0.79 | -0.62 | -0.27 | -0.65 | -0.75 | -0.52 | 0.61 | 0.44 | 0.48 | 0.81 | 0.77 | 0.89 | 0.69 | 0.41 | 0.38 |
| HD 49385 | E | 6127 | 3.98 | 1.66 | 0.29 | 0.20 | 0.31 | 0.20 | 0.12 | 0.27 | 0.34 | 0.22 | 0.23 | 0.22 | 0.15 | 0.15 | 0.28 | 0.16 | -0.09 | 0.09 | 0.60 | 0.38 | 0.65 | 0.29 | 0.46 | 0.49 | 0.46 | 0.36 | 0.56 |
| HD 50206 | E | 6459 | 3.88 | 2.12 | 0.37 | 0.28 | 0.16 | 0.22 | 0.12 | 0.24 | 0.27 | 0.22 | 0.30 | 0.22 | 0.14 | 0.14 | 0.34 | 0.19 | 0.05 | -0.05 | 0.06 | 0.15 | 0.30 | 0.14 | 0.45 | 0.28 | 0.16 | 0.24 | 0.34 |
| HD 50867 | E | 6256 | 4.30 | 1.52 | 0.03 | 0.07 | -0.03 | -0.01 | -0.05 | 0.06 | 0.10 | 0.07 | 0.04 | 0.04 | -0.14 | -0.05 | 0.08 | -0.05 | -0.36 | -0.22 | -0.05 | 0.09 | 0.23 | 0.08 | 0.49 | 0.28 | 0.25 | 0.24 | 0.23 |
| HD 51219 | E | 5626 | 4.41 | 0.81 | 0.06 | 0.22 | 0.11 | 0.11 | 0.36 | 0.14 | 0.14 | 0.16 | 0.14 | 0.16 | 0.09 | 0.10 | 0.09 | 0.12 | 0.19 | 0.12 | 0.03 | 0.26 | 0.27 | 0.01 | 0.50 | 0.53 | 0.64 | 0.59 | 0.66 |
| HD 51419 | E | 5723 | 4.46 | 0.64 | -0.36 | -0.11 | -0.19 | -0.21 | -0.10 | -0.18 | -0.18 | -0.16 | -0.22 | -0.28 | -0.45 | -0.29 | -0.20 | -0.30 | -0.35 | -0.26 | -0.18 | -0.18 | -0.05 | -0.16 | 0.08 | -0.09 | 0.12 | 0.02 | 0.46 |
| HD 51866 | E | 4860 | 4.59 | 0.31 | 0.43 | 0.35 | 0.29 | 0.30 | 0.88 | 0.26 | 0.34 | 0.34 | 0.57 | 0.33 | 0.30 | 0.24 | 0.35 | 0.36 | 0.54 | 0.48 | 0.42 | 0.43 | 0.35 | 0.22 | 0.98 | 1.19 | 1.34 | 1.55 | 0.91 |
| HD 52634 | E | 5999 | 4.07 | 2.77 | 0.14 | -0.06 | 0.65 | 0.08 | 0.35 | 0.05 | 0.05 | 0.23 | 0.11 | 0.13 | 0.09 | -0.03 | 0.29 | 0.02 | -0.35 | -0.61 | | 0.42 | 1.02 | -0.40 | 0.71 | 0.52 | 0.38 | 0.33 | 0.12 |
| HD 5294 | E | 5749 | 4.47 | 0.82 | -0.08 | -0.03 | -0.11 | -0.04 | 0.09 | 0.05 | -0.03 | -0.01 | -0.03 | 0.04 | -0.06 | -0.03 | -0.01 | -0.06 | -0.25 | -0.12 | 0.18 | 0.27 | 0.17 | 0.27 | 0.39 | 0.44 | 0.34 | 0.26 | 0.44 |
| HD 5351 | E | 4608 | 4.65 | 0.15 | -0.30 | -0.05 | -0.10 | 0.04 | 1.06 | -0.07 | 0.00 | -0.02 | 0.12 | -0.12 | -0.38 | -0.33 | 0.01 | -0.14 | -0.10 | -0.30 | 0.00 | -0.08 | -0.16 | -0.47 | 0.46 | 0.88 | 1.01 | 0.84 | 0.63 |
| HD 53927 | E | 4941 | 4.62 | 0.21 | -0.19 | -0.18 | -0.13 | -0.13 | 0.65 | -0.16 | -0.16 | -0.09 | 0.04 | -0.13 | -0.26 | -0.18 | -0.11 | -0.15 | 0.04 | -0.18 | 0.05 | -0.14 | 0.03 | -0.18 | 0.30 | 0.63 | 0.80 | 0.58 | 0.48 |
| HD 54371 | E | 5562 | 4.43 | 1.36 | 0.18 | 0.11 | 0.14 | 0.11 | 0.54 | 0.21 | 0.08 | 0.07 | 0.10 | 0.17 | 0.09 | 0.08 | 0.07 | 0.08 | -0.14 | -0.16 | 0.28 | 0.20 | 0.34 | 0.20 | 0.39 | 0.35 | 0.47 | 0.58 | 0.77 |
| HD 5600 | E | 6378 | 3.77 | 2.28 | -0.19 | -0.18 | -0.13 | -0.10 | -0.11 | -0.05 | -0.04 | -0.11 | 0.06 | -0.17 | -0.30 | -0.23 | 0.10 | -0.24 | -0.45 | -0.44 | -0.43 | 0.05 | 0.34 | -0.07 | 0.49 | 0.24 | 0.18 | 0.01 | 0.34 |
| HD 56303 | E | 5912 | 4.30 | 1.19 | 0.30 | 0.28 | 0.26 | 0.20 | 0.28 | 0.28 | 0.29 | 0.22 | 0.23 | 0.26 | 0.22 | 0.20 | 0.24 | 0.22 | 0.15 | 0.09 | 0.38 | 0.31 | 0.58 | 0.19 | 0.35 | 0.42 | 0.44 | 0.41 | 0.96 |
| HD 56515 | E | 5993 | 4.07 | 1.41 | 0.09 | 0.14 | 0.13 | 0.11 | 0.07 | 0.20 | 0.06 | 0.06 | 0.11 | 0.11 | 0.02 | 0.06 | 0.12 | 0.06 | -0.09 | -0.14 | 0.10 | 0.18 | 0.20 | 0.09 | 0.18 | 0.28 | 0.23 | 0.20 | 0.54 |
| HD 57707 | E | 5021 | 3.32 | 1.40 | 0.32 | 0.29 | 0.35 | 0.16 | 0.47 | 0.26 | 0.12 | 0.26 | 0.44 | 0.22 | 0.25 | 0.11 | 0.14 | 0.14 | 0.35 | 0.07 | 0.23 | 0.13 | 0.19 | 0.14 | 0.43 | 0.49 | 0.49 | 0.49 | 0.22 |
| HD 59090 | E | 6428 | 3.92 | 2.43 | 0.40 | 0.36 | 0.16 | 0.29 | 0.28 | 0.31 | 0.34 | 0.37 | 0.36 | 0.29 | 0.14 | 0.19 | 0.55 | 0.25 | -0.14 | -0.37 | | 0.29 | 0.84 | 0.04 | 0.51 | 0.16 | 0.40 | 0.46 | 0.35 |
| HD 59747 | E | 5099 | 4.57 | 0.43 | 0.13 | 0.17 | 0.10 | 0.12 | 0.46 | 0.22 | 0.16 | 0.26 | 0.35 | 0.28 | 0.18 | 0.16 | 0.14 | 0.19 | 0.34 | 0.25 | 0.38 | 0.41 | 0.51 | 0.35 | 0.65 | 0.78 | 0.91 | 0.74 | 0.58 |
| HD 59984 | E | 6005 | 4.03 | 1.38 | -0.63 | -0.48 | -0.70 | -0.50 | -0.40 | -0.50 | -0.59 | -0.52 | -0.63 | -0.68 | -0.93 | -0.70 | -0.42 | -0.70 | -0.88 | -0.87 | -0.84 | -0.73 | -0.68 | -0.58 | -0.48 | -0.44 | -0.49 | -0.49 | -0.24 |
| HD 62161 | E | 6449 | 4.25 | 3.42 | | 0.07 | | 0.27 | 0.18 | 0.00 | 0.39 | 0.61 | 0.56 | 0.46 | 0.23 | 0.16 | 0.67 | 0.36 | | -0.52 | | 0.85 | 1.76 | -0.17 | 1.13 | 0.71 | 1.22 | 1.06 | |
| HD 62323 | E | 6113 | 4.06 | 1.60 | 0.21 | 0.15 | 0.12 | 0.10 | 0.05 | 0.15 | 0.13 | 0.10 | 0.09 | 0.11 | 0.01 | 0.04 | 0.14 | 0.04 | -0.15 | -0.12 | -0.11 | 0.22 | 0.12 | 0.06 | 0.42 | 0.21 | 0.20 | 0.29 | 0.51 |
| HD 62613 | E | 5539 | 4.54 | 0.55 | -0.01 | 0.03 | 0.03 | 0.00 | 0.17 | 0.05 | 0.06 | 0.07 | 0.07 | 0.06 | 0.00 | 0.03 | 0.00 | 0.01 | 0.03 | 0.17 | 0.16 | 0.16 | 0.18 | 0.16 | 0.44 | 0.43 | 0.42 | 0.45 | 0.51 |
| HD 63433 | E | 5673 | 4.50 | 1.53 | 0.06 | 0.08 | 0.08 | 0.04 | 0.22 | 0.11 | 0.03 | 0.06 | 0.07 | 0.14 | -0.05 | 0.03 | -0.02 | 0.00 | -0.22 | -0.31 | 0.33 | 0.22 | 0.25 | 0.26 | 0.44 | 0.58 | 0.45 | 0.42 | 0.25 |
| HD 64021 | E | 6864 | 3.88 | 3.80 | | | 0.38 | 0.12 | 0.25 | 0.21 | 1.27 | 0.82 | 0.99 | 0.17 | 0.31 | 2.07 | 0.36 | 0.69 | | 1.52 | 0.96 | 1.35 | | 0.42 | 1.44 | 1.26 | 0.57 | 1.67 | |
| HD 64090 | E | 5528 | 4.62 | 2.03 | -1.80 | -1.25 | | -1.29 | | -1.40 | -1.49 | -1.41 | -1.17 | -1.60 | -1.92 | -1.64 | -0.93 | -1.70 | -2.05 | -1.81 | -1.69 | -1.16 | -0.45 | -1.68 | -1.11 | -0.41 | -0.32 | -0.38 | -1.60 |
| HD 64468 | E | 4914 | 4.52 | 0.85 | 0.38 | 0.29 | 0.34 | 0.33 | 0.94 | 0.28 | 0.36 | 0.31 | 0.53 | 0.33 | 0.31 | 0.28 | 0.39 | 0.34 | 0.59 | 0.59 | 0.16 | 0.49 | 0.28 | 0.05 | 0.50 | 0.77 | 0.99 | 0.92 | 1.11 |

| | | | | | | | | | | | | | | | | | | | | | | | | | | | |
|---|---|---|---|---|---|---|---|---|---|---|---|---|---|---|---|---|---|---|---|---|---|---|---|---|---|---|---|
| HD 64606 | E | 5250 | 4.66 | 0.15 | -0.64 | -0.41 | -0.48 | -0.40 | | -0.53 | -0.45 | -0.45 | -0.50 | -0.66 | -0.91 | -0.76 | -0.55 | -0.68 | -0.73 | -0.44 | -0.47 | -0.28 | -0.29 | -0.75 | -0.11 | 0.02 | 0.00 | -0.04 | 0.04 |
| HD 64815 | E | 5722 | 3.86 | 1.19 | -0.19 | -0.04 | -0.12 | -0.09 | 0.04 | -0.09 | -0.14 | -0.13 | -0.19 | -0.29 | -0.51 | -0.35 | -0.20 | -0.32 | -0.37 | -0.25 | 0.07 | -0.20 | -0.03 | -0.23 | 0.06 | -0.14 | -0.05 | -0.08 | 0.44 |
| HD 65583 | E | 5342 | 4.55 | 0.41 | -0.48 | -0.32 | -0.30 | -0.39 | | -0.35 | -0.46 | -0.27 | -0.34 | -0.52 | -0.77 | -0.61 | -0.35 | -0.58 | -0.64 | -0.46 | -0.63 | -0.39 | -0.16 | -0.67 | -0.25 | -0.26 | 0.03 | -0.11 | 0.10 |
| HD 65874 | E | 5911 | 3.88 | 1.49 | 0.26 | 0.15 | 0.17 | 0.14 | 0.16 | 0.16 | 0.15 | 0.11 | 0.09 | 0.14 | 0.10 | 0.08 | 0.12 | 0.11 | 0.02 | 0.03 | 0.13 | 0.12 | 0.30 | 0.04 | 0.16 | 0.12 | 0.11 | 0.11 | 0.69 |
| HD 68168 | E | 5750 | 4.40 | 0.97 | 0.32 | 0.29 | 0.22 | 0.19 | 0.33 | 0.27 | 0.30 | 0.24 | 0.26 | 0.25 | 0.22 | 0.20 | 0.22 | 0.23 | 0.27 | 0.20 | 0.36 | 0.45 | 0.28 | 0.23 | 0.48 | 0.49 | 0.59 | 0.50 | 0.71 |
| HD 68284 | E | 5945 | 3.92 | 1.40 | -0.39 | -0.30 | -0.37 | -0.35 | -0.31 | -0.33 | -0.26 | -0.31 | -0.33 | -0.43 | -0.62 | -0.48 | -0.16 | -0.48 | -0.67 | -0.55 | -0.57 | -0.42 | -0.07 | -0.39 | -0.27 | -0.21 | -0.07 | -0.15 | 0.08 |
| HD 68380 | E | 6538 | 4.33 | 1.86 | -0.05 | -0.02 | 0.06 | -0.01 | 0.01 | 0.06 | 0.02 | 0.08 | 0.11 | 0.00 | -0.20 | -0.04 | 0.12 | -0.04 | -0.18 | -0.33 | | 0.13 | 0.54 | 0.19 | 0.26 | 0.24 | 0.28 | 0.31 | |
| HD 68638 | E | 5364 | 4.35 | 1.48 | -0.08 | -0.09 | -0.09 | -0.14 | 0.28 | -0.15 | -0.19 | -0.15 | -0.10 | -0.11 | -0.27 | -0.22 | -0.17 | -0.21 | -0.28 | -0.36 | -0.13 | -0.12 | 0.08 | -0.26 | 0.16 | 0.10 | 0.24 | 0.10 | 0.25 |
| HD 69611 | E | 5846 | 4.25 | 1.06 | -0.40 | -0.15 | -0.30 | -0.29 | -0.30 | -0.27 | -0.31 | -0.23 | -0.34 | -0.46 | -0.80 | -0.55 | -0.36 | -0.51 | -0.59 | -0.46 | -0.52 | -0.31 | 0.11 | -0.55 | -0.25 | -0.26 | -0.21 | -0.19 | 0.52 |
| HD 70088 | E | 5566 | 4.54 | 0.99 | -0.11 | -0.01 | -0.03 | 0.05 | 0.11 | 0.05 | 0.01 | 0.02 | 0.01 | 0.07 | -0.05 | 0.01 | -0.08 | -0.04 | -0.20 | -0.16 | -0.01 | 0.32 | 0.17 | 0.30 | 0.48 | 0.49 | 0.49 | 0.48 | 0.53 |
| HD 70516 | E | 5744 | 4.42 | 2.16 | 0.16 | 0.27 | 0.19 | 0.17 | 0.21 | 0.19 | 0.12 | 0.16 | 0.13 | 0.24 | 0.07 | 0.09 | 0.26 | 0.14 | -0.24 | -0.16 | 0.30 | 0.33 | 0.41 | 0.17 | 0.53 | 0.22 | 0.41 | 0.51 | 0.55 |
| HD 70923 | E | 6021 | 4.22 | 1.33 | 0.35 | 0.19 | 0.15 | 0.17 | 0.24 | 0.19 | 0.26 | 0.15 | 0.18 | 0.18 | 0.20 | 0.14 | 0.22 | 0.18 | 0.05 | 0.14 | 0.18 | 0.18 | 0.28 | 0.10 | 0.41 | 0.20 | 0.22 | 0.22 | 0.74 |
| HD 71431 | E | 5887 | 3.91 | 1.42 | 0.14 | 0.13 | 0.00 | 0.08 | 0.11 | 0.09 | 0.11 | 0.06 | 0.03 | 0.09 | 0.00 | 0.01 | 0.10 | 0.05 | 0.01 | 0.05 | 0.17 | 0.14 | 0.14 | 0.03 | 0.22 | 0.21 | 0.24 | 0.16 | 0.29 |
| HD 71595 | E | 6682 | 4.02 | 1.95 | 0.00 | 0.05 | -0.10 | 0.01 | 0.04 | 0.09 | 0.08 | 0.07 | 0.12 | 0.01 | -0.12 | -0.02 | 0.27 | -0.06 | -0.31 | -0.22 | 0.40 | 0.14 | 0.50 | 0.28 | 0.22 | 0.15 | 0.18 | 0.21 | 0.29 |
| HD 71640 | E | 6098 | 4.31 | 1.29 | -0.06 | -0.07 | -0.17 | -0.12 | -0.06 | 0.00 | -0.04 | -0.04 | 0.16 | -0.03 | -0.23 | -0.13 | -0.03 | -0.21 | -0.46 | -0.30 | 0.25 | 0.08 | 0.39 | 0.15 | 0.18 | 0.27 | 0.17 | 0.10 | 0.56 |
| HD 73393 | E | 5703 | 4.47 | 0.88 | 0.15 | 0.15 | 0.19 | 0.12 | 0.21 | 0.16 | 0.16 | 0.15 | 0.11 | 0.14 | 0.10 | 0.10 | 0.09 | 0.10 | -0.02 | 0.14 | 0.27 | 0.32 | 0.28 | 0.17 | 0.46 | 0.47 | 0.54 | 0.36 | 0.77 |
| HD 74011 | E | 5751 | 4.10 | 1.11 | -0.45 | -0.28 | -0.33 | -0.33 | -0.21 | -0.30 | -0.33 | -0.32 | -0.35 | -0.52 | -0.84 | -0.59 | -0.42 | -0.57 | -0.68 | -0.52 | -0.71 | -0.50 | -0.57 | -0.52 | -0.25 | -0.33 | -0.28 | -0.33 | 0.04 |
| HD 75318 | E | 5422 | 4.40 | 0.78 | -0.03 | -0.01 | 0.05 | -0.05 | 0.20 | 0.03 | -0.04 | -0.01 | -0.02 | -0.05 | -0.12 | -0.09 | -0.06 | -0.07 | 0.03 | 0.05 | -0.01 | 0.00 | -0.03 | -0.20 | 0.31 | 0.35 | 0.39 | 0.33 | 0.39 |
| HD 75767 | E | 5782 | 4.40 | 0.91 | 0.02 | 0.07 | 0.00 | 0.02 | 0.14 | 0.06 | 0.02 | -0.01 | 0.02 | 0.02 | -0.03 | -0.01 | -0.01 | -0.01 | -0.09 | 0.05 | 0.02 | 0.12 | 0.31 | 0.11 | 0.27 | 0.27 | 0.34 | 0.17 | 0.55 |
| HD 7590 | E | 5979 | 4.45 | 1.38 | -0.06 | -0.01 | 0.08 | -0.01 | 0.06 | 0.08 | 0.05 | 0.03 | 0.04 | 0.07 | -0.11 | -0.02 | -0.01 | -0.07 | -0.26 | -0.28 | 0.05 | 0.21 | 0.24 | 0.25 | 0.34 | 0.34 | 0.40 | 0.15 | 0.27 |
| HD 75935 | E | 5451 | 4.50 | 1.45 | 0.11 | 0.11 | 0.07 | 0.05 | 0.26 | 0.11 | 0.02 | 0.07 | 0.12 | 0.13 | 0.07 | 0.05 | 0.05 | 0.02 | 0.00 | -0.18 | 0.31 | 0.23 | 0.22 | 0.11 | 0.41 | 0.39 | 0.43 | 0.51 | 0.49 |
| HD 76932 | E | 5966 | 4.18 | 1.26 | -0.81 | -0.47 | -0.77 | -0.51 | -0.36 | -0.52 | -0.61 | -0.52 | -0.41 | -0.76 | -1.09 | -0.82 | -0.36 | -0.80 | -1.10 | -0.64 | -0.06 | -0.76 | -0.07 | -0.64 | -0.37 | -0.23 | -0.15 | -0.21 | -0.71 |
| HD 77407 | E | 5989 | 4.44 | 1.70 | 0.10 | 0.10 | 0.06 | 0.07 | 0.14 | 0.20 | 0.17 | 0.13 | 0.15 | 0.18 | 0.07 | 0.10 | 0.15 | 0.08 | -0.18 | -0.16 | 0.06 | 0.32 | 0.40 | 0.42 | 0.33 | 0.45 | 0.47 | 0.50 | 0.31 |
| HD 7924 | E | 5172 | 4.60 | 0.83 | -0.07 | -0.08 | -0.03 | -0.05 | 0.40 | -0.10 | -0.04 | -0.05 | 0.07 | -0.07 | -0.16 | -0.14 | -0.07 | -0.13 | -0.06 | -0.08 | 0.11 | 0.00 | 0.09 | -0.20 | 0.44 | 0.35 | 0.45 | 0.24 | 0.29 |
| HD 82443 | E | 5315 | 4.58 | 1.60 | 0.08 | 0.09 | 0.07 | 0.08 | 0.50 | 0.20 | -0.01 | 0.05 | 0.11 | 0.13 | -0.03 | 0.02 | 0.01 | -0.01 | -0.07 | -0.12 | 0.25 | 0.23 | 0.23 | 0.19 | 0.41 | 0.61 | 0.53 | 0.35 | 0.49 |
| HD 84937 | E | 6383 | 4.48 | 2.76 | -2.19 | -1.72 | | -0.99 | | -1.76 | -1.52 | -1.66 | -1.21 | -1.57 | -2.29 | -2.02 | -0.96 | -1.86 | | -1.86 | | -1.76 | 0.03 | -2.21 | 0.28 | 0.26 | -0.43 | | | |
| HD 87883 | E | 4940 | 4.56 | 0.70 | 0.38 | 0.31 | 0.22 | 0.30 | 0.70 | 0.24 | 0.29 | 0.28 | 0.47 | 0.32 | 0.21 | 0.29 | 0.28 | 0.31 | 0.40 | 0.44 | 0.21 | 0.30 | 0.31 | 0.17 | 0.94 | 1.18 | 1.31 | 0.90 | 0.90 |
| HD 88725 | E | 5753 | 4.45 | 0.52 | -0.48 | -0.24 | -0.39 | -0.38 | -0.06 | -0.32 | -0.36 | -0.27 | -0.37 | -0.47 | -0.68 | -0.52 | -0.36 | -0.51 | -0.65 | -0.40 | -0.58 | -0.30 | 0.19 | -0.39 | 0.09 | -0.10 | -0.13 | 0.02 | 0.17 |
| HD 89269 | E | 5635 | 4.49 | 0.63 | -0.11 | -0.02 | -0.04 | -0.07 | 0.12 | -0.04 | -0.03 | -0.04 | -0.05 | -0.06 | -0.13 | -0.09 | -0.07 | -0.09 | -0.06 | -0.11 | -0.05 | 0.08 | 0.03 | 0.05 | 0.22 | 0.24 | 0.36 | 0.24 | 0.45 |
| HD 90875 | E | 4567 | 4.56 | 0.64 | 0.73 | 0.56 | 0.54 | 0.63 | 1.12 | 0.66 | 0.62 | 0.51 | 0.70 | 0.61 | 0.46 | 0.46 | 0.60 | 0.60 | 0.75 | 1.00 | 0.62 | 0.55 | 0.49 | 0.20 | 0.95 | 1.49 | 1.46 | 1.73 | 0.58 |
| HD 93215 | E | 5786 | 4.40 | 1.23 | 0.43 | 0.37 | 0.30 | 0.25 | 0.29 | 0.34 | 0.32 | 0.28 | 0.33 | 0.34 | 0.30 | 0.27 | 0.26 | 0.30 | 0.24 | 0.28 | 0.44 | 0.40 | 0.39 | 0.21 | 0.48 | 0.54 | 0.47 | 0.51 | 0.98 |
| HD 9407 | E | 5652 | 4.44 | 0.69 | 0.22 | 0.15 | 0.19 | 0.10 | 0.22 | 0.11 | 0.13 | 0.15 | 0.15 | 0.15 | 0.13 | 0.10 | 0.14 | 0.12 | 0.23 | 0.32 | 0.17 | 0.21 | 0.18 | 0.03 | 0.48 | 0.42 | 0.34 | 0.36 | 0.57 |
| HD 94765 | E | 5033 | 4.58 | 1.00 | 0.20 | 0.15 | -0.02 | 0.15 | 0.73 | 0.17 | 0.11 | 0.14 | 0.24 | 0.21 | 0.09 | 0.12 | 0.12 | 0.12 | 0.33 | -0.08 | 0.33 | 0.27 | 0.26 | 0.16 | 0.55 | 0.57 | 0.81 | 0.62 | 0.48 |
| HD 97503 | E | 4451 | 4.65 | 0.54 | 0.19 | 0.22 | 0.00 | 0.43 | 1.24 | 0.36 | 0.30 | 0.13 | 0.31 | 0.25 | 0.10 | 0.22 | 0.25 | 0.20 | 0.45 | -0.02 | 0.64 | 0.22 | 0.20 | -0.01 | 0.95 | 1.03 | 1.32 | 1.00 | 0.12 |
| HD 97658 | E | 5157 | 4.57 | 0.90 | -0.18 | -0.21 | -0.14 | -0.13 | 0.35 | -0.17 | -0.22 | -0.18 | -0.09 | -0.16 | -0.28 | -0.26 | -0.17 | -0.24 | -0.19 | -0.24 | -0.03 | -0.04 | 0.00 | -0.27 | 0.15 | 0.29 | 0.41 | 0.29 | 0.31 |
| HD 98630 | E | 6033 | 3.86 | 1.83 | 0.72 | 0.48 | 0.33 | 0.36 | 0.50 | 0.33 | 0.27 | 0.29 | 0.28 | 0.34 | 0.35 | 0.25 | 0.32 | 0.34 | 0.34 | 0.43 | 0.09 | 0.31 | 0.31 | 0.03 | 0.24 | 0.31 | 0.27 | 0.25 | 0.64 |
| HD 98800 | E | 4213 | 3.82 | 1.43 | -0.07 | 0.25 | -0.09 | 0.59 | 1.88 | 0.16 | 0.04 | -0.19 | -0.03 | 0.11 | -0.05 | 0.17 | 0.14 | 0.17 | 0.00 | 0.36 | -0.18 | 0.28 | -0.30 | -0.01 | 0.40 | 1.04 | 0.69 | 0.57 | 0.52 |
| HD 99747 | E | 6676 | 4.15 | 2.22 | -0.32 | -0.30 | | | -0.33 | -0.39 | -0.27 | -0.25 | -0.35 | -0.33 | -0.41 | -0.64 | -0.47 | 0.10 | -0.47 | -0.74 | -0.62 | | -0.05 | 0.16 | -0.16 | 0.20 | 0.03 | -0.09 | 0.19 | |

|  |  |  |  |  |  |  |  |  |  |  |  |  |  |  |  |  |  |  |  |  |  |  |  |  |  |  |  |  |  |
|---|---|---|---|---|---|---|---|---|---|---|---|---|---|---|---|---|---|---|---|---|---|---|---|---|---|---|---|---|---|
| HR 1687 | E | 6540 | 4.13 | 2.53 | 0.53 | 0.52 | 0.53 | 0.38 | 0.43 | 0.36 | 0.44 | 0.40 | 0.52 | 0.37 | 0.30 | 0.28 | 0.51 | 0.39 | 0.00 | 0.20 | | 0.46 | 0.82 | 0.27 | | 0.53 | 0.47 | 0.97 | 0.66 |
| HR 3144 | E | 6064 | 3.70 | 2.49 | 0.48 | 0.44 | 0.43 | 0.33 | 0.35 | 0.27 | 0.38 | 0.34 | 0.28 | 0.27 | 0.18 | 0.22 | 0.32 | 0.27 | -0.07 | 0.02 | | 0.44 | 0.76 | 0.08 | 0.74 | 0.46 | 0.32 | 0.50 | 0.33 |
| HR 4657 | E | 6316 | 4.42 | 1.50 | -0.46 | -0.19 | -0.41 | -0.34 | -0.41 | -0.37 | -0.41 | -0.35 | -0.25 | -0.59 | -0.84 | -0.64 | -0.05 | -0.59 | -0.81 | -0.54 | | -0.34 | 0.19 | -0.59 | -0.10 | 0.12 | -0.22 | 0.05 | 0.11 |
| HR 4867 | E | 6293 | 4.32 | 3.19 | -0.32 | | | 0.26 | | 0.56 | 0.03 | 0.35 | 0.81 | 0.50 | -0.03 | 0.14 | 0.59 | 0.45 | | -0.59 | 1.35 | 1.31 | 0.81 | -0.07 | 0.78 | 0.82 | 1.80 | 0.79 | |
| HR 5307 | E | 6461 | 4.16 | 1.72 | 0.88 | 0.53 | 0.36 | 0.57 | 0.82 | 0.62 | 0.66 | 0.96 | 0.72 | 0.78 | 0.51 | 0.42 | 0.56 | 0.56 | 0.02 | 0.16 | | 1.20 | 2.72 | 0.12 | 1.11 | 1.13 | 0.94 | 1.61 | 0.70 |
| HR 7438 | E | 6726 | 4.28 | 5.02 | 1.84 | | | 0.46 | 0.34 | 0.11 | 1.15 | 0.75 | 1.40 | 0.59 | 0.74 | 0.23 | 1.35 | 0.38 | | | | 1.16 | 0.30 | | | 1.30 | 1.09 | 0.95 | |
| HR 784 | E | 6245 | 4.37 | 1.40 | 0.04 | 0.05 | -0.01 | 0.07 | 0.16 | 0.14 | 0.18 | 0.09 | 0.12 | 0.15 | -0.02 | 0.05 | 0.14 | 0.02 | -0.23 | 0.06 | 0.00 | 0.26 | 0.58 | 0.26 | 0.57 | 0.31 | 0.33 | 0.19 | 0.49 |
| HR 7955 | E | 6205 | 3.78 | 1.89 | 0.40 | 0.32 | 0.19 | 0.23 | 0.19 | 0.27 | 0.25 | 0.24 | 0.21 | 0.22 | 0.20 | 0.17 | 0.26 | 0.23 | 0.04 | 0.16 | 0.03 | 0.24 | 0.28 | 0.03 | 0.27 | 0.24 | 0.29 | 0.18 | 0.50 |
| V* BW Ari | E | 5186 | 4.58 | 1.40 | 0.39 | 0.20 | 0.18 | 0.27 | 0.61 | 0.17 | 0.23 | 0.17 | 0.25 | 0.26 | 0.24 | 0.20 | 0.26 | 0.26 | 0.33 | 0.22 | 0.25 | 0.30 | 0.12 | 0.18 | 0.60 | 0.55 | 0.82 | 0.68 | 0.79 |
| V* BZ Cet | E | 5014 | 4.52 | 1.60 | 0.33 | 0.23 | 0.22 | 0.31 | 0.90 | 0.27 | 0.20 | 0.16 | 0.28 | 0.27 | 0.19 | 0.21 | 0.24 | 0.24 | 0.34 | 0.32 | 0.23 | 0.32 | 0.19 | 0.11 | 0.49 | 0.60 | 0.82 | 0.70 | 0.70 |
| V* DI Cam | E | 6337 | 3.89 | 2.28 | 0.07 | 0.00 | -0.59 | -0.05 | 0.03 | 0.03 | -0.06 | 0.24 | 0.28 | 0.04 | -0.13 | -0.11 | 0.28 | -0.05 | -0.58 | -0.26 | | 0.31 | 0.88 | -0.20 | | 0.37 | 0.45 | 0.39 | 0.87 |
| V* EI Eri | E | 5494 | 3.82 | 2.67 | 0.47 | | 0.37 | 0.54 | | 0.31 | 0.62 | 0.13 | 0.19 | 0.59 | 0.01 | -0.06 | 0.40 | 0.34 | | | 1.93 | 0.94 | 0.70 | | | 0.39 | 0.21 | 1.56 | 0.77 |
| V* GM Com | E | 6709 | 4.29 | 2.86 | 0.00 | 0.07 | -0.09 | 0.02 | 0.01 | 0.06 | 0.00 | 0.11 | 0.43 | -0.06 | -0.02 | -0.06 | 0.38 | 0.00 | -0.40 | -0.48 | | 0.35 | 0.77 | 0.17 | 0.73 | 0.50 | 0.37 | 0.32 | 1.51 |
| V* KX Cnc | E | 6008 | 4.09 | 1.60 | 0.08 | 0.12 | 0.13 | 0.07 | 0.12 | 0.18 | 0.01 | 0.04 | 0.07 | 0.14 | 0.06 | 0.06 | 0.06 | 0.06 | -0.14 | -0.14 | 0.28 | 0.20 | 0.18 | 0.02 | 0.27 | 0.14 | 0.23 | 0.17 | 0.19 |
| V* MV UMa | E | 4665 | 4.62 | 0.57 | -0.12 | 0.10 | 0.02 | 0.07 | 0.74 | -0.07 | -0.02 | -0.03 | 0.06 | -0.10 | -0.25 | -0.13 | -0.01 | -0.08 | 0.00 | -0.05 | -0.09 | 0.01 | -0.18 | -0.34 | 0.35 | 0.36 | 0.72 | 0.45 | 0.54 |
| V* NX Aqr | E | 5656 | 4.50 | 1.21 | -0.06 | 0.07 | 0.00 | 0.03 | 0.24 | 0.13 | 0.04 | 0.07 | 0.11 | 0.12 | -0.01 | 0.02 | 0.02 | -0.04 | -0.26 | -0.08 | 0.38 | 0.18 | 0.72 | 0.25 | 0.60 | 0.76 | 0.82 | 0.63 | 0.80 |
| V* V1309 Tau | E | 5791 | 4.47 | 1.51 | 0.28 | 0.23 | 0.21 | 0.23 | 0.39 | 0.25 | 0.22 | 0.16 | 0.24 | 0.27 | 0.21 | 0.17 | 0.17 | 0.18 | -0.03 | 0.25 | 0.18 | 0.35 | 0.29 | 0.30 | 0.38 | 0.47 | 0.47 | 0.58 | 0.55 |
| V* V1386 Ori | E | 5294 | 4.56 | 1.14 | 0.11 | 0.10 | 0.16 | 0.10 | 0.48 | 0.18 | 0.03 | 0.12 | 0.18 | 0.18 | 0.05 | 0.10 | 0.07 | 0.08 | 0.00 | -0.22 | 0.36 | 0.28 | 0.25 | 0.27 | 0.51 | 0.59 | 0.64 | 0.72 | 0.54 |
| V* V1709 Aql | E | 6913 | 4.13 | 3.36 | -0.19 | -0.07 | | 0.26 | 0.54 | 0.21 | 0.47 | 0.64 | 1.17 | 0.53 | 0.17 | 0.19 | 0.75 | 0.64 | 0.06 | -0.53 | | 0.62 | 1.90 | -0.24 | 1.16 | 0.53 | 0.96 | 0.92 | |
| V* V401 Hya | E | 5836 | 4.48 | 1.42 | 0.16 | 0.16 | 0.20 | 0.15 | 0.16 | 0.23 | 0.17 | 0.18 | 0.18 | 0.21 | 0.18 | 0.14 | 0.15 | 0.14 | -0.04 | -0.04 | 0.47 | 0.37 | 0.94 | 0.25 | 0.53 | 0.55 | 0.60 | 0.46 | 0.34 |
| V* V457 Vul | E | 5433 | 4.50 | 1.16 | 0.05 | 0.13 | 0.07 | 0.10 | 0.21 | 0.18 | 0.06 | 0.14 | 0.11 | 0.19 | 0.01 | 0.06 | 0.04 | 0.07 | -0.02 | -0.12 | 0.32 | 0.36 | 0.24 | 0.34 | 0.58 | 0.61 | 0.51 | 0.43 | 0.41 |
| V* V566 Oph | E | 6358 | 4.05 | 2.80 | -0.92 | | -0.19 | | 0.48 | 1.87 | 1.05 | 0.84 | 0.82 | 0.28 | -0.71 | 0.77 | 0.39 | | | | 1.20 | 4.57 | | 0.31 | 0.84 | 0.19 | 0.36 | 0.60 | |

| Column | Units | Designation | Description |
| --- | --- | --- | --- |
| Sp | N/A | N/A | Source for spectroscopic material |
| T | K | $T_{eff}$ | Effective Temperature |
| G | cm s$^{-2}$ | log g | log of the surface acceleration due to gravity |
| Vt | km s$^{-1}$ | $V_t$ | Microturbulent velocity |
| Na | Solar | [Na/H] | The abundance of sodium given logarithmically with respect to the solar value. |
| Mg | Solar | [Mg/H] | The abundance of magnesium given logarithmically with respect to the solar value. |
| Al | Solar | [Al/H] | The abundance of aluminum given logarithmically with respect to the solar value. |
| Si | Solar | [Si/H] | The abundance of silicon given logarithmically with respect to the solar value. |
| S | Solar | [S/H] | The abundance of sulfur given logarithmically with respect to the solar value. |
| Ca | Solar | [Ca/H] | The abundance of calcium given logarithmically with respect to the solar value. |
| Sc | Solar | [Sc/H] | The abundance of scandium given logarithmically with respect to the solar value. |
| Ti | Solar | [Ti/H] | The abundance of titanium given logarithmically with respect to the solar value. |
| V | Solar | [V/H] | The abundance of vanadium given logarithmically with respect to the solar value. |
| Cr | Solar | [Cr/H] | The abundance of chromium given logarithmically with respect to the solar value. |
| Mn | Solar | [Mn/H] | The abundance of manganese given logarithmically with respect to the solar value. |
| Fe | Solar | [Fe/H] | The abundance of iron given logarithmically with respect to the solar value. |
| Co | Solar | [Co/H] | The abundance of cobalt given logarithmically with respect to the solar value. |
| Ni | Solar | [Ni/H] | The abundance of nickel given logarithmically with respect to the solar value. |
| Cu | Solar | [Cu/H] | The abundance of copper given logarithmically with respect to the solar value. |
| Zn | Solar | [Zn/H] | The abundance of zinc given logarithmically with respect to the solar value. |
| Sr | Solar | [Sr/H] | The abundance of strontium given logarithmically with respect to the solar value. |
| Y | Solar | [Y/H] | The abundance of yttrium given logarithmically with respect to the solar value. |
| Zr | Solar | [Zr/H] | The abundance of zirconium given logarithmically with respect to the solar value. |
| Ba | Solar | [Ba/H] | The abundance of barium given logarithmically with respect to the solar value. |
| La | Solar | [La/H] | The abundance of lanthanum given logarithmically with respect to the solar value. |
| Ce | Solar | [Ce/H] | The abundance of cerium given logarithmically with respect to the solar value. |
| Nd | Solar | [Nd/H] | The abundance of neodymium given logarithmically with respect to the solar value. |
| Sm | Solar | [Sm/H] | The abundance of samarium given logarithmically with respect to the solar value. |
| Eu | Solar | [Eu/H] | The abundance of europium given logarithmically with respect to the solar value. |

Table 5

Lithium, Carbon, and Oxygen Data

| Primary | Sp | T (K) | G (cm s$^{-2}$) | $V_t$ (km s$^{-1}$) | $V_r$ (km s$^{-1}$) | Fe (log ε) | Li (log ε) | NLTE | L | 505.2 (log ε) | 538.0 (log ε) | $C_2$ (log ε) | 615.5 (log ε) | 630.0 (log ε) | C (log ε) | O (log ε) |
|---|---|---|---|---|---|---|---|---|---|---|---|---|---|---|---|---|
| 10 CVn | S | 5987 | 4.44 | 1.08 | 3.9 | 7.01 | 2.04 | 0.04 | | 7.78 | 7.86 | 8.03 | 8.64 | 8.66 | 7.89 | 8.65 |
| 10 Tau | S | 6013 | 4.05 | 1.48 | 6.0 | 7.41 | 2.42 | 0.04 | | 8.27 | 8.29 | 8.43 | 8.79 | 8.72 | 8.31 | 8.75 |
| 107 Psc | S | 5259 | 4.58 | 0.56 | 2.5 | 7.48 | 0.52 | 0.09 | L | 8.21 | 8.42 | 8.51 | 8.73 | 8.82 | 8.38 | 8.80 |
| 109 Psc | S | 5604 | 3.94 | 1.22 | 5.1 | 7.59 | 1.94 | 0.07 | | 8.41 | 8.54 | 8.49 | 8.70 | 8.69 | 8.48 | 8.69 |
| 11 Aql | S | 6144 | 3.57 | 2.83 | 28.2 | 7.50 | 1.88 | 0.03 | L | 8.18 | 8.28 | | 8.80 | 8.70 | 8.23 | 8.75 |
| 11 Aqr | S | 5929 | 4.24 | 1.40 | 5.7 | 7.73 | 2.43 | 0.04 | | 8.61 | 8.61 | 8.71 | 9.00 | 8.88 | 8.64 | 8.91 |
| 11 LMi | S | 5498 | 4.43 | 1.32 | 4.7 | 7.74 | 0.60 | 0.07 | L | 8.55 | 8.66 | 8.67 | 9.04 | 8.91 | 8.62 | 8.94 |
| 110 Her | S | 6457 | 3.94 | 2.37 | 18.0 | 7.56 | 1.14 | 0.01 | L | 8.21 | 8.30 | | 8.88 | | 8.25 | 8.88 |
| 111 Tau | S | 6184 | 4.38 | 2.04 | 17.0 | 7.57 | 2.95 | 0.03 | | 8.31 | 8.29 | 8.33 | 8.91 | 8.79 | 8.31 | 8.85 |
| 111 Tau B | S | 4576 | 4.66 | 0.80 | 4.4 | 7.65 | -0.02 | 0.14 | | | | 8.94 | | 9.10 | 8.94 | 9.10 |
| 112 Psc | S | 6031 | 4.03 | 1.68 | 7.7 | 7.75 | 1.05 | 0.04 | L | 8.55 | 8.64 | 8.73 | 9.06 | 8.87 | 8.62 | 8.97 |
| 12 Oph | S | 5262 | 4.57 | 1.02 | 2.4 | 7.50 | 0.26 | 0.09 | L | 8.32 | 8.52 | 8.47 | 8.70 | 8.69 | 8.44 | 8.69 |
| 13 Cet | S | 6080 | 4.07 | 0.70 | 12.0 | 7.40 | 2.18 | 0.03 | | 7.78 | 8.48 | 8.54 | 8.65 | 8.45 | 8.21 | 8.55 |
| 13 Ori | S | 5800 | 4.07 | 1.22 | 5.3 | 7.30 | 1.96 | 0.05 | | 8.14 | 8.17 | 8.26 | 8.85 | 8.75 | 8.19 | 8.78 |
| 13 Tri | S | 5957 | 3.95 | 1.55 | 6.1 | 7.31 | 2.61 | 0.04 | | 8.16 | 8.19 | 8.30 | 8.64 | 8.54 | 8.22 | 8.56 |
| 14 Cet | S | 6512 | 3.83 | 1.80 | 7.3 | 7.35 | 1.90 | 0.00 | | 8.07 | 8.08 | | 8.66 | | 8.08 | 8.66 |
| 14 Her | S | 5248 | 4.41 | 0.93 | 3.3 | 7.94 | 0.91 | 0.09 | | | | 8.82 | | 8.93 | 8.82 | 8.93 |
| 15 LMi | S | 5916 | 4.06 | 1.42 | 5.2 | 7.56 | 2.30 | 0.04 | | 8.39 | 8.42 | 8.47 | 8.85 | 8.76 | 8.43 | 8.78 |
| 15 Sge | S | 5946 | 4.40 | 1.21 | 5.6 | 7.52 | 2.33 | 0.04 | | 8.33 | 8.41 | 8.47 | 8.89 | 8.54 | 8.40 | 8.63 |
| 16 Cyg A | S | 5800 | 4.28 | 1.13 | 5.2 | 7.57 | 0.93 | 0.05 | L | 8.43 | 8.42 | 8.44 | 8.79 | 8.63 | 8.43 | 8.67 |
| 16 Cyg B | S | 5753 | 4.34 | 0.93 | 5.1 | 7.56 | 1.30 | 0.06 | | 8.39 | 8.51 | 8.45 | 8.85 | 8.70 | 8.45 | 8.74 |
| 17 Crt A | S | 6240 | 4.17 | 1.80 | 10.0 | 7.50 | 1.83 | 0.02 | | 8.17 | 8.24 | 8.50 | 8.78 | 8.84 | 8.27 | 8.81 |
| 17 Crt B | S | 6269 | 4.20 | 1.76 | 9.6 | 7.52 | 2.19 | 0.02 | | 8.17 | 8.23 | 8.39 | 8.74 | 8.77 | 8.24 | 8.75 |
| 17 Vir | S | 6146 | 4.33 | 1.60 | 7.8 | 7.60 | 2.84 | 0.03 | | 8.38 | 8.40 | 8.52 | 8.78 | 8.68 | 8.41 | 8.73 |
| 18 Cam | S | 5958 | 3.93 | 1.57 | 6.0 | 7.50 | 2.02 | 0.04 | | 8.29 | 8.38 | 8.39 | 8.78 | 8.70 | 8.36 | 8.72 |
| 18 Cet | S | 5861 | 3.99 | 1.29 | 5.3 | 7.28 | 2.43 | 0.05 | | 8.10 | 8.17 | 8.30 | 8.63 | 8.61 | 8.19 | 8.61 |
| 18 Sco | S | 5791 | 4.42 | 1.23 | 4.4 | 7.48 | 1.52 | 0.05 | | 8.26 | 8.33 | 8.42 | 8.65 | 8.54 | 8.34 | 8.57 |
| 20 LMi | S | 5771 | 4.32 | 1.33 | 4.7 | 7.67 | 1.46 | 0.05 | | 8.41 | 8.46 | 8.62 | 8.72 | 8.63 | 8.50 | 8.65 |
| 21 Eri | S | 5150 | 3.66 | 1.31 | 4.5 | 7.51 | 1.16 | 0.10 | | | | 8.47 | | 8.76 | 8.47 | 8.76 |
| 23 Lib | S | 5717 | 4.26 | 1.38 | 4.5 | 7.71 | 1.14 | 0.06 | | 8.61 | 8.65 | 8.74 | 8.94 | 8.86 | 8.67 | 8.88 |
| 24 LMi | S | 5760 | 4.03 | 1.35 | 5.7 | 7.47 | 1.96 | 0.06 | | 8.41 | 8.44 | 8.46 | 8.75 | 8.61 | 8.44 | 8.64 |
| 26 Dra | S | 5925 | 4.37 | 1.39 | 6.0 | 7.45 | 2.52 | 0.04 | | 8.30 | 8.36 | 8.31 | 8.75 | 8.58 | 8.32 | 8.62 |
| 33 Sex | S | 5124 | 3.72 | 1.24 | 4.8 | 7.47 | -0.03 | 0.10 | L | | | 8.44 | | 8.73 | 8.44 | 8.73 |

| Name | | HR | | | | | | | | | | | | | |
|---|---|---|---|---|---|---|---|---|---|---|---|---|---|---|---|
| 35 Leo | S | 5736 | 3.93 | 1.39 | 5.5 | 7.50 | 2.30 | 0.06 | | 8.41 | 8.40 | 8.48 | 8.83 | 8.66 | 8.43 | 8.70 |
| 36 And | S | 4809 | 3.19 | 1.50 | 5.8 | 7.58 | 0.54 | 0.12 | L | | 8.46 | | 8.79 | 8.46 | 8.79 |
| 36 Oph A | S | 5103 | 4.64 | 0.92 | 4.3 | 7.23 | -0.03 | 0.10 | L | | | 8.17 | | 8.40 | 8.17 | 8.40 |
| 36 Oph B | S | 5199 | 4.62 | 1.01 | 4.4 | 7.24 | 0.07 | 0.09 | L | | | 8.16 | | 8.31 | 8.16 | 8.31 |
| 36 UMa | S | 6173 | 4.40 | 1.33 | 5.3 | 7.38 | 2.67 | 0.03 | | 8.14 | 8.27 | 8.30 | 8.67 | 8.53 | 8.22 | 8.60 |
| 37 Gem | S | 5932 | 4.40 | 1.02 | 5.3 | 7.32 | 1.94 | 0.04 | | 8.16 | 8.22 | 8.31 | 8.70 | 8.58 | 8.23 | 8.61 |
| 38 LMi | S | 6090 | 3.71 | 2.66 | 15.9 | 7.82 | 2.98 | 0.03 | | 8.56 | 8.52 | 8.69 | 9.06 | 9.00 | 8.57 | 9.03 |
| 39 Gem | S | 6112 | 3.81 | 1.78 | 6.8 | 7.06 | 1.21 | 0.03 | L | 7.96 | 8.03 | 7.97 | 8.61 | 8.66 | 7.99 | 8.63 |
| 39 Leo | S | 6187 | 4.29 | 1.55 | 6.1 | 7.11 | 2.51 | 0.03 | | 7.97 | 8.04 | 8.11 | 8.41 | 8.55 | 8.03 | 8.48 |
| 39 Ser | S | 5830 | 4.45 | 1.02 | 4.2 | 7.04 | 1.23 | 0.05 | | 7.97 | 8.04 | 8.06 | 8.21 | 8.21 | 8.02 | 8.21 |
| 39 Tau | S | 5836 | 4.44 | 1.33 | 5.6 | 7.50 | 2.43 | 0.05 | | 8.33 | 8.36 | 8.38 | 8.70 | 8.65 | 8.36 | 8.66 |
| 4 Equ | S | 6086 | 3.76 | 1.74 | 7.9 | 7.46 | 1.08 | 0.03 | L | 8.35 | 8.39 | 8.37 | 8.86 | 8.67 | 8.37 | 8.77 |
| 42 Cap | S | 5706 | 3.65 | 1.62 | 8.1 | 7.45 | 2.24 | 0.06 | | 8.27 | 8.36 | 8.43 | 8.92 | 8.71 | 8.36 | 8.76 |
| 44 And | S | 5951 | 3.57 | 1.96 | 12.7 | 7.50 | 0.93 | 0.04 | L | 8.26 | 8.27 | 8.40 | 8.87 | 8.70 | 8.31 | 8.74 |
| 47 UMa | S | 5960 | 4.34 | 1.20 | 5.6 | 7.54 | 1.83 | 0.04 | | 8.35 | 8.35 | 8.53 | 8.80 | 8.72 | 8.41 | 8.74 |
| 49 Lib | S | 6297 | 3.91 | 2.04 | 11.0 | 7.48 | 0.45 | 0.02 | L | 8.28 | 8.41 | 8.57 | 8.96 | 8.95 | 8.39 | 8.96 |
| 49 Per | S | 4984 | 3.45 | 1.06 | 4.8 | 7.57 | 0.44 | 0.11 | | | | 8.38 | | 8.78 | 8.38 | 8.78 |
| 5 Ser | S | 6134 | 3.95 | 1.69 | 6.5 | 7.46 | 0.81 | 0.03 | L | 8.35 | 8.38 | 8.42 | 8.81 | 8.72 | 8.38 | 8.76 |
| 50 Per | S | 6313 | 4.33 | 2.21 | 19.0 | 7.73 | 3.01 | 0.02 | | 8.38 | 8.60 | 8.38 | 9.03 | 8.89 | 8.47 | 8.96 |
| 51 Boo Bn | S | 5821 | 4.43 | 1.11 | 5.3 | 7.60 | 2.19 | 0.05 | | 8.59 | 8.56 | 8.40 | 8.75 | 8.82 | 8.51 | 8.80 |
| 51 Boo Bs | S | 5990 | 4.33 | 1.77 | 7.1 | 7.58 | 2.70 | 0.04 | | 8.33 | 8.44 | 8.61 | 8.80 | 8.62 | 8.46 | 8.67 |
| 51 Peg | S | 5799 | 4.34 | 1.28 | 5.6 | 7.67 | 1.30 | 0.05 | | 8.47 | 8.54 | 8.65 | 9.01 | 8.89 | 8.55 | 8.92 |
| 54 Cnc | S | 5824 | 3.94 | 1.46 | 6.0 | 7.59 | 2.52 | 0.05 | | 8.48 | 8.51 | 8.56 | 8.85 | 8.80 | 8.51 | 8.82 |
| 54 Psc | S | 5289 | 4.52 | 0.72 | 4.1 | 7.70 | 0.67 | 0.09 | | 8.32 | 8.55 | 8.63 | 8.82 | 8.83 | 8.50 | 8.82 |
| 55 Vir | S | 5054 | 3.29 | 1.42 | 4.8 | 7.12 | 0.19 | 0.10 | L | | | 8.11 | | 8.54 | 8.11 | 8.54 |
| 58 Eri | S | 5839 | 4.47 | 1.29 | 6.4 | 7.48 | 2.39 | 0.05 | | 8.21 | 8.33 | 8.31 | 8.75 | 8.70 | 8.28 | 8.71 |
| 59 Eri | S | 6275 | 3.90 | 1.99 | 9.4 | 7.61 | 1.18 | 0.02 | L | 8.30 | 8.33 | 8.58 | 8.79 | 8.83 | 8.37 | 8.81 |
| 61 Cyg A | S | 4481 | 4.67 | 0.15 | 4.0 | 7.33 | -0.45 | 0.14 | L | | | 8.89 | | 9.02 | 8.89 | 9.02 |
| 61 Cyg B | S | 4171 | 4.70 | 0.15 | 5.5 | 7.39 | 0.06 | 0.16 | | | | | | | | |
| 61 Psc | S | 6273 | 3.92 | 2.11 | 10.6 | 7.65 | 2.89 | 0.02 | | 8.62 | 8.63 | 8.52 | 8.89 | 8.95 | 8.60 | 8.92 |
| 61 UMa | S | 5507 | 4.54 | 1.08 | 3.6 | 7.41 | 0.91 | 0.07 | | 8.22 | 8.34 | 8.38 | 8.68 | 8.58 | 8.31 | 8.61 |
| 61 Vir | S | 5578 | 4.44 | 0.92 | 4.6 | 7.46 | 0.26 | 0.07 | L | 8.29 | 8.27 | 8.39 | 8.78 | 8.73 | 8.32 | 8.74 |
| 63 Eri | S | 5422 | 3.33 | 1.69 | 6.8 | 7.30 | 1.02 | 0.08 | | 7.85 | 7.92 | 8.09 | 8.83 | 8.61 | 7.95 | 8.67 |
| 64 Aql | S | 4736 | 3.14 | 1.17 | 4.9 | 7.47 | 0.07 | 0.13 | | | | 8.49 | | 8.86 | 8.49 | 8.86 |
| 66 Cet | S | 6102 | 3.83 | 1.68 | 6.8 | 7.51 | 2.58 | 0.03 | | 8.37 | 8.34 | 8.55 | 8.79 | 8.68 | 8.39 | 8.74 |
| 66 Cet B | S | 5722 | 4.22 | 1.31 | 4.6 | 7.56 | 1.55 | 0.06 | | 8.38 | 8.42 | 8.50 | 8.99 | 8.68 | 8.43 | 8.76 |
| 70 Oph A | S | 5244 | 4.47 | 0.91 | 4.3 | 7.48 | -0.60 | 0.09 | | | | 8.33 | | 8.46 | 8.33 | 8.46 |
| 70 Vir | S | 5538 | 3.90 | 1.27 | 5.1 | 7.39 | 1.83 | 0.07 | | 8.17 | 8.21 | 8.33 | 8.83 | 8.70 | 8.24 | 8.73 |

| Name | | Teff | log g | vt | vsini | | | | | | | | | |
|---|---|---|---|---|---|---|---|---|---|---|---|---|---|---|
| 79 Cet | S | 5765 | 4.03 | 1.40 | 5.2 | 7.60 | 1.47 | 0.05 | | 8.35 | 8.40 | 8.57 | 8.80 | 8.92 | 8.44 | 8.89 |
| 83 Leo | S | 5380 | 4.42 | 1.04 | 5.1 | 7.76 | 0.44 | 0.08 | L | 8.66 | 8.76 | 8.66 | 8.95 | 8.82 | 8.69 | 8.85 |
| 83 Leo B | S | 4973 | 4.58 | 0.75 | 4.3 | 7.79 | 0.42 | 0.11 | | | | 8.85 | | 9.09 | 8.85 | 9.09 |
| 84 Her | S | 5810 | 3.77 | 1.76 | 7.7 | 7.71 | 0.90 | 0.05 | L | 8.58 | 8.58 | 8.70 | 9.10 | 9.13 | 8.62 | 9.13 |
| 85 Peg | S | 5454 | 4.54 | 0.58 | 3.2 | 6.72 | -0.09 | 0.08 | L | 7.85 | 8.03 | 7.86 | | 8.14 | 7.91 | 8.14 |
| 88 Leo | S | 6030 | 4.39 | 1.28 | 5.7 | 7.47 | 2.54 | 0.04 | | 8.29 | 8.32 | 8.38 | 8.86 | 8.67 | 8.32 | 8.76 |
| 9 Cet | S | 5822 | 4.46 | 1.62 | 8.2 | 7.69 | 2.58 | 0.05 | | 8.44 | 8.50 | 8.52 | 8.96 | 8.81 | 8.49 | 8.85 |
| 9 Com | S | 6239 | 3.91 | 2.06 | 9.8 | 7.68 | 1.30 | 0.02 | L | 8.57 | 8.55 | 8.65 | 9.09 | 8.96 | 8.58 | 9.03 |
| 94 Aqr | S | 5379 | 3.82 | 1.40 | 5.9 | 7.54 | 1.93 | 0.08 | | 8.59 | 8.60 | 8.50 | 8.81 | 8.77 | 8.56 | 8.78 |
| 94 Aqr B | S | 5219 | 4.35 | 1.29 | 3.9 | 7.66 | 0.67 | 0.09 | | | | 8.62 | | 8.76 | 8.62 | 8.76 |
| 94 Cet | S | 6064 | 4.04 | 1.82 | 10.3 | 7.66 | 1.89 | 0.03 | L | 8.52 | 8.54 | 8.57 | 9.05 | 8.94 | 8.54 | 9.00 |
| alf Aql | S | 7377 | 3.95 | 5.68 | 90.8 | 7.94 | 2.64 | -0.06 | L | 8.45 | | | 9.00 | | 8.45 | 9.00 |
| alf Cep | S | 7217 | 3.69 | 3.82 | 68.6 | 7.94 | 3.18 | -0.04 | L | 8.45 | | | 8.90 | | 8.45 | 8.90 |
| alf CMi | S | 6654 | 3.95 | 2.25 | 7.4 | 7.44 | 1.37 | -0.01 | L | 8.32 | 8.36 | | 8.79 | | 8.34 | 8.79 |
| alf Com A | S | 6391 | 4.09 | 2.29 | 20.5 | 7.39 | 2.13 | 0.01 | L | 8.15 | 8.17 | | 8.74 | | 8.16 | 8.74 |
| alf Crv | S | 7019 | 4.27 | 2.77 | 23.4 | 7.44 | 3.19 | -0.03 | | 8.29 | 8.28 | | 8.80 | | 8.29 | 8.80 |
| alf For A | S | 6195 | 3.95 | 1.77 | 7.3 | 7.26 | 1.99 | 0.03 | | 8.13 | 8.17 | 8.29 | 8.65 | 8.45 | 8.18 | 8.55 |
| alf PsA | S | 7671 | 4.03 | 3.90 | 95.4 | 7.76 | 2.97 | -0.08 | L | 8.45 | 8.38 | | 8.83 | | 8.42 | 8.83 |
| b Aql | S | 5466 | 4.10 | 1.45 | 6.1 | 7.76 | 0.89 | 0.08 | L | 8.71 | 8.70 | 8.70 | | 8.90 | 8.71 | 8.90 |
| b01 Cyg | S | 5090 | 3.69 | 1.38 | 4.8 | 7.44 | 0.63 | 0.10 | | | | 8.32 | | 8.75 | 8.32 | 8.75 |
| BD+04 701A | S | 6056 | 4.50 | 1.45 | 10.0 | 7.48 | 2.05 | 0.04 | | 8.28 | 8.41 | 8.35 | | 8.80 | 8.34 | 8.80 |
| BD+04 701B | S | 5642 | 4.50 | 0.58 | 10.0 | 7.09 | 1.78 | 0.06 | | 8.86 | 8.57 | 7.85 | | 8.71 | 8.42 | 8.71 |
| BD+18 2776 | S | 3644 | 4.46 | 0.67 | 6.9 | 7.89 | -0.49 | 0.20 | | | | | | | | |
| BD+27 4120 | S | 3899 | 4.80 | 0.35 | 8.5 | 7.65 | | 0.18 | | | | | | | | |
| BD+29 2963 | S | 5569 | 4.15 | 1.07 | 6.4 | 7.09 | 1.13 | 0.07 | | 8.39 | 8.25 | 8.10 | 8.70 | 8.64 | 8.25 | 8.66 |
| BD+30 2512 | S | 4313 | 4.68 | 0.15 | 4.1 | 7.68 | -0.10 | 0.15 | | | | | | | | |
| BD+33 529 | S | 3896 | 4.51 | 0.15 | 5.2 | 6.83 | -0.54 | 0.18 | | | | | | | | |
| BD+61 195 | S | 3799 | 4.84 | 0.15 | 8.0 | 7.93 | 0.09 | 0.19 | | | | | | | | |
| BD+67 1468A | S | 6680 | 4.13 | 2.89 | 23.0 | 7.80 | 2.19 | -0.01 | L | 8.46 | 8.60 | | 8.90 | | 8.53 | 8.90 |
| BD+67 1468B | S | 6642 | 4.15 | 2.78 | 18.6 | 7.77 | 0.90 | 0.00 | L | 8.48 | 8.58 | | 8.85 | | 8.53 | 8.85 |
| BD-04 782 | S | 4342 | 4.68 | 0.15 | 4.6 | 7.72 | -0.07 | 0.15 | | | | 9.07 | | 9.25 | 9.07 | 9.25 |
| BD-10 3166 | S | 5314 | 4.66 | 0.89 | 5.4 | 7.90 | 1.04 | 0.09 | | 8.51 | 8.56 | 8.94 | | 9.22 | 8.67 | 9.22 |
| bet Aql | S | 5144 | 3.58 | 1.22 | 4.6 | 7.28 | 0.43 | 0.10 | L | | | 8.25 | | 8.60 | 8.25 | 8.60 |
| bet Com | S | 6022 | 4.40 | 1.31 | 6.6 | 7.53 | 2.57 | 0.04 | | 8.36 | 8.34 | 8.37 | 8.73 | 8.59 | 8.36 | 8.66 |
| bet CVn | S | 5865 | 4.40 | 1.04 | 5.1 | 7.25 | 1.58 | 0.05 | | 8.11 | 8.27 | 8.21 | 8.68 | 8.73 | 8.20 | 8.72 |
| bet Vir | S | 6159 | 4.08 | 1.60 | 6.4 | 7.61 | 2.00 | 0.03 | | 8.45 | 8.48 | 8.59 | 8.89 | 8.91 | 8.49 | 8.90 |
| c Eri | S | 7146 | 4.23 | 4.98 | 81.2 | 7.70 | 2.59 | -0.04 | L | 8.34 | 8.49 | | 8.80 | | 8.41 | 8.80 |
| c UMa | S | 5995 | 4.12 | 1.48 | 6.5 | 7.52 | 2.72 | 0.04 | | 8.45 | 8.35 | 8.51 | 8.85 | 8.83 | 8.44 | 8.84 |

| Name | | Teff | logg | | | | | | | | | | | | |
|---|---|---|---|---|---|---|---|---|---|---|---|---|---|---|---|
| CCDM J14534+1542AB | S | 6135 | 4.11 | 1.82 | 10.3 | 7.72 | 2.26 | 0.03 | | 8.58 | 8.56 | 8.67 | 9.01 | 8.99 | 8.59 | 9.00 |
| chi Cnc | S | 6274 | 4.25 | 1.63 | 7.1 | 7.19 | 2.70 | 0.02 | | 8.13 | 8.15 | 8.31 | 8.73 | 8.62 | 8.17 | 8.67 |
| chi Dra | S | 6083 | 4.20 | 1.12 | 5.5 | 6.82 | 2.32 | 0.03 | | 7.83 | 7.88 | 8.25 | 8.59 | 8.31 | 7.93 | 8.45 |
| chi Her | S | 5890 | 3.96 | 1.44 | 5.0 | 6.97 | 2.47 | 0.05 | | 7.86 | 7.90 | 8.03 | 8.41 | 8.45 | 7.93 | 8.44 |
| chi01 Ori | S | 5983 | 4.44 | 1.83 | 10.5 | 7.49 | 2.89 | 0.04 | | 8.17 | 8.26 | 8.38 | 8.75 | 8.64 | 8.27 | 8.67 |
| del Cap | S | 7021 | 4.00 | 4.40 | 76.2 | 8.21 | 2.29 | -0.03 | L | 8.50 | 8.45 | | 8.90 | | 8.48 | 8.90 |
| del Eri | S | 5076 | 3.77 | 1.19 | 4.4 | 7.55 | 1.28 | 0.10 | | | | 8.55 | | 8.78 | 8.55 | 8.78 |
| del Tri | S | 5796 | 4.37 | 1.22 | 5.5 | 7.00 | 2.20 | 0.05 | | 7.98 | 8.12 | 8.12 | 8.43 | 8.32 | 8.08 | 8.35 |
| e Vir | S | 5999 | 4.17 | 1.75 | 9.0 | 7.60 | 2.76 | 0.04 | | 8.39 | 8.45 | 8.60 | 8.98 | 8.93 | 8.48 | 8.94 |
| eps Eri | S | 5123 | 4.57 | 0.90 | 4.0 | 7.42 | 0.52 | 0.10 | | | | 8.33 | | 8.48 | 8.33 | 8.48 |
| eps For | S | 5129 | 3.57 | 1.04 | 3.8 | 6.89 | 0.73 | 0.10 | | | | 8.01 | | 8.56 | 8.01 | 8.56 |
| eta Ari | S | 6485 | 3.98 | 1.84 | 9.4 | 7.33 | 2.93 | 0.01 | | 8.09 | 8.15 | | 8.49 | | 8.12 | 8.49 |
| eta Boo | S | 6050 | 3.75 | 2.38 | 14.4 | 7.76 | 1.12 | 0.04 | L | 8.59 | 8.55 | 8.63 | 9.07 | 9.07 | 8.58 | 9.07 |
| eta Cas | S | 5937 | 4.37 | 1.14 | 5.4 | 7.22 | 2.11 | 0.04 | | 8.15 | 8.24 | 8.28 | 8.64 | 8.64 | 8.22 | 8.64 |
| eta Cep | S | 5057 | 3.42 | 1.33 | 4.8 | 7.35 | 0.97 | 0.10 | | | | 8.36 | | 8.81 | 8.36 | 8.81 |
| eta CrB A | S | 6060 | 4.45 | 1.40 | 6.6 | 7.44 | 2.66 | 0.03 | | 8.25 | 8.27 | 8.36 | 8.81 | 8.82 | 8.28 | 8.82 |
| eta CrB B | S | 5948 | 4.51 | 1.18 | 7.0 | 7.42 | 2.56 | 0.04 | | 8.41 | 8.42 | 8.34 | 8.87 | 8.66 | 8.39 | 8.71 |
| eta Ser | S | 4985 | 3.15 | 1.37 | 4.7 | 7.30 | 0.15 | 0.11 | L | | | 8.14 | | 8.69 | 8.14 | 8.69 |
| gam Cep | S | 4850 | 3.21 | 1.50 | 5.0 | 7.61 | 0.40 | 0.12 | | | | 8.57 | | 8.95 | 8.57 | 8.95 |
| gam Lep | S | 6352 | 4.31 | 1.70 | 10.4 | 7.45 | 2.87 | 0.01 | | 8.24 | 8.26 | | 8.70 | | 8.25 | 8.70 |
| gam Ser | S | 6286 | 4.13 | 1.86 | 11.9 | 7.35 | 2.18 | 0.02 | | 8.17 | 8.18 | 8.13 | 8.65 | 8.77 | 8.16 | 8.71 |
| gam Vir A | S | 6922 | 4.27 | 2.83 | 29.7 | 7.54 | 3.19 | -0.02 | | 8.32 | 8.35 | | 8.80 | | 8.33 | 8.80 |
| gam01 Del | S | 6194 | 3.72 | 1.91 | 7.8 | 7.51 | 1.66 | 0.03 | | 8.37 | 8.42 | 8.43 | 8.95 | 8.82 | 8.40 | 8.88 |
| GJ 282 C | S | 3866 | 4.76 | 0.95 | 5.0 | 7.84 | 0.87 | 0.18 | | | | | | | | |
| GJ 528 A | S | 4471 | 4.42 | 0.53 | 3.1 | 7.54 | -0.28 | 0.14 | | | | 8.84 | | 8.94 | 8.84 | 8.94 |
| GJ 528 B | S | 4384 | 4.59 | 0.15 | 3.6 | 7.66 | -0.35 | 0.15 | | | | 9.00 | | 9.19 | 9.00 | 9.19 |
| HD 101177 | S | 5964 | 4.43 | 1.10 | 4.5 | 7.30 | 2.20 | 0.04 | | 8.12 | 8.17 | 8.22 | 8.62 | 8.54 | 8.17 | 8.56 |
| HD 101563 | S | 5868 | 3.89 | 1.57 | 6.3 | 7.43 | 1.08 | 0.05 | L | 8.44 | 8.47 | 8.58 | 8.77 | 8.59 | 8.50 | 8.64 |
| HD 101959 | S | 6095 | 4.42 | 1.37 | 6.0 | 7.38 | 2.60 | 0.03 | | 8.19 | 8.24 | 8.31 | 8.75 | 8.79 | 8.23 | 8.77 |
| HD 102158 | S | 5781 | 4.31 | 0.86 | 4.6 | 7.05 | -0.59 | 0.05 | L | 8.03 | 8.22 | 8.15 | 8.82 | 8.89 | 8.13 | 8.87 |
| HD 10307 | S | 5976 | 4.36 | 1.20 | 4.9 | 7.56 | 1.96 | 0.04 | | 8.29 | 8.35 | 8.53 | 8.70 | 8.61 | 8.39 | 8.63 |
| HD 103095 | S | 5178 | 4.72 | 1.30 | 3.1 | 6.22 | 0.71 | 0.10 | | | | 7.50 | | 7.66 | 7.50 | 7.66 |
| HD 103932 | S | 4585 | 4.58 | 0.60 | 5.0 | 7.77 | 0.08 | 0.14 | | | | 9.15 | | 9.31 | 9.15 | 9.31 |
| HD 104304 | S | 5555 | 4.42 | 0.97 | 5.2 | 7.77 | 0.79 | 0.07 | L | 8.63 | 8.68 | 8.77 | 8.97 | 8.85 | 8.69 | 8.88 |
| HD 10436 | S | 4393 | 4.72 | 0.15 | 5.1 | 7.33 | -0.14 | 0.15 | | | | 8.93 | | 9.18 | 8.93 | 9.18 |
| HD 106252 | S | 5890 | 4.37 | 1.17 | 5.3 | 7.39 | 1.75 | 0.05 | | 8.30 | 8.32 | 8.43 | 8.89 | 8.81 | 8.35 | 8.83 |
| HD 106640 | S | 6003 | 4.30 | 1.34 | 5.0 | 7.26 | 2.50 | 0.04 | | 8.05 | 8.10 | 8.19 | 8.60 | 8.78 | 8.10 | 8.69 |
| HD 108799 | S | 5878 | 4.33 | 1.46 | 9.1 | 7.31 | 2.75 | 0.05 | | 8.31 | 8.37 | 8.20 | 8.65 | 8.48 | 8.29 | 8.52 |
| HD 108874 | S | 5555 | 4.32 | 1.26 | 4.9 | 7.65 | 0.79 | 0.07 | L | 8.55 | 8.61 | 8.64 | 8.74 | 8.88 | 8.60 | 8.85 |
| HD 108954 | S | 6024 | 4.43 | 1.24 | 5.8 | 7.39 | 2.72 | 0.04 |   | 8.19 | 8.26 | 8.37 | 8.79 | 8.70 | 8.25 | 8.74 |
| HD 109057 | S | 6060 | 4.32 | 2.05 | 15.1 | 7.51 | 2.91 | 0.03 |   | 8.18 | 8.37 | 8.36 | 8.86 | 8.64 | 8.29 | 8.75 |
| HD 11007 | S | 5994 | 4.01 | 1.47 | 5.7 | 7.27 | 2.49 | 0.04 |   | 8.14 | 8.17 | 8.35 | 8.66 | 8.55 | 8.22 | 8.58 |
| HD 110315 | S | 4505 | 4.61 | 0.15 | 3.7 | 7.38 | -0.49 | 0.14 | L |   |   |   |   |   |   |   |
| HD 11038 | S | 6044 | 4.36 | 1.12 | 4.1 | 7.21 | 2.41 | 0.04 |   |   |   |   | 8.61 | 8.63 |   | 8.62 |
| HD 110745 | S | 6127 | 4.22 | 1.59 | 6.5 | 7.52 | 2.53 | 0.03 |   | 8.37 | 8.44 | 8.59 | 8.68 | 8.48 | 8.44 | 8.58 |
| HD 110833 | S | 4972 | 4.56 | 1.36 | 4.1 | 7.61 | 0.43 | 0.11 |   |   |   | 8.72 |   | 9.01 | 8.72 | 9.01 |
| HD 110869 | S | 5783 | 4.41 | 1.15 | 5.3 | 7.58 | 1.82 | 0.05 |   | 8.39 | 8.49 | 8.46 | 8.65 | 8.57 | 8.44 | 8.59 |
| HD 111513 | S | 5822 | 4.33 | 1.14 | 5.4 | 7.58 | 1.49 | 0.05 |   | 8.47 | 8.54 | 8.45 | 8.62 | 8.41 | 8.48 | 8.46 |
| HD 111540 | S | 5729 | 4.37 | 1.25 | 5.7 | 7.57 | 1.87 | 0.06 |   | 8.67 | 8.65 | 8.56 | 8.90 | 8.77 | 8.63 | 8.80 |
| HD 111799 | S | 5842 | 4.38 | 1.33 | 4.9 | 7.59 | 2.20 | 0.05 |   | 8.71 | 8.61 | 8.62 | 8.89 | 8.75 | 8.65 | 8.79 |
| HD 112068 | S | 5745 | 4.26 | 1.09 | 5.5 | 6.96 | 1.32 | 0.06 |   | 8.10 | 8.07 | 8.10 | 8.63 | 8.65 | 8.09 | 8.65 |
| HD 112257 | S | 5659 | 4.31 | 1.10 | 5.0 | 7.42 | 0.57 | 0.06 | L | 8.35 | 8.45 | 8.45 | 8.70 | 8.58 | 8.42 | 8.61 |
| HD 112758 | S | 5190 | 4.58 | 0.49 | 3.2 | 7.03 | 0.04 | 0.09 | L |   |   | 8.25 |   | 8.63 | 8.25 | 8.63 |
| HD 113470 | S | 5839 | 4.27 | 1.32 | 5.6 | 7.26 | 1.92 | 0.05 |   | 8.25 | 8.24 | 8.19 | 8.52 | 8.58 | 8.23 | 8.56 |
| HD 113713 | S | 6316 | 3.99 | 1.86 | 7.8 | 7.09 | 2.01 | 0.02 |   | 8.00 | 8.05 | 8.20 | 8.50 | 8.51 | 8.06 | 8.51 |
| HD 114762 | S | 5921 | 4.27 | 1.26 | 4.9 | 6.74 | 2.04 | 0.04 |   | 7.87 | 7.95 | 7.88 | 8.52 | 8.36 | 7.90 | 8.40 |
| HD 114783 | S | 5118 | 4.55 | 1.01 | 4.7 | 7.58 | 0.24 | 0.10 | L |   |   | 8.73 |   | 9.02 | 8.73 | 9.02 |
| HD 11507 | S | 4011 | 4.72 | 0.65 | 7.2 | 7.88 | 0.07 | 0.17 |   |   |   |   |   |   |   |   |
| HD 115404A | S | 5030 | 4.60 | 0.96 | 3.1 | 7.29 | -0.34 | 0.11 | L |   |   | 8.33 |   | 8.64 | 8.33 | 8.64 |
| HD 115404B | S | 3903 | 4.90 | 0.35 | 9.6 | 7.57 | 0.32 | 0.18 |   |   |   |   |   |   |   |   |
| HD 115953 | S | 3751 | 4.36 | 1.15 | 8.0 | 7.66 | -0.04 | 0.19 |   |   |   |   |   |   |   |   |
| HD 117043 | S | 5558 | 4.36 | 1.20 | 3.8 | 7.63 | 0.66 | 0.07 | L | 8.57 | 8.57 | 8.69 | 8.93 | 8.89 | 8.61 | 8.90 |
| HD 117845 | S | 5856 | 4.46 | 1.07 | 5.7 | 7.18 | 2.42 | 0.05 |   | 8.13 | 8.18 | 7.99 | 8.70 | 8.75 | 8.10 | 8.74 |
| HD 118203 | S | 5768 | 3.92 | 1.68 | 7.0 | 7.62 | 2.58 | 0.05 |   | 8.47 | 8.52 | 8.62 | 8.90 | 8.81 | 8.54 | 8.84 |
| HD 120066 | S | 5903 | 4.11 | 1.43 | 5.4 | 7.54 | 2.77 | 0.05 |   | 8.23 | 8.28 | 8.39 | 8.80 | 8.75 | 8.30 | 8.76 |
| HD 120467 | S | 4369 | 4.60 | 0.35 | 5.7 | 7.92 | 0.09 | 0.15 |   |   |   | 9.28 |   | 9.44 | 9.28 | 9.44 |
| HD 120690 | S | 5550 | 4.39 | 0.84 | 5.4 | 7.42 | 1.28 | 0.07 |   | 8.37 | 8.43 | 8.32 | 8.65 | 8.44 | 8.37 | 8.49 |
| HD 120730 | S | 5300 | 4.49 | 1.16 | 4.5 | 7.46 | 0.93 | 0.09 |   | 8.46 | 8.68 | 8.49 | 8.79 | 8.92 | 8.54 | 8.89 |
| HD 122064 | S | 4811 | 4.57 | 0.15 | 4.6 | 7.79 | 0.09 | 0.12 | L |   |   | 8.95 |   | 9.13 | 8.95 | 9.13 |
| HD 122303 | S | 4067 | 5.00 | 0.25 | 4.7 | 7.47 | 0.31 | 0.17 |   |   |   |   |   |   |   |   |
| HD 122742 | S | 5459 | 4.39 | 0.92 | 4.5 | 7.44 | 0.45 | 0.08 | L | 8.38 | 8.46 | 8.41 | 8.73 | 8.69 | 8.42 | 8.70 |
| HD 122967 | S | 7008 | 4.16 | 4.32 | 70.0 | 8.10 | 1.64 | -0.03 | L | 8.45 | 8.45 |   | 8.63 |   | 8.45 | 8.63 |
| HD 123A | S | 5823 | 4.43 | 1.26 | 6.1 | 7.56 | 2.43 | 0.05 |   | 8.35 | 8.36 | 8.37 | 8.90 | 8.81 | 8.36 | 8.83 |
| HD 123B | S | 5230 | 4.47 | 0.23 | 4.7 | 7.56 | 1.34 | 0.09 |   |   |   | 8.37 |   | 8.70 | 8.37 | 8.70 |
| HD 124553 | S | 6060 | 4.00 | 1.65 | 6.1 | 7.65 | 2.07 | 0.03 |   | 8.40 | 8.49 | 8.58 | 8.85 | 8.79 | 8.47 | 8.82 |
| HD 125040 | S | 6357 | 4.28 | 3.24 | 38.0 | 7.48 | 2.63 | 0.01 |   | 8.39 | 8.45 |   | 8.75 |   | 8.42 | 8.75 |

| Star | | Teff | log g | | | | | | | | | | | | |
|---|---|---|---|---|---|---|---|---|---|---|---|---|---|---|---|
| HD 125184 | S | 5637 | 4.06 | 1.37 | 5.6 | 7.72 | 0.61 | 0.06 | L | 8.58 | 8.65 | 8.68 | 8.94 | 8.92 | 8.64 | 8.92 |
| HD 126053 | S | 5727 | 4.46 | 0.94 | 3.8 | 7.14 | 1.04 | 0.06 | L | 7.88 | 8.09 | 8.20 | 8.33 | 8.43 | 8.06 | 8.41 |
| HD 12661 | S | 5651 | 4.35 | 1.06 | 5.0 | 7.81 | 1.01 | 0.06 | L | 8.65 | 8.81 | 8.73 | 8.97 | 8.87 | 8.73 | 8.90 |
| HD 127334 | S | 5643 | 4.25 | 1.32 | 5.1 | 7.64 | 0.88 | 0.06 | L | 8.60 | 8.64 | 8.64 | 8.97 | 8.98 | 8.63 | 8.98 |
| HD 128165 | S | 4793 | 4.59 | 0.76 | 2.9 | 7.47 | -0.02 | 0.12 | L | | | 8.74 | | 8.92 | 8.74 | 8.92 |
| HD 128311 | S | 4903 | 4.58 | 0.77 | 5.6 | 7.58 | 0.04 | 0.11 | L | | | 8.46 | | 8.70 | 8.46 | 8.70 |
| HD 129132 | S | 6609 | 3.41 | 2.80 | 25.3 | 7.31 | 1.85 | 0.00 | L | 8.20 | 8.15 | | 8.67 | | 8.17 | 8.67 |
| HD 129290 | S | 5837 | 4.19 | 1.26 | 5.0 | 7.34 | 1.67 | 0.05 | | 8.21 | 8.26 | 8.32 | 8.69 | 8.76 | 8.26 | 8.74 |
| HD 129829 | S | 6085 | 4.41 | 1.37 | 6.1 | 7.24 | 2.56 | 0.03 | | 8.14 | 8.17 | 8.10 | 8.81 | 8.76 | 8.15 | 8.78 |
| HD 130322 | S | 5422 | 4.54 | 0.88 | 4.3 | 7.51 | 0.46 | 0.08 | L | 8.26 | 8.44 | 8.43 | 8.81 | 8.78 | 8.38 | 8.78 |
| HD 13043 | S | 5877 | 4.15 | 1.33 | 5.6 | 7.54 | 2.00 | 0.05 | | 8.38 | 8.45 | 8.43 | 8.95 | 8.77 | 8.42 | 8.81 |
| HD 130948 | S | 5983 | 4.43 | 1.50 | 7.8 | 7.48 | 2.86 | 0.04 | | 8.29 | 8.29 | 8.37 | 8.68 | 8.49 | 8.32 | 8.53 |
| HD 131976 | S | 4077 | 5.00 | 0.15 | 5.0 | 7.42 | 0.49 | 0.17 | L | | | | | | | |
| HD 131977 | S | 4625 | 4.59 | 1.31 | 3.3 | 7.59 | -0.18 | 0.13 | L | | | 9.04 | | 9.22 | 9.04 | 9.22 |
| HD 132375 | S | 6336 | 4.18 | 1.90 | 9.4 | 7.56 | 2.21 | 0.02 | | 8.40 | 8.37 | 8.55 | 8.79 | 8.66 | 8.42 | 8.73 |
| HD 133161 | S | 5946 | 4.31 | 1.51 | 6.3 | 7.63 | 2.50 | 0.04 | | 8.61 | 8.62 | 8.63 | 8.95 | 8.92 | 8.62 | 8.93 |
| HD 135101 | S | 5637 | 4.24 | 0.87 | 4.4 | 7.54 | 0.65 | 0.06 | L | 8.54 | 8.53 | 8.52 | 8.90 | 8.95 | 8.53 | 8.94 |
| HD 135101B | S | 5529 | 4.07 | 1.27 | 4.3 | 7.49 | 0.70 | 0.07 | L | 8.40 | 8.34 | 8.52 | 8.85 | 8.73 | 8.42 | 8.76 |
| HD 135145 | S | 5852 | 4.05 | 1.47 | 5.2 | 7.34 | 1.74 | 0.05 | | 8.16 | 8.28 | 8.39 | 8.89 | 8.39 | 8.28 | 8.51 |
| HD 136064 | S | 6144 | 3.98 | 1.77 | 7.1 | 7.48 | 1.84 | 0.03 | | 8.42 | 8.39 | 8.49 | 8.73 | 8.70 | 8.42 | 8.72 |
| HD 136118 | S | 6143 | 4.08 | 1.66 | 8.5 | 7.41 | 2.27 | 0.03 | | 8.28 | 8.32 | 8.27 | 8.73 | 8.70 | 8.29 | 8.72 |
| HD 136231 | S | 5881 | 4.17 | 1.27 | 5.6 | 7.16 | 1.99 | 0.05 | | | | | 8.75 | 8.68 | | 8.69 |
| HD 1388 | S | 5884 | 4.31 | 1.20 | 5.6 | 7.44 | 2.02 | 0.05 | | 8.24 | 8.26 | 8.39 | 8.84 | 8.81 | 8.30 | 8.82 |
| HD 140913 | S | 5946 | 4.46 | 1.70 | 9.7 | 7.52 | 2.67 | 0.04 | | 8.35 | 8.49 | 8.36 | 9.04 | 8.97 | 8.40 | 8.99 |
| HD 141715 | S | 5836 | 4.25 | 2.04 | 10.3 | 7.37 | 2.71 | 0.05 | | 8.15 | 8.26 | 8.28 | 8.65 | 8.59 | 8.23 | 8.61 |
| HD 141937 | S | 5885 | 4.44 | 1.26 | 6.0 | 7.55 | 2.38 | 0.05 | | 8.44 | 8.45 | 8.45 | 8.85 | 8.79 | 8.45 | 8.80 |
| HD 14412 | S | 5492 | 4.62 | 0.48 | 3.6 | 7.11 | 0.88 | 0.07 | L | 7.81 | 8.03 | 8.16 | | 8.41 | 8.00 | 8.41 |
| HD 144579 | S | 5308 | 4.67 | 0.15 | 2.0 | 6.84 | -0.10 | 0.09 | L | 7.84 | 8.22 | 7.95 | | 8.37 | 8.00 | 8.37 |
| HD 144585 | S | 5850 | 4.22 | 1.41 | 5.1 | 7.75 | 1.61 | 0.05 | | 8.62 | 8.56 | 8.64 | 8.85 | 8.87 | 8.61 | 8.87 |
| HD 145148 | S | 4923 | 3.68 | 1.30 | 4.8 | 7.62 | 0.43 | 0.11 | | | | 8.69 | | 9.03 | 8.69 | 9.03 |
| HD 1461 | S | 5765 | 4.38 | 1.26 | 4.8 | 7.65 | 1.12 | 0.05 | L | 8.48 | 8.52 | 8.67 | 8.80 | 8.69 | 8.56 | 8.71 |
| HD 14624 | S | 5599 | 4.07 | 1.31 | 4.6 | 7.58 | 0.98 | 0.07 | L | 8.57 | 8.53 | 8.67 | 8.95 | 9.06 | 8.59 | 9.03 |
| HD 147681 | S | 6139 | 4.44 | 1.97 | 14.3 | 7.58 | 2.67 | 0.03 | | 8.32 | 8.32 | 8.43 | 8.82 | 8.80 | 8.34 | 8.81 |
| HD 149026 | S | 6003 | 4.13 | 1.73 | 8.3 | 7.70 | 2.34 | 0.04 | | 8.61 | 8.63 | 8.64 | 9.03 | 8.81 | 8.62 | 8.92 |
| HD 149143 | S | 5768 | 4.07 | 1.50 | 6.5 | 7.67 | 1.81 | 0.05 | | 8.68 | 8.69 | 8.63 | 9.08 | 8.87 | 8.67 | 8.92 |
| HD 15069 | S | 5722 | 4.11 | 1.22 | 4.7 | 7.48 | 1.81 | 0.06 | | 8.31 | 8.36 | 8.39 | 8.65 | 8.54 | 8.35 | 8.57 |
| HD 150706 | S | 5900 | 4.46 | 1.34 | 6.1 | 7.39 | 2.52 | 0.05 | | 8.17 | 8.29 | 8.29 | 8.72 | 8.79 | 8.25 | 8.77 |
| HD 150933 | S | 6087 | 4.28 | 1.37 | 5.5 | 7.59 | 2.56 | 0.03 | | 8.50 | 8.55 | 8.62 | 8.91 | 8.88 | 8.54 | 8.90 |

| | | | | | | | | | | | | | | |
|---|---|---|---|---|---|---|---|---|---|---|---|---|---|---|
| HD 151288 | S | 4181 | 4.70 | 0.15 | 4.7 | 7.85 | 0.06 | 0.16 | | | | | | |
| HD 151426 | S | 5711 | 4.28 | 1.07 | 4.9 | 7.37 | 1.05 | 0.06 | L | 8.26 | 8.38 | 8.43 | 8.80 | 8.61 | 8.35 | 8.66 |
| HD 151450 | S | 6108 | 4.38 | 1.31 | 6.0 | 7.45 | 2.49 | 0.03 | | 8.24 | 8.29 | 8.38 | 8.80 | 8.79 | 8.29 | 8.80 |
| HD 15189 | S | 6051 | 4.47 | 0.93 | 6.0 | 7.48 | 2.41 | 0.04 | | | | | 8.90 | 8.70 | | 8.80 |
| HD 152391 | S | 5431 | 4.51 | 1.32 | 5.2 | 7.42 | 1.13 | 0.08 | | 8.23 | 8.22 | 8.34 | 8.64 | 8.51 | 8.26 | 8.54 |
| HD 154160 | S | 5380 | 3.90 | 1.36 | 5.1 | 7.71 | 1.27 | 0.08 | | 8.83 | 8.88 | 8.71 | 9.07 | 9.00 | 8.81 | 9.02 |
| HD 154363 | S | 4373 | 4.66 | 0.15 | 2.5 | 7.15 | -0.30 | 0.15 | | | | | | | | |
| HD 154578 | S | 6294 | 4.14 | 1.91 | 12.9 | 7.26 | 2.18 | 0.02 | | 8.07 | 8.25 | 8.30 | 8.54 | 8.40 | 8.19 | 8.47 |
| HD 156062 | S | 5995 | 4.37 | 1.22 | 5.5 | 7.45 | 2.37 | 0.04 | | 8.30 | 8.33 | 8.49 | 8.75 | 8.62 | 8.37 | 8.65 |
| HD 156826 | S | 5155 | 3.52 | 1.19 | 4.7 | 7.22 | 0.30 | 0.10 | L | | | 8.16 | | 8.55 | 8.16 | 8.55 |
| HD 156846 | S | 6069 | 3.94 | 1.74 | 7.1 | 7.64 | 0.97 | 0.03 | L | 8.32 | 8.46 | 8.57 | 9.00 | 8.76 | 8.43 | 8.88 |
| HD 156968 | S | 5948 | 4.37 | 1.18 | 5.4 | 7.38 | 1.83 | 0.04 | | 8.23 | 8.27 | 8.36 | 8.72 | 8.68 | 8.28 | 8.69 |
| HD 157881 | S | 4161 | 4.67 | 0.15 | 5.6 | 7.72 | -0.03 | 0.16 | | | | | | | | |
| HD 158633 | S | 5313 | 4.57 | 0.55 | 3.4 | 7.06 | -0.11 | 0.09 | L | 7.69 | 8.08 | 8.10 | | 8.46 | 7.96 | 8.46 |
| HD 159222 | S | 5788 | 4.39 | 1.12 | 5.4 | 7.58 | 1.92 | 0.05 | | 8.47 | 8.53 | 8.45 | 8.97 | 8.87 | 8.48 | 8.90 |
| HD 160346 | S | 4864 | 4.52 | 0.29 | 3.0 | 7.50 | -0.08 | 0.12 | L | | | 8.47 | | 8.67 | 8.47 | 8.67 |
| HD 160933 | S | 5829 | 3.81 | 1.55 | 6.5 | 7.16 | 2.34 | 0.05 | | 8.07 | 8.16 | 8.24 | 8.68 | 8.58 | 8.16 | 8.60 |
| HD 16160 | S | 4886 | 4.63 | 0.15 | 2.4 | 7.43 | -0.44 | 0.12 | L | | | 8.70 | | 8.87 | 8.70 | 8.87 |
| HD 163492 | S | 5834 | 4.37 | 1.27 | 5.1 | 7.57 | 1.86 | 0.05 | | 8.55 | 8.50 | 8.53 | 8.59 | 8.70 | 8.53 | 8.67 |
| HD 163840 | S | 5760 | 4.10 | 1.26 | 5.6 | 7.48 | 1.98 | 0.06 | | 8.74 | 8.61 | 8.48 | 9.22 | 8.60 | 8.61 | 8.76 |
| HD 164595 | S | 5728 | 4.42 | 1.00 | 4.2 | 7.38 | 1.10 | 0.06 | L | 8.14 | 8.26 | 8.32 | 8.66 | 8.63 | 8.24 | 8.63 |
| HD 166 | S | 5481 | 4.52 | 1.27 | 6.2 | 7.55 | 2.27 | 0.07 | | 8.27 | 8.42 | 8.37 | 8.77 | 8.51 | 8.35 | 8.58 |
| HD 166601 | S | 6289 | 4.01 | 1.89 | 9.5 | 7.39 | 1.27 | 0.02 | L | 8.22 | 8.27 | 8.20 | 8.75 | 8.78 | 8.23 | 8.76 |
| HD 167588 | S | 5909 | 3.91 | 1.58 | 6.7 | 7.10 | 2.33 | 0.05 | | 8.05 | 8.08 | 8.22 | 8.60 | 8.63 | 8.12 | 8.63 |
| HD 167665 | S | 6179 | 4.25 | 1.55 | 7.1 | 7.33 | 2.53 | 0.03 | | 8.21 | 8.18 | 8.47 | 8.80 | 8.80 | 8.25 | 8.80 |
| HD 168443 | S | 5565 | 4.04 | 1.19 | 4.6 | 7.51 | 1.10 | 0.07 | | 8.47 | 8.51 | 8.51 | 9.08 | 8.82 | 8.50 | 8.88 |
| HD 168746 | S | 5576 | 4.33 | 0.96 | 4.1 | 7.39 | 0.48 | 0.07 | L | 8.20 | 8.37 | 8.35 | 8.89 | 8.61 | 8.31 | 8.68 |
| HD 170657 | S | 5133 | 4.59 | 0.89 | 4.2 | 7.34 | -0.02 | 0.10 | L | | | 8.07 | | 7.70 | 8.07 | 7.70 |
| HD 171706 | S | 5935 | 4.09 | 1.35 | 5.1 | 7.40 | 2.53 | 0.04 | | 8.22 | 8.26 | 8.41 | 8.70 | 8.59 | 8.29 | 8.62 |
| HD 172051 | S | 5669 | 4.49 | 0.95 | 4.0 | 7.24 | 1.28 | 0.06 | | 7.94 | 8.01 | 8.20 | 8.81 | 8.74 | 8.05 | 8.76 |
| HD 173818 | S | 4245 | 4.68 | 0.15 | 5.1 | 7.43 | -0.12 | 0.16 | | | | | | | | |
| HD 175225 | S | 5281 | 3.76 | 1.46 | 4.8 | 7.63 | 1.42 | 0.09 | | 8.54 | 8.60 | 8.67 | 8.86 | 8.89 | 8.60 | 8.88 |
| HD 175290 | S | 6359 | 4.02 | 1.84 | 12.1 | 7.17 | 2.81 | 0.01 | | 8.08 | 8.20 | | 8.65 | | 8.14 | 8.65 |
| HD 176051 | S | 5980 | 4.34 | 1.44 | 5.7 | 7.39 | 2.43 | 0.04 | | 8.16 | 8.23 | 8.37 | 8.71 | 8.62 | 8.25 | 8.65 |
| HD 176367 | S | 6087 | 4.47 | 2.58 | 22.0 | 7.58 | 3.49 | 0.03 | | 8.41 | 8.42 | | 8.75 | 8.70 | 8.41 | 8.72 |
| HD 177830 | S | 4813 | 3.49 | 1.30 | 4.8 | 7.83 | 0.49 | 0.12 | | | | 8.71 | | 9.01 | 8.71 | 9.01 |
| HD 178428 | S | 5646 | 4.21 | 1.16 | 5.1 | 7.58 | 0.02 | 0.06 | L | 8.54 | 8.54 | 8.53 | 8.92 | 8.82 | 8.54 | 8.85 |
| HD 178911 | S | 5825 | 3.86 | 1.44 | 7.0 | 7.47 | 2.34 | 0.05 | | 8.51 | 8.63 | 8.45 | 8.76 | 8.55 | 8.53 | 8.60 |

| Star | Type | Teff | log g | vt | [Fe/H] | ? | ? | L | ? | ? | ? | ? | ? | ? | ? |
|---|---|---|---|---|---|---|---|---|---|---|---|---|---|---|---|
| HD 178911B | S | 5602 | 4.37 | 1.29 | 4.6 | 7.69 | 0.47 | 0.07 | L | 8.58 | 8.56 | 8.71 | 8.96 | 9.07 | 8.62 | 9.04 |
| HD 179957 | S | 5771 | 4.38 | 1.09 | 4.7 | 7.49 | 0.63 | 0.05 | L | 8.27 | 8.35 | 8.45 | 8.84 | 8.77 | 8.36 | 8.79 |
| HD 179958 | S | 5807 | 4.33 | 1.13 | 4.9 | 7.53 | 0.36 | 0.05 | L | 8.28 | 8.36 | 8.45 | 8.85 | 8.79 | 8.37 | 8.81 |
| HD 181655 | S | 5673 | 4.17 | 1.43 | 4.2 | 7.48 | 1.95 | 0.06 | | 8.20 | 8.24 | 8.41 | 8.85 | 8.79 | 8.28 | 8.81 |
| HD 182488 | S | 5362 | 4.45 | 0.88 | 4.1 | 7.64 | 0.63 | 0.08 | L | 8.41 | 8.58 | 8.62 | 8.86 | 8.77 | 8.54 | 8.79 |
| HD 183263 | S | 5911 | 4.28 | 1.46 | 5.5 | 7.72 | 2.30 | 0.04 | | 8.55 | 8.62 | 8.69 | 9.18 | 9.05 | 8.62 | 9.08 |
| HD 184151 | S | 6430 | 3.77 | 2.21 | 15.3 | 7.21 | 2.27 | 0.01 | | 7.99 | 8.08 | | 8.64 | | 8.04 | 8.64 |
| HD 184152 | S | 5578 | 4.37 | 0.97 | 4.2 | 7.22 | 1.00 | 0.07 | | 8.02 | 8.19 | 8.29 | | 8.34 | 8.17 | 8.34 |
| HD 18445 | S | 4874 | 4.36 | 1.13 | 7.9 | 7.36 | -0.55 | 0.12 | L | | | 8.46 | | 8.61 | 8.46 | 8.61 |
| HD 184489 | S | 4133 | 4.74 | 0.15 | 6.1 | 7.31 | -0.04 | 0.17 | | | | | | | | |
| HD 184509 | S | 6069 | 4.33 | 1.30 | 6.1 | 7.35 | 2.49 | 0.03 | | 8.17 | 8.27 | 8.27 | 8.74 | 8.74 | 8.23 | 8.74 |
| HD 18455 | S | 5105 | 4.44 | 1.62 | 8.4 | 7.38 | 0.40 | 0.10 | L | | | 8.21 | | 8.18 | 8.21 | 8.18 |
| HD 184700 | S | 5745 | 4.23 | 1.02 | 4.4 | 7.31 | 0.59 | 0.06 | L | 8.31 | 8.30 | 8.36 | 8.80 | 8.70 | 8.32 | 8.73 |
| HD 187123 | S | 5796 | 4.30 | 1.25 | 5.0 | 7.55 | 0.75 | 0.05 | L | 8.41 | 8.45 | 8.41 | 8.90 | 8.67 | 8.43 | 8.73 |
| HD 188015 | S | 5667 | 4.26 | 1.35 | 5.0 | 7.68 | 0.35 | 0.06 | L | 8.61 | 8.64 | 8.71 | 8.75 | 8.66 | 8.65 | 8.68 |
| HD 188088 | S | 4818 | 4.20 | 1.45 | 10.2 | 7.59 | 0.41 | 0.12 | | | | 8.85 | | 9.14 | 8.85 | 9.14 |
| HD 188169 | S | 6509 | 4.25 | 1.91 | 12.0 | 7.37 | 1.31 | 0.00 | L | 8.24 | 8.39 | | 8.77 | | 8.32 | 8.77 |
| HD 189712 | S | 6363 | 3.85 | 2.21 | 9.5 | 6.94 | 1.56 | 0.01 | L | 7.91 | 7.95 | | 8.60 | | 7.93 | 8.60 |
| HD 189733 | S | 5044 | 4.60 | 0.83 | 4.5 | 7.48 | 0.29 | 0.10 | L | | | 8.44 | | 8.65 | 8.44 | 8.65 |
| HD 190007 | S | 4596 | 4.57 | 0.88 | 4.4 | 7.73 | -1.00 | 0.13 | L | | | 9.07 | | 9.21 | 9.07 | 9.21 |
| HD 190228 | S | 5264 | 3.71 | 1.16 | 4.4 | 7.19 | 1.14 | 0.09 | | 7.98 | 8.09 | 8.18 | 8.53 | 8.54 | 8.08 | 8.54 |
| HD 190360 | S | 5564 | 4.31 | 1.24 | 4.4 | 7.67 | 0.13 | 0.07 | L | 8.61 | 8.64 | 8.72 | 8.94 | 9.03 | 8.65 | 9.01 |
| HD 190771 | S | 5751 | 4.44 | 1.29 | 6.1 | 7.58 | 2.36 | 0.06 | | 8.41 | 8.47 | 8.40 | 8.85 | 8.68 | 8.43 | 8.72 |
| HD 192145 | S | 6069 | 3.98 | 1.59 | 5.8 | 7.16 | 2.28 | 0.03 | | 8.11 | 8.18 | 8.28 | 8.71 | 8.58 | 8.17 | 8.64 |
| HD 192263 | S | 4974 | 4.61 | 0.93 | 3.3 | 7.50 | 0.00 | 0.11 | L | | | 8.33 | | 8.49 | 8.33 | 8.49 |
| HD 192310 | S | 5044 | 4.51 | 0.25 | 3.7 | 7.52 | 0.19 | 0.10 | L | | | 8.52 | | 8.77 | 8.52 | 8.77 |
| HD 193555 | S | 6133 | 3.72 | 2.83 | 24.5 | 7.73 | 1.72 | 0.03 | L | 8.50 | 8.62 | | 9.11 | 9.08 | 8.56 | 9.09 |
| HD 193664 | S | 5945 | 4.44 | 1.08 | 5.0 | 7.40 | 2.28 | 0.04 | | 8.11 | 8.15 | 8.41 | 8.82 | 8.71 | 8.22 | 8.74 |
| HD 19383 | S | 6396 | 4.32 | 2.14 | 14.2 | 7.61 | 2.19 | 0.01 | | 8.26 | 8.41 | | 8.90 | | 8.34 | 8.90 |
| HD 195019 | S | 5660 | 4.05 | 1.32 | 5.1 | 7.42 | 1.40 | 0.06 | | 8.41 | 8.47 | 8.38 | 8.96 | 8.76 | 8.42 | 8.81 |
| HD 195564 | S | 5619 | 3.95 | 1.31 | 4.6 | 7.48 | 1.94 | 0.06 | | 8.43 | 8.38 | 8.47 | 8.90 | 8.85 | 8.43 | 8.87 |
| HD 19617 | S | 5682 | 4.43 | 1.17 | 3.8 | 7.63 | 0.55 | 0.06 | L | 8.48 | 8.53 | 8.65 | 8.86 | 9.14 | 8.56 | 9.07 |
| HD 196761 | S | 5483 | 4.53 | 0.82 | 3.9 | 7.21 | 0.40 | 0.07 | L | 8.05 | 8.17 | 8.21 | 8.68 | 8.48 | 8.14 | 8.53 |
| HD 197076 | S | 5844 | 4.46 | 1.09 | 5.1 | 7.38 | 2.20 | 0.05 | | 8.14 | 8.22 | 8.19 | 8.65 | 8.56 | 8.18 | 8.58 |
| HD 198387 | S | 5056 | 3.51 | 1.21 | 4.3 | 7.28 | 0.96 | 0.10 | | | | 8.21 | | 8.56 | 8.21 | 8.56 |
| HD 198483 | S | 5986 | 4.30 | 1.66 | 7.9 | 7.64 | 2.78 | 0.04 | | 8.55 | 8.57 | 8.60 | 8.97 | 8.86 | 8.57 | 8.89 |
| HD 198802 | S | 5736 | 3.81 | 1.46 | 5.5 | 7.43 | 1.87 | 0.06 | | 8.32 | 8.33 | 8.43 | 8.74 | 8.77 | 8.36 | 8.76 |
| HD 199604 | S | 5887 | 4.33 | 1.04 | 5.0 | 6.88 | 1.65 | 0.05 | | 7.86 | 7.90 | 8.07 | 8.73 | 8.50 | 7.94 | 8.56 |

| Star | | Teff | log g | ξ | [Fe/H] (or col) | | | | | | | | | | |
|------|---|------|------|-----|------|------|------|---|------|------|------|------|------|------|------|
| HD 200779 | S | 4389 | 4.61 | 0.15 | 4.1 | 7.72 | -0.12 | 0.15 | | | | 9.26 | | 9.42 | 9.26 | 9.42 |
| HD 201456 | S | 6201 | 4.20 | 1.55 | 8.1 | 7.55 | 2.69 | 0.03 | | 8.41 | 8.46 | 8.48 | 8.98 | 8.95 | 8.44 | 8.96 |
| HD 201496 | S | 5944 | 4.37 | 1.19 | 5.2 | 7.42 | 1.92 | 0.04 | | 8.30 | 8.36 | 8.44 | 8.76 | 8.65 | 8.37 | 8.68 |
| HD 202206 | S | 5693 | 4.40 | 1.27 | 5.7 | 7.75 | 1.17 | 0.06 | L | 8.49 | 8.54 | 8.61 | 8.95 | 9.01 | 8.55 | 9.00 |
| HD 202282 | S | 5796 | 4.30 | 1.31 | 5.1 | 7.51 | 1.53 | 0.05 | | 8.27 | 8.42 | 8.50 | 8.70 | 8.55 | 8.39 | 8.59 |
| HD 20339 | S | 5918 | 4.34 | 1.28 | 5.1 | 7.45 | 1.87 | 0.04 | | 8.29 | 8.33 | 8.42 | 8.75 | 8.64 | 8.34 | 8.67 |
| HD 20367 | S | 6050 | 4.36 | 1.48 | 5.5 | 7.50 | 2.89 | 0.04 | | 8.27 | 8.39 | 8.35 | 8.80 | 8.60 | 8.33 | 8.70 |
| HD 204153 | S | 6920 | 4.20 | 2.15 | 96.7 | 7.15 | 2.91 | -0.02 | L | 7.78 | 8.34 | | 8.70 | | 8.06 | 8.70 |
| HD 204485 | S | 7125 | 4.13 | 2.62 | 9.3 | 7.64 | 2.45 | -0.04 | | 8.48 | 8.49 | | 8.93 | | 8.49 | 8.93 |
| HD 205027 | S | 5752 | 4.31 | 0.89 | 4.4 | 7.11 | -1.00 | 0.06 | L | 8.25 | 8.28 | 8.20 | 8.67 | 8.57 | 8.24 | 8.59 |
| HD 205700 | S | 6656 | 4.11 | 1.98 | 8.3 | 7.32 | 2.78 | -0.01 | | 8.17 | 8.20 | | 8.79 | | 8.19 | 8.79 |
| HD 206282 | S | 6345 | 4.01 | 2.06 | 10.9 | 7.68 | 1.48 | 0.02 | | 8.44 | 8.48 | 8.59 | 8.84 | 8.70 | 8.48 | 8.77 |
| HD 207858 | S | 6290 | 3.97 | 2.14 | 11.1 | 7.59 | 2.65 | 0.02 | | 8.36 | 8.42 | 8.40 | 8.73 | 8.61 | 8.39 | 8.67 |
| HD 208801 | S | 4918 | 3.59 | 1.10 | 4.7 | 7.57 | 0.19 | 0.11 | L | | | 8.57 | | 8.91 | 8.57 | 8.91 |
| HD 21019 | S | 5514 | 3.79 | 1.21 | 4.8 | 7.04 | 1.35 | 0.07 | | 7.81 | 7.87 | 8.07 | 8.46 | 8.55 | 7.92 | 8.53 |
| HD 210277 | S | 5533 | 4.36 | 1.12 | 4.8 | 7.67 | 0.83 | 0.07 | L | 8.53 | 8.61 | 8.72 | 8.92 | 8.84 | 8.62 | 8.86 |
| HD 210460 | S | 5529 | 3.52 | 1.39 | 5.7 | 7.17 | 1.13 | 0.07 | | 8.02 | 7.96 | 8.14 | 8.76 | 8.51 | 8.04 | 8.57 |
| HD 210483 | S | 5842 | 4.10 | 1.34 | 5.1 | 7.34 | 1.91 | 0.05 | | 8.26 | 8.26 | 8.36 | 8.51 | 8.31 | 8.29 | 8.36 |
| HD 210855 | S | 6255 | 3.78 | 2.20 | 12.4 | 7.63 | 1.45 | 0.02 | L | 8.41 | 8.44 | 8.62 | 8.90 | 8.80 | 8.47 | 8.85 |
| HD 211476 | S | 5829 | 4.35 | 1.04 | 4.6 | 7.30 | 1.80 | 0.05 | | 8.11 | 8.23 | 8.25 | 8.62 | 8.52 | 8.20 | 8.54 |
| HD 211575 | S | 6589 | 4.23 | 2.62 | 20.6 | 7.68 | 2.02 | 0.00 | L | 8.39 | 8.47 | | 8.94 | | 8.43 | 8.94 |
| HD 213338 | S | 5558 | 4.51 | 0.84 | 4.6 | 7.54 | 0.79 | 0.07 | L | 8.37 | 8.47 | 8.43 | 8.83 | 8.74 | 8.42 | 8.76 |
| HD 214385 | S | 5711 | 4.47 | 0.74 | 3.8 | 7.18 | 0.92 | 0.06 | L | 8.08 | 8.05 | 8.15 | 8.90 | 8.74 | 8.09 | 8.78 |
| HD 214749 | S | 4531 | 4.63 | 0.80 | 4.8 | 7.57 | 0.06 | 0.14 | | | | 8.23 | | 8.20 | 8.23 | 8.20 |
| HD 215243 | S | 6393 | 4.12 | 1.94 | 9.5 | 7.53 | 0.35 | 0.01 | L | 8.35 | 8.39 | | 8.79 | | 8.37 | 8.79 |
| HD 21531 | S | 4231 | 4.67 | 0.15 | 5.0 | 7.89 | 0.11 | 0.16 | | | | | | | | |
| HD 215625 | S | 6228 | 4.39 | 1.42 | 6.6 | 7.55 | 2.73 | 0.02 | | 8.45 | 8.43 | 8.53 | 8.80 | 8.87 | 8.46 | 8.84 |
| HD 216133 | S | 3973 | 4.92 | 0.15 | 6.8 | 7.58 | 0.21 | 0.18 | | | | | | | | |
| HD 216770 | S | 5411 | 4.48 | 0.95 | 4.7 | 7.85 | 0.89 | 0.08 | | 8.59 | 8.69 | 8.80 | 9.06 | 8.82 | 8.69 | 8.88 |
| HD 217107 | S | 5541 | 4.29 | 1.19 | 5.2 | 7.75 | -0.12 | 0.07 | L | 8.68 | 8.67 | 8.72 | 8.98 | 8.90 | 8.69 | 8.92 |
| HD 217357 | S | 4192 | 4.82 | 0.15 | 5.8 | 7.61 | 0.01 | 0.16 | | | | | | | | |
| HD 217577 | S | 5762 | 4.11 | 1.23 | 4.8 | 7.31 | 0.67 | 0.06 | L | 8.10 | 8.10 | 8.19 | 8.56 | 8.47 | 8.13 | 8.49 |
| HD 217877 | S | 6015 | 4.35 | 1.29 | 5.5 | 7.39 | 2.29 | 0.04 | | 8.26 | 8.28 | 8.45 | 8.87 | 8.82 | 8.31 | 8.84 |
| HD 217958 | S | 5771 | 4.23 | 1.38 | 6.5 | 7.70 | 2.23 | 0.05 | | 8.73 | 8.77 | 8.67 | 9.00 | 8.89 | 8.72 | 8.92 |
| HD 218101 | S | 5217 | 3.81 | 1.26 | 4.3 | 7.54 | 1.22 | 0.09 | | | | 8.52 | | 8.79 | 8.52 | 8.79 |
| HD 219428 | S | 5930 | 4.35 | 1.43 | 7.4 | 7.55 | 2.55 | 0.04 | | 8.48 | 8.53 | 8.41 | 8.85 | 8.77 | 8.47 | 8.79 |
| HD 220008 | S | 5653 | 3.82 | 1.35 | 4.9 | 7.26 | 2.15 | 0.06 | | 8.05 | 8.19 | 8.24 | 8.81 | 8.67 | 8.16 | 8.71 |
| HD 220689 | S | 5921 | 4.37 | 1.17 | 5.5 | 7.45 | 1.93 | 0.04 | | 8.40 | 8.36 | 8.46 | 8.50 | 8.47 | 8.41 | 8.48 |

| Star | | Teff | log g | ξ | v sin i | log N(Li) | [Fe/H] | σ | L | C | N | O | Na | Mg | Al | Si |
|---|---|---|---|---|---|---|---|---|---|---|---|---|---|---|---|---|
| HD 22072 | S | 4974 | 3.48 | 1.13 | 4.6 | 7.15 | -0.09 | 0.11 | L | | | 8.30 | | 8.76 | 8.30 | 8.76 |
| HD 221356 | S | 6137 | 4.39 | 1.39 | 5.6 | 7.25 | 2.59 | 0.03 | | 8.04 | 8.02 | 8.29 | 8.64 | 8.77 | 8.08 | 8.70 |
| HD 221445 | S | 6230 | 3.77 | 1.87 | 8.3 | 7.32 | 1.13 | 0.02 | L | 8.10 | 8.15 | 8.37 | 8.64 | 8.59 | 8.17 | 8.61 |
| HD 221503 | S | 4312 | 4.66 | 0.15 | 5.9 | 7.75 | 0.09 | 0.15 | | | | | | | | |
| HD 222582 | S | 5796 | 4.37 | 1.09 | 5.0 | 7.49 | 0.69 | 0.05 | L | 8.30 | 8.34 | 8.40 | 8.84 | 8.77 | 8.35 | 8.79 |
| HD 222645 | S | 6211 | 4.33 | 1.47 | 6.6 | 7.38 | 2.68 | 0.02 | | 8.16 | 8.28 | 8.41 | 8.69 | 8.56 | 8.26 | 8.63 |
| HD 22292 | S | 6483 | 4.31 | 2.64 | 27.4 | 7.51 | 1.89 | 0.01 | L | 8.34 | 8.55 | | 8.70 | | 8.44 | 8.70 |
| HD 223084 | S | 5958 | 4.33 | 1.19 | 7.3 | 7.28 | 2.51 | 0.04 | | 8.23 | 8.30 | 8.16 | 8.60 | 8.50 | 8.23 | 8.52 |
| HD 223110 | S | 6516 | 3.96 | 2.60 | 17.8 | 7.61 | 2.74 | 0.00 | | 8.32 | 8.41 | | 8.64 | | 8.36 | 8.64 |
| HD 223238 | S | 5889 | 4.30 | 1.20 | 5.0 | 7.52 | 1.33 | 0.05 | | 8.33 | 8.39 | 8.51 | 8.83 | 8.83 | 8.41 | 8.83 |
| HD 22455 | S | 5886 | 4.37 | 1.22 | 5.1 | 7.51 | 1.75 | 0.05 | | 8.39 | 8.41 | 8.45 | 8.77 | 8.80 | 8.42 | 8.79 |
| HD 22468A | S | 4736 | 3.35 | 1.78 | 34.5 | 7.31 | 1.35 | 0.13 | | | | | | | | |
| HD 22468B | S | 4631 | 4.47 | 0.15 | 4.1 | 7.56 | -0.56 | 0.13 | L | | | 8.26 | | 8.37 | 8.26 | 8.37 |
| HD 225239 | S | 5647 | 3.76 | 1.30 | 5.0 | 7.00 | 1.95 | 0.06 | | 7.93 | 7.95 | 8.06 | 8.65 | 8.58 | 7.98 | 8.60 |
| HD 231701 | S | 6240 | 4.17 | 1.65 | 7.1 | 7.49 | 2.74 | 0.02 | | 8.23 | 8.31 | 8.48 | 8.69 | 8.70 | 8.31 | 8.70 |
| HD 232979 | S | 3893 | 4.62 | 0.15 | 5.0 | 7.71 | -0.10 | 0.18 | | | | | | | | |
| HD 23349 | S | 6018 | 4.26 | 1.42 | 5.5 | 7.59 | 2.47 | 0.04 | | 8.41 | 8.43 | 8.60 | 8.90 | 8.84 | 8.46 | 8.87 |
| HD 23356 | S | 4990 | 4.63 | 1.36 | 3.6 | 7.38 | -0.07 | 0.11 | L | | | 8.40 | | 8.60 | 8.40 | 8.60 |
| HD 234078 | S | 4157 | 4.68 | 0.15 | 4.0 | 7.61 | -0.05 | 0.16 | | | | | | | | |
| HD 23453 | S | 3748 | 4.37 | 0.25 | 8.3 | 7.92 | -0.06 | 0.19 | | | | | | | | |
| HD 23476 | S | 5662 | 4.48 | 0.81 | 3.9 | 7.09 | 0.97 | 0.06 | L | 8.07 | 8.01 | 8.21 | 8.69 | 8.60 | 8.10 | 8.62 |
| HD 23596 | S | 6008 | 4.15 | 1.50 | 6.1 | 7.69 | 2.71 | 0.04 | | 8.56 | 8.60 | 8.70 | 8.95 | 8.74 | 8.60 | 8.85 |
| HD 239928 | S | 5899 | 4.43 | 1.08 | 4.9 | 7.52 | 1.94 | 0.05 | | 8.40 | 8.37 | 8.46 | 8.75 | 8.65 | 8.41 | 8.68 |
| HD 2475 | S | 6012 | 4.19 | 1.66 | 8.8 | 7.56 | 2.69 | 0.04 | | 8.26 | 8.29 | 8.49 | 8.80 | 8.54 | 8.32 | 8.67 |
| HD 2582 | S | 5705 | 4.19 | 1.41 | 5.7 | 7.60 | 0.94 | 0.06 | L | 8.56 | 8.61 | 8.61 | 8.88 | 8.78 | 8.59 | 8.80 |
| HD 2589 | S | 5154 | 3.64 | 1.15 | 4.3 | 7.41 | 0.31 | 0.10 | L | | | 8.48 | | 8.79 | 8.48 | 8.79 |
| HD 2638 | S | 5120 | 4.54 | 1.33 | 5.5 | 7.66 | | 0.10 | | | | | | | | |
| HD 26505 | S | 5878 | 4.11 | 1.39 | 5.2 | 7.50 | 1.29 | 0.05 | L | 8.26 | 8.34 | 8.50 | 8.88 | 8.97 | 8.37 | 8.94 |
| HD 26913 | S | 5661 | 4.52 | 1.55 | 7.6 | 7.45 | 2.31 | 0.06 | | 8.09 | 8.20 | 8.26 | 8.77 | 8.74 | 8.18 | 8.75 |
| HD 26923 | S | 5989 | 4.43 | 1.20 | 4.8 | 7.44 | 2.79 | 0.04 | | 8.13 | 8.19 | 8.34 | 8.65 | 8.43 | 8.22 | 8.48 |
| HD 2730 | S | 6192 | 3.99 | 1.71 | 9.1 | 7.37 | 2.58 | 0.03 | | 8.06 | 8.13 | 8.19 | 8.68 | 8.65 | 8.11 | 8.66 |
| HD 27530 | S | 5926 | 4.38 | 1.49 | 6.3 | 7.65 | 2.23 | 0.04 | | 8.59 | 8.53 | 8.66 | 8.85 | 8.95 | 8.59 | 8.92 |
| HD 28185 | S | 5658 | 4.33 | 1.27 | 4.6 | 7.69 | 0.81 | 0.06 | L | 8.53 | 8.58 | 8.71 | 9.02 | 9.02 | 8.61 | 9.02 |
| HD 28343 | S | 4152 | 4.65 | 0.35 | 5.3 | 7.89 | 0.17 | 0.17 | | | | | | | | |
| HD 285660 | S | 6055 | 4.27 | 1.65 | 9.2 | 7.47 | 2.40 | 0.04 | | 8.28 | 8.37 | 8.33 | 8.79 | 8.89 | 8.33 | 8.84 |
| HD 28571 | S | 5736 | 4.22 | 1.07 | 3.8 | 7.24 | 0.30 | 0.06 | L | 8.18 | 8.26 | 8.32 | 8.86 | 8.84 | 8.25 | 8.85 |
| HD 28635 | S | 6140 | 4.35 | 1.54 | 6.4 | 7.57 | 2.98 | 0.03 | | 8.43 | 8.47 | 8.49 | 8.88 | 8.59 | 8.46 | 8.73 |
| HD 29587 | S | 5682 | 4.48 | 0.75 | 3.9 | 6.94 | 0.10 | 0.06 | L | 7.80 | 8.02 | 8.05 | 8.70 | 8.67 | 7.96 | 8.68 |

| Star | | T | log g | ξ | v sin i | [Fe/H] | [α/Fe] | σ | Flag | C | N | O | Na | Mg | Al | Si |
|---|---|---|---|---|---|---|---|---|---|---|---|---|---|---|---|---|
| HD 29645 | S | 6002 | 4.02 | 1.63 | 7.2 | 7.58 | 2.61 | 0.04 | | 8.52 | 8.47 | 8.52 | 8.82 | 8.75 | 8.50 | 8.79 |
| HD 29697 | S | 4421 | 4.64 | 0.97 | 10.8 | 7.68 | 0.87 | 0.15 | | | | 8.95 | | 9.14 | 8.95 | 9.14 |
| HD 31527 | S | 5908 | 4.36 | 1.08 | 5.2 | 7.30 | 1.89 | 0.05 | | 8.12 | 8.18 | 8.27 | 8.87 | 8.72 | 8.19 | 8.76 |
| HD 31949 | S | 6213 | 4.30 | 1.78 | 11.0 | 7.41 | 2.65 | 0.02 | | 8.23 | 8.31 | 8.30 | 8.73 | 8.67 | 8.28 | 8.70 |
| HD 32147 | S | 4790 | 4.57 | 0.15 | 3.1 | 7.86 | 0.19 | 0.12 | | | | 8.84 | | 9.01 | 8.84 | 9.01 |
| HD 32715 | S | 6615 | 4.25 | 2.45 | 42.0 | 7.62 | 2.30 | 0.00 | L | 8.09 | 8.50 | | 8.90 | | 8.29 | 8.90 |
| HD 332612 | S | 6124 | 4.10 | 1.74 | 8.9 | 7.47 | 1.96 | 0.03 | | 8.22 | 8.38 | 8.25 | 8.65 | 8.44 | 8.29 | 8.55 |
| HD 334372 | S | 5765 | 4.01 | 1.32 | 4.8 | 7.39 | 2.40 | 0.05 | | 8.14 | 8.23 | 8.33 | 9.07 | 8.91 | 8.23 | 8.95 |
| HD 33636 | S | 5953 | 4.45 | 1.11 | 6.3 | 7.38 | 2.46 | 0.04 | | 8.15 | 8.27 | 8.30 | 8.67 | 8.49 | 8.24 | 8.53 |
| HD 33866 | S | 5625 | 4.33 | 1.12 | 5.0 | 7.34 | 1.47 | 0.06 | | 8.29 | 8.40 | 8.26 | 8.98 | 8.88 | 8.32 | 8.91 |
| HD 34721 | S | 6004 | 4.14 | 1.45 | 6.4 | 7.40 | 2.37 | 0.04 | | 8.27 | 8.28 | 8.43 | 8.76 | 8.76 | 8.31 | 8.76 |
| HD 3556 | S | 6019 | 4.42 | 1.43 | 6.1 | 7.60 | 2.68 | 0.04 | | 8.34 | 8.40 | 8.47 | 8.79 | 8.62 | 8.39 | 8.71 |
| HD 35961 | S | 5740 | 4.37 | 0.90 | 5.3 | 7.24 | 1.72 | 0.06 | | 8.23 | 8.32 | 8.15 | 8.55 | 8.46 | 8.23 | 8.48 |
| HD 36003 | S | 4536 | 4.60 | 0.73 | 5.0 | 7.53 | −0.28 | 0.14 | L | | | 9.17 | | 9.34 | 9.17 | 9.34 |
| HD 37124 | S | 5561 | 4.39 | 0.70 | 3.6 | 7.06 | 0.68 | 0.07 | L | 7.98 | 8.28 | 8.17 | 8.53 | 8.52 | 8.14 | 8.52 |
| HD 37605 | S | 5318 | 4.47 | 0.69 | 4.5 | 7.78 | 0.72 | 0.09 | | 8.55 | 8.71 | 8.70 | 8.94 | 8.88 | 8.65 | 8.90 |
| HD 3795 | S | 5456 | 3.89 | 1.02 | 4.2 | 6.93 | 0.29 | 0.08 | L | 7.65 | 7.79 | 8.03 | 8.69 | 8.66 | 7.82 | 8.67 |
| HD 38529 | S | 5450 | 3.72 | 1.56 | 6.2 | 7.72 | 0.38 | 0.08 | L | 8.65 | 8.72 | 8.69 | 9.01 | 9.00 | 8.69 | 9.01 |
| HD 38700 | S | 6049 | 4.45 | 1.57 | 8.3 | 7.45 | 3.08 | 0.04 | | 8.20 | 8.34 | 8.38 | 8.68 | 8.58 | 8.29 | 8.63 |
| HD 38858 | S | 5798 | 4.48 | 1.01 | 4.2 | 7.28 | 1.57 | 0.05 | | 8.05 | 8.07 | 8.23 | 8.40 | 8.30 | 8.12 | 8.33 |
| HD 38A | S | 4065 | 4.71 | 0.15 | 6.4 | 7.66 | 0.10 | 0.17 | | | | | | | | |
| HD 38B | S | 4028 | 4.72 | 0.15 | 7.7 | 7.63 | 0.17 | 0.17 | | | | | | | | |
| HD 39881 | S | 5719 | 4.27 | 1.07 | 4.6 | 7.33 | 0.78 | 0.06 | L | 8.34 | 8.36 | 8.42 | 8.99 | 8.89 | 8.38 | 8.92 |
| HD 400 | S | 6240 | 4.13 | 1.64 | 7.3 | 7.28 | 2.36 | 0.02 | | 8.16 | 8.18 | 8.35 | 8.66 | 8.50 | 8.21 | 8.58 |
| HD 40979 | S | 6150 | 4.35 | 1.63 | 9.1 | 7.67 | 2.91 | 0.03 | | 8.42 | 8.50 | 8.64 | 8.78 | 8.65 | 8.50 | 8.71 |
| HD 41330 | S | 5876 | 4.13 | 1.28 | 4.8 | 7.30 | 1.97 | 0.05 | | 8.12 | 8.23 | 8.26 | 8.82 | 8.70 | 8.20 | 8.73 |
| HD 41708 | S | 5867 | 4.47 | 1.03 | 4.9 | 7.52 | 2.18 | 0.05 | | 8.36 | 8.45 | 8.41 | 8.74 | 8.64 | 8.41 | 8.67 |
| HD 4203 | S | 5471 | 4.10 | 1.21 | 5.6 | 7.80 | 1.05 | 0.08 | | 8.86 | 8.93 | 8.72 | 9.11 | 8.92 | 8.84 | 8.97 |
| HD 4208 | S | 5674 | 4.47 | 0.94 | 4.4 | 7.24 | 0.52 | 0.06 | L | 8.03 | 8.05 | 8.22 | 8.80 | 8.69 | 8.10 | 8.71 |
| HD 43162 | S | 5651 | 4.50 | 1.51 | 7.4 | 7.49 | 2.30 | 0.06 | | 8.21 | 8.31 | 8.33 | 8.75 | 8.61 | 8.28 | 8.65 |
| HD 43587 | S | 5859 | 4.27 | 1.24 | 4.9 | 7.40 | 2.04 | 0.05 | | 8.29 | 8.33 | 8.36 | 8.89 | 8.72 | 8.33 | 8.76 |
| HD 43745 | S | 6087 | 3.92 | 1.67 | 6.3 | 7.54 | 1.71 | 0.03 | | 8.40 | 8.48 | 8.55 | 8.81 | 8.80 | 8.46 | 8.80 |
| HD 45067 | S | 6049 | 3.95 | 1.70 | 8.0 | 7.41 | 2.44 | 0.04 | | 8.21 | 8.25 | 8.35 | 8.75 | 8.72 | 8.25 | 8.74 |
| HD 45088 | S | 4778 | 4.37 | 0.15 | 8.2 | 7.26 | −0.49 | 0.12 | L | | | 8.20 | | 8.57 | 8.20 | 8.57 |
| HD 45184 | S | 5852 | 4.41 | 1.22 | 5.4 | 7.51 | 2.01 | 0.05 | | 8.37 | 8.38 | 8.38 | 8.75 | 8.63 | 8.38 | 8.66 |
| HD 45205 | S | 5921 | 4.15 | 1.31 | 5.7 | 6.65 | 2.26 | 0.04 | | 7.80 | 7.76 | 7.93 | 8.25 | 8.37 | 7.83 | 8.34 |
| HD 45350 | S | 5567 | 4.22 | 1.33 | 4.7 | 7.69 | 0.65 | 0.07 | L | 8.50 | 8.59 | 8.61 | 8.93 | 8.95 | 8.57 | 8.94 |
| HD 45588 | S | 6214 | 4.24 | 1.70 | 8.8 | 7.51 | 2.77 | 0.02 | | 8.29 | 8.27 | 8.54 | 8.78 | 8.93 | 8.33 | 8.86 |

| Star | | Teff | log g | vt | vsini | [Fe/H] | ? | ? | L? | ? | ? | ? | ? | ? | ? | ? |
|---|---|---|---|---|---|---|---|---|---|---|---|---|---|---|---|---|
| HD 45759 | S | 6132 | 4.37 | 2.26 | 16.1 | 7.59 | 3.09 | 0.03 | | 8.42 | 8.35 | 8.30 | 8.80 | 8.60 | 8.37 | 8.70 |
| HD 4628 | S | 5044 | 4.61 | 0.35 | 3.2 | 7.25 | -0.41 | 0.10 | L | | | 8.32 | | 8.58 | 8.32 | 8.58 |
| HD 46375 | S | 5265 | 4.33 | 1.24 | 4.3 | 7.68 | 0.73 | 0.09 | | 8.51 | 8.59 | 8.74 | | 9.02 | 8.61 | 9.02 |
| HD 48938 | S | 6055 | 4.33 | 1.27 | 5.1 | 7.09 | 2.37 | 0.04 | | 7.90 | 7.93 | 8.16 | 8.57 | 8.45 | 7.96 | 8.51 |
| HD 49674 | S | 5621 | 4.39 | 1.37 | 4.7 | 7.70 | 0.57 | 0.06 | L | 8.41 | 8.44 | 8.60 | 8.89 | 8.72 | 8.49 | 8.76 |
| HD 50281 | S | 4708 | 4.64 | 0.15 | 5.5 | 7.56 | -0.11 | 0.13 | L | | | 8.39 | | 8.49 | 8.39 | 8.49 |
| HD 50554 | S | 6019 | 4.41 | 1.21 | 5.5 | 7.46 | 2.47 | 0.04 | | 8.25 | 8.27 | 8.41 | 8.88 | 8.78 | 8.29 | 8.83 |
| HD 50806 | S | 5640 | 4.06 | 1.16 | 5.1 | 7.52 | 0.61 | 0.06 | L | 8.43 | 8.52 | 8.55 | 8.89 | 8.82 | 8.50 | 8.84 |
| HD 52265 | S | 6086 | 4.29 | 1.51 | 6.7 | 7.65 | 2.72 | 0.03 | | 8.50 | 8.55 | 8.55 | 9.00 | 8.93 | 8.53 | 8.96 |
| HD 52698 | S | 5155 | 4.55 | 0.98 | 5.2 | 7.66 | 0.61 | 0.10 | | | | 8.55 | | 8.77 | 8.55 | 8.77 |
| HD 52711 | S | 5992 | 4.41 | 1.26 | 5.2 | 7.41 | 2.03 | 0.04 | | 8.15 | 8.21 | 8.38 | 8.89 | 8.79 | 8.25 | 8.82 |
| HD 5494 | S | 6044 | 4.03 | 1.67 | 14.3 | 7.39 | 1.80 | 0.04 | | 8.44 | 8.49 | 8.40 | 8.83 | 8.80 | 8.45 | 8.82 |
| HD 55054 | S | 6219 | 4.14 | 1.70 | 8.5 | 7.42 | 2.26 | 0.02 | | 8.29 | 8.33 | 8.46 | 8.68 | 8.60 | 8.34 | 8.64 |
| HD 55575 | S | 5937 | 4.32 | 1.26 | 5.2 | 7.15 | 1.71 | 0.04 | | 8.02 | 8.13 | 8.23 | 8.69 | 8.62 | 8.13 | 8.64 |
| HD 55693 | S | 5854 | 4.32 | 1.32 | 5.8 | 7.73 | 1.39 | 0.05 | | 8.54 | 8.66 | 8.67 | 8.92 | 8.79 | 8.62 | 8.82 |
| HD 57006 | S | 6264 | 3.77 | 2.10 | 8.3 | 7.51 | 1.79 | 0.02 | | 8.26 | 8.31 | 8.51 | 8.71 | 8.91 | 8.33 | 8.81 |
| HD 603 | S | 5967 | 4.37 | 1.15 | 5.6 | 7.31 | 2.31 | 0.04 | | 8.04 | 8.09 | 8.31 | 8.83 | 8.73 | 8.15 | 8.76 |
| HD 6064 | S | 6363 | 3.78 | 2.03 | 11.1 | 7.60 | 3.22 | 0.01 | | 8.32 | 8.36 | | 8.79 | | 8.34 | 8.79 |
| HD 61606 | S | 4932 | 4.61 | 1.00 | 3.6 | 7.47 | 0.01 | 0.11 | L | | | 8.19 | | 8.22 | 8.19 | 8.22 |
| HD 61632 | S | 5632 | 4.02 | 1.09 | 5.3 | 6.93 | 1.49 | 0.06 | | 7.95 | 8.12 | 7.96 | 8.82 | 8.80 | 8.01 | 8.80 |
| HD 63598 | S | 5828 | 4.38 | 0.93 | 4.8 | 6.63 | 1.62 | 0.05 | | 7.77 | 7.81 | 7.79 | 8.77 | 8.60 | 7.79 | 8.64 |
| HD 63754 | S | 6040 | 3.92 | 1.78 | 7.7 | 7.61 | 2.16 | 0.04 | | 8.55 | 8.54 | 8.70 | 8.98 | 8.79 | 8.57 | 8.89 |
| HD 68988 | S | 5878 | 4.39 | 1.37 | 6.4 | 7.76 | 2.12 | 0.05 | | 8.73 | 8.75 | 8.74 | 9.06 | 8.95 | 8.74 | 8.98 |
| HD 69830 | S | 5443 | 4.53 | 0.92 | 5.2 | 7.45 | 0.86 | 0.08 | L | 8.29 | 8.43 | 8.40 | 8.73 | 8.72 | 8.37 | 8.72 |
| HD 70889 | S | 6036 | 4.43 | 1.25 | 5.9 | 7.57 | 2.59 | 0.04 | | 8.37 | 8.34 | 8.37 | 8.89 | 8.65 | 8.36 | 8.77 |
| HD 7091 | S | 6109 | 4.29 | 1.16 | 7.1 | 7.21 | 2.33 | 0.03 | | | | | 8.75 | 8.73 | | 8.74 |
| HD 71148 | S | 5835 | 4.39 | 1.17 | 5.8 | 7.49 | 1.86 | 0.05 | | 8.36 | 8.38 | 8.36 | 8.69 | 8.51 | 8.37 | 8.56 |
| HD 71881 | S | 5863 | 4.30 | 1.14 | 5.3 | 7.43 | 1.57 | 0.05 | | 8.38 | 8.43 | 8.47 | 8.88 | 8.73 | 8.43 | 8.77 |
| HD 7230 | S | 6637 | 4.17 | 3.73 | 45.7 | 7.75 | 2.19 | 0.00 | L | 8.58 | 8.28 | | 8.82 | | 8.43 | 8.82 |
| HD 72659 | S | 5902 | 4.12 | 1.31 | 5.1 | 7.43 | 2.28 | 0.05 | | 8.32 | 8.34 | 8.38 | 8.70 | 8.67 | 8.35 | 8.68 |
| HD 72946 | S | 5674 | 4.49 | 1.23 | 5.3 | 7.57 | 1.41 | 0.06 | | 8.33 | 8.37 | 8.47 | 8.76 | 8.69 | 8.39 | 8.71 |
| HD 7352 | S | 6023 | 4.31 | 1.30 | 6.4 | 7.55 | 2.59 | 0.04 | | 8.42 | 8.43 | 8.39 | 8.80 | 8.67 | 8.42 | 8.74 |
| HD 73596 | S | 6744 | 3.41 | 2.68 | 26.1 | 7.49 | 1.68 | -0.01 | L | 7.85 | 8.13 | | 8.72 | | 7.99 | 8.72 |
| HD 73668 | S | 5922 | 4.37 | 1.24 | 5.8 | 7.45 | 2.41 | 0.04 | | 8.27 | 8.28 | 8.37 | 8.75 | 8.64 | 8.31 | 8.67 |
| HD 73752 | S | 5654 | 3.98 | 1.51 | 6.4 | 7.68 | 1.60 | 0.06 | | 8.65 | 8.68 | 8.70 | 8.95 | 8.93 | 8.68 | 8.93 |
| HD 7397 | S | 6029 | 4.15 | 1.37 | 8.2 | 7.36 | 2.44 | 0.04 | | | | | 8.70 | 8.78 | | 8.74 |
| HD 74156 | S | 6005 | 4.06 | 1.62 | 6.5 | 7.52 | 2.50 | 0.04 | | 8.34 | 8.39 | 8.49 | 8.92 | 8.95 | 8.39 | 8.93 |
| HD 7514 | S | 5682 | 4.24 | 1.22 | 4.6 | 7.31 | 0.68 | 0.06 | L | 8.41 | 8.38 | 8.43 | 9.02 | 8.94 | 8.41 | 8.96 |

| | | | | | | | | | | | | | | |
|---|---|---|---|---|---|---|---|---|---|---|---|---|---|---|
| HD 75488 | S | 5996 | 4.18 | 1.36 | 4.9 | 7.04 | 2.19 | 0.04 | | 7.88 | 7.97 | 8.20 | 8.51 | 8.47 | 8.02 | 8.48 |
| HD 76151 | S | 5780 | 4.43 | 1.10 | 3.9 | 7.57 | 1.85 | 0.05 | | 8.39 | 8.44 | 8.53 | 8.85 | 8.69 | 8.45 | 8.73 |
| HD 78366 | S | 5981 | 4.42 | 1.41 | 5.6 | 7.50 | 2.37 | 0.04 | | 8.25 | 8.31 | 8.34 | 8.87 | 8.77 | 8.30 | 8.80 |
| HD 79210 | S | 3973 | 4.69 | 0.15 | 5.8 | 7.64 | -0.07 | 0.18 | | | | | | | | |
| HD 79211 | S | 3763 | 4.49 | 0.15 | 6.2 | 7.81 | -0.25 | 0.19 | | | | | | | | |
| HD 80372 | S | 6039 | 4.36 | 1.43 | 7.7 | 7.49 | 2.51 | 0.04 | | 8.42 | 8.57 | 8.38 | 8.96 | 8.74 | 8.47 | 8.85 |
| HD 80606 | S | 5467 | 3.52 | 1.50 | 5.9 | 7.70 | 0.73 | 0.08 | L | 8.43 | 8.50 | 8.65 | 8.78 | 8.57 | 8.53 | 8.62 |
| HD 80607 | S | 5535 | 3.35 | 1.45 | 5.9 | 7.76 | 0.82 | 0.07 | L | 8.28 | 8.31 | 8.75 | | 8.27 | 8.45 | 8.27 |
| HD 81040 | S | 5753 | 4.48 | 1.17 | 5.3 | 7.38 | 1.98 | 0.06 | | 8.08 | 8.25 | 8.32 | 8.60 | 8.45 | 8.22 | 8.49 |
| HD 8173 | S | 6009 | 4.40 | 1.14 | 5.4 | 7.47 | 2.31 | 0.04 | | 8.23 | 8.23 | 8.44 | 8.90 | 8.91 | 8.27 | 8.90 |
| HD 81809 | S | 5666 | 3.72 | 1.27 | 5.5 | 7.14 | 1.21 | 0.06 | | 7.98 | 8.03 | 8.23 | 8.65 | 8.61 | 8.08 | 8.62 |
| HD 82106 | S | 4826 | 4.63 | 0.86 | 3.5 | 7.51 | 0.02 | 0.12 | L | | | 8.66 | | 8.95 | 8.66 | 8.95 |
| HD 82943 | S | 5919 | 4.35 | 1.36 | 6.5 | 7.67 | 2.42 | 0.04 | | 8.61 | 8.65 | 8.59 | 9.05 | 8.98 | 8.61 | 8.99 |
| HD 84117 | S | 6239 | 4.34 | 1.54 | 7.0 | 7.45 | 2.59 | 0.02 | | 8.22 | 8.30 | 8.52 | 8.76 | 8.75 | 8.31 | 8.75 |
| HD 84703 | S | 6125 | 4.14 | 1.60 | 7.1 | 7.50 | 2.64 | 0.03 | | 8.25 | 8.35 | 8.53 | 8.80 | 8.95 | 8.35 | 8.88 |
| HD 85725 | S | 5892 | 3.72 | 1.84 | 8.3 | 7.59 | 1.06 | 0.05 | L | 8.48 | 8.47 | 8.61 | 8.87 | 8.88 | 8.52 | 8.88 |
| HD 8574 | S | 6045 | 4.19 | 1.43 | 6.6 | 7.45 | 2.57 | 0.04 | | 8.32 | 8.36 | 8.39 | 8.79 | 8.84 | 8.35 | 8.82 |
| HD 8673 | S | 6409 | 4.25 | 3.13 | 30.7 | 7.63 | 2.10 | 0.01 | L | 8.46 | 8.61 | | 8.94 | | 8.54 | 8.94 |
| HD 87097 | S | 5993 | 4.40 | 1.88 | 11.1 | 7.50 | 2.65 | 0.04 | | 8.35 | 8.35 | 8.42 | 8.81 | 8.72 | 8.38 | 8.74 |
| HD 88133 | S | 5371 | 3.82 | 1.41 | 4.9 | 7.73 | 1.75 | 0.08 | | 8.60 | 8.66 | 8.69 | 8.93 | 8.93 | 8.65 | 8.93 |
| HD 88230 | S | 4215 | 4.70 | 0.15 | 4.3 | 7.68 | 0.07 | 0.16 | | | | | | | | |
| HD 88371 | S | 5673 | 4.27 | 1.08 | 4.4 | 7.17 | 0.59 | 0.06 | L | 8.13 | 8.21 | 8.26 | 8.70 | 8.74 | 8.20 | 8.73 |
| HD 88595 | S | 6322 | 4.12 | 1.91 | 9.0 | 7.54 | 1.57 | 0.02 | L | 8.36 | 8.48 | 8.51 | 8.94 | 8.84 | 8.44 | 8.89 |
| HD 88737 | S | 6129 | 3.73 | 2.20 | 12.0 | 7.69 | 1.40 | 0.03 | L | 8.47 | 8.45 | 8.59 | 8.94 | 8.72 | 8.49 | 8.83 |
| HD 89319 | S | 4942 | 3.35 | 1.61 | 5.4 | 7.70 | 0.64 | 0.11 | | | | 8.66 | | 8.99 | 8.66 | 8.99 |
| HD 89744 | S | 6207 | 3.92 | 2.11 | 10.4 | 7.65 | 2.03 | 0.02 | | 8.48 | 8.46 | 8.62 | 8.95 | 8.71 | 8.50 | 8.83 |
| HD 90508 | S | 5757 | 4.34 | 1.06 | 4.9 | 7.14 | 0.81 | 0.06 | L | 8.18 | 8.21 | 8.19 | 8.70 | 8.66 | 8.19 | 8.67 |
| HD 9224 | S | 5848 | 4.16 | 1.24 | 4.9 | 7.46 | 1.73 | 0.05 | | 8.37 | 8.41 | 8.45 | 8.79 | 8.73 | 8.41 | 8.74 |
| HD 92788 | S | 5710 | 4.32 | 1.20 | 5.5 | 7.73 | 0.96 | 0.06 | L | 8.60 | 8.63 | 8.69 | 8.98 | 9.00 | 8.64 | 9.00 |
| HD 9369 | S | 7237 | 4.09 | 3.50 | 48.9 | 8.19 | 3.78 | -0.05 | | 8.45 | 8.71 | | 9.12 | | 8.58 | 9.12 |
| HD 94132 | S | 4988 | 3.44 | 1.50 | 5.3 | 7.62 | 0.46 | 0.11 | | | | 8.63 | | 9.04 | 8.63 | 9.04 |
| HD 94915 | S | 5985 | 4.41 | 1.33 | 5.0 | 7.38 | 2.40 | 0.04 | | 8.18 | 8.22 | 8.20 | 8.81 | 8.89 | 8.20 | 8.87 |
| HD 95650 | S | 3655 | 4.56 | 0.25 | 6.0 | 7.99 | -0.24 | 0.20 | | | | | | | | |
| HD 96276 | S | 6040 | 4.34 | 1.38 | 5.4 | 7.41 | 2.45 | 0.04 | | 8.18 | 8.34 | 8.49 | 8.80 | 8.56 | 8.31 | 8.68 |
| HD 97100 | S | 5008 | 5.00 | 0.15 | 5.6 | 7.24 | 0.01 | 0.11 | L | | | | | | | |
| HD 97101 | S | 4189 | 4.67 | 0.70 | 6.0 | 8.10 | 0.10 | 0.16 | | | | | | | | |
| HD 97334 | S | 5906 | 4.44 | 1.57 | 7.8 | 7.56 | 2.69 | 0.05 | | 8.37 | 8.46 | 8.42 | 8.65 | 8.65 | 8.42 | 8.65 |
| HD 975 | S | 6405 | 4.27 | 1.66 | 11.7 | 7.43 | 1.55 | 0.01 | L | 8.33 | 8.41 | | 8.91 | | 8.37 | 8.91 |

| | | | | | | | | | | | | | | |
|---|---|---|---|---|---|---|---|---|---|---|---|---|---|---|
| HD 97584 | S | 4732 | 4.62 | 0.91 | 4.4 | 7.46 | -0.13 | 0.13 | L | | | 8.54 | | 8.68 | 8.54 | 8.68 |
| HD 98388 | S | 6338 | 4.34 | 2.03 | 12.9 | 7.63 | 2.93 | 0.02 | | 8.33 | 8.36 | 8.30 | 8.94 | 8.71 | 8.34 | 8.83 |
| HD 98712 | S | 4300 | 4.72 | 0.15 | 8.0 | 7.59 | 0.19 | 0.16 | | | | | | | | |
| HD 98823 | S | 6414 | 3.59 | 2.87 | 39.9 | 7.63 | 3.00 | 0.01 | | 7.78 | 8.32 | | 8.74 | | 8.05 | 8.74 |
| HR 1179 | S | 6292 | 4.17 | 1.77 | 10.6 | 7.35 | 2.36 | 0.02 | | 8.18 | 8.19 | 8.22 | 8.68 | 8.53 | 8.19 | 8.61 |
| HR 1232 | S | 4905 | 3.26 | 1.50 | 5.2 | 7.52 | 0.41 | 0.11 | | | | 8.45 | | 8.88 | 8.45 | 8.88 |
| HR 159 | S | 5501 | 4.28 | 1.08 | 3.9 | 7.26 | 0.38 | 0.07 | L | 8.07 | 8.18 | 8.24 | 8.60 | 8.51 | 8.16 | 8.53 |
| HR 1665 | S | 5935 | 3.92 | 1.63 | 7.6 | 7.48 | 2.23 | 0.04 | | 8.22 | 8.29 | 8.41 | 8.83 | 8.87 | 8.31 | 8.86 |
| HR 1685 | S | 4976 | 3.48 | 1.20 | 4.6 | 7.55 | 0.31 | 0.11 | L | | | 8.51 | | 8.98 | 8.51 | 8.98 |
| HR 1925 | S | 5243 | 4.53 | 1.00 | 5.4 | 7.54 | 0.20 | 0.09 | L | | | 8.26 | | 8.30 | 8.26 | 8.30 |
| HR 1980 | S | 6095 | 4.38 | 1.39 | 5.6 | 7.49 | 2.66 | 0.03 | | 8.33 | 8.36 | 8.47 | 8.82 | 8.62 | 8.37 | 8.72 |
| HR 2208 | S | 5762 | 4.50 | 1.42 | 6.1 | 7.45 | 2.07 | 0.06 | | 8.18 | 8.35 | 8.41 | 8.78 | 8.68 | 8.31 | 8.71 |
| HR 244 | S | 6201 | 4.09 | 1.58 | 8.1 | 7.62 | 1.21 | 0.03 | L | 8.21 | 8.33 | 8.54 | 8.80 | 8.92 | 8.33 | 8.86 |
| HR 2692 | S | 4995 | 3.47 | 1.09 | 4.5 | 7.08 | 0.25 | 0.11 | L | | | 8.31 | | 8.71 | 8.31 | 8.71 |
| HR 2866 | S | 6358 | 4.25 | 1.76 | 10.1 | 7.40 | 2.39 | 0.01 | | 8.29 | 8.31 | | 8.77 | | 8.30 | 8.77 |
| HR 3193 | S | 5989 | 3.91 | 1.68 | 7.4 | 7.50 | 2.28 | 0.04 | | 8.36 | 8.39 | 8.47 | 8.84 | 8.81 | 8.40 | 8.82 |
| HR 3271 | S | 5975 | 3.96 | 1.62 | 6.3 | 7.56 | 2.49 | 0.04 | | 8.38 | 8.36 | 8.49 | 8.89 | 8.81 | 8.41 | 8.83 |
| HR 3395 | S | 6222 | 4.32 | 1.52 | 7.1 | 7.52 | 3.06 | 0.02 | | 8.40 | 8.42 | 8.41 | 8.80 | 8.70 | 8.41 | 8.75 |
| HR 357 | S | 6491 | 3.72 | 2.76 | 44.5 | 7.52 | 3.12 | 0.01 | | 7.96 | 8.38 | | 8.60 | | 8.17 | 8.60 |
| HR 3762 | S | 5153 | 3.40 | 1.31 | 4.9 | 7.10 | -1.09 | 0.10 | L | | | 8.06 | | 8.47 | 8.06 | 8.47 |
| HR 3901 | S | 6081 | 4.01 | 1.66 | 7.1 | 7.57 | 1.44 | 0.03 | L | 8.38 | 8.47 | 8.53 | 8.95 | 8.81 | 8.44 | 8.88 |
| HR 4051 | S | 5978 | 4.18 | 1.44 | 6.6 | 7.45 | 2.43 | 0.04 | | 8.48 | 8.47 | 8.46 | 8.85 | 8.75 | 8.47 | 8.77 |
| HR 407 | S | 6520 | 3.72 | 3.15 | 33.9 | 7.41 | 3.25 | 0.00 | | 8.07 | 8.38 | | 8.54 | | 8.23 | 8.54 |
| HR 4285 | S | 5916 | 3.76 | 1.66 | 6.7 | 7.17 | 1.24 | 0.04 | L | 8.00 | 8.04 | 8.18 | 8.65 | 8.65 | 8.07 | 8.65 |
| HR 448 | S | 5861 | 3.99 | 1.51 | 6.9 | 7.66 | 2.50 | 0.05 | | 8.51 | 8.60 | 8.66 | 9.11 | 8.97 | 8.59 | 9.01 |
| HR 4864 | S | 5615 | 4.49 | 1.30 | 5.0 | 7.50 | 1.44 | 0.07 | | 8.30 | 8.34 | 8.38 | 8.69 | 8.62 | 8.34 | 8.64 |
| HR 495 | S | 4661 | 3.08 | 1.39 | 5.3 | 7.79 | 0.46 | 0.13 | | | | 8.64 | | 8.95 | 8.64 | 8.95 |
| HR 511 | S | 5422 | 4.55 | 0.91 | 4.3 | 7.53 | 0.48 | 0.08 | L | 8.06 | 8.34 | 8.41 | 8.85 | 8.73 | 8.27 | 8.76 |
| HR 5258 | S | 6417 | 3.71 | 2.57 | 22.1 | 7.58 | 2.25 | 0.01 | | 8.38 | 8.55 | | 8.78 | | 8.47 | 8.78 |
| HR 5317 | S | 6437 | 3.74 | 2.93 | 30.7 | 7.59 | 2.75 | 0.01 | | 8.32 | 8.37 | | 8.80 | | 8.35 | 8.80 |
| HR 5335 | S | 4862 | 3.29 | 1.68 | 5.3 | 7.76 | 0.80 | 0.12 | | | | 8.73 | | 9.09 | 8.73 | 9.09 |
| HR 5387 | S | 6753 | 4.29 | 2.17 | 10.7 | 7.46 | 1.54 | -0.01 | L | 8.27 | 8.28 | | 8.80 | | 8.27 | 8.80 |
| HR 5504 | S | 5990 | 4.01 | 1.78 | 8.1 | 7.65 | 2.49 | 0.04 | | 8.44 | 8.41 | 8.62 | 8.85 | 8.78 | 8.49 | 8.80 |
| HR 5630 | S | 6186 | 4.30 | 1.46 | 6.8 | 7.52 | 2.72 | 0.03 | | 8.46 | 8.45 | 8.50 | 8.94 | 8.84 | 8.46 | 8.89 |
| HR 5706 | S | 4785 | 3.52 | 1.02 | 5.0 | 7.86 | 0.44 | 0.12 | | | | 8.85 | | 9.19 | 8.85 | 9.19 |
| HR 5740 | S | 5956 | 3.97 | 1.87 | 9.6 | 7.78 | 1.26 | 0.04 | L | 8.71 | 8.68 | 8.71 | 9.18 | 8.98 | 8.70 | 9.03 |
| HR 6105 | S | 5934 | 4.16 | 1.38 | 5.8 | 7.57 | 2.42 | 0.04 | | 8.42 | 8.49 | 8.40 | 8.89 | 8.75 | 8.44 | 8.78 |
| HR 6106 | S | 6009 | 3.97 | 1.68 | 7.0 | 7.60 | 2.25 | 0.04 | | 8.38 | 8.43 | 8.52 | 8.94 | 8.87 | 8.42 | 8.91 |

| Name | | Teff | logg | v | | | | | | | | | | |
|---|---|---|---|---|---|---|---|---|---|---|---|---|---|---|
| HR 6269 | S | 5634 | 3.95 | 1.34 | 5.5 | 7.57 | 2.10 | 0.06 | | 8.55 | 8.55 | 8.46 | 8.80 | 8.66 | 8.52 | 8.69 |
| HR 6301 | S | 4955 | 3.31 | 1.39 | 5.0 | 7.46 | 0.87 | 0.11 | | | | 8.31 | | 8.83 | 8.31 | 8.83 |
| HR 6372 | S | 5744 | 4.02 | 1.49 | 5.2 | 7.73 | 2.26 | 0.06 | | 8.62 | 8.61 | 8.74 | 9.03 | 8.95 | 8.66 | 8.97 |
| HR 6465 | S | 5719 | 4.43 | 1.06 | 4.4 | 7.51 | 0.48 | 0.06 | L | 8.28 | 8.30 | 8.42 | 8.97 | 8.98 | 8.33 | 8.97 |
| HR 6516 | S | 5583 | 4.15 | 1.69 | 7.8 | 7.49 | 0.52 | 0.07 | L | 8.29 | 8.37 | 8.45 | 8.90 | 8.85 | 8.37 | 8.86 |
| HR 6669 | S | 6134 | 4.27 | 1.45 | 6.0 | 7.46 | 2.61 | 0.03 | | 8.45 | 8.35 | 8.47 | 8.83 | 8.76 | 8.41 | 8.80 |
| HR 672 | S | 6079 | 4.07 | 1.56 | 5.6 | 7.61 | 3.10 | 0.03 | | 8.45 | 8.46 | 8.68 | 8.93 | 8.89 | 8.50 | 8.91 |
| HR 6722 | S | 5539 | 3.72 | 1.54 | 5.0 | 7.50 | 0.00 | 0.07 | L | 8.40 | 8.42 | 8.44 | 9.16 | 8.82 | 8.42 | 8.90 |
| HR 6756 | S | 4907 | 3.39 | 1.30 | 4.7 | 7.49 | 0.41 | 0.11 | | | | 8.44 | | 8.82 | 8.44 | 8.82 |
| HR 6806 | S | 5042 | 4.58 | 0.98 | 3.8 | 7.30 | 0.32 | 0.10 | L | | | 8.66 | | 9.00 | 8.66 | 9.00 |
| HR 6847 | S | 5791 | 4.33 | 1.09 | 4.9 | 7.47 | 1.06 | 0.05 | L | 8.37 | 8.38 | 8.45 | 8.80 | 8.89 | 8.40 | 8.87 |
| HR 6907 | S | 6311 | 4.04 | 1.76 | 6.4 | 7.61 | 0.82 | 0.02 | L | 8.42 | 8.45 | 8.69 | 8.90 | 8.85 | 8.48 | 8.87 |
| HR 6950 | S | 5337 | 3.85 | 1.28 | 4.6 | 7.50 | 1.29 | 0.08 | | 8.47 | 8.55 | 8.49 | 8.71 | 8.70 | 8.50 | 8.70 |
| HR 7079 | S | 6319 | 4.31 | 1.73 | 8.3 | 7.41 | 2.71 | 0.02 | | 8.27 | 8.30 | 8.45 | 8.73 | 8.79 | 8.32 | 8.76 |
| HR 7291 | S | 6173 | 4.34 | 1.51 | 8.6 | 7.66 | 2.50 | 0.03 | | 8.44 | 8.55 | 8.55 | 8.80 | 8.83 | 8.51 | 8.82 |
| HR 7522 | S | 6026 | 3.89 | 1.86 | 8.0 | 7.60 | 1.05 | 0.04 | L | 8.37 | 8.45 | 8.63 | 8.83 | 8.80 | 8.45 | 8.81 |
| HR 7569 | S | 5774 | 4.10 | 1.26 | 4.9 | 7.35 | 0.28 | 0.05 | L | 8.28 | 8.27 | 8.46 | 8.88 | 8.79 | 8.34 | 8.81 |
| HR 761 | S | 6230 | 3.85 | 1.86 | 8.0 | 7.26 | 0.85 | 0.02 | L | 8.13 | 8.13 | 8.34 | 8.69 | 8.68 | 8.17 | 8.68 |
| HR 7637 | S | 5947 | 4.24 | 1.29 | 7.6 | 7.42 | 2.46 | 0.04 | | 8.16 | 8.27 | 8.37 | 8.65 | 8.49 | 8.27 | 8.53 |
| HR 7793 | S | 6261 | 4.36 | 1.55 | 9.9 | 7.41 | 2.64 | 0.02 | | 8.30 | 8.29 | 8.28 | 8.71 | 8.71 | 8.29 | 8.71 |
| HR 7855 | S | 6217 | 4.13 | 1.69 | 7.8 | 7.50 | 2.45 | 0.02 | | 8.24 | 8.26 | 8.51 | 8.73 | 8.80 | 8.30 | 8.77 |
| HR 7907 | S | 6189 | 4.25 | 1.74 | 9.1 | 7.66 | 2.65 | 0.03 | | 8.52 | 8.54 | 8.60 | 8.95 | 8.89 | 8.55 | 8.92 |
| HR 8133 | S | 5874 | 3.88 | 1.45 | 5.8 | 7.47 | 2.21 | 0.05 | | 8.26 | 8.31 | 8.46 | 8.65 | 8.55 | 8.34 | 8.58 |
| HR 8148 | S | 5433 | 4.43 | 0.39 | 5.2 | 7.21 | 0.74 | 0.08 | L | 8.18 | 8.28 | 8.11 | | 8.36 | 8.19 | 8.36 |
| HR 8170 | S | 5903 | 4.26 | 1.81 | 18.8 | 7.35 | 2.57 | 0.05 | | 8.38 | 8.47 | 8.31 | | 8.60 | 8.39 | 8.60 |
| HR 857 | S | 5201 | 4.59 | 1.56 | 6.3 | 7.54 | 2.66 | 0.09 | | | | 8.05 | | 7.42 | 8.05 | 7.42 |
| HR 8631 | S | 5453 | 3.69 | 1.29 | 5.3 | 7.29 | 1.53 | 0.08 | | 8.07 | 8.19 | 8.24 | 8.66 | 8.55 | 8.17 | 8.58 |
| HR 8832 | S | 4883 | 4.60 | 1.27 | 4.1 | 7.58 | 0.27 | 0.12 | | | | 8.76 | | 8.93 | 8.76 | 8.93 |
| HR 8924 | S | 4784 | 3.26 | 1.76 | 5.3 | 7.81 | 0.66 | 0.12 | | | | 8.75 | | 9.09 | 8.75 | 9.09 |
| HR 8964 | S | 5821 | 4.43 | 1.43 | 5.7 | 7.57 | 2.04 | 0.05 | | 8.33 | 8.41 | 8.38 | 8.83 | 8.79 | 8.37 | 8.80 |
| HR 9074 | S | 6227 | 4.36 | 1.54 | 9.0 | 7.50 | 2.77 | 0.02 | | 8.23 | 8.28 | 8.22 | 8.72 | 8.82 | 8.25 | 8.77 |
| HR 9075 | S | 6112 | 4.39 | 1.37 | 7.4 | 7.53 | 2.85 | 0.03 | | 8.23 | 8.28 | 8.36 | 8.78 | 8.75 | 8.28 | 8.76 |
| i Boo A | S | 5878 | 4.38 | 3.80 | 97.2 | 6.38 | 1.70 | 0.05 | L | | 8.65 | 7.69 | | 8.40 | 8.17 | 8.40 |
| i Boo B | S | 5240 | 4.50 | 0.96 | 4.2 | 6.94 | 1.46 | 0.09 | | | | 7.81 | | 8.15 | 7.81 | 8.15 |
| iot Peg | S | 6504 | 4.19 | 2.02 | 9.3 | 7.37 | 3.06 | 0.00 | | 8.24 | 8.26 | | 8.83 | | 8.25 | 8.83 |
| iot Per | S | 5995 | 4.15 | 1.49 | 6.4 | 7.59 | 2.47 | 0.04 | | 8.43 | 8.44 | 8.65 | 8.96 | 8.90 | 8.51 | 8.91 |
| iot Psc | S | 6177 | 4.08 | 1.60 | 8.1 | 7.36 | 2.26 | 0.03 | | 8.15 | 8.19 | 8.26 | 8.66 | 8.72 | 8.19 | 8.69 |
| kap Del | S | 5633 | 3.66 | 1.60 | 6.0 | 7.45 | 1.55 | 0.06 | | 8.27 | 8.32 | 8.39 | 8.85 | 8.77 | 8.33 | 8.79 |

| Name | | T | log g | [Fe/H] | v sin i | | | | | | | | | | |
|---|---|---|---|---|---|---|---|---|---|---|---|---|---|---|---|
| kap For | S | 5867 | 3.96 | 1.43 | 6.1 | 7.40 | 2.34 | 0.05 | | 8.33 | 8.34 | 8.44 | 8.83 | 8.71 | 8.37 | 8.74 |
| kap01 Cet | S | 5730 | 4.50 | 1.47 | 6.1 | 7.53 | 2.12 | 0.06 | | 8.28 | 8.33 | 8.42 | 8.85 | 8.74 | 8.35 | 8.76 |
| ksi Boo A | S | 5480 | 4.53 | 1.38 | 6.0 | 7.29 | 2.42 | 0.07 | | 8.17 | 8.16 | 8.17 | 8.59 | 8.45 | 8.17 | 8.49 |
| ksi Boo B | S | 4767 | 5.00 | 0.15 | 5.1 | 7.45 | 0.60 | 0.12 | | | | 9.15 | | 9.35 | 9.15 | 9.35 |
| ksi Peg | S | 6250 | 4.01 | 1.77 | 9.7 | 7.28 | 2.38 | 0.02 | | 8.04 | 8.15 | 8.27 | 8.63 | 8.52 | 8.13 | 8.58 |
| ksi UMa B | S | 5667 | 4.44 | 0.77 | 8.3 | 6.79 | 2.13 | 0.06 | | 8.14 | 8.29 | 7.73 | | 8.47 | 8.05 | 8.47 |
| lam Aur | S | 5931 | 4.28 | 1.28 | 5.7 | 7.56 | 2.09 | 0.04 | | 8.33 | 8.39 | 8.50 | 8.76 | 8.75 | 8.41 | 8.75 |
| lam Ser | S | 5908 | 4.13 | 1.45 | 5.3 | 7.44 | 1.86 | 0.05 | | 8.23 | 8.31 | 8.44 | 8.77 | 8.59 | 8.32 | 8.64 |
| LP 837-53 | S | 3666 | 4.89 | 1.25 | 7.0 | 7.82 | -0.05 | 0.20 | | | | | | | | |
| m Per | S | 6704 | 3.83 | 2.15 | 90.4 | 7.59 | 1.78 | -0.01 | L | 8.45 | 8.55 | | 8.80 | | 8.50 | 8.80 |
| m Tau | S | 5631 | 4.03 | 1.11 | 5.0 | 7.24 | 1.58 | 0.06 | | 8.32 | 8.33 | 8.23 | 8.82 | 8.68 | 8.29 | 8.71 |
| mu. Cas | S | 5434 | 4.57 | 0.53 | 2.4 | 6.72 | -0.03 | 0.08 | L | 7.40 | 7.92 | 7.97 | | 8.50 | 7.76 | 8.50 |
| mu. Her | S | 5560 | 3.99 | 1.36 | 5.2 | 7.70 | 1.21 | 0.07 | | 8.66 | 8.65 | 8.71 | 8.91 | 8.93 | 8.67 | 8.92 |
| mu.01 Cyg | S | 6354 | 3.93 | 1.96 | 11.6 | 7.32 | 1.23 | 0.01 | L | 8.21 | 8.11 | | 8.53 | | 8.16 | 8.53 |
| mu.02 Cnc | S | 5857 | 3.97 | 1.55 | 6.3 | 7.62 | 2.67 | 0.05 | | 8.49 | 8.48 | 8.68 | 8.96 | 8.93 | 8.55 | 8.94 |
| mu.02 Cyg | S | 5998 | 4.33 | 1.27 | 6.4 | 7.23 | 2.38 | 0.04 | | 8.22 | 8.17 | 8.21 | 8.64 | 8.46 | 8.20 | 8.50 |
| NAME 23 H. Cam | S | 6215 | 4.38 | 1.38 | 7.3 | 7.41 | 2.66 | 0.02 | | 8.25 | 8.31 | 8.45 | 8.67 | 8.53 | 8.31 | 8.60 |
| ome Leo | S | 5883 | 3.79 | 1.60 | 6.8 | 7.48 | 1.97 | 0.05 | | 8.32 | 8.37 | 8.44 | 8.75 | 8.56 | 8.38 | 8.61 |
| ome Sgr | S | 5455 | 3.61 | 1.42 | 5.6 | 7.42 | 2.17 | 0.08 | | 8.18 | 8.27 | 8.30 | 8.70 | 8.62 | 8.25 | 8.64 |
| omi Aql | S | 6173 | 4.20 | 1.53 | 6.1 | 7.64 | 2.66 | 0.03 | | 8.34 | 8.39 | 8.58 | 8.81 | 8.83 | 8.41 | 8.82 |
| omi02 Eri | S | 5202 | 4.59 | 0.33 | 3.0 | 7.22 | 0.29 | 0.09 | L | | | 8.36 | | 8.62 | 8.36 | 8.62 |
| phi02 Cet | S | 6218 | 4.39 | 1.32 | 5.4 | 7.38 | 2.82 | 0.02 | | 8.20 | 8.27 | 8.34 | 8.76 | 8.72 | 8.26 | 8.74 |
| pi.03 Ori | S | 6509 | 4.31 | 2.22 | 18.5 | 7.64 | 2.23 | 0.00 | | 8.33 | 8.40 | | 8.79 | | 8.36 | 8.79 |
| psi Cap | S | 6633 | 4.24 | 3.36 | 41.7 | 7.67 | 1.65 | 0.00 | L | 8.10 | 8.53 | | 8.99 | | 8.31 | 8.99 |
| psi Cnc | S | 5305 | 3.53 | 1.51 | 5.3 | 7.38 | 0.89 | 0.09 | | 8.10 | 8.17 | 8.34 | 8.90 | 8.79 | 8.20 | 8.82 |
| psi Ser | S | 5635 | 4.45 | 1.07 | 4.1 | 7.48 | 1.49 | 0.06 | | 8.24 | 8.28 | 8.36 | 8.70 | 8.63 | 8.29 | 8.65 |
| psi01 Dra A | S | 6573 | 3.97 | 2.55 | 13.9 | 7.47 | 2.69 | 0.00 | | 8.30 | 8.27 | | 8.76 | | 8.28 | 8.76 |
| psi01 Dra B | S | 6237 | 4.31 | 1.60 | 8.0 | 7.52 | 2.81 | 0.02 | | 8.30 | 8.36 | 8.37 | 8.91 | 8.88 | 8.34 | 8.89 |
| psi05 Aur | S | 6128 | 4.34 | 1.47 | 6.5 | 7.59 | 2.67 | 0.03 | | 8.34 | 8.42 | 8.60 | 8.84 | 8.72 | 8.42 | 8.78 |
| rho Cap | S | 6856 | 3.95 | 4.61 | 54.4 | 7.33 | 3.05 | -0.02 | L | 8.15 | 8.10 | | 8.90 | | 8.12 | 8.90 |
| rho CrB | S | 5850 | 4.17 | 1.11 | 4.9 | 7.27 | 1.49 | 0.05 | | 8.12 | 8.18 | 8.26 | 8.79 | 8.65 | 8.19 | 8.69 |
| rho01 Cnc | S | 5248 | 4.43 | 1.09 | 4.4 | 7.80 | 0.73 | 0.09 | | | | 8.88 | | 9.23 | 8.88 | 9.23 |
| sig Boo | S | 6781 | 4.29 | 2.13 | 10.2 | 7.11 | 1.27 | -0.01 | L | 8.02 | 7.99 | | 8.63 | | 8.00 | 8.63 |
| sig CrB A | S | 5923 | 4.12 | 1.52 | 29.4 | 7.47 | 2.28 | 0.04 | | 8.74 | 8.19 | 8.42 | 8.73 | 8.60 | 8.45 | 8.63 |
| sig CrB B | S | 5992 | 4.47 | 1.26 | 4.9 | 7.47 | 2.53 | 0.04 | | 8.15 | 8.21 | 8.41 | 8.79 | 8.70 | 8.26 | 8.72 |
| sig Dra | S | 5338 | 4.57 | 0.96 | 3.2 | 7.26 | 0.23 | 0.08 | L | 8.05 | 8.29 | 8.22 | 8.41 | 8.43 | 8.19 | 8.43 |
| tau Boo | S | 6447 | 4.26 | 2.38 | 16.8 | 7.80 | 1.74 | 0.01 | L | 8.51 | 8.52 | | 8.88 | | 8.51 | 8.88 |
| tau Cet | S | 5403 | 4.53 | 0.41 | 3.6 | 7.02 | 0.28 | 0.08 | L | 7.79 | 8.14 | 8.14 | 8.74 | 8.54 | 8.02 | 8.59 |

| Name | Type | Teff | logg | vt | vsini | [Fe/H] | ? | ? | L | ? | ? | ? | ? | ? | ? | ? |
|---|---|---|---|---|---|---|---|---|---|---|---|---|---|---|---|---|
| tau01 Eri | S | 6395 | 4.29 | 2.62 | 26.8 | 7.64 | 2.05 | 0.01 | L | 8.48 | 8.47 | | 8.93 | | 8.48 | 8.93 |
| tet Boo | S | 6294 | 4.07 | 3.04 | 30.4 | 7.49 | 2.23 | 0.02 | L | 8.44 | 8.44 | | 8.72 | 8.95 | 8.44 | 8.84 |
| tet Per | S | 6310 | 4.32 | 1.64 | 10.2 | 7.55 | 2.89 | 0.02 | | 8.38 | 8.34 | 8.41 | 8.76 | 8.79 | 8.37 | 8.77 |
| tet UMa | S | 6371 | 3.80 | 2.02 | 9.4 | 7.35 | 3.25 | 0.01 | | 8.07 | 8.12 | | 8.54 | | 8.09 | 8.54 |
| ups And | S | 6269 | 4.12 | 1.87 | 10.9 | 7.67 | 2.41 | 0.02 | | 8.37 | 8.43 | 8.60 | 8.87 | 8.79 | 8.44 | 8.83 |
| V* AK Lep | S | 4925 | 4.60 | 1.70 | 4.3 | 7.42 | -0.27 | 0.11 | L | | | 8.59 | | 8.95 | 8.59 | 8.95 |
| V* AR Lac | S | 5342 | 3.61 | 1.75 | 88.0 | 8.48 | 0.77 | 0.08 | L | 8.63 | 9.09 | | | | | |
| V* DE Boo | S | 5260 | 4.52 | 1.34 | 5.7 | 7.52 | 0.75 | 0.09 | | 8.42 | 8.49 | 8.46 | 8.75 | 8.75 | 8.46 | 8.75 |
| V* HN Peg | S | 5972 | 4.44 | 1.80 | 11.2 | 7.47 | 2.87 | 0.04 | | 8.25 | 8.26 | 8.27 | 8.68 | 8.69 | 8.26 | 8.69 |
| V* IL Aqr | S | 3935 | 5.00 | 1.75 | 5.0 | 6.49 | | | | | | | | | | |
| V* pi.01 UMa | S | 5921 | 4.46 | 1.98 | 10.4 | 7.46 | 2.87 | 0.04 | | 8.16 | 8.15 | 8.40 | 8.69 | 8.68 | 8.24 | 8.69 |
| V* V2213 Oph | S | 6030 | 4.39 | 1.48 | 7.1 | 7.50 | 2.75 | 0.04 | | 8.27 | 8.32 | 8.38 | 8.86 | 8.65 | 8.31 | 8.76 |
| V* V2215 Oph | S | 4463 | 4.68 | 0.15 | 3.7 | 7.37 | -0.56 | 0.14 | L | | | 8.48 | | 8.54 | 8.48 | 8.54 |
| V* V2502 Oph | S | 6969 | 4.02 | 2.59 | 11.1 | 7.39 | 3.00 | -0.03 | | 8.37 | 8.40 | | 8.77 | | 8.39 | 8.77 |
| V* V2689 Ori | S | 4073 | 4.72 | 0.15 | 6.3 | 7.92 | 0.11 | 0.17 | | | | | | | | |
| V* V376 Peg | S | 6066 | 4.33 | 1.33 | 6.5 | 7.50 | 2.69 | 0.03 | | 8.28 | 8.39 | 8.43 | 8.79 | 8.57 | 8.35 | 8.68 |
| V* V450 And | S | 5653 | 4.46 | 1.29 | 5.8 | 7.44 | 2.35 | 0.06 | | 8.21 | 8.37 | 8.34 | 8.70 | 8.75 | 8.31 | 8.73 |
| V* V819 Her | S | 5699 | 3.28 | 0.58 | 8.1 | 7.59 | 1.86 | 0.06 | | 7.56 | 7.39 | 8.44 | 8.72 | 8.69 | 7.80 | 8.70 |
| w Her | S | 5780 | 4.33 | 1.09 | 1.9 | 7.10 | 0.36 | 0.05 | L | 8.02 | 8.01 | 8.24 | 8.76 | 8.78 | 8.09 | 8.77 |
| Wolf 1008 | S | 5787 | 4.07 | 1.39 | 5.7 | 7.12 | 1.91 | 0.05 | | 8.05 | 8.08 | 8.17 | 8.65 | 8.45 | 8.10 | 8.50 |
| zet Her | S | 5759 | 3.69 | 1.69 | 6.2 | 7.50 | 0.89 | 0.06 | L | 8.44 | 8.45 | 8.46 | 8.85 | 8.75 | 8.45 | 8.77 |
| 10 Tau | H | 6013 | 4.05 | 1.61 | 3.2 | 7.42 | 2.41 | 0.04 | | 8.26 | 8.32 | 8.38 | 8.75 | 8.51 | 8.31 | 8.63 |
| 11 LMi | H | 5498 | 4.43 | 1.54 | 1.4 | 7.79 | 0.64 | 0.07 | L | 8.52 | 8.62 | 8.67 | 8.86 | 8.85 | 8.60 | 8.85 |
| 15 LMi | H | 5916 | 4.06 | 1.49 | 1.6 | 7.64 | 2.32 | 0.04 | | 8.37 | 8.45 | 8.59 | 8.92 | 8.89 | 8.47 | 8.90 |
| 16 Cyg B | H | 5753 | 4.34 | 1.22 | 1.0 | 7.58 | 0.76 | 0.06 | L | 8.35 | 8.44 | 8.47 | 8.94 | 8.81 | 8.42 | 8.84 |
| 37 Gem | H | 5932 | 4.40 | 1.20 | 1.6 | 7.37 | 1.93 | 0.04 | | 8.19 | 8.24 | 8.31 | 8.72 | 8.75 | 8.25 | 8.75 |
| 88 Leo | H | 6030 | 4.39 | 1.25 | 2.1 | 7.54 | 2.53 | 0.04 | | 8.32 | 8.37 | 8.39 | 8.72 | 8.72 | 8.35 | 8.72 |
| b Aql | H | 5466 | 4.10 | 1.12 | 1.8 | 7.84 | 0.44 | 0.08 | L | 8.79 | 8.86 | 8.72 | | 8.87 | 8.79 | 8.87 |
| bet CVn | H | 5865 | 4.40 | 1.07 | 1.2 | 7.29 | 1.61 | 0.05 | | 8.18 | 8.24 | 8.23 | 8.68 | 8.60 | 8.22 | 8.62 |
| HD 10086 | H | 5658 | 4.46 | 1.16 | 3.8 | 7.66 | 1.82 | 0.06 | | 8.35 | 8.48 | 8.58 | 8.95 | 9.09 | 8.47 | 9.06 |
| HD 101177 | H | 5964 | 4.43 | 1.17 | 1.4 | 7.37 | 2.21 | 0.04 | | 8.08 | 8.19 | 8.27 | 8.73 | 8.53 | 8.18 | 8.58 |
| HD 102158 | H | 5781 | 4.31 | 1.09 | 1.0 | 7.09 | 0.66 | 0.05 | L | 8.08 | 8.15 | 8.19 | 8.79 | 8.70 | 8.14 | 8.72 |
| HD 112257 | H | 5659 | 4.31 | 1.13 | 1.0 | 7.48 | 0.56 | 0.06 | L | 8.36 | 8.43 | 8.47 | 9.03 | 8.88 | 8.42 | 8.92 |
| HD 116442 | H | 5281 | 4.62 | 0.88 | 1.0 | 7.19 | 0.07 | 0.09 | L | 7.75 | 8.33 | 8.19 | 8.70 | 8.45 | 8.09 | 8.51 |
| HD 12051 | H | 5400 | 4.39 | 1.68 | 1.2 | 7.71 | 0.83 | 0.08 | | 8.45 | 8.65 | 8.66 | 9.18 | 8.92 | 8.59 | 8.98 |
| HD 130087 | H | 5991 | 4.12 | 1.55 | 3.2 | 7.75 | 1.00 | 0.04 | | | | | | | | |
| HD 13043 | H | 5877 | 4.15 | 1.43 | 1.9 | 7.59 | 2.00 | 0.05 | | 8.38 | 8.43 | 8.60 | 8.89 | 8.81 | 8.47 | 8.83 |
| HD 130948 | H | 5983 | 4.43 | 1.47 | 6.3 | 7.50 | 2.88 | 0.04 | | 8.28 | 8.36 | 8.30 | 8.66 | 8.55 | 8.31 | 8.58 |

|  | |  |  |  |  |  |  |  |  |  |  |  |  |  |  |
|---|---|---|---|---|---|---|---|---|---|---|---|---|---|---|---|
| HD 135101 | H | 5637 | 4.24 | 1.15 | 1.0 | 7.56 | 0.67 | 0.06 | L | 8.47 | 8.57 | 8.52 | 9.11 | 8.99 | 8.52 | 9.02 |
| HD 139323 | H | 5046 | 4.48 | 0.79 | 1.0 | 7.90 | 0.45 | 0.10 | | | | 8.66 | | 8.89 | 8.66 | 8.89 |
| HD 144579 | H | 5308 | 4.67 | 0.73 | 1.0 | 6.92 | 0.04 | 0.09 | L | 7.80 | 8.27 | 8.04 | 8.40 | 8.35 | 8.04 | 8.36 |
| HD 147044 | H | 5890 | 4.39 | 1.43 | 2.0 | 7.45 | 2.20 | 0.05 | | 8.35 | 8.30 | 8.37 | 8.66 | 8.70 | 8.34 | 8.69 |
| HD 152792 | H | 5675 | 3.84 | 1.33 | 1.2 | 7.15 | 2.19 | 0.06 | | 8.01 | 8.10 | 8.14 | 8.67 | 8.49 | 8.08 | 8.54 |
| HD 159222 | H | 5788 | 4.39 | 1.36 | 1.8 | 7.63 | 1.96 | 0.05 | | 8.46 | 8.53 | 8.55 | 9.03 | 8.96 | 8.52 | 8.98 |
| HD 170778 | H | 5932 | 4.46 | 1.64 | 7.1 | 7.47 | 2.88 | 0.04 | | 8.19 | 8.31 | 8.25 | 8.94 | 8.78 | 8.25 | 8.82 |
| HD 182488 | H | 5362 | 4.45 | 1.29 | 1.0 | 7.67 | 0.63 | 0.08 | L | 8.47 | 8.56 | 8.68 | 8.93 | 8.94 | 8.57 | 8.94 |
| HD 183341 | H | 5952 | 4.27 | 1.47 | 2.5 | 7.57 | 2.32 | 0.04 | | 8.36 | 8.38 | 8.55 | 8.90 | 8.95 | 8.43 | 8.94 |
| HD 193664 | H | 5945 | 4.44 | 1.25 | 1.2 | 7.41 | 2.26 | 0.04 | | 8.15 | 8.23 | 8.30 | 8.63 | 8.60 | 8.23 | 8.61 |
| HD 197076 | H | 5844 | 4.46 | 1.19 | 1.0 | 7.43 | 2.20 | 0.05 | | 8.11 | 8.24 | 8.31 | 8.83 | 8.73 | 8.22 | 8.76 |
| HD 210460 | H | 5529 | 3.52 | 1.45 | 2.3 | 7.22 | 1.16 | 0.07 | | 7.93 | 8.01 | 8.12 | 8.63 | 8.57 | 8.02 | 8.58 |
| HD 210640 | H | 6377 | 3.98 | 1.52 | 1.7 | 7.72 | 1.85 | 0.01 | | 7.60 | 7.70 | | | | 7.65 | |
| HD 223238 | H | 5889 | 4.30 | 1.35 | 1.0 | 7.56 | 1.43 | 0.05 | | 8.37 | 8.44 | 8.51 | 8.97 | 8.81 | 8.44 | 8.85 |
| HD 24213 | H | 6053 | 4.18 | 1.53 | 2.8 | 7.59 | 2.65 | 0.04 | | 8.29 | 8.32 | 8.41 | 8.84 | 8.88 | 8.33 | 8.86 |
| HD 28005 | H | 5727 | 4.27 | 1.38 | 1.3 | 7.78 | 1.43 | 0.06 | | 8.69 | 8.73 | 8.73 | 9.23 | 9.02 | 8.72 | 9.08 |
| HD 38858 | H | 5798 | 4.48 | 1.19 | 1.0 | 7.34 | 1.65 | 0.05 | | 8.05 | 8.13 | 8.31 | 8.66 | 8.62 | 8.17 | 8.63 |
| HD 39881 | H | 5719 | 4.27 | 1.08 | 1.2 | 7.41 | 0.75 | 0.06 | L | 8.34 | 8.40 | 8.44 | 8.93 | 8.91 | 8.39 | 8.91 |
| HD 47127 | H | 5616 | 4.32 | 1.12 | 1.2 | 7.65 | 0.72 | 0.07 | L | 8.40 | 8.52 | 8.62 | 9.00 | 8.92 | 8.51 | 8.94 |
| HD 5372 | H | 5847 | 4.37 | 1.44 | 1.9 | 7.73 | 1.22 | 0.05 | L | 8.56 | 8.60 | 8.65 | 9.03 | 8.83 | 8.60 | 8.88 |
| HD 58781 | H | 5576 | 4.41 | 1.20 | 1.3 | 7.64 | 0.71 | 0.07 | L | 8.47 | 8.58 | 8.63 | 9.02 | 8.93 | 8.56 | 8.95 |
| HD 64090 | H | 5528 | 4.62 | 2.34 | 8.0 | 5.78 | 1.45 | 0.07 | | 6.99 | 7.31 | 7.35 | 7.59 | 7.31 | 7.22 | 7.38 |
| HD 65583 | H | 5342 | 4.55 | 0.90 | 1.0 | 6.84 | 0.26 | 0.08 | L | 7.76 | 8.13 | 8.00 | 8.30 | 8.26 | 7.96 | 8.27 |
| HD 68017 | H | 5565 | 4.43 | 0.66 | 1.0 | 7.14 | -0.68 | 0.07 | L | 8.06 | 8.30 | 8.21 | 8.64 | 8.64 | 8.19 | 8.64 |
| HD 71148 | H | 5835 | 4.39 | 1.17 | 1.1 | 7.53 | 1.88 | 0.05 | | 8.28 | 8.37 | 8.37 | 8.85 | 8.70 | 8.34 | 8.73 |
| HD 72760 | H | 5328 | 4.56 | 1.41 | 1.0 | 7.53 | 0.81 | 0.08 | | 8.19 | 8.50 | 8.52 | 8.96 | 8.96 | 8.40 | 8.96 |
| HD 76752 | H | 5685 | 4.28 | 1.19 | 1.5 | 7.53 | 0.75 | 0.06 | L | 8.46 | 8.45 | 8.42 | 8.98 | 8.78 | 8.44 | 8.83 |
| HD 76909 | H | 5598 | 4.15 | 1.50 | 1.4 | 7.79 | 0.87 | 0.07 | L | 8.66 | 8.78 | 8.73 | 9.06 | 8.97 | 8.72 | 8.99 |
| HD 8648 | H | 5711 | 4.16 | 1.46 | 1.7 | 7.62 | 1.38 | 0.06 | | 8.47 | 8.56 | 8.60 | 9.04 | 8.82 | 8.55 | 8.87 |
| HD 89269 | H | 5635 | 4.49 | 1.07 | 1.1 | 7.36 | 0.86 | 0.06 | L | 8.09 | 8.27 | 8.29 | 8.68 | 8.54 | 8.22 | 8.57 |
| HD 98618 | H | 5735 | 4.35 | 1.13 | 1.6 | 7.52 | 1.61 | 0.06 | | 8.33 | 8.44 | 8.35 | 8.76 | 8.56 | 8.38 | 8.61 |
| HD 9986 | H | 5791 | 4.42 | 1.32 | 1.0 | 7.56 | 1.81 | 0.05 | | 8.40 | 8.47 | 8.41 | 8.70 | 8.50 | 8.43 | 8.55 |
| HR 2208 | H | 5762 | 4.50 | 1.50 | 2.2 | 7.48 | 2.09 | 0.06 | | 8.19 | 8.28 | 8.39 | 8.68 | 8.58 | 8.29 | 8.61 |
| HR 511 | H | 5422 | 4.55 | 0.76 | 1.1 | 7.64 | 0.30 | 0.08 | L | 8.04 | 8.37 | 8.49 | 8.92 | 8.68 | 8.30 | 8.74 |
| HR 6465 | H | 5719 | 4.43 | 1.16 | 1.0 | 7.56 | 0.64 | 0.06 | L | 8.27 | 8.44 | 8.43 | 8.96 | 8.78 | 8.38 | 8.82 |
| HR 6847 | H | 5791 | 4.33 | 1.21 | 1.0 | 7.52 | 1.01 | 0.05 | L | | | | | | | |
| HR 8964 | H | 5821 | 4.43 | 1.50 | 2.2 | 7.63 | 2.04 | 0.05 | | 8.32 | 8.38 | 8.50 | 8.80 | 8.65 | 8.40 | 8.69 |
| iot Psc | H | 6177 | 4.08 | 1.62 | 6.3 | 7.38 | 2.21 | 0.03 | | 8.19 | 8.25 | 8.21 | 8.80 | 8.64 | 8.22 | 8.72 |

| Name | | | | | | | | | | | | | | | |
|---|---|---|---|---|---|---|---|---|---|---|---|---|---|---|---|
| sig Dra | H | 5338 | 4.57 | 0.94 | 1.0 | 7.38 | 0.05 | 0.08 | L | 7.90 | 8.35 | 8.43 | 8.90 | 8.82 | 8.22 | 8.84 |
| V* BZ Cet | H | 5014 | 4.52 | 1.11 | 1.5 | 7.76 | 0.57 | 0.11 | | | | 8.56 | | 8.74 | 8.56 | 8.74 |
| 1 Hya | E | 6359 | 4.11 | 2.69 | 47.5 | 7.54 | 1.30 | 0.01 | L | 7.57 | 8.33 | | 8.88 | | 7.95 | 8.88 |
| 101 Tau | E | 6465 | 4.31 | 2.66 | 34.1 | 7.69 | 2.20 | 0.01 | L | 8.53 | 8.57 | | 8.86 | | 8.55 | 8.86 |
| 14 Boo | E | 6180 | 3.91 | 1.81 | 7.5 | 7.58 | 1.02 | 0.03 | L | 8.41 | 8.46 | 8.71 | 8.88 | 8.95 | 8.49 | 8.91 |
| 15 Peg | E | 6452 | 4.09 | 1.95 | 8.2 | 6.96 | 1.14 | 0.01 | L | 7.93 | 7.94 | | 8.54 | | 7.94 | 8.54 |
| 22 Lyn | E | 6395 | 4.34 | 1.65 | 9.7 | 7.27 | 2.41 | 0.01 | | 8.12 | 8.13 | | 8.81 | | 8.13 | 8.81 |
| 30 Ari B | E | 6257 | 4.34 | 3.36 | 43.5 | 7.69 | 2.53 | 0.02 | L | 7.96 | 8.54 | 8.40 | 8.90 | 8.91 | 8.28 | 8.90 |
| 34 Peg | E | 6258 | 3.92 | 1.94 | 9.8 | 7.52 | 1.66 | 0.02 | | 8.23 | 8.32 | 8.47 | 8.79 | 8.82 | 8.32 | 8.80 |
| 36 Dra | E | 6522 | 4.07 | 1.98 | 10.3 | 7.17 | 1.20 | 0.00 | L | 8.00 | 8.06 | | 8.57 | | 8.03 | 8.57 |
| 38 Cet | E | 6480 | 3.87 | 1.89 | 9.0 | 7.30 | 2.56 | 0.01 | | 8.13 | 8.09 | | 8.63 | | 8.11 | 8.63 |
| 4 Aqr | E | 6440 | 3.79 | 3.48 | 34.6 | 7.48 | 2.65 | 0.01 | L | 8.31 | 8.44 | | 8.70 | | 8.38 | 8.70 |
| 40 Leo | E | 6467 | 4.11 | 2.57 | 17.5 | 7.59 | 1.70 | 0.01 | L | 8.37 | 8.43 | | 8.89 | | 8.40 | 8.89 |
| 47 Ari | E | 6644 | 4.21 | 3.01 | 26.7 | 7.72 | 2.00 | -0.01 | L | 8.50 | 8.53 | | 9.08 | | 8.52 | 9.08 |
| 49 Peg | E | 6275 | 3.95 | 1.68 | 6.7 | 7.29 | 1.51 | 0.02 | L | 8.16 | 8.22 | 8.37 | 8.72 | 8.69 | 8.23 | 8.71 |
| 51 Ari | E | 5603 | 4.46 | 0.85 | 3.6 | 7.64 | 1.11 | 0.07 | | 8.44 | 8.52 | 8.59 | 8.95 | 8.98 | 8.52 | 8.97 |
| 6 And | E | 6338 | 4.23 | 2.43 | 20.7 | 7.34 | 1.78 | 0.02 | L | 8.25 | 8.34 | 8.27 | 8.65 | 8.56 | 8.29 | 8.60 |
| 6 Cet | E | 6242 | 4.09 | 1.53 | 6.3 | 7.17 | 2.46 | 0.02 | | 8.00 | 8.07 | 8.10 | 8.58 | 8.40 | 8.05 | 8.49 |
| 68 Eri | E | 6421 | 4.03 | 2.01 | 10.8 | 7.19 | 1.38 | 0.01 | L | 8.04 | 8.07 | | 8.67 | | 8.05 | 8.67 |
| 71 Ori | E | 6560 | 4.31 | 1.59 | 7.2 | 7.60 | 1.62 | 0.00 | L | 8.38 | 8.40 | | 8.77 | | 8.39 | 8.77 |
| 84 Cet | E | 6236 | 4.29 | 2.64 | 33.1 | 7.50 | 2.75 | 0.02 | | 8.07 | 8.45 | 8.25 | 8.80 | 8.70 | 8.26 | 8.75 |
| 89 Leo | E | 6538 | 4.30 | 2.22 | 15.1 | 7.63 | 2.85 | 0.00 | | 8.35 | 8.43 | | 8.86 | | 8.39 | 8.86 |
| b Her | E | 6000 | 4.22 | 1.26 | 4.5 | 6.90 | 2.38 | 0.04 | | 7.92 | 8.01 | 7.88 | 8.69 | 8.45 | 7.95 | 8.57 |
| BD+01 2063 | E | 4978 | 4.63 | 1.50 | 4.5 | 7.17 | 1.89 | 0.11 | | | | 8.42 | | 8.79 | 8.42 | 8.79 |
| BD+12 4499 | E | 4653 | 4.66 | 0.69 | 2.8 | 7.66 | -0.08 | 0.13 | L | | | 8.84 | | 9.02 | 8.84 | 9.02 |
| BD+17 4708 | E | 6180 | 4.16 | 1.54 | 9.0 | 5.99 | 2.30 | 0.03 | | 6.86 | 7.12 | 7.15 | 7.41 | 7.41 | 7.02 | 7.41 |
| BD+23 465 | E | 5240 | 4.49 | 1.22 | 3.9 | 7.68 | 0.85 | 0.09 | | | | 8.58 | | 8.86 | 8.58 | 8.86 |
| BD+29 366 | E | 5777 | 4.46 | 1.34 | 2.6 | 6.60 | 1.59 | 0.05 | | 7.81 | 7.68 | 7.91 | 8.00 | 7.98 | 7.80 | 7.99 |
| BD+33 99 | E | 4499 | 4.61 | 0.50 | 2.9 | 7.62 | -0.21 | 0.14 | | | | 8.92 | | 9.05 | 8.92 | 9.05 |
| BD+41 3306 | E | 5053 | 4.54 | 0.15 | 1.7 | 7.02 | -0.12 | 0.10 | L | | | 8.39 | | 8.78 | 8.39 | 8.78 |
| BD+43 699 | E | 4802 | 4.66 | 0.15 | 1.7 | 7.23 | -0.09 | 0.12 | L | | | 8.58 | | 8.83 | 8.58 | 8.83 |
| BD+46 1635 | E | 4215 | 4.65 | 1.36 | 5.0 | 7.90 | 0.05 | 0.16 | | | | | | | | |
| BD+52 2815 | E | 4203 | 4.66 | 0.15 | 4.2 | 7.60 | -0.19 | 0.16 | | | | | | | | |
| c Boo | E | 6560 | 4.27 | 3.22 | 44.1 | 7.64 | 2.10 | 0.00 | L | 8.07 | 8.53 | | 8.87 | | 8.30 | 8.87 |
| CCDM J20051-0418AB | E | 6370 | 4.07 | 1.78 | 10.1 | 7.30 | 2.26 | 0.01 | | 7.98 | 8.09 | | 8.58 | | 8.03 | 8.58 |
| CCDM J21031+0132AB | E | 6423 | 3.74 | 2.48 | 18.4 | 7.47 | 1.55 | 0.01 | L | 8.22 | 8.33 | | 8.80 | | 8.27 | 8.80 |
| CCDM J22071+0034AB | E | 6059 | 3.91 | 1.47 | 6.1 | 7.35 | 2.48 | 0.03 | | 8.00 | 8.09 | 8.45 | 8.60 | 8.64 | 8.13 | 8.62 |
| eta UMi | E | 6788 | 4.00 | 3.75 | 48.5 | 7.50 | 2.50 | -0.01 | L | 7.74 | 8.10 | | 8.72 | | 7.92 | 8.72 |

| Name | | T | log g | ξ | v sin i | A(Fe) | [Fe/H] | age | | A(C) | A(N) | A(O) | A(Na) | A(Mg) | A(Al) | A(Si) |
|---|---|---|---|---|---|---|---|---|---|---|---|---|---|---|---|---|
| GJ 1067 | E | 4378 | 4.68 | 0.57 | 3.8 | 7.72 | 0.08 | 0.15 | | | | 8.88 | | 9.04 | 8.88 | 9.04 |
| GJ 697 | E | 4881 | 4.64 | 1.03 | 3.7 | 7.44 | 0.30 | 0.12 | | | | 8.54 | | 8.75 | 8.54 | 8.75 |
| HD 10086 | E | 5658 | 4.46 | 1.04 | 3.8 | 7.57 | 1.70 | 0.06 | | 8.36 | 8.43 | 8.44 | 8.95 | 8.88 | 8.41 | 8.90 |
| HD 10145 | E | 5642 | 4.38 | 0.82 | 3.5 | 7.52 | 0.50 | 0.06 | L | 8.26 | 8.34 | 8.39 | 8.95 | 8.92 | 8.33 | 8.92 |
| HD 105631 | E | 5363 | 4.52 | 1.01 | 3.6 | 7.67 | 0.43 | 0.08 | L | 8.30 | 8.51 | 8.58 | 8.90 | 8.86 | 8.46 | 8.87 |
| HD 106116 | E | 5665 | 4.34 | 1.14 | 3.9 | 7.65 | 1.07 | 0.06 | | 8.38 | 8.45 | 8.56 | 8.95 | 9.04 | 8.47 | 9.02 |
| HD 106691 | E | 6647 | 4.13 | 3.08 | 38.6 | 7.58 | 2.20 | -0.01 | L | 7.86 | 8.24 | | 8.81 | | 8.05 | 8.81 |
| HD 107611 | E | 6391 | 4.21 | 2.54 | 19.3 | 7.45 | 2.65 | 0.01 | | 8.17 | 8.30 | | 8.87 | | 8.23 | 8.87 |
| HD 10853 | E | 4655 | 4.66 | 0.48 | 3.7 | 7.51 | 0.10 | 0.13 | | | | 8.60 | | 8.72 | 8.60 | 8.72 |
| HD 110463 | E | 4926 | 4.61 | 1.03 | 2.6 | 7.51 | 0.96 | 0.11 | | | | 8.48 | | 8.89 | 8.48 | 8.89 |
| HD 111069 | E | 5865 | 4.40 | 1.14 | 4.1 | 7.72 | 1.81 | 0.05 | | 8.46 | 8.52 | 8.65 | 8.95 | 8.92 | 8.54 | 8.92 |
| HD 11373 | E | 4759 | 4.61 | 0.85 | 2.8 | 7.67 | 0.07 | 0.12 | L | | | 8.80 | | 9.02 | 8.80 | 9.02 |
| HD 115274 | E | 6153 | 3.84 | 1.27 | 9.3 | 7.44 | 3.08 | 0.03 | | 8.25 | 8.29 | 8.50 | 8.83 | 8.66 | 8.32 | 8.74 |
| HD 116956 | E | 5308 | 4.52 | 1.26 | 6.5 | 7.65 | 1.62 | 0.09 | | 8.36 | 8.47 | 8.47 | 9.03 | 8.78 | 8.43 | 8.84 |
| HD 117635 | E | 5217 | 4.24 | 0.80 | 5.5 | 7.13 | 0.21 | 0.09 | L | | | 8.30 | | 8.80 | 8.30 | 8.80 |
| HD 118096 | E | 4568 | 4.73 | 0.15 | 3.8 | 7.15 | -0.15 | 0.14 | | | | 8.69 | | 8.84 | 8.69 | 8.84 |
| HD 119332 | E | 5213 | 4.53 | 1.04 | 2.3 | 7.44 | 0.72 | 0.09 | | | | 8.46 | | 8.71 | 8.46 | 8.71 |
| HD 119802 | E | 4716 | 4.64 | 1.19 | 2.5 | 7.64 | -0.06 | 0.13 | L | | | 8.77 | | 8.97 | 8.77 | 8.97 |
| HD 12051 | E | 5397 | 4.39 | 0.91 | 2.8 | 7.73 | 0.84 | 0.08 | | 8.42 | 8.64 | 8.66 | 9.05 | 8.94 | 8.57 | 8.97 |
| HD 122120 | E | 4486 | 4.62 | 0.49 | 2.5 | 7.79 | -0.31 | 0.14 | | | | 9.20 | | 9.39 | 9.20 | 9.39 |
| HD 124292 | E | 5497 | 4.49 | 0.89 | 2.9 | 7.42 | 0.77 | 0.07 | L | 8.22 | 8.36 | 8.43 | 8.85 | 8.80 | 8.34 | 8.81 |
| HD 124642 | E | 4664 | 4.60 | 1.26 | 3.8 | 7.59 | -0.09 | 0.13 | | | | 8.78 | | 8.95 | 8.78 | 8.95 |
| HD 128429 | E | 6456 | 4.26 | 2.26 | 17.2 | 7.49 | 1.23 | 0.01 | L | 8.38 | 8.46 | | 8.96 | | 8.42 | 8.96 |
| HD 12846 | E | 5733 | 4.48 | 0.66 | 2.9 | 7.30 | 0.89 | 0.06 | L | 8.09 | 8.20 | 8.23 | 8.75 | 8.54 | 8.17 | 8.60 |
| HD 130307 | E | 5043 | 4.59 | 0.66 | 2.7 | 7.40 | 0.05 | 0.10 | L | | | 8.35 | | 8.59 | 8.35 | 8.59 |
| HD 132142 | E | 5229 | 4.58 | 0.47 | 1.5 | 7.13 | 0.14 | 0.09 | L | | | 8.28 | | 8.55 | 8.28 | 8.55 |
| HD 132254 | E | 6279 | 4.21 | 1.75 | 9.6 | 7.59 | 2.47 | 0.02 | | 8.30 | 8.39 | 8.44 | 8.80 | 9.06 | 8.37 | 8.93 |
| HD 133002 | E | 5562 | 3.55 | 1.29 | 4.8 | 7.15 | 0.47 | 0.07 | L | 7.77 | 7.87 | 7.78 | 8.64 | 8.54 | 7.80 | 8.57 |
| HD 13403 | E | 5617 | 4.00 | 0.98 | 3.4 | 7.18 | 1.10 | 0.07 | | 8.08 | 8.18 | 8.23 | 8.75 | 8.63 | 8.16 | 8.66 |
| HD 135204 | E | 5457 | 4.49 | 0.93 | 2.5 | 7.47 | 0.59 | 0.08 | L | 8.22 | 8.38 | 8.55 | 9.05 | 8.97 | 8.38 | 8.99 |
| HD 135599 | E | 5270 | 4.55 | 1.03 | 3.8 | 7.50 | 0.83 | 0.09 | | 8.13 | 8.39 | 8.38 | 8.90 | 8.71 | 8.30 | 8.76 |
| HD 13579 | E | 5133 | 4.51 | 0.55 | 3.0 | 7.88 | 0.54 | 0.10 | | | | 8.71 | | 8.98 | 8.71 | 8.98 |
| HD 139777 | E | 5775 | 4.47 | 1.52 | 6.5 | 7.49 | 2.90 | 0.05 | | 8.29 | 8.28 | 8.34 | 8.79 | 8.37 | 8.30 | 8.48 |
| HD 139813 | E | 5394 | 4.56 | 1.45 | 5.8 | 7.51 | 2.47 | 0.08 | | 8.12 | 8.49 | 8.40 | 9.00 | 8.90 | 8.33 | 8.92 |
| HD 140283 | E | 5823 | 3.80 | 2.24 | 6.0 | 5.15 | 2.35 | 0.05 | | 6.46 | 7.11 | 6.47 | 6.57 | 6.57 | 6.68 | 6.57 |
| HD 14348 | E | 6036 | 4.03 | 1.62 | 7.1 | 7.65 | 1.45 | 0.04 | | 8.47 | 8.59 | 8.64 | 9.02 | 8.95 | 8.55 | 8.98 |
| HD 14374 | E | 5474 | 4.55 | 0.71 | 3.7 | 7.56 | 1.11 | 0.07 | | 8.24 | 8.39 | 8.35 | 8.57 | 8.50 | 8.32 | 8.52 |
| HD 144287 | E | 5315 | 4.42 | 0.88 | 1.7 | 7.40 | 0.50 | 0.09 | L | 8.24 | 8.38 | 8.43 | 9.10 | 8.90 | 8.35 | 8.95 |

|  | |  |  |  |  | | | | | | | | | |
|------|---|------|-------|-----|-----|------|------|---|------|------|------|------|------|------|------|
| HD 145435 | E | 6083 | 4.12 | 1.42 | 5.7 | 7.49 | 2.58 | 0.03 | | 8.39 | 8.39 | 8.36 | 8.79 | 8.69 | 8.38 | 8.74 |
| HD 145729 | E | 6028 | 4.39 | 1.09 | 4.8 | 7.44 | 2.50 | 0.04 | | 8.32 | 8.28 | 8.40 | 8.92 | 8.80 | 8.32 | 8.86 |
| HD 146946 | E | 5738 | 4.31 | 0.96 | 4.8 | 7.18 | 2.06 | 0.06 | | 8.16 | 8.35 | 8.06 | 8.50 | 8.09 | 8.19 | 8.19 |
| HD 153525 | E | 4820 | 4.63 | 1.75 | 3.5 | 7.37 | -0.01 | 0.12 | L | | | 8.40 | | 8.54 | 8.40 | 8.54 |
| HD 154931 | E | 5869 | 3.97 | 1.42 | 4.9 | 7.37 | 2.40 | 0.05 | | 8.23 | 8.21 | 8.46 | 8.50 | 8.40 | 8.30 | 8.43 |
| HD 155712 | E | 4936 | 4.58 | 0.24 | 2.4 | 7.48 | 0.21 | 0.11 | L | | | 8.52 | | 8.73 | 8.52 | 8.73 |
| HD 15632 | E | 5749 | 4.48 | 0.81 | 3.5 | 7.60 | 1.37 | 0.06 | | 8.34 | 8.37 | 8.56 | 8.85 | 8.61 | 8.43 | 8.67 |
| HD 156985 | E | 4778 | 4.59 | 0.51 | 1.3 | 7.47 | -0.49 | 0.12 | L | | | 8.31 | | 8.35 | 8.31 | 8.35 |
| HD 157089 | E | 5830 | 4.14 | 1.17 | 3.9 | 6.92 | 1.90 | 0.05 | | 7.85 | 8.03 | 7.99 | 8.58 | 8.51 | 7.96 | 8.52 |
| HD 159062 | E | 5385 | 4.48 | 0.27 | 2.1 | 7.19 | 0.30 | 0.08 | L | 8.06 | 8.38 | 8.36 | 8.91 | 8.72 | 8.27 | 8.77 |
| HD 159482 | E | 5805 | 4.36 | 0.86 | 3.6 | 6.75 | 0.65 | 0.05 | L | 7.97 | 7.93 | 7.62 | 8.45 | 8.45 | 7.84 | 8.45 |
| HD 160964 | E | 4589 | 4.67 | 0.78 | 2.9 | 7.45 | -0.09 | 0.14 | | | | 8.96 | | 9.21 | 8.96 | 9.21 |
| HD 161098 | E | 5637 | 4.50 | 0.69 | 3.1 | 7.33 | 1.11 | 0.06 | | 7.94 | 8.18 | 8.19 | 8.40 | 8.35 | 8.10 | 8.36 |
| HD 163183 | E | 5928 | 4.46 | 1.70 | 9.2 | 7.43 | 2.75 | 0.04 | | 8.25 | 8.29 | 8.33 | 8.75 | 8.71 | 8.29 | 8.72 |
| HD 16397 | E | 5821 | 4.37 | 0.92 | 2.6 | 7.00 | 1.03 | 0.05 | L | 8.01 | 8.03 | 8.03 | 8.75 | 8.69 | 8.03 | 8.70 |
| HD 164651 | E | 5599 | 4.45 | 0.76 | 2.7 | 7.52 | 0.11 | 0.07 | L | 8.18 | 8.36 | 8.38 | 8.75 | 8.81 | 8.30 | 8.79 |
| HD 165173 | E | 5484 | 4.54 | 0.74 | 2.9 | 7.59 | 0.89 | 0.07 | | 8.32 | 8.43 | 8.50 | 8.72 | 8.62 | 8.42 | 8.65 |
| HD 165401 | E | 5816 | 4.44 | 0.97 | 3.6 | 7.09 | 0.67 | 0.05 | L | 8.14 | 8.11 | 8.19 | 8.71 | 8.74 | 8.15 | 8.73 |
| HD 165476 | E | 5816 | 4.25 | 1.02 | 3.8 | 7.46 | 0.78 | 0.05 | L | 8.31 | 8.31 | 8.40 | 8.92 | 8.62 | 8.34 | 8.70 |
| HD 165590 | E | 5663 | 4.09 | 1.56 | 16.0 | 7.29 | 2.50 | 0.06 | | 8.30 | 8.35 | 8.23 | 8.70 | 8.70 | 8.29 | 8.70 |
| HD 165670 | E | 6285 | 4.32 | 1.88 | 12.1 | 7.47 | 1.51 | 0.02 | L | 8.44 | 8.37 | 8.32 | 9.00 | 9.05 | 8.39 | 9.03 |
| HD 165672 | E | 5866 | 4.37 | 1.26 | 5.2 | 7.66 | 2.28 | 0.05 | | 8.48 | 8.53 | 8.55 | 9.00 | 8.79 | 8.52 | 8.84 |
| HD 166183 | E | 6380 | 4.05 | 1.92 | 10.2 | 7.45 | 1.34 | 0.01 | L | 8.28 | 8.34 | | 8.85 | | 8.31 | 8.85 |
| HD 166435 | E | 5811 | 4.47 | 1.75 | 10.6 | 7.47 | 0.81 | 0.05 | L | 8.17 | 8.31 | 8.32 | 8.70 | 8.40 | 8.26 | 8.48 |
| HD 167278 | E | 6483 | 3.82 | 2.70 | 21.1 | 7.25 | 2.04 | 0.01 | L | 8.04 | 8.19 | | 8.76 | | 8.12 | 8.76 |
| HD 169822 | E | 5529 | 4.53 | 1.13 | 3.4 | 7.31 | 0.50 | 0.07 | L | 8.22 | 8.42 | 8.36 | 8.50 | 8.40 | 8.33 | 8.43 |
| HD 170008 | E | 4978 | 3.33 | 0.90 | 3.4 | 7.15 | -0.26 | 0.11 | L | | | 8.06 | | 8.50 | 8.06 | 8.50 |
| HD 170291 | E | 6327 | 4.22 | 1.90 | 11.5 | 7.55 | 2.78 | 0.02 | | 8.34 | 8.44 | 8.41 | 8.80 | 8.84 | 8.40 | 8.82 |
| HD 170512 | E | 6152 | 4.27 | 1.58 | 7.5 | 7.68 | 2.76 | 0.03 | | 8.53 | 8.57 | 8.65 | 8.95 | 8.93 | 8.57 | 8.94 |
| HD 170579 | E | 6417 | 4.22 | 2.03 | 13.6 | 7.25 | 1.32 | 0.01 | L | 8.24 | 8.28 | | 8.80 | | 8.26 | 8.80 |
| HD 171314 | E | 4530 | 4.60 | 0.69 | 3.3 | 7.75 | -0.05 | 0.14 | | | | 9.12 | | 9.32 | 9.12 | 9.32 |
| HD 171888 | E | 6095 | 4.00 | 1.56 | 6.5 | 7.54 | 1.22 | 0.03 | L | 8.40 | 8.45 | 8.51 | 8.85 | 8.75 | 8.44 | 8.80 |
| HD 171951 | E | 6094 | 4.03 | 1.39 | 4.7 | 7.21 | 2.49 | 0.03 | | 7.98 | 8.12 | 8.34 | 8.72 | 8.71 | 8.10 | 8.72 |
| HD 171953 | E | 6480 | 3.82 | 2.28 | 47.7 | 7.77 | 2.99 | 0.01 | L | 8.25 | 8.69 | | 9.06 | | 8.47 | 9.06 |
| HD 172675 | E | 6303 | 4.37 | 1.66 | 10.3 | 7.42 | 2.53 | 0.02 | | 8.21 | 8.28 | 8.15 | 8.65 | 8.74 | 8.23 | 8.70 |
| HD 172718 | E | 6132 | 3.95 | 1.58 | 6.7 | 7.38 | 2.61 | 0.03 | | 8.13 | 8.09 | 8.28 | 8.85 | 8.60 | 8.15 | 8.72 |
| HD 172961 | E | 6571 | 4.32 | 2.21 | 17.1 | 7.43 | 2.10 | 0.00 | L | 8.19 | 8.18 | | 8.70 | | 8.19 | 8.70 |
| HD 173174 | E | 6009 | 3.98 | 1.70 | 6.4 | 7.67 | 2.30 | 0.04 | | 8.29 | 8.40 | 8.56 | 8.82 | 8.58 | 8.39 | 8.70 |

| Star | | Teff | log g | ξ | EW | A | B | C | | D | E | F | G | H | I | J |
|---|---|---|---|---|---|---|---|---|---|---|---|---|---|---|---|---|
| HD 173605 | E | 5722 | 4.01 | 2.40 | 20.9 | 7.50 | 1.16 | 0.06 | L | 8.43 | 8.63 | 8.24 | 8.90 | 8.83 | 8.43 | 8.85 |
| HD 173634 | E | 6505 | 3.76 | 2.94 | 37.0 | 7.60 | 2.75 | 0.00 | | 8.30 | 8.50 | | 8.77 | | 8.40 | 8.77 |
| HD 17382 | E | 5245 | 4.49 | 1.06 | 4.0 | 7.54 | 0.77 | 0.09 | | | | 8.41 | | 8.76 | 8.41 | 8.76 |
| HD 174080 | E | 4676 | 4.60 | 1.12 | 3.1 | 7.67 | -0.04 | 0.13 | | | | 8.94 | | 9.15 | 8.94 | 9.15 |
| HD 174719 | E | 5647 | 4.50 | 0.88 | 3.5 | 7.34 | 1.15 | 0.06 | | 8.05 | 8.05 | 8.27 | 8.85 | 8.72 | 8.12 | 8.76 |
| HD 175272 | E | 6638 | 3.98 | 2.93 | 23.6 | 7.60 | 2.82 | 0.00 | | 8.36 | 8.43 | | 8.95 | | 8.39 | 8.95 |
| HD 175726 | E | 6069 | 4.43 | 2.00 | 13.6 | 7.43 | 2.85 | 0.03 | | 8.34 | 8.21 | 8.38 | 8.74 | 8.64 | 8.29 | 8.69 |
| HD 175805 | E | 6318 | 3.72 | 3.42 | 35.4 | 7.86 | 1.69 | 0.02 | L | 8.28 | 8.80 | 8.47 | 9.04 | 9.08 | 8.53 | 9.06 |
| HD 175806 | E | 6171 | 3.63 | 1.99 | 7.5 | 7.45 | 3.08 | 0.03 | | 8.25 | 8.33 | 8.21 | 8.70 | 8.40 | 8.27 | 8.55 |
| HD 176118 | E | 6665 | 4.04 | 2.14 | 10.5 | 7.69 | 2.98 | -0.01 | | 8.53 | 8.57 | | 8.97 | | 8.55 | 8.97 |
| HD 176377 | E | 5868 | 4.46 | 0.99 | 4.1 | 7.24 | 2.18 | 0.05 | | 8.02 | 8.15 | 8.15 | 8.52 | 8.39 | 8.11 | 8.42 |
| HD 177749 | E | 6403 | 3.97 | 1.92 | 6.0 | 7.48 | 2.45 | 0.01 | | 8.30 | 8.29 | | 8.83 | | 8.30 | 8.83 |
| HD 177904 | E | 6902 | 3.84 | 3.31 | 33.6 | 7.64 | 2.70 | -0.02 | | 7.90 | 8.18 | | 8.86 | | 8.04 | 8.86 |
| HD 178126 | E | 4541 | 4.66 | 0.15 | 2.8 | 7.13 | -0.22 | 0.14 | | | | 8.81 | | 9.02 | 8.81 | 9.02 |
| HD 180161 | E | 5400 | 4.53 | 1.34 | 3.9 | 7.62 | 0.58 | 0.08 | L | 8.51 | 8.46 | 8.62 | 9.05 | 9.04 | 8.53 | 9.04 |
| HD 180945 | E | 6415 | 4.04 | 2.36 | 17.3 | 7.56 | 1.96 | 0.01 | | 8.33 | 8.37 | | 8.91 | | 8.35 | 8.91 |
| HD 181096 | E | 6270 | 3.92 | 1.85 | 7.7 | 7.23 | 1.38 | 0.02 | L | 8.20 | 8.20 | 8.34 | 8.84 | 8.69 | 8.23 | 8.76 |
| HD 181420 | E | 6606 | 4.18 | 2.72 | 22.0 | 7.60 | 1.80 | 0.00 | L | 8.20 | 8.37 | | 8.87 | | 8.28 | 8.87 |
| HD 181806 | E | 6404 | 3.92 | 1.82 | 6.7 | 7.53 | 2.61 | 0.01 | | 8.38 | 8.34 | | 8.92 | | 8.36 | 8.92 |
| HD 182274 | E | 6307 | 4.32 | 1.47 | 9.0 | 7.28 | 1.29 | 0.02 | L | 8.33 | 8.41 | 8.35 | 8.70 | 8.40 | 8.36 | 8.55 |
| HD 182736 | E | 5237 | 3.66 | 1.08 | 3.8 | 7.34 | 0.11 | 0.09 | L | | | 8.19 | | 8.59 | 8.19 | 8.59 |
| HD 182905 | E | 5376 | 3.88 | 1.18 | 4.1 | 7.57 | 1.77 | 0.08 | | 8.42 | 8.43 | 8.43 | 9.00 | 8.80 | 8.42 | 8.85 |
| HD 183341 | E | 5952 | 4.27 | 1.22 | 4.7 | 7.58 | 2.29 | 0.04 | | 8.30 | 8.37 | 8.55 | 8.80 | 8.85 | 8.41 | 8.84 |
| HD 183658 | E | 5828 | 4.45 | 0.87 | 4.3 | 7.62 | 1.44 | 0.05 | | 8.35 | 8.43 | 8.64 | 9.00 | 9.10 | 8.47 | 9.07 |
| HD 183870 | E | 5015 | 4.62 | 0.51 | 3.5 | 7.55 | -0.05 | 0.11 | L | | | 8.47 | | 8.82 | 8.47 | 8.82 |
| HD 184499 | E | 5807 | 4.10 | 1.11 | 3.7 | 6.98 | 1.42 | 0.05 | | 7.89 | 8.11 | 8.15 | 8.81 | 8.71 | 8.05 | 8.73 |
| HD 184768 | E | 5635 | 4.29 | 0.85 | 3.4 | 7.45 | 0.74 | 0.06 | L | 8.44 | 8.51 | 8.48 | 9.00 | 8.90 | 8.48 | 8.93 |
| HD 184960 | E | 6290 | 4.23 | 1.63 | 9.8 | 7.45 | 2.76 | 0.02 | | 8.22 | 8.25 | 8.17 | 8.81 | 8.78 | 8.22 | 8.80 |
| HD 185269 | E | 5987 | 3.97 | 1.59 | 7.3 | 7.63 | 2.50 | 0.04 | | 8.41 | 8.51 | 8.55 | 8.80 | 8.54 | 8.49 | 8.61 |
| HD 185414 | E | 5806 | 4.45 | 0.73 | 3.9 | 7.40 | 2.12 | 0.05 | | 8.21 | 8.26 | 8.31 | 8.89 | 8.79 | 8.26 | 8.82 |
| HD 186104 | E | 5759 | 4.33 | 1.09 | 3.6 | 7.61 | 1.43 | 0.06 | | 8.30 | 8.48 | 8.56 | 9.03 | 9.03 | 8.45 | 9.03 |
| HD 186226 | E | 6371 | 3.93 | 2.33 | 14.3 | 7.69 | 1.31 | 0.01 | L | 8.41 | 8.48 | | 8.99 | | 8.45 | 8.99 |
| HD 186379 | E | 5923 | 3.98 | 1.33 | 4.2 | 7.15 | 2.37 | 0.04 | | 8.01 | 8.07 | 8.28 | 8.44 | 8.43 | 8.12 | 8.43 |
| HD 186413 | E | 5918 | 4.16 | 1.29 | 5.1 | 7.50 | 2.36 | 0.04 | | 8.33 | 8.39 | 8.35 | 8.89 | 8.79 | 8.36 | 8.82 |
| HD 18757 | E | 5674 | 4.33 | 0.70 | 2.6 | 7.23 | 0.76 | 0.06 | L | 8.22 | 8.28 | 8.25 | 8.85 | 8.62 | 8.25 | 8.68 |
| HD 18768 | E | 5815 | 3.85 | 1.37 | 3.6 | 6.94 | 2.46 | 0.05 | | 7.91 | 7.94 | 8.05 | 8.59 | 8.49 | 7.97 | 8.51 |
| HD 187897 | E | 5905 | 4.35 | 1.39 | 5.9 | 7.61 | 2.45 | 0.05 | | 8.42 | 8.46 | 8.57 | 8.90 | 8.99 | 8.48 | 8.97 |
| HD 188326 | E | 5342 | 3.85 | 1.00 | 3.5 | 7.41 | 0.93 | 0.08 | | 8.20 | 8.37 | 8.42 | 8.85 | 8.76 | 8.33 | 8.78 |

| Star | | Teff | log g | ξ | vsini | [Fe/H] | [α/Fe] | σ | | C | N | O | Na | Mg | Al | Si |
|---|---|---|---|---|---|---|---|---|---|---|---|---|---|---|---|---|
| HD 189509 | E | 6368 | 4.34 | 2.82 | 31.3 | 7.55 | 2.98 | 0.01 | | 8.29 | 8.40 | | 8.84 | | 8.35 | 8.84 |
| HD 189558 | E | 5773 | 3.93 | 1.40 | 4.0 | 6.39 | 2.35 | 0.05 | | 7.60 | 7.43 | 7.68 | 8.23 | 8.22 | 7.57 | 8.22 |
| HD 19019 | E | 6113 | 4.40 | 1.18 | 5.3 | 7.43 | 0.84 | 0.03 | L | 8.19 | 8.28 | 8.37 | 8.55 | 8.44 | 8.26 | 8.50 |
| HD 190404 | E | 5088 | 4.60 | 0.54 | 2.2 | 6.93 | -0.07 | 0.10 | L | | | 8.33 | | 8.73 | 8.33 | 8.73 |
| HD 190412 | E | 5388 | 4.29 | 0.97 | 4.2 | 7.16 | 0.03 | 0.08 | L | 8.14 | 8.18 | 8.11 | 8.72 | 8.53 | 8.14 | 8.58 |
| HD 190498 | E | 6415 | 3.93 | 2.69 | 25.9 | 7.67 | 1.76 | 0.01 | L | 8.31 | 8.42 | | 9.06 | | 8.36 | 9.06 |
| HD 191533 | E | 6254 | 3.84 | 1.98 | 9.3 | 7.47 | 2.91 | 0.02 | | 8.21 | 8.24 | 8.56 | 8.76 | 8.81 | 8.29 | 8.78 |
| HD 191785 | E | 5213 | 4.52 | 0.68 | 0.4 | 7.43 | 0.52 | 0.09 | L | | | 8.57 | | 8.90 | 8.57 | 8.90 |
| HD 19308 | E | 5767 | 4.21 | 1.31 | 3.8 | 7.60 | 1.10 | 0.05 | L | 8.50 | 8.52 | 8.66 | 8.95 | 8.87 | 8.56 | 8.89 |
| HD 193374 | E | 6522 | 3.91 | 2.73 | 20.6 | 7.62 | 1.60 | 0.00 | L | 8.30 | 8.46 | | 8.76 | | 8.38 | 8.76 |
| HD 194154 | E | 6456 | 4.30 | 2.93 | 39.1 | 7.57 | 2.10 | 0.01 | | 8.08 | 8.35 | | 8.70 | | 8.21 | 8.70 |
| HD 19445 | E | 6052 | 4.49 | 1.97 | 5.0 | 5.56 | 2.35 | 0.04 | | 6.97 | 7.12 | 7.75 | 6.98 | 6.98 | 7.19 | 6.98 |
| HD 194598 | E | 6126 | 4.37 | 1.77 | 4.1 | 6.41 | 2.36 | 0.03 | | 7.50 | 7.39 | 7.57 | 8.50 | 8.45 | 7.47 | 8.48 |
| HD 195005 | E | 6149 | 4.41 | 1.33 | 6.4 | 7.50 | 2.75 | 0.03 | | 8.35 | 8.37 | 8.39 | 8.78 | 8.72 | 8.37 | 8.75 |
| HD 195104 | E | 6226 | 4.38 | 1.36 | 6.8 | 7.40 | 2.69 | 0.02 | | 8.14 | 8.20 | 8.29 | 8.66 | 8.78 | 8.19 | 8.72 |
| HD 195633 | E | 6024 | 3.99 | 1.59 | 4.8 | 6.87 | 2.30 | 0.04 | | 7.92 | 7.94 | 8.14 | 8.35 | 8.29 | 7.97 | 8.32 |
| HD 196218 | E | 6204 | 4.19 | 1.46 | 5.1 | 7.36 | 2.36 | 0.02 | | 8.31 | 8.28 | 8.46 | 8.81 | 8.63 | 8.33 | 8.72 |
| HD 198061 | E | 6379 | 3.98 | 3.53 | 32.6 | 7.47 | 2.50 | 0.01 | L | 8.28 | 8.44 | | 8.78 | | 8.36 | 8.78 |
| HD 199598 | E | 5918 | 4.37 | 1.04 | 3.8 | 7.51 | 2.51 | 0.04 | | 8.40 | 8.48 | 8.39 | 8.75 | 8.63 | 8.43 | 8.66 |
| HD 20039 | E | 5331 | 3.68 | 0.40 | 3.5 | 6.99 | 1.27 | 0.08 | | 8.08 | 7.91 | 7.89 | 8.28 | 8.16 | 7.96 | 8.19 |
| HD 200391 | E | 5709 | 3.97 | 0.40 | 43.5 | 7.65 | 1.30 | 0.06 | L | 7.70 | 8.55 | 8.22 | 8.87 | 8.87 | 8.16 | 8.87 |
| HD 200560 | E | 4894 | 4.52 | 1.13 | 3.8 | 7.62 | 0.08 | 0.11 | L | | | 8.31 | | 8.35 | 8.31 | 8.35 |
| HD 200580 | E | 5870 | 4.00 | 1.36 | 5.2 | 6.90 | 2.14 | 0.05 | | 7.80 | 8.01 | 8.20 | 8.50 | 8.45 | 8.00 | 8.46 |
| HD 201099 | E | 5947 | 4.22 | 1.16 | 4.1 | 7.06 | 2.11 | 0.04 | | 8.00 | 8.19 | 7.98 | 8.74 | 8.64 | 8.06 | 8.66 |
| HD 20165 | E | 5137 | 4.58 | 0.15 | 0.8 | 7.55 | 0.42 | 0.10 | L | | | 8.45 | | 8.77 | 8.45 | 8.77 |
| HD 201891 | E | 5998 | 4.39 | 1.39 | 7.1 | 6.46 | 2.17 | 0.04 | | 7.63 | 7.48 | 7.85 | 8.20 | 8.13 | 7.65 | 8.14 |
| HD 202575 | E | 4737 | 4.65 | 1.15 | 3.7 | 7.48 | 0.06 | 0.13 | L | | | 8.69 | | 8.91 | 8.69 | 8.91 |
| HD 203235 | E | 6242 | 4.16 | 1.62 | 6.3 | 7.64 | 2.73 | 0.02 | | 8.38 | 8.44 | 8.68 | 8.90 | 8.81 | 8.46 | 8.85 |
| HD 204426 | E | 5658 | 3.98 | 1.00 | 3.7 | 7.11 | 1.58 | 0.06 | | 8.07 | 8.17 | 8.15 | 9.04 | 9.01 | 8.13 | 9.01 |
| HD 204734 | E | 5262 | 4.58 | 1.15 | 5.1 | 7.56 | 0.99 | 0.09 | | 8.14 | 8.38 | 8.40 | 8.85 | 8.71 | 8.31 | 8.75 |
| HD 20512 | E | 5270 | 3.56 | 1.31 | 3.1 | 7.34 | 1.50 | 0.09 | | 8.08 | 8.14 | 8.31 | 8.90 | 8.63 | 8.18 | 8.70 |
| HD 205434 | E | 4454 | 4.68 | 0.60 | 5.2 | 7.53 | 0.64 | 0.14 | | | | 9.24 | | 9.41 | 9.24 | 9.41 |
| HD 205702 | E | 6060 | 4.19 | 1.46 | 6.4 | 7.57 | 2.56 | 0.03 | | 8.39 | 8.45 | 8.61 | 8.90 | 8.91 | 8.46 | 8.90 |
| HD 206374 | E | 5579 | 4.53 | 0.83 | 3.1 | 7.43 | 1.22 | 0.07 | | 8.14 | 8.31 | 8.32 | 8.75 | 8.65 | 8.26 | 8.68 |
| HD 208038 | E | 4995 | 4.63 | 0.53 | 3.2 | 7.49 | 1.10 | 0.11 | | | | 8.35 | | 8.62 | 8.35 | 8.62 |
| HD 208313 | E | 5030 | 4.61 | 0.52 | 2.3 | 7.52 | 0.41 | 0.11 | | | | 8.43 | | 8.64 | 8.43 | 8.64 |
| HD 209472 | E | 6480 | 4.18 | 2.16 | 20.5 | 7.38 | 1.50 | 0.01 | L | 8.35 | 8.23 | | 8.79 | | 8.29 | 8.79 |
| HD 210752 | E | 6024 | 4.42 | 1.09 | 4.0 | 6.94 | 2.34 | 0.04 | | 7.79 | 7.80 | 7.95 | 8.65 | 8.58 | 7.83 | 8.61 |

| Star | | T | log g | ξ | vsini | [Fe/H] | [α/Fe] | [Eu/Fe] | | | | | | | |
|---|---|---|---|---|---|---|---|---|---|---|---|---|---|---|---|
| HD 21197 | E | 4562 | 4.56 | 0.82 | 3.0 | 7.87 | 0.01 | 0.14 | | | | 9.01 | | 9.15 | 9.01 | 9.15 |
| HD 214683 | E | 4893 | 4.65 | 0.56 | 2.7 | 7.26 | 0.81 | 0.11 | | | | 8.45 | | 8.69 | 8.45 | 8.69 |
| HD 216259 | E | 5002 | 4.67 | 0.95 | 1.0 | 6.87 | 0.25 | 0.11 | L | | | 8.23 | | 8.42 | 8.23 | 8.42 |
| HD 216520 | E | 5103 | 4.56 | 1.11 | 2.5 | 7.32 | 0.31 | 0.10 | L | | | 8.40 | | 8.64 | 8.40 | 8.64 |
| HD 217813 | E | 5849 | 4.42 | 1.24 | 4.5 | 7.50 | 2.64 | 0.05 | | 8.20 | 8.19 | 8.27 | 8.75 | 8.53 | 8.22 | 8.58 |
| HD 218059 | E | 6382 | 4.31 | 1.52 | 7.8 | 7.21 | 2.46 | 0.01 | | 8.06 | 8.18 | | 8.69 | | 8.12 | 8.69 |
| HD 218209 | E | 5623 | 4.46 | 0.56 | 2.2 | 7.05 | 0.51 | 0.06 | L | 7.97 | 8.06 | 8.03 | 8.43 | 8.33 | 8.02 | 8.36 |
| HD 218566 | E | 4846 | 4.53 | 0.95 | 3.3 | 7.84 | 0.12 | 0.12 | L | | | 8.98 | | 9.29 | 8.98 | 9.29 |
| HD 218687 | E | 5902 | 4.39 | 1.83 | 11.4 | 7.42 | 1.05 | 0.05 | L | 8.27 | 8.35 | 8.31 | 8.80 | 8.74 | 8.31 | 8.76 |
| HD 218868 | E | 5509 | 4.41 | 1.03 | 3.2 | 7.71 | 0.66 | 0.07 | L | 8.44 | 8.46 | 8.59 | 9.00 | 8.99 | 8.49 | 8.99 |
| HD 219396 | E | 5649 | 4.03 | 1.06 | 4.2 | 7.41 | 0.83 | 0.06 | L | 8.24 | 8.35 | 8.44 | 9.06 | 9.04 | 8.34 | 9.04 |
| HD 219420 | E | 6175 | 4.19 | 1.49 | 7.1 | 7.49 | 1.18 | 0.03 | L | 8.35 | 8.45 | 8.59 | 8.75 | 8.46 | 8.44 | 8.60 |
| HD 219538 | E | 5078 | 4.60 | 0.15 | 2.5 | 7.58 | 0.34 | 0.10 | L | | | 8.43 | | 8.69 | 8.43 | 8.69 |
| HD 219623 | E | 6177 | 4.31 | 1.43 | 6.8 | 7.57 | 2.75 | 0.03 | | 8.33 | 8.40 | 8.51 | 8.95 | 8.66 | 8.39 | 8.80 |
| HD 220140 | E | 5075 | 4.58 | 2.77 | 17.1 | 7.55 | 2.70 | 0.10 | | | | 8.29 | | 8.97 | 8.29 | 8.97 |
| HD 220182 | E | 5335 | 4.56 | 1.16 | 5.6 | 7.54 | 1.83 | 0.08 | | 8.15 | 8.45 | 8.36 | 8.82 | 8.69 | 8.32 | 8.72 |
| HD 220221 | E | 4783 | 4.58 | 1.16 | 2.8 | 7.69 | -0.08 | 0.12 | L | | | 8.90 | | 9.18 | 8.90 | 9.18 |
| HD 221354 | E | 5282 | 4.49 | 0.89 | 0.5 | 7.62 | 0.67 | 0.09 | L | 8.43 | 8.51 | 8.68 | 8.97 | 8.90 | 8.54 | 8.92 |
| HD 221585 | E | 5530 | 3.93 | 1.37 | 3.9 | 7.75 | 1.55 | 0.07 | | 8.59 | 8.69 | 8.68 | 9.11 | 9.01 | 8.65 | 9.04 |
| HD 221851 | E | 5192 | 4.58 | 1.00 | 2.6 | 7.43 | 0.32 | 0.09 | L | | | 8.47 | | 8.88 | 8.47 | 8.88 |
| HD 222155 | E | 5694 | 3.94 | 1.23 | 3.8 | 7.34 | 1.03 | 0.06 | L | 8.18 | 8.24 | 8.36 | 8.90 | 8.83 | 8.26 | 8.85 |
| HD 224465 | E | 5770 | 4.42 | 0.98 | 3.6 | 7.58 | 0.83 | 0.05 | L | 8.38 | 8.39 | 8.48 | 8.95 | 8.76 | 8.42 | 8.81 |
| HD 22879 | E | 5932 | 4.36 | 1.22 | 3.1 | 6.65 | 1.51 | 0.04 | | 7.68 | 7.78 | 7.76 | 8.40 | 8.37 | 7.74 | 8.38 |
| HD 232781 | E | 4679 | 4.68 | 0.73 | 2.5 | 7.16 | -0.19 | 0.13 | L | | | 8.64 | | 8.84 | 8.64 | 8.84 |
| HD 23439A | E | 5192 | 4.67 | 0.15 | 0.3 | 6.57 | 0.03 | 0.09 | L | | | 7.84 | | 8.35 | 7.84 | 8.35 |
| HD 238087 | E | 4213 | 4.64 | 0.98 | 4.4 | 7.77 | -0.15 | 0.16 | | | | | | | | |
| HD 24040 | E | 5756 | 4.17 | 1.26 | 4.6 | 7.67 | 0.69 | 0.06 | L | 8.53 | 8.55 | 8.60 | 8.95 | 8.96 | 8.56 | 8.95 |
| HD 24238 | E | 5031 | 4.60 | 0.15 | 0.0 | 7.06 | -0.04 | 0.11 | L | | | 8.42 | | 8.83 | 8.42 | 8.83 |
| HD 24409 | E | 5568 | 4.34 | 0.84 | 3.8 | 7.35 | 0.72 | 0.07 | L | 8.36 | 8.46 | 8.31 | 8.65 | 8.56 | 8.38 | 8.58 |
| HD 24451 | E | 4547 | 4.64 | 0.82 | 3.9 | 7.67 | -0.13 | 0.14 | | | | 8.96 | | 9.16 | 8.96 | 9.16 |
| HD 24496 | E | 5429 | 4.44 | 0.97 | 2.9 | 7.41 | 0.20 | 0.08 | L | 8.38 | 8.41 | 8.36 | 8.60 | 8.57 | 8.38 | 8.58 |
| HD 245 | E | 5805 | 4.35 | 0.66 | 3.6 | 6.96 | 1.21 | 0.05 | L | 8.01 | 8.00 | 7.96 | 8.85 | 8.71 | 7.99 | 8.75 |
| HD 24552 | E | 5916 | 4.43 | 1.09 | 4.7 | 7.53 | 2.23 | 0.04 | | 8.15 | 8.31 | 8.32 | 9.00 | 8.87 | 8.26 | 8.91 |
| HD 25329 | E | 4855 | 4.73 | 1.63 | 2.1 | 5.89 | -0.32 | 0.12 | L | | | 7.97 | | 8.33 | 7.97 | 8.33 |
| HD 25457 | E | 6268 | 4.33 | 2.41 | 18.7 | 7.58 | 3.09 | 0.02 | | 8.27 | 8.36 | 8.42 | 8.80 | 8.60 | 8.34 | 8.70 |
| HD 25621 | E | 6307 | 3.86 | 2.54 | 18.0 | 7.57 | 1.36 | 0.02 | L | 8.29 | 8.39 | 8.43 | 8.91 | 8.73 | 8.36 | 8.82 |
| HD 25825 | E | 5976 | 4.41 | 1.24 | 8.6 | 7.58 | 2.63 | 0.04 | | 8.40 | 8.46 | 8.51 | 8.98 | 8.99 | 8.45 | 8.99 |
| HD 26345 | E | 6676 | 4.28 | 3.02 | 32.5 | 7.67 | 2.20 | -0.01 | L | 8.44 | 8.52 | | 8.89 | | 8.48 | 8.89 |

| Star | | Teff | log g | ξ | vsini | [Fe/H] | [X/Fe] | err | flag | A | B | C | D | E | F | G |
|---|---|---|---|---|---|---|---|---|---|---|---|---|---|---|---|---|
| HD 26784 | E | 6261 | 4.29 | 2.33 | 17.3 | 7.70 | 2.95 | 0.02 | | 8.55 | 8.61 | 8.49 | 9.01 | 8.90 | 8.56 | 8.96 |
| HD 27808 | E | 6230 | 4.30 | 2.03 | 13.1 | 7.63 | 2.98 | 0.02 | | 8.42 | 8.49 | 8.56 | 8.81 | 8.61 | 8.47 | 8.71 |
| HD 28005 | E | 5727 | 4.27 | 1.22 | 4.2 | 7.76 | 1.45 | 0.06 | | 8.66 | 8.72 | 8.72 | 9.15 | 9.12 | 8.70 | 9.13 |
| HD 28099 | E | 5738 | 4.42 | 1.56 | 5.0 | 7.62 | 2.33 | 0.06 | | 8.53 | 8.58 | 8.52 | 8.91 | 8.61 | 8.54 | 8.69 |
| HD 28344 | E | 5921 | 4.39 | 1.82 | 8.1 | 7.62 | 2.74 | 0.04 | | 8.38 | 8.50 | 8.49 | 9.06 | 9.06 | 8.46 | 9.06 |
| HD 283704 | E | 5524 | 4.51 | 1.30 | 4.5 | 7.68 | 1.47 | 0.07 | | 8.46 | 8.50 | 8.59 | 8.95 | 8.89 | 8.52 | 8.90 |
| HD 283750 | E | 4405 | 4.51 | 1.99 | 8.0 | 7.96 | 0.24 | 0.15 | | | | 9.23 | | 9.36 | 9.23 | 9.36 |
| HD 284248 | E | 6157 | 4.32 | 2.00 | 3.8 | 5.88 | 2.38 | 0.03 | | 7.10 | 6.90 | 7.04 | 7.70 | 7.50 | 7.01 | 7.60 |
| HD 284574 | E | 5370 | 4.47 | 0.98 | 6.0 | 7.75 | 1.23 | 0.08 | | 8.47 | 8.59 | 8.57 | 8.95 | 8.88 | 8.54 | 8.89 |
| HD 284930 | E | 4667 | 4.63 | 1.15 | 4.3 | 7.74 | -0.25 | 0.13 | L | | | 8.91 | | 9.08 | 8.91 | 9.08 |
| HD 285690 | E | 4975 | 4.45 | 0.78 | 2.9 | 7.67 | 0.57 | 0.11 | | | | 8.47 | | 8.69 | 8.47 | 8.69 |
| HD 28946 | E | 5366 | 4.57 | 0.78 | 2.4 | 7.40 | 0.02 | 0.08 | L | 8.07 | 8.47 | 8.27 | 8.62 | 8.40 | 8.27 | 8.46 |
| HD 28992 | E | 5865 | 4.40 | 1.54 | 7.2 | 7.57 | 2.60 | 0.05 | | 8.45 | 8.53 | 8.40 | 8.95 | 8.93 | 8.46 | 8.94 |
| HD 29310 | E | 5841 | 4.14 | 1.98 | 8.5 | 7.54 | 2.69 | 0.05 | | 8.43 | 8.54 | 8.32 | 8.80 | 8.63 | 8.43 | 8.67 |
| HD 29419 | E | 6058 | 4.37 | 1.51 | 4.9 | 7.63 | 2.84 | 0.03 | | 8.43 | 8.42 | 8.57 | 9.01 | 9.08 | 8.45 | 9.04 |
| HD 30562 | E | 5894 | 4.08 | 1.54 | 5.8 | 7.70 | 2.64 | 0.05 | | 8.43 | 8.50 | 8.59 | 8.96 | 9.06 | 8.51 | 9.04 |
| HD 3268 | E | 6232 | 4.10 | 1.41 | 6.7 | 7.39 | 1.57 | 0.02 | L | 8.14 | 8.27 | 8.51 | 8.60 | 8.40 | 8.27 | 8.50 |
| HD 32850 | E | 5226 | 4.56 | 1.20 | 2.2 | 7.31 | -0.51 | 0.09 | L | | | 8.21 | | 8.41 | 8.21 | 8.41 |
| HD 330 | E | 5932 | 3.79 | 1.44 | 5.7 | 7.37 | 2.62 | 0.04 | | 8.12 | 8.23 | 8.39 | 8.84 | 8.84 | 8.25 | 8.84 |
| HD 332518 | E | 4386 | 4.69 | 0.15 | 3.5 | 7.47 | -0.25 | 0.15 | | | | 9.16 | | 9.33 | 9.16 | 9.33 |
| HD 33564 | E | 6393 | 4.22 | 2.21 | 14.3 | 7.61 | 2.16 | 0.01 | | 8.41 | 8.45 | | 9.02 | | 8.43 | 9.02 |
| HD 345957 | E | 5943 | 4.08 | 1.55 | 5.0 | 6.16 | 2.66 | 0.04 | | 7.32 | 7.09 | 7.90 | 7.70 | 7.58 | 7.44 | 7.61 |
| HD 3628 | E | 5810 | 4.03 | 1.21 | 3.9 | 7.39 | 1.21 | 0.05 | | 8.08 | 8.18 | 8.47 | 8.85 | 8.86 | 8.24 | 8.85 |
| HD 37008 | E | 5091 | 4.60 | 0.62 | 1.6 | 7.13 | -0.10 | 0.10 | L | | | | | | | |
| HD 3765 | E | 4987 | 4.55 | 0.89 | 2.6 | 7.69 | 0.42 | 0.11 | | | | 8.80 | | 9.05 | 8.80 | 9.05 |
| HD 38230 | E | 5212 | 4.47 | 0.36 | 2.2 | 7.53 | 0.47 | 0.09 | L | | | 8.33 | | 8.38 | 8.33 | 8.38 |
| HD 40512 | E | 6471 | 4.28 | 1.80 | 47.4 | 7.95 | 2.00 | 0.01 | L | 8.10 | 8.44 | | 9.17 | | 8.27 | 9.17 |
| HD 40616 | E | 5769 | 3.99 | 1.17 | 4.0 | 7.23 | 2.05 | 0.05 | | 8.12 | 8.17 | 8.20 | 8.88 | 8.78 | 8.16 | 8.81 |
| HD 42250 | E | 5410 | 4.48 | 0.71 | 2.3 | 7.59 | 0.30 | 0.08 | L | 8.26 | 8.59 | 8.63 | 8.95 | 8.92 | 8.50 | 8.93 |
| HD 4256 | E | 4904 | 4.58 | 0.79 | 3.4 | 7.82 | 0.40 | 0.11 | | | | 8.84 | | 9.04 | 8.84 | 9.04 |
| HD 42618 | E | 5775 | 4.46 | 0.79 | 3.5 | 7.44 | 1.36 | 0.05 | | 8.20 | 8.28 | 8.38 | 8.90 | 8.83 | 8.29 | 8.84 |
| HD 42983 | E | 4918 | 3.61 | 1.07 | 3.2 | 7.62 | 0.57 | 0.11 | | | | 8.62 | | 8.97 | 8.62 | 8.97 |
| HD 43318 | E | 6241 | 3.87 | 1.70 | 6.2 | 7.34 | 0.56 | 0.02 | L | 8.17 | 8.28 | 8.43 | 8.74 | 8.56 | 8.26 | 8.65 |
| HD 43856 | E | 6150 | 4.19 | 1.31 | 5.7 | 7.31 | 2.50 | 0.03 | | 8.21 | 8.20 | 8.32 | 8.90 | 8.77 | 8.23 | 8.84 |
| HD 44966 | E | 6349 | 4.25 | 1.86 | 11.7 | 7.60 | 2.64 | 0.02 | | 8.38 | 8.43 | 8.54 | 8.89 | 8.95 | 8.43 | 8.92 |
| HD 45282 | E | 5422 | 3.23 | 1.54 | 3.6 | 6.12 | 1.27 | 0.08 | | 7.28 | 7.43 | 7.44 | 7.78 | 7.69 | 7.38 | 7.71 |
| HD 46090 | E | 5575 | 4.36 | 1.04 | 4.6 | 7.40 | 2.67 | 0.07 | | 8.49 | 8.36 | 8.38 | 9.38 | 9.45 | 8.41 | 9.43 |
| HD 46301 | E | 6396 | 3.74 | 3.57 | 43.7 | 7.77 | 2.51 | 0.01 | | 8.10 | 8.50 | | 8.75 | | 8.30 | 8.75 |

| | | | | | | | | | | | | | | | |
|---|---|---|---|---|---|---|---|---|---|---|---|---|---|---|---|
| HD 4635 | E | 5036 | 4.55 | 0.81 | 0.7 | 7.67 | 0.50 | 0.10 | | | | 8.64 | | 8.85 | 8.64 | 8.85 |
| HD 46780 | E | 5793 | 4.25 | 1.95 | 11.0 | 7.56 | 2.57 | 0.05 | | 8.32 | 8.44 | 8.28 | 9.06 | 8.97 | 8.35 | 8.99 |
| HD 47157 | E | 5689 | 4.40 | 0.42 | 3.8 | 8.02 | 1.02 | 0.06 | L | 8.64 | 8.77 | 8.90 | 9.00 | 8.95 | 8.77 | 8.96 |
| HD 47309 | E | 5792 | 4.37 | 1.01 | 3.7 | 7.55 | 1.34 | 0.05 | | 8.41 | 8.42 | 8.44 | 8.90 | 8.81 | 8.42 | 8.83 |
| HD 47752 | E | 4698 | 4.62 | 0.71 | 2.8 | 7.48 | 0.04 | 0.13 | | | | 8.73 | | 8.96 | 8.73 | 8.96 |
| HD 48565 | E | 6024 | 4.07 | 1.33 | 4.8 | 6.83 | 0.91 | 0.04 | L | 8.21 | 8.31 | 8.24 | 8.67 | 8.66 | 8.26 | 8.66 |
| HD 49385 | E | 6127 | 3.98 | 1.66 | 5.9 | 7.62 | 1.95 | 0.03 | | 8.36 | 8.43 | 8.53 | 9.07 | 9.03 | 8.42 | 9.05 |
| HD 50206 | E | 6459 | 3.88 | 2.12 | 7.6 | 7.61 | 1.20 | 0.01 | L | 8.48 | 8.56 | | 8.92 | | 8.52 | 8.92 |
| HD 50867 | E | 6256 | 4.30 | 1.52 | 7.6 | 7.43 | 2.73 | 0.02 | | 8.20 | 8.25 | 8.38 | 8.78 | 8.63 | 8.26 | 8.70 |
| HD 51219 | E | 5626 | 4.41 | 0.81 | 3.6 | 7.55 | 0.51 | 0.06 | L | 8.32 | 8.47 | 8.49 | 8.90 | 8.82 | 8.43 | 8.84 |
| HD 51419 | E | 5723 | 4.46 | 0.64 | 0.7 | 7.17 | 1.06 | 0.06 | L | 8.04 | 8.10 | 8.22 | 8.65 | 8.55 | 8.12 | 8.58 |
| HD 51866 | E | 4860 | 4.59 | 0.31 | 0.7 | 7.70 | 0.06 | 0.12 | L | | | 8.73 | | 8.90 | 8.73 | 8.90 |
| HD 52634 | E | 5999 | 4.07 | 2.77 | 23.0 | 7.45 | 3.00 | 0.04 | | 8.31 | 8.45 | 8.30 | 8.75 | 8.50 | 8.35 | 8.56 |
| HD 5294 | E | 5749 | 4.47 | 0.82 | 4.2 | 7.42 | 2.11 | 0.06 | | 8.21 | 8.22 | 8.32 | 8.85 | 8.78 | 8.25 | 8.80 |
| HD 5351 | E | 4608 | 4.65 | 0.15 | 3.6 | 7.11 | -0.53 | 0.13 | L | | | 8.84 | | 9.02 | 8.84 | 9.02 |
| HD 53927 | E | 4941 | 4.62 | 0.21 | 0.0 | 7.26 | -0.51 | 0.11 | L | | | 8.37 | | 8.63 | 8.37 | 8.63 |
| HD 54371 | E | 5562 | 4.43 | 1.36 | 6.3 | 7.54 | 1.94 | 0.07 | | 8.38 | 8.49 | 8.39 | 8.35 | 8.07 | 8.42 | 8.14 |
| HD 5600 | E | 6378 | 3.77 | 2.28 | 10.6 | 7.25 | 1.75 | 0.01 | L | 8.12 | 8.17 | | 8.67 | | 8.15 | 8.67 |
| HD 56303 | E | 5912 | 4.30 | 1.19 | 4.2 | 7.66 | 2.50 | 0.04 | | 8.39 | 8.47 | 8.54 | 8.84 | 8.74 | 8.47 | 8.77 |
| HD 56515 | E | 5993 | 4.07 | 1.41 | 5.9 | 7.54 | 2.52 | 0.04 | | 8.21 | 8.26 | 8.36 | 8.85 | 8.73 | 8.28 | 8.76 |
| HD 57707 | E | 5021 | 3.32 | 1.40 | 4.0 | 7.57 | 0.35 | 0.11 | | | | 8.33 | | 8.84 | 8.33 | 8.84 |
| HD 59090 | E | 6428 | 3.92 | 2.43 | 16.5 | 7.68 | 1.47 | 0.01 | L | 8.31 | 8.42 | | 8.93 | | 8.37 | 8.93 |
| HD 59747 | E | 5099 | 4.57 | 0.43 | 0.8 | 7.62 | 1.46 | 0.10 | | | | 8.22 | | 8.37 | 8.22 | 8.37 |
| HD 59984 | E | 6005 | 4.03 | 1.38 | 4.2 | 6.77 | 2.42 | 0.04 | | 7.76 | 7.75 | 7.91 | 8.64 | 8.52 | 7.79 | 8.58 |
| HD 62161 | E | 6449 | 4.25 | 3.42 | 42.8 | 7.64 | 2.81 | 0.01 | L | 7.85 | 8.40 | | 8.80 | | 8.13 | 8.80 |
| HD 62323 | E | 6113 | 4.06 | 1.60 | 6.8 | 7.50 | 1.08 | 0.03 | L | 8.38 | 8.39 | 8.44 | 8.95 | 8.85 | 8.40 | 8.90 |
| HD 62613 | E | 5539 | 4.54 | 0.55 | 1.6 | 7.48 | 0.46 | 0.07 | L | 8.09 | 8.30 | 8.37 | 8.82 | 8.71 | 8.25 | 8.74 |
| HD 63433 | E | 5673 | 4.50 | 1.53 | 7.7 | 7.49 | 2.49 | 0.06 | | 8.23 | 8.35 | 8.20 | 8.70 | 8.56 | 8.26 | 8.60 |
| HD 64021 | E | 6864 | 3.88 | 3.80 | 49.3 | 7.77 | 3.00 | -0.02 | | 7.50 | 8.30 | | 8.90 | | 7.90 | 8.90 |
| HD 64090 | E | 5528 | 4.62 | 2.03 | 3.1 | 5.82 | 1.50 | 0.07 | | 6.98 | 7.19 | 7.40 | 8.30 | 8.25 | 7.19 | 8.26 |
| HD 64468 | E | 4914 | 4.52 | 0.85 | 1.6 | 7.71 | 0.38 | 0.11 | | | | 8.81 | | 9.04 | 8.81 | 9.04 |
| HD 64606 | E | 5250 | 4.66 | 0.15 | 0.0 | 6.69 | 0.55 | 0.09 | L | 7.59 | 8.03 | 8.09 | 8.80 | 8.83 | 7.90 | 8.82 |
| HD 64815 | E | 5722 | 3.86 | 1.19 | 4.3 | 7.11 | 1.08 | 0.06 | L | 8.15 | 8.18 | 8.20 | 8.95 | 8.77 | 8.17 | 8.81 |
| HD 65583 | E | 5342 | 4.55 | 0.41 | 1.0 | 6.85 | 0.28 | 0.08 | L | 7.88 | 8.06 | 8.13 | 8.83 | 8.75 | 8.02 | 8.77 |
| HD 65874 | E | 5911 | 3.88 | 1.49 | 4.5 | 7.54 | 3.21 | 0.04 | | 8.37 | 8.42 | 8.48 | 8.90 | 8.78 | 8.42 | 8.81 |
| HD 68168 | E | 5750 | 4.40 | 0.97 | 3.6 | 7.65 | 1.55 | 0.06 | | 8.33 | 8.44 | 8.54 | 8.85 | 8.87 | 8.44 | 8.86 |
| HD 68284 | E | 5945 | 3.92 | 1.40 | 4.9 | 6.98 | 2.49 | 0.04 | | 7.90 | 7.91 | 8.04 | 8.72 | 8.58 | 7.95 | 8.61 |
| HD 68380 | E | 6538 | 4.33 | 1.86 | 11.2 | 7.43 | 1.75 | 0.00 | L | 8.27 | 8.30 | | 8.86 | | 8.28 | 8.86 |
| HD 68638 | E | 5364 | 4.35 | 1.48 | 6.2 | 7.24 | 0.57 | 0.08 | L | 8.08 | 8.30 | 8.15 | 8.67 | 8.39 | 8.17 | 8.46 |
| HD 69611 | E | 5846 | 4.25 | 1.06 | 2.8 | 6.92 | 0.55 | 0.05 | L | 8.00 | 8.03 | 8.05 | 8.75 | 8.82 | 8.03 | 8.81 |
| HD 70088 | E | 5566 | 4.54 | 0.99 | 4.5 | 7.45 | 1.93 | 0.07 | | 8.28 | 8.42 | 8.23 | 8.90 | 8.77 | 8.31 | 8.80 |
| HD 70516 | E | 5744 | 4.42 | 2.16 | 13.2 | 7.56 | 2.87 | 0.06 | | 8.38 | 8.46 | 8.28 | 9.00 | 8.85 | 8.37 | 8.88 |
| HD 70923 | E | 6021 | 4.22 | 1.33 | 4.1 | 7.60 | 2.45 | 0.04 | | 8.41 | 8.49 | 8.58 | 8.93 | 9.03 | 8.47 | 8.98 |
| HD 71431 | E | 5887 | 3.91 | 1.42 | 5.5 | 7.47 | 2.42 | 0.05 | | 8.28 | 8.36 | 8.39 | 8.65 | 8.71 | 8.34 | 8.69 |
| HD 71595 | E | 6682 | 4.02 | 1.95 | 9.9 | 7.45 | 3.20 | -0.01 | | 8.27 | 8.29 | | 8.79 | | 8.28 | 8.79 |
| HD 71640 | E | 6098 | 4.31 | 1.29 | 5.6 | 7.32 | 2.50 | 0.03 | | 8.23 | 8.32 | 8.34 | 8.83 | 8.82 | 8.29 | 8.82 |
| HD 73393 | E | 5703 | 4.47 | 0.88 | 2.9 | 7.56 | 1.05 | 0.06 | L | 8.33 | 8.40 | 8.42 | 8.90 | 8.90 | 8.38 | 8.90 |
| HD 74011 | E | 5751 | 4.10 | 1.11 | 3.6 | 6.87 | 1.69 | 0.06 | | 7.99 | 8.07 | 7.94 | 8.70 | 8.55 | 8.00 | 8.59 |
| HD 75318 | E | 5422 | 4.40 | 0.78 | 0.8 | 7.36 | 0.45 | 0.08 | L | 8.09 | 8.30 | 8.38 | 8.80 | 8.63 | 8.26 | 8.67 |
| HD 75767 | E | 5782 | 4.40 | 0.91 | 4.4 | 7.44 | 1.38 | 0.05 | | 8.36 | 8.37 | 8.41 | 8.88 | 8.78 | 8.38 | 8.80 |
| HD 7590 | E | 5979 | 4.45 | 1.38 | 8.2 | 7.45 | 2.82 | 0.04 | | 8.20 | 8.27 | 8.24 | 8.75 | 8.56 | 8.24 | 8.61 |
| HD 75935 | E | 5451 | 4.50 | 1.45 | 5.9 | 7.50 | 1.49 | 0.08 | | 8.18 | 8.37 | 8.31 | 8.78 | 8.56 | 8.29 | 8.61 |
| HD 76932 | E | 5966 | 4.18 | 1.26 | 3.2 | 6.66 | 2.12 | 0.04 | | 7.66 | 7.77 | 7.87 | 8.22 | 7.92 | 7.77 | 8.00 |
| HD 77407 | E | 5989 | 4.44 | 1.70 | 7.7 | 7.57 | 3.28 | 0.04 | | 8.37 | 8.31 | 8.48 | 8.96 | 9.13 | 8.38 | 9.09 |
| HD 7924 | E | 5172 | 4.60 | 0.83 | 0.9 | 7.32 | 0.17 | 0.10 | L | | | 8.20 | | 8.42 | 8.20 | 8.42 |
| HD 82443 | E | 5315 | 4.58 | 1.60 | 6.7 | 7.47 | 2.69 | 0.09 | | 8.17 | 8.44 | 8.32 | 8.80 | 8.67 | 8.31 | 8.70 |
| HD 84937 | E | 6383 | 4.48 | 2.76 | 4.6 | 5.45 | 2.24 | 0.01 | | 6.50 | 7.03 | | 6.87 | | 6.76 | 6.87 |
| HD 87883 | E | 4940 | 4.56 | 0.70 | 2.4 | 7.70 | -0.72 | 0.11 | L | | | 8.71 | | 9.07 | 8.71 | 9.07 |
| HD 88725 | E | 5753 | 4.45 | 0.52 | 1.9 | 6.94 | 0.67 | 0.06 | L | 7.91 | 8.01 | 8.03 | 8.75 | 8.65 | 7.98 | 8.68 |
| HD 89269 | E | 5635 | 4.49 | 0.63 | 2.7 | 7.37 | 1.08 | 0.06 | | 8.10 | 8.19 | 8.31 | 8.85 | 8.80 | 8.20 | 8.81 |
| HD 90875 | E | 4567 | 4.56 | 0.64 | 3.2 | 7.92 | 0.28 | 0.14 | | | | 9.12 | | 9.22 | 9.12 | 9.22 |
| HD 93215 | E | 5786 | 4.40 | 1.23 | 4.6 | 7.73 | 1.94 | 0.05 | | 8.38 | 8.45 | 8.60 | 9.05 | 8.94 | 8.48 | 8.96 |
| HD 9407 | E | 5652 | 4.44 | 0.69 | 3.0 | 7.55 | 0.96 | 0.06 | L | 8.31 | 8.43 | 8.46 | 8.92 | 8.87 | 8.40 | 8.88 |
| HD 94765 | E | 5033 | 4.58 | 1.00 | 3.3 | 7.55 | 0.33 | 0.10 | L | | | 8.49 | | 8.86 | 8.49 | 8.86 |
| HD 97503 | E | 4451 | 4.65 | 0.54 | 3.7 | 7.62 | -0.08 | 0.14 | | | | 8.95 | | 9.17 | 8.95 | 9.17 |
| HD 97658 | E | 5157 | 4.57 | 0.90 | 1.3 | 7.20 | 0.19 | 0.10 | L | | | 8.27 | | 8.57 | 8.27 | 8.57 |
| HD 98630 | E | 6033 | 3.86 | 1.83 | 7.6 | 7.72 | 1.30 | 0.04 | L | 8.60 | 8.59 | 8.64 | 9.14 | 8.94 | 8.60 | 9.04 |
| HD 98800 | E | 4213 | 3.82 | 1.43 | 9.7 | 7.53 | 2.20 | 0.16 | | | | | | | | |
| HD 99747 | E | 6676 | 4.15 | 2.22 | 12.4 | 7.00 | 1.60 | -0.01 | L | 7.96 | 7.97 | | 8.53 | | 7.96 | 8.53 |
| HR 1687 | E | 6540 | 4.13 | 2.53 | 16.1 | 7.75 | 2.00 | 0.00 | L | 8.51 | 8.61 | | 8.98 | | 8.56 | 8.98 |
| HR 3144 | E | 6064 | 3.70 | 2.49 | 15.3 | 7.69 | 1.36 | 0.03 | L | 8.39 | 8.51 | 8.48 | 8.90 | 8.91 | 8.45 | 8.90 |
| HR 4657 | E | 6316 | 4.42 | 1.50 | 8.4 | 6.83 | 0.91 | 0.02 | L | 7.89 | 7.99 | 8.10 | 8.60 | 8.42 | 7.97 | 8.51 |
| HR 4867 | E | 6293 | 4.32 | 3.19 | 44.1 | 7.64 | 2.41 | 0.02 | L | 7.80 | 8.30 | 8.30 | 8.85 | 8.86 | 8.10 | 8.85 |
| HR 5307 | E | 6461 | 4.16 | 1.72 | 32.3 | 7.90 | 2.00 | 0.01 | L | 8.45 | 8.63 | | 9.05 | | 8.54 | 9.05 |
| HR 7438 | E | 6726 | 4.28 | 5.02 | 49.3 | 7.70 | 2.10 | -0.01 | L | 8.34 | 8.31 | | 8.90 | | 8.33 | 8.90 |
| HR 784 | E | 6245 | 4.37 | 1.40 | 8.3 | 7.51 | 3.17 | 0.02 | | 8.30 | 8.37 | 8.41 | 8.80 | 8.50 | 8.35 | 8.65 |

| Name | Sp | T | G | Vt | Vr | Fe | Li | NLTE | L | 505.2 | 538.0 | C2 | 615.5 | 630.0 | C | O |
|---|---|---|---|---|---|---|---|---|---|---|---|---|---|---|---|---|
| HR 7955 | E | 6205 | 3.78 | 1.89 | 8.7 | 7.66 | 1.00 | 0.02 | L | 8.33 | 8.43 | 8.66 | 8.89 | 8.98 | 8.44 | 8.94 |
| iot Vir | E | 6217 | 3.75 | 2.41 | 17.1 | 7.41 | 1.42 | 0.02 | L | 8.23 | 8.35 | 8.20 | 8.81 | 8.84 | 8.27 | 8.82 |
| kap CrB | E | 4863 | 3.18 | 1.25 | 4.2 | 7.67 | 0.29 | 0.12 | | | | 8.44 | | 8.88 | 8.44 | 8.88 |
| phi Vir | E | 5551 | 3.44 | 2.33 | 16.0 | 7.43 | 2.49 | 0.07 | | 8.34 | 8.35 | 8.23 | 8.75 | 8.62 | 8.31 | 8.65 |
| tau01 Hya | E | 6507 | 4.21 | 2.49 | 32.2 | 7.63 | 2.10 | 0.00 | L | 8.31 | 8.44 | | 8.82 | | 8.38 | 8.82 |
| tet Dra | E | 6208 | 3.79 | 3.03 | 30.7 | 7.66 | 1.89 | 0.02 | L | 8.29 | 8.57 | 8.25 | 8.99 | 8.69 | 8.39 | 8.84 |
| V* BW Ari | E | 5186 | 4.58 | 1.40 | 5.1 | 7.65 | 0.68 | 0.09 | | | | 8.58 | | 8.84 | 8.58 | 8.84 |
| V* BZ Cet | E | 5014 | 4.52 | 1.60 | 3.6 | 7.65 | 0.58 | 0.11 | | | | 8.66 | | 9.04 | 8.66 | 9.04 |
| V* DI Cam | E | 6337 | 3.89 | 2.28 | 20.3 | 7.35 | 2.29 | 0.02 | | 8.11 | 8.26 | 8.30 | 8.60 | 8.57 | 8.21 | 8.58 |
| V* EI Eri | E | 5494 | 3.82 | 2.67 | 46.8 | 7.42 | 1.75 | 0.07 | | 8.00 | 8.60 | 8.30 | 8.72 | 8.64 | 8.30 | 8.66 |
| V* GM Com | E | 6709 | 4.29 | 2.86 | 20.8 | 7.42 | 1.80 | -0.01 | L | 8.23 | 8.31 | | 8.70 | | 8.27 | 8.70 |
| V* KX Cnc | E | 6008 | 4.09 | 1.60 | 6.9 | 7.54 | 3.00 | 0.04 | | 8.26 | 8.28 | 8.37 | 8.85 | 8.74 | 8.29 | 8.79 |
| V* MV UMa | E | 4665 | 4.62 | 0.57 | 2.7 | 7.28 | -0.54 | 0.13 | L | | | 8.57 | | 8.78 | 8.57 | 8.78 |
| V* NX Aqr | E | 5656 | 4.50 | 1.21 | 4.3 | 7.47 | 2.80 | 0.06 | | 8.07 | 8.33 | 8.38 | 8.80 | 8.69 | 8.26 | 8.72 |
| V* V1309 Tau | E | 5791 | 4.47 | 1.51 | 7.1 | 7.62 | 2.52 | 0.05 | | 8.49 | 8.53 | 8.50 | 9.05 | 8.95 | 8.51 | 8.98 |
| V* V1386 Ori | E | 5294 | 4.56 | 1.14 | 4.6 | 7.54 | 1.30 | 0.09 | | 8.29 | 8.42 | 8.35 | 8.97 | 8.84 | 8.35 | 8.87 |
| V* V1709 Aql | E | 6913 | 4.13 | 3.36 | 35.6 | 7.66 | 3.21 | -0.02 | | 8.58 | 8.42 | | 8.85 | | 8.50 | 8.85 |
| V* V401 Hya | E | 5836 | 4.48 | 1.42 | 5.2 | 7.60 | 2.30 | 0.05 | | 8.31 | 8.37 | 8.51 | 8.90 | 8.75 | 8.40 | 8.78 |
| V* V457 Vul | E | 5433 | 4.50 | 1.16 | 7.4 | 7.53 | 1.91 | 0.08 | | 8.26 | 8.34 | 8.23 | 8.90 | 8.74 | 8.27 | 8.78 |
| V* V566 Oph | E | 6358 | 4.05 | 2.80 | 48.7 | 6.66 | 2.20 | 0.01 | L | 6.96 | 7.47 | | 8.69 | | 7.21 | 8.69 |

| | | |
|---|---|---|
| Sp | | Source for spectroscopic material |
| T | K | Effective Temperature |
| G | cm s-2 | Logarithium of the surface acceleration (gravity) computed from average mass, temperature, and luminosity. |
| Vt | km s-1 | Microturbulent velocity |
| Vr | km s-1 | Rotational velocity |
| Fe | log ε | Iron abundance. The solar iron abundance is 7.47 relative to H = 12. |
| Li | log ε | Lithium abundance. The solar lithium abundance is 1.0 dex |
| NLTE | | Correction for non local thermodynamic equilibrium |
| L | | L = Upper limit on lithium abundance |
| 505.2 | log ε | Carbon abundance from C I 505.2 nm line. |
| 538.0 | log ε | Carbon abundance from C I 538.0 nm line. |
| C2 | log ε | Carbon abundance from C2 Swan lines - primary indicator at 513.5 nm |
| 615.5 | log ε | Oxygen abundance from [O I] 630.0 nm line |
| 630.0 | log ε | Oxygen abundance from O I 615.5 triplet |
| C | log ε | Mean carbon abundance - weights discussed in text |
| O | log ε | Mean oxygen abundance - weights discussed in text |

Table 6

Parameter and Iron Abundance Comparison

| Study | N | T type | G type | <dT> | s_T | <dG1> | s_G1 | <dG2> | s_G2 | <dFe1> | s_Fe1 | <dFe2> | s_Fe2 |
|---|---|---|---|---|---|---|---|---|---|---|---|---|---|
| This vs LH2006 | 207 | E | Ion | -74 | 116 | -0.09 | 0.27 | -0.17 | 0.22 | 0.00 | 0.10 | -0.04 | 0.12 |
| This vs C+2010 | 74 | IRFM | Var | 2 | 52 | 0.02 | 0.19 | | | | | | |
| This vs C+2011 | 612 | Var | Var | -23 | 97 | -0.03 | 0.09 | -0.08 | 0.24 | 0.02 | 0.12 | 0.00 | 0.13 |
| This vs R+2013 | 211 | E | Ion | 22 | 76 | -0.06 | 0.18 | 0.04 | 0.10 | 0.02 | 0.22 | -0.02 | 0.25 |
| This vs R+2014 | 30 | E | Ion | -8 | 57 | -0.07 | 0.05 | -0.07 | 0.15 | 0.01 | 0.03 | 0.01 | 0.03 |
| This vs B+2014 | 100 | E | Ion | -21 | 100 | -0.04 | 0.13 | -0.06 | 0.17 | 0.00 | 0.07 | -0.01 | 0.08 |
| B+2014 vs C+2011 | 642 | | | -24 | 105 | -0.01 | 0.14 | | | -0.02 | 0.14 | | |
| This vs MM2012 | 15 | Var | A | 17 | 33 | -0.02 | 0.06 | -0.02 | 0.11 | 0.03 | 0.04 | 0.03 | 0.05 |
| This vs C+2013 | 10 | Var | A | 26 | 61 | -0.02 | 0.04 | -0.03 | 0.15 | 0.00 | 0.08 | -0.01 | 0.09 |

N = number of common stars

T Type = Method of effective temperature determination in comparison study. E = Excitation analysis  IRFM = Infrared Flux Method  Var = combination or literature derived

G Type = Method of gravity determination in comparison study. Ion = spectroscopic ionization balance.  Var = Combination (line wings/isochrones / literature) .  A = astroseimology

dT = New effective temperature - Source
dG1 = New Mass Gravity - Source
dG2 = New Ionization Gravity - Source
dFe1 = New Mass Gravity [Fe/H] - Source
dFe2 = New Ionization Gravity [Fe/H] - Source

s_X = Standard deviation of differences

Comparison Studies

| | | | |
|---|---|---|---|
| LH2006 | Luck & Heiter 2006 | R+2014 | Ramírez et al. 2014 |
| C+2010 | Casagrande et al. 2010 | B+2014 | Bensby et al. 2014 |
| C+2011 | Casagrande et al. 2011 | MM2012 | Morel & Miglio. 2012 |
| R+2013 | Ramírez et al. 2013 | C+2013 | Creevey et al. 2013 |

Figures

Figure 1: Top Panel: The difference in Fe I and Fe II total iron abundances as a function of effective temperature as determined from the mass-derived gravity. Bottom Panel: The difference in gravity from the mass-derived value minus the ionization balance determination. The limit at a difference of 1.0 dex reflects the maximum change allowed. The low-temperature differences of 0.0 dex signify the lack of Fe II data.

Figure 2: Top Panel: [S/Fe], [Si/Fe], and [Ni/Fe] versus effective temperature for the program stars. The data shown are derived using the mass determined gravities. It is immediately evident that below 4500 K, the S and Si data are not reliable as the ratios rapidly increase as a function of temperature. The most likely explanation is the intrusion of significant blends with decreasing temperature. Bottom Panel: Fe I and Fe II total iron abundances as a function of temperatures. The Fe II data is also affected below 4500 K by significant blend problems.

Figure 3: The HR diagram for the program stars. The sample is broken into subsets based on rotational velocity and proximity to the main sequence. The dividing line for rotational velocity is 20 km s$^{-1}$. The proximity division is the solid line indicated. Stars below the line are considered dwarfs while those above are except for one, subgiants.

Figure 4: <[x/Fe]> versus element. The offsets from [x/Fe] are discussed in §4.2.

Figure 5: [Si/Fe], [Ca/Fe], [Mn/Fe], [Ni/Fe], and [Zn/Fe] versus [Fe/H]. The trends exhibited in these panels are discussed in §4.2.

Figure 6: Top Two Panels: [Nd/Fe] versus [Fe/H] and effective temperature. Bottom Two Panels: [Eu/Fe] versus [Fe/H] and effective temperature. While the Nd data appear influenced by the cooler stars (lower panel of Nd data), eliminating the outlying points has little effect upon the mean. The same is true for Eu.

Figure 7: Histogram for [Fe/H] of "non-hosts" and hosts. The smooth curves are 4 parameter Weibull fits. The peaks are offset by 0.12 dex, and a Mann-Whitney test indicates that the two distributions are different.

Figure 8: Top Panel: Lithium abundance versus effective temperature. Bottom Panel: Lithium abundance versus mass. Both show the dwarfs only. Astration as a function of temperature / mass is most pronounced in the data.

Figure 9: Lithium versus temperature with both dwarfs and "non-dwarfs." The only non-dwarf that is a young star is HD 98800. The other non-dwarfs have lithium abundances consistent with the dwarfs and are subgiants.

Figure 10: Lithium versus temperature with the planet-hosts segregated from the non-hosts. Both classes appear to have equivalent lithium abundances at the same temperature.

Figure 11:     Mean lithium abundances for non-host and host dwarfs versus effective temperature. The bin size is 100 K and both the temperature and lithium abundance shown are the average value within the bin. The error bars on the lithium abundance are one standard deviation. Note that the host abundances tend to be lower than the corresponding non-host lithium abundance.

Figure 12:     Histograms of the lithium abundances in three temperature bins. Astration effects are obvious in the abundances: the decline in mean abundance and the broadening of the Gaussian fit width going to lower temperature. A Mann-Whitney test indicates that only the lower temperature set shows a significant difference between the hosts and non-hosts. In that temperature range, the hosts show a lower abundance than do the non-hosts.

Figure 13:     Top Panel:     [C/Fe] versus [Fe/H]. Middle Panel: [O/Fe] versus [Fe/H], Bottom Panel: C/O versus [Fe/H]. The effects of galactic chemical evolution dominate the data with the influence of Type Ia supernovae increasing as solar [Fe/H] is approached. What is curious is that the C/O ratio at [Fe/H] of 0 is about 0.4 and not the solar ratio of 0.55.

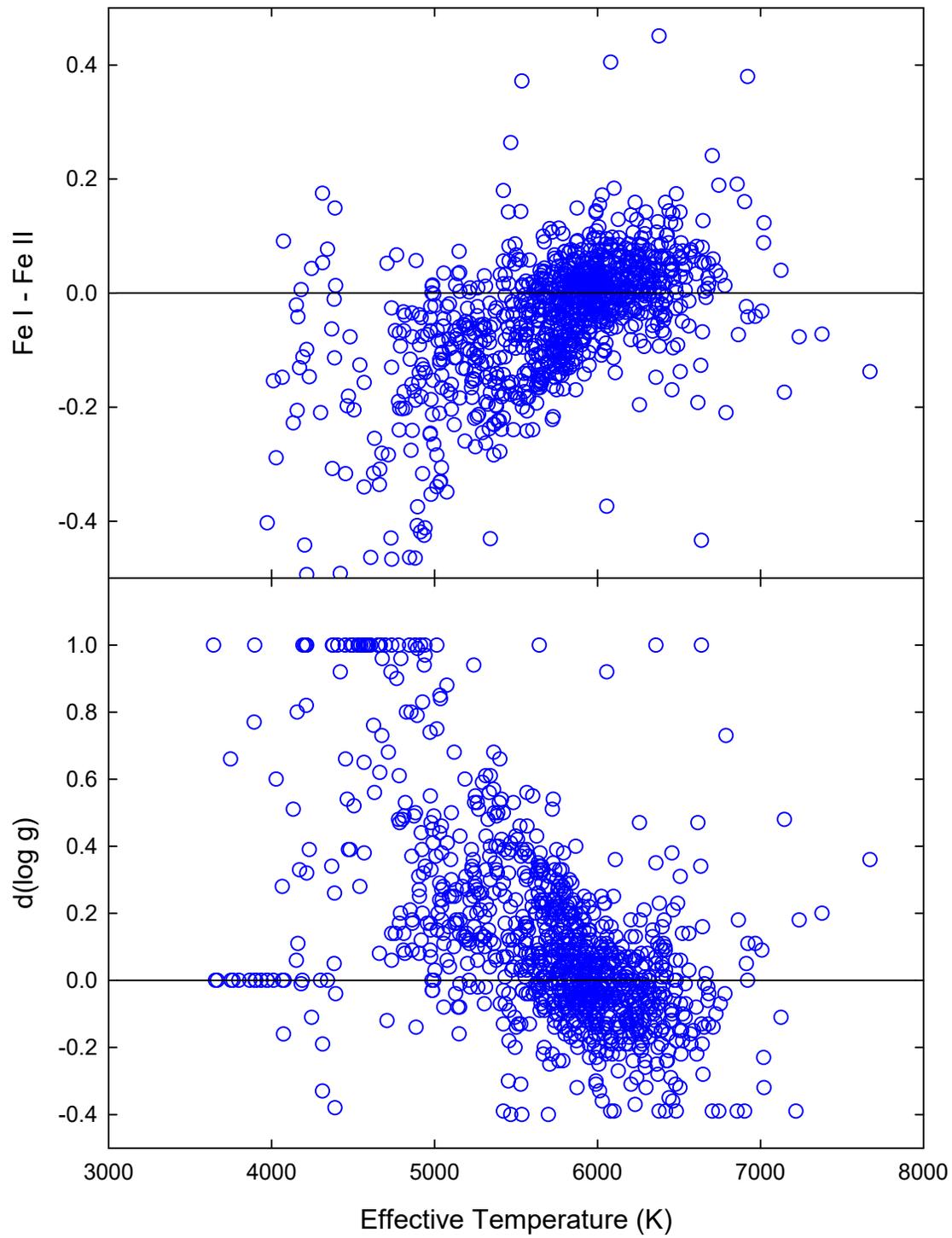

Figure 1

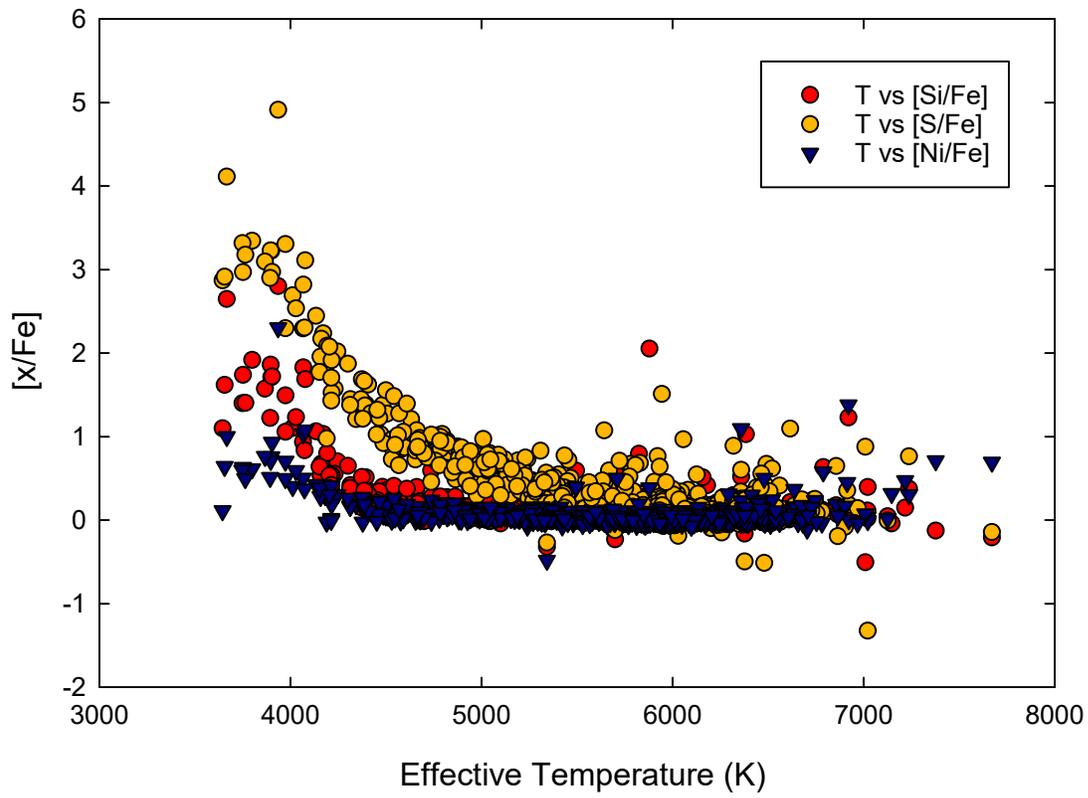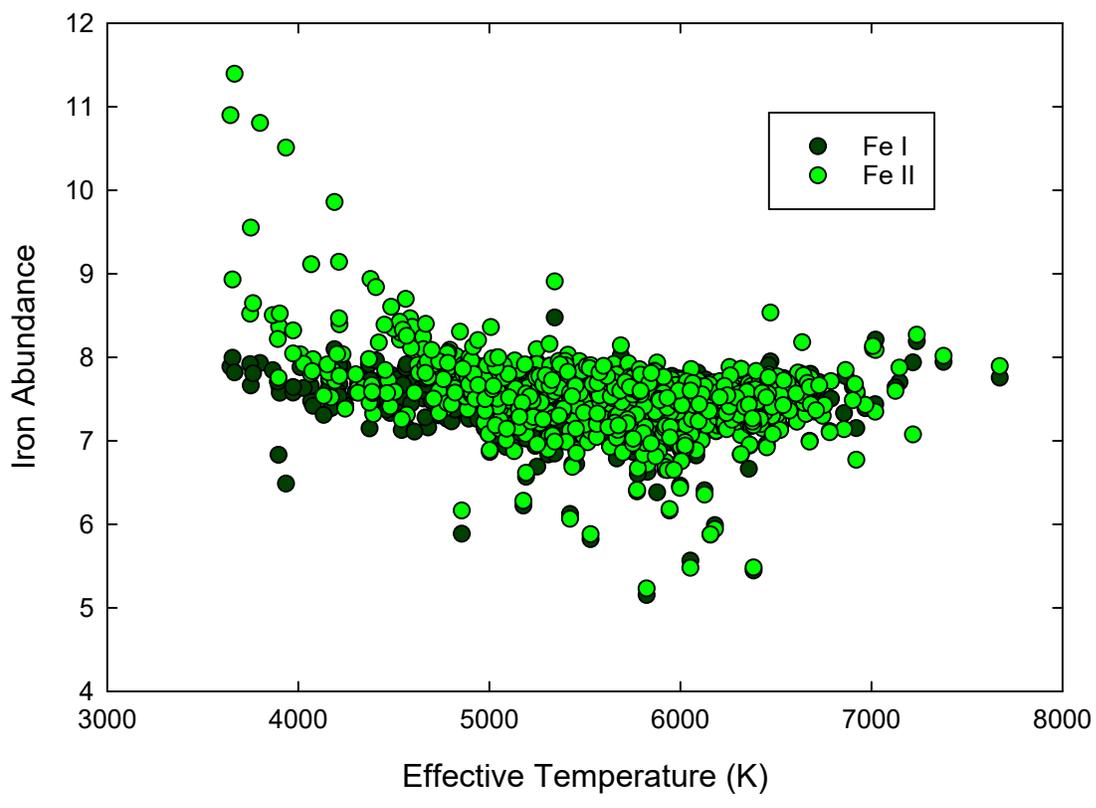

Figure 2

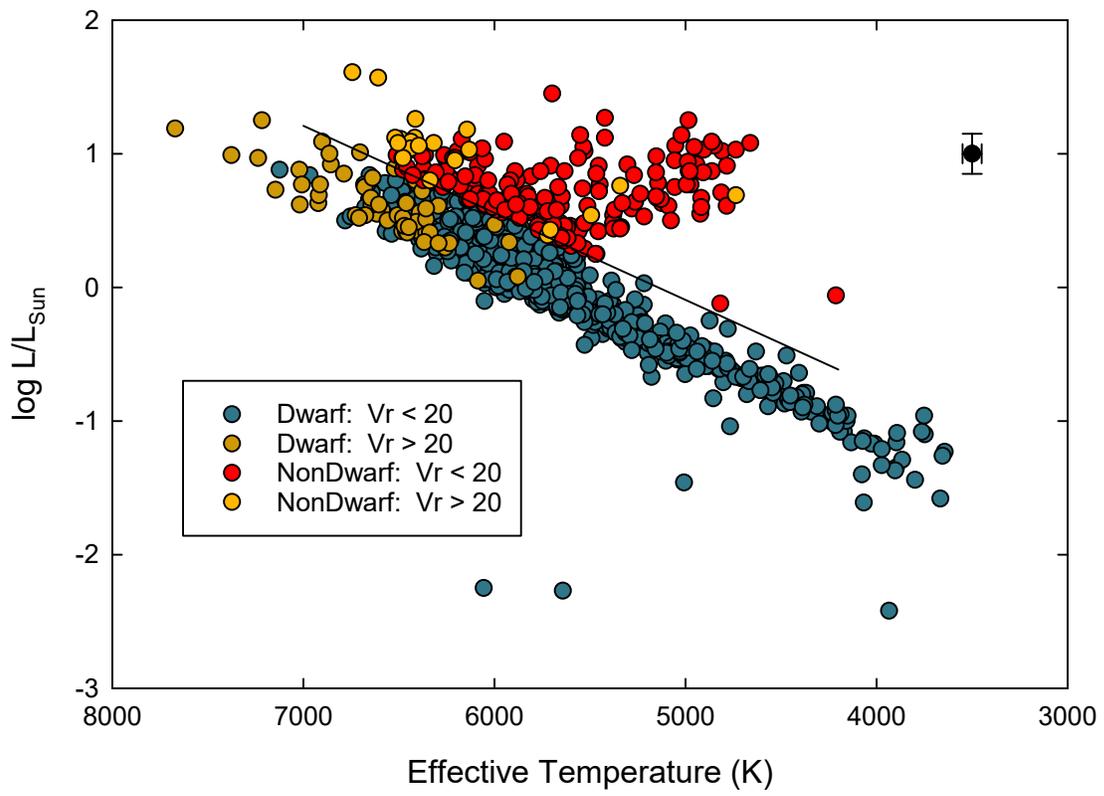

Figure 3

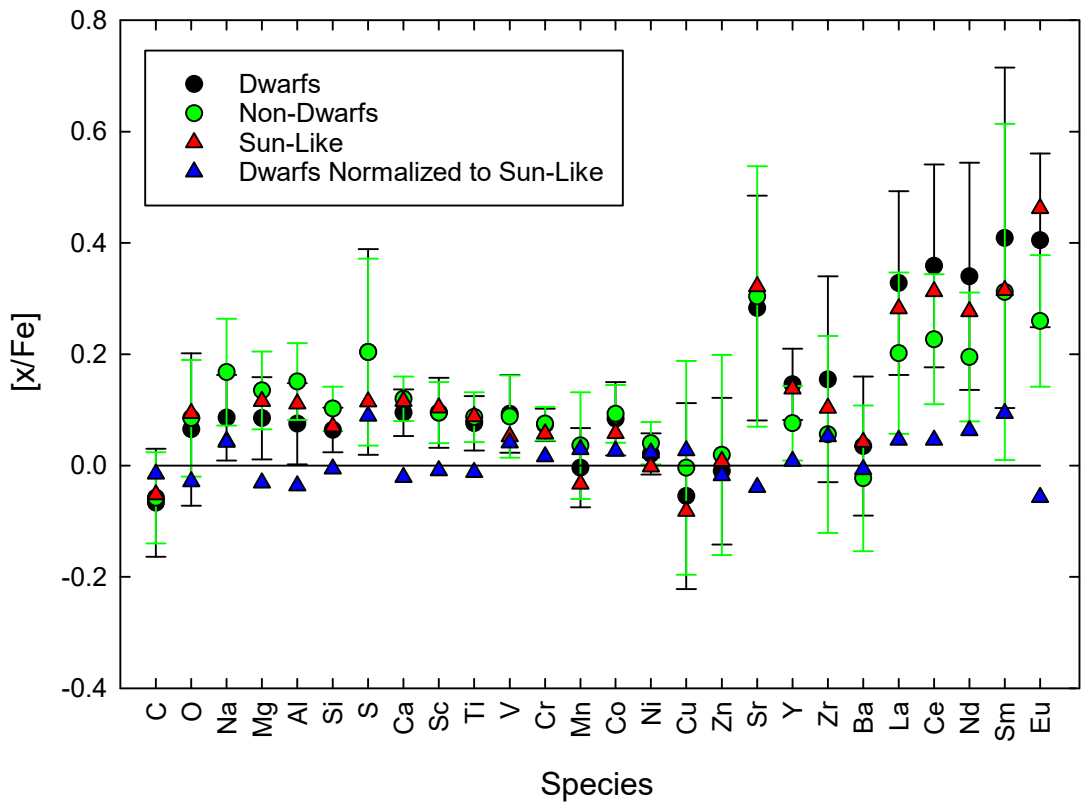

Figure 4

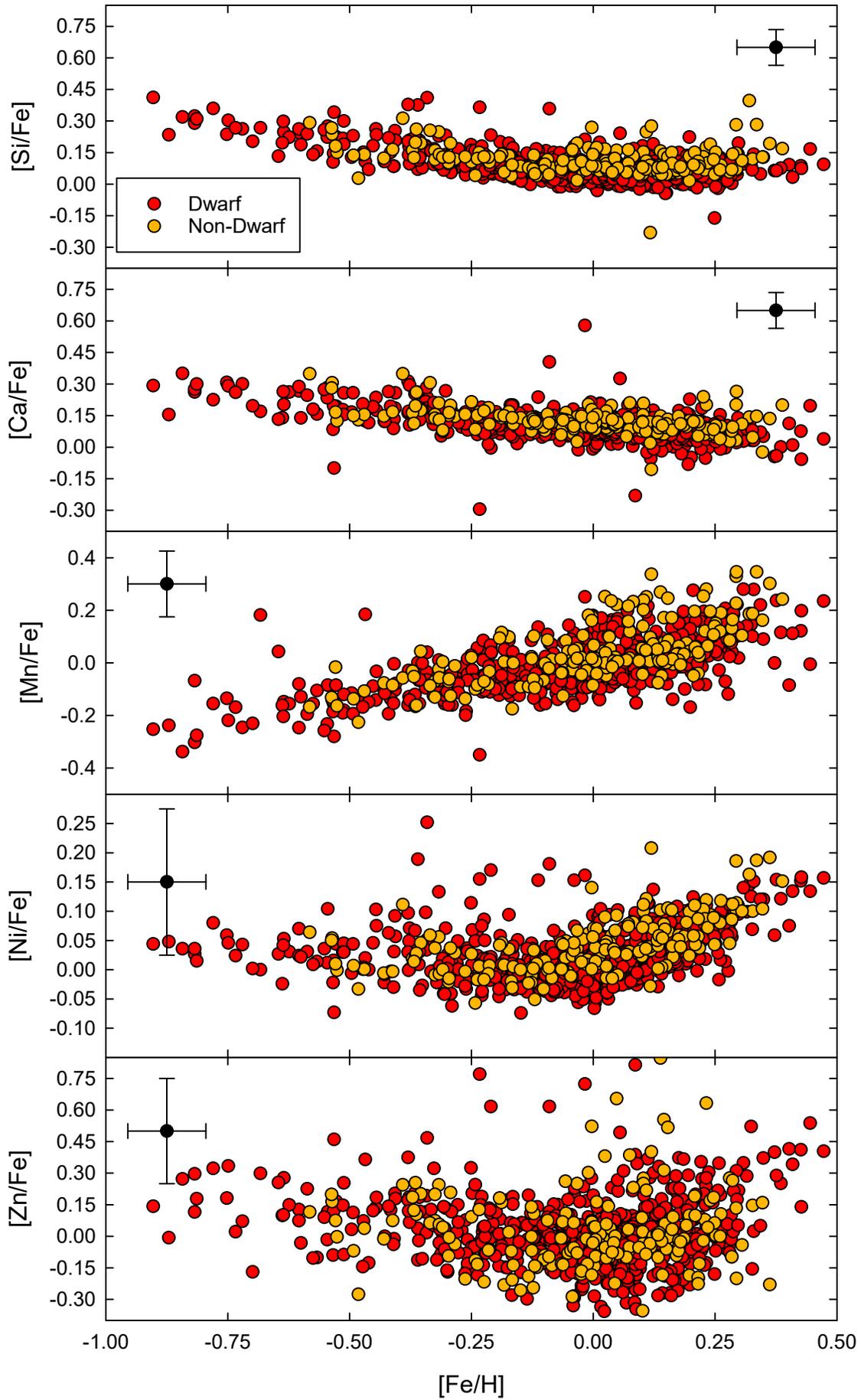

Figure 5

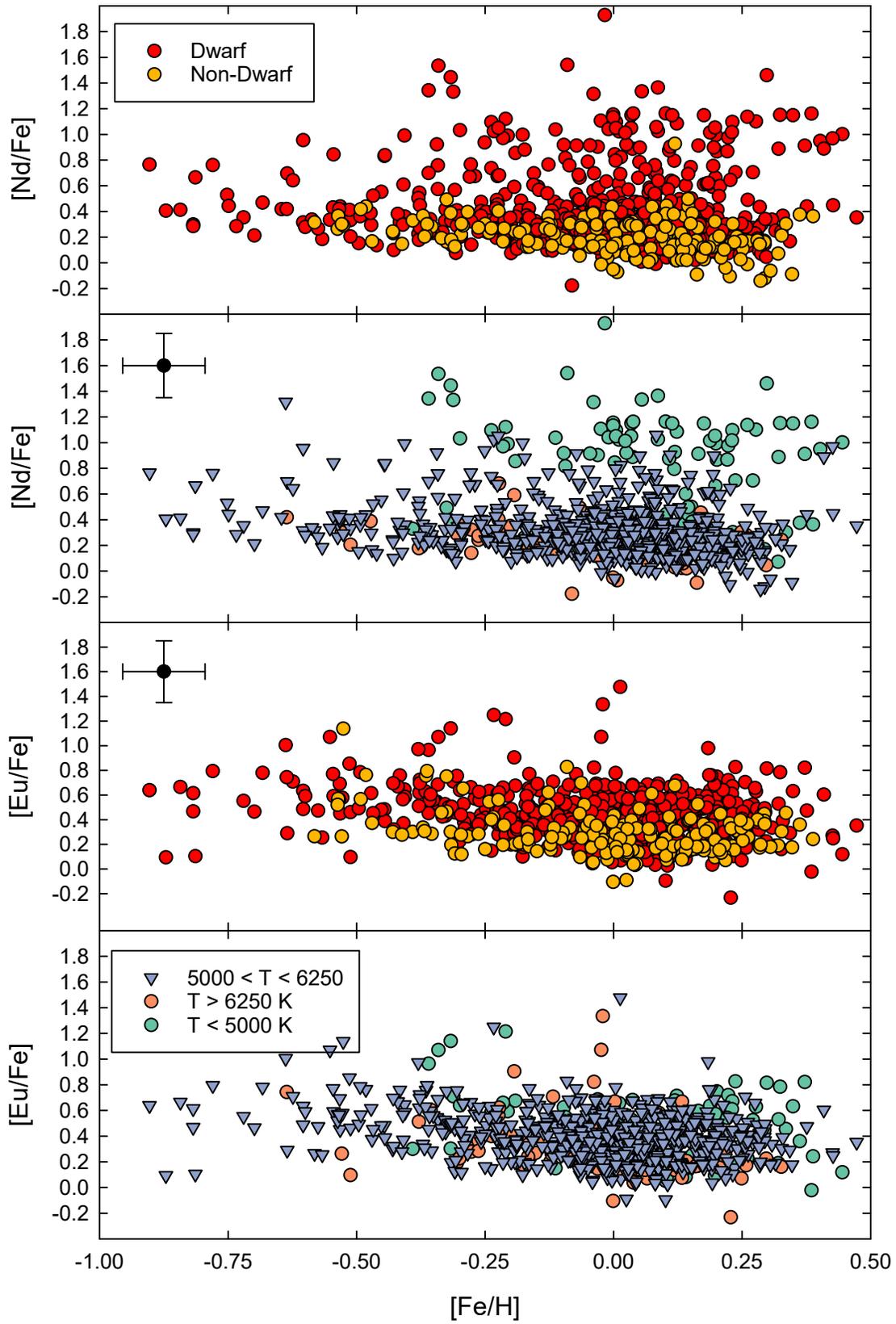

Figure 6

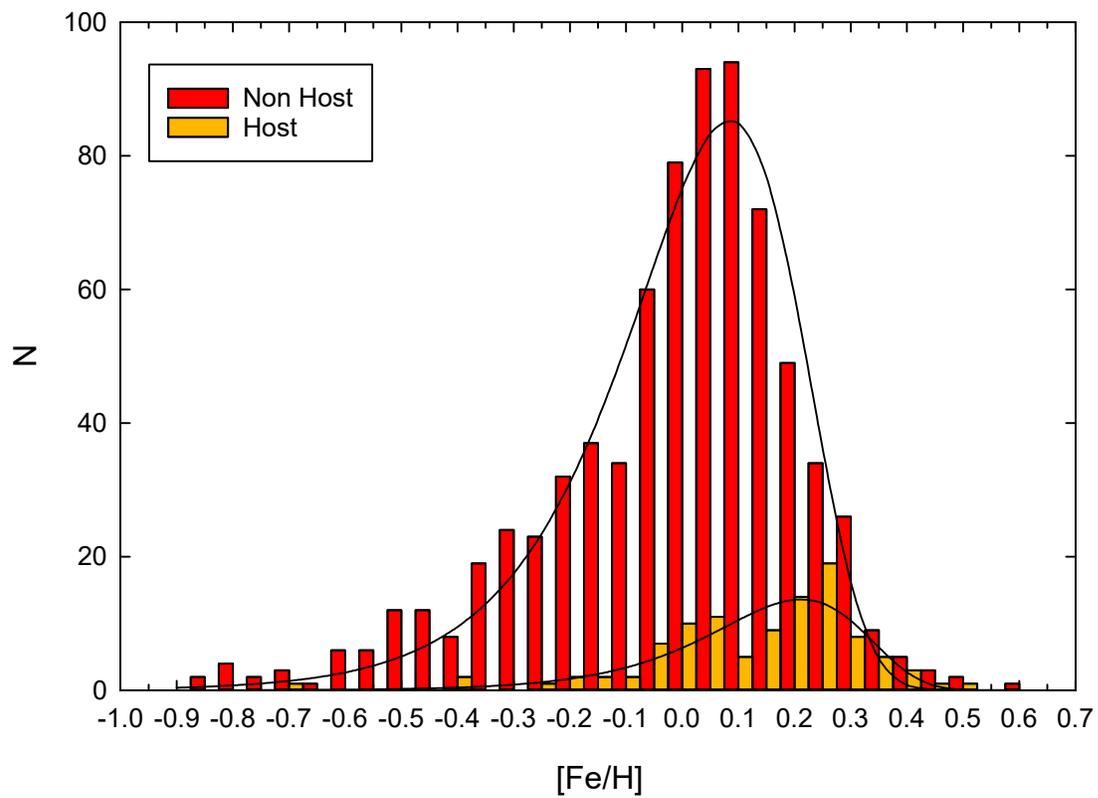

Figure 7

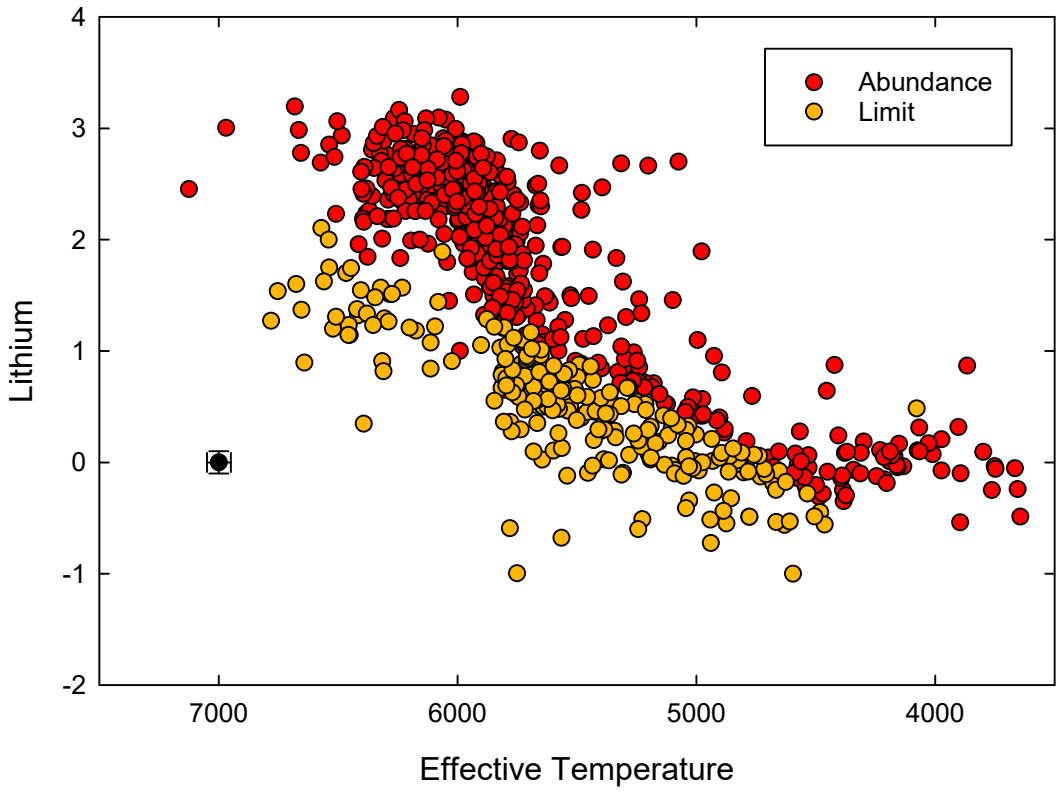

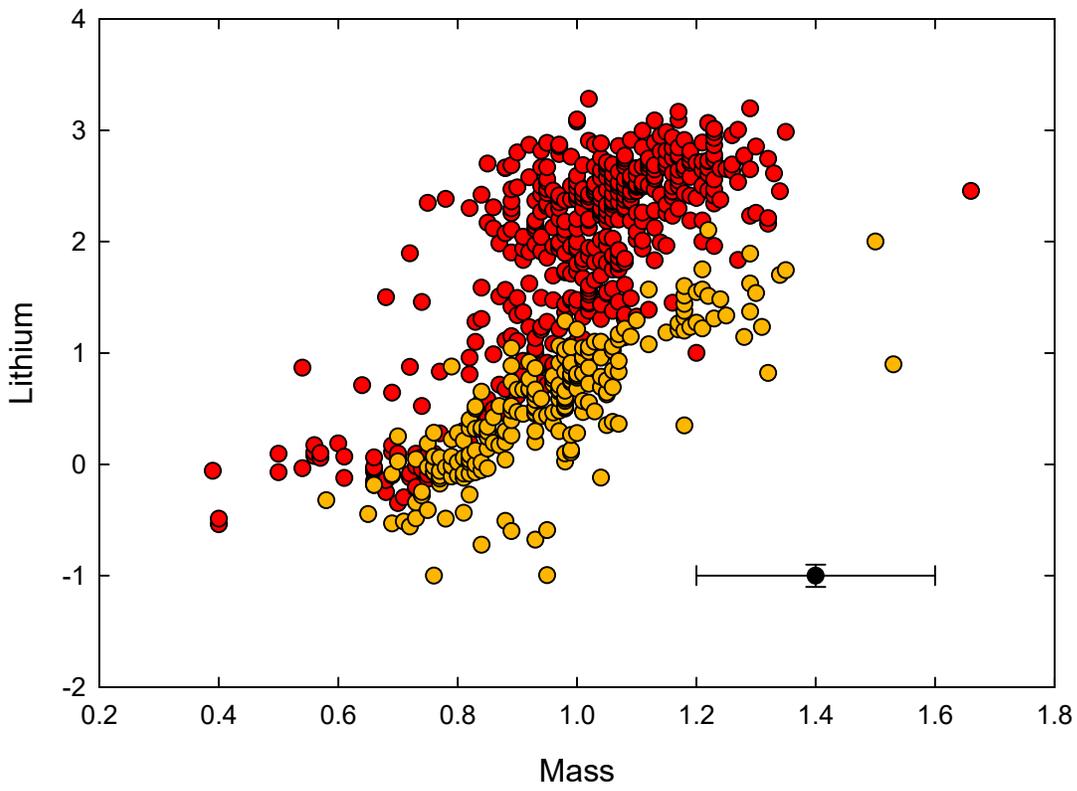

Figure 8

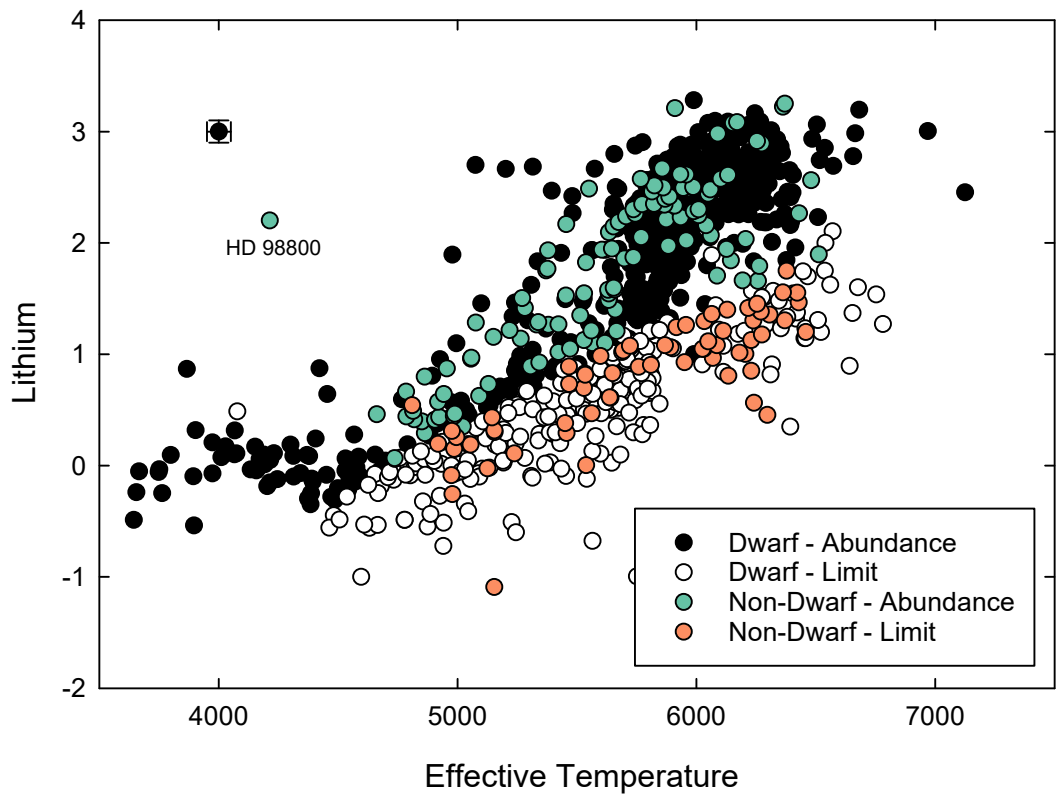

Figure 9

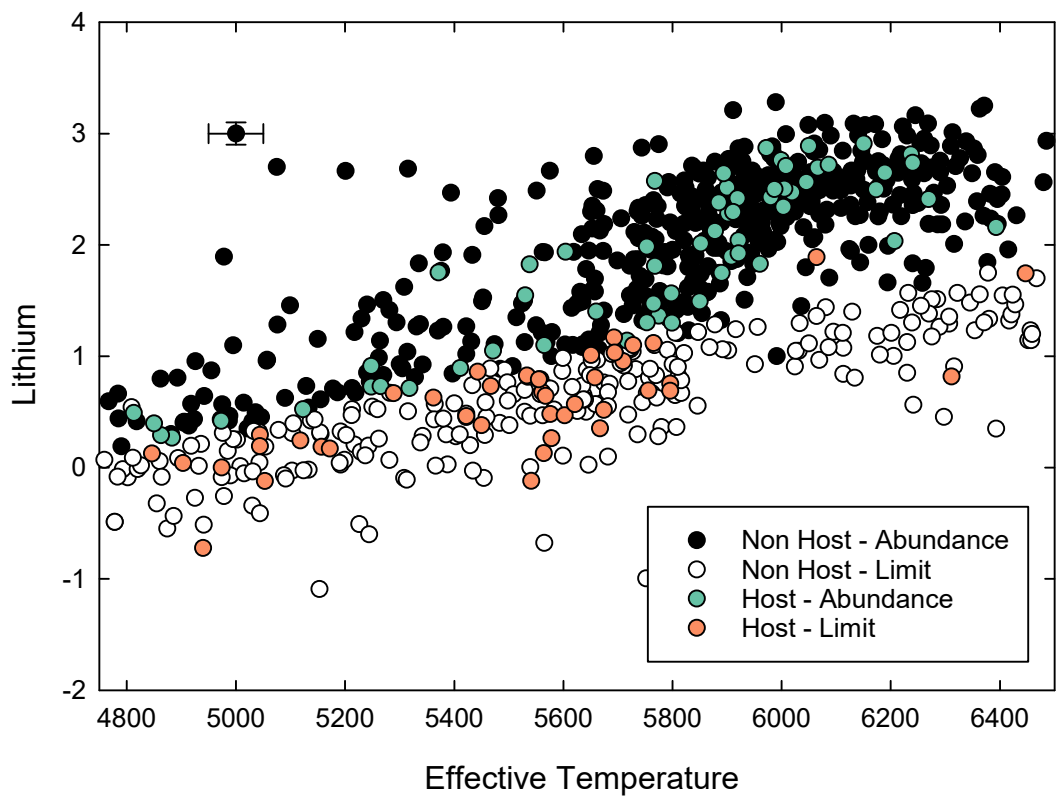

Figure 10

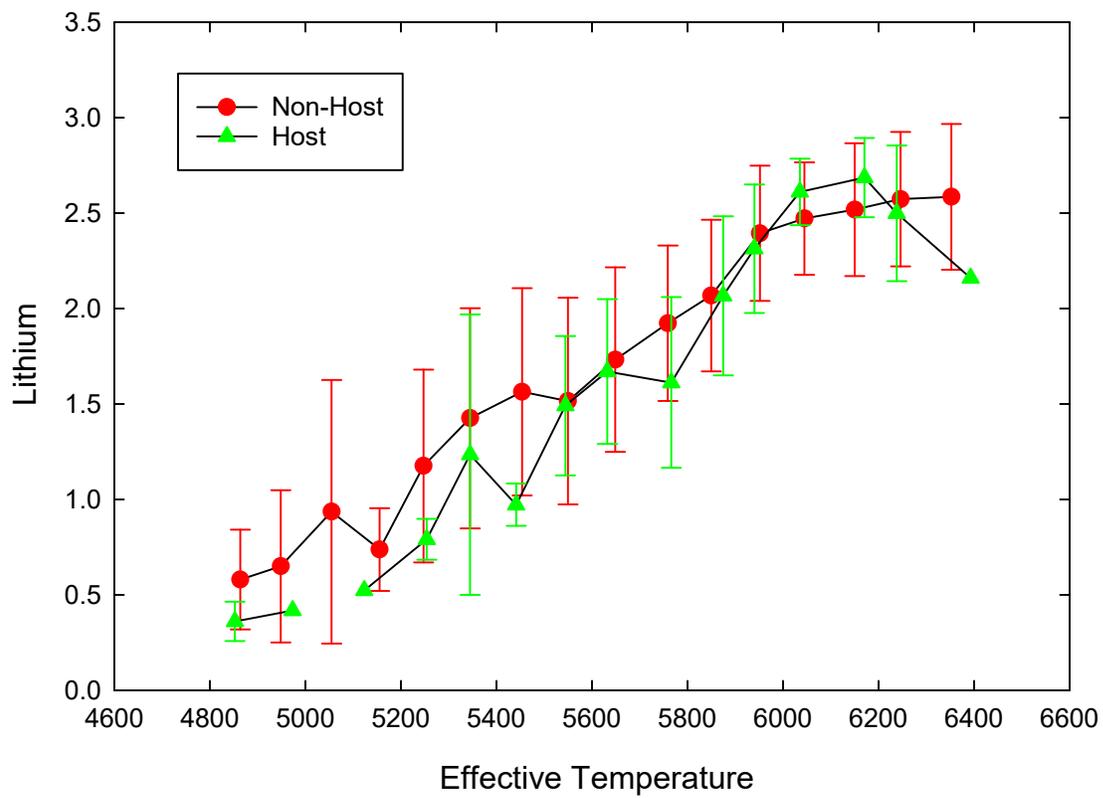

Figure 11

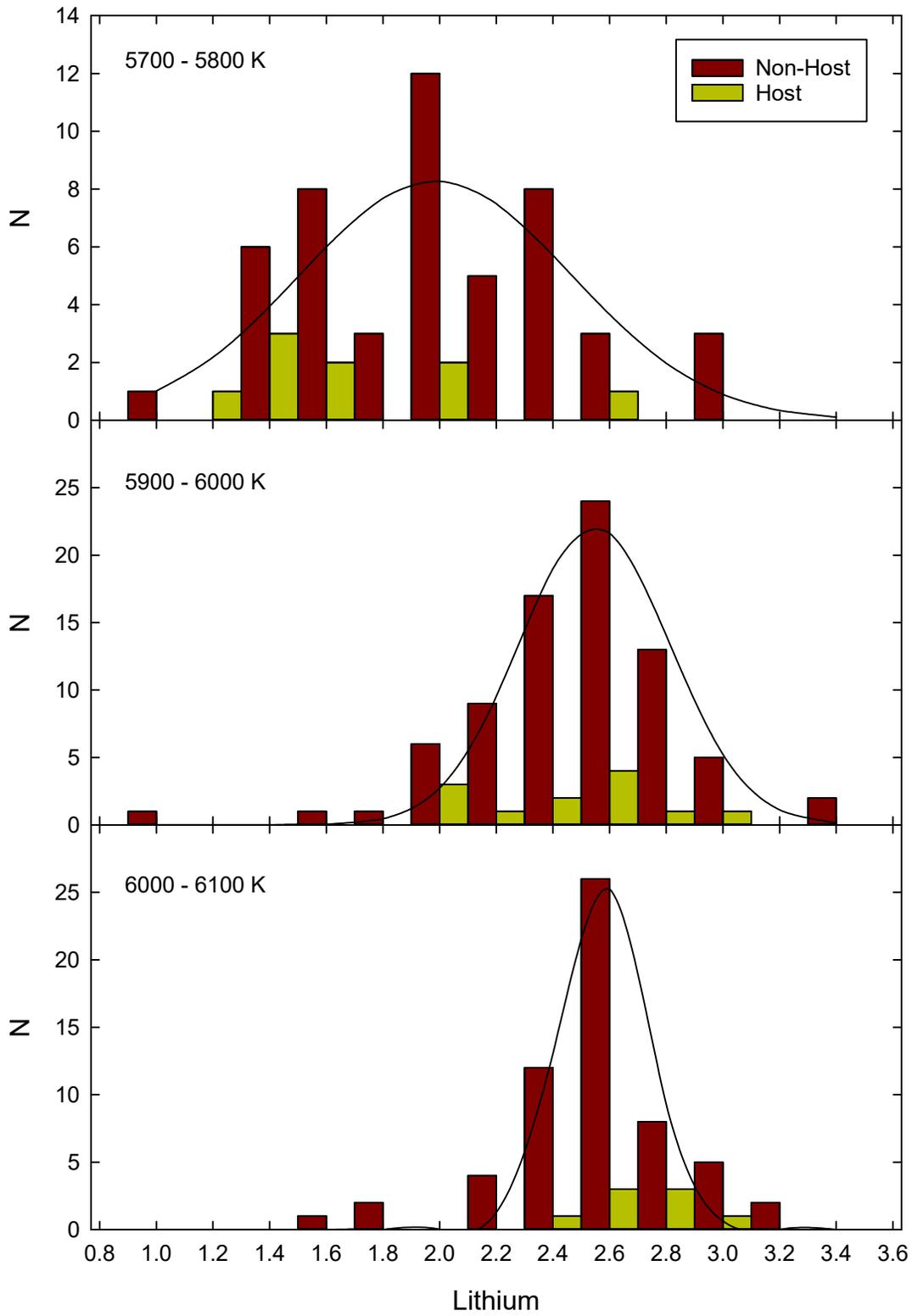

Figure 12

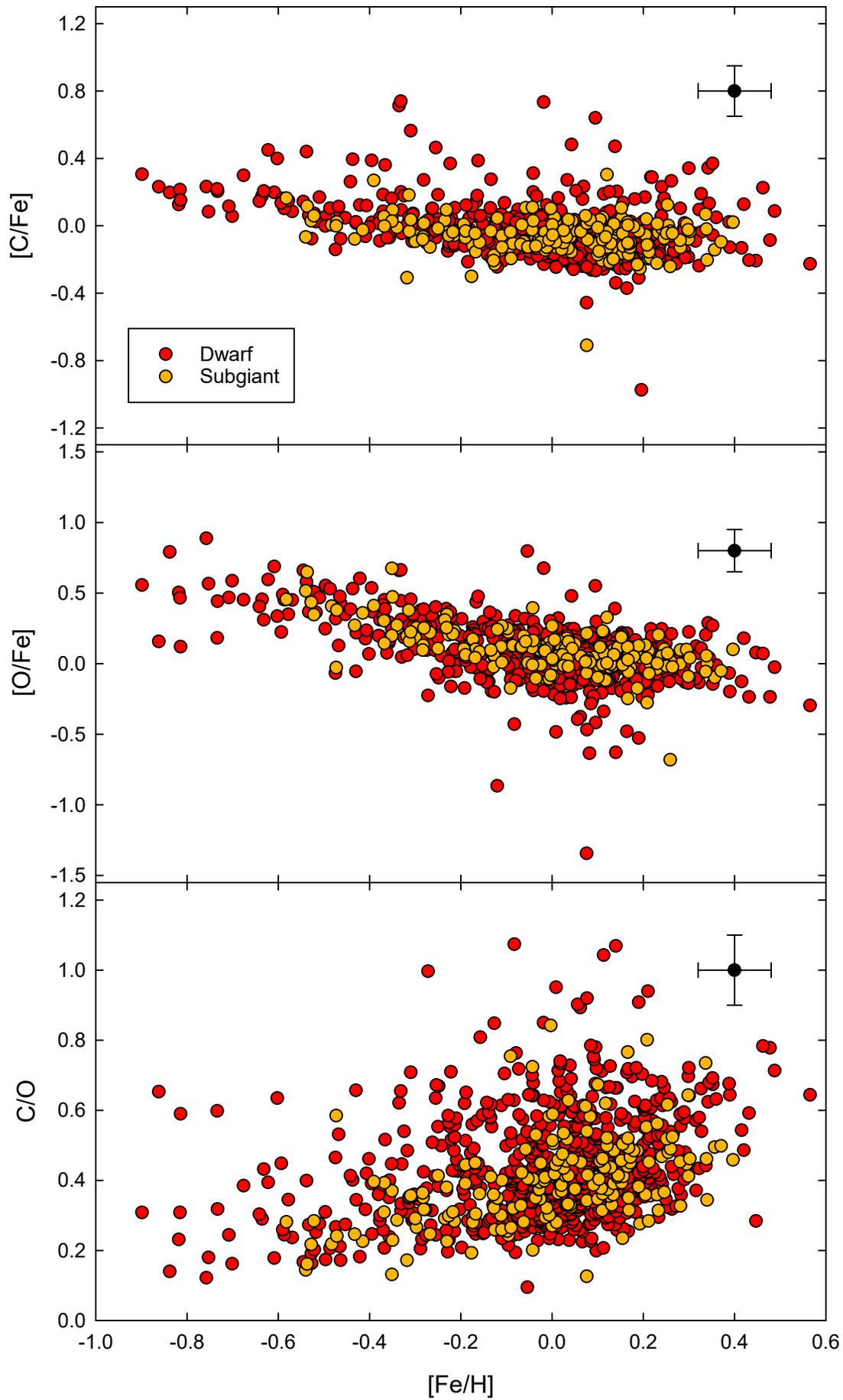

Figure 13